%% file: CMI_decay_main_arxiv.tex
\documentclass[aps,a4paper,showkeys,nofootinbib,longbibliography,twocolumn]{revtex4-2}
\usepackage{braket}
\usepackage[utf8]{inputenc}
\usepackage{filecontents}
\usepackage{natbib}
\usepackage{amsmath,amssymb,bm,amsthm}
\usepackage{subfigure}
\usepackage{graphicx}
\usepackage{dcolumn}
\usepackage{bm}
\usepackage[mathlines]{lineno}
\usepackage{booktabs}
\usepackage{color}
\newcounter{one}
\usepackage{url}

\usepackage{textcase}

\usepackage{colortbl}
\usepackage{tabularx}
\usepackage{verbatim}
\usepackage{multirow}

\usepackage[T1]{fontenc} 
\usepackage{lmodern}
\usepackage{bbm}
\usepackage[utf8]{inputenc}
\usepackage{amsfonts}
\usepackage{array}

\usepackage{bbm}
\usepackage{enumitem}
\usepackage{umoline}
\usepackage[usenames,svgnames]{xcolor}
\usepackage{natbib}
\usepackage[hyperindex,breaklinks]{hyperref}
\hypersetup{
     colorlinks=true,       		
     linkcolor=Navy,          	
     citecolor=Navy,            
     filecolor=Navy,      		
     urlcolor=Navy,           	
    runcolor=cyan,
 }
\usepackage{graphicx}
\usepackage{amsfonts}
\usepackage{amssymb}
\usepackage{amsmath}
\usepackage{bbm}
\usepackage{enumitem}

\newcommand{\tr}[0]{ {\rm tr}}

\newcommand{\ceil}[1]{ \lceil #1 \rceil}
\newcommand{\half}[1]{{ \rm h}}
\newcommand{\Oorderof}{\mathcal{O}}
\newcommand{\orderof}[1]{\Oorderof(#1)} 

\newcommand{\for}[0]{\quad \textrm{for} \quad}

\newcommand{\dist}{d}

\newcommand{\co}{{\rm c}}

\newcommand{\diam}{{\rm diam}}

\newcommand{\poly}{{\rm poly}}

\newcommand{\ad}{{\rm ad}}

\newcommand{\Or}{\quad {\rm or} \quad}
\newcommand{\AND}{\quad {\rm and} \quad}

\def\beq{\begin{equation}}
\def\eeq{\end{equation}}
\def\nbeq{\begin{equation*}}
\def\neeq{\end{equation*}}
\def\<{\langle}
\def\>{\rangle}

\def\tr{{\rm tr}}
\renewcommand{\d}{\partial}

\newcommand{\Er}{\mathcal{E}}

\newcommand{\mF}{{\mathcal{F}}}

\newcommand{\Poly}{{\rm Poly}}

\newcommand{\mI}{{\mathcal{I}}}

\newcommand{\mP}{{\mathcal{P}}}
\newcommand{\mQ}{{\mathcal{Q}}}

\newcommand{\mA}{{\mathcal{A}}}
\newcommand{\mB}{{\mathcal{B}}}
\newcommand{\mC}{{\mathcal{C}}}
\newcommand{\mD}{{\mathcal{D}}}
\newcommand{\mG}{{\mathcal{G}}}

\newtheorem{theorem}{Theorem}
\newtheorem{subtheorem}{Subtheorem}
\newtheorem{lemma}{Lemma}
\newtheorem{corol}[lemma]{Corollary}
 
\newtheorem{definition}{Definition}  
\newtheorem{prop}[lemma]{Proposition} 
\newtheorem{claim}[lemma]{Claim} 
\newtheorem{conj}[theorem]{Conjecture} 

\newcommand{\bal}[2]{#1[#2]}

\renewcommand{\d}{\partial}

\newcommand{\br}[1]{\left( #1 \right)}
\newcommand{\brr}[1]{\left[ #1 \right]}
\newcommand{\brrr}[1]{\left\{ #1 \right\}}
 \newcommand{\norm}[1]{\left \|  #1 \right \|}

\newcommand{\abs}[1]{\left| #1 \right|}
\newcommand{\tO}[0]{\tilde{\mathcal{O}}}

\usepackage{mathtools}
\def\multiset#1#2{\ensuremath{\left(\kern-.3em\left(\genfrac{}{}{0pt}{}{#1}{#2}\right)\kern-.3em\right)}}

\renewcommand\thefootnote{*\arabic{footnote}}

\newcommand{\mfL}[0]{\mathfrak{L}}

\begin{document}

\title{Clustering of conditional mutual information and quantum Markov structure \\ at arbitrary temperatures}


\author{Tomotaka Kuwahara$^{1,2}$}
\affiliation{$^{1}$ 
Analytical quantum complexity RIKEN Hakubi Research Team, RIKEN Cluster for Pioneering Research (CPR)/ RIKEN Center for Quantum Computing (RQC), Wako, Saitama 351-0198, Japan
}
\affiliation{$^{2}$
PRESTO, Japan Science and Technology (JST), Kawaguchi, Saitama 332-0012, Japan}

\begin{abstract}


Recent investigations have unveiled exotic quantum phases that elude characterization by simple bipartite correlation functions. In these phases, long-range entanglement arising from tripartite correlations plays a central role. Consequently, the study of multipartite correlations has become a focal point in modern physics.
In these, Conditional Mutual Information (CMI) is one of the most well-established information-theoretic measures, adept at encapsulating the essence of various exotic phases, including topologically ordered ones. Within the realm of quantum many-body physics, it has been a long-sought goal to establish a quantum analog to the Hammersley--Clifford theorem that bridges the two concepts of the Gibbs state and the Markov network. This theorem posits that the correlation length of CMI remains short-range across all thermal equilibrium quantum phases.
In this work, we demonstrate that CMI exhibits exponential decay with respect to distance, with its correlation length increasing polynomially with respect to the inverse temperature. While this clustering theorem has previously been believed to hold for high temperatures devoid of thermal phase transitions, it has remained elusive at low temperatures, where genuine long-range entanglement is corroborated to exist by the quantum topological order.
Our findings unveil that, even at low temperatures, a broad class of tripartite entanglement cannot manifest in the long-range regime. To achieve the proof, we establish a comprehensive formalism for analyzing the locality of effective Hamiltonians on subsystems, commonly known as the `entanglement Hamiltonian' or `Hamiltonian of mean force.' 
As one outcome of our analyses, we enhance the prior clustering theorem concerning bipartite entanglement. In essence, this means that we investigate genuine bipartite entanglement that extends beyond the limitations of the Positive Partial Transpose (PPT) class.

\end{abstract}

\maketitle


\section{Introduction}


In quantum many-body physics, a fundamental challenge is uncovering structures that apply universally, regardless of specific system details. 
One of the simplest ways to characterize this is by examining the correlation function between two observables, which reveals a distinct short-range behavior in non-critical phases both at finite~\cite{Araki1969,Park1995,ueltschi2004cluster,PhysRevX.4.031019,frohlich2015some} and zero temperatures~\cite{PhysRevB.69.104431,ref:Hastings2004-Markov,ref:Hastings2006-ExpDec,ref:Nachtergaele2006-LR}. 
Recent advances in quantum information science have introduced various methods for comprehending the complexity of quantum phases of matter from an information-theoretic perspective~\cite{RevModPhys.80.517,Latorre_2009,LAFLORENCIE20161,RevModPhys.93.045003}. Among these, one particularly renowned and elegant concept is the area law of quantum entanglement~\cite{RevModPhys.82.277}. 
While the area law has been a conjecture in high dimensions at absolute zero temperature~\cite{Hastings_2007,arad2013area,10.1145/3519935.3519962}, it has been rigorously confirmed to hold at non-zero temperatures~\cite{PhysRevLett.100.070502,PhysRevX.11.011047,Bernigau_2015,Lemm2023thermalarealaw}.
The area law has had a profound impact on various areas of research. It has dramatically influenced numerical techniques employing the tensor network formalism~\cite{PhysRevLett.93.207204,RevModPhys.77.259}. 
Additionally, it has paved the way for developing efficiency-guaranteed algorithms that compute physical observables~\cite{landau2015polynomial,Arad2017}.
In more recent developments, the study of bipartite quantum entanglement between distant subsystems has emerged as a promising avenue for universally revealing short-range characteristics, even at thermal critical point~\cite{PhysRevX.12.021022}. So far, quantum long-range entanglement is believed to manifest primarily in the form of tripartite (or higher-order multipartite) correlations at non-zero temperatures.

Over the past two decades, our understanding of quantum phases has evolved, revealing that they cannot be fully characterized solely through bipartite correlation measures. A prominent example of this complexity is found in topologically ordered phases, which exhibit genuinely multipartite correlations~\cite{PhysRevB.82.155138,RevModPhys.89.041004}.
Efforts in the field have been directed towards devising comprehensive information-theoretic measures to capture multipartite correlations in quantum many-body systems. Thus far, one of the most established measures for quantifying tripartite correlations is Conditional Mutual Information (CMI)~\cite{doi:10.1063/1.1643788,berta2015renyi,sutter2018approximate}.
The CMI has found versatile applications, including the definition of topological entanglement entropy~\cite{PhysRevLett.96.110404,PhysRevLett.96.110405,PhysRevA.93.022317}. Recent studies have uncovered additional uses for the CMI, such as characterizing information scrambling~\cite{Ding2016,PhysRevA.97.042330,PhysRevB.102.064305,LoMonaco_2023}, identifying measurement-induced quantum phase transitions~\cite{PhysRevX.10.041020,PhysRevX.11.011030,Koh2023}, and studying entanglement in conformal field theory~\cite{PhysRevD.87.046003,Rota2016,PhysRevB.108.L161116,Maric2023}, among others.
In different contexts, researchers have extensively investigated the operational significance of the CMI. Remarkably, they have rigorously clarified the relationship between the CMI and the error of the recovery map~\cite{Petz1986,Fawzi2015,PhysRevLett.115.050501,doi101098rspa20150338,Junge2018,Sutter2018}.


 \begin{figure}[tt]
\centering
\includegraphics[clip, scale=0.43]{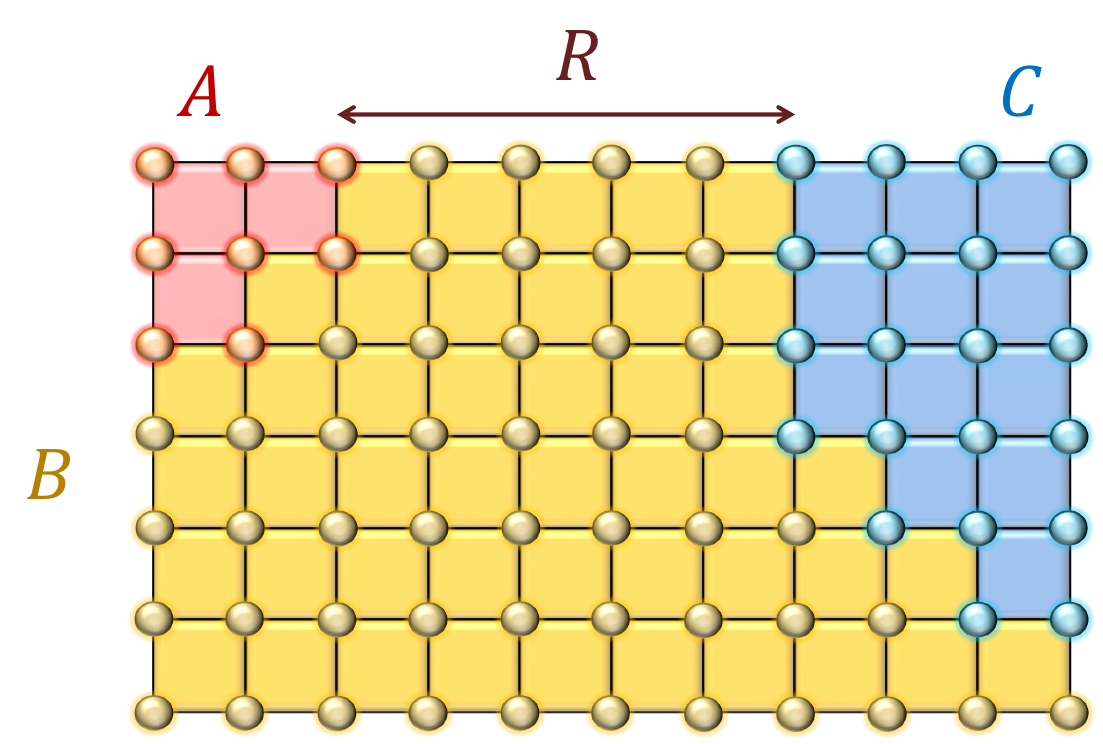}
\caption{Illustration of the main problem. 
We consider a finite-dimensional lattice system, depicted in 2D. The system is partitioned into three subsystems: $A$, $B$, and $C$, and we examine the conditional mutual information (CMI) between $A$ and $C$, conditioned on $B$.
When the system Hamiltonian is classical or commuting, the Hammersley--Clifford theorem indicates that subsystems $A$ and $C$ are conditionally independent. This independence implies that the CMI is zero when the distance between regions $A$ and $C$ exceeds the interaction length. The quantum analog suggests a rapid decay of the CMI with respect to the distance, as conjectured in Eq.~\eqref{quantum_Markov_Conj_ineq}.
A positive resolution of this conjecture would imply that generic quantum Gibbs states form approximate Markov networks, which, in turn, implies the existence of a local recovery map capable of reconstructing the global Gibbs state from the reduced density matrices~\cite{Fawzi2015}. Our main result offers a partial solution to this conjecture, encapsulated in the primary statements Eqs.~\eqref{main_high_dim_CMI} and~\eqref{main_1D_dim_CMI}. }
\label{fig:Fig1}
\end{figure}


Here, our fundamental question is ``\textit{Can we establish a universal theorem on the tripartite correlation based on the CMI?}''
To provide context for this query, we draw upon the well-known Hammersley--Clifford theorem~\cite{HammersleyClifford1971} within classical many-body systems, 
which establishes an equivalence between classical Gibbs states and the concept of a Markov network (or Markov random field). 
This network's characteristics are a finite correlation length of the CMI.
Then, one might envision a similar relationship in quantum systems in the analogy of classical Gibbs states. 
Although it breaks down in a strict sense, 
conjectures have emerged suggesting that a modified form of the theorem could hold approximately, 
with CMI decaying rapidly as the distance [See Ineq.~\eqref{quantum_Markov_Conj_ineq} below]. 
To date, mathematical proofs have been restricted in two scenarios: i) commuting Hamiltonians~\cite{Leifer2008,brown2012quantum,Jouneghani2014}, and ii) 1D Gibbs states at arbitrary temperatures~\cite{kato2016quantum}. Beyond one dimension, the problem was still open even at high temperatures, where the absolute convergence of the cluster expansion breaks down~\cite{PhysRevLett.124.220601,Erratum2025}.
When temperatures are low, the structure of entanglement becomes exceptionally complex, exhibiting long-range entanglement even at $\orderof{1}$ temperatures~\cite{PhysRevB.78.155120,PhysRevLett.107.210501,8104078,10.1145/3564246.3585114}.
Up to this point, no theoretical efforts have overcome this challenge, and the clustering of the CMI remains an inaccessible problem.

The resolution of this conjecture holds substantial significance in various ways. In the realm of quantum many-body physics, it enables us to impose stricter constraints on the nature of long-range entanglement that persists at non-zero temperatures.
Additionally, it assures us that the recovery map for the quantum Gibbs state can be constructed using local quantum channels, even when dealing with low temperatures.
From a practical standpoint, the Markov property is one of the foundational assumptions when dealing with unknown probability distributions in classical theory~\cite{ACKLEY1985147,pmlr-v5-salakhutdinov09a}. 
In the quantum world, the concept of the Markov property plays a fundamental role in various quantum technologies. Some notable applications include quantum Hamiltonian learning~\cite{Anshu_2021,anshu2021efficient}, efficiency-guaranteed quantum Gibbs sampling~\cite{PhysRevLett.103.220502,Brandao2019}, quantum marginal problems~\cite{kim2016markovian,PhysRevX.11.021039}, and more.
The establishment of the quantum version of the Hammersley--Clifford theorem has long been an ambitious goal in quantum information science.

In the present paper, we report an unconditional proof of the conjecture regarding the decay of Conditional Mutual Information (CMI) at arbitrary temperatures and in arbitrary finite dimensions. At this stage, it is important to note that our result does not provide a complete resolution of the quantum version of the Hammersley--Clifford theorem, which will be remarked after the main theorem. Nevertheless, it furnishes strong evidence that the Markov property generally holds, even at extremely low temperatures.

\section{Results}

\subsection{Setup and notations}

We consider a quantum system on a $D$-dimensional lattice, where $\Lambda$ represents the set of all sites.
For any subset $L \subset \Lambda$, let $|L|$ denote the number of sites in $L$.
For two disjoint subsets $L$ and $L'$, we define the distance $\dist_{L,L'}$ as the length of the shortest path connecting $L$ and $L'$.
We also denote the Hilbert space dimension of $L$ by $\mathcal{D}_L$.  

We then define the Hamiltonian as follows:
\begin{align}
&H = \sum_{i,i' \in \Lambda} h_{i,i'} + \sum_{i \in \Lambda} h_i, \notag \\
&\|h_{i,i'}\| \le \bar{J}_{\ell} := \bar{J}_0 e^{-\mu \ell}, \label{def:Hamiltonian_main}
\end{align}
for $\dist_{i,i'} = \ell$,
where the operator $h_{i,i'}$ represents an interaction term acting on the two sites $i$ and $i'$. The condition for $\|h_{i,i'}\|$ implies that the interaction is short-range (or exponentially decaying).
Here, $\|\cdots\|$ denotes the operator norm, which is equal to the largest singular value of the given operators. 
We can generalize all the results to generic $k$-local Hamiltonians, i.e., $H = \sum_{Z \subset \Lambda : |Z| \le k} h_Z$ where $h_Z$ is an interaction term acting on the subset $Z$ ($|Z| \le k$).
For an arbitrary subset $L\subseteq \Lambda$, we define the subset Hamiltonian as 
\begin{align}
&H_L = \sum_{i,i' \in L} h_{i,i'} + \sum_{i \in L} h_i.  \label{def:Hamiltonian_main_subset}
\end{align}

Throughout this paper, we consider the quantum Gibbs state defined as 
\begin{align}
\rho_\beta := \frac{e^{\beta H}}{Z_\beta},\quad Z_\beta=1,  \label{Quantum_Gibbs}
\end{align}
where $Z_\beta := \tr(e^{\beta H})$, and we shift the energy origin so that $Z_\beta=1$.
For simplicity, we omit the minus sign, so $e^{-\beta H}$ becomes $e^{\beta H}$ without loss of generality.
For a given quantum state $\rho$, we denote the reduced density matrix on the subregion $L$ by $\rho_{L}$:
\begin{align}
\rho_{L} = \tr_{L^\co} (\rho), \quad L^\co := \Lambda \setminus L,
\end{align}
where $ \tr_{L^\co}$ means the partial trace for the Hilbert space on the complementary region $L^\co$. 
In particular, for the Gibbs state, we denote the reduced density matrix by $\rho_{\beta,L}$.

For any three sets $A$, $B$, and $C$ within $\Lambda$, we start by defining the conditional mutual information (CMI) $\mI_{\rho}(A:C|B)$ between $A$ and $C$, conditioned on $B$, for a density matrix $\rho$:
\begin{align}
&\mI_{\rho}(A:C|B) \notag \\
&:= S_{\rho}(AB) + S_{\rho}(BC) - S_{\rho}(ABC) - S_{\rho}(B),
\end{align}
where, for $\forall L \subseteq \Lambda$, we define $S_{\rho}(L)$ as the von Neumann entropy of the reduced density matrix $\rho_L$.
Using the mutual information $\mI_\rho(A:B) := S_{\rho}(A) + S_{\rho}(B) - S_{\rho}(AB)$, we can also describe the CMI by 
\begin{align}
\mI_{\rho}(A:C|B) = \mI_\rho(A:BC) - \mI_\rho(A:B).
\end{align}

The quantum Markov conjecture regarding the CMI is now presented as follows:
For an arbitrary quantum Gibbs state $\rho_\beta$, the CMI $\mI_{\rho_\beta}(A:C|B)$, where $\Lambda = A \sqcup B \sqcup C$, rapidly decays with the distance between $A$ and $C$:
\begin{align}
\textrm{[{\bf Conjecture}]} \quad 
&\mI_{\rho_\beta}(A:C|B) \le  \mG_\mI(R),
\label{quantum_Markov_Conj_ineq}
\end{align}
with $R = \dist_{A,C}$, where $\mG_\mI(R)$ is a super-polynomially decaying function depending on $\beta$, $\{A,B,C\}$, and the system details.
It is important to note that the condition $\Lambda = A \sqcup B \sqcup C$ is crucial. If $A \sqcup B \sqcup C \subset \Lambda$, even for commuting Hamiltonians, there exists a counterexample to~\eqref{quantum_Markov_Conj_ineq} at low temperatures~\cite{PhysRevB.78.155120,PhysRevLett.107.210501}.

In accordance with the subsystem-size dependence (e.g., $|A|$ or $|C|$), the Markov property can be classified into three levels~\cite{lauritzen1996graphical,koller2009probabilistic}:
\begin{enumerate}
    \item{} Global Markov Property: This is the strongest level, implying that the CMI decays even when $A$ and $C$ are macroscopic, i.e., $|A|, |C| = \orderof{|\Lambda|}$.
    \item{} Local Markov Property: At this level, a small subsystem size for either $A$ or $C$ is required, i.e., $\min(|A|, |C|) = \orderof{1}$.
    \item{} Pairwise Markov Property: The weakest level, where both subsystems $A$ and $C$ must be small, i.e., $|A|, |C| = \orderof{1}$.
\end{enumerate}
In classical theory, provided the probability distribution is strictly positive, these three concepts are equivalent. However, in quantum scenarios, the conditions for such equivalence remain unclear. To establish the global Markov property in a quantum system, the decay rate $\mG_\mI(R)$ of the CMI should depend on the subsystem sizes $|A|$ and $|C|$ at most polynomially.

\subsection{Exponential clustering of the CMI}

The central achievement of this work lies in the unconditional proof of the conjecture~\eqref{quantum_Markov_Conj_ineq}, which is summarized in the following statement~(see \cite[Supplementary Theorems~4 and 5]{Supplement_CMI}):\\

\noindent
{\bf Theorem.}
For an arbitrary quantum Gibbs state, the conditional mutual information is upper-bounded as in~\eqref{quantum_Markov_Conj_ineq} with the following expression for $\mG_\mI(R)$:
\begin{align}
\label{main_high_dim_CMI}
\mG_\mI(R) = \mathcal{D}_{AC} e^{-c_1 R / \left( \beta^{D+1} \log(R) \right) + c_2 \log(\beta |AC|)},
\end{align}
where $\mathcal{D}_{AC}$ is the Hilbert space dimension on the subsystems $A \sqcup C$.
In particular, for one-dimensional systems, the upper bound is improved to
\begin{align}
\label{main_1D_dim_CMI}
\mG_\mI(R) = e^{-c_3 R / \beta + c_4 \beta \log(\beta R)} \quad (D=1).
\end{align}
Here, the parameters $c_1$, $c_2$, $c_3$, and $c_4$ are $\orderof{1}$ constants that depend only on the fundamental parameters of the system (see 
\cite[Supplementary Table~I]{Supplement_CMI}).
These results ensure the (almost) exponential decay of the conditional mutual information at arbitrary temperatures.

{~}

We summarize several key points. 
For $D=1$, the result does not depend on the subsystem sizes $|A|$ and $|C|$, and hence they can be macroscopically large as $|A|, |C| = \mathcal{O}(|\Lambda|)$.
We thus conclude that the global Markov property holds in one-dimensional quantum Gibbs states at arbitrary temperatures.   
This result significantly improves upon the previous one in Ref.~\cite{kato2016quantum}, which relies on the exponential clustering for bipartite correlations and yielded subexponential decay of $\mG_\mI(R) = \exp \left( -e^{-\Omega(\beta)} \sqrt{R} \right)$.
Additionally, for some classes of matrix product density operators (MPDO) with translation invariance, 
the exponential decay of the CMI has been proven under the assumption of the gap of the transfer matrix~\cite{10.1063/5.0085358,svetlichnyy2022matrix}. 
Our result leads to the unconditional proof of the CMI decay of MPDOs which have a quasi-local parent Hamiltonian~\cite{chen2023matrix}. 

On the other hand, for $D \ge 2$, we have achieved the exponential decay of the CMI at arbitrary temperatures. However, the current bound is insufficient to prove the global Markov property due to the growth of the coefficient $\mathcal{D}_{AC}$, which increases as $e^{\Omega(|A| + |C|)}$. Essentially, if we compare two large subregions (with sizes on the order of $|\Lambda|$), the influence of $\mathcal{D}_{AC}$ dominates over the decay factor $e^{-\Omega\brr{R/\log(R)}}$.
In this sense, the statement~\eqref{main_high_dim_CMI} implies the pairwise Markov property at this stage, and the complete solution of the conjecture~\eqref{quantum_Markov_Conj_ineq} is still open in high dimensions.

We further mention the temperature dependence of the CMI. The CMI correlation length increases at most polynomially as $\mathcal{O}(\beta)$ and $\mathcal{O}(\beta^{D+1})$ for $D=1$ and $D \ge 2$, respectively. 
The different scaling for $\beta$ between 1D and higher dimensions arises from the use of different analytical techniques for high-dimensional cases (see Sec.~\ref{subsec:First step}).
Notably, our results imply that the correlation length of the CMI is much smaller than that of the bipartite correlation function, which can be infinitely large at critical points in high dimensions or at least exponentially increases with $\beta$ as $e^{\mathcal{O}(\beta)}$ in one dimension~\cite{Araki1969,kimura2024clustering}.

Finally, we discuss the possibility of a qualitative improvement in the CMI decay rate. Based on numerical tests~\cite{DK_com}, it appears that the decay rate does not improve beyond exponential decay, even in 1D systems with finite-range interactions. This observation contrasts with the findings in Ref.~\cite{gondolf2024conditional}, where the authors reported a super-exponential decay of conditional information based on the Belavkin-Staszewski relative entropy~\cite{Belavkin1982}, an alternative metric for characterizing conditional independence in quantum Gibbs states.



\subsubsection{Bipartite vs. Tripartite Measures.}

Our findings demonstrate a fundamental difference between conditional mutual information (CMI) and standard bipartite correlation measures such as mutual information (MI). In high-dimensional quantum systems, it is well known that two-point correlations typically decay as a power law near criticality. Even in one dimension, the correlation length associated with MI often grows exponentially with inverse temperature, i.e., as $e^{\mathcal{O}(\beta)}$~\cite{Araki1969,kimura2024clustering}.

By contrast, the CMI remains short-ranged even at or near criticality. As we have shown in the main theorem, the CMI decays exponentially with distance at all positive temperatures, and its correlation length grows at most polynomially with $\beta$. This locality stems from the structural nature of CMI: it quantifies correlations between $A$ and $C$ that remain after conditioning on an intermediate region $B$. Because $B$ acts as an information ``shield,''  it effectively blocks any long-range correlations between $A$ and $C$ unless information about $B$ is available.
This type of locality is even more transparent in classical thermal equilibrium states, where it can be rigorously shown that $I(A:C|B) = 0$ whenever $A$ and $C$ are separated beyond the interaction length. In other words, classical CMI exhibits strict locality.

It is also important to emphasize that CMI and MI are fundamentally incomparable quantities: there is no general monotonicity relation between them. In some scenarios, the MI between two regions $A$ and $C$ can vanish while the CMI conditioned on $B$ remains nonzero, indicating correlations that only emerge upon conditioning. Conversely, there are cases where the CMI vanishes but the MI remains finite, typically reflecting indirect correlations mediated by $B$. These contrasting behaviors highlight that CMI captures distinct and often more subtle aspects of correlation structure than MI.

The robustness of short-range behavior in CMI at low temperatures reflects a general physical phenomenon: certain types of quantum correlations remain fragile even within strongly correlated regimes at low temperatures. This feature highlights the ability of the CMI to detect structural simplicity in quantum correlations: it tends to vanish even in regimes where standard correlation measures remain large. As such, the CMI serves as a sensitive probe for uncovering quasi-local structures or disentangling patterns that remain hidden in bipartite measures such as mutual information.
 A similar contrast emerges in the context of bipartite entanglement measures such as the entanglement of formation (EoF). 
 As we discuss in the subsequent section, the EoF also tends to decay rapidly even when MI remains large.

 \section{Entanglement clustering beyond the PPT class}
 
Regarding the applications of the main inequalities~\eqref{main_high_dim_CMI} and \eqref{main_1D_dim_CMI}, they allow us to address the clustering of genuine bipartite entanglement. 
Previous work~\cite{PhysRevX.12.021022} demonstrated that exponential clustering for bipartite entanglement holds at arbitrary temperatures. 
However, a severe drawback still remained that it could not exclude the possibility of the existence of bound entanglement~\cite{PhysRevLett.80.5239,PhysRevLett.82.1056}, which has been a primary open problem. 
More precisely, the statement was limited to the positive-partial-transpose (PPT) relative entanglement~\cite{PhysRevLett.87.217902,PhysRevA.66.032310,PhysRevA.78.032310,Girard_2014}, which has similar properties to entanglement negativity~\cite{PhysRevA.65.032314}. 

To capture bound entanglement, we need to analyze a faithful entanglement measure, meaning that the measure should be zero if and only if the target quantum state is separable. We call a bipartite quantum state $\rho_{AB}$ separable if it can be decomposed into a mixed state of product states, i.e.,
\begin{align}
\rho_{AB} = \sum_{s} p_s \rho_{A,s} \otimes \rho_{B,s}.
\end{align}
Conveniently, the CMI serves as an upper bound for bipartite entanglement in the form of squashed entanglement, which is a faithful measure~\cite{Brandao2011,Li2014}.
For an arbitrary quantum state $\rho_{AB}$ composed of systems $A$ and $B$, squashed entanglement is defined as follows:
\begin{align}
\label{Squashed_entanglement_def_00}
&E_{\rm sq}(\rho_{AB}) \notag \\
&:= \inf_E \left\{ \frac{1}{2} \mI_{\rho_{ABE}}(A:B|E) \biggl| \tr_E (\rho_{ABE}) = \rho_{AB} \right\},
\end{align}
where $\inf_E$ is taken over all extensions of $\rho_{AB}$ such that $\tr_E (\rho_{ABE}) = \rho_{AB}$.
Assuming that inequality~\eqref{quantum_Markov_Conj_ineq} holds, we can derive a clustering theorem for squashed entanglement as follows:
\begin{align}
\label{Sq_exp_decay_coj_00}
E_{\rm sq}(\rho_{\beta, AB}) &\le \frac{1}{2} \mI_{\rho_\beta}(A:B|C) \le \frac{1}{2} \mG_\mI(R),
\end{align}
with $R = \dist_{A,B}$.
In the inequality, the ancilla system $E$ is chosen as the residual system to $AB$, i.e., $E = C= \Lambda \setminus (AB)$.
This aspect enables us to prove entanglement clustering for genuine bipartite entanglement, including Entanglement of Formation (EoF).

To capture the broadest class of bipartite entanglement measures, we adopt the EoF since it provides an upper bound for other entanglement measures~\cite{PhysRevA.99.042304}, such as the relative entanglement~\cite{PhysRevLett.78.2275}, the entanglement cost~\cite{PhysRevA.54.3824}, and the squashed entanglement~\cite{doi:10.1063/1.1643788}.
The EoF for an arbitrary bipartite quantum state $\rho_{AB}$ is defined as follows:
\begin{align}
\label{def_Entanglement_form_corr_00}
E_F(\rho_{AB}) &:= \inf_{\{p_s, \ket{\psi_{s,AB}}\}} 
\sum_{s} p_s S_{\ket{\psi_{s,AB}}}(A),
\end{align}
where $S_{\ket{\psi_{s,AB}}}(A)$ is the von Neumann entropy for the reduced density matrix onto the subsystem $A$.
The convex roof $\inf_{\{p_s, \ket{\psi_{s,AB}}\}}$ is taken over all decompositions $\rho_{AB} = \sum_s p_s \ket{\psi_{s,AB}} \bra{\psi_{s,AB}}$ with $p_s > 0$.

In general, we can prove the following statement as a consequence of the main results~\eqref{main_high_dim_CMI} and \eqref{main_1D_dim_CMI}
(see \cite[Supplementary Corollary~42 and Proposition~43]{Supplement_CMI}):\\

\noindent
{\bf Clustering theorem for the EoF.}
Let $A$ and $B$ be subsystems that are separated by a distance $R$.
Then, for the reduced density matrix $\rho_{\beta,AB}$, the entanglement of formation obeys the following clustering theorem:
\begin{align}
\label{corol:ent_F_high_ineq_main_00}
E_F(\rho_{\beta,AB}) \le \mathcal{D}_{AB} e^{-c_1' R / \left( \beta^{D+1} \log(R) \right) + c_2' \log(R)},
\end{align}
which holds in arbitrary dimensions. In particular, for one-dimensional systems, we can improve the upper bound to 
\begin{align}
\label{corol:ent_F_one_ineq_main_00}
E_F(\rho_{\beta,AB}) \le e^{c_3' \beta \log(\beta) - c_4' R / \beta^2}. 
\end{align}
Here, $c'_1$, $c'_2$, $c'_3$, and $c'_4$ are $\orderof{1}$ constants. 

{~}

\noindent

The general upper bound~\eqref{corol:ent_F_high_ineq_main_00} has the same limitation as~\eqref{main_high_dim_CMI} in that it is meaningful only when the subsystem sizes of $A$ and $B$ are much smaller than the length $R$. The dependence on the Hilbert space dimension $\mathcal{D}_{AB}$ has been removed in the one-dimensional case. 
We emphasize that the inequality~\eqref{corol:ent_F_one_ineq_main_00} provides a non-trivial upper bound even in thermodynamic limit (i.e., $|A|, |B|\to \infty$).
This point is a significant advantage compared to the similar result in Ref.~\cite[Theorem~12 therein]{PhysRevX.12.021022} for the PPT relative entanglement, which becomes meaningless in the thermodynamic limit.

\section{Quasi-locality of the 1D entanglement Hamiltonian} \label{sec:Quasi-locality of the 1D entanglement Hamiltonian}

Analyzing the effective Hamiltonian on subsystems poses the most significant challenge in our work, which is also known as the entanglement Hamiltonian~\cite{PhysRevLett.101.010504,Peschel_2009,PhysRevB.81.054106,PhysRevLett.121.200602,Eisler_2020,Kokail2021,Zache2022entanglement} or Hamiltonian of mean force~\cite{RevModPhys.92.041002,PhysRevLett.127.250601,10.1116/5.0073853,burke2023structure}. 
The latter terminology is more commonly used in the fields of statistical mechanics and non-equilibrium physics.
When the subsystem is given by $L \subset \Lambda$, we define the entanglement Hamiltonian by
\begin{align}
\rho_{\beta,L} = e^{\beta H^\ast_L}, \quad 
H^\ast_L = \frac{1}{\beta} \log(\rho_{\beta,L}).
\label{def_rho_L_H^ast_L_main}
\end{align}
The quasi-locality of the entanglement Hamiltonian implies the decay of the CMI [see Eq.~\eqref{CMI_form_by_H_A|B} below], while the converse does not hold in general~\cite{Foot2}.
Hence, it is a stronger concept than the conjecture~\eqref{quantum_Markov_Conj_ineq}.
The entanglement Hamiltonian of the quantum Gibbs state attracts much attention in modern quantum technologies~\cite{anshu2021efficient,10.1116/5.0073853,hakoshima2024localized}.

 \begin{figure}[tt]
\centering
\includegraphics[clip, scale=0.33]{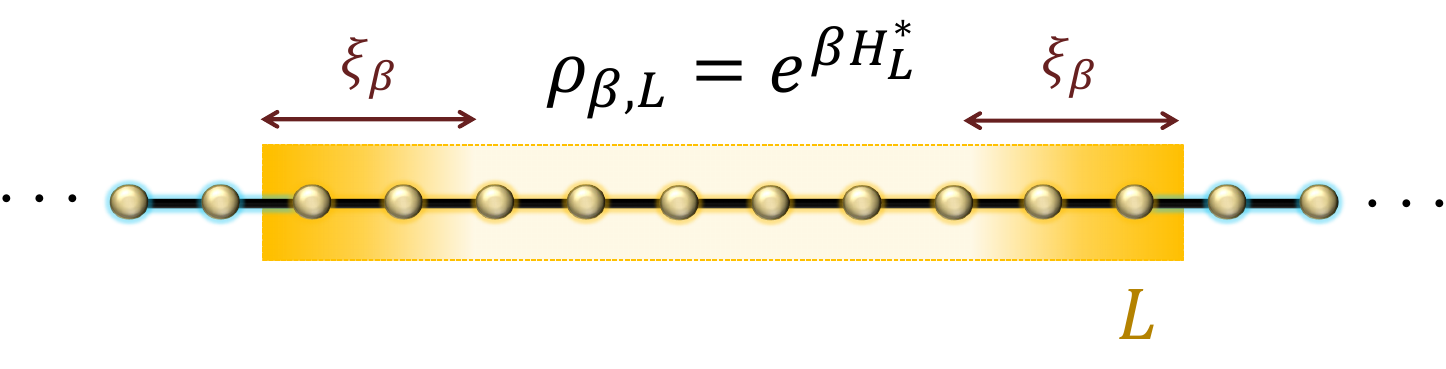}
\caption{Illustration of 1D entanglement Hamiltonian. 
For a contiguous region $L\subset \Lambda$, we define the entanglement Hamiltonian $H^\ast_L$ by $\beta^{-1} \log(\rho_{\beta,L})$ as in Eq.~\eqref{def_rho_L_H^ast_L_main}. 
The effective interaction terms in $H^\ast_L$ (i.e., $H^\ast_L-H_L$) are expected to be quasi-local around the boundary of $L$, with the length scale of the quasi-locality given by $\xi_\beta$. 
The quasi-locality of the entanglement Hamiltonian imposes a much stronger structural constraint than the CMI decay~\eqref{quantum_Markov_Conj_ineq}.
In one-dimensional systems with finite-range interactions, we rigorously prove the error bound~\eqref{Hamiltonian_eff_approx_main}, which yields $\xi_\beta = e^{e^{\mathcal{O}(\beta)}}$.
}
\label{fig:Fig2}
\end{figure}

{~}

\noindent
{\bf Quasi-locality of entanglement Hamiltonian.}
Let $\tilde{H}^\ast_{L,R}$ be the approximate entanglement Hamiltonian such that the effective interaction terms are localized around the boundary within a distance $R$ (see Fig.~\ref{fig:Fig2}).  
Then, if the interaction length is finite, i.e., $\bar{J}_\ell = 0$ for $\ell > {\rm const.}$ in \eqref{def:Hamiltonian_main}, the approximation error is given by  
\begin{align}
\label{Hamiltonian_eff_approx_main}
\norm{H^\ast_L - \tilde{H}^\ast_{L,R}} \le e^{\xi_\beta - c_5 R/\beta},
\end{align}
where $\xi_\beta = e^{e^{\mathcal{O}(\beta)}}$ and $c_5$ is a constant of order $\mathcal{O}(1)$. 
 
{~}

From this theorem, we can ensure that the non-locality of the effective interaction terms is limited to a distance of at most $e^{e^{\mathcal{O}(\beta)}}$.
In general, it is much larger than the correlation length of $e^{\mathcal{O}(\beta)}$~\cite{kimura2024clustering}. 
As a reason, the result~\eqref{Hamiltonian_eff_approx_main} relies on analyses based on imaginary time evolution, while the 1D CMI decay~\eqref{main_1D_dim_CMI} is based on the quantum belief propagation technique~\cite{PhysRevB.76.201102,PhysRevB.86.245116} (see Secs.~\ref{subsec:First step} and~\ref{subsec:entanglement_Ham_analyses}).
The assumption of the finite-range interactions also arises from this approach; for exponentially decaying interactions, 
Ref.~\cite{Perez-Garcia2023} shows that the imaginary-time Lieb--Robinson bound intrinsically diverges below a temperature threshold in 1D.
We conjecture that the quasi-locality of the entanglement Hamiltonian follows a similar inequality to that of the CMI~\eqref{main_1D_dim_CMI}.

One straightforward application of this inequality~\eqref{Hamiltonian_eff_approx_main} is to establish the logarithmic sample complexity of Hamiltonian learning with respect to the system size $|\Lambda|$.
In Hamiltonian learning~\cite{Anshu_2021,bak2023,Haah2024}, given $N$ copies of the same quantum Gibbs state, we reconstruct the Hamiltonian using $N$ measurement data sets.
Using the inequality~\eqref{Hamiltonian_eff_approx_main} along with the method described in Ref.~\cite{anshu2021efficient}---which is designed for commuting Hamiltonians---we derive the following expression for sample complexity (see \cite[Supplementary Corollary~48]{Supplement_CMI}): 
\begin{align}
N = e^{\xi_\beta} (1/\epsilon)^{c_6 \beta^2} \log(|\Lambda|)
\end{align}
with $c_6=\orderof{1}$ to ensure the estimation error of $\epsilon = \max_{Z \subset \Lambda} \norm{h_Z - \tilde{h}_Z}$, where $\{\tilde{h}_Z\}_{Z \subset \Lambda}$ are the reconstructed interaction terms.  
This is the first result achieving logarithmic sample complexity for 1D Hamiltonian learning at arbitrary temperatures. 
For the Hamiltonian learning problem, the recent flat polynomial approximation techniques~\cite{bak2023,narayanan2024} might be more effective than using the quasi-locality of the effective Hamiltonian. 
Nevertheless, we emphasize that the quasi-local structure of the entanglement Hamiltonian reveals more fundamental aspects of quantum many-body physics beyond its application to Hamiltonian learning.

The fundamental difficulty in accessing the entanglement Hamiltonian, i.e., the inequality~\eqref{Hamiltonian_eff_approx_main}, stems from the analytical instability of the operator logarithm. In particular, even exponentially small features in the spectrum of the reduced density matrix (e.g., on the order of $e^{-\Omega(|L|)}$) can contribute significantly to the structure of the entanglement Hamiltonian. To faithfully capture such fine structure, one must approximate the reduced density matrix with extremely high precision, typically up to errors of order $e^{-\Omega(|L|)}$. However, achieving this level of precision generally requires using global information or non-local methods, which in turn destroys any quasi-locality structure in the approximating Hamiltonian. 
This difficulty reflects a deeper conceptual gap between the nature of the CMI decay and the quasi-local structure of entanglement Hamiltonians.

To make this more concrete, we show that one can construct a quasi-local Hamiltonian $\tilde{H}^\ast_L$ supported around the subsystem $L$ such that
\begin{align}
\label{Hamiltonian_eff_approx}
e^{\beta \tilde{H}^\ast_L} \approx \rho_{\beta,L},
\end{align}
but this approximation does not imply that $\tilde{H}^\ast_L$ is close to the true entanglement Hamiltonian $H^\ast_L$. While such an approximation is sufficient to establish the exponential decay of the CMI (see Sec.~\ref{subsec:Third step}), it does not grant us access to the exact structure of $H^\ast_L$. This limitation presents an obstacle for applications such as achieving polylogarithmic sample complexity in Hamiltonian learning~\cite{anshu2021efficient}. Nevertheless, in the one-dimensional case (see \cite[Supplementary Theorem~6]{Supplement_CMI}), it is possible to partially bridge this gap under more favorable conditions.

\section{Proof techniques}

\subsection{Overviews}\label{subsec:Overview/}

As outlined in Section~\ref{sec:Quasi-locality of the 1D entanglement Hamiltonian}, our main objective is to demonstrate the quasi-local nature of the entanglement Hamiltonian as suggested in Eq.~\eqref{Hamiltonian_eff_approx}. Using the notation introduced in Eq.~\eqref{def_rho_L_H^ast_L_main}, the CMI is represented as:
\begin{align}
&\mI_{\rho_\beta}(A:C|B) = \tr \left[ \rho_\beta H_{\rho_\beta}(A:C|B) \right], \notag \\
&H_{\rho_\beta}(A:C|B) = -\beta \left( H^\ast_{AB} + H^\ast_{BC} - H - H^\ast_B \right),
\label{CMI_form_by_H_A|B}
\end{align}
where we have set $H^\ast_{ABC} = H$. 
Our significant technical contribution lies in developing a systematic methodology to analyze the entanglement Hamiltonian, denoted as $H^\ast_{L}$.  

In the following, we choose to work with the effective Hamiltonian on $L^{\co}$ (i.e., $H^\ast_{L^\co}$), since taking the partial trace over $L$ leads to a simpler notation. This notational simplification is made solely for presentation purposes and does not affect the core of the argument.

Our approach is structured into three steps:
\begin{enumerate}
\item{} Reduction to Connected Exponential Operators: We first approximate the reduced density matrix $\rho_{\beta,L^\co}$ as an exponential operator:
\begin{align}
\label{approx_exponential_reduced}
\quad  e^{\beta H^\ast_{L^\co}} \approx \mathcal{T} e^{\int_0^\tau V_x dx} e^{\beta H_0} \br{\mathcal{T} e^{\int_0^\tau V_x dx}}^\dagger,
\end{align}
where $V_x$ ($0 \le x \le \tau$) is a quasi-local operator around region $L$, $\mathcal{T}$ denotes the time-ordering operator, and $H_0$ is a $k$-local Hamiltonian that may differ from the original $H$. The precision of this approximation is dependent upon $\tau$ and the quasi-local nature of $V_x$.

The motivation for this step lies in the perturbative treatment of the reduced density matrix via the formalism of connected exponential operators. Rather than working directly with $\log (\rho_{\beta,L^\co})$, we consider an iterative update from
\begin{align}
\quad \quad   e^{\beta H^\ast_{L^\co,x_0}}:= \mathcal{T} e^{\int_0^{x_0} V_x dx} e^{\beta H_0} \br{\mathcal{T} e^{\int_0^{x_0} V_x dx}}^\dagger, \notag 
\end{align}
to
\begin{align}
\quad \quad  e^{\beta H^\ast_{L^\co,x_0+dx}} := e^{V_{x_0+dx} dx}  e^{\beta H^\ast_{L^\co,x_0}}e^{V^\dagger_{x_0+dx} dx}, \notag 
\end{align}
in a perturbative manner. This representation allows us to handle the operator logarithm indirectly while keeping the evolution quasi-local at each infinitesimal step.

\item{} Logarithm of Exponential Operators: We next consider the logarithm of the approximated exponential operators as in Eq.~\eqref{approx_exponential_reduced}:
\begin{align}
\label{approx_exponential_reduced_log}
\quad \quad \hat{H}_\tau = \log \left[ \mathcal{T} e^{\int_0^\tau V_x dx} e^{\beta H_0} \left(\mathcal{T} e^{\int_0^\tau V_x dx}\right)^\dagger \right].
\end{align}
Our analysis needs to establish that the effective interaction terms $\hat{H}_\tau - H_0$ are localized around the region $L$. Additionally, we need to demonstrate that these effective interactions are predominantly determined by the interaction terms within the neighboring region of $L$
\item{} Linking Approximations to CMI: To derive the CMI decay, we connect the approximation in Eq.~\eqref{approx_exponential_reduced} with the quasi-locality of $\hat{H}_\tau$. The challenge lies in that a close approximation between two operators cannot translate to the closeness of their logarithms. The qualitatively optimal bound between $H^\ast_{L^\co}$ and $\hat{H}_\tau$ is expressed as:
\begin{align}
\label{two_effective_Ham_distance}
\norm{H^\ast_{L^\co} - \hat{H}_\tau} \le \frac{c \delta_\tau}{\lambda_{\min}(\rho_{\beta,L^\co})}
\end{align}
with $c$ a constant of $\orderof{1}$, where $\delta_\tau$ represents the approximation error for Eq.~\eqref{approx_exponential_reduced}, and $\lambda_{\min}(\rho_{\beta,L^\co})$ is the minimum eigenvalue of $\rho_{\beta,L^\co}$. 
In general, $\lambda_{\min}(\rho_{\beta,L^\co}) = e^{-\orderof{|L^\co|}}$, and hence we need to make the error $\delta_\tau$ exponentially small with the system size, i.e., $\delta_\tau = e^{-\Omega(|\Lambda|)}$. 
Achieving an exponentially small error $\delta_\tau$ spoils the quasi-locality of the effective interactions. Hence, we approach the estimation of the CMI without directly analyzing the true entanglement Hamiltonian.
\end{enumerate}

In the following, we demonstrate how these steps are taken.
To explain the essence of the analyses, we omit most of the detailed calculations.
Further details and specific calculations are reserved for the supplementary materials (see \cite[Supplementary sections~III, IV, V, and VI]{Supplement_CMI}).

\subsection{First step}   \label{subsec:First step}

In the initial phase of our methodology, we adapt two distinct formalisms based on the dimensional characteristics of the system:
\begin{enumerate}
    \item[(i)] the Belief-Propagation (BP) formalism for one-dimensional systems, as detailed in \cite[Supplementary section~III B]{Supplement_CMI},
    \item[(ii)] the Partial-Trace-Projection (PTP) formalism for higher-dimensional systems, described in \cite[Supplementary section~III C]{Supplement_CMI}.
\end{enumerate}
We emphasize that the BP formalism cannot be directly applied to high-dimensional systems in our setting. Although the quantum belief propagation itself is defined for arbitrary dimensions, in higher dimensions the error grows exponentially with the size of the boundary between regions (see the discussion below), which prevents us from using this approach beyond 1D.

We begin with the belief-propagation formalism to derive the exponential form shown in Eq.~\eqref{approx_exponential_reduced}, which is used to approximate the reduced density matrix. We briefly review the quantum belief propagation~\cite{PhysRevB.76.201102,PhysRevB.86.245116}.  
In essence, it provides a mechanism to connect two exponential operators, $e^{\beta H_0}$ and $e^{\beta(H_0 + V)}$, for any operators $H_0$ and $V$, as follows:
\begin{align}
\label{BP_eq_A_B}
e^{\beta(H_0 +V)} = \Phi_V e^{\beta H_0} \Phi^\dagger_V .
\end{align}
The operator $\Phi_V$, known as the quantum-belief-propagation operator, is defined by:
\begin{align}
\label{BP_eq_A_B_operator}
&\Phi_V = \mathcal{T} e^{\int_0^1 \phi_\tau d\tau} ,\quad  \phi_\tau= \int_{-\infty}^\infty f_\beta (t) V(H_\tau, t) dt  ,\notag \\
& f_\beta (t):= \frac{2}{\beta \pi} \log\br{\frac{e^{\pi |t| /\beta }+1}{e^{\pi |t| /\beta }-1}} ,
\end{align} 
where $V(H_\tau, t) = e^{iH_\tau t} V e^{-iH_\tau t}$ and $H_\tau = H_0 + \tau V$. The filter function $f_\beta (t)$ exhibits a decay of $e^{-\orderof{|t|/\beta}}$, emphasizing the dominance of the time integral within $\orderof{\beta}$. Specifically, when $H_0$ represents a short-range interacting Hamiltonian, the Lieb--Robinson bound~\cite{ref:LR-bound72,PhysRevLett.97.050401} can be applied to the time evolution $V(H_\tau, t)$. By selecting $V$ as a local operator within region $L \subset \Lambda$, we can accurately approximate $V(H_\tau, t)$ onto the neighboring region around $L$.

 \begin{figure}[tt]
\centering
\includegraphics[clip, scale=0.33]{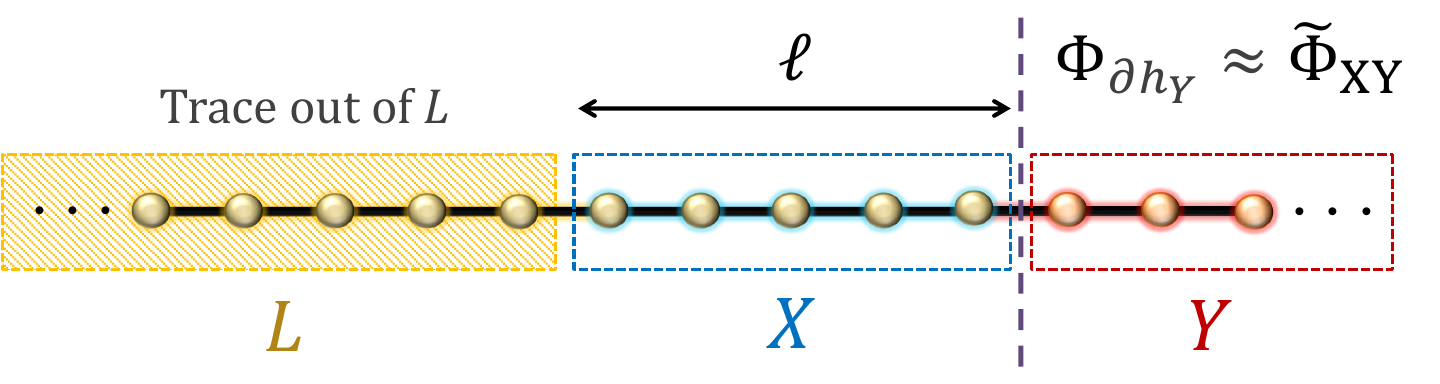}
\caption{Belief propagation formalism for the entanglement Hamiltonian. 
By tracing out the subsystem $L$, we aim to derive the connected exponential form given in Eq.~\eqref{approx_exponential_reduced} to approximate the reduced density matrix $\rho_{L^\co}$. 
For this purpose, we apply the quantum belief propagation method to segregate the boundary interaction term $\partial h_Y$ from the quantum Gibbs state, which is represented as $e^{\beta H} = e^{\beta (H_{LX} + H_Y + \partial h_Y)}$.
In this setup, the quantum-belief-propagation operator $\Phi_{\partial h_Y}$ serves as a bridge linking $e^{\beta (H_{LX} + H_Y)}$ to $e^{\beta H}$. By approximating this operator within the region $XY$ with $\tilde{\Phi}_{XY}$, we can manage the partial trace $\tr_{L}(\cdots)$ independently from the quantum belief propagation operation. This approximation leads to the desired form of Eq.~\eqref{BP_approx_reduce_den}.}
\label{fig:Fig3}
\end{figure}

In what follows, we partition the one-dimensional system into three parts: $\Lambda = L \cup X \cup Y$, where $L$ is the target region, and $X$ is a subsystem of length $\ell$ that connects to $L$ (see Fig.~\ref{fig:Fig3}). We then decompose the Hamiltonian as $H = H_{LX} + H_Y + \partial h_Y$, with $\partial h_Y$ representing the boundary interaction between the subsystems $LX$ and $Y$.
Employing the quantum belief propagation described in Eq.~\eqref{BP_eq_A_B} with $H_0 = H_{LX} + H_Y$ and $V = \partial h_Y$, we derive:
\begin{align}
\label{BP_eq_A_B_apply1D}
e^{\beta H} = \Phi_{\partial h_Y} e^{\beta H_{LX}} \otimes e^{\beta H_Y} \Phi^\dagger_{\partial h_Y} .
\end{align}
Applying the Lieb--Robinson bound to the time evolution described in Eq.~\eqref{BP_eq_A_B_operator}, we obtain the following approximation (see \cite[Supplementary Lemmas~9 and 10]{Supplement_CMI}):
\begin{align}
\label{BP_eq_A_B_error_bound}
&\norm{ \Phi_{\partial h_Y}^{-1} \tilde{\Phi}_{XY} -1 } \le e^{c \beta \norm{\partial h_Y} - c' \ell /\beta} , \notag \\
&\tilde{\Phi}_{XY}= \mathcal{T} e^{\int_0^1 \tilde{\phi}_{XY,\tau} d\tau}  ,
\end{align}
where $c$ and $c'$ are constants of $\orderof{1}$, and $\tilde{\phi}_{XY,\tau}$ is supported on $XY$.
By applying the approximation $\Phi_{\partial h_Y} \approx \tilde{\Phi}_{XY}$ to Eq.~\eqref{BP_eq_A_B_apply1D}, we approximate the reduced density matrix $\rho_{\beta,L^\co}$ as
\begin{align}
\rho_{\beta,L^\co} 
&\approx \tilde{\Phi}_{XY}  \tr_{L} \br{e^{\beta H_{LX}}} \otimes e^{\beta H_Y} \tilde{\Phi}^\dagger_{XY} \notag \\
&=\mathcal{T} e^{\int_0^1 \tilde{\phi}_{XY,\tau} d\tau} e^{\beta (\tilde{H}_{X}^\ast+ H_Y)}\br{\mathcal{T} e^{\int_0^1 \tilde{\phi}_{XY,\tau} d\tau}}^\dagger  ,\label{BP_approx_reduce_den} 
\end{align}
which is the desired form of Eq.~\eqref{approx_exponential_reduced},
where $\tilde{H}_{X}^\ast$ is defined such that $\tr_{L} \left(e^{\beta H_{LX}}\right) = e^{\beta \tilde{H}_{X}^\ast}$. 
The approximation error is dependent on the length $\ell$ of $X$ and aligns with the error bound in Eq.~\eqref{BP_eq_A_B_error_bound}.

In the application of the BP formalism to derive the effective Hamiltonian, it is necessary to select the length $\ell$ to be larger than $\beta^2 \norm{\partial h_Y}$ to ensure a small approximation error from the bound~\eqref{BP_eq_A_B_error_bound}. In higher dimensions, the norm $\norm{\partial h_Y}$ scales with the surface size of $X$, which is $\orderof{\ell^{D-1}}$. However, for $D \geq 2$ and $\beta \gg 1$, it becomes challenging to satisfy the condition $\ell \gtrsim \beta^2 \orderof{\ell^{D-1}}$.
To address this limitation, we develop an alternative approach. We introduce an ancillary system $L_a$, which is a copy of the subsystem $L$, and express the partial trace operation as follows:
\begin{align}
&\tr_L (\rho_\beta) =\mD_L \bra{\mP_L} \rho_\beta \otimes \hat{1}_{L_a}  \ket{\mP_L} , \notag \\
& \ket{\mP_L} := \frac{1}{\sqrt{\mD_L}} \sum_{j=1}^{\mD_L} \ket{j_L} \otimes \ket{j_{L_a}} ,
\end{align}
where $\ket{\mP_L}$ represents a maximally entangled state between the subsystems $L$ and $L_a$. We define the Partial-Trace-Projection (PTP) operator $\mP_L$ as the projection onto the state $\ket{\mP_L}$, which results in:
\begin{align}
\rho_{\beta,L^\co}\otimes \mP_L=\mD_L \mP_L  \rho_\beta \otimes \hat{1}_{L_a}   \mP_L ,
\label{exact_PTP_form} 
\end{align}
where the operator $\mP_L$ is supported on the doubled Hilbert space on $L \cup L_a$.

To approximate Eq.~\eqref{exact_PTP_form} in the exponential form as Eq.~\eqref{approx_exponential_reduced}, we define the approximate PTP operator $\mP_{L,\tau}$ as:
\begin{align}
\mP_{L,\tau} := e^{-\tau \mQ_L}, \quad \mQ_L := 1 - \mP_L.
\label{approx_PTP_form} 
\end{align}
Using the approximate PTP operator, we obtain the desired exponential form of the reduced density matrix:
\begin{align}
\label{approx_reduced_density_PTP}
\rho_{\beta,L^\co} \otimes \mP_L
&\approx \mD_L e^{-\tau \mQ_L} e^{\beta H} e^{-\tau \mQ_L},
\end{align}
where the approximation error is proved to be $\mD_L e^{-\orderof{\tau}}$. Due to the coefficient $\mD_L$, which scales as $e^{\Omega(|L|)}$, it is required to make $\tau = \Omega(|L|)$. This requirement results in a problematic dependence on the Hilbert space dimension for the CMI decay as detailed in Eq.~\eqref{main_high_dim_CMI}.

In summary, Step~1 allows us to approximate the reduced density matrix  in a tractable form, namely the connected exponential form given in 
Eq.~\eqref{approx_exponential_reduced_log}. 
The next challenge is to analyze the operator logarithm of this form in order to establish that the effective Hamiltonian of the approximate reduced state has a quasi-local structure.

\subsection{Second step}   \label{subsec:Second step}

The logarithm of connected exponential operators, such as $\log\left(e^{H_0} e^{V}\right)$, is commonly analyzed using the Baker-Campbell-Hausdorff formula or the Magnus expansion~\cite{BLANES2009151}. If a small order truncation of the Magnus expansion is employed, it can ensure the quasi-locality of the effective Hamiltonian $\hat{H}_\tau$ in Eq.~\eqref{approx_exponential_reduced_log}. However, a significant challenge arises due to the fact that the Magnus expansion is not absolutely convergent~\cite{doi:10.1080/00018732.2015.1055918, KUWAHARA201696, PhysRevLett.116.120401}.

To establish the quasi-locality of the effective Hamiltonian, we utilize the following theoretical framework outlined in~\cite[Supplementary Lemma~18 and Corollary~19]{Supplement_CMI}:\\

\noindent
{\bf Effective Hamiltonian Theory}.
Let $H_0$ and $V_{\tau_1}$ ($0 \leq \tau_1 \leq \tau$) be arbitrary operators (possibly interaction operators). By defining the effective operators $\hat{H}_\tau$ as 
\begin{align}
\label{effecetve_mA_tau}
\hat{H}_\tau := \log\left[\mathcal{T} e^{\int_0^\tau V_{\tau_1} d\tau_1} e^{\beta H_0} \left(\mathcal{T} e^{\int_0^\tau V_{\tau_1} d\tau_1}\right)^\dagger\right],
\end{align}
we can simplify it to the form:
\begin{align}
\hat{H}_\tau =  U_\tau (\beta H_0 + \hat{V}_\tau) U_\tau^\dagger
\end{align}
with
\begin{align}
\label{eff_Ham_U_B_mC}
&U_\tau := \mathcal{T} e^{-i\int_0^\tau \mC_{\tau_1} d\tau_1}, \notag \\
&\hat{V}_\tau :=2 \int_0^\tau U_{\tau_1}^\dagger V_{\tau_1} U_{\tau_1} d\tau_1, \notag \\
&\mC_{\tau_1} := \frac{2}{\beta}\int_{-\infty}^\infty g_\beta(t) e^{i H_{\tau_1} t } V_{\tau_1} e^{- i H_{\tau_1} t } dt,
\end{align}
where $g_\beta(t) := -{\rm sign}(t) (e^{2\pi |t|/\beta} - 1)^{-1}$. 

{~}

In the Supplementary Materials~\cite{Supplement_CMI}, we use the symbols $\mA$ and $\mB$ for the corresponding operators $H_0$ and $V$.

This expression significantly simplifies the form of the effective Hamiltonian. For simplicity, we assume $V_\tau = V$ without $\tau$ dependence in the following analysis. We then point out that from the definition~\eqref{effecetve_mA_tau}, the operator $\hat{H}_\tau$ originates from the differential equation:
\begin{align}
\label{derivative_mA_tau}
\frac{d}{d\tau}\hat{H}_\tau =2 V - i [\mC_\tau, \hat{H}_\tau].
\end{align}
Assuming the effective operator $H_\tau$ satisfies quasi-locality, the Lieb--Robinson bound can be applied to ensure the quasi-locality of $\mC_\tau$ in Eq.~\eqref{eff_Ham_U_B_mC}, which further implies the quasi-locality of $dH_\tau/d\tau$. Note that $g_\beta(t)$ approximately decays as $e^{-\mathcal{O}(|t|/\beta)}$, making the time integral in the range of $|t| = \mathcal{O}(\beta)$ dominant in the definition of $\mC_\tau$.

The challenge emerges from the fact that Eq.~\eqref{derivative_mA_tau} includes double-bracket flows, i.e., terms like $[[V, \hat{H}_\tau], \hat{H}_\tau]$. According to Ref.~\cite{hastings2022doublebracket}, the Lieb--Robinson bound for double-bracket flow does not generally preserve quasi-locality, unlike the standard Heisenberg equation. To illustrate this, consider:
\begin{align}
-i [\mC_\tau, \hat{H}_\tau] = \frac{-2i}{\beta} \int_{-\infty}^\infty g_\beta(t) \left[e^{i H_\tau t} V e^{-i H_\tau t}, \hat{H}_\tau\right] dt,
\end{align}
applying the definition of $\mC_\tau$ from Eq.~\eqref{eff_Ham_U_B_mC}. For small $\delta t \ll 1$, we approximate $g_\beta(\delta t) \approx -\beta/(2\pi \delta t)$, resulting in the double-bracket flow appearing as:
\begin{align}
\label{eq_integrand_g_beta}
\frac{-i g_\beta(\delta t)}{\beta} \left[e^{i H_\tau \delta t} V e^{-i H_\tau \delta t}, \hat{H}_\tau\right] \approx \frac{[[V, \hat{H}_\tau], \hat{H}_\tau]}{2\pi},
\end{align}
up to an $\mathcal{O}(\delta t)$ error. This analysis does not necessarily imply a breakdown in the quasi-locality of the effective Hamiltonian. First, for sufficiently small $\delta t$, although double-bracket terms appear in the integral of $\mC_\tau$ in Eq.~\eqref{eff_Ham_U_B_mC}, their contribution (i.e., the integral for $|t| \leq \delta t$) is proportional to $\delta t$. Moreover, it is observed that the integrand~\eqref{eq_integrand_g_beta} deviates from exact double-bracket flows for large $t$. By analyzing this deviation more carefully, we can still ensure the quasi-locality of the effective interaction terms.

In the present analyses, we take $V = V_{i_0}$ for example, where the Hamiltonian $H_0$ is a short-range interacting Hamiltonian as specified in Eq.~\eqref{def:Hamiltonian_main}, and $V_{i_0}$ is a local operator located at site $i_0 \in \Lambda$. We focus on estimating the effective Hamiltonian $\hat{H}_\tau = \log(e^{\tau V_{i_0}} e^{\beta H_0} e^{\tau V_{i_0}})$. Notably, $V_{i_0}$ can also be generalized to a quasi-local operator around any subsystem, as discussed in~\cite[Supplementary Sec. V]{Supplement_CMI}.
A crucial aspect we examine is the quasi-locality of the unitary operator $U_\tau$, particularly through the commutator norm $\norm{[U_\tau, u_i]}$, where $u_i$ is any local unitary operator on a site $i \in \Lambda$. The dynamic property of this commutator is given by:
\begin{align}
\label{dtau_U_tau_u_i_upp}
\frac{d}{d\tau} \norm{[U_\tau, u_i]} \le \norm{[\mC_\tau, u_i]}.
\end{align}

\begin{widetext}
To estimate $\norm{[\mC_\tau, u_i]}$, we upper-bound the  commutator norm between the time-evolved $V_{i_0}$ and $u_i$~\cite[Supplementary Lemma~22]{Supplement_CMI}:
\begin{align}
\label{time_evo_V_upper_bound_hat}
\norm{\left[e^{i \hat{H}_\tau t} V_{i_0} e^{-i \hat{H}_\tau t}, u_i\right]} &\le \norm{[V_{i_0}, u_i(-t)]} \notag \\
&\quad + 2 \norm{V_{i_0}} \left( \norm{[U_\tau, u_i]} + \norm{[U_\tau, u_i(-t)]} + \int_0^t \norm{\left[\hat{V}_\tau, u_i(-t_1)\right]} dt_1 \right),
\end{align}
where we denote $u_i(-t) = e^{-i H_0 t} u_i e^{i H_0 t}$.
In analyzing the growth of the commutator norm, we employ the upper bound $\norm{[U_\tau, u_i]} \le Q(\tau, r)$, where $r = \dist_{i_0, i}$, and $Q(\tau, r)$ is defined as:
\begin{align}
\label{Q_tau_r_anzats}
Q(\tau, r) = e^{\kappa_0 \tau + \kappa_1 \tau \log(r + \tau + e) - \kappa_\beta r},
\end{align}
with $\kappa_0$, $\kappa_1$, and $\kappa_\beta$ being free parameters chosen afterward.
\end{widetext}

To ensure consistency between the upper bounds in Eq.~\eqref{dtau_U_tau_u_i_upp} and $\norm{[U_\tau, u_i]} \le Q(\tau, r)$, we can prove that it is sufficient to satisfy the condition:
\begin{align}
\label{condition_mC_comm_u_i}
\norm{[\mC_\tau, u_i]} \le \left[\kappa_0 + \kappa_1 \log(r + \tau + e)\right] Q(\tau, r).
\end{align}
Applying the Lieb--Robinson bound to $u_i(-t)$ and utilizing the upper bound~\eqref{Q_tau_r_anzats} for $\norm{[U_\tau, u_i]}$, we can upper-bound the right-hand side of the inequality~\eqref{time_evo_V_upper_bound_hat}. After integrating Eq.~\eqref{time_evo_V_upper_bound_hat} with the filter function $g_\beta(t)$, we further upper-bound $\norm{[\mC_\tau, u_i]}$ using the parameters $\kappa_0$, $\kappa_1$, and $\kappa_\beta$. 
Under appropriate choices of parameters such as:
\begin{align}
\kappa_0 &\propto \beta^D \norm{V_{i_0}} \log(\beta \norm{V_{i_0}}), \notag \\
\kappa_1 &\propto \beta^D \norm{V_{i_0}}, \notag \\
\kappa_\beta &\propto 1/\beta,
\end{align}
we can satisfy the condition~\eqref{condition_mC_comm_u_i}. This estimation shows that the unitary operator $U_{\tau}$ maintains quasi-locality around the site $i_0$ with exponentially decaying tails, expressed as $e^{\mathcal{O}(\beta^D \norm{V_{i_0}}) - r/\beta}$. The detailed calculations are complex, and we defer all specifics to the supplementary materials~\cite[Supplementary Subtheorem~1]{Supplement_CMI}.
In the one-dimensional case, the dependence of $\kappa_0$ and $\kappa_1$ on $\beta$ disappears entirely (i.e., they are $\mathcal{O}(1)$ with respect to $\beta$), as shown in~\cite[Supplementary Theorem~3]{Supplement_CMI}. We conjecture that a similar improvement to a $\beta^{D-1}$ dependence may also be possible in higher dimensions.

 \begin{figure}[tt]
\centering
\includegraphics[clip, scale=0.8]{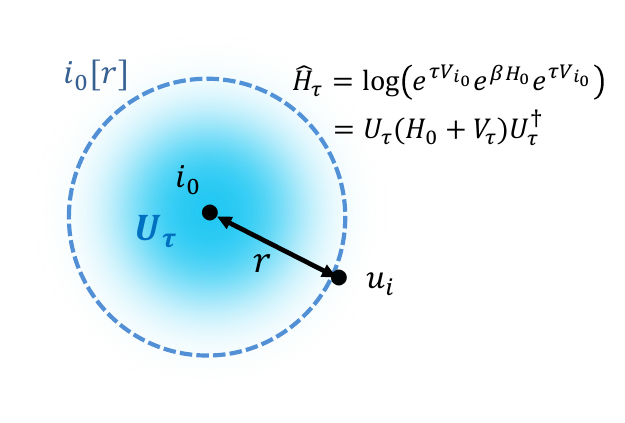}
\caption{Quasi-locality of the effective interaction terms. 
We are examining the operator logarithm given by $\log\br{e^{\tau V_{i_0}} e^{\beta H} e^{\tau V_{i_0}}}$ and begin by establishing the quasi-locality of the unitary operator $U_\tau$ in Eq.~\eqref{eff_Ham_U_B_mC}, which is instrumental in defining the effective Hamiltonian presented in Eq.~\eqref{effecetve_mA_tau}. The quasi-locality of $U_\tau$ is evaluated based on the norm $\norm{\brr{U_\tau, u_i}}$, where $u_i$ represents any arbitrary unitary operator located on a site $i$ ($\dist_{i,i_0} = r$).  
However, confirming the quasi-locality of $U_\tau$ alone does not fulfill all our requirements. We further stipulate that $U_\tau$ should primarily be influenced by the neighboring region around the site $i_0$. This requirement leads us to approximate $U_\tau$ by $U_{\tau, i_0[r]}$, as expressed in Eq.~\eqref{U_tau_ball_region_approx}. Here, $U_{\tau, i_0[r]}$ is specifically determined by the ball region centered at the site $i_0$ with radius $r$ (inside the dashed circle).
}
\label{fig:Fig4}
\end{figure}

At this stage, we have established only the quasi-locality of the unitary operator $ U_{\tau} $. However, this does not necessarily imply that $ U_{\tau} $ is determined solely by the neighboring region around the site $ i_0 $. To address this mathematically, we consider the subset Hamiltonian $ H_{0, i_0[r]} $ on the ball region centered at the site $ i_0 $ with radius $ r $, and define:
\begin{align}
\label{U_tau_ball_region_approx}
&\log(e^{\tau V_{i_0}} e^{\beta H_{0, i_0[r]}} e^{\tau V_{i_0}})  \notag \\
&= \beta U_{\tau, i_0[r]} (H_{0, i_0[r]} + \hat{V}_{\tau, i_0[r]}) U_{\tau, i_0[r]}^\dagger,
\end{align}
where the unitary $ U_{\tau, i_0[r]} $ is determined only by the subset Hamiltonian $ H_{0, i_0[r]} $. If $ U_\tau \approx U_{\tau, i_0[r]} $, we can infer that the unitary $ U_\tau $ is approximately constructed from $ H_{0, i_0[r]} $.
Utilizing the quasi-locality from Eq.~\eqref{Q_tau_r_anzats}, we can also derive the approximation error between $ U_\tau $ and $ U_{\tau, i_0[r]} $, which qualitatively follows the same upper bound as in Eq.~\eqref{Q_tau_r_anzats} (see \cite[Supplementary Theorem~2]{Supplement_CMI}).

So far, we have shown that when the connected exponential form approximates the reduced density matrix, its operator logarithm can be controlled and exhibits a quasi-local structure. 
In Step~3, we connect this quasi-locality to the decay of the conditional mutual information (CMI), thereby deriving our main results on CMI decay, Eqs.~\eqref{main_high_dim_CMI} and~\eqref{main_1D_dim_CMI}.

\subsection{Third step}   \label{subsec:Third step}
Here, we introduce $\tilde{\rho}_{\beta,L^\co}$,
which denotes an approximate reduced density matrix for
$\rho_{\beta,L^\co}$. Concretely, $\tilde{\rho}_{\beta,L^\co}$ is
constructed via the connected exponential: in one dimension, using the
BP formalism~\eqref{BP_eq_A_B_error_bound}, and in higher dimensions
using the PTP formalism~\eqref{approx_reduced_density_PTP}. 
Throughout this section, the tilde notation consistently indicates a
quasi-local approximation, in contrast to the exact reduced states
used earlier.

%
%
 Leveraging the quasi-locality of the effective Hamiltonian described in Step 2, we can estimate the operator norm of
\begin{align}
\tilde{H}(A:C|B): = -\beta \bigl[ & \log (\tilde{\rho}_{\beta,AB}) +\log (\tilde{\rho}_{\beta,BC})  \notag \\
 &- \log (\tilde{\rho}_{\beta,ABC}) - \log (\tilde{\rho}_{\beta,B}) \bigr].
\end{align}
Employing the quasi-locality function from Eq.~\eqref{Q_tau_r_anzats}, we are able to approximate $\norm{\tilde{H}(A:C|B)} \lesssim Q(\tau, R)$, where $R = \dist_{A, C}$ and $\tau$ follows from the approximation in Eq.~\eqref{approx_exponential_reduced}. From the definition of conditional mutual information (CMI) in Eq.~\eqref{CMI_form_by_H_A|B}, we derive:
\begin{align}
\label{error_CMI_relative_ent}
&\mI_{\rho_\beta}(A:C|B) - \tr \left( \rho_\beta \tilde{H}(A:C|B) \right) \notag \\
&= -S(\rho_{\beta, AB} \| \tilde{\rho}_{\beta, AB}) - S(\rho_{\beta, BC} \| \tilde{\rho}_{\beta, BC}) \notag \\
&\quad + S(\rho_{\beta, ABC} \| \tilde{\rho}_{\beta, ABC}) + S(\rho_{\beta, B} \| \tilde{\rho}_{\beta, B}),
\end{align}
where $S(\rho \| \tilde{\rho})$ is the relative entropy, defined as $S(\rho \| \tilde{\rho}) := \tr \left[ \rho \log(\rho) - \rho \log(\tilde{\rho}) \right]$.

In order to complete the argument, it is necessary to upper-bound the right-hand side of Eq.~\eqref{error_CMI_relative_ent}. At this stage, however,
the techniques required in the BP formalism and in the PTP formalism differ from each
other. Since the notation becomes rather cumbersome in this part of the proof, we refer
interested readers to the detailed statements in \cite[Supplementary Lemma~14 and Corollary~17]{Supplement_CMI}.
The approaches allow us to demonstrate the decay of CMI without directly accessing the true entanglement Hamiltonian $H^\ast_L$, addressing the more complex challenge described in Eq.~\eqref{two_effective_Ham_distance}.

This leads to the proof of our main results on CMI decays~\eqref{main_high_dim_CMI} and \eqref{main_1D_dim_CMI} (see \cite[Supplementary Theorems~4 and~5]{Supplement_CMI}).

As a supplementary remark, the following simple upper bound cannot be used:
\begin{align}
&\mI_{\rho_\beta}(A:C|B) \notag \\
&\le \norm{\tilde{H}(A:C|B)} + 4 \max_{L \subseteq \Lambda} \left[ S(\rho_{\beta, L^\co} \| \tilde{\rho}_{\beta, L^\co}) \right].
\label{simple_bound_for_CMI}
\end{align}
Even for $\rho_{\beta, L^\co} \approx \tilde{\rho}_{\beta, L^\co}$, the continuity inequality does not suffice for estimating $S(\rho_{\beta, L^\co} \| \tilde{\rho}_{\beta, L^\co})$. Utilizing the Alicki-Fannes inequality~\cite{Alicki_2004}, we find the following inverse Pinsker's inequality:
\begin{align}
S(\rho_{\beta, L^\co} \| \tilde{\rho}_{\beta, L^\co}) \lesssim |L^\co| \cdot \norm{ \rho_{\beta, L^\co} - \tilde{\rho}_{\beta, L^\co}}_1.
\end{align}
However, the dependency on the size of $|L^\co|$ suggests that the simple inequality~\eqref{simple_bound_for_CMI} is not applicable to the thermodynamic limit as $|\Lambda| \to \infty$. Notably, our main results, encapsulated in Eqs.~\eqref{main_high_dim_CMI} and~\eqref{main_1D_dim_CMI}, exclude the $|\Lambda|$ dependency. Since the continuity inequality is already optimal, we must explore alternative methods to establish the independence of the relative entropy from $|\Lambda|$. This is why we consider a refined approach in~\cite[Supplementary Lemma~14 and Corollary~17]{Supplement_CMI}.

As an outlook, it remains an interesting question to identify cases where the inverse Pinsker's inequality holds without dependence on the logarithm of the Hilbert space dimension, possibly by extending the current analytical techniques.

\subsection{Analyses of the 1D entanglement Hamiltonian}   \label{subsec:entanglement_Ham_analyses}

Finally, we provide a technical overview of accessing the true entanglement Hamiltonian in a one-dimensional case. The central approach relies on the continuity of the operator logarithm concerning the relative error, as stated in \cite[Supplementary Theorem~7]{Supplement_CMI}:\\

\noindent
{\bf Continuity of the Operator Logarithm in Terms of the Relative Error:}
Consider two density operators $\rho$ and $\sigma$. We define the relative error $\delta_{\rm R}(\rho,\sigma)$ as:
\begin{align}
\delta_{\rm R}(\rho,\sigma) := \sup_{\ket{\psi}} \frac{|\bra{\psi} (\rho - \sigma) \ket{\psi}|}{\bra{\psi} \rho \ket{\psi}},
\label{relative/error_main_text}
\end{align}
where $\sup_{\ket{\psi}}$ is taken over all quantum states. The continuity inequality for $\log(\rho) - \log(\sigma)$ is then given by:
\begin{align}
\norm{\log(\sigma) - \log(\rho)} \lesssim \delta_{\rm R}(\rho,\sigma) \log \left[ \lambda^{-1}_{\min}(\rho) \right],
\label{relative/error_main_text/ineq}
\end{align}
where $\lambda_{\min}(\rho)$ is the minimum eigenvalue of $\rho$.

{~}

With this framework, by analyzing the relative error between $\rho_{\beta,L}$ and its approximation $\tilde{\rho}_{\beta,L}$, we establish the quasi-locality of the true entanglement Hamiltonian $\log(\rho_{\beta,L})$ compared to $\log(\tilde{\rho}_{\beta,L})$. Nevertheless, two primary challenges arise: (i) estimating the minimum eigenvalue of the reduced density matrix, and (ii) estimating the relative error $\delta_{\rm R}(\rho_{\beta,L}, \tilde{\rho}_{\beta,L})$.

While naively straightforward, the estimation of $\lambda^{-1}_{\min}(\rho_{\beta,L})$, expected to be proportional to $\beta |L|$, presents difficulties. This proportionality is established in commuting cases, as shown in Ref.~\cite{anshu2021efficient}. Extending this to non-commuting cases requires strict $k$-locality of the Hamiltonian, utilizing technical methods from Refs.~\cite{Arad_2016, Anshu_2021}. Generally, it can be shown (see \cite[Supplementary Proposition~12]{Supplement_CMI}):
\begin{align}
\log\brr{ \lambda^{-1}_{\min}(\rho_{\beta,L})} \le \beta \bar{J}_0 |L| + \log(16 \bar{J}_0 |L| \mathcal{D}_{L})
\end{align}
for two-body interacting Hamiltonians,
where we can choose $\bar{J}_0 = 3J_0 + 4J_0 |L^\co|^{-1} \log(8 \mathcal{D}_{L^\co})+1$ using the Hamiltonian parameter $J_0$ in Eq.~\eqref{def:Hamiltonian_main}. This approach can be generalized to any $k$-local Hamiltonians.

To address the second challenge, i.e., the estimation of $\delta_{\rm R}(\rho_{\beta,L}, \tilde{\rho}_{\beta,L})$, we first note that the relative error for the reduced density matrices is always less than that for the global ones, i.e., $\delta_{\rm R}(\rho_{\beta,L}, \tilde{\rho}_{\beta,L}) \le \delta_{\rm R}(\rho_{\beta}, \tilde{\rho}_{\beta})$. In one-dimensional cases, this comparison involves two density matrices derived using the quantum belief propagation, as specified in Eqs.~\eqref{BP_eq_A_B_apply1D} and \eqref{BP_approx_reduce_den}:
\begin{align}
&\rho_{\beta} = \Phi_{\partial h_Y} e^{\beta H_{LX}} \otimes e^{\beta H_Y} \Phi^\dagger_{\partial h_Y}, \notag \\
&\tilde{\rho}_{\beta} = \Phi_{XY} e^{\beta H_{LX}} \otimes e^{\beta H_Y} \Phi^\dagger_{XY}.
\end{align}
The upper bound for the relative error can then be derived as:
\begin{align}
\delta_{\rm R}(\rho_{\beta}, \tilde{\rho}_{\beta}) \le \norm{1 - \mathcal{W} \mathcal{W}^\dagger},
\end{align}
where $H_0 = H_{LX} + H_Y$ and 
\begin{align}
\mathcal{W} := e^{-\beta H_0/2} \Phi_{\partial h_Y}^{-1} \Phi_{XY} e^{\beta H_0/2}.
\end{align}
From the inequality~\eqref{BP_eq_A_B_error_bound}, we have $\Phi_{\partial h_Y}^{-1} \Phi_{XY} \approx 1$. Hence, the primary issue is the amplification by the imaginary time evolution of $e^{-\beta H_0/2}$.

The analysis of this imaginary time evolution reveals that the difficulty stems from the potential exponential amplification of norms, as expressed by:
\begin{align}
\norm{e^{-\beta H_0/2} O_X e^{\beta H_0/2} } \le e^{c_7 \beta |X| + e^{c_8\beta} }\norm{O_X},  \notag 
\end{align}
with $c_7,c_8=\orderof{1}$ independent of $|\Lambda|$, where $O_X$ is an operator supported on $X \subset \Lambda$. On the other hand, utilizing the Lieb--Robinson bound, we can decompose:
\begin{align}
\Phi_{\partial h_Y}^{-1} \Phi_{XY} = 1 + \sum_{\ell=1}^\infty \Phi_\ell,
\end{align}
where each $\Phi_\ell$ is an $\ell$-local operator with a norm that decays exponentially with $\ell/\beta$. However, the amplification by the imaginary time evolution for each decomposed term $\Phi_\ell$ can be significant, leading to:
$
\norm{e^{-\beta H_0/2} \Phi_\ell e^{\beta H_0/2}} = e^{\Omega(\beta \ell)+ e^{\Omega(\beta)}} \norm{\Phi_\ell}.
$
Such amplification substantially breaks down the approximation $\mathcal{W} \approx 1$ for large $\beta$. 
To derive a better bound, we need to fully utilize the structure of the Hamiltonian in more elaborate ways (see~\cite[Supplementary Subtheorem~2]{Supplement_CMI}).

\section{Outlook}

In this work, we have addressed the resolution of the conjecture stated in Eq.~\eqref{quantum_Markov_Conj_ineq}, proposing that every quantum Gibbs state approximates a Markov network at any temperature. Our principal findings are encapsulated in the inequalities presented in Eqs.~\eqref{main_high_dim_CMI} and \eqref{main_1D_dim_CMI}. Due to the dependency on the Hilbert space dimension $\mathcal{D}_{AC}$, our analysis only concludes that quantum Gibbs states constitute approximately pairwise Markov networks, rather than forming a global Markov network. This study significantly advances the understanding of the entanglement Hamiltonian by systematically approximating its properties.
As applications of our results, we have identified a clustering property of genuine bipartite entanglement beyond the PPT class and have established the logarithmic sample complexity required for learning 1D Hamiltonians.

Despite these advancements, various open questions remain. One of the most pressing challenges is improving the subsystem-size dependence (i.e., $|A|, |C|$) of the CMI decay to a polynomial form $\poly(|A|, |C|)$, which would enable its application to macroscopically large subsystems. The current analyses are highly intricate, whereas point-by-point calculations have been optimized. Hence, we believe that further improvement is likely to require a theoretical leap. This could involve leveraging the quasi-locality of the Hamiltonians more effectively. One potential approach is to impose additional constraints on the quantum Gibbs states, such as clustering of bipartite correlations, as discussed in Ref.~\cite{kato2016quantum}. Even under these constraints, the improvement is still beneficial for applications like quantum Gibbs sampling, as detailed in Ref.~\cite{Brandao2019}.

The second crucial problem we address is identifying the types of long-range entanglement that robustly persist at finite temperatures. Despite extensive studies, computing entanglement in quantum Gibbs states remains a significant challenge, even at the numerical level~\cite{PhysRevB.76.184442, PhysRevB.78.155120, MAZAC20122096, PhysRevA.88.042318, Gabbrielli2018, Shapourian_2019, PhysRevLett.125.116801, PhysRevLett.125.140603, PhysRevResearch.2.043345,kim2024thermal,PhysRevLett.134.190404}. Our approaches are twofold. First, we aim to refine the subsystem-size dependence as described in Eq.~\eqref{corol:ent_F_high_ineq_main_00} for bipartite-entanglement decay in high dimensions.
From the relation~\eqref{Sq_exp_decay_coj_00}, this is closely linked to the first open problem regarding the CMI decay. 
The second approach considers the presence of multipartite entanglement.
So far, understanding multipartite entanglement is challenging due to the need for established methods for its characterization~\cite{PhysRevA.67.012108, RevModPhys.81.865, PhysRevA.92.042329, PhysRevLett.127.040403, PhysRevLett.127.140501, MA2024}. 
We still anticipate that certain types of tripartite entanglement may be absent in general many-body systems, as the CMI effectively describes tripartite correlations.

The third open problem focuses on improving the quasi-locality analyses of the entanglement Hamiltonian. In our studies of the CMI decay, we have only needed to address the approximate entanglement Hamiltonian in terms of Eq.~\eqref{Hamiltonian_eff_approx}. However, more than this approximation is required in order to access the true entanglement Hamiltonian.
In the 1D case, this issue is partially resolved, as demonstrated in Eq.~\eqref{Hamiltonian_eff_approx_main}, although the temperature dependence remains doubly exponential.
By improving the temperature dependence of the polynomial forms, we can access the 1D entanglement Hamiltonian at the ground state, where $\beta=\log(n)$ is taken for approximating the Gibbs state to the ground state. 
This may further lead to the solution of the Li--Haldane--Poilblanc conjecture~\cite{PhysRevLett.101.010504,PhysRevLett.105.077202,PhysRevB.109.195169} in a generic setup beyond the CFT class~\cite{Zache2022entanglement,PhysRevB.98.134403}. 
Our present approach relies on a new continuity inequality for the operator logarithm, indicated by Eq.~\eqref{relative/error_main_text/ineq}, suggesting that a small relative error between two operators implies their logarithms are similarly close. Future improvements could involve developing a better bound by introducing an alternative measure of operator distance, distinct from the relative error.
The essential obstacle to accessing the entanglement Hamiltonian stems from the difficulty of connecting the closeness of density matrices and the closeness of the logarithms of those matrices. This issue has been a central challenge in the Hamiltonian learning problem~\cite{Anshu_2021,bak2023,chen2025_learn}. We expect that recent technical advances in that area will contribute significantly to resolving the quasi-locality of the entanglement Hamiltonian in more general settings.

%

Finally, the primary method used to analyze the behavior of exponential operators extends beyond studying the quantum Gibbs states. This technique is particularly valuable for examining the effective Hamiltonians in systems undergoing open quantum dynamics, which involve interactions with the environment, as also reviewed in~\cite{RevModPhys.92.041002}. A thorough analysis of quantum effects arising from environmental interactions is essential for a better understanding of quantum thermalization processes as well as for characterizing mixed quantum phases~\cite{lee2024universalspreadingconditionalmutual,zhang2024nonlocalgrowthquantumconditional,sang2024stabilitymixedstatequantumphases}. Our current work will pave the way for new analytical approaches to tackle these complex issues.

\section{Acknowledgment}

T. K. acknowledges the Hakubi projects of RIKEN.
T. K. was supported by JST PRESTO (Grant No. JPMJPR2116), ERATO (Grant No. JPMJER2302), and JSPS Grants-in-Aid for Scientific Research (No. JP23H01099, JP24H00071), Japan. 
We are grateful to Angela Capel, Chi-Fang Chen, Shang Cheng, Donghoon Kim, Isaac Kim, Kohtaro Kato, Quynh The Nguyen, Shengqi Sang, Samuel Scalet, and Chao Yin for helpful discussions and comments on related topics.

\textit{Note added.}
After submission of this manuscript, the CMI decay was revisited using a completely different approach, constructing the recovery map~\cite{kato2025,chen2025} with the combination of Fawzi--Renner inequality~\cite{Fawzi2015}. In particular, the result in Ref.~\cite{chen2025} partially overcomes one of the critical open problems regarding subsystem-size (i.e., $|A|$, $|C|$) dependence at any temperatures.
Still, the method presented here has a strong advantage in that it provides a route to access the structure of the entanglement Hamiltonian.

\bibliography{Quantum_Markov}

\clearpage

\widetext{

\input{CMI_decay_supple_arxiv.tex}

}

\end{document}

%% file: CMI_decay_supple_arxiv.tex


\renewcommand{\theequation}{S.\arabic{equation}}

\renewcommand{\thesection}{S.\Roman{section}}

\begin{center}
{\large \bf Supplementary Materials for  ``Clustering of conditional mutual information and quantum Markov structure at arbitrary temperatures''}\\
\vspace*{0.3cm}
\author{Tomotaka Kuwahara$^{1,2,3}$}
\affiliation{$^{1}$ 
Analytical Quantum Complexity RIKEN Hakubi Research Team, RIKEN Center for Quantum Computing (RQC), Wako, Saitama 351-0198, Japan
}
\affiliation{$^{2}$ 
RIKEN Pioneering Research Institute (PRI), Wako, Saitama 351-0198, Japan
}

\affiliation{$^{3}$
PRESTO, Japan Science and Technology (JST), Kawaguchi, Saitama 332-0012, Japan}

\end{center}

\renewcommand\thefootnote{*\arabic{footnote}}


\setcounter{equation}{0}
\setcounter{section}{0}

\tableofcontents

\clearpage
\newpage

\section*{Roadmap of the materials}

The supplementary materials are mainly devoted to providing detailed proofs 
of the statements presented in the main text. 
Since the material is rather long, 
we provide in Figs.~\ref{fig_Overview_1} and \ref{fig_Overview_1} a roadmap that outlines 
the logical flow and structure of the supplement.
Throughout the Supplementary Materials, we use the symbols $\mA$ and $\mB$ (corresponding to $H_0$ and $V$ in the main text) to simplify the notation.

 \begin{figure}[bb]
\centering
\includegraphics[clip, width=\textwidth]{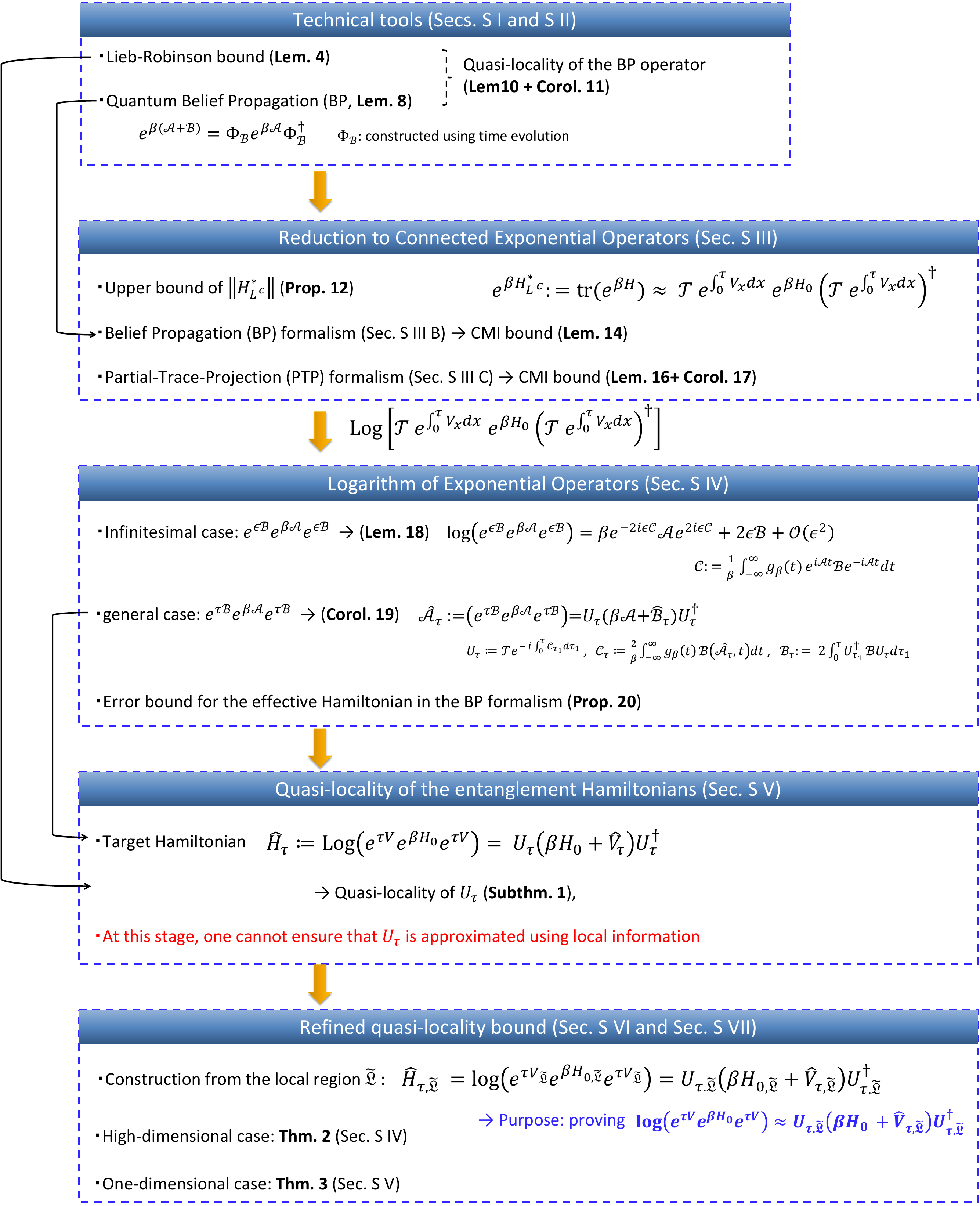}
  \caption{Roadmap (from Sec. S I to Sec. S VII)}
  \label{fig_Overview_1}
\end{figure}

\clearpage

 \begin{figure}[bb]
\centering
\includegraphics[clip, width=\textwidth]{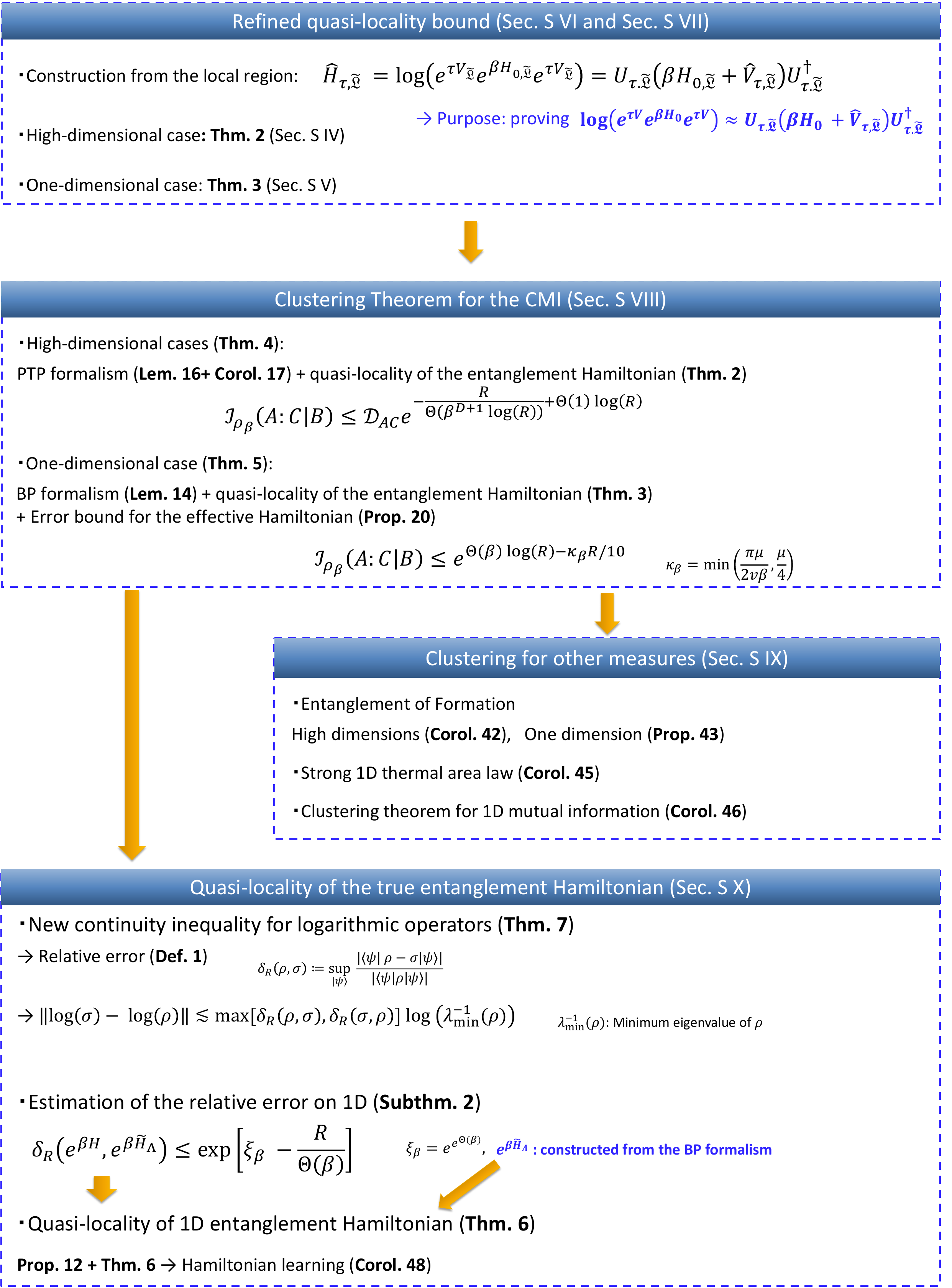}
  \caption{Roadmap (from Sec. S VI to Sec. S X)}
  \label{fig_Overview_2}
  \end{figure}

\clearpage

\section{Precise system setup and notations}

\begin{figure}[bb]
\centering
\includegraphics[clip, scale=0.5]{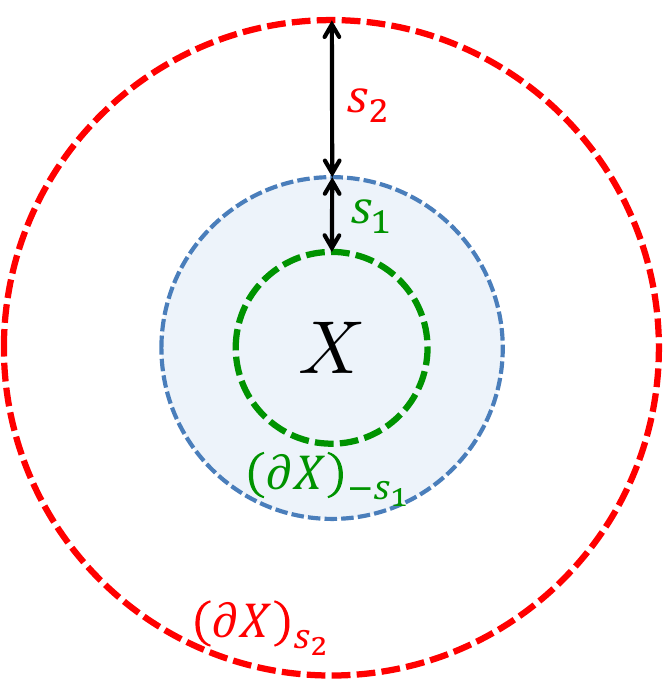}
\caption{Schematic picture to describe the definition of $(\partial X)_s$ for positive and negative $s$.
}
\label{fig_partial_X_s}
\end{figure}

Consider a quantum system on a $D$-dimensional lattice with $n$ sites, with $\Lambda$ representing the set of all sites. 
For any arbitrary partial set $X\subseteq \Lambda$, we denote the cardinality (number of sites in $X$) as $|X|$. The complementary subset of $X$ is denoted by $X^\co := \Lambda\setminus X$. We define $\mathcal{D}_X$ as the dimension of the Hilbert space on $X$.
Also, we often denote $X\cup Y$ by $XY$ for simplicity. 
For subsets $X$ and $Y$ of $\Lambda$, the distance $\dist_{X,Y}$ is defined as the shortest path length on the graph connecting $X$ and $Y$, with $\dist_{X,Y}=0$ if $X\cap Y \neq \emptyset$. When $X$ comprises only one element (i.e., $X=\{i\}$), we use $\dist_{i,Y}$ to represent $\dist_{\{i\},Y}$ for simplicity. 
We also define $\diam(X)$ as follows: 
$
\diam(X):  =1+ \max_{i,i'\in X} (\dist_{i,i'}).
$
The surface subset of $X$ is denoted by 
\begin{align}
\partial X:=\{ i\in X| \dist_{i,X^\co}=1\} . \label{surface_definition_partial_X}
\end{align}
Moreover, we define $(\partial X)_s$ as follows (see Fig.~\ref{fig_partial_X_s}): 
\begin{align}
\label{notation_partial_X_s}
(\partial X)_s := 
\begin{cases}
\{i\in X| \dist_{i,\partial X} =s\} &\for s\le 0, \\
\{i\in X^\co | \dist_{i,\partial X} =s\} &\for s> 0. 
\end{cases}
\end{align}
where $(\partial X)_0= \partial X$ and we have 
\begin{align}
X = \bigcup_{s=0}^\infty (\partial X)_{-s} ,\quad \Lambda= \bigcup_{s=-\infty}^\infty  (\partial X)_s .
\end{align}

For a subset $X\subseteq \Lambda$, the extended subset $\bal{X}{r}$ is defined as
\begin{align}
\bal{X}{r}:= \{i\in \Lambda| \dist_{X,i} \le r \} , \label{def:bal_X_r}
\end{align}
where $\bal{X}{0}=X$, and $r$ is an arbitrary positive number (i.e., $r\in \mathbb{R}^+$). We introduce a geometric parameter $\gamma$ determined solely by the lattice structure, satisfying $\gamma \ge 1$ as a constant of $\orderof{1}$, which fulfills the inequalities:
\begin{align}
\max_{i\in \Lambda} ( |\partial i[r] | ) \le \gamma r^{D-1} ,\quad \max_{i\in \Lambda} ( | i[r] | )\le \gamma r^{D} ,\label{def:parameter_gamma}
\end{align}
where $r\ge 1$.

We consider a Hamiltonian with short-range (or exponentially decaying) interactions on an arbitrary finite-dimensional graph:
\begin{align}
H = \sum_{Z} h_Z, 
\quad \max_{i\in \Lambda} \sum_{Z:Z\ni i}\|h_Z\| \le \bar{J}_0 \label{def:Hamiltonian}
\end{align}
with the interaction decays such that 
\begin{align}
\label{def_short_range_long_range}
J_{i,i'}:= \sum_{Z:Z\ni \{i,i'\}}\norm{ h_Z} \le \bar{J}_{\dist_{i,i'}}:=
\bar{J}_0 e^{-\mu \dist_{i,i'}} ,
\end{align}
where we define $\norm{\cdots}$ as the operator norm, i.e., the maximum singular value of a target operator. 
We often utilize the trace norm $\norm{\cdots}_1$ which means $\norm{O}_1 := \tr\br{\sqrt{O^\dagger O}}$. 
We note that for $i=i'$, the condition~\eqref{def_short_range_long_range} reduces to the condition in Eq.~\eqref{def:Hamiltonian}. 
We denote the subset Hamiltonian on a region $L$ by $H_L$ and all the interaction terms acting on $L$ by $\widehat{H_L}$, i.e., 
\begin{align}
H_L := \sum_{Z: Z\subset L} h_Z, \quad \widehat{H_L}:= \sum_{Z: Z\cup L\neq \emptyset} h_Z = H- H_{L^\co} . 
\label{def:Hamiltonian_subset_L}
\end{align}
Also, we define the boundary interaction terms on the region $L$ as follows:
\begin{align}
\partial h_{L}\coloneqq  H-H_{L} - H_{L^\co}=\sum_{Z: Z\cap L\neq \emptyset ,\ Z\cap L^\co\neq \emptyset} h_Z \Or \widehat{H_L}= H_{L}+\partial h_{L}
 \label{def:Ham_surface}
\end{align}

For an arbitrary operator $O_1$, we define the time evolution of $O_1$ by another operator $O_2$ as 
\begin{align}
O_1(O_2,t) := e^{iO_2t} O_1 e^{-iO_2t} .
\end{align}
For simplicity, we often denote $O_1(H,t)$ by $O_1(t)$.

We consider the quantum Gibbs state $\rho_\beta$ with a fixed inverse temperature $\beta$:
\begin{align}
\rho_\beta := e^{\beta H}/Z_\beta ,\quad Z_\beta= \tr (e^{\beta H}). 
\end{align}
Throughout, we use $\rho_\beta$ to denote the Gibbs state at inverse temperature $\beta$, and reserve $\rho$ without a subscript for general density matrices.
We also define the reduced density matrix and the corresponding effective Hamiltonian (a.k.a entanglement Hamiltonian) as 
\begin{align}
\rho_{\beta,L} = \tr_{L^\co} (\rho_\beta), \quad 
H^\ast_L =(1/\beta) \log(\rho_{\beta,L}),
\label{def_rho_L_H^ast_L}
\end{align}
where $\tr_{L^\co}(\cdots)$ denotes the partial trace operation with respect to $L^\co$.
We define normalized partial trace $\tilde{\tr}_X(O)$ as follows:
\begin{align}
\label{definition_of_tilde_tr_partial}
\tilde{\tr}_X(O) : =\frac{1}{\tr_X(\hat{1})} \tr_X(O).
\end{align}
We note that $\tilde{\tr}_X(O)$ is supported on $X^\co$ and commutes with arbitrary operators supported on $X$, i.e., $[\tilde{\tr}_X(O),O_X]=0$. 
Also, the norm $\norm{\tilde{\tr}_X(O)}$ is always smaller than or equal to $\norm{O}$, i.e., $\norm{\tilde{\tr}_X(O)}\le \norm{O}$.

Here, we introduce the notation $\Theta(x_1,x_2,\ldots,x_s)$ that means a function with respect to $\{x_1,x_2,\ldots,x_s\}$ as follows:
\begin{align}
\label{Theta_notation_def}
\Theta(x_1,x_2,\ldots,x_s) = \sum_{\sigma_1,\sigma_2,\ldots,\sigma_s=0,1} c_{\sigma_1,\sigma_2,\ldots,\sigma_s} x_1^{\sigma_1}x_2^{\sigma_2} \cdots x_s^{\sigma_s} \quad (0<c_{\sigma_1,\sigma_2,\ldots,\sigma_s}<\infty )  ,
\end{align}
where the coefficients $\{c_{\sigma_1,\sigma_2,\ldots,\sigma_s}\}$ only depend on the fundamental parameters listed in Table~\ref{tab:fund_para}. 
Also, we use the notation of $\tO(x)$ in the following sense:
\begin{align}
\tO(x)= \mathcal{O} \brr{x \cdot {\rm polylog} (x)} . 
\end{align}

\begin{table*}[tt]
 \caption{Fundamental parameters in our statements}
  \label{tab:fund_para} 
\begin{ruledtabular}
\begin{tabular}{lr}
\textrm{\textbf{Definition}}&\textrm{\textbf{Parameters}} 
\\
\colrule
Spatial dimension    
&  $D$ \\
Constant for spatial structure [see Ineq.~\eqref{def:parameter_gamma}]    
& $\gamma$  \\
Interaction strength and decay rate of short-range interactions [see the definition~\eqref{def_short_range_long_range}].
&    $\bar{J}_0$, $\mu$ \\
 Parameters in the Lieb--Robinson bound [see Ineq.~\eqref{Lieb--Robinson_main_short}]
 &   $v$, $C$  
\end{tabular}
\end{ruledtabular}
\end{table*}

\subsection{Convenient lemma for interactions}

\begin{lemma} \label{sum_interaction_terms}
For arbitrary subsets $X$ and $Y$,  we obtain the following upper bound: 
\begin{align}
\label{sum_interaction_terms_main_eq}
\sum_{Z: Z\cap X\neq \emptyset,\ Z\cap Y\neq \emptyset} \norm{h_Z} \le 
|X| \cdot |Y| \bar{J}_{\dist_{X,Y}}.
\end{align}
In particular, by choosing $Y=X[r]^\co$, we have 
\begin{align}
\sum_{Z: Z\cap X\neq \emptyset,\ Z\cap X[r]^\co \neq \emptyset} \norm{h_Z} \le  
 |\partial X| \mathcal{J}_r,\quad  \mathcal{J}_r := 
\tilde{J}_0 e^{-\mu r/2}  
\label{sum_interaction_terms_main_eq_short_long}
\end{align}
with
\begin{align}
\tilde{J}_0:= \frac{\bar{J}_0 \gamma e^{\mu}(2/\mu)^D D!  }{e^{\mu/2}-1},
\end{align}
where we assume the surface subset $(\partial X)_{-s}$ decreases with $s$. Note that we adopt the definition of~\eqref{notation_partial_X_s} for $(\partial X)_{-s}$. 
\end{lemma}

\textit{Proof of Lemma~\ref{sum_interaction_terms}.}
We first use 
\begin{align}
\sum_{Z: Z\cap X\neq \emptyset,\ Z\cap Y\neq \emptyset} \norm{h_Z} \le 
\sum_{i: i\in X} \sum_{i': i'\in Y} \sum_{Z:Z\ni \{i,i'\}} \norm{h_Z}  \le
 \sum_{i: i\in X} \sum_{i': i'\in Y} J_{i,i'}  .
\end{align}
By applying the condition~\eqref{def_short_range_long_range} to the above inequality, we obtain the first main inequality~\eqref{sum_interaction_terms_main_eq}.
We then consider $Y=X[r]^\co$.
We then obtain  
  \begin{align}
  \label{short_rage_case_integral_Ham}
\sum_{i\in X} \sum_{i'\in X[r]^\co} J_{i,i'}   &
\le \bar{J}_0 \sum_{s=0}^\infty \sum_{i\in (\partial X)_{-s}} \sum_{i'\in \Lambda: \dist_{i,i'} > r+s} e^{-\mu \dist_{i,i'}}    .
\end{align}
using the notation of $(\partial X)_s$ in Eq.~\eqref{notation_partial_X_s}.
As in Ref.~\cite[(S.6) in the supplemental]{kuwahara2020absence}, we can derive 
\begin{align}
\label{ineq_sum_decay_func}
\max_i \sum_{i'\in \Lambda: \dist_{i,i'}>r+s}e^{-\mu \dist_{i,i'}} &= \max_i \sum_{l=r+s+1}^\infty \sum_{i'\in \Lambda: \dist_{i,i'}=l}e^{-\mu l} \notag \\
&\le \gamma \sum_{l=r+s+1}^\infty l^{D-1}e^{-\mu l}  \le \gamma \int_{r+s+1}^\infty x^{D-1} e^{-\mu (x-1)}dx  \notag \\
&\le  \gamma  e^{-\mu (r+s)/2} \int_{0}^\infty x^{D-1} e^{-\mu (x-1)/2}dx
= \gamma e^{\mu/2} (2/\mu)^D D! e^{-\mu (r+s)/2} 
\end{align}
using the constant $\gamma$, where we use $\{i'\in \Lambda: \dist_{i,i'}=s\} = \partial i[s]$ and $|\partial i[s]|\le \gamma s^{D-1}$ for $\forall i\in \Lambda$ and $s\ge r+1$. 
Applying the inequality~\eqref{ineq_sum_decay_func} to \eqref{short_rage_case_integral_Ham}, we have 
 \begin{align}
 \label{short_rage_case_integral_Ham_2}
\sum_{i\in X} \sum_{i'\in X[r]^\co} J_{i,i'}   
&\le \bar{J}_0 \gamma e^{\mu/2} (2/\mu)^D D!  \sum_{s=0}^\infty \sum_{i\in (\partial X)_{-s}}e^{-\mu (r+s)/2}    \notag\\
&\le \bar{J}_0 \gamma e^{\mu/2} (2/\mu)^D D!  |\partial X| \frac{1}{1-e^{-\mu/2}}e^{-\mu r/2} ,
\end{align}
which gives the main inequality~\eqref{sum_interaction_terms_main_eq_short_long}. 
This completes the proof of Lemma~\ref{sum_interaction_terms}. $\square$

{~} \\

{\bf Remark.}
We can refine the upper bound~\eqref{sum_interaction_terms_main_eq_short_long} as 
\begin{align}
\sum_{Z: Z\cap X\neq \emptyset,\ Z\cap X[r]^\co \neq \emptyset} \norm{h_Z} \le  
 |\partial X| \bar{J}_0 \gamma(2D/\mu^2)^{D} (r+1)^{D-1} e^{-\mu r}, 
\label{sum_interaction_terms_main_eq_short_long_refine}
\end{align}
but we mainly utilize the upper bound~\eqref{sum_interaction_terms_main_eq_short_long} for simplicity. 
In the inequality~\eqref{ineq_sum_decay_func}, we can re-estimate
\begin{align}
\label{ineq_sum_decay_func___2}
\max_i \sum_{i'\in \Lambda: \dist_{i,i'}>r+s}e^{-\mu \dist_{i,i'}}\le \gamma \int_{r+s+1}^\infty x^{D-1} e^{-\mu (x-1)}dx  \le  \gamma (1/\mu)^D (r+s+D)^{D-1} e^{-\mu (r+s)} ,
\end{align}
where we use 
\begin{align}
\int_{z}^\infty x^{m} e^{-\mu x}dx \le (z+m)^m (1/\mu)^{m+1} e^{-\mu z}.
\end{align}
Applying the inequality~\eqref{ineq_sum_decay_func___2} to \eqref{short_rage_case_integral_Ham}, we have 
 \begin{align}
 \label{short_rage_case_integral_Ham_2//2}
\sum_{i\in X} \sum_{i'\in X[r]^\co} J_{i,i'}   
&\le \bar{J}_0 \gamma \sum_{s=0}^\infty \sum_{i\in (\partial X)_{-s}} (r+s+D)^{D-1} e^{-\mu (r+s)}    \le \bar{J}_0 \gamma(1/\mu)^D |\partial X| e^{-\mu r} \sum_{s=0}^\infty (r+s+D)^{D-1} e^{-\mu s} \notag \\
&\le \bar{J}_0 \gamma(1/\mu)^D |\partial X| e^{-\mu r} \int_{r+D}^\infty x^{D-1} e^{-\mu (x-1)} dx 
\le  \bar{J}_0 \gamma(1/\mu)^{2D} |\partial X| (r+2D-1)^{D-1} e^{-\mu (r+D-1)} \notag \\
&\le \bar{J}_0 \gamma(2D/\mu^2)^{D} |\partial X| (r+1)^{D-1} e^{-\mu r},
\end{align}
which gives the improved inequality~\eqref{sum_interaction_terms_main_eq_short_long_refine}.

%
%

\subsection{Conditional mutual information and Quantum Markov structure} \label{sec:Conditional mutual information and Quantum Markov conjecture}

In this subsection, we introduce the quantum conditional mutual information $\mI_{\rho_{ABC}}(A:C|B)$ and the related conjecture on it.
First, the definition of the conditional mutual information between the subsets $A$ and $C$ conditioned on the subset $B$ is defined as follows:
\begin{align}
\mI_{\rho}(A:C|B) := &S_{\rho} (AC) +S_{\rho}(BC)-S_{\rho}(ABC)  -S_{\rho}(B) ,  \label{def:cond_info}
\end{align}
where we define $S_\rho(L)$ for $\forall L\subseteq \Lambda$ as the von Neumann entropy for the reduced density matrix on the subset $L$:
\begin{align}
S_\rho(L):=- \tr \brr{\rho_L \log(\rho_L)}.
\end{align}
Note that we have referred to the definition~\eqref{def_rho_L_H^ast_L}.

The conditional mutual information is deeply connected to the quantum Markov property.
For an arbitrary density matrix $\rho$, the quantum Markov property implies the following equation for an arbitrary tripartition of the total system ($\Lambda=A\sqcup B \sqcup C$):
 \begin{align}
 \label{Quantum_Markov_st_def}
&\mI_{\rho}(A:C|B) = 0 \for \dist_{A,B} \ge r_0 ,
\end{align}
where $r_0$ is a constant of $\orderof{1}$. 
That is, when the subset $A$ (or $C$) is shielded by the subset $B$, the two subsets are conditionally independent. 
The quantum Markov property is known to have a proper operational meaning regarding the recovery map~\cite{sutter2018approximate}. 
The existence of the recovery map is utilized in improving the entanglement clustering in one-dimensional systems (see Proposition~\ref{corol:ent_F_one}). 

The Hammersley--Clifford theorem gives an interesting relation between the quantum Markov structure and the quantum Gibbs state.
It says that for the classical probability distribution, the quantum Markov structure and the classical Ising model are equivalent; that is, 
any probability distribution with the Markov structure~\eqref{Quantum_Markov_st_def} is expressed in the form of the Gibbs state by the classical Ising Hamiltonian. Conversely, any Gibbs state by the classical Ising Hamiltonian has the Markov structure. 
On the other hand, it has been a long-standing open question whether there exists a similar relationship between the quantum Markov structure and the quantum Gibbs states. 

As a partial solution, when the Hamiltonian is short-range and commuting, the above Markov property strictly holds for quantum Gibbs states at arbitrary temperatures~\cite{brown2012quantum,Jouneghani2014}. 
On the other hand, when the Hamiltonian is non-commuting, the quantum Markov property~\eqref{Quantum_Markov_st_def} breaks down in the exact sense. 
Still, even for non-commuting Hamiltonians, it has been conjectured~\cite{Brandao2019,kato2016quantum} that the quantum Markov property generally holds in an approximate sense:
\begin{conj} \label{conj:Quantum_CMI}
For arbitrary quantum Gibbs states, the conditional mutual information $\mI_{\rho_\beta}(A:C|B)$ ($\Lambda=A\sqcup B \sqcup C$) rapidly decays with the distance between $A$ and $C$:
 \begin{align}
 \label{quantum_Markov_Conj_ineq_supp}
&\mI_{\rho_\beta}(A:C|B) \le  \mG_\mI(R)  ,\quad R=\dist_{A,C} , 
\end{align}
where $ \mG_\mI(R) $ is a super-polynomially decaying function which depend on $\beta$, $\{A,B,C\}$ and fundamental parameters as in Table~\ref{tab:fund_para}.  
\end{conj}

This approximate version of the quantum Markov property also possesses a similar operational meaning to the exact Markov property~\cite{Fawzi2015}, and it plays a critical role in preparing the quantum Gibbs states on a quantum computer~\cite{Kastoryano2016,kim2017markovian,Brandao2019}.
In one-dimensional cases, the conjecture has been proved at arbitrary $\orderof{1}$ temperatures~\cite{kato2016quantum}, where the decay rate is given by a subexponential form of $e^{e^{-\orderof{\beta}}\orderof{R^{1/2}}}$.
In dimensions greater than 1, Conjecture~\ref{conj:Quantum_CMI} is known to be true only in high-temperature regimes, i.e., above a threshold temperature~\cite{PhysRevLett.124.220601}.
The primary goal of this paper is to prove the conjecture in arbitrary temperature regimes. 

\subsubsection{Relation to entanglement clustering}

We also mention a connection between the quantum Markov structure and the bipartite entanglement.
As has been mentioned in Ref.~\cite{PhysRevX.12.021022}, the solution of the quantum Markov conjecture immediately includes the exponential clustering of the quantum squashed entanglement. 
For an arbitrary quantum state $\rho_{AB}$ that is composed of the systems $A$ and $B$, the squashed entanglement is defined as follows:
\begin{align}
\label{Squashed_entanglement_def}
&E_{\rm sq}(\rho_{AB}) := \inf_E \left\{\frac{1}{2}\mI_{\rho_{ABE}}(A:B|E) \biggl |  \tr_E (\rho_{ABE}) = \rho_{AB}  \right\} ,
\end{align}
where $\inf_E$ is taken over all extensions of $\rho_{AB}$ such that $\tr_E (\rho_{ABE}) = \rho_{AB}$. 
As one of the convenient properties, the squashed entanglement satisfies the faithfulness; that is, it is equal to zero if and only if the quantum state is not entangled~\cite{Brandao2011,Li2014}. 

Under the assumption that the inequality~\eqref{quantum_Markov_Conj_ineq_supp} holds, we can derive the exponential clustering for the squashed entanglement in the following way:
\begin{align}
\label{Sq_exp_decay_coj}
E_{\rm sq}(\rho_{\beta, AB})  &\le \frac{1}{2} \mI_{\rho_\beta}(A:B|C) \notag \\
&\le \frac{1}{2} \mG_\mI(R)   . 
\end{align}
Now, the ancilla system $E$ is chosen as the residual system to $AB$, i.e., $E\to C= \Lambda\setminus (AB)$.

To utilize the quantum squashed entanglement to upper-bound other entanglement measures, we use the quantity $\delta_{\rho_{AB}}$, which characterizes the minimum distance between the target state and separable (i.e., non-entangled) states:
\begin{align}
\label{minimum_distance_SEP}
\delta_{\rho_{AB}} := \inf_{\sigma_{AB} \in {\rm SEP}(A:B) } \norm{ \rho_{AB} - \sigma_{AB}}_1 ,
\end{align}
where ${\rm SEP}(A:B)$ is a set of quantum states defined on the subsets $A$ and $B$ that have no entanglement. 
One can derive the following upper bound for $\delta_{\rho_{AB}}$ in terms of the squashed entanglement as shown in Ref.~\cite[Corollary~3.13]{lami2018nonclassical} (see also Ref.~\cite{berta2023entanglement}): 
\begin{align}
\label{minimum_distance_SEP_squash_relation}
\delta_{\rho_{AB}} \le 4 \min(\mathcal{D}_A,\mathcal{D}_B) \sqrt{E_{\rm sq}(\rho_{AB})  }  \le 2 \min(\mathcal{D}_A,\mathcal{D}_B) \sqrt{2\mI_{\rho_\beta}(A:B|E)  },
\end{align}
where $\mathcal{D}_{A}$ and $\mathcal{D}_{B}$ are the dimensions of the Hilbert spaces of $A$ and $B$, respectively. 
Therefore, if $E_{\rm sq}(\rho_{AB})  \ll 1/\mathcal{D}_{AB}$, we can conclude that $\delta_{\rho_{AB}}$ is as small as $\mG_\mI(R)$ from~\eqref{Sq_exp_decay_coj}.
In the case where $|AB|$ is macroscopically large (i.e., $|AB|=\orderof{|\Lambda|}$),  
$\mathcal{D}_{AB}$ is exponentially large with $|\Lambda|$, and hence the inequality~\eqref{Sq_exp_decay_coj} cannot be used to provide an upper bound on other entanglement measures such as the relative entanglement entropy and the EoF. 
In the case of one-dimensional systems, we can overcome the problem by employing the prescription introduced in Ref.~\cite[Theorem~12]{PhysRevX.12.021022} (see also Sec.~\ref{sec:Relation to other information measures}).

\subsection{Upper bound on the norm of commutators}

In this section, we show several convenient lemmas on the norm of commutators that are often used in our analysis. 

\begin{lemma} \label{lemma_commutator_generalization}
Let $O$ be an arbitrary operator. Also, for an arbitrary unitary operator $u_i$ acting on a site $i\in \Lambda$, we assume the following inequality:
\begin{align}
\label{lemma_commutator_generalization/assumption}
\sup_{u_i} \norm{[O,u_i]} \le \delta_i ,
\end{align} 
where $\sup_{u_i}$ is taken for the set of all the unitary operators acting on the site $i$. 
Then, we obtain 
\begin{align}
\label{first_ineq:lemma_commutator_generalization}
\norm{O - \tilde{\tr} (O)} \le \sum _{i\in \Lambda} \delta_i. 
\end{align} 
Also, for an arbitrary operator $O_X$ which is supported on subset $X\subset \Lambda$, we obtain 
\begin{align}
\label{sec_ineq:lemma_commutator_generalization}
\norm{[O,O_X]} \le 2\norm{O_X} \sum_{i\in X} \delta_i .
\end{align} 
\end{lemma}

{\bf Remark.} The inequality~\eqref{first_ineq:lemma_commutator_generalization} gives upper bound for the norm of $\norm{O}$ when $O$ is traceless (i.e., $\tr(O)=0$):
\begin{align}
\label{first_ineq:lemma_commutator_generalization_tr=0}
\norm{O} \le \sum _{i\in \Lambda} \delta_i  \for  \tr(O)=0.
\end{align} 
Also, when the initial condition is generalized as 
\begin{align}
\sup_{u_X} \norm{[O,u_X]} \le \delta_X ,
\end{align} 
one can derive 
\begin{align}
\label{sec_ineq:lemma_commutator_generalization_2}
\norm{[O,O_X]} \le 2\norm{O_X} \delta_X .
\end{align} 
In the above inequality, we upper-bound $\delta_X$ by $\sum _{i\in \Lambda} \delta_i$ when we begin with the assumption~\eqref{lemma_commutator_generalization/assumption}.

{~}\\
\textit{Proof of Lemma~\ref{lemma_commutator_generalization}.}
For the proof, we consider
\begin{align}
\tilde{\tr}_{X}(O) = \frac{\hat{1}_X \otimes  \tr_{X}(O)}{\tr_X(\hat{1})} ,
\end{align}
where we use the definition~\eqref{definition_of_tilde_tr_partial}. 
Note that $\tilde{\tr}_{X}(O)$ is supported on $X^\co$, and hence $[\tilde{\tr}_{X}(O) ,O_X]=0$. 
We now label the sites in $X$ by $X=\{1,2,\ldots,|X|\}$ without loss of generality. 
Using the discussion in Ref.~\cite{PhysRevLett.97.050401}, we have
\begin{align}
\tilde{\tr}_{X}(O) :=\int d\mu(u_{1}) d\mu(u_{2}) \cdots d\mu(u_{|X|}) (u_1 u_2\cdots u_{|X|})^\dagger O  (u_1u_2 \cdots u_{|X|}) ,
\end{align}
and hence we obtain 
\begin{align}
\norm{ O-\tilde{\tr}_{X}(O)}& \le \int d\mu(u_{1}) d\mu(u_{2}) \cdots d\mu(u_{|X|}) \sum_{i\in X}
\norm{ [O, u_i ]} \le   \sum _{i\in X} \sup_{u_i} \norm{ [O, u_i ]}\le   \sum_{i\in X} \delta_i    ,
\label{basic_ansatz_m_general_X_start}
\end{align}
where $\mu(u_i)$ is the Haar measure for the unitary operator on the site $i\in \Lambda$.
Therefore, by choosing $X=\Lambda$, we obtain the first main inequality~\eqref{first_ineq:lemma_commutator_generalization}. 
Also, by using the inequality of  
\begin{align}
\norm{ [  O , O_X] } = \norm{ [ O-\tilde{\tr}_{X}(O), O_X] } \le 2\norm{O_X}\cdot  \norm{ O-\tilde{\tr}_{X}(O)},
\end{align}
we obtain the second main inequality~\eqref{sec_ineq:lemma_commutator_generalization}.
This completes the proof of Lemma~\ref{lemma_commutator_generalization}. $\square$

{~}\\

\begin{lemma} \label{norm_local_approx}
Let $O_X$ be an arbitrary operator supported on a subset $X\subset \Lambda$. 
If $O_X(t)$ satisfies the Lieb--Robinson bound as 
\begin{align}
\label{Lieb_Robinson_assump_O_X_O_Y_t}
\norm{[O_X(t),O_Y]} \le \mathcal{G}(X,Y,t,\dist_{X,Y})
\end{align}
with $\mathcal{G}(X,Y,t,\dist_{X,Y})$ an appropriate function, 
we have 
\begin{align}
\label{main_ineq1:norm_local_approx}
\norm{O_X(t)- O_{X[r]}^{(t)}} \le \mathcal{G}(X,X[r]^\co,t,r+1), \quad   O_{X[r]}^{(t)}:= \tilde{\tr}_{X[r]^\co}\brr{O_X(t)} .
\end{align}
Similarly, we also obtain 
\begin{align}
\label{main_ineq2:norm_local_approx}
\norm{O_{X[r+\delta r]}^{(t)} - O_{X[r]}^{(t)}  } \le \mathcal{G}(X,X[r+\delta r]\setminus X[r],t,r+1) , 
\end{align}
where the integers $r$ and $\delta r$ can be arbitrarily chosen.
\end{lemma} 
{~}\\
\textit{Proof of Lemma~\ref{norm_local_approx}.}
The inequality~\eqref{main_ineq1:norm_local_approx} is immediately proved by using 
\begin{align}
\tilde{\tr}_{X[r]^\co}\brr{O_X(t)} 
=\int d\mu(U_{X[r]^\co})  U_{X[r]^\co}^\dagger O_X(t) U_{X[r]^\co} ,
\end{align}
which yields 
\begin{align}
\norm{O_X(t)- \tilde{\tr}_{X[r]^\co}\brr{O_X(t)}} \le 
\norm{ \brr{ O_X(t) ,U_{X[r]^\co}  } } \le \mathcal{G}(X,X[r]^\co,t,r) 
\label{local_approxiamtion_O_X_X_brr_r}
\end{align}
from the inequality~\eqref{Lieb_Robinson_assump_O_X_O_Y_t} and $\dist_{X,X[r]^\co}=r+1$.

We then let $X'=X[r+\delta r]\setminus X[r]$. 
The second inequality~\eqref{main_ineq2:norm_local_approx} is derived in the same way as follows:
\begin{align}
\norm{O_{X[r+\delta r]}^{(t)} - O_{X[r]}^{(t)}  } 
&=\norm{ \tilde{\tr}_{X[r+\delta r]^\co}\brr{O_X(t)} - \tilde{\tr}_{X[r]^\co}\brr{O_X(t)}}  \notag\\
&=\norm{\int d\mu(U_{X[r+\delta r]^\co})  
U_{X[r+\delta r]^\co}^\dagger \br{ O_X(t) -  \int d\mu(U_{X'})  U_{X'}^\dagger O_X(t)  U_{X'} }  U_{X[r+\delta r]^\co}} \notag \\
&\le  \norm{ O_X(t) - \int d\mu(U_{X'})  U_{X'}^\dagger O_X(t)U_{X'}} \notag \\
&\le  \int d\mu(U_{X'}) \norm{\brr{ O_X(t) , U_{X'} }} \le  \mathcal{G}(X,X',t,r+1)  .
\label{Proof_norm_local_approx_fin}
\end{align}
This completes the proof of Lemma~\ref{norm_local_approx}. $\square$

\subsection{Lieb--Robinson bound and locality analyses}

Here, we introduce the Lieb--Robinson bound, which characterizes the quasi-locality of interactions under time evolution.
It plays a central role in our analyses alongside the quantum belief propagation. 
One can prove the following statement in general~\cite{ref:LR-bound72,ref:Hastings2006-ExpDec,Nachtergaele2006,ref:Nachtergaele2006-LR}:
\begin{lemma}[Lieb--Robinson bound~\cite{ref:Nachtergaele2006-LR}] \label{Lieb--Robinson_lemma_corol}
For arbitrary operators $O_X$ and $O_Y$ with unit norm and $\dist_{X,Y}=R$, the norm of the commutator $[O_X(t), O_Y]$ satisfies the following inequality:
\begin{align}
\label{Lieb--Robinson_main_short}
\norm{ [O_X(t), O_Y]} \le  C \min(|\partial X|,|\partial Y|) \br{e^{v|t|}-1} e^{-\mu R } ,
\end{align}
where we have adopted $C$ and $v$ as fundamental parameters in Table~\ref{tab:fund_para}, but they can be expressed using the parameters, i.e., $D$, $\gamma$, $\bar{J}_0$ and $\mu$.   
\end{lemma}

\noindent
{\bf Remark.} 
Originally derived inequality in Ref.~\cite{ref:Nachtergaele2006-LR} is slightly weaker than~\eqref{Lieb--Robinson_main_short}:  
\begin{align}
\label{Lieb--Robinson_main_short_i,j}
\norm{ [O_X(t), O_Y]} \le C_0 \min(|X|,|Y|)  \br{e^{v|t|}-1} e^{-\mu R } .
\end{align}
Using the inequality~\eqref{sec_ineq:lemma_commutator_generalization} in Lemma~\ref{lemma_commutator_generalization}, we improve the bound~\eqref{Lieb--Robinson_main_short_i,j}  with $O\to O_X(t)$, $O_X \to O_Y$ and $\delta_i \to C \br{e^{v|t|}-1} e^{-\mu \dist_{i,X}}$, we have 
\begin{align}
\label{Lieb--Robinson_main_short_i,j_XY}
\norm{ [O_X(t), O_Y]} \le  C \br{e^{v|t|}-1} \sum_{i\in Y} e^{-\mu \dist_{i,X}} ,
\end{align}
and 
\begin{align}
\label{Lieb--Robinson_main_short_i,j_XY/sum}
\sum_{i\in Y} e^{-\mu \dist_{i,X}} \le \sum_{s=0}^\infty \sum_{i\in (\partial Y)_{-s}} e^{-\mu (\dist_{X,Y}+s)} 
\le  |\partial Y| e^{-\mu R } \sum_{s=0}^\infty  e^{-\mu s}
=\frac{|\partial Y| e^{-\mu R }}{1-e^{-\mu}} . 
\end{align}
Combining~\eqref{Lieb--Robinson_main_short_i,j_XY} and \eqref{Lieb--Robinson_main_short_i,j_XY/sum} yields the inequality~\eqref{Lieb--Robinson_main_short} by letting $C=C_0/(1-e^{-\mu})$.

We define $O_X(H,t,\tilde{X})$ as an approximation of $O_X(H,t)$ on the subset $\tilde{X}$, 
\begin{align}
O_X(H,t,\tilde{X}) := \frac{1}{\tr_{\tilde{X}^\co}(\hat{1})} \tr_{\tilde{X}^\co} \left[O_X(H,t)\right] \otimes \hat{1}_{\tilde{X}^\co},
\label{supp_def:W_X_local_approx}
\end{align}
where $\tr_{\tilde{X}^\co}(\cdots)$ is the partial trace with respect to the subset $\tilde{X}^\co$.
We then obtain from Lemma~\ref{norm_local_approx}
\begin{align}
\label{local_approximation_Lieb_Robinson}
\norm{ O_X(H,t)  - O_X(H,t,X[r]) } \le C \|O_X\|\cdot |\partial X| \br{e^{v|t|}-1} e^{-\mu r}, 
\end{align}
where the notation $X[r]$ has been defined in Eq.~\eqref{def:bal_X_r}.

Throughout the proof, we consider the Lieb--Robinson bound for short-range interacting (or with exponentially decaying interactions) systems, but the Lieb--Robinson bound holds for more general classes of Hamiltonians with power-law decaying (i.e., long-range) interactions~\cite{PhysRevLett.114.157201,chen2019finite,PhysRevX.10.031010,PhysRevLett.127.160401}.  
As long as the Lieb--Robinson bound holds, we expect that our main results will be generalized to the long-range interacting systems.

\subsection{Commutator analysis of different quasi-local operators}

\begin{figure}[tt]
\centering
\includegraphics[clip, scale=0.4]{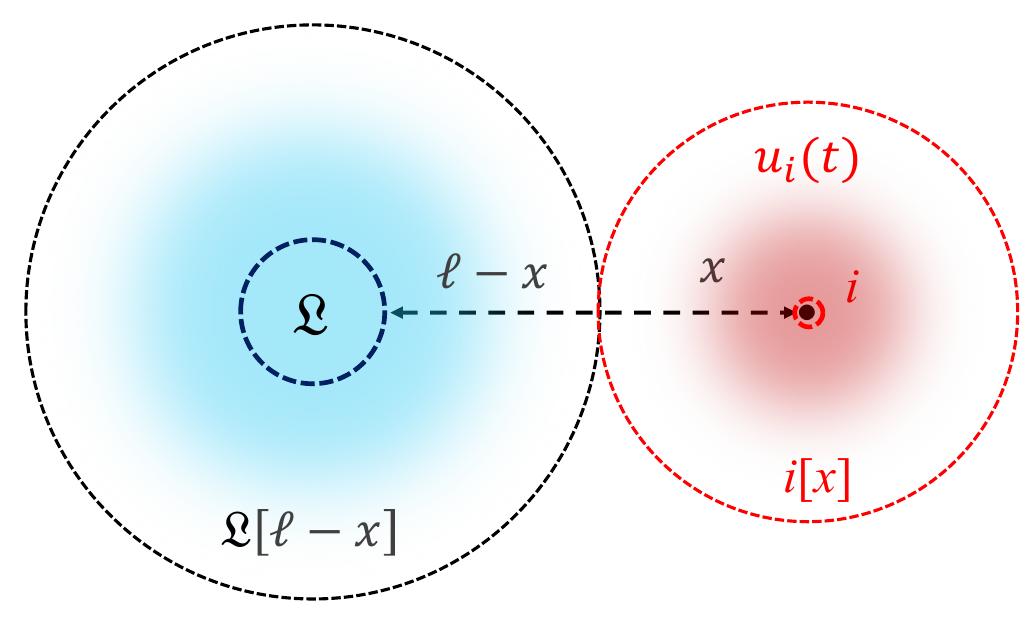}
\caption{Schematic picture of the setup of Lemma~\ref{lem:Quasi-local_commutator_bound}. 
We consider two quasi-local operators; the former one is $O$ that satisfies the quasi-locality described by~\eqref{lem:Quasi-local_commutator_bound_cond}, while the latter one is given by $u_i(t)$ [or $u_i(\lambda t)$ with $0\le \lambda\le 1$]. 
The commutator between $O$ and $\int_{-\infty}^\infty f(t) u_i(\lambda t) dt$ is upper-bounded as in~\eqref{main_ineq:lem:Quasi-local_commutator_bound}. 
In the proof, we decompose the operator $u_i(\lambda t)$ as in Eq.~\eqref{decomp_u_i_lambda_t_0} and take the commutator with each of the decomposed terms as in~\eqref{ineq:U_tau_decomp_s}. 
}
\label{fig:Fig_commute}
\end{figure}

In this section, we address the following question. 
Consider two quasi-local operators, $O_1$ and $O_2$. 
A fundamental problem is to estimate the degree of quasi-locality of their commutator, $\brr{O_1,O_2}$. 
Such an analysis of locality for distinct quasi-local operators plays a central role in our study. 
Related techniques have also been developed in the context of the Lieb--Robinson bound~\cite{PhysRevX.10.031010,kuwahara2020absence}.
The following lemma characterizes the commutator norm of $\int_{-\infty}^{-\infty} f(t)  \norm{\brr{O, u_i(t)}} dt$ (see also Fig.~\ref{fig:Fig_commute}), where $u_i(t)$ is characterized by the time evolution for a unitary operator at the site $i\in \Lambda$ and $f(t)$ is a filter function. 
This also characterizes the quasi-locality of the operator $ \int_{-\infty}^{-\infty} f(t) O(-t) dt$.

\begin{lemma} \label{lem:Quasi-local_commutator_bound}
Let $O$ be a quasi-local operator around a subset $\mfL$ in the sense that
\begin{align}
\label{lem:Quasi-local_commutator_bound_cond}
\norm{[O,u_i]} \le  \mF(\ell) ,\quad \ell=\dist_{i,\mfL} 
\end{align}
for an arbitrary unitary operator on a site $i$, where $\mF(\ell) $ is a monotonically decaying function. Then, under the Lieb--Robinson bound in short-range interacting systems, we obtain
\begin{align}
\label{main_ineq:lem:Quasi-local_commutator_bound}
\int_{-\infty}^\infty f(t) \norm{[O,u_i(\lambda t)]}dt \le  2 f_1 \mathcal{F}(\ell)  
+2\gamma(\Delta \ell)^D\sum_{s=1}^\infty s^D\mathcal{F}(\ell-s\Delta \ell) \brr{ C f_2 v e^{-\mu s \Delta \ell/2} + f_{t_0}(s)}
\end{align}
for $0\le \lambda \le 1$, where we define
\begin{align}
\label{definition_lem:Quasi-local_commutator_bound}
f_{t_0}(s):= \int_{|t| \ge t_0} f(t)dt ,\quad f_1:=\int_{-\infty}^\infty f(t) dt,\quad f_2:=\int_{|t|\le t_0}|t|f(t)dt\le\int_{-\infty}^\infty|t| f(t) dt,
\end{align}
where $t_0=\max(0,\mu s \Delta \ell/(2v) - \mu (\Delta \ell-1)/v)$, and $ f(t)$ is an arbitrary positive function. 
In particular, by taking $\Delta \ell=1$, the inequality~\eqref{main_ineq:lem:Quasi-local_commutator_bound} reduces to 
\begin{align}
\label{main_ineq:lem:Quasi-local_commutator_bound_Delta_eLL=1}
\int_{-\infty}^\infty f(t) \norm{[O,u_i(\lambda t)]}dt \le 2f_1 \mathcal{F}(\ell)  
+2\gamma \sum_{s=1}^\infty s^D\mathcal{F}(\ell-s) \brr{ C f_2 v e^{-\mu s/2} + f_{t_0}(s)} ,
\end{align}
where $t_0=\mu s/(2v)$. 
\end{lemma}

{\bf Remark.} 
The lemma is utilized in Propositions~\ref{prop:Quasi-local_commutator_bound_tau} and \ref{prop:Quasi-local_commutator_bound_tau_2} for example, where we let the filter function as the function $g_\beta(t)$ in the quantum belief propagation~\eqref{sup_Def:Phi_0_phi_Hastings}, which plays an essential role in Lemma~\ref{connection_of_exponential_operator} for connection of exponential operators.

{~}\\
\textit{Proof of Lemma~\ref{lem:Quasi-local_commutator_bound}.} 
We first decompose 
\begin{align}
\label{decomp_u_i_lambda_t_0}
u_{i}(\lambda t) =u_{i}^{(t)} + \sum_{s=1}^\infty  \br{ u_{i[\ell_s]}^{(t)}  - u_{i[\ell_{s-1}]}^{(t)} } ,\quad u_{i[\ell_s]}^{(t)}:=  \tilde{\tr}_{i[\ell_s]^\co}\brr{u_i(\lambda t)},
\end{align}
where we use $u_{i[\infty]}^{(t)}= u_{i}(\lambda t)$ and define $\ell_s$ as 
\begin{align}
\ell_s := s \Delta \ell. 
\end{align}
Using the inequality~\eqref{sec_ineq:lemma_commutator_generalization} in Lemma~\ref{lemma_commutator_generalization}, we obtain 
\begin{align}
\label{ineq:U_tau_decomp_s}
\norm{\brr{O,  u_{i[\ell_s]}^{(t)}  - u_{i[\ell_{s-1}]}^{(t)} }} 
&\le 2 \norm{u_{i[\ell_s]}^{(t)}  - u_{i[\ell_{s-1}]}^{(t)} } \sum_{i'\in i[\ell_s]} \sup_{u_{i'}} \norm{[O,u_{i'}]}  \notag \\
& \le  2\norm{u_{i[\ell_s]}^{(t)}  - u_{i[\ell_{s-1}]}^{(t)} } \cdot |i[\ell_s]|\mathcal{F}(\ell -\ell_s)  \notag \\
& \le  2\gamma \ell_s^D \mathcal{F}(\ell -\ell_s) \norm{u_{i[\ell_s]}^{(t)}  - u_{i[\ell_{s-1}]}^{(t)} } ,
\end{align}
where we use $\dist_{i',\mfL}\ge \dist_{i,\mfL} -\ell_s = \ell -\ell_s$ for $i'\in i[\ell_s]$ and $\dist_{i,\mfL} =\ell$.
In the same way, we have 
\begin{align}
\label{ineq:U_tau_decomp_s__2}
\norm{\brr{O, u_{i}^{(t)}  }} 
&\le 2 \norm{ u_{i}^{(t)}  } \mathcal{F}(\ell) =  2  \mathcal{F}(\ell),
\end{align}
where we use $\norm{\tilde{\tr}_{i^\co}\brr{u_i(\lambda t)}} \le \norm{u_i(\lambda t)}=1$. 

Moreover, by combining the Lieb--Robinson bound and the inequality~\eqref{main_ineq2:norm_local_approx} in Lemma~\ref{norm_local_approx}, we have 
 \begin{align}
 \label{ineq:u_is_norm}
\norm{u_{i[\ell_s]}^{(t)}  - u_{i[\ell_{s-1}]}^{(t)} }
\le\min\brr{1,C\br{e^{v|\lambda t|} -1} e^{-\mu (\ell_{s-1}+1)}} \le \min\brr{1,C\br{e^{v|t|} -1}e^{-\mu(\ell_s-\Delta \ell+1) }},
\end{align}
where we use $0\le \lambda \le1$.
Combining the inequalities~\eqref{ineq:U_tau_decomp_s} and \eqref{ineq:u_is_norm}, we obtain
\begin{align}
\label{ineq:U_tau_decomp_s_reduce}
\norm{\brr{O,  u_{i[\ell_s]}^{(t)}  - u_{i[\ell_{s-1}]}^{(t)}  }} 
&\le  2\gamma \ell_s^D \mathcal{F}(\ell -\ell_s)  \min\brr{1,C\br{e^{v|t|} -1} e^{-\mu(\ell_s-\Delta \ell+1) }} .
\end{align}
Therefore, we can derive for $\norm{\brr{O,  u_i(t) }}$
\begin{align}
\norm{\brr{O,  u_i( \lambda t) }}
\le 2\mathcal{F}(\ell) +2\gamma \sum_{s=1}^\infty \ell_s^D\mathcal{F}(\ell-\ell_s) \min\brr{1,C\br{e^{v|t|} -1}e^{-\mu(\ell_s-\Delta \ell+1) } } ,
\label{norm_O_u_i_t_clear_form}
\end{align}
whose integral then reads 
\begin{align}
\label{integral_commutator_O_u_it}
&\int_{-\infty}^\infty f(t) \norm{\brr{O,  u_i( \lambda t) }} dt \notag \\
&\le 2\mathcal{F}(\ell)  \int_{-\infty}^\infty  f(t)dt  +2\gamma \sum_{s=1}^\infty \ell_s^D\mathcal{F}(\ell-\ell_s) \int_{-\infty}^\infty f(t)\min\brr{1,C\br{e^{v|t|} -1}e^{-\mu(\ell_s-\Delta \ell+1) } } dt . 
\end{align}

Let us set $t_0=\max\br{ 0, \mu \ell_s/(2v) - \mu (\Delta \ell-1)/v}$ and define $f_{t_0}(s)$ as in Eq.~\eqref{definition_lem:Quasi-local_commutator_bound}.
We then obtain
\begin{align}
\label{integral_ft_ge_delta_t}
\int_{-\infty}^\infty f(t)\min\brr{1,C\br{e^{v|t|} -1}e^{-\mu(\ell_s-\Delta \ell+1) } } dt 
&\le  \int_{|t|\le t_0}  f(t) C\br{e^{v|t|} -1}e^{-\mu(\ell_s-\Delta \ell+1) }   dt + \int_{|t| \ge t_0} f(t)dt  \notag \\
&\le Cv e^{-\mu \ell_s/2}   \int_{|t|\le t_0} |t| f(t) dt + f_{t_0}(s)  \notag \\
&\le Cv e^{-\mu \ell_s/2}   \int_{|t|\le t_0} |t| f(t) dt + f_{t_0}(s)  ,
\end{align}
where we use $e^{v|t|} -1\le v|t| e^{v|t|} \le v|t| e^{\mu \ell_s/2 + \mu (-\Delta \ell+1)}$ for $|t|\le t_0$.

By applying the inequality~\eqref{integral_ft_ge_delta_t} to Eq.~\eqref{integral_commutator_O_u_it}, we obtain
\begin{align}
\label{integral_ft_ge_delta_t_final}
\int_{-\infty}^\infty  f(t) \norm{\brr{O,  u_i(t) }} dt
&\le 2f_1\mathcal{F}(\ell) +2C f_2 v\gamma  \sum_{s=1}^\infty \ell_s^D\mathcal{F}(\ell-\ell_s) e^{-\mu \ell_s/2} 
+2\gamma  \sum_{s=1}^\infty \ell_s^D\mathcal{F}(\ell-\ell_s) f_{t_0}(s),
\end{align}
which reduces to the desired inequality~\eqref{main_ineq:lem:Quasi-local_commutator_bound} by letting $\ell_s=s\Delta \ell$. 
This completes the proof.  $\square$

{~}

\hrulefill{\bf [ End of Proof of Lemma~\ref{lem:Quasi-local_commutator_bound}]}

{~}

We can also prove a similar statement using the same proof technique:
\begin{corol} \label{corol:Quasi-local_commutator_bound}
Under the same setup as in Lemma~\ref{lem:Quasi-local_commutator_bound}, we consider
\begin{align}
\label{eq:corol:Quasi-local_commutator}
\int_{-\infty}^\infty f(t) \int_0^t \norm{[O,u_i(t_1)]} dt_1 dt. 
\end{align}
Then, the same inequality as~\eqref{main_ineq:lem:Quasi-local_commutator_bound} holds for~\eqref{eq:corol:Quasi-local_commutator} by replacing $f(t)$ by $|t|f(t)$, where the definitions of $f_{t_0}(s)$, $f_1$ and $f_2$ are replaced correspondingly. 
\end{corol}
{~}\\
\textit{Proof of Corollary~\ref{corol:Quasi-local_commutator_bound}.}
We start from the inequality~\eqref{norm_O_u_i_t_clear_form}.
By integrating it from $t_1=0$ to $t_1=t$ in \eqref{norm_O_u_i_t_clear_form}, we have 
\begin{align}
 \int_0^t \norm{\brr{O,  u_i(t_1) }}dt_1  
\le2 |t| \mathcal{F}(\ell) +2\gamma \sum_{s=1}^\infty \ell_s^D\mathcal{F}(\ell-\ell_s) \min\brr{|t|,C \br{ \frac{e^{v|t|} -1}{v} -|t|} e^{-\mu(\ell_s-\Delta \ell+1) } } ,
\label{norm_O_u_i_t_clear_form_corol}
\end{align}
Then, the inequality~\eqref{integral_ft_ge_delta_t} is replaced by 
\begin{align}
\label{integral_ft_ge_delta_t__1}
&\int_{-\infty}^\infty f(t)\min\brr{|t|,C \br{ \frac{e^{v|t|} -1}{v} -|t|} e^{-\mu(\ell_s-\Delta \ell+1) } }dt  \notag\\
&\le  \int_{|t|\le t_0}  f(t) C \br{ \frac{e^{v|t|} -1}{v} -|t|} e^{-\mu(\ell_s-\Delta \ell+1) } dt + \int_{|t| \ge t_0} |t| f(t)dt .
\end{align}
Using $e^{v|t|} -1\le v|t| e^{v|t|}$, we have 
\begin{align}
\label{integral_ft_ge_delta_t/2}
\frac{e^{v|t|} -1}{v} -|t| \le  |t|\br{ e^{v|t|}    -1} \le v|t|^2 .  
\end{align}
By integrating \eqref{norm_O_u_i_t_clear_form_corol} by multiplying $f(t)$ and applying the inequalities~\eqref{integral_ft_ge_delta_t__1} and~\eqref{integral_ft_ge_delta_t/2}, 
we obtain the similar inequality to~\eqref{main_ineq:lem:Quasi-local_commutator_bound} with $f(t)$ replaced by $|t|f(t)$.
This completes the proof of Corollary~\ref{corol:Quasi-local_commutator_bound}. $\square$

%
%
%
%

\subsection{1D version}

From the inequality~\eqref{main_ineq:lem:Quasi-local_commutator_bound} in Lemma~\ref{lem:Quasi-local_commutator_bound}, we can roughly estimate 
\begin{align}
\int_{-\infty}^\infty f(t) \norm{[O,u_i(\lambda t)]}dt \sim \Theta(\xi^D) \mathcal{F}(\ell)   ,
\end{align}
if $f(t)$ decays exponentially with $t/\xi$, i.e., $f(t) \sim e^{-\orderof{t/\xi}}$, where we choose $\Delta \ell$ such that $\Delta \ell \propto \xi$. 
Using the dependence, we cannot obtain the best upper bound for the conditional mutual information as in Theorem~\ref{CMI_decay_PTP_general_1D_improve}.  
We will also show the point in Sec.~\ref{sec:Improved bound for 1D cases}. 
We, in the following, derive the improved statement by adopting a slightly different ansatz for the quasi-locality of the operator [see~\eqref{initial_Quasi-local_commutator_bound}]: 

\begin{lemma} \label{lem:Quasi-local_commutator_bound2}
Let $O$ be a quasi-local operator around a subset $\mfL$ in the sense that
\begin{align}
\label{initial_Quasi-local_commutator_bound}
\norm{[O,u_X] } \le  \mF(\ell) ,  \quad \dist_{X,\mfL}=\ell, 
\end{align}
where the unitary operator $u_X$ is arbitrarily chosen, and $\mF(\ell) $ is a monotonically decaying function. Then, under the Lieb--Robinson bound in short-range interacting systems, we obtain
\begin{align}
\int_{-\infty}^\infty f(t) \norm{[O,u_X(t)]} dt\le 2f_1 \mathcal{F}(\ell)  +2\sum_{s=1}^\infty \mathcal{F}(\ell-s\Delta \ell) \brr{ C|\partial (\mfL[\ell])| f_2 v e^{-\mu s \Delta \ell /2} + f_{t_0}(s)}
\label{main_ineq:lem:Quasi-local_commutator_bound2}
\end{align}
for an arbitrary choice of $\Delta \ell$,
where we adopt the same definitions for $f_1$, $f_2$ and $f_{t_0}(s)$ as in Eq.~\eqref{definition_lem:Quasi-local_commutator_bound} with $t_0=\max(0,\mu s \Delta \ell/(2v) - \mu (\Delta \ell-1)/v)$. 
\end{lemma}

{\bf Remark.} We note that the condition~\eqref{initial_Quasi-local_commutator_bound} implies the local approximation of 
\begin{align}
\norm{O -\tilde{\tr}_{\mfL[\ell-1]^\co}(O) } \le  \mF(\ell) ,
\end{align}
where we use the upper bound \eqref{main_ineq1:norm_local_approx}.
The term $|\partial (\mfL[\ell])|$ scales as $\ell^{D-1}$ concerning $\ell$, and hence, in dimensions larger than 1, we have an upper bound like $\orderof{\ell^{D-1}}\mathcal{F}(\ell)$ in~\eqref{main_ineq:lem:Quasi-local_commutator_bound2}, while the inequality~\eqref{main_ineq:lem:Quasi-local_commutator_bound} gives an upper bound of $\orderof{\ell^{0}}\mathcal{F}(\ell)$. 
Hence, the upper bound~\eqref{main_ineq:lem:Quasi-local_commutator_bound2} yields a rather weaker bound in high-dimensional systems.
However, in the 1D case, this lemma leads to a qualitatively better estimation of the correlation length of the conditional mutual information (see Sec.~\ref{sec:Improved bound for 1D cases}).


{~}\\
\textit{Proof of Lemma~\ref{lem:Quasi-local_commutator_bound2}.} 
We choose $X=\mfL[\ell-1]^\co$ since any unitary operator $u_X$ with $\dist_{X,\mfL}=\ell$ can be expressed in the form of $u_{\mfL[\ell-1]^\co}$.
For simplicity of notation, we here introduce $X_s$ as 
\begin{align}
\label{def_X_s_lemm}
X_s = \mfL[\ell-1-\ell_s]^\co ,\quad \ell_s=s \Delta \ell,
\end{align}
where $X=\mfL[\ell-1]^\co=X_0$. We then decompose 
\begin{align}
u_{X_0}(t) =u_{X_0}^{(t)} + \sum_{s=1}^\infty  \br{ u_{X_s}^{(t)}  - u_{X_{s-1}}^{(t)} } ,\quad u_{X_s}^{(t)}:=  \tilde{\tr}_{X_s^\co}\brr{u_X(t)}.
\end{align}

Using the initial condition~\eqref{initial_Quasi-local_commutator_bound}, we obtain 
\begin{align}
\label{ineq:U_tau_decomp_s__1D}
\norm{\brr{O,  u_{X_s}^{(t)}  - u_{X_{s-1}}^{(t)}}} 
&\le 2\mathcal{F}(\ell -\ell_s)\norm{u_{X_s}^{(t)}  - u_{X_{s-1}}^{(t)} }  ,
\end{align}
where we use the inequality~\eqref{sec_ineq:lemma_commutator_generalization_2} with $O_X \to u_{X_s}^{(t)}  - u_{X_{s-1}}^{(t)} $ and $\dist_{X_s, \mfL}= \ell -\ell_s$.
Moreover, by applying the Lieb--Robinson bound to the inequality~\eqref{main_ineq2:norm_local_approx} in Lemma~\ref{norm_local_approx}, we have 
 \begin{align}
 \label{ineq:u_is_norm__1D}
\norm{u_{X_s}^{(t)}  - u_{X_{s-1}}^{(t)} } 
\le\min\brr{1,C |\partial X| \br{e^{v|t|} -1} e^{-\mu (\ell_s-\Delta \ell+1) }},
\end{align}
where we use $\ell_{s-1}= \ell_s-\Delta \ell$.
Combining the inequalities~\eqref{ineq:U_tau_decomp_s__1D} and \eqref{ineq:u_is_norm__1D}, we obtain
\begin{align}
\label{ineq:U_tau_decomp_s_reduce__1D}
\norm{\brr{O,  u_{X_s}^{(t)}-u_{X_{s-1}}^{(t)}}} 
&\le 2\mathcal{F}(\ell-\ell_s)\min\brr{1,C |\partial (\mfL[\ell])| \br{e^{v|t|} -1} e^{-\mu (\ell_s-\Delta \ell+1) }}  ,
\end{align}
where we use the definition of $X=\mfL[\ell-1]^\co$, which gives $\partial X =\partial (\mfL[\ell])$.
We then rely on similar analyses to the derivation of~\eqref{integral_ft_ge_delta_t}.
We set $t_0=\max\br{ 0, \mu \ell_s/(2v) - \mu (\Delta \ell-1)/v}$ and define $f_{t_0}(s)$ as in Eq.~\eqref{definition_lem:Quasi-local_commutator_bound}. 
Under the choice, we obtain
\begin{align}
\label{integral_ft_ge_delta_t_1D}
\int_{-\infty}^\infty f(t)\min\brr{1,C\br{e^{v|t|} -1}e^{-\mu (\ell_s -\Delta \ell+1)} } dt 
&\le  \int_{|t|\le t_0}  f(t) C\br{e^{v|t|} -1}e^{-\mu \ell_s + \mu(\Delta \ell-1)}   dt + \int_{|t| \ge t_0} f(t)dt  \notag \\
&\le Cv e^{-\mu \ell_s/2}   \int_{|t|\le t_0} |t| f(t) dt + f_{t_0}(s)  \notag \\
&\le Cv e^{-\mu \ell_s/2}   \int_{|t|\le t_0} |t| f(t) dt + f_{t_0}(s) ,
\end{align}
where we use $e^{v|t|} -1\le v|t| e^{v|t|} \le v|t| e^{\mu \ell_s/2 - \mu (\Delta \ell-1)}$ for $t\le t_0$.
By combining the inequalities~\eqref{ineq:U_tau_decomp_s_reduce__1D} and~\eqref{integral_ft_ge_delta_t_1D}, we reach the inequality of 
\begin{align}
\label{integral_ft_ge_delta_t_final_1D}
\int_{-\infty}^\infty  f(t) \norm{\brr{O,  u_X(t) }} dt
\le 2f_1 \mathcal{F}(\ell)  +2\sum_{s=1}^\infty \mathcal{F}(\ell-s\Delta \ell) \brr{ C|\partial (\mfL[\ell])| f_2 v e^{-\mu s \Delta \ell /2} + f_{t_0}(s)} . 
\end{align}
This completes the proof of Lemma~\ref{lem:Quasi-local_commutator_bound2}. $\square $

\section{Quantum Belief propagation}\label{Sec:Quantum Belief propagation}

Here, we introduce the quantum belief propagation~\cite{PhysRevB.76.201102,PhysRevB.86.245116,kato2016quantum}:
\begin{lemma} \label{belief_propagation_lemma}
For arbitrary operators $\mA$ and $\mB$, we consider the following decomposition:
\begin{align}
\label{eq_belief_propagation_operator}
e^{\beta(\mA+\mB)} = \Phi_{\mB} e^{\beta \mA}\Phi_{\mB}^\dagger
\end{align}
for a fixed $\beta$. Then, the quantum belief propagation provides the explicit form of $\Phi_{\mB}$ as follows:
\begin{align}
\label{sup_Def:Phi_0_phi}
&\Phi_{\mB}:= \mathcal{T} e^{\int_0^{1} \phi_{\mB,\tau} d\tau} , \notag \\
&\phi_{\mB,\tau}:= \frac{\beta}{2}  \int_{-\infty}^\infty f_\beta(t)\mB(\mA_\tau,t) dt, 
\end{align}
where $\mathcal{T}$ is the time ordering operator, $\mA_\tau= \mA+ \tau \mB$, and $f_\beta(t)$ is defined as
\begin{align}
\label{sup_Def:Phi_0_phi_def_f_beta_tilde_f}
f_\beta(t) = \frac{1}{2\pi}  \int_{-\infty}^\infty 
\tilde{f}_\beta(\omega) e^{-i\omega t} d\omega ,\quad 
\tilde{f}(\omega):= \frac{\tanh(\beta \omega/2)}{\beta\omega/2} .
\end{align}
Note that $\phi_{\mB,\tau}$ and $\Phi_{\mB}$ are Hermitian operators. 
\end{lemma}

\noindent
{\bf Remark.}
The explicit form of $f_\beta(t)$ can be calculated as~\cite{Anshu_2021}
\begin{align}
\label{expolicit_form_of_f_beta_t}
f_\beta(t) = \frac{2}{\beta \pi} \log \left (\frac{e^{\pi |t|/\beta }+1}{e^{\pi |t|/\beta }-1} \right) 
\le \frac{2}{\beta \pi}\cdot \frac{2}{e^{\pi |t|/\beta }-1}.
\end{align}
Note that $f_\beta(t)$ is a positive real function that decays exponentially with $t$.
Also, the form of \eqref{sup_Def:Phi_0_phi} is different from the original one by Hastings~\cite{PhysRevB.76.201102}.
In fact, we will also utilize an equivalent but different version by Hastings, which has been defined as follows:
\begin{align}
\label{eq_belief_propagation_operator2}
e^{\beta(\mA+\mB)} = \hat{\Phi}_{\mB} e^{\beta \mA}\hat{\Phi}_{\mB}^\dagger,
\end{align}
where the quantum belief propagation operator has the form of 
\begin{align}
& \hat{\Phi}_{\mB}:= \mathcal{T} e^{\int_0^{1}  \hat{\phi}_{\mB,\tau} d\tau} , \notag \\
& \hat{\phi}_{\mB,\tau}:= \frac{\beta \mB}{2} + i \int_{-\infty}^\infty g_\beta(t)\mB(\mA_\tau,t) dt, 
\label{sup_Def:Phi_0_phi_Hastings}
\end{align}
where $g_\beta(t)$ is defined as
\begin{align}
 \label{def_g_t_exp_decay}
g_\beta(t)  := -\sum_{m=1}^\infty {\rm sign}(t) e^{-2\pi m |t|/\beta}= - {\rm sign}(t)  \frac{e^{-2\pi |t|/\beta}}{1-e^{-2\pi |t|/\beta}}.
\end{align}
The two forms~\eqref{sup_Def:Phi_0_phi} and \eqref{sup_Def:Phi_0_phi_Hastings} for the belief propagation are equivalent. 
One advantage of the form~\eqref{sup_Def:Phi_0_phi} is that the integral of $f_\beta(t)$ is convergent, i.e., 
$\int_{-\infty}^\infty f_\beta(t) dt=\tilde{f}_\beta(\omega=0)=1$, while the integral of $|g_\beta(t)|$ is not convergent 
because of $|g_\beta(t)| \propto 1/|t|$ for $|t|\ll 1$. 
However, the form~\eqref{eq_belief_propagation_operator2} from the belief propagation plays a critical role in estimating the quasi-locality due to the connection of exponential operators as in Lemma~\ref{connection_of_exponential_operator} below.

\subsubsection{Proof of Lemma~\ref{belief_propagation_lemma}}

We give a simpler proof for the belief propagation based on the Baker-Campbell-Hausdorff (BCH) formula.
Our purpose is to derive the Hermitian operator $\phi$ such that 
\begin{align}
\label{B_omega_B_eq_starting/point}
e^{\epsilon \phi} e^{\beta \mA}e^{\epsilon \phi} = e^{\beta (\mA + \epsilon \mB)} 
\end{align}
for infinitesimally small $\epsilon$.
From the BCH formula (see Ref.~\cite[Eq.~(2.7)]{Scharf_1988} for example), we have in general
\begin{align}
\label{BCH_formula_starting_point}
e^{\epsilon \phi} e^{\beta \mA}e^{\epsilon \phi^\dagger} = 
\exp\left[\beta \mA+ \epsilon \br{ \frac{\beta\ad_{\mA}}{e^{\beta\ad_{\mA}}-1} \phi+ {\rm h.c.} } +\orderof{\epsilon^2}\right] .
\end{align}
Hence, for $\phi=\phi^\dagger$, we have 
\begin{align}
\label{connection_B_and_phi}
\beta \mB = \frac{\beta \ad_{\mA} }{e^{\beta\ad_{\mA}}-1} \phi  + {\rm h.c.} 
\end{align}
up to an error of $\orderof{\epsilon^2}$.
We define $\mB_\omega = \sum_{i,j} \bra{a_i} \mB \ket{a_j} \delta(a_i-a_j-\omega) \ket{a_i}\bra{a_j}$ and 
$\phi_\omega = \sum_{i,j} \bra{a_i} \phi \ket{a_j} \delta(a_i-a_j-\omega) \ket{a_i}\bra{a_j}$ with $\{\ket{a_i}\}$ the eigenstates of $\mA$. 
We notice that
\begin{align}
\label{B_omega_B_eq_}
\mB_\omega=\frac{1}{2\pi} \int_{-\infty}^\infty \mB(\mA,t) e^{-i\omega t} dt,\quad 
\mB= \int_{-\infty}^\infty  \mB_\omega d\omega. 
\end{align}
Then, we obtain $\ad_{\mA}(\phi_\omega)=\omega \phi_\omega$ ($[\ad_{\mA}(\phi_\omega)]^\dagger=\omega \phi_{-\omega}$), 
and hence the equation~\eqref{connection_B_and_phi} yields
\begin{align}
\label{express_mB_omega_fourier}
&\mB_\omega = \left(\frac{\omega }{e^{\beta \omega}-1}+ \frac{\omega }{1-e^{-\beta \omega}} 
\right) \phi_\omega= \frac{\omega}{\tanh(\beta \omega/2)} \phi_\omega \to \phi_\omega =   \frac{\tanh(\beta \omega/2)}{\omega} \mB_\omega . 
\end{align}
Therefore, we have from $\phi=\int_{-\infty}^\infty \phi_\omega d\omega$ [see Eq.~\eqref{B_omega_B_eq_}]
\begin{align}
\phi=\frac{1}{2\pi} \int_{-\infty}^\infty  \int_{-\infty}^\infty  \frac{\tanh(\beta \omega/2)}{\omega} \mB(\mA,t) e^{-i\omega t} dt d\omega
=\frac{\beta}{2}   \int_{-\infty}^\infty  
f_\beta(t)\mB(\mA,t)  dt  ,
\end{align}
where we used
\begin{align}
f_\beta(t) = \frac{1}{2\pi} \int_{-\infty}^\infty  
 \frac{\tanh(\beta \omega/2)}{\beta\omega/2} e^{-i\omega t} d\omega =
 \frac{1}{2\pi} \int_{-\infty}^\infty 
\tilde{f}_\beta(\omega) e^{-i\omega t} d\omega  .
\end{align}
By repeating the process of Eq.~\eqref{B_omega_B_eq_starting/point} iteratively, we can prove Lemma~\ref{belief_propagation_lemma}.  

{~}

In deriving Eq.~\eqref{eq_belief_propagation_operator2}, we start from a more general form as
\begin{align}
e^{\epsilon (\phi_1+ i\phi_2)} e^{\beta \mA}e^{\epsilon  (\phi_1- i\phi_2)} = 
\exp\left[\beta \mA+ \epsilon \br{ \frac{\beta\ad_{\mA}}{e^{\beta\ad_{\mA}}-1} \phi_1  + {\rm h.c.} } +i \epsilon \br{ \frac{\beta\ad_{\mA}}{e^{\beta\ad_{\mA}}-1} \phi_2  - {\rm h.c.} } +\orderof{\epsilon^2}\right] .
\end{align}
We then need to choose $\phi_1$ and $\phi_2$ such that
\begin{align}
 \br{ \frac{\omega}{e^{\beta\omega}-1}+ \frac{\omega}{1-e^{-\beta\omega}}   } \phi_{1,\omega} 
 +  i \br{ \frac{\omega}{e^{\beta\omega}-1}-\frac{\omega}{1-e^{-\beta\omega}}   } \phi_{2,\omega}
 = \frac{\omega}{\tanh(\beta \omega/2)} \phi_{1,\omega}  -i\omega \phi_{2,\omega} = \mB_\omega.
\end{align}
Then, choosing $\phi_{1,\omega}=\beta \mB_\omega/2$ gives $\phi_{2,\omega}$ as 
\begin{align}
\phi_{2,\omega} =\frac{1}{i\omega} \br{\frac{\beta\omega/2}{\tanh(\beta \omega/2)}  -1}  \mB_\omega.
\end{align}
Thus, the integral of $\int_{-\infty}^\infty \br{\phi_{1,\omega} \pm i \phi_{2,\omega}} d\omega$ gives 
\begin{align}
&\phi_{1}\pm i\phi_2 = \frac{\beta \mB}{2} \pm \frac{1}{2\pi} \int_{-\infty}^\infty  \int_{-\infty}^\infty \frac{1}{\omega} \br{\frac{\beta\omega/2}{\tanh(\beta \omega/2)}  -1} \mB(\mA,t) e^{-i\omega t} dt d\omega \notag\\
&\longrightarrow  i\phi_2= i \int_{-\infty}^\infty dt \mB(\mA,t)  \frac{1}{2\pi i} \int_{-\infty}^\infty \frac{1}{\omega} \br{\frac{\beta\omega/2}{\tanh(\beta \omega/2)}  -1} e^{-i\omega t}d\omega  ,
\end{align}
which yields Eq.~\eqref{sup_Def:Phi_0_phi_Hastings}. Here, $g_\beta(t)$ is given by
\begin{align}
\label{definition_of_g_beta_t}
g_\beta(t) = \frac{1}{2\pi i} \int_{-\infty}^\infty \frac{1}{\omega} \br{\frac{\beta\omega/2}{\tanh(\beta \omega/2)}  -1} e^{-i\omega t} d\omega.
\end{align}
This completes the proof\footnote{The form of Eq.~\eqref{def_g_t_exp_decay} can be derived by using the decomposition of 
\begin{align}
(x/2)\coth (x/2)= \frac{(x/2)}{\tanh(x/2)} = 1 + \sum_{m\neq 0} \frac{x}{x-2\pi i m }
\end{align}
}. $\square$

\subsection{Error bound for approximate quantum Belief propagation}

Here, we consider the approximation of the belief propagation operator $\Phi_{\mB}$ in Eq.~\eqref{eq_belief_propagation_operator} as 
$\tilde{\Phi}_{\mB}$ as  
\begin{align}
\label{approx_bp_propagation}
&\tilde{\Phi}_{\mB}:= \mathcal{T} e^{\int_0^{1} \tilde{\phi}_{\mB,\tau} d\tau} . 
\end{align}
The question here is whether or not we can obtain  
\begin{align}
\Phi_{\mB} e^{\beta \mA}\Phi_{\mB}^\dagger   \overset{\text{?}}{=}\tilde{\Phi}_{\mB} e^{\beta \mA}\tilde{\Phi}_{\mB}^\dagger   .
\end{align}
This kind of approximation is often critical in our analyses.
We prove the following lemma:
\begin{lemma} \label{lema:bp_error_est}
Let us define $\tilde{\phi}_{\mB,\tau}$ as in Eq.~\eqref{approx_bp_propagation} such that $\| \phi_{\mB,\tau}- \tilde{\phi}_{\mB,\tau} \| \le \delta 
$ for $\forall \tau$.
Then, the norm difference between $\Phi_{\mB} e^{\beta \mA}\Phi_{\mB}^\dagger$ and $\tilde{\Phi}_{\mB} e^{\beta \mA}\tilde{\Phi}_{\mB}$ is upper-bounded by
 \begin{align}
\label{main_ineq:lema:bp_error_est}
\norm{\Phi_{\mB} e^{\beta \mA}\Phi_{\mB}^\dagger  -\tilde{\Phi}_{\mB} e^{\beta \mA}\tilde{\Phi}_{\mB}^\dagger}_1
\le  13 e^{2\beta \norm{\mB}} \delta  \norm{\Phi_{\mB} e^{\beta \mA} \Phi_{\mB}^\dagger}_1
\end{align}
as long as $\delta\le 1$. 
\end{lemma}


\textit{Proof of Lemma~\ref{lema:bp_error_est}.}
We first note that the norm of operator $\phi_{\mB}$ is estimated by Eq.~\eqref{sup_Def:Phi_0_phi} as 
\begin{align}
\label{sup_Def:Phi_0_phi_norm_up}
\norm{\phi_{\mB,\tau}}\le \frac{\beta}{2}  \int_{-\infty}^\infty f_\beta(t) \norm{\mB(\mA_\tau,t)} dt\le  
 \frac{\beta \norm{\mB} }{2}  \int_{-\infty}^\infty f_\beta(t) dt=  \frac{\beta \norm{\mB} }{2},
\end{align}
where we use $f_\beta(t) \ge 0$ and $ \int_{-\infty}^\infty f_\beta(t) dt= \tilde{f}(\omega=0)=1$ from Eq.~\eqref{sup_Def:Phi_0_phi_def_f_beta_tilde_f}.
We thus obtain 
\begin{align}
\label{sup_Def:Phi_0_phi_norm_up_tilde}
\norm{\tilde{\phi}_{\mB,\tau}}\le \norm{\phi_{\mB,\tau}-\tilde{\phi}_{\mB,\tau}} + \norm{\phi_{\mB,\tau}}
\le \frac{\beta \norm{\mB} }{2} +\delta,
\end{align}
where we use the condition $\| \phi_{\mB,\tau}- \tilde{\phi}_{\mB,\tau} \| \le \delta$.

To estimate the error in~\eqref{main_ineq:lema:bp_error_est}, we start from 
\begin{align}
\label{simple_est_QBP_norm}
\norm { \Phi_{\mB} e^{\beta \mA}\Phi_{\mB}^\dagger - \tilde{\Phi}_{\mB} e^{\beta \mA}\tilde{\Phi}_{\mB}^\dagger }_1 
&= \norm { \Phi_{\mB} e^{\beta \mA}\Phi_{\mB}^\dagger  
- \tilde{\Phi}_{\mB}\Phi_{\mB}^{-1} \Phi_{\mB} e^{\beta \mA} \Phi_{\mB}^\dagger  (\Phi_{\mB}^{-1})^\dagger \tilde{\Phi}_{\mB}^\dagger  }_1  \notag \\
&\le \norm{\Phi_{\mB} e^{\beta \mA} \Phi_{\mB}^\dagger}_1\br{ 2 \norm{1-\tilde{\Phi}_{\mB}\Phi_{\mB}^{-1}} +\norm{1-\tilde{\Phi}_{\mB}\Phi_{\mB}^{-1}}^2} .
\end{align} 
The norm of $1-\tilde{\Phi}_{\mB}\Phi_{\mB}^{-1}$ can be upper-bounded as 
\begin{align}
\label{simple_est_QBP_norm_norm_bound}
\norm{1-\tilde{\Phi}_{\mB}\Phi_{\mB}^{-1}} \le \norm{\Phi_{\mB}^{-1}} \cdot \norm{\Phi_{\mB}-\tilde{\Phi}_{\mB}}
&\le  e^{\beta \norm{\mB} /2} \cdot  \delta  \cdot  e^{\delta+ \beta \norm{\mB} /2} ,
\end{align}
where we use
\begin{align}
\label{sup_Def:Phi_0_phi_norm_Phi_norm}
\norm{\Phi_{\mB}^{-1}}\le e^{\beta \norm{\mB} /2} ,\quad \norm{\Phi_{\mB}}\le e^{\beta \norm{\mB} /2}
\end{align}
and the inequality as
\begin{align}
\label{upp_norm_Delta_U_tau_derr}
\norm{\mathcal{T} e^{\int_0^t A(x)dx}- \mathcal{T} e^{\int_0^t B(x)dx}} \le \int_0^t\norm{\mathcal{T} e^{\int_0^x A(x)dx}} \cdot \norm{A(x)-B(x)} \cdot \norm{\mathcal{T} e^{\int_x^t B(x)dx}}dx 
\end{align}
for arbitrary operators $A(x)$ and $B(x)$ [see Ref.~\cite[Eq.~(32)]{KUWAHARA201696} for example]. 
Using the condition $ \delta \le 1$, we reduce the inequality~\eqref{simple_est_QBP_norm_norm_bound} to
\begin{align}
\norm{1-\tilde{\Phi}_{\mB}\Phi_{\mB}^{-1}} \le \delta e^{\beta \norm{\mB}+1} ,
\end{align}
which also yields 
\begin{align}
\norm{1-\tilde{\Phi}_{\mB}\Phi_{\mB}^{-1}}^2 \le  \delta^2 e^{2\beta \norm{\mB}+2} \le \delta e^{2\beta \norm{\mB}+2}  .
\end{align}
By applying the above two inequalities to~\eqref{simple_est_QBP_norm}, we obtain 
\begin{align}
\norm { \Phi_{\mB} e^{\beta \mA}\Phi_{\mB}^\dagger - \tilde{\Phi}_{\mB} e^{\beta \mA}\tilde{\Phi}_{\mB}^\dagger }_1 
&\le \norm{\Phi_{\mB} e^{\beta \mA} \Phi_{\mB}^\dagger}_1 \delta \br{2e \cdot e^{\beta \norm{\mB}}+e^2e^{2\beta \norm{\mB}}  }
\le  13 e^{2\beta \norm{\mB}} \delta \norm{\Phi_{\mB} e^{\beta \mA} \Phi_{\mB}^\dagger}_1, \notag 
\end{align} 
which gives the main inequality~\eqref{main_ineq:lema:bp_error_est}. This completes the proof. $\square$

{~}

\hrulefill{\bf [ End of Proof of Lemma~\ref{lema:bp_error_est}]}

{~}

In various contexts, we often consider the decomposition of the Hamiltonian as 
\begin{align}
H= H_L +H_{L^\co} + \partial h_L,\label{def:Ham_surface_decomposition_BP}
\end{align}
where $\partial h_L$ has been defined as the surface interaction term as in Eq.~\eqref{def:Ham_surface}.
We then set $\mA=H_L +H_{L^\co}$ and $\mB=\partial h_L$ in Eq.~\eqref{eq_belief_propagation_operator}. 
In this case, $\phi_{\partial h_L,\tau}$ is constructed by using the time evolution $\partial h_L( H_L +H_{L^\co} + \tau \partial h_L ,t)$. 
Hence, we can utilize the Lieb--Robinson bound~\eqref{local_approximation_Lieb_Robinson} to approximate $\phi_{\partial h_L,\tau}$ onto a subset $(\partial L)[r]$.
We prove the following statement: 
\begin{lemma} \label{lem:BP_error_approx_local}
Under the decomposition~\eqref{def:Ham_surface_decomposition_BP}, we define
\begin{align}
\label{tilde_phi_definitoon_R_tau}
\tilde{\phi}_{\partial h_L,\tau}^{(r)} :=\tilde{\tr}_{(\partial L)[r]^\co} \brr{\phi_{\partial h_L,\tau}}.
\end{align}
Then, for an arbitrary $\tau\in [0,1]$, we can prove 
\begin{align}
\label{main_ineq::lem:BP_error_approx_local___0}
\norm{\phi_{\partial h_L,\tau}-\tilde{\phi}_{\partial h_L,\tau}^{(r)} }\le \bar{\phi}_{\beta, |\partial L|} e^{-\kappa_\beta r} 
\end{align}
with
\begin{align}
\label{def:bar_phi_beta_partial_L}
\bar{\phi}_{\beta, |\partial L|} := 4\beta \gamma\tilde{J}_0  |\partial L|   e^{\mu/2} 
 \brr{1+\frac{2\beta \gamma Cv  |\partial L|}{7} \br{\frac{4D}{e\mu}}^{D} +\frac{8}{\pi^2}  \log\br{e+\frac{e}{\kappa_\beta}} \frac{e^{\kappa_\beta }}{e^{\kappa_\beta }-1}},
\end{align}
where we define $\kappa_\beta$ as follows:
\begin{align}
\label{definition_kappa_beta_1}
\kappa_\beta := \min\br{ \frac{\pi \mu}{2v\beta} , \frac{\mu}{4}}.
\end{align}
\end{lemma}

\textit{Proof of Lemma~\ref{lem:BP_error_approx_local}.}
For an arbitrary unitary operator $u_X$ acting on $X \subset \Lambda$, we consider 
\begin{align}
\label{start_pro_lem:BP_error_approx_local}
\norm{[ \phi_{\partial h_L,\tau}, u_X]} \le 
\frac{\beta}{2}  \int_{-\infty}^\infty f_\beta(t) \norm{[ \partial h_L(H_\tau,t),u_X]} dt ,\quad H_\tau:=H_L +H_{L^\co} + \tau \partial h_L.
\end{align}
By using Lemma~\ref{norm_local_approx}, we can obtain the upper bound for the LHS of Eq.~\eqref{main_ineq::lem:BP_error_approx_local___0} by choosing $X=(\partial L)[r]^\co$.
To estimate the quantity~\eqref{start_pro_lem:BP_error_approx_local}, we utilize Lemma~\ref{lem:Quasi-local_commutator_bound2} by choosing 
\begin{align}
\mfL\to \partial L,\quad O \to \partial h_L,\quad f(t) \to \frac{\beta}{2}   f_\beta(t),\quad \Delta \ell \to 1.
\end{align}
Note that under the choice of $\mfL\to \partial L$ the summation in~\eqref{main_ineq:lem:Quasi-local_commutator_bound2} with respect to $s$ can be truncated up to $s=\ell$ since $X_s$ in Eq.~\eqref{def_X_s_lemm} becomes the total set $\Lambda$ for $s=\ell$, i.e., $X_\ell=\br{\partial L[-1]}^\co=\Lambda$.

To utilize the inequality~\eqref{main_ineq:lem:Quasi-local_commutator_bound2}, we have to estimate an upper bound of
\begin{align}
\label{F_ell_est_bp_approx}
\norm{ [\partial h_L,u_Y} \le 
\begin{cases} 
\displaystyle 
2\sum_{Z:Z\cap L \neq \emptyset,\ Z\cap Y\neq \emptyset} \norm{h_Z}  &\for Y \subset L^\co, \\
\displaystyle 
2\sum_{Z:Z\cap L^\co \neq \emptyset,\ Z\cap Y\neq \emptyset} \norm{h_Z}     &\for Y \subset L , 
\end{cases}
\end{align}
for unitary operators supported on $Y$ such that $\dist_{Y,\partial L}=\ell$, 
where we use the form of~\eqref{def:Ham_surface}.
Using Lemma~\ref{sum_interaction_terms} with $X \to L$ ($X\to L^\co$) and $Y\to L[\ell-1]^\co$ ($Y\to L[-\ell]$), we obtain
\begin{align}
\label{F_ell_est_bp_approx_h_Z_sum}
\norm{ [\partial h_L,u_{(\partial L)[\ell-1]^\co}} &\le 
2\sum_{Z:Z\cap L \neq \emptyset,\ Z\cap L[\ell-1]^\co\neq \emptyset} \norm{h_Z} 
+2\sum_{Z:Z\cap L^\co \neq \emptyset,\ Z\cap L[-\ell] \neq \emptyset} \norm{h_Z} \notag \\
&\le 2(|\partial L| + |\partial L^\co| ) \tilde{J}_0 e^{-\mu (\ell-1)/2} 
\le 4 \gamma\tilde{J}_0  |\partial L| e^{-\mu (\ell-1)/2},
\end{align}
where we use $|\partial L^\co| \le|\partial L[1]|\le \gamma |\partial L|$. 
Therefore, we can choose the quasi-locality function $\mathcal{F}(\ell)$ in~\eqref{initial_Quasi-local_commutator_bound} as follows:
\begin{align}
\mathcal{F}(\ell)=4 \gamma\tilde{J}_0  |\partial L| e^{-\mu (\ell-1)/2}  =: \tilde{J}_{L} e^{-\mu (\ell-1)/2} ,
\end{align}
where we use the fact that $\dist_{\partial L, (\partial L)[\ell-1]^\co}=\ell$.

We now have all the ingredients to estimate the RHS of the inequality~\eqref{main_ineq:lem:Quasi-local_commutator_bound2}, where we choose $\Delta \ell=1$. 
We first estimate $f_1$ and $f_2$, which are now calculated as 
\begin{align}
&f_1= \int_{-\infty}^\infty \frac{\beta}{2}   f_\beta(t)dt= \frac{\beta}{2} ,\quad 
f_2\le  \int_{-\infty}^\infty \frac{\beta}{2}  |t| f_\beta(t)dt=
\frac{7\beta^2 \zeta(3)}{2\pi^3} \le \frac{\beta^2}{7}, 
\end{align}
where $\zeta(x)$ is the Riemann zeta function, and we use the inequality of 
\begin{align}
\int_{-\infty}^\infty   |t| f_\beta(t)dt= \frac{2}{\pi \beta} \int_{-\infty}^\infty |t| \log \left (\frac{e^{\pi |t|/\beta }+1}{e^{\pi |t|/\beta }-1} \right) dt =\frac{2}{\pi \beta}\cdot \frac{7\beta^2 \zeta(3)}{2\pi^2} .
\end{align}
Also, for  $f_{t_0}(s)$, we obtain by using~\eqref{expolicit_form_of_f_beta_t}
\begin{align}
f_{t_0}(s)= 2\int_{t_0}^\infty \frac{\beta}{2}   f_\beta(t)dt
\le  \frac{4}{\pi}\int_{t_0}^\infty  \frac{1}{e^{\pi |t|/\beta }-1}dt 
=\frac{4}{\pi} \cdot \frac{-1}{\pi/\beta} \log\br{1-e^{-\pi t_0 /\beta}}
\le \frac{4\beta}{\pi^2}\log\br{e+\frac{e\beta}{\pi t_0}} e^{-\pi t_0 /\beta} ,
\end{align}
where we use $-\log(1-e^{-x})\le \log(e+e/x) e^{-x}$. Note that $t_0$ is chosen as $t_0=\mu s /(2v)$ for $\Delta \ell=1$ in Lemma~\ref{lem:Quasi-local_commutator_bound2}. 
We thus calculate the RHS of~\eqref{main_ineq:lem:Quasi-local_commutator_bound2} as 
\begin{align}
\label{proof_pro_lem:BP_error_approx_local_fin}
&2f_1 \mathcal{F}(\ell)  +2 \sum_{s=1}^{\ell} \mathcal{F}(\ell-s) \brr{ C|\partial (\partial L[\ell])| f_2 v e^{-\mu s/2} + f_{t_0}(s)} \notag \\
&\le  4\beta  \gamma\tilde{J}_0  |\partial L| e^{-\mu (\ell-1)/2} + 
\frac{2\beta^2 \gamma Cv  |\partial L| \ell^{D-1}}{7}4 \gamma\tilde{J}_0  |\partial L| \ell e^{-\mu (\ell-1)/2} 
+8 \gamma\tilde{J}_0  |\partial L|   e^{\mu/2}     \cdot \frac{4\beta}{\pi^2}  \log\br{e+\frac{e}{\kappa_\beta}} \frac{e^{\kappa_\beta }}{e^{\kappa_\beta }-1} e^{-\kappa_\beta \ell} \notag \\
 &\le  4\beta \gamma\tilde{J}_0  |\partial L|   e^{\mu/2} e^{-\kappa_\beta \ell}  
 \br{1+\frac{2\beta \gamma Cv  |\partial L| }{7}  \ell^{D}e^{-\mu \ell/4} +\frac{8}{\pi^2}  \log\br{e+\frac{e}{\kappa_\beta}} \frac{e^{\kappa_\beta }}{e^{\kappa_\beta }-1}} 
 ,
\end{align}
where we use $|\partial (X[\ell])| \le \gamma \ell^{D-1} |X|$ for $\forall X\subset \Lambda$, the definition~\eqref{definition_kappa_beta_1} for $\kappa_\beta$, 
$\mu/2 \ge 2\kappa_\beta$ ($\mu/4 \ge \kappa_\beta$) and $\pi t_0 /\beta=\pi \mu s /(2v\beta) \ge \kappa_\beta s$ to derive the inequalities of
\begin{align}
 \sum_{s=1}^{\ell}\mathcal{F}(\ell-s)  e^{-\mu s/2} \le 4 \gamma\tilde{J}_0  |\partial L|  
 \sum_{s=1}^{\ell}e^{-\mu (\ell-s-1)/2} \cdot e^{-\mu s/2}
 = 4 \gamma\tilde{J}_0  |\partial L|   \ell e^{-\mu (\ell-1)/2}  ,
\end{align}
and 
\begin{align}
 \sum_{s=1}^{\ell} \mathcal{F}(\ell-s) f_{t_0}(s)
 &\le 
 4 \gamma\tilde{J}_0  |\partial L|   e^{\mu/2} \sum_{s=1}^{\ell}e^{-2\kappa_\beta (\ell-s)} \cdot \frac{4\beta}{\pi^2}\log\br{e+\frac{e}{\kappa_\beta s}} e^{-\kappa_\beta  s} \notag \\
 &\le 4 \gamma\tilde{J}_0  |\partial L|   e^{\mu/2}     \cdot \frac{4\beta}{\pi^2}  \log\br{e+\frac{e}{\kappa_\beta}} \frac{e^{\kappa_\beta }}{e^{\kappa_\beta }-1} e^{-\kappa_\beta \ell}  .
\end{align}
Now, Eq.~\eqref{start_pro_lem:BP_error_approx_local} is upper-bounded by~\eqref{proof_pro_lem:BP_error_approx_local_fin} by letting $\ell=r$.
Therefore, by applying 
\begin{align}
\ell^{D} e^{-\mu \ell/4}\le  \br{\frac{4D}{e\mu}}^{D} 
\end{align}
to~\eqref{proof_pro_lem:BP_error_approx_local_fin}, we prove the main inequality~\eqref{main_ineq::lem:BP_error_approx_local___0}, 
where we use $x^m e^{-\mu x/4}\le \brr{4m/(e\mu)}^m$ for $m\in \mathbb{N}$.
This completes the proof. $\square$

{~}

\hrulefill{\bf [ End of Proof of Lemma~\ref{lem:BP_error_approx_local}]}

{~}

From Lemmas~\ref{belief_propagation_lemma}, \ref{lema:bp_error_est} and~\ref{lem:BP_error_approx_local}, we immediately prove the following corollary:
\begin{corol}\label{corol:high_dimensional_applicaton_bp}
Let $L$ be a connected region and $\tilde{\Phi}_{\partial h_L}^{(r)}$ be defined as  
\begin{align}
\label{tilde_Phi_definition_R}
\tilde{\Phi}_{\partial h_L}^{(r)}:= \mathcal{T} e^{\int_0^1 \tilde{\phi}_{\partial h_L,\tau}^{(r)} d\tau} .
\end{align}
Then, using $\tilde{\Phi}_{\partial h_L}^{(r)}$ as the belief propagation operator instead of $\Phi_{\partial h_L}$, we obtain the error as follows:
 \begin{align}
\label{main_ineq:corol:high_dimensional_applicaton_bp}
\frac{1}{Z_\beta}\norm{e^{\beta H} -\tilde{\Phi}_{\partial h_L}^{(r)} e^{\beta \br{H_L +H_{L^\co}}}  \tilde{\Phi}_{\partial h_L}^{(r)\dagger} }_1
\le  13  \bar{\phi}_{\beta, |\partial L|} e^{2\beta \norm{\partial h_L}-\kappa_\beta r} .
\end{align}
\end{corol}

{~} \\

{\bf Remark.} By using the error bound of \eqref{main_ineq:corol:high_dimensional_applicaton_bp}, the error is sufficiently small under the condition that
 \begin{align}
 \label{corol:high_dimensional_applicaton_bp_caveat}
r\gtrsim \frac{2\beta}{\kappa_\beta} \norm{\partial h_L} = \orderof{\beta^2} |\partial L|,
\end{align}
where we use $\kappa_\beta=\orderof{1/\beta}$ from Eq.~\eqref{definition_kappa_beta_1} and 
$\norm{\partial h_L} \le \sum_{Z: Z\cap \mfL\neq \emptyset,\ Z\cap L^\co \neq \emptyset}  \norm{h_Z} 
\le |\partial L| \mathcal{J}_0= \tilde{J}_0  |\partial L| $ from the inequality~\eqref{sum_interaction_terms_main_eq_short_long} in Lemma~\ref{sum_interaction_terms}.
In one-dimensional systems, the cardinality $|\partial L|$ is $\orderof{1}$ and hence the condition reduces to $r\gtrsim \orderof{\beta^2}$. 
On the other hand, in higher dimensions, the cardinality $|\partial L|$ is as large as $[\diam(L)]^{D-1}$, i.e., 
$r\gtrsim \beta^2 |\partial L|\approx \beta^2[\diam(L)]^{D-1}$. 
This point prohibits us from utilizing the belief propagation in estimating the conditional mutual information as in Sec.~\ref{sec:BP_formalism} in dimensions greater than one~\footnote{This point might be improved by cleverly employing the quasi-locality of the Hamiltonian. However, as far as we know, there are no results that improve the bound as in~\eqref{main_ineq:corol:high_dimensional_applicaton_bp} at arbitrary temperatures.}.

 \begin{figure}[tt]
\centering
\includegraphics[clip, scale=0.4]{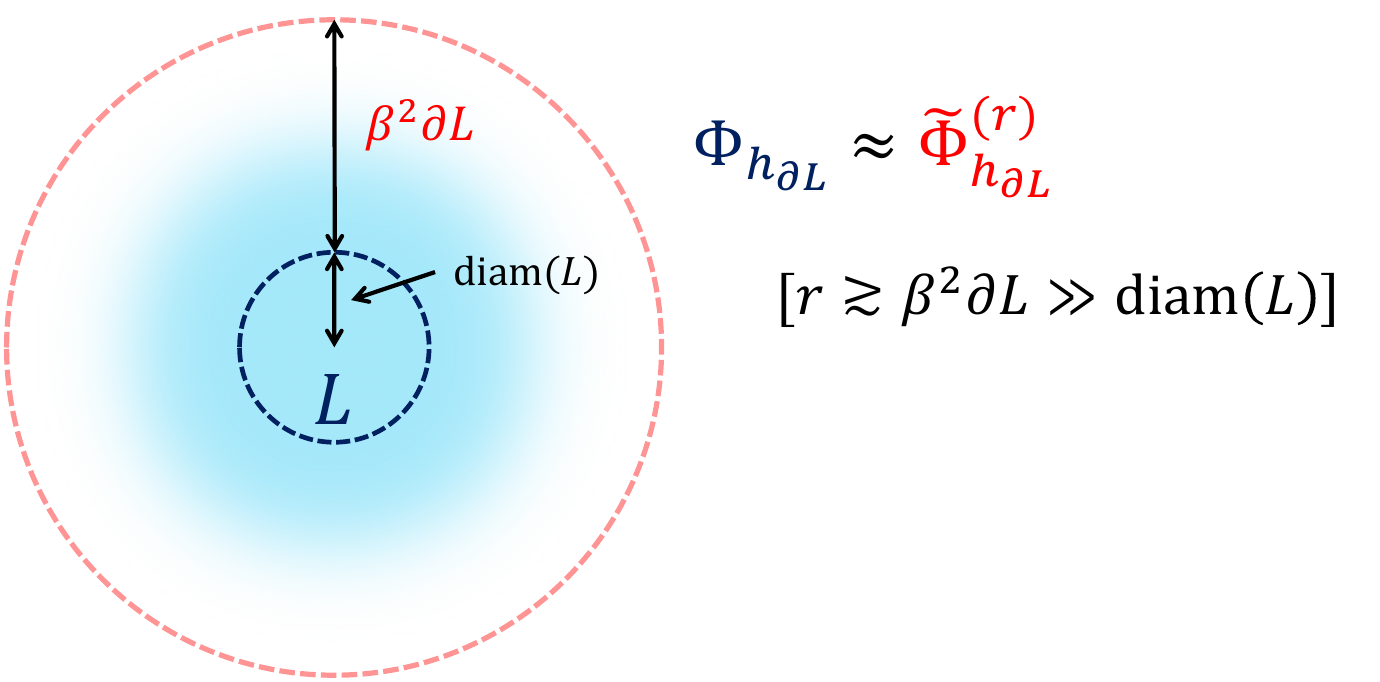}
\caption{Approximate belief propagation operator in high dimensions (2D picture). When we consider the belief propagation operator $\Phi_{\partial h_L}$ for the decomposition of $H=H_L + H_{L^\co}$, the local approximation by $\Phi_{\partial h_L}\approx \tilde{\Phi}_{\partial h_L}^{(r)}$ holds for $r \gtrsim \beta^2 |\partial L|$ from the inequality~\eqref{main_ineq:corol:high_dimensional_applicaton_bp}. 
Therefore, the approximate belief propagation operator $\tilde{\Phi}_{\partial h_L}^{(r)}$ is completely non-local regardless of the size of $L$ as long as we look at it in the region $L$. 
This becomes a primary bottleneck in applying the belief propagation technique in a high-dimensional Gibbs state. 
}
\label{fig_BP_highD}
\end{figure}

\section{Effective Hamiltonian theory on subsystem} \label{sec:effective Hamiltonian theory on subsystem}

In this section, we consider the entanglement Hamiltonian that is defined by the partial trace operation for a subset $L\subset \Lambda$:
\begin{align}
\tr_{L} \br{e^{\beta H}} = e^{\beta H^\ast_{L^\co}}. 
\end{align}
When considering the effective interaction term $H^\ast_{L^\co}- H_{L^\co}$, the problem is whether such interaction approximately localizes around the boundary of $L$. 
This kind of analysis is the crucial ingredient to treating the quantum conditional mutual information (see also Lemma~\ref{lem:entropy_bound_1D} and Corollary~\ref{corol:error_est_APTP_continuity_bound_CMI} below). 
In the case where the Hamiltonian is commuting and has an interaction length up to $k$, it is clear that the effective interaction is strictly localized around $\partial L$.
The proof is immediately given as follows:
\begin{align}
\tr_{L} \br{ e^{\beta H} } = \tr_{L} \br{ e^{\beta H_{L^\co}} e^{\beta \br{H_L + \partial h_L} }  } =  
 e^{\beta H_{L^\co}}  \tr_{L} \br{e^{\beta \br{H_L + \partial h_L} }  } = e^{\beta H_{L^\co} +\log \brrr{\tr_{L} \brr{e^{\beta \br{H_L + \partial h_L} }} }}   ,
\end{align}
which gives $H^\ast_{L^\co}= H_{L^\co} + (1/\beta)\log \brrr{\tr_{L} \brr{e^{\beta \br{H_L + \partial h_L} }} }$.
Note that the support of $\partial h_L$ is included in $L[k]$ from the assumption on the interaction length. Therefore, $\tr_{L} \brr{e^{\beta \br{H_L + \partial h_L} }}$ as well as 
$\log\brrr{\tr_{L} \brr{e^{\beta \br{H_L + \partial h_L} }}}$ is also supported on $L[k]\setminus L$.
Thus, we conclude that the effective interaction is localized up to a distance $k$.

The challenge here is to generalize the statement to non-commuting Hamiltonians.
So far, no established theoretical framework exists to analyze the entanglement Hamiltonian after partial trace. 
In the following, we show two formalisms that rely on quantum belief propagation (BP formalism, see Sec.~\ref{sec:BP_formalism}) and partial-trace projection (PTP formalism, see Sec.~\ref{sec:PTP_formalism}).
The former is utilized in one-dimensional systems, while the latter is applied to high-dimensional systems. 
We note that in the 1D case, the result becomes better when we use the BP formalism.

\subsection{Norm of the entanglement Hamiltonian} \label{subsec:norm_effective Hamiltonian theory on subsystem}

Before going to the quasi-locality, we first derive a fundamental statement on the norm of $\beta H^\ast_{L^\co}=\log\brr{\tr_L \br{\rho_\beta}}=\log(\rho_{\beta,L^\co} )$, where we use the notation~\eqref{def_rho_L_H^ast_L}. 
At first glance, it is a trivial problem to prove
\begin{align}
\label{desired_ineq_e^_beta_H_L}
\norm{\beta H^\ast_{L^\co}}= \orderof{ \beta |L^\co|} .
\end{align}
Indeed, in the commuting cases, one can easily derive it by using 
\begin{align}
\label{e^beta_H_decomp}
e^{\beta H}=e^{\beta H_{L} +\beta \widehat{H_{L^\co}}}  \succeq e^{\beta H_{L}  - \beta \norm{\widehat{H_{L^\co}} }} ,
\end{align}
which yields 
\begin{align}
\label{e^beta_H_decomp__2_proof}
\frac{1}{Z_\beta}\tr_L \br{e^{\beta H}} \succeq  \frac{\tr_L\br{e^{\beta H_{L}} }}{Z_\beta} e^{- \beta \norm{\widehat{H_{L^\co}} }}
 \succeq e^{- 2\beta \norm{\widehat{H_{L^\co}} } - \log (\mathcal{D}_{L^\co})} ,
\end{align}
where we use the notation~\eqref{def:Hamiltonian_subset_L}, i.e., $\widehat{H_{L^\co}}= H_{L^\co}+\partial h_{L}$, and the last inequality is derived from the Golden-Thompson inequality as   
\begin{align}
Z_\beta=\tr \br{e^{\beta H_{L} +\beta \widehat{H_{L^\co}}} } \le \tr \br{e^{\beta H_{L}} e^{\beta \widehat{H_{L^\co}}} } 
&\le e^{\beta \norm{\widehat{H_{L^\co}}}}  \tr \br{ e^{\beta H_{L}} \otimes \hat{1}_{L^\co}}  =e^{\beta \norm{\widehat{H_{L^\co}}}} \mathcal{D}_{L^\co} \tr_L \br{ e^{\beta H_L} }.
\end{align}
Note that $\mathcal{D}_{L^\co}$ is the Hilbert space dimension on $L^\co$. 
From the upper bound $\norm{\widehat{H_{L^\co}}} \le \bar{J}_0 |L^\co|$ which is derived from the condition~\eqref{def:Hamiltonian}, we reduce the inequality~\eqref{e^beta_H_decomp__2_proof} to the desired one~\eqref{desired_ineq_e^_beta_H_L} as follows:
 \begin{align}
 \label{eff_Ham_commuting_bound}
\norm{\beta H^\ast_{L^\co}}\le 2\beta \norm{\widehat{H_{L^\co}}} + \log (\mathcal{D}_{L^\co}) \le  2\beta \bar{J}_0 |L^\co|+ \log (\mathcal{D}_{L^\co})  .
\end{align}

The difficulty in the non-commuting case originates from the fact that the operator inequality~\eqref{e^beta_H_decomp} does not generally hold. 
Indeed, from Ref.~\cite[Lemma~12]{kimura2024clustering}, we ensure 
\begin{align} 
\label{e^beta_H_decomp_non_comm}
e^{\beta H}=e^{\beta H_{L} +\beta \widehat{H_{L^\co}}  }  \succeq e^{\beta H_{L} - \zeta} 
\end{align}
only when $\zeta \ge \beta \norm{\widehat{H_{L^\co}} }+\beta \norm{ H_{L}} =\Omega(\beta |\Lambda|)$. 
This condition is proven to be qualitatively tight. 
Hence, the inequality~\eqref{e^beta_H_decomp_non_comm} leads to a rather weaker bound of $\log\brr{\tr_L \br{e^{\beta H}}/Z_\beta} = \beta \orderof{|\Lambda|}$. 
By employing the techniques in Ref.~\cite{Anshu_2021} based on \cite{Arad_2016}, we can prove the desired bound~\eqref{desired_ineq_e^_beta_H_L} for general quantum Gibbs states.
\begin{prop} \label{prop:Norm of the effective Hamiltonian}
Under the condition that the Hamiltonian is $k$-local as 
\begin{align}
H = \sum_{Z:|Z|\le k} h_Z,  \quad \max_{i\in \Lambda} \sum_{Z:Z\ni i}\|h_Z\| \le \bar{J}_0 \label{def:Hamiltonian_k-local},
\end{align}
we have 
\begin{align}
\label{main_ineq_prop:Norm of the effective Hamiltonian} 
\norm{\beta H^\ast_{L^\co}} \le \beta \bar{J}_0 |L^\co| + \log(16 \bar{J}_0 |L^\co|  \mathcal{D}_{L^\co}) = \orderof{ \beta |L^\co| }, 
\end{align}
where $\bar{J}_0$ is defined by the inequality of
\begin{align}
\label{def_bar_J_0_1}
3J_0 |L^\co|+ (2J_0 k) \log(8\mathcal{D}_{L^\co})+1 \le \bar{J}_0 |L^\co| .
\end{align}
\end{prop}

{\bf Remark.} From the statement, as long as the Hamiltonian is $k$-local, we obtain qualitatively the same upper bound as the one in the commuting cases~\eqref{eff_Ham_commuting_bound}. On the other hand, as the $k$ increases, the upper bound linearly increases with $k$. 
For $L^\co=\orderof{1}$, our result implies that the entanglement Hamiltonian on $L^\co$ has an amplitude of $\orderof{1}$. Hence, only a small portion of the global interaction terms in $H_L$ contributes to the entanglement Hamiltonian. 

In our definition of the Hamiltonian in Eq.~\eqref{def:Hamiltonian}, we do not assume the strict $k$-locality.
Without the strict $k$-locality, we can prove a similar but weaker inequality to~\eqref{main_ineq_prop:Norm of the effective Hamiltonian} as long as the bound~\eqref{AKL:2016_bound} in Ref.~\cite{Arad_2016} is given (see below).
Using Ref.~\cite[Supplementary Lemma~9]{Anshu_2021}, we can obtain 
\begin{align}
g_{\Delta,X} \le \Theta(|X|) e^{-\Theta(1) (\Delta/|X|)^{\mathfrak{u}/(2-\mathfrak{u})}} 
\end{align}
under the similar condition to~\eqref{def_short_range_long_range}:
\begin{align}
\sum_{Z:Z\ni \{i,i'\}}\norm{ h_Z} \le  \bar{J}_0 e^{-\mu \dist_{i,i'}^{\mathfrak{u}}} \quad (0\le \mathfrak{u} \le 1) ,
\end{align}
where the case of $\mathfrak{u}=1$ corresponds to the condition~\eqref{def_short_range_long_range}.
Then, in the inequality~\eqref{lambda_min_P_L^co_lower}, we need to choose the parameter $\Delta$ as 
\begin{align}
\Delta = \Theta(|L^\co|) \brr{\log(\mathcal{D}_{L^\co})}^{(2-\mathfrak{u})/\mathfrak{u}}  ,
\end{align}
which yields 
\begin{align}
\label{main_ineq_prop:Norm of the effective Hamiltonian_general} 
\norm{\beta H^\ast_{L^\co}} = \mathcal{O} \br{ \beta  |L^\co|^{2/\mathfrak{u}} }    .
\end{align}
This is still polynomial concerning the subsystem size $|L^\co|$. 

As a simple application, we consider the following entanglement Hamiltonian learning on a small subsystem, which will also be utilized for the global Hamiltonian learning in one-dimensional systems (see Sec.~\ref{sec:1D Hamiltonian learning with poly-log sample complexity}):
\begin{corol} \label{corol:effective_Ham_Learning}
Let us adopt the same setup as in Proposition~\ref{prop:Norm of the effective Hamiltonian}. 
Then, given $N$ copies of the quantum Gibbs state $\rho_\beta$, one can reconstruct an entanglement Hamiltonian $\log(\sigma_X)$ up to the following error with the success probability larger than $0.99$: 
\begin{align}
\label{main_corol:effective_Ham_Learning}
\norm{\log(\sigma_X)- \log(\rho_{\beta,X})}
&\le  \frac{\log (N/20)}{N} 920\mathcal{D}_X^2 (16 \bar{J}_0 |X|  \mathcal{D}_{X})^{3/2} e^{3\beta \bar{J}_0 |X|/2}
\notag \\
&\le  \frac{\log (N)}{N}e^{\Theta(\beta) |X|},
\end{align}
where we assume $N\ge 40$. 
\end{corol}

\textit{Proof of Corollary~\ref{corol:effective_Ham_Learning}}.
Using the result in Ref.~\cite[Theorem~1]{10.1145/2897518.2897585}, the sufficient number of copies to reconstruct $\rho_{\beta,X}$ up to an error $\epsilon$ is 
upper-bounded by 
\begin{align}
\label{N_epsilon_tomography}
N \le 20 (\mathcal{D}_X^2/\epsilon) \log (\mathcal{D}_X^2/\epsilon)  ,
\end{align}
where the success probability is larger than $0.99$. 
Here, the error is estimated by using the trace norm; that is, the reconstructed state $\sigma_X$ satisfies 
\begin{align}
\norm{\sigma_X - \rho_{\beta,X}}_1 \le \epsilon. 
\end{align}
By solving the inequality~\eqref{N_epsilon_tomography}, we have 
\begin{align}
\label{main_effective_Ham_Learning_1}
(\mathcal{D}_X^2/\epsilon) \ge \frac{(N/20)}{W(N/20)} \longrightarrow 
\epsilon \le \frac{40\mathcal{D}_X^2 \log (N/20)}{N}  ,
\end{align}
where we use $N\ge 40$ and $W(x) \le 2\log(x)$ for $x\ge 2$. 
Note that the function $W(x)$ is the Lambert W function and is defined by $W(x) e^{W(x)}=x$.

Also, by combining the inequalities~\eqref{ineq:delta_rho_sigma_norm} and~\eqref{main_ineq:thm:refined continuity_0} in Theorem~\ref{thm:refined continuity} below, we prove 
 \begin{align}
 \label{main_effective_Ham_Learning_2}
\norm{\log(\sigma_X)- \log(\rho_{\beta,X})}\le  \frac{\norm{\sigma_X - \rho_{\beta,X}}}{\lambda^{-1}_{\min}}  \brr{ \frac{4\log (2\lambda^{-1}_{\min}) }{\pi} \log\br{\frac{e\log (2\lambda^{-1}_{\min})}{2\pi }} +23 } \le \frac{23 \epsilon}{\lambda^{-3/2}_{\min}}  , 
\end{align} 
where we let $\lambda_{\rm min}$ be the minimum eigenvalue of $\rho_{\beta,X}$, and the operator norm is smaller than the trace norm, i.e., $\norm{\sigma_X - \rho_{\beta,X}} \le \norm{\sigma_X - \rho_{\beta,X}}_1 \le  \epsilon$.
By choosing $L^\co$ as $X$ in the upper bound~\eqref{main_ineq_prop:Norm of the effective Hamiltonian}, we prove 
\begin{align}
\label{main_effective_Ham_Learning_3}
\lambda^{-3/2}_{\min}\le e^{(3/2)\norm{\beta H^\ast_{X}}} \le (16 \bar{J}_0 |X|  \mathcal{D}_{X})^{3/2} e^{3\beta \bar{J}_0 |X|/2}.
\end{align}
By combining the inequalities~\eqref{main_effective_Ham_Learning_1}, \eqref{main_effective_Ham_Learning_2} and \eqref{main_effective_Ham_Learning_3}, we prove the main inequality~\eqref{main_corol:effective_Ham_Learning}.
This completes the proof of Corollary~\ref{corol:effective_Ham_Learning}. $\square$

\subsubsection{Proof of Proposition~\ref{prop:Norm of the effective Hamiltonian}}
For the proof, we aim to derive the lower bound on the minimum eigenvalue of $ \rho_{\beta, L^\co}$, which we denote by $\lambda_{\min}$ 
Because of $\rho_{\beta, L^\co} \preceq \hat{1}_{L^\co}$, we have 
\begin{align}
\norm{ \log(\rho_{\beta,L^\co} )} = \log(1/\lambda_{\min}) ,
\end{align}
and it also gives the upper bound of $\norm{\beta H^\ast_{L^\co}}$.
To derive the minimum eigenvalue $\lambda_{\min}$, we consider 
\begin{align}
\label{lambda_min_P_L^co}
\lambda_{\min}= \inf_{P_{L^\co}} \brr{ \tr_{L^\co} (P_{L^\co}  \rho_{\beta, L^\co}) } =  \inf_{P_{L^\co}} \brr{  \tr \br{P_{L^\co} \rho_{\beta} }}  .
\end{align}

For the estimation of $\lambda_{\min}$, we consider an arbitrary projector $P$ and estimate 
\begin{align} 
\delta = \tr (P \rho_\beta) = \norm{P\sqrt{\rho_\beta}}_F .
\end{align}
To apply the techniques in Ref.~\cite{Anshu_2021}, we define the projection onto the energy eigenspace with the energies within $[ E, \infty)$ and $(-\infty, E]$ as $\Pi_{\ge E}$ and $\Pi_{\le E}$, respectively.
We then denote the parameter $g_{\Delta,X}$ which satisfies the following inequality:
\begin{align} 
\label{AKL:2016_bound}
\sup_{O_X: \norm{O_X}=1} \norm{\Pi_{\ge E+\Delta} O_X \Pi_{\le E} } \le g_{\Delta,X} \for  \forall E\in \mathbb{R} .
\end{align}
The parameter $g_{\Delta,X}$ characterizes the robustness of the energy spectrum to local operators and has been given by~\cite{Arad_2016} 
 \begin{align} 
g_{\Delta,X} = e^{-(\Delta -3J_0 |X|)/(4J_0k)} ,
\end{align}
where the above expression was given in \cite[Theorem 3.2]{Phd_kuwahara}.   

Using $g_{\Delta,X}$, we generally obtain the following inequality on the robustness of the expectation of $P$ to the local unitary operator $U_X$:
\begin{align} 
\label{U_X_rotation}
 \norm{PU_X\sqrt{\rho_\beta}}_F  \le (4 \Delta +2) e^{\beta (\Delta +1)}\delta + 2 g_{\Delta,X}^2  \for  \forall \Delta >0   ,
\end{align}
where the proof was given in Ref.~\cite[Supplementary inequality~(116)]{Anshu_2021}.
The interpretation of the inequality~\eqref{U_X_rotation} is that local unitary operators do not significantly influence the expectation value of $P$. 

Moreover, if $P$ is supported on $X$, i.e., $P=P_X$, there exists a unitary operator $\check{U}_X$ to reduce $ \norm{P\check{U}_X\sqrt{\rho_\beta}}_F$ to the infinite-temperature average such that~\cite[Claim~36]{Anshu_2021} 
 \begin{align} 
 \label{U_X_rotation_2}
 \norm{P\check{U}_X\sqrt{\rho_\beta}}_F  \ge \frac{1}{2}  \norm{P_X\sqrt{\rho_{\beta=0}}}_F  \ge  \frac{\tr_X(P_X)}{2\mathcal{D}_X}  .
\end{align}
By combining~\eqref{U_X_rotation} and \eqref{U_X_rotation_2}, we arrive at the inequality of
\begin{align} 
\label{U_X_rotation_delta_result}
\delta\ge \frac{e^{-\beta (\Delta +1)}}{4( \Delta +1)} \br{\frac{\tr_X(P_X)}{2\mathcal{D}_X} -  2 g_{\Delta,X}^2 } \for  \forall \Delta >0   .
\end{align}

Therefore, by applying the inequality~\eqref{U_X_rotation_delta_result} to Eq.~\eqref{lambda_min_P_L^co} with $X=L^\co$ and $\tr_{L^\co}\br{P_{L^\co}}=1$, we prove the lower bound of 
\begin{align}
\label{lambda_min_P_L^co_lower}
\lambda_{\min}
&\ge  \frac{e^{-\beta (\Delta +1)}}{4 (\Delta +1)} \br{\frac{1}{2\mathcal{D}_{L^\co}} -  2 e^{-(\Delta -3J_0 |L^\co|)/(2J_0k)} }\ge \frac{e^{-\beta  \bar{J}_0 |L^\co| }}{16 \mathcal{D}_{L^\co}  \bar{J}_0 |L^\co| } ,
\end{align}
where we choose $\Delta =3J_0 |L^\co|+ (2J_0 k) \log(8\mathcal{D}_{L^\co})$ and use $\Delta +1 \le \bar{J}_0 |L^\co|$ from the definition of $\bar{J}_0$ in Eq.~\eqref{def_bar_J_0_1}. 
We thus prove the main inequality~\eqref{main_ineq_prop:Norm of the effective Hamiltonian}.
This completes the proof of Proposition~\ref{prop:Norm of the effective Hamiltonian}. $\square$

\subsection{Belief propagation formalism} \label{sec:BP_formalism}

 \begin{figure}[tt]
\centering
\includegraphics[clip, scale=0.4]{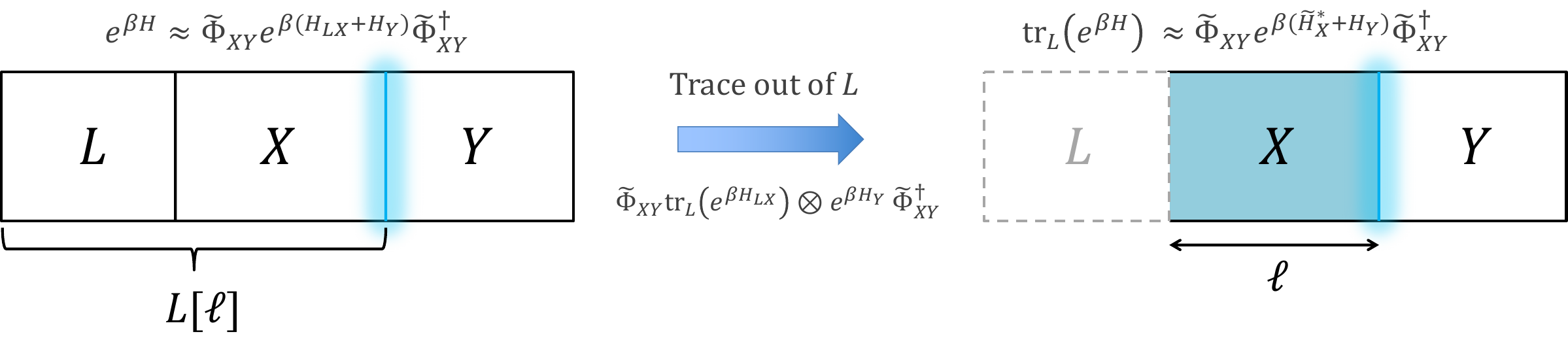}
\caption{Schematic picture of the Belief Propagation (BP) formalism. Let's decompose the total system into $\Lambda = L \cup X \cup Y$ as given in Eq.~\eqref{X:L_brr_r_Y:L_brr_r_^co}. We begin with the approximation of the belief propagation operator between the regions $LX$ and $Y$ so that it can be supported on $X$ and $Y$, which is denoted by $\tilde{\Phi}_{XY}$. Then, for the approximated Gibbs state $\tilde{\Phi}_{XY}  e^{\beta (H_{LX}+H_{L[\ell]^\co})}\tilde{\Phi}_{XY}^\dagger$, we can perform the partial trace with respect to $L$ without influencing the region $Y$, i.e.,
$\tr_{L} \brr{ \tilde{\Phi}_{XY}  e^{\beta (H_{LX}+H_{L[\ell]^\co})}\tilde{\Phi}_{XY}^\dagger}
= \tilde{\Phi}_{XY} \tr_{L} \brr{e^{\beta H_{LX}} } \otimes e^{\beta H_{L[\ell]^\co}}\tilde{\Phi}_{XY}^\dagger$. 
 Therefore, we need to calculate the entanglement Hamiltonian of $\tilde{\Phi}_{XY} e^{\beta (\tilde{H}^\ast_{X}+H_{L[\ell]^\co})} \tilde{\Phi}_{XY}^\dagger$, where $\tilde{H}^\ast_{X}$ is defined by the partial trace for $e^{\beta H_{LX}}$. Subsequently, the problem now reduces to estimating the quasi-locality for connections of exponential operators (see Sec.~\ref{Sec:Connection of the exponential operators}).
  }
\label{fig:partial_trace_belief}
\end{figure}

From this section, we consider the construction of the entanglement Hamiltonian.  
We first consider a tripartition of the total system as $\Lambda= L\sqcup X \sqcup Y$ (see Fig.~\ref{fig:partial_trace_belief}), where we define 
 \begin{align}
 \label{X:L_brr_r_Y:L_brr_r_^co}
X=L[\ell] \setminus L ,\quad Y=L[\ell]^\co ,\quad X\sqcup Y=L^\co.
\end{align}
Then, the quantum belief propagation for the bipartition of the Hamiltonian as $H=H_{L[\ell]}+H_{L[\ell]^\co} + \partial h_{L[\ell]}$ gives 
\begin{align}
 \label{X:L_brr_r_Y:L_brr_r_^co_belief_prop}
e^{\beta H} = \Phi_{\partial h_{L[\ell]}}e^{\beta \br{H_{LX}+H_{L[\ell]^\co}}}\Phi_{\partial h_{L[\ell]}}^\dagger ,
\end{align}
where we denote $L[\ell]=LX$. 
We consider the approximation of $\Phi_{\partial h_{L[\ell]}}$ onto the $XY$, which we denote by $\tilde{\Phi}_{XY}$. 
By the notation of Corollary~\ref{corol:high_dimensional_applicaton_bp}, $\tilde{\Phi}_{XY}$ is described by $\tilde{\Phi}_{\partial h_{L[\ell]}}^{(\ell-1)}$, and hence we obtain
\begin{align}
\label{norm_ineq:tr_e_beta_H/00}
\frac{1}{Z_\beta}\norm{e^{\beta H} - \tilde{\Phi}_{XY} e^{\beta \br{H_{LX}+H_{L[\ell]^\co}}}\tilde{\Phi}_{XY}^\dagger}_1 \le
13  \bar{\phi}_{\beta,|\partial L[\ell]|} \norm{\partial h_{L[\ell]}}e^{2\beta \norm{\partial h_{L[\ell]}}-\kappa_\beta (\ell-1)} ,
\end{align}
where we use the inequality~\eqref{main_ineq:corol:high_dimensional_applicaton_bp}. 
The above inequality immediately yields  
\begin{align}
\label{norm_ineq:tr_e_beta_H}
\frac{1}{Z_\beta}\norm{\tr_{L} \br{e^{\beta H}} - \tilde{\Phi}_{XY} \tr_L\brr{ e^{\beta \br{H_{LX}+H_{L[\ell]^\co}}}} \tilde{\Phi}_{XY}^\dagger}_1 \le
13  \bar{\phi}_{\beta,|\partial L[\ell]|} \norm{\partial h_{L[\ell]}}e^{2\beta \norm{\partial h_{L[\ell]}}-\kappa_\beta (\ell-1)} .
\end{align}
Hence, by defining 
\begin{align}
\label{partial_trace_tr_L_H_LX}
\tr_L\brr{ e^{\beta \br{H_{LX}+H_{L[\ell]^\co}}}} = \tr_L\br{ e^{\beta H_{LX}}} e^{\beta H_{L[\ell]^\co}}=: e^{\beta \tilde{H}^\ast_{X}} e^{\beta H_{L[\ell]^\co}} ,
\end{align}
we reduce the inequality~\eqref{norm_ineq:tr_e_beta_H} to
\begin{align}
\label{approx_BP_formalism}
\frac{1}{Z_\beta}\norm{e^{\beta H^\ast_{L^\co}} - \tilde{\Phi}_{XY} e^{\beta \br{\tilde{H}^\ast_{X}+H_{L[\ell]^\co}}} \tilde{\Phi}_{XY}^\dagger}_1 \le
13  \bar{\phi}_{\beta, |\partial L[\ell]|}  \norm{\partial h_{L[\ell]}}e^{2\beta \norm{\partial h_{L[\ell]}}-\kappa_\beta (\ell-1)} ,
\end{align}
where we use $[H_{LX},H_{L[\ell]^\co}]=0$ in defining the effective Hamiltonian $\tilde{H}^\ast_{X}$.
Note that we cannot obtain any information on the structure of $\tilde{H}^\ast_{X}$ except that it is supported on $X$. 
Subsequently, we need to consider the entanglement Hamiltonian of 
\begin{align}
\label{BP_formalism_exponential_connection_eq}
 \tilde{\Phi}_{XY} e^{\beta \br{\tilde{H}^\ast_{X}+H_{L[\ell]^\co}}} \tilde{\Phi}_{XY}^\dagger
 = 
\mathcal{T} e^{\int_0^1 \tilde{\phi}_{XY,\tau}d\tau} e^{\beta \br{\tilde{H}^\ast_{X}+H_{L[\ell]^\co}}} \br{\mathcal{T} e^{\int_0^1 \tilde{\phi}_{XY,\tau}d\tau} }^\dagger ,
\end{align}
with
\begin{align}
 \tilde{\Phi}_{XY} := 
\mathcal{T} e^{\int_0^1 \tilde{\phi}_{XY,\tau}d\tau} ,
\end{align}
which necessitates us to estimate the quasi-locality of the entanglement Hamiltonian due to the connection of exponential operators (see Sec.~\ref{Sec:Connection of the exponential operators}).
Note that $\norm{\partial h_{L[\ell]}}\propto \ell^{D-1}$, and hence we cannot obtain a good approximation in~\eqref{approx_BP_formalism} for dimensions higher than $2$.

As the second question, we are interested in whether we can efficiently approximate the conditional mutual information.
The straightforward way is to utilize the continuity inequality~\cite{10129917}.
By denoting 
\begin{align}
\label{def_tilde_rho_beta_delta}
\tilde{\rho}_{\beta}:=\frac{\tilde{\Phi}_{XY} e^{\beta \br{H_{LX}+H_{L[\ell]^\co}}} \tilde{\Phi}_{XY}}{\tr\br{\tilde{\Phi}_{XY} e^{\beta \br{H_{LX}+H_{L[\ell]^\co}}} \tilde{\Phi}_{XY}}} , \quad  \delta_\beta :=\norm{\rho_\beta -\tilde{\rho}_{\beta}  }_1, 
\end{align}
one can derive 
\begin{align}
\label{continuity_CMI_general}
\abs{\mI_{\rho_\beta}(A:C|B) -  \mI_{\tilde{\rho}_{\beta}}(A:C|B)}
\le \delta_\beta \log \brr{\min(\mathcal{D}_A,\mathcal{D}_C)} + \br{1+ \frac{\delta_\beta}{2}} \br{\frac{\delta_\beta}{2+\delta_\beta}} ,
\end{align}
where $h(x)=-x\log (x) - (1-x)\log(1-x)$ with $0\le x\le 1$. 
However, when we consider the thermodynamic limit as $|\Lambda|\to \infty$, the above upper bound may not be utilized since $\min(\mathcal{D}_A,\mathcal{D}_C)$ can be infinitely large.
To resolve this issue, we prove the following lemma:
\begin{lemma} \label{lem:entropy_bound_1D}
Let us define $H_{\rho}(A:C|B)$ as 
\begin{align}
H_{\rho}(A:C|B) =-\log(\rho_{AB}) -\log(\rho_{BC})  + \log(\rho_{ABC}) + \log(\rho_{B})
\end{align}
for an arbitrary quantum state $\rho$. Here, we do not assume that $A$, $B$ and $C$ constitute the total system, i.e., 
 \begin{align}
A\cup B\cup C \subseteq \Lambda.
\end{align}
Then, we obtain 
\begin{align}
\label{main_ineq;lem:entropy_bound_1D}
\mI_{\rho_\beta}(A:C|B) 
\le \norm{ H_{\tilde{\rho}_{\beta}}(A:C|B) }+4\norm{\beta H - \log \br{\tilde{\Phi}_{XY} e^{\beta \br{H_{LX}+H_{L[\ell]^\co}}}\tilde{\Phi}_{XY}^\dagger}}+4\delta_{\beta,\ell,L}  ,
\end{align}
where we define $\delta_{\beta,\ell,L}$ by~\eqref{norm_ineq:tr_e_beta_H/00} as follows:
\begin{align}
\label{def_delta_beta_ell}
\delta_{\beta,\ell,L}:=13  \bar{\phi}_{\beta,|\partial L[\ell]|} \norm{\partial h_{L[\ell]}}e^{2\beta \norm{\partial h_{L[\ell]}}-\kappa_\beta (\ell-1)} .
\end{align}
\end{lemma} 

{\bf Remark.} 
The second term in the RHS of~\eqref{main_ineq;lem:entropy_bound_1D} is close to the norm difference of 
$\norm{ \log(\rho_\beta) - \log(\tilde{\rho}_{\beta})}$. 
Although we can ensure $\rho_\beta \approx \tilde{\rho}_{\beta}$, the logarithm of the operator is upper-bounded by using the minimum eigenvalue $\lambda_{\min} $ of $\rho_\beta$ [see the inequality~\eqref{main_effective_Ham_Learning_2}]: 
\begin{align}
\norm{ \log(\rho_\beta) - \log(\tilde{\rho}_{\beta})} \lesssim \norm{ \rho_\beta -\tilde{\rho}_{\beta}}/\lambda_{\min} =e^{\orderof{|\Lambda|}} \norm{ \rho_\beta -\tilde{\rho}_{\beta}} ,
\end{align}
which necessitates the exponentially accurate error between $\rho_\beta$ and $\tilde{\rho}_{\beta}$ to ensure the good approximation for $\norm{ \log(\rho_\beta) - \log(\tilde{\rho}_{\beta})}$. This has been a main bottleneck to the quantum Hamiltonian learning~\cite{Anshu_2021}. 
In Proposition~\ref{prop:bp_error_est_effective}, we will give a much better error bound between $\log(\rho_\beta)$ and $\log(\tilde{\rho}_{\beta})$.

\subsubsection{Proof of Lemma~\ref{lem:entropy_bound_1D}}
We start from the following inequality that holds for arbitrarily reduced density matrices $\rho_{L^\co}$ and $\tilde{\rho}_{L^\co}$:
\begin{align}
\label{relative_entropy_upp_monotonicity}
 S\br{\rho_{L^\co}| \tilde{\rho}_{L^\co}}
 :=\tr\brr{\rho_{L^\co} \log(\rho_{L^\co})- \rho_{L^\co}\log(\tilde{\rho}_{L^\co})}
 \le S\br{\rho| \tilde{\rho}} =\tr\brr{ \rho \log (\rho) -\rho \log (\tilde{\rho}) }         ,
\end{align}
which is derived from the monotonicity of the quantum relative entropy~\cite{Lindblad1974,Uhlmann1977,10.5555/2871378.2871383}.
Then, for the conditional mutual information $\mI_{\rho_\beta}(A:C|B) = \tr\brr{\rho_\beta H_{\rho_\beta}(A:C|B)}$, we obtain 
\begin{align}
&\tr\brr{\rho_\beta H_{\rho_\beta}(A:C|B)} - \tr\brr{\rho_\beta  H_{\tilde{\rho}_{\beta}}(A:C|B)} \notag \\
&= -  S\br{\rho_{\beta,AB}| \tilde{\rho}_{\beta,AB}} -  S\br{\rho_{\beta,BC}| \tilde{\rho}_{\beta,BC}} + S\br{\rho_{\beta,ABC}| \tilde{\rho}_{\beta,ABC}} + S\br{\rho_{\beta,B}| \tilde{\rho}_{\beta,B}} .
\label{conditional_mutual_information_relative_ent_rho_diff}
\end{align}
Therefore, by applying~\eqref{relative_entropy_upp_monotonicity} to the above equation, we have 
\begin{align}
&\abs{\tr\brr{\rho_\beta H_{\rho_\beta}(A:C|B)} - \tr\brr{\rho_\beta  H_{\tilde{\rho}_{\beta}}(A:C|B)}} \le  4 S\br{\rho_\beta| \tilde{\rho}_{\beta}} \notag \\
&\longrightarrow \mI_{\rho_\beta}(A:C|B) \le  \norm{H_{\tilde{\rho}_{\beta}}(A:C|B)} + 4 S\br{\rho_\beta| \tilde{\rho}_{\beta}} . 
\label{mI_rho_beta_A:C_B_uup}
\end{align}
From the definitions of $\rho_\beta$ and $\tilde{\rho}_\beta$, we can immediately obtain 
\begin{align}
\label{relative_ent_mI_rho_beta_A:C_B_uup}
S\br{\rho_\beta| \tilde{\rho}_{\beta}} 
&= \tr \brrr{\rho_\beta \brr{\beta H - \log \br{\tilde{\Phi}_{XY} e^{\beta \br{H_{LX}+H_{L[\ell]^\co}}}\tilde{\Phi}_{XY}^\dagger}}  } -\log(Z_\beta) + \log\brr{\tr\br{\tilde{\Phi}_{XY} e^{\beta \br{H_{LX}+H_{L[\ell]^\co}}} \tilde{\Phi}_{XY}^\dagger}}   \notag \\
&\le \norm{\beta H - \log \br{\tilde{\Phi}_{XY} e^{\beta \br{H_{LX}+H_{L[\ell]^\co}}}\tilde{\Phi}_{XY}^\dagger}} +\delta_{\beta,\ell,L} ,
\end{align}
where we use the inequality~\eqref{norm_ineq:tr_e_beta_H/00} to obtain 
\begin{align}
\tr \br{\tilde{\Phi}_{XY} e^{\beta \br{H_{LX}+H_{L[\ell]^\co}}}\tilde{\Phi}_{XY}^\dagger}=\norm{\tilde{\Phi}_{XY} e^{\beta \br{H_{LX}+H_{L[\ell]^\co}}}\tilde{\Phi}_{XY}^\dagger}_1
&\le \norm{e^{\beta H} }_1+ \norm{e^{\beta H} - \tilde{\Phi}_{XY} e^{\beta \br{H_{LX}+H_{L[\ell]^\co}}}\tilde{\Phi}_{XY}^\dagger}_1 \notag \\
&\le Z_\beta (1+\delta_{\beta,\ell,L} ) .
\end{align}
By applying the inequality~\eqref{relative_ent_mI_rho_beta_A:C_B_uup} to the inequality~\eqref{mI_rho_beta_A:C_B_uup}, we reach the upper bound of~\eqref{main_ineq;lem:entropy_bound_1D}.
This completes the proof of Lemma~\ref{lem:entropy_bound_1D}. $\square$

%
%

\subsection{Partial-trace-projection formalism} \label{sec:PTP_formalism}

As has been shown, we cannot utilize the belief propagation technique to derive the entanglement Hamiltonian in high dimensions.
We thus take another route to obtain it.
The statements in this section hold for arbitrary states $\rho$, and are not restricted to thermal states $\rho_\beta$.

Let us adopt an ancilla system $L_a$ and define the maximally entangled states $\ket{\mP_L}$ between $L$ and $L_a$ as follows: 
\begin{align}
\ket{\mP_L} := \sum_{j=1}^{\mathcal{D}_L} \frac{1}{\sqrt{\mathcal{D}_L}} \ket{j_L} \otimes \ket{j_{L_a}} ,
\end{align}
which gives the partial transpose as 
\begin{align}
\label{PPT_equation_D_L}
\bra{\mP_L}\rho \otimes \hat{1}_{L_a} \ket{\mP_L} = \tilde{\tr}_L (\rho) = \frac{1}{\mathcal{D}_L}  \tr_L (\rho), 
\end{align}
where $\{\ket{j_L}\}$ and $\{\ket{j_{L_a}}\}$ are the arbitrary orthonormal bases of the Hilbert spaces on the subsystems $L$ and $L_a$, respectively. 
In the following, we omit the notation of $\hat{1}_{L_a}$ and simply denote $\rho \otimes \hat{1}_{L_a}$ by $\rho$. 

We thus define 
\begin{align}
\mP_L := \ket{\mP_L}\bra{\mP_L} , 
\end{align}
as the partial-trace projection (PTP) onto the subsystem $L$. 
Then, we approximate the PTP by the following exponential form:
\begin{align}
\label{PTP_operator_def}
\mP_{L,\tau}:= e^{-\tau \mQ_L} ,\quad \mQ_L:= 1- \mP_L .
\end{align}
By making $\tau \to \infty$, we have $\mP_{L,\tau}\to \mP_{L}$ with an exponentially small error with $\tau$. 
We indeed prove the following lemma:
\begin{lemma} \label{lem:error_est_APTP}
Let $\rho$ be an arbitrary quantum state.
Then, we obtain the norm bound of 
\begin{align}
\label{main:eq:lem:error_est_APTP}
\norm{\mP_{L,\tau} \rho \mP_{L,\tau} - \mP_{L} \rho \mP_{L} }_1 \le 2 e^{-\tau} ,
\end{align}
which also yields 
\begin{align}
\label{main2:eq:lem:error_est_APTP__2}
\norm{ \frac{ \mP_{L,\tau} \rho \mP_{L,\tau}}{\tr\br{ \mP_{L,\tau} \rho \mP_{L,\tau}}} - \rho_{L^\co} \otimes \mP_{L} }_1 \le 
4\mathcal{D}_L e^{-\tau}.
\end{align}
\end{lemma}

\textit{Proof of Lemma~\ref{lem:error_est_APTP}.}
We start with the inequality of 
\begin{align}
\norm{\mP_{L,\tau} \rho \mP_{L,\tau} - \mP_{L} \rho \mP_{L} }_1 
&= \norm{\br{ \mP_{L,\tau} - \mP_{L} }  \rho \mP_{L,\tau} + \mP_{L} \rho \br{\mP_{L,\tau} -  \mP_{L} }  }_1  \notag \\
&\le  \norm{\mP_{L,\tau} - \mP_{L}} \cdot \norm{\rho}_1 \cdot \norm{\mP_{L,\tau}} 
+ \norm{ \mP_{L} } \cdot \norm{\rho}_1 \cdot \norm{ \mP_{L,\tau} -  \mP_{L} }  \le 2 \norm{ \mP_{L,\tau} -  \mP_{L} } .
\end{align}
Using $\mP_{L,\tau}=\mP_{L,\tau}(  \mP_{L}+  \mQ_{L}) =\mP_{L}+  e^{-\tau }\mQ_{L}$, we reduce the above inequality to the main inequality~\eqref{main:eq:lem:error_est_APTP}. 
The second inequality~\eqref{main2:eq:lem:error_est_APTP__2} is immediately derived from Eq.~\eqref{PPT_equation_D_L} as follows:
\begin{align}
\label{main2:eq:lem:error_est_APTP}
\norm{ \frac{ \mP_{L,\tau} \rho \mP_{L,\tau}}{\tr\br{ \mP_{L,\tau} \rho \mP_{L,\tau}}} - \rho_{L^\co} \otimes \mP_{L} }_1
&\le \norm{ \frac{ \mP_{L,\tau} \rho \mP_{L,\tau}}{\tr\br{ \mP_{L,\tau} \rho \mP_{L,\tau}}} -  \mathcal{D}_L\mP_{L,\tau} \rho \mP_{L,\tau}}_1
+ \norm{ \mathcal{D}_L\mP_{L,\tau} \rho \mP_{L,\tau}- \rho_{L^\co} \otimes \mP_{L} }_1\notag \\
&= \abs{ \frac{1}{\tr\br{ \mP_{L,\tau} \rho \mP_{L,\tau}}} -  \mathcal{D}_L} \cdot \norm{ \mP_{L,\tau} \rho \mP_{L,\tau}}_1
+ \norm{ \mathcal{D}_L\mP_{L,\tau} \rho \mP_{L,\tau}- \mathcal{D}_L \mP_{L} \rho \mP_{L}  }_1 \notag \\
&=  \abs{1-  \mathcal{D}_L \norm{\mP_{L,\tau} \rho \mP_{L,\tau}}_1} + 2\mathcal{D}_L e^{-\tau} 
\le 4\mathcal{D}_L  e^{-\tau},
\end{align}
where, in the last inequality, we use $ \mathcal{D}_L \norm{\mP_{L,\tau} \rho \mP_{L,\tau}}_1\le \mathcal{D}_L \norm{\mP_{L} \rho \mP_{L} }_1 + 2\mathcal{D}_L  e^{-\tau}=1+2\mathcal{D}_L  e^{-\tau}$ from the inequality~\eqref{main:eq:lem:error_est_APTP}.
This completes the proof. $\square$

{~}

\hrulefill{\bf [ End of Proof of Lemma~\ref{lem:error_est_APTP}]}

{~}

We then consider the von Neumann entropy for the reduced density matrix $\rho_{L^\co}$. 
From Eq.~\eqref{PPT_equation_D_L}, we can immediately obtain
\begin{align}
S\br{\mathcal{D}_L \mP_L \rho \mP_L} = S\br{\rho_{L^\co} \otimes \mP_L} =  S\br{\rho_{L^\co}} + S\br{\mP_L}= S\br{\rho_{L^\co}} . 
\end{align}
Note that $\ket{\mP_L}$ is pure state, and hence $S\br{\mP_L}=0$. 
Now, the problem is to estimate the error of the relative entropy of
\begin{align}
\abs{ S\br{  \frac{\mP_{L,\tau} \rho \mP_{L,\tau}}{\tr\br{\mP_{L,\tau} \rho \mP_{L,\tau}}} \biggr | \mathcal{D}_L \mP_{L} \rho \mP_{L}}}  ,
\end{align}
which is connected to the conditional mutual information using Eq.~\eqref{conditional_mutual_information_relative_ent_rho_diff}.
As in the case of one-dimensional systems, we have to avoid the factor of $\log(\mathcal{D}_{L^\co})$, which appears from the continuity inequality of the relative entropy~\cite{doi:10.1063/1.2044667}.
As in the following lemma, we can obtain a better continuity bound using the property of $\mP_{L}$ (see Sec.~\ref{sec:Proof of Lemma_lem:error_est_APTP_continuity_bound} for the proof). 
\begin{lemma}\label{lem:error_est_APTP_continuity_bound}
Let $\rho$ be an arbitrary density matrix and $\rho_{L^\co,\tau}$ be defined as 
 \begin{align}
 \label{definition_rho_L^co_tau_PTP}
\rho_{L^\co,\tau}:=  \frac{\mP_{L,\tau} \rho \mP_{L,\tau}}{\tr\br{\mP_{L,\tau} \rho \mP_{L,\tau}}} 
=\frac{e^{-\tau \mQ_L}\rho e^{-\tau \mQ_L}}{\tr\br{e^{-\tau \mQ_L} \rho e^{-\tau \mQ_L} }}  .
\end{align}
Then, we obtain the upper bound of
\begin{align}
\label{ineq:lem:error_est_APTP_continuity_bound}
S\br{\rho_{L^\co,\infty}|\rho_{L^\co,\tau}}\le 16\mathcal{\chi}_{\tau,\rho,L^\co} \mathcal{D}_L e^{-\tau }  , 
\end{align}
where $\mathcal{\chi}_{\tau,\rho,L^\co}$ is defined as 
 \begin{align}
\mathcal{\chi}_{\tau,\rho,L^\co}:=  \sup_{u_L} \norm{\brr{\log\br{\rho_{L^\co,\tau}}, u_L}} + \int_{\tau}^\infty   e^{\tau-\tau_1}\sup_{u_L} \norm{\brr{\log\br{\rho_{L^\co,\tau_1}}, u_L}} d\tau_1 .
\label{mathcal_chi_tau_rho_L^_co}
\end{align}
\end{lemma}

We now combine this lemma with a similar inequality to Eq.~\eqref{conditional_mutual_information_relative_ent_rho_diff}, i.e.,
\begin{align}
\label{ineq_on_rho_A_C_B_cond_M}
\tr\brr{\rho H_{\rho}(A:C|B)} 
&= -  S\br{\rho_{AB,\infty}|\rho_{AB,\tau}} -  S\br{\rho_{BC,\infty}|\rho_{BC,\tau}} + S\br{\rho_{ABC,\infty}| \rho_{ABC,\tau}} + S\br{\rho_{B,\infty}|\rho_{B,\tau}} \notag \\ 
&\quad + \tr\brr{\rho  H_{\rho,\tau}(A:C|B)},
\end{align}
where $H_{\rho,\tau}(A:C|B)$ is defined as 
 \begin{align}
H_{\rho,\tau}(A:C|B):=-  \mP_C \log(\rho_{AB,\tau}) \mP_C - \mP_A \log(\rho_{BC,\tau}) \mP_A +\log(\rho_{ABC,\tau})+ \mP_{AC}\log(\rho_{B,\tau})\mP_{AC} .
\end{align}
Note that from $\rho_{AB,\infty} =\rho_{C^\co,\infty}=  \mP_C \rho  \mP_C$ [see Eq.~\eqref{definition_rho_L^co_tau_PTP}], we have
 \begin{align}
S\br{\rho_{AB}} 
= S\br{\rho_{AB,\infty}} &= - S\br{\rho_{AB,\infty}|\rho_{AB,\tau}} - \tr \brr{ \rho_{AB,\infty} \log\br{ \rho_{AB,\tau}}} \notag \\
&= - S\br{\rho_{AB,\infty}|\rho_{AB,\tau}} - \tr \brr{\rho \mP_C \log\br{ \rho_{AB,\tau}} \mP_C } .
\end{align}
From Lemma~\ref{lem:error_est_APTP_continuity_bound} and the inequality~\eqref{ineq_on_rho_A_C_B_cond_M}, we can also upper-bound the conditional mutual information, which is given as follows:
\begin{corol}\label{corol:error_est_APTP_continuity_bound_CMI}
For an arbitrary tripartition of $\Lambda=A\sqcup B \sqcup C$, the quantum conditional mutual information $\mI_{\rho}(A:C|B)$ is upper-bounded as follows:
\begin{align}
\label{ineq:corol:error_est_APTP_continuity_bound_CMI}
\mI_{\rho}(A:C|B) \le  16 e^{-\tau } \br{\mathcal{D}_C\mathcal{\chi}_{\tau,\rho,AB} + \mathcal{D}_A\mathcal{\chi}_{\tau,\rho,BC} +\mathcal{D}_{AC} \mathcal{\chi}_{\tau,\rho,B}} 
 + \norm{H_{\rho, \tau}(A:C|B)} .
\end{align}
For the proof, we use $\rho_{L^\co,\tau}= \rho$ for $L=\emptyset$ (or $L^\co=\Lambda$), which gives $S\br{\rho_{ABC}|\rho_{ABC,\tau}}=0$. 
\end{corol}

{\bf Remark.}
To estimate the RHS of the inequality~\eqref{ineq:corol:error_est_APTP_continuity_bound_CMI}, it is crucial to estimate $\mathcal{\chi}_{\tau,\rho,L^\co}$ in Eq.~\eqref{mathcal_chi_tau_rho_L^_co}. 
When we consider the quantum Gibbs state, 
as will be shown in Corollary~\ref{connection_of_exponential_operator_repeat} [see also Eq.~\eqref{main_eq:connection_of_exponential_operator_corol}] and Subtheorem~\ref{sub_thm:U_tau_u_i_commun},
the logarithm of $\rho_{\beta,L^\co,\tau}$ is expressed in the form of $ U_\tau( \beta H_0 + \hat{V}_{\tau} )U_\tau^\dagger $ with $V=Q_L$.
The theorem implies that the non-locality of the unitary operator $U_\tau$ linearly increases with $\tau$, and hence the commutator norm $\norm{\brr{\log\br{\rho_{L^\co,\tau}}, u_L}}=\norm{\brr{\beta H_0 + \hat{V}_{\tau} , U_\tau^\dagger  u_L U_\tau}}$ is roughly upper-bounded
by $\Theta(\beta |L[\tau]|) =\Theta(\beta|L| \tau^D)$. 
Thus, the integral with respect to $\tau_1$ converges as $\int_{\tau}^\infty e^{\tau -\tau_1} \Theta(\beta|L| \tau_1^D)d\tau_1$, which makes $\mathcal{\chi}_{\tau,\rho,L^\co}$ of order of $\beta|L| \tau^D$~\footnote{The actual estimation of $\mathcal{\chi}_{\tau,\rho,L^\co}$ is provided in Lemma~\ref{lemma:upp_bound_chi/tau/rho_beta_Lco}, which is worse than $\orderof{\beta|L| \tau^D}$ but is still $\poly(\beta, |L|,\tau)$}.

Thus, the first term in the RHS of~\eqref{ineq:corol:error_est_APTP_continuity_bound_CMI}, i.e., $ 16 e^{-\tau } \br{\mathcal{D}_C\mathcal{\chi}_{\tau,\rho,AB} + \mathcal{D}_A\mathcal{\chi}_{\tau,\rho,BC} +\mathcal{D}_{AC} \mathcal{\chi}_{\tau,\rho,B}} $, becomes sufficiently small by choosing $\tau\gtrsim |AC|$. 
However, when $|A|$ or $|C|$ is macroscopically large (i.e., $|A|,|C|=\orderof{|\Lambda|}$), we cannot obtain a meaningful upper bound for the second term~$\norm{H_{\rho, \tau}(A:C|B)} $ (see Lemma~\ref{lemma:upp_bound_norm_CMI_Hamiltonian} below).

\subsubsection{Proof of Lemma~\ref{lem:error_est_APTP_continuity_bound}} \label{sec:Proof of Lemma_lem:error_est_APTP_continuity_bound}
For an arbitrary base $\{\ket{\phi_s}\}_s$, the von Neumann entropy $S(\rho)$ is upper-bounded by~\cite{schur1923uber,Horn1954,10.5555/2011326.2011331}
\begin{align}
\label{ineq:majorization_entropy_diagonal}
S(\rho) \le - \sum_{s} p_{\phi,s} \log(p_{\phi,s} ) ,\quad \sum_s p_{\phi,s}=1,
\end{align}
where $p_{\phi,s}:= \bra{\phi_s} \rho \ket{\phi_s}$.
Note that the equation is achieved iff $\{\ket{\phi_s}\}_s$ are given by the eigenbase of $\rho$. 

We start with the spectral decomposition of $\rho_{L^\co,\tau}$ as 
\begin{align}
\label{rho_tau_spectral_decomp_s}
\rho_{L^\co,\tau} = \sum_s \lambda_s \ket{s} \bra{s} ,
\end{align}
where $\{\ket{s}\}$ are the eigenstate states of $\rho_{L^\co,\tau}$. 
We consider 
\begin{align}
\label{rho_tau_+_dtau_eq}
\rho_{L^\co,\tau +d\tau} = \frac{e^{-d\tau \mQ_L}  e^{-\tau \mQ_L} \rho  e^{-\tau \mQ_L}e^{-d\tau \mQ_L} }{\tr\br{ e^{-d\tau \mQ_L}  e^{-\tau \mQ_L} \rho  e^{-\tau \mQ_L}e^{-d\tau \mQ_L} } }.
\end{align}
We obtain 
\begin{align}
\tr\br{ e^{-d\tau \mQ_L}  e^{-\tau \mQ_L} \rho  e^{-\tau \mQ_L}e^{-d\tau \mQ_L} } 
=& \tr\br{ e^{-\tau \mQ_L} \rho  e^{-\tau \mQ_L} }-  \tr\br{\mQ_L e^{-\tau \mQ_L} \rho  e^{-\tau \mQ_L} }2d\tau +\orderof{d\tau^2} \notag \\
=& \tr\br{ e^{-\tau \mQ_L} \rho  e^{-\tau \mQ_L} } \brr{1 -2 \tr\br{\mQ_L \rho_{L^\co,\tau}} d\tau } +\orderof{d\tau^2}  ,
\end{align}
which reduces Eq.~\eqref{rho_tau_+_dtau_eq} to 
\begin{align}
\label{rho_tau_+_dtau_eq_reduce}
\rho_{L^\co,\tau +d\tau} = \rho_{L^\co,\tau} -  \{\rho_{L^\co,\tau},  \mQ_L\} d \tau + 2\tr\br{\mQ_L \rho_{L^\co,\tau}}  \rho_{L^\co,\tau} d\tau +\orderof{d\tau^2} .
\end{align}

Using Eq.~\eqref{rho_tau_spectral_decomp_s}, we calculate
\begin{align}
\bra{s}\rho_{L^\co,\tau +d\tau} \ket{s}
&=\lambda_s- 2\lambda_s \bra{s}\mQ_L \ket{s} d\tau + 2 \lambda_s \tr\br{\mQ_L \rho_{L^\co,\tau}}  d\tau  +\orderof{d\tau^2}  ,
\end{align}
and 
\begin{align}
\log \br{\bra{s}\rho_{L^\co,\tau +d\tau} \ket{s}} 
&=\log \br{\lambda_s} + \log\br{1- 2\bra{s}\mQ_L \ket{s} d\tau + 2 \tr\br{\mQ_L \rho_{L^\co,\tau}}  d\tau  +\orderof{d\tau^2} }  \notag \\
&= \log \br{\lambda_s} +  2\bra{s}\mQ_L \ket{s} d\tau - 2 \tr\br{\mQ_L \rho_{L^\co,\tau}}  d\tau  +\orderof{d\tau^2}  .
\end{align}
Then, by using the inequality~\eqref{ineq:majorization_entropy_diagonal}, we upper-bound $S(\rho_{L^\co,\tau +d\tau})$ as 
\begin{align}
\label{inq_S_tau_d_tau_^2}
&S(\rho_{L^\co,\tau +d\tau}) \le - \sum_s  \bra{s}\rho_{L^\co,\tau +d\tau} \ket{s} \log\br{\bra{s}\rho_{L^\co,\tau +d\tau} \ket{s}} \notag \\
&=  - \sum_s \brr{\lambda_s- 2\lambda_s \bra{s}\mQ_L \ket{s} d\tau + 2 \lambda_s \tr\br{\mQ_L \rho_{L^\co,\tau}}  d\tau }
\cdot \brr{\log \br{\lambda_s} +  2\bra{s}\mQ_L \ket{s} d\tau - 2 \tr\br{\mQ_L \rho_{L^\co,\tau}}  d\tau  }  +\orderof{d\tau^2}   \notag \\
&= S(\rho_{L^\co,\tau}) -2 d\tau  \sum_s \lambda_s\brr{\bra{s}\mQ_L \ket{s} -  \tr\br{\mQ_L \rho_{L^\co,\tau}} }   +
2 d\tau \sum_s  \lambda_s  \brr{\bra{s}\mQ_L \ket{s} -  \tr\br{\mQ_L \rho_{L^\co,\tau}}} \log(\lambda_s) +\orderof{d\tau^2}  \notag \\
&= S(\rho_{L^\co,\tau})  +
2 d\tau \sum_s  \lambda_s  \brr{\bra{s}\mQ_L \ket{s} -  \tr\br{\mQ_L \rho_{L^\co,\tau}}} \log(\lambda_s) +\orderof{d\tau^2} , 
\end{align}
where we use $ \sum_s \lambda_s\brr{\bra{s}\mQ_L \ket{s} -  \tr\br{\mQ_L \rho_{L^\co,\tau}} }= 0$
from $ \sum_s \lambda_s\bra{s}\mQ_L \ket{s}= \tr \br{\mQ_L \sum_s \lambda_s\ket{s}\bra{s} }= \tr \br{\mQ_L \rho_{L^\co,\tau}}$
 in the last equation. 
For the second term in the RHS of~\eqref{inq_S_tau_d_tau_^2}, we have 
\begin{align}
\label{inq_S_tau_d_tau_^2_seconed:term}
\sum_s  \lambda_s  \brr{\bra{s}\mQ_L \ket{s} -  \tr\br{\mQ_L \rho_{L^\co,\tau}}} \log(\lambda_s) 
&= \sum_s \bra{s}\mQ_L \rho_{L^\co,\tau} \log\br{\rho_{L^\co,\tau}}\ket{s} -  \tr\br{\mQ_L \rho_{L^\co,\tau}} \cdot  \tr\brr{ \rho_{L^\co,\tau} \log(\rho_{L^\co,\tau})} \notag \\
&=\tr\brr{\mQ_L \rho_{L^\co,\tau} \log\br{\rho_{L^\co,\tau}}}  -  \tr\br{\mQ_L \rho_{L^\co,\tau}} \cdot  \tr\brr{ \rho_{L^\co,\tau} \log(\rho_{L^\co,\tau})}.
\end{align}
Combining~\eqref{inq_S_tau_d_tau_^2} and~\eqref{inq_S_tau_d_tau_^2_seconed:term}, we obtain 
\begin{align}
\label{inq_S_tau_d_tau_^2_upp}
\abs{ \frac{S(\rho_{L^\co,\tau})}{d\tau}} \le
 2 \abs{ \tr\brrr{ \brr{\mQ_L -\tr\br{\mQ_L \rho_{L^\co,\tau}} } \rho_{L^\co,\tau} \log\br{\rho_{L^\co,\tau}}}  }.
\end{align}
The upper bound is expressed as a bipartite correlation between $\log(\rho_{L^\co,\tau})$ and $\mathcal{Q}_L$, which is upper-bounded by $\mathcal{O}(|L^\co|)$ in general, as the entanglement Hamiltonian for $\rho_{L^\co,\tau}$ is supported on $L^\co$. If the clustering theorem holds, only the surface term around the region $L$ of $\log(\rho_{L^\co,\tau})$ contributes to the upper bound~\eqref{inq_S_tau_d_tau_^2_upp}. Unfortunately, we cannot exploit the clustering property, as high-dimensional Gibbs states may exhibit long-range correlations at thermal critical points.

To estimate the RHS of the inequality~\eqref{inq_S_tau_d_tau_^2_upp}, we first prove
\begin{align}
\label{O_Z_exp_mQL_eqq_0}
\tr\brrr{ \brr{\mQ_L -\tr\br{\mQ_L \rho_{L^\co,\tau}} } \rho_{L^\co,\tau} O_{L^\co}}  =0
\end{align}
for an arbitrary operator $O_{L^\co}$ that is supported on $L^\co$. 
Because of $\tr_L \br{ \cdots  O_{L^\co}} =O_{L^\co} \tr_L \br{ \cdots }$, we have 
\begin{align}
 \tr \br{\mQ_L \rho_{L^\co,\tau} O_{L^\co}} 
 &=  \tr_{L^\co} \brrr{ O_{L^\co}\tr_L \brr{\br{1-\mP_L} \rho_{L^\co,\tau}}  }  \notag \\
 &=  \tr_{L^\co} \brr{ O_{L^\co}\tr_L \br{ \rho_{L^\co,\tau} - \mP_L \rho_{L^\co,\tau} \mP_L}  } \notag \\
  &=\tr \br{\rho_{L^\co,\tau} - \mP_L \rho_{L^\co,\tau} \mP_L } \tr_{L^\co}\brr{O_{L^\co} \tr_L\br{\rho_{L^\co,\tau}}} 
  = \tr \br{\mQ_L \rho_{L^\co,\tau} } \tr\br{ O_{L^\co} \rho_{L^\co,\tau}} ,
\end{align}
where we use $\tr_L\br{\mP_L \rho_{L^\co,\tau} \mP_L}\propto \tr_L\br{\rho_{L^\co,\tau}}$ in the third equation from Eq.~\eqref{PPT_equation_D_L} and 
$\norm{\rho_{L^\co,\tau} - \mP_L \rho_{L^\co,\tau} \mP_L }_1=\tr\br{\mQ_L\rho_{L^\co,\tau} }$ in the fourth equation. 
The above equation immediately yields Eq.~\eqref{O_Z_exp_mQL_eqq_0}. 

By using the fact that the operator $\tilde{\tr}_{L} \brr{ \log\br{\rho_{L^\co,\tau}}}$ is supported on $L^\co$, we obtain from Eq.~\eqref{O_Z_exp_mQL_eqq_0} 
\begin{align}
\label{inq_S_tau_d_tau_^2_upp_abs_RHS}
 2 \abs{ \tr\brrr{ \brr{\mQ_L -\tr\br{\mQ_L \rho_{L^\co,\tau}} } \rho_{L^\co,\tau} \log\br{\rho_{L^\co,\tau}}}  }
 &= 2 \abs{ \tr\brrr{ \brr{\mQ_L -\tr\br{\mQ_L \rho_{L^\co,\tau}} } \rho_{L^\co,\tau}  \br{ \log\br{\rho_{L^\co,\tau}} -\tilde{\tr}_{L} \brr{ \log\br{\rho_{L^\co,\tau}}}   }  }  } \notag \\
 &\le 2 \norm{ \brr{\mQ_L -\tr\br{\mQ_L \rho_{L^\co,\tau}} } \rho_{L^\co,\tau}}_1 \cdot \norm{\log\br{\rho_{L^\co,\tau}} -\tilde{\tr}_{L} \brr{ \log\br{\rho_{L^\co,\tau}}}   } \notag \\
 &\le 4 \tr\br{\mQ_L \rho_{L^\co,\tau}} \norm{ \log\br{\rho_{L^\co,\tau}} -\tilde{\tr}_{L} \brr{ \log\br{\rho_{L^\co,\tau}}} } \notag \\
 &\le 16\mathcal{D}_L e^{-\tau}\norm{ \log\br{\rho_{L^\co,\tau}} -\tilde{\tr}_{L} \brr{ \log\br{\rho_{L^\co,\tau}}} } , 
\end{align}
where, in the last inequality, we use 
\begin{align}
\tr\br{\mQ_L \rho_{L^\co,\tau}}= 1- \tr\br{\mP_L \rho_{L^\co,\tau} \mP_L} = 1- \norm{\mP_L \rho_{L^\co,\tau} \mP_L}_1  
&\le 1- \br{\norm{\rho_{L^\co,\infty}}_1 -  \norm{\mP_L \rho_{L^\co,\tau} \mP_L-\rho_{L^\co,\infty}}_1 }   \notag \\
&= \norm{\mP_L \rho_{L^\co,\tau} \mP_L-\rho_{L^\co,\infty}}_1 \le 4\mathcal{D}_L e^{-\tau},
\end{align}
which is derived from $\norm{\mP_L \rho_{L^\co,\tau} \mP_L
-\rho_{L^\co,\infty}}_1 \le \norm{\rho_{L^\co,\tau}-\rho_{L^\co,\infty}}_1  \le 4\mathcal{D}_L e^{-\tau}$ by~\eqref{main2:eq:lem:error_est_APTP}. 
Note that $\rho_{L^\co,\infty}=\rho_{L^\co} \otimes \mP_L = \mP_L \rho_{L^\co,\infty} \mP_L $.
By applying the inequality~\eqref{inq_S_tau_d_tau_^2_upp_abs_RHS} to~\eqref{inq_S_tau_d_tau_^2_upp} and integrating it with $\tau$, we obtain 
\begin{align}
\label{inq_S_tau_d_tau_^2_upp_fin}
\abs{S(\rho_{L^\co,\infty}) - S(\rho_{L^\co,\tau})} 
\le \int_{\tau}^\infty \abs{ \frac{S(\rho_{L^\co,\tau_1})}{d\tau_1}} d\tau_1
&\le 16\mathcal{D}_L \int_{\tau}^\infty   e^{-\tau_1}\norm{ \log\br{\rho_{L^\co,\tau_1}} -\tilde{\tr}_{L} \brr{ \log\br{\rho_{L^\co,\tau_1}}} } d\tau_1  \notag \\
&\le 16\mathcal{D}_L \int_{\tau}^\infty   e^{-\tau_1}\sup_{u_L} \norm{\brr{\log\br{\rho_{L^\co,\tau_1}}, u_L}}  d\tau_1  .
\end{align}

To connect the upper bound~\eqref{inq_S_tau_d_tau_^2_upp_fin} to $S(\rho_{L^\co,\infty}|\rho_{L^\co,\tau})$, we need to estimate 
\begin{align}
\label{tr_rho_inf_tau_rho_tau_rho_tau_tr_dif1}
\tr \brr{\rho_{L^\co,\infty} \log \br{\rho_{L^\co,\tau}} - \rho_{L^\co,\tau} \log \br{\rho_{L^\co,\tau}}  } 
= \int_\tau^\infty  \tr \brr{ \br{ \rho_{L^\co,\tau_1+d\tau_1} -   \rho_{L^\co,\tau_1} } \log \br{\rho_{L^\co,\tau}}}   .
\end{align}
By using Eq.~\eqref{rho_tau_+_dtau_eq_reduce}, we obtain 
\begin{align}
\label{tr_rho_inf_tau_rho_tau_rho_tau_tr_dif2}
\tr \brr{ \br{ \rho_{L^\co,\tau_1+d\tau_1} -   \rho_{L^\co,\tau_1} } \log \br{\rho_{L^\co,\tau}}}
=\tr \brr{ \br{ -  \{\rho_{L^\co,\tau_1},  \mQ_L\} + 2\tr\br{\mQ_L \rho_{L^\co,\tau_1}}  \rho_{L^\co,\tau_1} } \log \br{\rho_{L^\co,\tau}}}  d\tau_1 .
\end{align}
Using the same analyses as in the inequality~\eqref{inq_S_tau_d_tau_^2_upp_abs_RHS}, we have 
\begin{align}
\label{tr_rho_inf_tau_rho_tau_rho_tau_tr_dif3}
\abs{\tr \brr{ \br{ -  \{\rho_{L^\co,\tau_1},  \mQ_L\} + 2\tr\br{\mQ_L \rho_{L^\co,\tau_1}}  \rho_{L^\co,\tau_1} } \log \br{\rho_{L^\co,\tau}}}   } 
&\le  2 \abs{\tr \brrr{ \brr{  \mQ_L -\tr\br{\mQ_L \rho_{L^\co,\tau_1}}}  \rho_{L^\co,\tau_1}  \log \br{\rho_{L^\co,\tau}}}   }  \notag \\
&\le16\mathcal{D}_L e^{-\tau_1}\norm{ \log\br{\rho_{L^\co,\tau}} -\tilde{\tr}_{L} \brr{ \log\br{\rho_{L^\co,\tau}}} }  .
\end{align}
By applying Eq.~\eqref{tr_rho_inf_tau_rho_tau_rho_tau_tr_dif2} and the inequality~\eqref{tr_rho_inf_tau_rho_tau_rho_tau_tr_dif3} to Eq.~\eqref{tr_rho_inf_tau_rho_tau_rho_tau_tr_dif1}, we obtain 
\begin{align}
\label{tr_rho_inf_tau_rho_tau_rho_tau_tr_dif__fin}
\abs{\tr \brr{\rho_{L^\co,\infty} \log \br{\rho_{L^\co,\tau}} - \rho_{L^\co,\tau} \log \br{\rho_{L^\co,\tau}}  } } 
&\le 16\mathcal{D}_L\norm{ \log\br{\rho_{L^\co,\tau}} -\tilde{\tr}_{L} \brr{ \log\br{\rho_{L^\co,\tau}}} } \int_\tau^\infty   e^{-\tau_1}d\tau_1   \notag \\
&\le 16\mathcal{D}_L e^{-\tau} \sup_{u_L} \norm{\brr{\log\br{\rho_{L^\co,\tau}}, u_L}}  . 
\end{align}
By combining the inequalities~\eqref{inq_S_tau_d_tau_^2_upp_fin} and~\eqref{tr_rho_inf_tau_rho_tau_rho_tau_tr_dif__fin}, 
we prove the main inequality~\eqref{ineq:lem:error_est_APTP_continuity_bound} from
\begin{align}
S\br{\rho_{L^\co,\infty}|\rho_{L^\co,\tau}}
&= \tr\brr{\rho_{L^\co,\infty} \log\br{\rho_{L^\co,\infty}} - \rho_{L^\co,\infty} \log\br{\rho_{L^\co,\tau}}} \notag \\
&= \tr\brr{\rho_{L^\co,\infty} \log\br{\rho_{L^\co,\infty}} -  \rho_{L^\co,\tau} \log\br{\rho_{L^\co,\tau}} } 
-\tr\brr{\rho_{L^\co,\infty} \log\br{\rho_{L^\co,\tau}}- \rho_{L^\co,\tau} \log\br{\rho_{L^\co,\tau}}}  \notag \\
&\le  \abs{S(\rho_{L^\co,\infty}) - S(\rho_{L^\co,\tau})} + \abs{\tr \brr{\rho_{L^\co,\infty} \log \br{\rho_{L^\co,\tau}} - \rho_{L^\co,\tau} \log \br{\rho_{L^\co,\tau}}  } } .
\end{align}
This completes the proof. $\square$

{~}

\hrulefill{\bf [ End of Proof of Lemma~\ref{lem:error_est_APTP_continuity_bound}]}

{~}


\section{Connection of exponential operators}\label{Sec:Connection of the exponential operators}

In the practical use of the BP formalism or the PTP formalism for the entanglement Hamiltonian as shown in Sec.~\ref{sec:effective Hamiltonian theory on subsystem}, we have to estimate the logarithm of the product of exponential operators. 
To make the problem clearer, we here consider a set of operators $\{\mB_j\}_{j=1}^N$ and calculate the logarithm of
\begin{align}
\label{def:psi_N_connect}
\Psi_N := e^{\epsilon \mB_N} \cdots  e^{\epsilon \mB_1} e^{\beta \mA} e^{\epsilon  \mB_1}\cdots  e^{\epsilon \mB_N} \quad (N \propto 1/\epsilon),
\end{align}
where we choose $\epsilon$ infinitesimally small, which makes $N$ infinitely large (i.e., $N \to \infty$).

In the BP formalism for one-dimensional systems (see Sec.~\ref{sec:BP_formalism}), we need to consider the logarithm of Eq.~\eqref{BP_formalism_exponential_connection_eq}, that is,
\begin{align}
\label{BP_formalism_exponential_connection_eq_log}
 \log\br{\tilde{\Phi}_{XY} e^{\beta \br{\tilde{H}^\ast_{X}+H_{L[\ell]^\co}}} \tilde{\Phi}_{XY}^\dagger}
 =\log\brr{ 
\mathcal{T} e^{\int_0^1 \tilde{\phi}_{XY,\tau}d\tau} e^{\beta \br{\tilde{H}^\ast_{X}+H_{L[\ell]^\co}}} \br{\mathcal{T} e^{\int_0^1 \tilde{\phi}_{XY,\tau}d\tau} }^\dagger} .
\end{align}
Therefore, we choose 
\begin{align}
\label{BP_formalism_exponential_connection_eq_log_choice}
\mA \to \tilde{H}^\ast_{X}+H_{L[\ell]^\co} , \quad \{\mB_j\}_{j=1}^N := \{\tilde{\phi}_{XY,j/N}\}_{j=1}^N,\quad \epsilon=\frac{1}{N}  \for N\to \infty .
\end{align}
where we define $\{\mB_j\}_{j=1}^N$ as the discretization of the function $\tilde{\phi}_{XY,\tau}$.

On the other hand, in the PTP formalism for high-dimensional systems (see Sec.~\ref{sec:PTP_formalism}), we need to consider the logarithm of 
Eq.~\eqref{definition_rho_L^co_tau_PTP} with the choice of $\rho=e^{\beta H}$:
\begin{align}
\label{PTP_formalism_exponential_connection_eq_log}
 \log\br{\mP_{L,\tau} e^{\beta H} \mP_{L,\tau} }
 =\log\br{e^{-\tau \mQ_L}  e^{\beta H} e^{-\tau \mQ_L}} ,
\end{align}
where we use the definition of Eq.~\eqref{PTP_operator_def}, i.e., $\mP_{L,\tau}:= e^{-\tau \mQ_L}$ and $\mQ_L:= 1- \mP_L $.
We thus choose 
\begin{align}
\mA \to H , \quad \{\mB_j\}_{j=1}^N := \mQ_L,\quad \epsilon=\frac{\tau}{N}  \for N\to \infty .
\end{align}

Our purpose here is to investigate the property of $\log(\Psi_N)$, which is derived from the sequential estimations of the logarithm of  
\begin{align}
\label{def:psi_N_connect_m_to_m+1}
\Psi_{m+1} = e^{\epsilon \mB_{m+1}} \Psi_{m} e^{\epsilon \mB_{m+1}} =e^{\epsilon \mB_{m+1}} e^{\log(\Psi_{m})} e^{\epsilon \mB_{m+1}}  .
\end{align}
For this purpose, we prove the following convenient lemma, which plays a key role in our analyses:
\begin{lemma} \label{connection_of_exponential_operator}
For arbitrary operators in the form of 
\begin{align}
 e^{\epsilon  \mB} e^{\beta \mA} e^{\epsilon  \mB} ,
\end{align}
we obtain the logarithm as 
\begin{align}
\label{main_eq:connection_of_exponential_operator}
&\log( e^{\epsilon \mB} e^{\beta \mA} e^{\epsilon \mB} ) =\beta  e^{-2i \epsilon \mathcal{C}} \mA e^{2i \epsilon \mathcal{C}} + 2\epsilon  \mB + \orderof{\epsilon^2}, \\
&\mathcal{C} := \frac{1}{\beta}\int_{-\infty}^\infty g_\beta(t) e^{i\mA t} \mB e^{-i\mA t}dt.
\end{align}
where $g_\beta(t)$ has been defined in the context of the belief propagation~\eqref{def_g_t_exp_decay}.
\end{lemma}

\subsubsection{Proof of Lemma~\ref{connection_of_exponential_operator}}
For the proof, we start from the belief propagation~\eqref{sup_Def:Phi_0_phi_Hastings} as follows:
\begin{align}
\label{sup_Def:Phi_0_phi_Hastings_starting point}
e^{\beta \mA+ 2 \epsilon \mB} = \hat{\Phi}_{\mB,\epsilon} e^{\beta \mA} \hat{\Phi}_{\mB,\epsilon}^\dagger,
\end{align}
where $\hat{\Phi}_{\mB,\epsilon} $ is obtained from Eqs.~\eqref{eq_belief_propagation_operator2} and \eqref{sup_Def:Phi_0_phi_Hastings}
\begin{align}
\hat{\Phi}_{\mB,\epsilon} = \exp \left[  \epsilon \mB  + \frac{2 i\epsilon}{\beta} \int_{-\infty}^\infty g_\beta(t) e^{i\mA t} \mB e^{-i\mA t}dt \right] 
=  e^{2 i \epsilon \mathcal{C}} e^{\epsilon \mB}  + \orderof{\epsilon^2} ,
\end{align}
where we use the definition of the operator $\mC$.
By applying the explicit form of $\hat{\Phi}_{\mB,\epsilon}$ to Eq.~\eqref{sup_Def:Phi_0_phi_Hastings_starting point}, we have 
\begin{align}
 e^{\epsilon  \mB} e^{\beta \mA} e^{\epsilon \mB} &=  e^{-2i \epsilon \mathcal{C}} e^{\beta \mA+ 2 \epsilon \beta \mB} e^{2i \epsilon \mathcal{C}} + \orderof{\epsilon^2} \notag \\
 &=\exp\left[ \beta e^{-2i \epsilon \mathcal{C}} \mA e^{2i \epsilon \mathcal{C}} + 2\epsilon \mB  + \orderof{\epsilon^2}\right] .
\end{align}
This completes the proof. $\square$

{~}

\hrulefill{\bf [ End of Proof of Lemma~\ref{connection_of_exponential_operator}]}
{~}

{~}\\

\noindent
{\bf Remark.}
We emphasize that the same analyses cannot be applied to $e^{\beta \mA} e^{\epsilon  \mB}$. 
Usually, the Baker-Campbell-Hausdorff formula~\eqref{BCH_formula_starting_point} gives 
\begin{align}
\log(e^{\beta \mA} e^{\epsilon  \mB})= 
\beta \mA  +\epsilon \br{ \frac{\beta \ad_{\mA}}{1-e^{-\beta \ad_{\mA}}} \mB }  + \orderof{\epsilon^2}.
\end{align}
Here, the Taylor expansion of the operator $\frac{\ad_{\mA}}{1-e^{-\beta \ad_{\mA}}} B$ gives
\begin{align}
\frac{\ad_{\mA}}{1-e^{-\beta \ad_{\mA}}} \mB = \sum_{m=0}^\infty \frac{\mB_m}{m!} (\beta \ad_{\mA})^m B,
\end{align}
where $\{\mB_m\}_{m=0}^\infty$ are the Bernoulli numbers, which grow as $(c m)^{\orderof{m}}$ with $c$ a positive constant.
Unfortunately, we cannot utilize the expansion because it is not a convergent series unless $\|\mA\|$ and $\|\mB\|$ are sufficiently small.
Due to this fact, theoretical analyses of the quasi-locality estimation based on the  BCH expansion are notoriously challenging.
On the other hand, if we apply the BCH formula to $\log(e^{\epsilon  \mB} e^{\beta \mA} e^{\epsilon  \mB})$, we have 
\begin{align}
\log(e^{\epsilon \mB} e^{\beta \mA} e^{\epsilon  \mB})= 
\beta \mA  +\epsilon \br{ \frac{\beta \ad_{\mA}}{1-e^{-\beta \ad_{\mA}}} \mB +{\rm h.c.}}  + \orderof{\epsilon^2},
\end{align}
where we use the same equation in Eq.~\eqref{BCH_formula_starting_point} for the derivation of the belief propagation.
By applying the spectral expressions as in Eq.~\eqref{B_omega_B_eq_} and  \eqref{express_mB_omega_fourier}, we can derive 
\begin{align}
\frac{\beta \ad_{\mA}}{1-e^{-\beta \ad_{\mA}}} \mB +{\rm h.c.} = 
\int_{-\infty}^\infty  \mB(\mA,t) \br{ \frac{1}{2\pi} \int_{-\infty}^\infty  \frac{\beta\omega}{\tanh(\beta \omega/2)} e^{-i\omega t} d\omega } dt ,
\end{align}
which gives the equivalent expression to Eq.~\eqref{main_eq:connection_of_exponential_operator}. 
The primary difference in considering $\log(e^{\beta \mA} e^{\epsilon  \mB})$ is that we need to take 
the Fourier transform of  $\frac{\beta \omega}{1-e^{-\beta\omega}}$ instead of $\frac{\beta\omega}{\tanh(\beta \omega/2)}$, which is not well defined. 
We also show another proof of Lemma~\ref{connection_of_exponential_operator} in Appendix~\ref{Another proof for Lemma_connection_of_exponential_operator} using the perturbation theory for the operator logarithm.

By connecting Lemma~\ref{connection_of_exponential_operator} iteratively, we obtain the following corollary:
\begin{corol} \label{connection_of_exponential_operator_repeat}
The logarithm of the operator 
\begin{align}
 e^{\tau \mB} e^{\beta \mA} e^{\tau  \mB} 
\end{align}
is expressed as 
\begin{align}
\label{main_eq:connection_of_exponential_operator_corol}
&\hat{\mA}_\tau:=\log(  e^{\tau \mB} e^{\beta \mA} e^{\tau  \mB}  ) = U_\tau ( \beta \mA + \hat{\mB}_\tau ) U_\tau^\dagger ,
\end{align}
where we define
\begin{align}
U_\tau :=\mathcal{T} e^{- i\int_0^\tau \mC_{\tau_1} d\tau_1}  ,\quad   \hat{\mB}_\tau := 2\int_0^\tau U_{\tau_1}^\dagger  \mB   U_{\tau_1} d\tau_1 ,\quad 
\mC_\tau :=\frac{2}{\beta}\int_{-\infty}^\infty g_\beta(t)  \mB(\hat{\mA}_\tau, t)dt.
\label{connection_of_exponential_operator_repeat_def}
\end{align}
\end{corol}

{\bf Remark.} When we consider $\mathcal{T} e^{\int_0^\tau \mB_{\tau_1} d\tau_1}$ instead of $ e^{\tau \mB}$, we have the same statement, but the operators $\hat{\mB}_\tau$ and $\mC_\tau$ are replaced by
\begin{align}
\hat{\mB}_\tau := 2\int_0^\tau U_{\tau_1}^\dagger  \mB_{\tau_1}   U_{\tau_1} d\tau_1 ,\quad 
\mC_\tau :=\frac{2}{\beta}\int_{-\infty}^\infty g_\beta(t)  \mB_\tau(\hat{\mA}_\tau, t)dt.
\label{connection_of_exponential_operator_repeat_def_tau_dependent}
\end{align}

\textit{Proof of Corollary~\ref{connection_of_exponential_operator_repeat}.}
We assume the form of Eq.~\eqref{main_eq:connection_of_exponential_operator_corol} for $\tau \le x$ and prove the case of $\tau=x+dx$.
Here, we denote 
\begin{align}
e^{\beta \hat{\mA}_x}= e^{x \mB} e^{\beta \mA} e^{x \mB} ,\quad e^{\beta \hat{\mA}_{x+dx}}= e^{(x+dx) \mB} e^{\beta \mA} e^{(x+dx) \mB} e^{dx \mB}
= e^{dx \mB}  e^{\beta\hat{\mA}_x} e^{dx \mB}.
\end{align}
Then, Lemma~\ref{connection_of_exponential_operator} gives the form of $\hat{\mA}_{x+dx}$ as 
\begin{align}
\beta \hat{\mA}_{x+dx} &=  \beta  e^{-i dx \mathcal{C}_x} \hat{\mA}_x e^{i dx \mathcal{C}_x} + 2dx  \mB + \orderof{dx^2} \notag  \\
&=e^{-i dx \mathcal{C}_x} \br{\beta   \hat{\mA}_x  + 2dx  \mB}e^{i dx \mathcal{C}_x} + \orderof{dx^2} \notag \\
&=e^{-i dx \mathcal{C}_x} \br{ U_x( \beta \mA + \hat{\mB}_x ) U_x^\dagger   + 2dx  \mB}e^{i dx \mathcal{C}_x} + \orderof{dx^2} 
\notag \\
&= U_{x+dx} \br{ \beta \mA + \hat{\mB}_x + 2dxU_x^\dagger \mB U_x } U_{x+dx}^\dagger + \orderof{dx^2} ,
\end{align}
which yields the desired equation~\eqref{main_eq:connection_of_exponential_operator_corol} for $\tau=x+dx$. 
This completes the proof of Corollary~\ref{connection_of_exponential_operator_repeat}. $\square$

\subsection{Refined error bound for approximate quantum Belief propagation}

In this subsection, we consider the approximation of the belief propagation operator $\Phi_{\mB}$ in Eq.~\eqref{eq_belief_propagation_operator} as 
$\tilde{\Phi}_{\mB}$ as  
\begin{align}
\label{approx_bp_propagation__2}
&\tilde{\Phi}_{\mB}:= \mathcal{T} e^{\int_0^{1} \tilde{\phi}_{\mB,\tau} d\tau} ,
\end{align}
where $\tilde{\phi}_{\mB,\tau}$ is Hermitian. 
In Lemma~\ref{lema:bp_error_est}, we have proved the error bound for the approximation of 
\begin{align}
\Phi_{\mB} e^{\beta \mA}\Phi_{\mB}^\dagger\approx \tilde{\Phi}_{\mB} e^{\beta \mA}\tilde{\Phi}_{\mB}^\dagger .
\end{align}
However, we cannot usually ensure that the logarithms of $\Phi_{\mB} e^{\beta \mA}\Phi_{\mB}$ and $\tilde{\Phi}_{\mB} e^{\beta \mA}\tilde{\Phi}_{\mB}$ are close to each other. 
Using Lemma~\ref{connection_of_exponential_operator}, we aim to estimate the error for the approximation of 
\begin{align}
\log\br{\Phi_{\mB} e^{\beta \mA}\Phi_{\mB}^\dagger} \approx \log\br{\tilde{\Phi}_{\mB} e^{\beta \mA}\tilde{\Phi}_{\mB}^\dagger}  .
\end{align}

This kind of approximation is critical in estimating the conditional mutual information (see Lemma~\ref{lem:entropy_bound_1D}).
We prove the following lemma:
\begin{prop} \label{prop:bp_error_est_effective}
Let us define $\tilde{\phi}_{\mB,\tau}$ as in Eq.~\eqref{approx_bp_propagation__2} such that $\| \phi_{\mB,\tau}- \tilde{\phi}_{\mB,\tau} \| \le  \delta \norm{\mB}/2$ for $\forall \tau$.
Then, the norm difference between $\Phi_{\mB} e^{\beta \mA}\Phi_{\mB}^\dagger$ and $\tilde{\Phi}_{\mB} e^{\beta \mA}\tilde{\Phi}_{\mB}$ is upper-bounded by
 \begin{align}
\label{main_ineq:prop:bp_error_est_effective}
\norm{ \log\br{\Phi_{\mB} e^{\beta \mA}\Phi_{\mB}^\dagger}-\log\br{\tilde{\Phi}_{\mB} e^{\beta \mA}\tilde{\Phi}_{\mB}^\dagger} } 
\le  3  \mathcal{N}_{\mA,\mB} (\beta \nu_1 +1) e^{\beta \nu_2}  \delta ,
\end{align}
where we define $\mathcal{N}_{\mA,\mB}$ as 
\begin{align}
\mathcal{N}_{\mA,\mB} := \max(4\pi, \beta \norm{\mA} + \beta \norm{\mB}) , 
\end{align} 
and choose $\nu_1$ and $\nu_2$ as 
\begin{align}
 \label{prop:bp_error_est_effective_choice_nu_1_nu_2}
\nu_1=4\norm{\mB} \log\br{\mathcal{N}_{\mA,\mB}},\quad \nu_2=\norm{\mB}\br{ 14\log\br{\mathcal{N}_{\mA,\mB}} + 1}.
\end{align}
\end{prop}

{\bf Remark.} 
From the leading term in the inequality~\eqref{main_ineq:prop:bp_error_est_effective}, we can upper-bound
 \begin{align}
\label{main_ineq:lema:bp_error_est_rougth}
\norm{ \log\br{\Phi_{\mB} e^{\beta \mA}\Phi_{\mB}^\dagger}-\log\br{\tilde{\Phi}_{\mB} e^{\beta \mA}\tilde{\Phi}_{\mB}^\dagger} } 
\le e^{\Theta(\beta) \norm{\mB}\log\br{\beta \norm{\mA}}  } \delta 
\end{align}
by using the $\Theta(1)$ notation in Eq.~\eqref{Theta_notation_def}. 
Compared with the upper bound in~\eqref{main_ineq:lema:bp_error_est}, the additional logarithmic term $\log\br{\beta \norm{\mA}}$ appears in the exponential. 
In applying the proposition to many-body systems, the norm $\norm{\mA}$ becomes as large as the system size $|\Lambda|$, which makes the upper bound meaningless.
To avoid it, we need to use 
\begin{align}
\label{key_decomposition_for_norm_eff_Ham_dif}
&\norm{\log\br{\Phi_{\mB} e^{\beta \mA}\Phi_{\mB}^\dagger} - \log\br{\tilde{\Phi}_{\mB} e^{\beta \mA}\tilde{\Phi}_{\mB}^\dagger}}  
\le 
\norm{\log\br{\Phi_{\mB} e^{\beta \mA_L}\Phi_{\mB}^\dagger} - \log\br{\tilde{\Phi}_{\mB} e^{\beta \mA_L}\tilde{\Phi}_{\mB}^\dagger}} \notag\\
&\quad\quad\quad\quad\quad\quad\quad\quad\quad\quad +
\norm{\log\br{\Phi_{\mB} e^{\beta \mA}\Phi_{\mB}^\dagger} - \log\br{\tilde{\Phi}_{\mB} e^{\beta \mA}\tilde{\Phi}_{\mB}^\dagger}
-\brr{\log\br{\Phi_{\mB} e^{\beta \mA_L}\Phi_{\mB}^\dagger} - \log\br{\tilde{\Phi}_{\mB} e^{\beta \mA_L}\tilde{\Phi}_{\mB}^\dagger}}
 } ,
\end{align}
where $\mA_L$ consists of interaction terms that act on the subsets $L \subset \Lambda$ [see Eq.~\eqref{def:Hamiltonian_subset_L}]. 

For the first term, we apply Proposition~\ref{prop:bp_error_est_effective}, which now yields $\mathcal{N}_{\mA_L,\mB}
=\orderof{\beta |L|}$. 
More precisely, for $\Phi_{\mB} e^{\beta \mA_L}\Phi_{\mB}$, we can no longer ensure $\Phi_{\mB} e^{\beta \mA_L}\Phi_{\mB}= e^{\beta \br{\mA_L+\mB}}$.  Hence, we need to consider 
 \begin{align}
e^{\beta \br{\mA_L+\mB}} = \Phi'_{\mB} e^{\beta \mA_L}\Phi^{'\dagger}_{\mB},
\end{align}
and decompose as 
\begin{align}
\label{key_decomposition_for_norm_eff_Ham_dif__alt}
&\norm{\log\br{\Phi_{\mB} e^{\beta \mA_L}\Phi_{\mB}^\dagger} - \log\br{\tilde{\Phi}_{\mB} e^{\beta \mA_L}\tilde{\Phi}_{\mB}^\dagger}} \notag \\
&\le\norm{\log\br{\Phi'_{\mB} e^{\beta \mA_L}\Phi^{'\dagger}_{\mB}} -\log\br{\Phi_{\mB} e^{\beta \mA_L}\Phi_{\mB}^\dagger} }
+  \norm{\log\br{\Phi'_{\mB} e^{\beta \mA_L}\Phi^{'\dagger}_{\mB}} - \log\br{\tilde{\Phi}_{\mB} e^{\beta \mA_L}\tilde{\Phi}_{\mB}^\dagger  }} ,
\end{align}
where each of the above terms can be treated by Proposition~\ref{prop:bp_error_est_effective}. 
Regarding the second term in the RHS of the above inequality~\eqref{key_decomposition_for_norm_eff_Ham_dif}, it is generally uncontrollable by the choice of $L$. 
If we assume the quasi-locality of the operators $\mA$ (and $\mB$), we can ensure that those terms exponentially decay with the size of $L$ (see Theorem~\ref{thm:Refined locality estimation for effective Hamiltonian} and Lemma~\ref{lem:beta_H_log_tilde..._up}).

\subsubsection{Proof of Proposition~\ref{prop:bp_error_est_effective}}
For the proof, we adopt the ansatz of
\begin{align}
\log\br{\tilde{\Phi}_{\mB,\tau} e^{\beta \mA}\tilde{\Phi}^\dagger_{\mB,\tau}} :=   u_\tau^\dagger \br{\beta\mA +\tau \mB + \Delta \mA_\tau} u_\tau,
\end{align}
with
\begin{align}
\label{tilde_Phi_mB_tau/def}
\tilde{\Phi}_{\mB,\tau} :=
 \mathcal{T} e^{\int_0^\tau  \tilde{\phi}'_{\mB,\tau} d\tau} , \quad 
\tilde{\phi}'_{\mB,\tau}:= \frac{\tilde{\phi}_{\mB,\tau/\beta} }{\beta} \quad (0\le \tau \le \beta), 
\end{align}
where $u_\tau$ is appropriately chosen in an iterative way [see Eq.~\eqref{def_u/tau/dtau_*} below].
Note that $\tilde{\Phi}_{\mB,\beta}= \mathcal{T} e^{\int_0^{\beta}  \tilde{\phi}_{\mB,\tau/\beta} (d\tau/\beta)}=
 \mathcal{T} e^{\int_0^1  \tilde{\phi}_{\mB,x} dx}=\tilde{\Phi}_{\mB}$. 
We now aim to estimate the norms of $\norm{\Delta \mA_\tau}$ and $\norm{u_\tau-1}$ in the form of 
\begin{align}
\label{Delta_mA_tau_u_tau-1}
\norm{\Delta \mA_\tau} \le Q(\tau) \delta ,\quad \norm{u_\tau-1} \le Q(\tau) \delta , \quad  Q(\tau):= (\tau \nu_1 +1) e^{\tau \nu_2} ,
\end{align}
where we choose $\nu_1$ and $\nu_2$ appropriately, and it will be eventually set as in Eq.~\eqref{prop:bp_error_est_effective_choice_nu_1_nu_2}. 
For the proof of the above inequalities, it is enough to prove 
\begin{align}
\label{dtau_Delta_mA_tau_nu_1_nu2}
\abs{\frac{d\norm{\Delta \mA_\tau}}{d\tau}} \le \frac{\nu_1  + \nu_2 Q(\tau) }{2} \delta  ,\quad 
\abs{\frac{d\norm{u_\tau-1} }{d\tau}} \le \frac{\nu_1  + \nu_2 Q(\tau) }{2} \delta  ,
\end{align}
since $\delta \int_0^{\tau} \brr{ \nu_1  + \nu_2 Q(\tau_1) } d\tau_1  \le  \tau \nu_1 \delta + (\tau \nu_1 +1)  \delta\int_0^\tau \nu_2 e^{\tau_1 \nu_2}  d\tau 
\le 2 Q(\tau) \delta $.
Under the inequality~\eqref{Delta_mA_tau_u_tau-1}, we obtain 
 \begin{align}
 \label{desrived_inequality_phi_tilde_phi_log}
\norm{ \log\br{\Phi_{\mB} e^{\beta \mA}\Phi_{\mB}^\dagger}-\log\br{\tilde{\Phi}_{\mB} e^{\beta \mA}\tilde{\Phi}_{\mB}^\dagger} }
&\le  \norm{\beta\br{\mA+\mB} - u_\beta^\dagger \br{\beta\mA +\beta \mB + \Delta \mA_\beta} u_\beta} \notag \\
&\le 2\norm{u_\beta -1}\cdot \beta\br{\norm{\mA}+\norm{\mB}} + \norm{\Delta \mA_\beta}
\le3 \mathcal{N}_{\mA,\mB} Q(\beta) \delta . 
\end{align}

In the following, we aim to prove the inequality~\eqref{Delta_mA_tau_u_tau-1}. 
First of all, for $\tau=0$, we have $\norm{\Delta \mA_\tau}=\norm{u_\tau-1}=0$, and the inequalities in~\eqref{Delta_mA_tau_u_tau-1} are trivially satisfied. 
We then assume the inequalities~\eqref{Delta_mA_tau_u_tau-1} up to a fixed $\tau$ and consider the case of $\tau+d\tau$:
\begin{align}
\log \brr{e^{\tilde{\phi}'_{\mB,\tau}d\tau} e^{ u_\tau^\dagger \br{\beta\mA+\tau \mB + \Delta \mA_{\tau}} u_\tau}  e^{\tilde{\phi}'_{\mB,\tau}d\tau}} = 
u_{\tau+d\tau}^\dagger \br{ \beta \mA+\tau \mB +d\tau \mB + \Delta \mA_{\tau+d\tau}}  u_{\tau+d\tau},
\end{align}
where the unitary operator $u_{\tau+d\tau}$ is defined as in Eq.~\eqref{def_u/tau/dtau_*} using $u_\tau$
We now need to prove the inequality~\eqref{dtau_Delta_mA_tau_nu_1_nu2}.
We estimate the upper bounds of $\norm{\Delta \mA_{\tau+d\tau}}$ and $\norm{u_{\tau+d\tau}-1}$ in terms of $\norm{\Delta \mA_{\tau}}$ and $\norm{u_{\tau}-1}$, respectively.  
By using Lemma~\ref{connection_of_exponential_operator}, we obtain 
\begin{align}
\log \brr{e^{\tilde{\phi}'_{\mB,\tau}d\tau} e^{u_\tau^\dagger \br{\beta \mA+\tau \mB + \Delta \mA_{\tau}} u_\tau }e^{\tilde{\phi}'_{\mB,\tau}d\tau}} 
=e^{-2i \tilde{\mathcal{C}}_{\tilde{\phi}',\tau} d\tau  } u_\tau^\dagger \br{ \beta \mA+\tau \mB + \Delta \mA_{\tau}} u_\tau e^{2i \tilde{\mathcal{C}}_{\tilde{\phi}',\tau}d\tau  } + 2d\tau \tilde{\phi}'_{\mB,\tau}  ,
\label{error_ver_dtau_connect_effective_term}
\end{align}
where we have ignored the terms of order of $\orderof{d\tau^2}$ and define $\tilde{\mathcal{C}}_{\tilde{\phi}',\tau}$ as in Eq.~\eqref{main_eq:connection_of_exponential_operator}:
\begin{align}
\label{tilde_mathcal_C_tilde_phi_tau/def}
\tilde{\mathcal{C}}_{\tilde{\phi}',\tau}:= \frac{1}{\beta}\int_{-\infty}^\infty g_\beta(t)  \cdot
\tilde{\phi}'_{\mB,\tau}\brr{u_\tau^\dagger \br{\mA+\tau \mB/\beta  + \Delta \mA_{\tau}/\beta } u_\tau ,t } dt,
\end{align}
where we have defined $\tilde{\phi}'_{\mB,\tau}\brr{u_\tau^\dagger \br{\mA+\tau \mB/\beta  + \Delta \mA_{\tau}/\beta } u_\tau ,t }=
e^{i t u_\tau^\dagger \br{\mA+\tau \mB/\beta  + \Delta \mA_{\tau}/\beta } u_\tau}
\tilde{\phi}'_{\mB,\tau} 
e^{-i t u_\tau^\dagger \br{\mA+\tau \mB/\beta  + \Delta \mA_{\tau}/\beta } u_\tau}$. 

Because the following equation exactly holds, 
\begin{align}
\label{exct_ver_dtau_connect_effective_term}
\log \brr{e^{\phi'_{\mB,\tau}d\tau} e^{\beta  \mA+\tau \mB}e^{\phi'_{\mB,\tau}d\tau}} = 
\beta \mA+\tau \mB +d\tau \mB  ,  
\end{align}
we have 
\begin{align}
e^{-2i \mathcal{C}_{\phi',\tau} d\tau  } \br{\beta \mA+\tau \mB } e^{2i \mathcal{C}_{\phi',\tau}d\tau  } + 2d\tau \phi'_{\mB,\tau} = 
\beta \mA+\tau \mB +d\tau \mB , 
\end{align}
where $\phi'_{\mB,\tau}$ and $ \mathcal{C}_{\phi',\tau} $ are defined in the same ways as in Eqs.~\eqref{tilde_Phi_mB_tau/def} and \eqref{tilde_mathcal_C_tilde_phi_tau/def}, respectively. 
By applying Eq.~\eqref{exct_ver_dtau_connect_effective_term} to \eqref{error_ver_dtau_connect_effective_term}, we have 
\begin{align}
&e^{-2i \tilde{\mathcal{C}}_{\tilde{\phi}',\tau} d\tau} u_\tau^\dagger \br{\beta \mA+\tau \mB + \Delta \mA_{\tau}} u_\tau e^{2i \tilde{\mathcal{C}}_{\tilde{\phi}',\tau}d\tau } + 2d\tau \tilde{\phi}'_{\mB,\tau}  \notag \\
=& e^{-2i \tilde{\mathcal{C}}_{\tilde{\phi}',\tau} d\tau} u_\tau^\dagger e^{2i \mathcal{C}_{\phi',\tau} d\tau  } 
\br{ e^{-2i \mathcal{C}_{\phi',\tau} d\tau  } \br{ \beta\mA+\tau \mB + \Delta \mA_{\tau}} e^{2i \mathcal{C}_{\phi',\tau} d\tau  }}
  e^{-2i \mathcal{C}_{\phi',\tau} d\tau}  u_\tau e^{2i \tilde{\mathcal{C}}_{\tilde{\phi}',\tau}d\tau } + 2d\tau \tilde{\phi}'_{\mB,\tau}  \notag \\
=& e^{-2i \tilde{\mathcal{C}}_{\tilde{\phi}',\tau} d\tau} u_\tau^\dagger e^{2i \mathcal{C}_{\phi',\tau} d\tau  } 
\br{ \beta\mA+\tau \mB +d\tau \mB  - 2d\tau \phi'_{\mB,\tau} 
+e^{-2i \mathcal{C}_{\phi',\tau} d\tau  }  \Delta \mA_{\tau} e^{2i \mathcal{C}_{\phi',\tau} d\tau  }   }
 e^{-2i \mathcal{C}_{\phi',\tau} d\tau  }  u_\tau e^{2i \tilde{\mathcal{C}}_{\tilde{\phi}',\tau}d\tau } + 2d\tau \tilde{\phi}'_{\mB,\tau}  \notag \\
 =&u_{\tau+d\tau}^\dagger 
\br{ \beta \mA+\tau \mB +d\tau \mB  - 2d\tau \phi'_{\mB,\tau} 
+e^{-2i \mathcal{C}_{\phi',\tau} d\tau  }  \Delta \mA_{\tau} e^{2i \mathcal{C}_{\phi',\tau} d\tau  } 
+2d\tau u_{\tau+d\tau} \tilde{\phi}'_{\mB,\tau} u_{\tau+d\tau}^\dagger   } 
u_{\tau+d\tau} ,
\end{align}
where we define 
\begin{align}
\label{def_u/tau/dtau_*}
u_{\tau+d\tau}:= e^{-2i \mathcal{C}_{\phi',\tau} d\tau  }  u_\tau e^{2i \tilde{\mathcal{C}}_{\tilde{\phi}',\tau}d\tau }  .
\end{align} 
Using the notation of $u_{\tau+d\tau}$, the above equation gives $\Delta A_{\tau+d\tau}$ as 
\begin{align}
\Delta A_{\tau+d\tau}= e^{-2i \mathcal{C}_{\phi',\tau} d\tau  }  \Delta \mA_{\tau} e^{2i \mathcal{C}_{\phi',\tau} d\tau  } 
+ 2d\tau \br{u_{\tau+d\tau} \tilde{\phi}'_{\mB,\tau} u_{\tau+d\tau}^\dagger  -\phi'_{\mB,\tau} } ,
\end{align}
whose upper bound is given by
\begin{align}
\norm{\Delta A_{\tau+d\tau}} 
&\le  \norm{\Delta \mA_{\tau}}+ 2d\tau \norm{\tilde{\phi}'_{\mB,\tau}- \phi'_{\mB,\tau}} +2d\tau \norm{u_{\tau+d\tau} \phi'_{\mB,\tau}u_{\tau+d\tau}^\dagger  -\phi'_{\mB,\tau} } \notag \\
&\le \norm{\Delta \mA_{\tau}}+ 2d\tau \norm{\tilde{\phi}'_{\mB,\tau}- \phi'_{\mB,\tau}} + 4d\tau \norm{\phi'_{\mB,\tau}} \cdot \norm{u_{\tau}-1} ,
\end{align}
where we use $ u_{\tau+d\tau}d\tau=u_{\tau}d\tau + \orderof{d\tau^2}$. 
We therefore obtain using the assumption~\eqref{Delta_mA_tau_u_tau-1} 
\begin{align}
\abs{\frac{\norm{\Delta A_{\tau+d\tau}} - \norm{\Delta \mA_{\tau}} }{d\tau}}
\le \norm{\mB} \br{\delta  + 2\norm{u_\tau-1} } 
\le \norm{\mB} \brr{1 + 2Q(\tau) } \delta  ,
\label{final_ineq_for_Delta_A}
\end{align}
where we use $\norm{\phi'_{\mB,\tau}}\le\norm{\phi_{\mB,\tau/\beta}}/\beta\le  \norm{\mB}/2$ from Eq.~\eqref{tilde_Phi_mB_tau/def}
 and the inequality~\eqref{sup_Def:Phi_0_phi_norm_up} and $\norm{\tilde{\phi}'_{\mB,\tau}- \phi'_{\mB,\tau}}\le 
 \delta \norm{\mB}/2$ from the initial condition.

We then estimate $\norm{u_{\tau+d\tau}-1}$ as 
\begin{align}
\norm{e^{-2i \mathcal{C}_{\phi',\tau} d\tau  }  u_\tau e^{2i \tilde{\mathcal{C}}_{\tilde{\phi}',\tau}d\tau } -1} 
&= \norm{e^{2i \tilde{\mathcal{C}}_{\tilde{\phi}',\tau}d\tau }\br{u_\tau -1} e^{-2i \mathcal{C}_{\phi',\tau} d\tau  } + e^{2i \tilde{\mathcal{C}}_{\tilde{\phi}',\tau}d\tau } e^{-2i \mathcal{C}_{\phi',\tau} d\tau  }   -1}\notag \\
&\le \norm{u_\tau-1} + \norm{e^{2i \tilde{\mathcal{C}}_{\tilde{\phi}',\tau}d\tau } e^{-2i \mathcal{C}_{\phi',\tau} d\tau  }   -1} 
\le  \norm{u_\tau-1} + 2d\tau \norm{\tilde{\mathcal{C}}_{\tilde{\phi}',\tau} - \mathcal{C}_{\phi',\tau} },
\end{align}
which gives 
\begin{align}
\label{ineq_for_uta_dtau_variance}
\frac{\norm{u_{\tau+d\tau}-1} - \norm{u_\tau-1} }{d\tau}
\le 2\norm{\tilde{\mathcal{C}}_{\tilde{\phi}',\tau} - \mathcal{C}_{\phi',\tau} } .
\end{align}
To estimate the RHS of the above inequality, we prove the lemma as follows:
\begin{lemma} \label{lemma:upp_mathcal_C_tilde_df}
For an arbitrary $\tau$, we prove the upper bound of 
\begin{align}
\label{main_ineq:lemma:upp_mathcal_C_tilde_df}
\norm{\tilde{\mathcal{C}}_{\tilde{\phi}',\tau} - \mathcal{C}_{\phi',\tau} }
\le \frac{\norm{\mB}}{\pi}\log\br{\mathcal{N}_{\mA,\mB}} \br{2\delta +10\norm{u_\tau-1} }  
+ \frac{1}{4}  \norm{\mB} \cdot \norm{ \Delta \mA_{\tau}}  .
\end{align}
We show the proof in Sec.~\ref{proof_lemma:upp_mathcal_C_tilde_df}. 
\end{lemma}

By applying the inequality~\eqref{main_ineq:lemma:upp_mathcal_C_tilde_df} to~\eqref{ineq_for_uta_dtau_variance} with the use the assumption~\eqref{Delta_mA_tau_u_tau-1}, we obtain 
\begin{align}
\label{ineq_for_uta_dtau_variance_fin}
&\frac{\norm{u_{\tau+d\tau}-1} - \norm{u_\tau-1} }{d\tau}
\le \frac{2\norm{\mB}}{\pi}\log\br{\mathcal{N}_{\mA,\mB}} \br{2 +10 Q(\tau) }\delta  
+ \frac{1}{2} \norm{\mB} Q(\tau) \delta     \notag \\
&\quad \quad \quad \quad \le  2\norm{\mB}\log\br{\mathcal{N}_{\mA,\mB}} \delta   
+ \norm{\mB}\cdot\brr{ 7\log\br{\mathcal{N}_{\mA,\mB}} + \frac{1}{2} }        Q(\tau) \delta .
\end{align}
Therefore, the inequalities~\eqref{final_ineq_for_Delta_A} and \eqref{ineq_for_uta_dtau_variance_fin} reduce to the form of~\eqref{dtau_Delta_mA_tau_nu_1_nu2} under the choices in Eq.~\eqref{prop:bp_error_est_effective_choice_nu_1_nu_2}, that is, 
$\nu_1=4\norm{\mB} \log\br{\mathcal{N}_{\mA,\mB}}$ and $\nu_2=\norm{\mB} \br{14\log\br{\mathcal{N}_{\mA,\mB}} + 1}$. 
We thus prove the inequality~\eqref{Delta_mA_tau_u_tau-1}, which also proves the inequality~\eqref{desrived_inequality_phi_tilde_phi_log}. 
This completes the proof of Proposition~\ref{prop:bp_error_est_effective}. $\square$

\subsubsection{Proof of Lemma~\ref{lemma:upp_mathcal_C_tilde_df}} \label{proof_lemma:upp_mathcal_C_tilde_df}
For the norm of $ \norm{\tilde{\mathcal{C}}_{\tilde{\phi}',\tau} - \mathcal{C}_{\phi',\tau} }$, we have to treat the integral with the filter function $g_\beta(t)$, which diverges to infinity as $1/|t|$. 
So, as in Eq.~\eqref{int_g_beta_t_V} below, we need to decompose the time evolution to $|t|\le \Delta_t$ and $|t|>\Delta_t$, where $\Delta_t$ is chosen appropriately afterward.
We, in the following, treat the two integrals: 
\begin{align}
\label{t_>_delta_t_phi_g_beta}
&\norm{\frac{1}{\beta}\int_{|t|> \Delta_t} g_\beta(t) 
\brrr{  \tilde{\phi}'_{\mB,\tau} \brr{ u_\tau^\dagger  \br{\mA+\tau \mB/\beta + \Delta \mA_{\tau}/\beta}   u_\tau , t} 
- \phi'_{\mB,\tau} \br{\mA+\tau \mB/\beta , t}
}dt
}
\end{align}
and 
\begin{align}
\label{t_le_delta_t_phi_g_beta}
&\norm{\frac{1}{\beta}\int_{|t|\le \Delta_t} t g_\beta(t) \int_0^1 
\brrr{ \tilde{\psi}'_{\mB,\tau}\brr{ u_\tau^\dagger  \br{\mA+\tau \mB/\beta + \Delta \mA_{\tau}/\beta}   u_\tau , \lambda t} 
-\psi'_{\mB,\tau} \br{\mA+\tau \mB/\beta , \lambda t}
}d\lambda  dt 
},
\end{align}
where we denote $\psi'_{\mB,\tau} = \ad_{\mA+\tau \mB/\beta}\br{ \phi'_{\mB,\tau}} $ and 
$\tilde{\psi}'_{\mB,\tau}= \ad_{u_\tau^\dagger  \br{\mA+\tau \mB/\beta + \Delta \mA_{\tau}/\beta}u_\tau} \br{\tilde{\phi}'_{\mB,\tau}}$.

Before analyzing~\eqref{t_>_delta_t_phi_g_beta} and \eqref{t_le_delta_t_phi_g_beta}, 
we obtain the following upper bound for arbitrary operators $\tilde{O}$ and $O$:
\begin{align}
\label{tilde_O_O_time_ev_diff}
&\norm{\tilde{O} \brr{ u_\tau^\dagger  \br{\mA+\tau \mB/\beta + \Delta \mA_{\tau}/\beta}   u_\tau , t} 
-O \br{\mA+\tau \mB/\beta , t}}  \notag \\
&\le  \norm{\tilde{O} - O}+\br{4\norm{u_\tau-1} +2t \norm{ \Delta \mA_{\tau}/\beta}}\cdot \norm{O} , 
\end{align}
which is derived from the following two inequalities:
\begin{align}
&\norm{\tilde{O} \brr{ u_\tau^\dagger  \br{\mA+\tau \mB/\beta + \Delta \mA_{\tau}/\beta}   u_\tau , t} 
-O \brr{ u_\tau^\dagger  \br{\mA+\tau \mB/\beta + \Delta \mA_{\tau}/\beta}   u_\tau , t}}  \le  \norm{\tilde{O} - O}. 
\end{align}
and 
\begin{align}
&\norm{O \brr{ u_\tau^\dagger  \br{\mA+\tau \mB /\beta + \Delta \mA_{\tau} /\beta}   u_\tau , t} 
-O \br{\mA+\tau \mB/\beta , t}} \notag \\
= &\norm{u_\tau^\dagger  e^{i(\mA+\tau \mB/\beta + \Delta \mA_{\tau}/\beta) t}  u_\tau  O u_\tau^\dagger   e^{-i(\mA+\tau \mB/\beta + \Delta \mA_{\tau}/\beta) t}  u_\tau - e^{i(\mA+\tau \mB/\beta) t}O e^{-i(\mA+\tau \mB/\beta ) t}  } \notag \\
\le &\br{4\norm{u_\tau-1} +2 \norm{ e^{i(\mA+\tau \mB/\beta + \Delta \mA_{\tau}/\beta) t}-  e^{i(\mA+\tau \mB/\beta) t}}}\cdot \norm{O} 
\le \br{4\norm{u_\tau-1} +2|t|\cdot \norm{ \Delta \mA_{\tau}/\beta}}\cdot \norm{O} ,
\end{align}
where we use 
\begin{align}
 e^{i(\mA+\tau \mB/\beta + \Delta \mA_{\tau}/\beta) t} =  e^{i(\mA+\tau \mB/\beta) t} \mathcal{T} e^{i \beta^{-1}\int_0^t  \Delta \mA_{\tau}(\mA+\tau \mB/\beta,-x)dx} .
\end{align}
We apply the inequality~\eqref{tilde_O_O_time_ev_diff} to~\eqref{t_>_delta_t_phi_g_beta} and~\eqref{t_le_delta_t_phi_g_beta} to prove the inequality~\eqref{main_ineq:lemma:upp_mathcal_C_tilde_df}.

Using the inequality~\eqref{main_eq:connection_of_exponential_operator_trans_1} in Lemma~\ref{belief_norm_lemma}, we have 
\begin{align}
\label{main_eq:operator_trans_1_g_beta_re}
&\int_{|t| > \Delta_t}  |g_\beta (t)|  dt \le  \frac{2\beta}{\pi}\log\br{\frac{\beta}{2\pi \Delta_t}} ,\quad \int_{|t| > \Delta_t}  |t g_\beta (t)|  dt \le  2\br{\frac{\beta}{2\pi}}^{2} \zeta(2) = \frac{\beta^2}{12} \for \Delta_t \le \frac{\beta}{4\pi},
\end{align}
and hence, we obtain
\begin{align}
\label{upp_:t_>_delta_t_phi_g_beta}
\eqref{t_>_delta_t_phi_g_beta} 
\le &\frac{1}{\beta}\int_{|t|> \Delta_t} |g_\beta(t) | \brr{
\norm{\tilde{\phi}'_{\mB,\tau}  - \phi'_{\mB,\tau} }+\br{4\norm{u_\tau-1} +2|t|\cdot \norm{ \Delta \mA_{\tau}/\beta}}\cdot \norm{\phi'_{\mB,\tau}}}   dt \notag \\
\le &\frac{2}{\pi}\log\br{\frac{\beta}{2\pi \Delta_t}} \br{\norm{\tilde{\phi}'_{\mB,\tau}  - \phi'_{\mB,\tau} }+4\norm{u_\tau-1} \cdot \norm{\phi'_{\mB,\tau}}}  
+  \frac{1}{6}  \norm{ \Delta \mA_{\tau}}\cdot \norm{\phi'_{\mB,\tau}} \notag \\
\le &\frac{\norm{\mB}}{\pi}\log\br{\frac{\beta}{2\pi \Delta_t}} 
\br{\delta +4\norm{u_\tau-1} }  
+  \frac{\norm{\mB}}{12}  \norm{ \Delta \mA_{\tau}},
\end{align}
where in the last inequality we use $\norm{\phi'_{\mB,\tau}}\le  \norm{\mB}/2$ and $\norm{\tilde{\phi}'_{\mB,\tau}- \phi'_{\mB,\tau}}\le 
 \delta \norm{\mB}/2$ as in the inequality~\eqref{final_ineq_for_Delta_A}. 

In the same way, using the inequality~\eqref{main_eq:connection_of_exponential_operator_trans_2} in Lemma~\ref{belief_norm_lemma},
we have 
\begin{align}
\label{main_eq:operator_trans_2_g_beta_re}
\int_{|t| \le \Delta_t} | t g_\beta (t) | dt  \le  \frac{\beta}{\pi} \Delta_t , \quad  
\int_{|t| \le \Delta_t} | t^2 g_\beta (t) | dt  \le  \frac{\beta}{2\pi} \Delta_t^{2} , 
\end{align}
which yields 
\begin{align}
\label{upp_:t_le_delta_t_phi_g_beta}
\eqref{t_le_delta_t_phi_g_beta} 
\le &\frac{1}{\beta}\int_{|t|\le \Delta_t} |t g_\beta(t)| \int_0^1 \brr{
\norm{\tilde{\psi}'_{\mB,\tau}  - \psi'_{\mB,\tau} }+\br{4\norm{u_\tau-1} +2|\lambda t|\cdot \norm{ \Delta \mA_{\tau}/\beta}}\cdot \norm{\psi'_{\mB,\tau}}}   dt \notag \\
\le &\frac{\Delta_t}{\pi}\br{\norm{\tilde{\psi}'_{\mB,\tau}  - \psi'_{\mB,\tau} }+4\norm{u_\tau-1} \cdot \norm{\psi'_{\mB,\tau}}}  
+ \frac{\Delta_t^2}{\pi \beta}  \norm{ \Delta \mA_{\tau}}\cdot \norm{\psi'_{\mB,\tau}}.
\end{align}
To estimate the norm of $\psi'_{\mB,\tau} - \tilde{\psi}'_{\mB,\tau} $, we calculate as 
\begin{align}
\psi'_{\mB,\tau} - \tilde{\psi}'_{\mB,\tau} 
&= \ad_{\mA+\tau \mB/\beta}\br{ \phi'_{\mB,\tau}}  - \ad_{u_\tau^\dagger  \br{\mA+\tau \mB/\beta + \Delta \mA_{\tau}/\beta}u_\tau} \br{\tilde{\phi}'_{\mB,\tau}} \notag \\
&=\ad_{\mA+\tau \mB/\beta}\br{ \phi'_{\mB,\tau}-\tilde{\phi}'_{\mB,\tau} }
+ \ad_{\mA+\tau \mB/\beta} \br{\tilde{\phi}'_{\mB,\tau}}
  - \ad_{u_\tau^\dagger  \br{\mA+\tau \mB/\beta + \Delta \mA_{\tau}/\beta}u_\tau} \br{\tilde{\phi}'_{\mB,\tau}} ,
\end{align}
which yields 
\begin{align}
\label{upp_:t_le_delta_t_phi_g_beta_der1}
\norm{\psi'_{\mB,\tau} - \tilde{\psi}'_{\mB,\tau} } 
&\le 
2  \norm{\mA+\tau \mB/\beta} \cdot \norm{\tilde{\phi}'_{\mB,\tau}-\phi'_{\mB,\tau}}
+ 
2 \norm{\tilde{\phi}'_{\mB,\tau}} \cdot \norm{\mA+\tau \mB/\beta - u_\tau^\dagger  \br{\mA+\tau \mB/\beta + \Delta \mA_{\tau}/\beta}u_\tau} \notag \\
&\le 2  \br{ \norm{\mA} +\norm{\mB}}  \cdot \norm{\tilde{\phi}'_{\mB,\tau}-\phi'_{\mB,\tau}}
+ 2 \norm{\tilde{\phi}'_{\mB,\tau}} \cdot \brr{2 \norm{u_\tau-1} \cdot  \br{ \norm{\mA} +\norm{\mB}}+  \norm{\Delta \mA_{\tau}/\beta}} \notag \\
&\le  \norm{\mB} \br{ \norm{\mA} +\norm{\mB}} \br{\delta + 2 \norm{u_\tau-1}} + \norm{\mB} \cdot \norm{\Delta \mA_{\tau}/\beta} ,
\end{align}
where we use $\norm{\mA+\tau \mB/\beta} \le \norm{\mA} +\norm{\mB}$ from $\tau \le \beta$. 
Using a similar analysis, we also obtain 
\begin{align}
\label{upp_:t_le_delta_t_phi_g_beta_der2}
\norm{\psi'_{\mB,\tau}}
&= \norm{ \ad_{\mA+\tau \mB/\beta}\br{ \phi'_{\mB,\tau}} }  \le 
2\norm{\mA+\tau \mB/\beta} \cdot \norm{\phi'_{\mB,\tau}}\le 
\norm{\mB} \br{ \norm{\mA} +\norm{\mB}} .
\end{align}
By applying the upper bounds~\eqref{upp_:t_le_delta_t_phi_g_beta_der1} and \eqref{upp_:t_le_delta_t_phi_g_beta_der2} to the inequality~\eqref{upp_:t_le_delta_t_phi_g_beta}, we finally obtain 
\begin{align}
\label{upp_:t_le_delta_t_phi_g_beta_fin}
\eqref{t_le_delta_t_phi_g_beta} 
\le &\frac{\Delta_t}{\pi} \norm{\mB} \br{ \norm{\mA} +\norm{\mB}} \br{\delta + 6 \norm{u_\tau-1} +\frac{\Delta_t}{\beta} \norm{ \Delta \mA_{\tau}} }
+\frac{\Delta_t}{\pi \beta} \norm{\mB} \cdot \norm{\Delta \mA_{\tau}} .
\end{align}

We now choose $\Delta_t=\beta/\mathcal{N}_{\mA,\mB}=\min\brr{\beta/4\pi,(\norm{\mA} +\norm{\mB})^{-1}}$, which satisfies the condition~\eqref{cond_delta_t_1/2}. 
By combining the inequalities~\eqref{upp_:t_>_delta_t_phi_g_beta} and~\eqref{upp_:t_le_delta_t_phi_g_beta_fin} with the above choice of $\Delta_t$, we obtain
\begin{align}
&\eqref{t_>_delta_t_phi_g_beta} + \eqref{t_le_delta_t_phi_g_beta}  \notag \\
&\le \frac{\norm{\mB}}{\pi}\log\br{\frac{\beta}{2\pi \Delta_t}} 
\br{\delta +4\norm{u_\tau-1} }  
+  \frac{\norm{\mB}}{12}  \norm{ \Delta \mA_{\tau}}  \notag \\
&\quad + \frac{\Delta_t}{\pi} \norm{\mB} \br{ \norm{\mA} +\norm{\mB}} \br{\delta + 6 \norm{u_\tau-1} +\frac{\Delta_t}{\beta} \norm{ \Delta \mA_{\tau}} }+\frac{\Delta_t}{\pi\beta} \norm{\mB} \cdot \norm{\Delta \mA_{\tau}} \notag \\
&\le \frac{\norm{\mB}}{\pi}\log\br{\frac{\mathcal{N}_{\mA,\mB}}{2\pi }} 
\br{\delta +4\norm{u_\tau-1} }  
+  \frac{\norm{\mB}}{12}  \norm{ \Delta \mA_{\tau}}   + \frac{1}{\pi} \norm{\mB}\br{\delta + 6 \norm{u_\tau-1} +\frac{2\Delta_t}{\beta} \norm{ \Delta \mA_{\tau}} }\notag \\
&\le \frac{\norm{\mB}}{\pi}\log\br{\mathcal{N}_{\mA,\mB}} \br{2\delta +10\norm{u_\tau-1} }  
+ \br{\frac{1}{12} + \frac{1}{2\pi } } \norm{\mB} \cdot \norm{ \Delta \mA_{\tau}} ,
\end{align}
which reduces to the main inequality~\eqref{main_ineq:lemma:upp_mathcal_C_tilde_df}, where we use $1/12 + 1/(2\pi)\le 1/4$.
This completes the proof of Lemma~\ref{lemma:upp_mathcal_C_tilde_df}. $\square$

\section{Quasi-locality of the entanglement Hamiltonian}

In this section, we estimate quasi-locality after the connection of exponential operators as in Eq.~\eqref{def:psi_N_connect} using Lemma~\ref{connection_of_exponential_operator} (or Corollary~\ref{connection_of_exponential_operator_repeat}).
In general, the forms of $\{\Psi_j\}_{j=1}^N$ are extremely complicated. 
To see this point, we first consider
 \begin{align}
\Psi_1 =e^{\epsilon \beta \mB_1} e^{\beta \mA} e^{\epsilon \beta \mB_1}= \exp\br{\beta e^{-2i \epsilon \mC_1}  \mA e^{2i \epsilon \mC_1} +2\epsilon \beta \mB_1},
\end{align}
where $\mC_1$ is a quasi-local operator, which is defined by Eq.~\eqref{main_eq:connection_of_exponential_operator}.
Second, we consider 
 \begin{align}
\Psi_2 &=e^{\epsilon \beta \mB_2} e^{\beta e^{-2i \epsilon \mC_1}A e^{2i \epsilon \mC_1} + 2\epsilon \beta \mB_1} e^{ \epsilon  \beta \mB_2} \notag \\
&=  \exp \br{\beta e^{-2i \epsilon \mC_2}e^{-2i\epsilon \mC_1}A e^{2i \epsilon \mC_1}e^{2i \epsilon  \mC_2} +2 \beta \epsilon e^{-2i \epsilon \mC_2}B_1e^{2i \epsilon \mC_2}+2 \beta \epsilon \mB_2}  ,
\end{align}
where 
\begin{align}
\mC_2:=\int_{-\infty}^\infty g_\beta (t)e^{i \br{ e^{-2i \epsilon \mC_1}Ae^{2i\epsilon \mC_1}+2 \epsilon \mB_1}t}B_2e^{-i\br{e^{-2i \epsilon \mC_1}Ae^{2i\epsilon \mC_1}+2 \epsilon \mB_1}t}dt.
\end{align}
Thus, the operator $\mC_2$ includes the double exponential operators.  
In general, $\mC_j$ is characterized by an exponential tower of operators; in the limit of $N\to \infty$, the number of tower layers becomes infinite.
This point makes analyses for the logarithm of $\Psi_N$ quite challenging.

In the following, we show our main mathematical theorems to estimate the quasi-locality after the connection of exponential operators.
As the basic setup, we let $H_0$ be an arbitrary Hamiltonian with short-range interaction with the condition~\eqref{def_short_range_long_range}.
Then, we start with the logarithm of 
\begin{align}
\label{H_tau_express_V_tau_H0}
\hat{H}_\tau =\log\br{e^{\tau V} e^{\beta H_0} e^{\tau V} },
\end{align}
Corollary~\ref{connection_of_exponential_operator_repeat} gives the form of 
\begin{align}
\label{H_tau_explicit_form_0}
\hat{H}_\tau = U_\tau( \beta H_0 + \hat{V}_\tau )U_\tau^\dagger  ,
\end{align}
where 
\begin{align}
U_\tau =\mathcal{T} e^{-i\int_0^\tau \mC_{\tau_1} d\tau_1}  ,\quad   \hat{V}_\tau = 2\int_0^\tau U_{\tau_1}^\dagger  V   U_{\tau_1} d\tau_1 ,\quad 
\mC_\tau =\frac{2}{\beta}\int_{-\infty}^\infty g_\beta(t)  V(\hat{H}_\tau, t)dt.
\label{U_tau_V_tau_mu_tau_def}
\end{align}
We aim to prove the following statement on the quasi-locality of the unitary operator $U_\tau$: 
\begin{subtheorem} \label{sub_thm:U_tau_u_i_commun}
Let $V$ be quasi-local around the subset $\mfL$ ($\mfL\subset\Lambda$) in the sense of
\begin{align}
\label{assumption_proof_le_0V}
\norm{[V,u_i]} \le \norm{V} Q(0,\ell),  \quad \ell=\dist_{i,\mfL} ,  
\end{align}
where $u_i$ is an arbitrary unitary operator defined on the site $i$ that is separated from $\mfL$ by distance $\ell$.  
Then, we obtain 
\begin{align}
\label{upper_bound_U_tau_u_i}
\norm{[U_\tau,u_i]}  \le Q (\tau,r) \for \forall \tau >0 
\end{align}
with 
\begin{align}
\label{assumption_proof_le_tau}
Q (\tau,r) = e^{ \kappa_0\tau+\kappa_1 \tau \log(r+\tau+e)-\kappa_\beta r },
\end{align}
where we define $\kappa_1$ and $\kappa_\beta$ as 
\begin{align}
&\kappa_0= \Theta(\beta^D)\norm{V} \log (\beta\norm{V}\cdot |\mfL|)  , \quad \kappa_1=\Theta(\beta^D) \norm{V}  , 
\label{kappa_0_kappa_1_definition}\\
&\kappa_\beta := \min\br{ \frac{\pi \mu}{2v\beta} , \frac{\mu}{4}} . \label{definition_kappa_beta_1_again}
\end{align}
\end{subtheorem}

{\bf Remark.} In Eq.~\eqref{H_tau_express_V_tau_H0}, we consider the case where the operator $V$ does not depend on $\tau$, but the proof technique can be straightforwardly generalized to the logarithm of 
\begin{align}
e^{V_N \tau/N}  \cdots e^{V_2 \tau/N}  e^{V_1 \tau/N}   e^{\beta H_0}e^{V_1 \tau/N}     e^{V_2 \tau/N} \cdots  e^{V_N \tau/N}  ,
\end{align}
where $N$ is arbitrarily large. 
We also note that from the inequality $\norm{[U_\tau,u_i]} \le 2$, we can replace $Q(\tau,r)$ in~\eqref{upper_bound_U_tau_u_i} as 
\begin{align}
Q (\tau,r)  \to \min \brr{2,Q (\tau,r) },
\label{trivial_bound_/for_Q_tau_r}
\end{align}
but we omit the $\min(\cdots)$ notation.



\subsection{Proof of Subtheorem~\ref{sub_thm:U_tau_u_i_commun}}

For the proof, we use the inductive method. 
For $\tau=0$, we trivially obtain $U_\tau=\hat{1}$, and hence $\norm{[U_\tau,u_i]}=0$.
We assume the inequality~\eqref{assumption_proof_le_tau} up to a certain $\tau$ and prove the case of $\tau+d\tau$ with $d\tau$ infinitesimally small.
We have 
\begin{align}
\norm{[U_{\tau+d\tau},u_i]} = \norm{[e^{-i\mC_\tau d\tau}U_{\tau},u_i]}\le \norm{[U_{\tau},u_i]}+ d\tau \norm{[\mC_\tau,u_i]},
\end{align}
which yields $(d/d\tau)\norm{[U_\tau,u_i]} \le  \norm{[\mC_\tau,u_i]}$. 
Then, our task is to prove 
\begin{align}
 \norm{[\mC_\tau,u_i]} 
&\le \frac{d}{d\tau} e^{\kappa_0\tau+\kappa_1 \tau \log(r+\tau+e)-\kappa_\beta r } = Q (\tau,r)\brr{\kappa_0+\kappa_1 \log(r+\tau+e) + \frac{\kappa_1\tau}{r+\tau+e} } 
\label{aim_to_prove_statement_mC_pree}
\end{align}
under the assumption of \eqref{upper_bound_U_tau_u_i} for $[0,\tau]$. 
The inequality~\eqref{aim_to_prove_statement_mC_pree} and $\norm{[U_0,u_i]}=0$ yields the desired inequality of
\begin{align}
\norm{[U_\tau,u_i]}= \int_0^\tau \frac{d}{d\tau_1}\norm{[U_{\tau_1},u_i]} d\tau_1 
&\le \int_0^\tau \norm{[\mC_{\tau_1},u_i]}  d\tau_1 
\le \int_0^\tau  \frac{d}{d\tau_1} e^{\kappa_0\tau_1+\kappa_1 \tau_1 \log(r+\tau_1+e)-\kappa_\beta r }  d\tau_1   \notag \\
&\le e^{\kappa_0\tau+\kappa_1 \tau \log(r+\tau+e)-\kappa_\beta r }=Q (\tau,r). 
\end{align}
In proving the inequality~\eqref{aim_to_prove_statement_mC_pree}, it is enough to prove 
\begin{align}
\label{aim_to_prove_statement_mC}
 \norm{[\mC_\tau,u_i]} \le  \brr{\kappa_0+ \kappa_1  \log(r+\tau+e) } Q (\tau,r)
\end{align}
under the choices of $\kappa_0$ and $\kappa_1$ as in Eq.~\eqref{kappa_0_kappa_1_definition} and \eqref{definition_kappa_beta_1_again}, respectively.

In the following, the primary task is to estimate the norm of
\begin{align}
\label{aim_to_prove_statement_mC_stating}
\brr{\mC_\tau , u_i}=\frac{2}{\beta} \int_{-\infty}^\infty g_\beta(t)  \brr{ V(\hat{H}_\tau, t),u_i} dt.
\end{align}
In the above equation, we need to consider the integral, which includes the function $g_\beta (t)$. 
A challenging problem that may occur is that the integral 
\begin{align}
\int_{-\infty}^\infty |g_\beta (t)| dt 
\end{align}
is divergent since $|g_\beta (t)| \propto 1/|t|$ for $t\ll1$ [see also Eq.~\eqref{def_g_t_exp_decay}]. 
To avoid the divergence, we utilize the following equation as in Ref.~\cite[(D. 83) and (D.84)]{PhysRevX.12.021022}:
\begin{align}
O(H,t) = O + \int_0^1 \frac{d}{d\lambda}O(H,\lambda t) d\lambda = 
 O +it \int_0^1\ad_{H} [O(H,\lambda t)] d\lambda  ,
\end{align}
which yields the following decomposition:
\begin{align}
\int_{-\infty}^\infty g_\beta(t) V(\hat{H}_\tau, t)dt
&= \int_{|t|> \delta t} g_\beta(t) V(\hat{H}_\tau, t) dt
+ \int_{|t|\le \delta t} g_\beta(t) \br{V+ i t \int_0^1 \ad_{\hat{H}_\tau} [V(\hat{H}_\tau,\lambda t)] d\lambda}  dt \notag \\
&= \int_{|t|> \delta t} g_\beta(t) V(\hat{H}_\tau, t)dt
+ i\int_{|t|\le \delta t} t g_\beta(t) \int_0^1 \ad_{\hat{H}_\tau} [V(\hat{H}_\tau,\lambda t)] d\lambda  dt ,
\label{int_g_beta_t_V}
\end{align}
where in the second equation, we use $\int_{|t|\le \delta t} g_\beta(t)dt =0$ since $g_\beta(t)$ is an odd function, and the parameter $\delta t$ is appropriately chosen afterward so that it can satisfy
\begin{align}
\label{cond_delta_t_1/2_00}
\delta t \le \frac{\beta}{4\pi}.
\end{align}

To analyze the commutator norm of $\norm{[\mC_\tau,u_i]}$ in~\eqref{aim_to_prove_statement_mC_pree}, we prove the following general statement: 
\begin{lemma} \label{lemma:commutator_u_i}
For arbitrary local unitary operator $u_i$ and operator $O$, we have the following upper bounds:
\begin{align}
\label{first_ineq_lem_V_u_i}
\norm { [ O (\hat{H}_\tau ,t), u_i] } \le&  \norm {\brr{ O  ,u_i(H_0,-t)}} \notag \\
&+2\norm{O}\br{ \norm{[U_\tau,u_i]} + \norm {\brr{  U_\tau ,u_i(H_0,-t)}}  + \int_0^t \norm{\brr{\hat{V}_\tau, u_i (H_0,-t_1) } }dt_1 } . 
\end{align}
\end{lemma}

{~}\\
\textit{Proof of Lemma~\ref{lemma:commutator_u_i}.}
Using the explicit form of $\hat{H}_\tau$ in Eq.~\eqref{H_tau_explicit_form_0}, we start from 
\begin{align}
\label{start_time_O_H_tau_t}
O(\hat{H}_\tau ,t)=e^{i\hat{H}_\tau t}Oe^{-i\hat{H}_\tau t}= 
 U_\tau e^{i \br{ H_0 + \hat{V}_\tau } t }  U_\tau^\dagger  O U_\tau  e^{-i \br{ H_0 + \hat{V}_\tau } t } U_\tau^\dagger ,
\end{align}
which immediately yields 
\begin{align}
\label{starting_V_H_atu_uv}
\norm{ \brr{O(\hat{H}_\tau ,t),u_i}}\le 
2\norm{O} \cdot \norm{[U_\tau,u_i]} + 
\norm { \brr{ e^{i \br{ H_0 + \hat{V}_\tau } t }  U_\tau^\dagger O U_\tau  e^{-i \br{ H_0 + \hat{V}_\tau } t } ,u_i }}.
\end{align}

The second term is upper-bounded by
\begin{align}
&\norm { \brr{ e^{i \br{ H_0 + \hat{V}_\tau } t }  U_\tau^\dagger  O U_\tau e^{-i \br{ H_0 + \hat{V}_\tau } t } ,u_i}}
= \norm {\brr{  U_\tau^\dagger  O U_\tau,e^{-i \br{ H_0 + \hat{V}_\tau } t }   u_i   e^{i \br{ H_0 + \hat{V}_\tau } t }}}  \notag \\
&= \norm {\brr{  U_\tau^\dagger  O U_\tau,e^{-i H_0t } \tilde{U}_\tau u_i  \tilde{U}_\tau^\dagger     e^{i H_0  t }}}   
=\norm {\brr{  U_\tau^\dagger  O U_\tau,e^{-i H_0t } \br{ [\tilde{U}_\tau, u_i]  \tilde{U}_\tau^\dagger  +u_i}   e^{i H_0  t }}}   
\notag \\
&\le 2\norm{O} \cdot \norm{[\tilde{U}_\tau, u_i]} + \norm {\brr{  U_\tau^\dagger  O U_\tau,u_i(H_0,-t)}}    \notag \\
&\le 2\norm{O} \cdot \norm{[\tilde{U}_\tau, u_i]} + 2\norm{O} \cdot \norm {\brr{  U_\tau ,u_i(H_0,-t)}} +\norm {\brr{ O ,u_i(H_0,-t)}}  ,
\label{ineq_U_tau_V_U_taudagger}
\end{align}
where $ \tilde{U}_\tau := e^{i H_0t } e^{-i \br{ H_0 + \hat{V}_\tau } t } $. 
Finally, we have 
\begin{align}
\norm{\brr{ \tilde{U}_\tau,u_i}} 
&=\norm{ \brr{ \mathcal{T} e^{-\int_0^t e^{iH_0t_1} \hat{V}_\tau e^{-iH_0t_1} dt_1},u_i}} \notag \\
&\le \int_0^t \norm{\brr{\hat{V}_\tau, e^{-iH_0t_1} u_i e^{iH_0t_1}} }dt_1
= \int_0^t \norm{\brr{\hat{V}_\tau,  u_i (H_0,-t_1)} }dt_1,
\label{upp_tilde_U_tau_u_i}
\end{align}
which reduces the inequality~\eqref{ineq_U_tau_V_U_taudagger} to 
\begin{align}
&\norm { \brr{ e^{i \br{ H_0 + \hat{V}_\tau } t }  U_\tau^\dagger O U_\tau e^{-i \br{ H_0 + \hat{V}_\tau } t } ,u_i}} \notag \\
& \le  \norm {\brr{ O ,u_i(H_0,-t)}}+  2\norm{O} \cdot \norm {\brr{  U_\tau ,u_i(H_0,-t)}} +2\norm{O}  \int_0^t \norm{\brr{\hat{V}_\tau,  u_i (H_0,-t_1)} }dt_1 .
\label{ineq_U_tau_V_U_taudagger2}
\end{align}
Combining the above inequality with the inequality~\eqref{starting_V_H_atu_uv}, we obtain the main inequality~\eqref{first_ineq_lem_V_u_i}. 
This completes the proof. $\square$

{~}

\hrulefill{\bf [ End of Proof of Lemma~\ref{lemma:commutator_u_i}]}

{~}

Using Lemma~\ref{lemma:commutator_u_i} in Eq.~\eqref{int_g_beta_t_V} with $O=V$ or $O=\ad_{\hat{H}_\tau(V)}$, the norm of Eq.~\eqref{aim_to_prove_statement_mC_stating}, i.e., 
\begin{align}
\frac{2}{\beta}\norm{\int_{-\infty}^\infty g_\beta(t)  \brr{ V(\hat{H}_\tau, t),u_i} dt}  , 
\end{align}
is upper-bounded by using the summation of the following three terms: 
\begin{align}
&\int_{|t|> \delta t} dt  |g_\beta(t)| 
\Bigl( \norm {\brr{ V  ,u_i(H_0,-t)}} + 2\norm{V} \cdot \norm{[U_\tau,u_i]}+ 2\norm{V} \cdot  \norm {\brr{  U_\tau ,u_i(H_0,-t)}} \Bigr) , \label{norm_V_tau_t_u_i_com_1} \\
&\int_{|t|\le \delta t} dt\int_0^1d\lambda  |tg_\beta(t)| \Bigl( \norm {\brr{ \ad_{\hat{H}_\tau} (V)  ,u_i(H_0,-\lambda t)}}  
 \notag \\
&\quad \quad\quad\quad\quad\quad\quad \quad\quad\quad + 2 \norm{\ad_{\hat{H}_\tau} (V)} \cdot  \norm{[U_\tau,u_i]}+ 2 \norm{\ad_{\hat{H}_\tau} (V)} \cdot \norm {\brr{  U_\tau ,u_i(H_0,-\lambda t)}}   \Bigr)   ,  \label{norm_V_tau_t_u_i_com_2} \\
&2\norm{V} \int_{|t|> \delta t} dt  |g_\beta(t)|  \int_0^t dt_1\norm{\brr{\hat{V}_\tau, u_i (H_0,-t_1) } }
+2\norm{\ad_{\hat{H}_\tau} (V)} \int_{|t|\le \delta t}dt |t g_\beta(t)| \int_0^1  \int_0^{\lambda t} dt_1\norm{\brr{\hat{V}_\tau, u_i (H_0,-t_1) } }  d\lambda  . \label{norm_V_tau_t_u_i_com_3}
\end{align}

In analyzing the above terms, we have to treat the norms like $\norm {\brr{  U_\tau ,u_i(H_0,-t)}}$. 
For this purpose, based on Lemma~\ref{lem:Quasi-local_commutator_bound}, we prove the following proposition (see Sec.~\ref{proof_prop:Quasi-local_commutator_bound_tau} for the proof):
\begin{prop}\label{prop:Quasi-local_commutator_bound_tau}
Let us adopt the same setup as in Lemma~\ref{lem:Quasi-local_commutator_bound}. 
We also choose $\mathcal{F}(\ell)$ as 
\begin{align}
\label{choice_of_nathcal_F_ell}
\mathcal{F}(\ell)= \exp\br{ K_\ell- \kappa_\beta \ell} , 
\end{align}
where $K_\ell$ monotonically increases with $\ell$. 
Then, we consider two kinds of functions as 
\begin{align}
g_1(t) =  \begin{cases}  
| g_\beta(t)| &\for |t| \ge \delta t,  \\
0 & \for |t| < \delta t, 
\end{cases} 
\quad \quad 
g_2(t) =  \begin{cases}  
0&\for |t| \ge \delta t,  \\
|tg_\beta(t)|  & \for |t| < \delta t.  
\end{cases} 
\label{definition_g_1t_g_2t}
\end{align}
We then obtain
\begin{align}
\label{main:eq:prop:Quasi-local_commutator_bound_tau}
\int_{-\infty}^\infty g_1(t) \norm{[O,u_i(H_0,t)]}dt \le  \mathfrak{g}_1  \mathcal{F}(\ell)  , 
\AND 
\int_{-\infty}^\infty g_2(t) \norm{[O,u_i(H_0,\lambda t)]}dt \le  \mathfrak{g}_2   \delta t  \mathcal{F}(\ell)  , 
\end{align}
where $0\le \lambda \le 1$, and $\mathfrak{g}_1$ and $\mathfrak{g}_2$ are defined as  
\begin{align}
\label{main:eq:prop:Quasi-local_commutator_bound_tau_parameter_g_1and_g_2}
&\mathfrak{g}_{1}:= \frac{4\beta}{\pi} \br{1+ \gamma (\Delta \ell)^D  e^{\kappa_\beta \Delta \ell } }\log\br{\frac{\beta}{2\pi \delta t}}+  \frac{8\gamma (4D \Delta \ell)^D\beta}{\pi}  e^{2\kappa_\beta\Delta \ell} +\frac{\gamma Cv\beta^2}{3}\br{\frac{8D^2}{e\mu}}^D    , \notag \\
&\mathfrak{g}_{2}:= \frac{2\beta}{\pi}  \brr{1+ 2^{D} D! \gamma C v (1+4/\mu)^D  \delta t  },
\end{align}
where $\Delta \ell$ is an arbitrary integer that satisfies 
\begin{align}
\kappa_\beta \Delta \ell \ge 1 .
\label{kappa_beta_Detla_ell_condition}
\end{align}
\end{prop}

{\bf Remark.}
By choosing $\Delta \ell$ such that $\kappa_\beta \Delta \ell=\orderof{1}$ [e.g., $\Delta \ell =\ceil{ 1/\kappa_\beta}=\orderof{\beta}$], we have $\kappa_\beta \Delta \ell\le 2$, and hence
\begin{align}
&\mathfrak{g}_{1}\le \Theta(\beta^{D+1})\log (\beta/\delta t)  , \notag \\
&\mathfrak{g}_{2}\le  \Theta(\beta) ,
\label{definition_g_1_g_2_mathfrak}
\end{align}
using the $\Theta$ notation in Eq.~\eqref{Theta_notation_def}, where we use $\kappa_\beta =\orderof{\beta}$ and $\delta t =\orderof{1}$ as has been given in~\eqref{definition_kappa_beta_1} and \eqref{cond_delta_t_1/2_00} [see also Eq.~\eqref{choice_fo_delta_t}], respectively.

By using Proposition~\ref{prop:Quasi-local_commutator_bound_tau}, we are able to upper-bound \eqref{norm_V_tau_t_u_i_com_1} 
under the conditions of  
\begin{align}
\norm{[V,u_i]}\le \norm{V} Q(0,r),\quad \norm{[U_\tau,u_i]}\le Q(\tau,r), 
\end{align}
where $Q(\tau,r)$ in Eq.~\eqref{assumption_proof_le_tau} can be expressed by the form of Eq.~\eqref{choice_of_nathcal_F_ell}.
We then obtain 
\begin{align}
&\int_{|t|> \delta t} dt  |g_\beta(t)| \cdot \norm {\brr{ V  ,u_i(H_0,-t)}} \le  \mathfrak{g}_1\norm{V}  Q(0,r) \le \mathfrak{g}_1\norm{V}  Q(\tau,r) 
,\notag \\
& \int_{|t|> \delta t} dt  |g_\beta(t)| \cdot 2\norm{V} \cdot \norm{[U_\tau,u_i]}
\le 2\norm{V} Q(\tau,r) \cdot  \frac{2\beta}{\pi}\log\br{\frac{\beta}{2\pi \delta t}} \le \mathfrak{g}_1\norm{V}  Q(\tau,r),  \notag \\
&\int_{|t|> \delta t} dt  |g_\beta(t)| \cdot 2\norm{V} \cdot  \norm {\brr{  U_\tau ,u_i(H_0,-t)}} \le2 \mathfrak{g}_1\norm{V}  Q(\tau,r) ,
 \end{align}
where, in the second inequality, we use the inequality~\eqref{main_eq:connection_of_exponential_operator_trans_1} in Lemma~\ref{belief_norm_lemma} below.
We therefore obtain 
\begin{align}
\textrm{Eq.~\eqref{norm_V_tau_t_u_i_com_1}} 
\le 4\mathfrak{g}_1  \norm{V}  Q(\tau,r)  . 
\label{norm_V_tau_t_u_i_com_1_fin}
\end{align}

To apply Proposition~\ref{prop:Quasi-local_commutator_bound_tau} to the terms of \eqref{norm_V_tau_t_u_i_com_2}, 
we need to consider the quasi-locality of $\ad_{\hat{H}_\tau} (V)$, which can be estimated by the following general statement (see Sec.~\ref{proof_prop:quasi-locality of_ad__H__tau_V} for the proof):
\begin{prop}[{\bf Analysis of \mbox{\boldmath $\ad_{\hat{H}_\tau} (O)$}}] \label{prop:quasi-locality of_ad__H__tau_V}
Let $O$ be an arbitrary quasi-local operator around $\mfL$ such that 
\begin{align}
&\norm{\brr{O, u_i}} \le \mathcal{N}_O\check{Q}(\tau,r) \quad {\rm with} \quad  \mathcal{N}_O\ge \norm{O}, 
\label{assum_quasi_locality_of_O_first_0}
\end{align}
with
\begin{align}
\check{Q}(\tau,r) := \min\br{ 2, e^{\check{\kappa}_0\tau+\check{\kappa}_1 \tau \log(r+\tau+e)-\check{\kappa}_\beta r }} \ge Q(\tau,r),
\label{assum_quasi_locality_of_O_first}
\end{align}
where $\dist_{i,\mfL}=r$ and $u_i$ is an arbitrary local unitary operator on the site $i$.
Note that the condition has been imposed for $Q(\tau,r)$ as in~\eqref{trivial_bound_/for_Q_tau_r}. 
Under the condition~\eqref{assumption_proof_le_tau}, $\ad_{\hat{H}_\tau} (V)$ is quasi-local in the sense that 
\begin{align}
\norm{\brr{\ad_{\hat{H}_\tau} (O) ,u_i }}\le \mathcal{N}_O g_{\tau,r}\check{Q}(\tau,r)
\label{main_eq:prop:quasi-locality of_ad__H__tau_V}
\end{align}
with
\begin{align}
\label{def_C_V_tau_r}
 g_{\tau,r}:=24\norm{V} \tau +2^{2D+6}\gamma^2 |\mfL|^2 \bar{J}_0\br{\check{C}_{2,\tau}^{2D} +r^{2D} + 8\gamma |\mfL| \check{C}_{2,\tau}^{2D}  \check{C}_{3,\tau}^{3D} }\Bigr]  , 
\end{align}
where we define $\check{C}_{2,\tau}$ and $\check{C}_{3,\tau}$ as in Eq.~\eqref{upper-bound:tilde_Q_tau_r_lemma}, i.e., $\check{C}_{\nu,\tau}:= \frac{\tau+e}{2} + \frac{32}{\check{\kappa}_\beta ^2}  \br{\nu^2 D^2 + \check{\kappa}_1^2 \tau^2+\frac{\check{\kappa}_\beta \check{\kappa}_0 \tau}{8}}$ for $\nu>0$. 
Also, we have 
\begin{align}
\label{main_eq:prop:quasi-locality of_ad__H__tau_V_2} 
\norm{\ad_{\hat{H}_\tau} (O)} \le \mathcal{N}_O  g'_{\tau}, \quad g'_{\tau}:=
2 \tau \norm{V} + 2^{2D+5}  \gamma^2 |\mfL|^2 \check{C}_{2,\tau}^{2D} \bar{J}_0 \br{ 1 +  2\gamma |\mfL|   \check{C}_{3,\tau}^{3D}  }   .
\end{align}
\end{prop}

{\bf Remark.}
In applying the proposition to the case of $O=V$, we can choose as $\check{Q}(\tau,r)=Q(\tau,r)$ and $\mathcal{N}_V=\norm{V}$ since $\norm{\brr{V, u_i}}\le \norm{V} Q(0,r)\le \norm{V} Q(\tau,r)$.
In this application, we should replace $\check{C}_{\nu, \tau}$ in $ g_{\tau,r}$ and $g'_{\tau}$ with $C_{\nu, \tau}$:
 \begin{align}
\label{def_C_V_tau_r_without_check}
C_{\nu,\tau}:=\frac{\tau+e}{2} + \frac{32}{\kappa_\beta^2}  \br{\nu^2 D^2 + \kappa_1^2 \tau^2+\frac{\kappa_\beta\kappa_0\tau}{8}} \quad  (\nu\in \mathbb{N}) .
\end{align}
By using the $\Theta$ notation in Eq.~\eqref{Theta_notation_def}, we obtain 
\begin{align}
\label{upp_t_tau_r_g'}
&  g_{\tau,r}\le \Theta(\norm{V}, |\mfL|^3,  r^{2D}, \beta^{10D}, \kappa_0^{5D}, \kappa_1^{10D},\tau^{10D}) , \notag \\
& g'_{\tau} \le  \Theta(\norm{V}, |\mfL|^3, \beta^{10D}, \kappa_0^{5D}, \kappa_1^{10D},\tau^{10D}).
\end{align}
As shown in the following section (Sec.~\ref{sec:Completing the proof_quasi_locality}), for our purpose, it is enough to ensure that $g_{\tau,r}$ and $g'_{\tau} $ are upper-bounded by finite-degree polynomials of $\{ \norm{V}, |\mfL|, r, \beta, \kappa_0, \kappa_1,\tau$\}.

By applying Propositions~\ref{prop:Quasi-local_commutator_bound_tau} and \ref{prop:quasi-locality of_ad__H__tau_V} to Eq.~\eqref{norm_V_tau_t_u_i_com_2},
we have 
\begin{align}
&\int_{|t|\le \delta t} dt\int_0^1d\lambda  |tg_\beta(t)|  \cdot \norm {\brr{ \ad_{\hat{H}_\tau} (V)  ,u_i(H_0,-\lambda t)}}\le 
g_{\tau,r}\norm{V} \mathfrak{g}_2   \delta t  Q(\tau,r) ,   \notag \\
& \int_{|t|\le \delta t} dt\int_0^1d\lambda  |tg_\beta(t)| \cdot  2 \norm{\ad_{\hat{H}_\tau} (V)} \cdot \norm{[U_\tau,u_i]}  
\le   2g'_{\tau} \norm{V}  Q(\tau,r)\int_{|t|\le \delta t} dt\int_0^1d\lambda  |tg_\beta(t)|   \le  \frac{2\beta g'_{\tau}}{\pi}  \delta t\norm{V}    Q(\tau,r),
 \notag \\
&\int_{|t|\le \delta t} dt\int_0^1d\lambda  |tg_\beta(t)| \cdot  2\norm{\ad_{\hat{H}_\tau} (V)} \cdot \norm {\brr{  U_\tau ,u_i(H_0,-\lambda t)}}  
\le 2g'_{\tau} \norm{V} \cdot  \mathfrak{g}_2 \delta t    Q(\tau,r)  ,
\end{align}
where we use the inequality~\eqref{main_eq:connection_of_exponential_operator_trans_2} in the second inequality.
We thus upper-bound Eq.~\eqref{norm_V_tau_t_u_i_com_2} as follows:
\begin{align}
\textrm{Eq.~\eqref{norm_V_tau_t_u_i_com_2}} 
\le \br{\mathfrak{g}_2  g_{\tau,r}  +\frac{2\beta}{\pi}g'_{\tau} + 2 \mathfrak{g}_2  g'_{\tau} } \norm{V} \delta t Q(\tau,r)  
\le \br{g_{\tau,r}  + 3g'_{\tau} } \norm{V} \mathfrak{g}_2 \delta t Q(\tau,r)   ,
\label{norm_V_tau_t_u_i_com_2_fin}
\end{align}
where we use $\frac{2\beta}{\pi}\le \mathfrak{g}_2$ from the definition~\eqref{main:eq:prop:Quasi-local_commutator_bound_tau_parameter_g_1and_g_2}.

Finally, to analyse the commutator $\brr{\hat{V}_\tau, u_i (H_0,-t_1) }$ in \eqref{norm_V_tau_t_u_i_com_3}, we use the relation
 \begin{align}
\norm{\brr{\hat{V}_\tau, u_i (H_0,-t_1) } }
&= 
\norm{\brr{ \int_0^\tau U_{\tau_1}V   U_{\tau_1}^\dagger  d\tau_1 , u_i (H_0,-t_1) } }  \notag \\
&\le \int_0^\tau \br{ 2 \norm{V} \cdot  \norm{\brr{  U_{\tau_1}, u_i (H_0,-t_1) } }
+  \norm{ [ V , u_i (H_0,-t_1) }} d\tau_1 ,
\label{norm_V_tau_t_u_i_com_3_pre}
 \end{align}
Then, to estimate Eq.~\eqref{norm_V_tau_t_u_i_com_3}, we prove the following proposition using Corollary~\ref{corol:Quasi-local_commutator_bound}.
(see Sec.~\ref{proof_prop:Quasi-local_commutator_bound_tau_2} for the proof):
\begin{prop}\label{prop:Quasi-local_commutator_bound_tau_2}
Let us adopt the same setup as in Proposition~\ref{prop:Quasi-local_commutator_bound_tau}. 
We also choose $\mathcal{F}(\ell)$ as in Eq.~\eqref{choice_of_nathcal_F_ell}. 
Then,  we have 
\begin{align}
&\int_{-\infty}^\infty g_1(t) \int_0^t \norm{[O,u_i(H_0,t_1)]} dt_1 dt\le  \mathfrak{g}_3  \mathcal{F}(\ell)  ,  \notag\\
&\int_{-\infty}^\infty g_2(t) \int_0^{\lambda t} \norm{[O,u_i(H_0,t_1)]} dt_1 dt\le  \mathfrak{g}_4 \delta t^2 \mathcal{F}(\ell)  , 
\label{main:eq:prop:Quasi-local_commutator_bound_tau__2}
\end{align}
with
$\mathfrak{g}_3$ and $\mathfrak{g}_4$ defined as  
\begin{align}
&\mathfrak{g}_3:=  \frac{\beta^2}{6} \brr{1+ \gamma (\Delta \ell)^D e^{\kappa_\beta \Delta \ell }} 
 + 2\gamma  (4D \Delta \ell)^D e^{2\kappa_\beta \Delta \ell }
\br {\frac{16D\beta \kappa_\beta \Delta \ell}{\pi}+\frac{\beta^2}{6} }+ \frac{ 4\gamma Cv\beta^3}{\pi^3}\br{\frac{8D^2}{e\mu}}^D
 \le  \Theta(\beta^{D+2}) , \notag \\
&\mathfrak{g}_4 = \frac{\beta}{\pi} \brr{1+ \frac{2}{3} \cdot 2^{D} D! C v (1+4/\mu)^D \delta t} \le \mathfrak{g}_2,
\label{definition_g_3_g_4_mathfrak}
\end{align}
respectively, where $g_1(t)$ and $g_2(t)$ have been defined in Eq.~\eqref{definition_g_1t_g_2t}. 
Note that the inequality for $\mathfrak{g}_3$ can be derived by choosing $\Delta \ell =\ceil{ 1/\kappa_\beta}$, for example. 
\end{prop}

By applying Proposition~\ref{prop:Quasi-local_commutator_bound_tau_2} to the integral of Eq.~\eqref{norm_V_tau_t_u_i_com_3_pre}, the first term in Eq.~\eqref{norm_V_tau_t_u_i_com_3} is upper-bounded by
 \begin{align}
 \label{norm_V_tau_t_u_i_com_3_ca1}
&2\norm{V}\int_{|t|> \delta t} dt |g_\beta(t)| \int_0^t  \norm{\brr{\hat{V}_\tau, u_i (H_0,-t_1) } }dt_1\notag \\
&\le 2\norm{V}\int_0^\tau d\tau_1 \int_{|t|> \delta t}dt |g_\beta(t)| \int_0^t  \br{ 2 \norm{V} \cdot  \norm{\brr{  U_{\tau_1}, u_i (H_0,-t_1) } }
+  \norm{ [ V , u_i (H_0,-t_1) }}  dt_1  \notag \\
&\le2 \mathfrak{g}_3  \norm{V}^2 \int_0^\tau d\tau_1  \brr{ 2  Q(\tau_1,r) +Q(0,r) } 
\le 6 \mathfrak{g}_3  \norm{V}^2 \int_0^\tau d\tau_1 Q(\tau_1,r),
 \end{align}
 where we use $Q(0,r)\le Q(\tau,r)$. 
Also, the second term in Eq.~\eqref{norm_V_tau_t_u_i_com_3} is upper-bounded by 
\begin{align}
&2\norm{\ad_{\hat{H}_\tau} (V)} \int_{|t|\le \delta t}dt  |t g_\beta(t)| \int_0^1  \int_0^{\lambda t} dt_1\norm{\brr{\hat{V}_\tau, u_i (H_0,-t_1) } }  d\lambda  \notag \\
&\le 2g'_{\tau}\norm{V} \int_{|t|\le \delta t}dt  |t g_\beta(t)|   \int_0^{t} dt_1\norm{\brr{\hat{V}_\tau, u_i (H_0,-t_1) } }   \notag \\
&\le 2g'_{\tau}\norm{V} \int_0^\tau d\tau_1 \int_{|t|\le \delta t}dt  |t g_\beta(t)|   \int_0^{t}  \br{ 2 \norm{V} \cdot  \norm{\brr{  U_{\tau_1}, u_i (H_0,-t_1) } }
+  \norm{ [ V , u_i (H_0,-t_1) }}  dt_1 \notag \\
&\le 6\mathfrak{g}_4  g'_{\tau} \delta t^2 \norm{V}^2 \int_0^\tau d\tau_1 Q(\tau_1,r), 
 \label{norm_V_tau_t_u_i_com_3_ca1//22}
\end{align}
where we use the inequality~\eqref{main_eq:prop:quasi-locality of_ad__H__tau_V_2} to derive $\norm{\ad_{\hat{H}_\tau} (V)}\le g'_{\tau}\norm{V}$.
 
From the condition~\eqref{assumption_proof_le_tau}, we have 
\begin{align}
\label{norm_V_tau_t_u_i_com_3_ca2}
 \int_0^\tau d\tau_1Q (\tau_1,r) &= e^{-\kappa_\beta r } \int_0^\tau d\tau_1  e^{\kappa_0\tau_1+\kappa_1 \tau_1 \log(r+\tau_1+e)} 
 \le  e^{-\kappa_\beta r +\kappa_1 \tau \log(r+\tau+e)} \int_0^\tau d\tau_1  e^{\kappa_0\tau_1}  \notag \\
& = e^{-\kappa_\beta r  +\kappa_1 \tau \log(r+\tau+e)}  \frac{e^{\kappa_0\tau}- 1}{\kappa_0} 
\le \frac{Q (\tau,r)}{\kappa_0} . 
\end{align}
By combining the inequalities~\eqref{norm_V_tau_t_u_i_com_3_ca1} and~\eqref{norm_V_tau_t_u_i_com_3_ca1//22} with~\eqref{norm_V_tau_t_u_i_com_3_ca2}, we obtain 
\begin{align}
\textrm{Eq.~\eqref{norm_V_tau_t_u_i_com_3}} 
\le  \frac{6\norm{V}^2}{\kappa_0}   Q(\tau,r) \br{\mathfrak{g}_3 +  \mathfrak{g}_4  g'_{\tau}\delta t^2}  . 
\label{norm_V_tau_t_u_i_com_3_up_fin}
\end{align}

\subsubsection{Completing the proof of Subtheorem~\ref{sub_thm:U_tau_u_i_commun}} \label{sec:Completing the proof_quasi_locality}

Collecting~\eqref{norm_V_tau_t_u_i_com_1_fin}, \eqref{norm_V_tau_t_u_i_com_2_fin} and \eqref{norm_V_tau_t_u_i_com_3_up_fin} together, 
we obtain
\begin{align}
 \norm{[\mC_\tau,u_i]} 
 &\le \frac{2}{\beta} \int_{-\infty}^\infty |g_\beta(t)|\cdot \norm{ [ V(\hat{H}_\tau, t), u_i ]}dt \notag \\
 &\le \frac{2\norm{V} }{\beta}\brr{ 4\mathfrak{g}_1  +\br{ g_{\tau,r}  + 3g'_{\tau} }  \mathfrak{g}_2 \delta t +
 \frac{6\norm{V}}{\kappa_0}    \br{\mathfrak{g}_3 +  \mathfrak{g}_4  g'_{\tau}\delta t^2} }
 Q(\tau,r)  ,
 \label{upper_bpound_/mc_tau_ui}
\end{align}
We now choose $\delta t$ as  
\begin{align}
\label{choice_fo_delta_t}
\delta t = \frac{1}{g_{\tau,r}  + 3g'_{\tau} }.   
\end{align}
Because of~\eqref{choice_fo_delta_t} and 
\begin{align}
& g_{\tau,r}\le \Theta(\norm{V},|\mfL|^3, r^{2D}, \beta^{10D}, \kappa_0^{5D}, \kappa_1^{10D},\tau^{10D}) , \quad  g'_{\tau} \le  \Theta(\norm{V},|\mfL|^3, \beta^{10D}, \kappa_0^{5D}, \kappa_1^{10D},\tau^{10D}) ,
\end{align}
we have 
\begin{align}
\label{upper_bound_de_ta_t_-1}
\log(\delta t^{-1}) \le  \Theta(1) \brr{ \log(r+\tau+e)  + \log(\kappa_0 \kappa_1\beta \norm{V}\cdot |\mfL|) } .
\end{align}

Then, the inequality~\eqref{upper_bpound_/mc_tau_ui} imposes the following condition for $\kappa_0$ and $\kappa_1$ from~\eqref{aim_to_prove_statement_mC}:
\begin{align}
\label{condition_kappa_1and_kpappa_2}
 \frac{2\norm{V}}{\beta}\brr{ 4\mathfrak{g}_1   + \mathfrak{g}_2 +  \frac{6\norm{V}}{\kappa_0}    \br{\mathfrak{g}_3 +  \frac{\mathfrak{g}_4}{9g'_{\tau}}  }} \le \kappa_0+ \kappa_1  \log(r+\tau+e) .
\end{align}
To determine $\kappa_0$ and $\kappa_1$, we first recall 
\begin{align}
\label{ineq:mathfrak_g_1_2_3_4_g}
&\mathfrak{g}_{1}\le \Theta(\beta^{D+1})\log (\beta/\delta t)  , 
\quad\mathfrak{g}_4\le \mathfrak{g}_{2}\le  \Theta(\beta) , \quad \mathfrak{g}_3 \le  \Theta(\beta^{D+2})
\end{align}
from the definitions of $\{\mathfrak{g}_{1},\mathfrak{g}_{2},\mathfrak{g}_{3},\mathfrak{g}_{4}\}$ in~\eqref{definition_g_1_g_2_mathfrak} and \eqref{definition_g_3_g_4_mathfrak}. 
We thus obtain
\begin{align}
 & \frac{2\norm{V}}{\beta}\brr{ 4\mathfrak{g}_1   + \mathfrak{g}_2 +  \frac{6}{\kappa_0}    \br{\mathfrak{g}_3\norm{V}   +  \frac{\mathfrak{g}_4}{9g'_{\tau}}  }}
 \le \norm{V} \brr{ \Theta(\beta^D)\log (\beta/\delta t))  + \frac{\norm{V}}{\kappa_0} \Theta(\beta^{D+1})}\notag \\
 &\le \norm{V} \brr{ \Theta(\beta^D) \log(r+\tau+e)  + \Theta(\beta^D)\log(\kappa_0\kappa_1 \beta\norm{V}\cdot |\mfL|)  + \frac{\norm{V}}{\kappa_0} \Theta(\beta^{D+1}) } .\label{upper_bound_g_1_g_2_g_3_f_2}
 \end{align}

By applying the inequalities~\eqref{upper_bound_g_1_g_2_g_3_f_2} to \eqref{condition_kappa_1and_kpappa_2}, we have to satisfy the inequalities of
\begin{align}
\kappa_0\ge  \norm{V}\brr{  \Theta(\beta^D)\log(\kappa_0\kappa_1 \beta\norm{V}\cdot |\mfL|) + \frac{\norm{V}}{\kappa_0} \Theta(\beta^{D+1}) \log(\beta) }
\end{align}
and 
\begin{align}
\kappa_1  \log(r+\tau+e) \ge\norm{V}  \Theta(\beta^D) \log(r+\tau+e) . 
\end{align}
The above condition is satisfied by choosing 
\begin{align}
\kappa_0= \Theta(\beta^D)\norm{V} \log (\beta\norm{V}\cdot |\mfL|)  , \quad \kappa_1=\Theta(\beta^D) \norm{V} .
\end{align}
This completes the proof of Subtheorem~\ref{sub_thm:U_tau_u_i_commun}.  $\square$

\subsection{Proof of Proposition~\ref{prop:Quasi-local_commutator_bound_tau}} \label{proof_prop:Quasi-local_commutator_bound_tau}

We start from upper-bounding the integrals of $g_1(t)$ [or $g_2(t)$] and $|t|g_1(t)$ [or $|t|g_2(t)$]. 
We can prove the following lemma: 
\begin{lemma} \label{belief_norm_lemma}
For an arbitrary $\delta t >0$ such that
\begin{align}
\label{cond_delta_t_1/2}
\delta t \le \frac{\beta}{4\pi},
\end{align}
the integrals of $\{g_1(t), |t| g_1(t)\}$ and $\{g_2(t), |t| g_2(t)\}$ are upper-bounded by
\begin{align}
\label{main_eq:connection_of_exponential_operator_trans_1}
&\int_{-\infty}^\infty g_1(t) dt
= \int_{|t| > \delta t}  |g_\beta (t)|  dt \le  \frac{2\beta}{\pi}\log\br{\frac{\beta}{2\pi \delta t}} ,\notag \\
&\int_{-\infty}^\infty  |t|^m  g_1(t) dt
=\int_{|t| > \delta t}  |t^m g_\beta (t)|  dt \le  2\br{\frac{\beta}{2\pi}}^{m+1} m! \zeta(m+1)  \quad (m \ge 1).
\end{align}
Also, for an arbitrary $\delta t >0$, we have 
\begin{align}
\label{main_eq:connection_of_exponential_operator_trans_2}
&\int_{-\infty}^\infty |t|^m g_2(t) dt=\int_{|t| \le \delta t} | t^{m+1} g_\beta (t) | dt  \le  \frac{\beta}{(m+1)\pi} \delta t^{m+1} \quad (m\ge 0), 
\end{align}
where $\zeta(x)$ is the Riemann zeta function, i.e., $\zeta(x)=\sum_{s=1}^\infty (1/s^x)$,  which is given by $\zeta(2)=\pi^2x/6$, $\zeta(3)=1.20205\cdots$, $\zeta(4)=\pi^4/90$, and so on. 
\end{lemma}
{~}\\
\textit{Proof of Lemma~\ref{belief_norm_lemma}.}
From the definition~\eqref{def_g_t_exp_decay}, i.e., $$g_\beta(t)= -{\rm sign}(t)  \frac{e^{-2\pi |t|/\beta}}{1-e^{-2\pi |t|/\beta}},$$ 
the function $g_\beta(t)$ is an odd function, $g_\beta(-t)=-g_\beta(t)$.
Thus, from $|g_\beta(-t)|=|g_\beta(t)|$, we obtain by letting $z=2\pi t/\beta$ and $\delta z=2\pi \delta t/\beta$ ($\le 1/2$)
\begin{align}
\label{integral_g_beta_t_t_ge_delta_t}
\int_{|t| > \delta t}  |g_\beta (t)|  dt 
&=2 \int_{\delta t}^\infty  |g_\beta (t)|  dt 
= 2\int_{\delta z}^\infty   \frac{e^{-z}}{1-e^{-z}} \frac{\beta dz}{2\pi} =\frac{\beta}{\pi}\int_{\delta z}^\infty  \frac{1}{e^{z}-1}dz \notag \\
&=- \frac{\beta}{\pi} \log\br{1 - e^{-\delta z} } \le  \frac{\beta}{\pi} \brr{\delta z+ \log(1/\delta z) } ,
\end{align}
and 
\begin{align}
\label{integral_t_g_beta_t_ge_delta__1}
\int_{|t| > \delta t}  |t^m g_\beta (t)|  dt 
&= 2\int_{\delta t}^\infty \br{\frac{\beta z}{2\pi}}^m \cdot  \frac{e^{-z}}{1-e^{-z}} \frac{\beta dz}{2\pi}
 =2\br{\frac{\beta}{2\pi}}^{m+1}\int_{\delta z}^\infty  \frac{z^m}{e^{z}-1}dz \notag \\
&\le 2\br{\frac{\beta}{2\pi}}^{m+1}\int_{0}^\infty  \frac{z^m}{e^{z}-1}dz  
=  2\br{\frac{\beta}{2\pi}}^{m+1} m! \zeta(m+1)  ,
\end{align}
where $\zeta(x)$ has been defined as the Riemann zeta function. 
By substituting $\delta z$ by $2\pi \delta t/\beta$, we prove the first inequality in~\eqref{main_eq:connection_of_exponential_operator_trans_1}, where we use the inequalities of $x+\log(1/x)\le 2\log(1/x)$  for $x\le 1/2$. 

In the same way, we can derive 
\begin{align}
\int_{|t| \le \delta t}  |t^{m+1} g_\beta (t)|  dt 
 &= 2\int_{0}^{\delta z} \br{\frac{\beta z}{2\pi}}^{m+1} \frac{e^{-z}}{1-e^{-z}} \frac{\beta dz}{2\pi} \notag \\
 &=2 \br{\frac{\beta}{2\pi}}^{m+2}\int_{0}^{\delta z}  \frac{z^{m+1}}{e^{z}-1}dz  \notag \\
 &\le \frac{2}{m+1} \br{\frac{\beta}{2\pi}}^{m+2} \delta z^{m+1} = 
 \frac{\beta}{(m+1)\pi} \delta t^{m+1},
  \label{integral_t__g_beta_t_ineq}
\end{align}
where we use $ \int_{0}^{z_0}\frac{z^m}{e^{z}-1}\le z_0^m/m$ for arbitrary $z_0>0$ and $m\in \mathbb{N}$ and substitute $\delta z$ by $2\pi \delta t/\beta$ in the last equation.   
We thus prove the second main inequality~\eqref{main_eq:connection_of_exponential_operator_trans_2}.
This completes the proof. $\square$


{~}

\hrulefill{\bf [ End of Proof of Lemma~\ref{belief_norm_lemma}]}

{~}

We then rely on Lemma~\ref{belief_norm_lemma} for the proof. 
We start from the case of $f(t)=g_1(t)$ in the statement to derive the first inequality in~\eqref{main:eq:prop:Quasi-local_commutator_bound_tau}.  
Using the inequality~\eqref{main_eq:connection_of_exponential_operator_trans_1} in Lemma~\ref{belief_norm_lemma}, the parameters $f_1$ and $f_2$ in~\eqref{definition_lem:Quasi-local_commutator_bound} immediately given by
\begin{align}
\label{est_f_1andf_2_g_1case}
&f_1= \int_{-\infty}^\infty g_1(t) dt \le  \frac{2\beta}{\pi}\log\br{\frac{\beta}{2\pi \delta t}}, \notag \\
&f_2\le \int_{-\infty}^\infty  |t| g_1(t) dt \le \frac{\beta^2}{2\pi^2} \zeta(2) = \frac{\beta^2}{12}.
\end{align}
Also, because $t_0=\max(0,\mu s \Delta \ell/(2v) - \mu (\Delta \ell-1)/v)=[\mu/(2v)]\Delta \ell(s-2) +\mu/v$ for $s\ge 2$ from the definition,  we have
\begin{align}
\label{est_f_t_0_s_g_1case}
f_{t_0}(s)= \int_{|t| \ge t_0}g_1(t)dt &= -\frac{\beta}{\pi}  \log\br{1-e^{-2\pi t_0/\beta}  } 
\le \frac{\beta}{\pi}  \log\br{e+ \frac{e\beta}{2\pi t_0}}e^{-2\pi t_0/\beta} \notag \\
&\le \frac{\beta}{\pi}  \log\br{e+ \frac{e}{2\kappa_\beta \Delta \ell}}e^{-2\kappa_\beta (s-2)  \Delta \ell } 
\le   \frac{2\beta}{\pi} e^{-2\kappa_\beta (s-2) \Delta \ell }   
\end{align}
for $s\ge 2$, 
where we use Eq.~\eqref{integral_g_beta_t_t_ge_delta_t}, the inequality $\log(1-e^{-x}) \le \log(e+e/x) e^{-x}$,  
$2\pi t_0/\beta\ge  \pi \mu (s-2) \Delta \ell/(v\beta) \ge 2\kappa_\beta (s-2) \Delta \ell$, and 
$\kappa_\beta \Delta \ell \ge 1$ from the condition~\eqref{kappa_beta_Detla_ell_condition}. 
Recall that $2\kappa_\beta \le \min\br{ \frac{\pi \mu}{v\beta} , \frac{\mu}{2},2} \le \frac{\pi \mu}{v\beta}$ from the definition~\eqref{definition_kappa_beta_1}.
For $s\le 1$, we can adopt $f_{t_0}(s)\le f_1$ by letting $t_0=0$. 
Note that $\log(e+e/x) e^{-x}$ monotonically decreases with $x$ for $x\ge 0$.

Using the above estimations, we calculate the summation in the RHS of~\eqref{main_ineq:lem:Quasi-local_commutator_bound}  from the choice of  $\mathcal{F}(\ell)$ as in Eq.~\eqref{choice_of_nathcal_F_ell}.
First, we obtain the following upper bound from the inequalities~\eqref{est_f_1andf_2_g_1case} and $\kappa_\beta\le \mu/4$ from the definition of $\kappa_\beta$:
\begin{align}
\label{upper_bound_s^D_mathcal_F_ell-s_f_2term_0}
&2\gamma (\Delta \ell)^D\sum_{s=1}^\infty s^D\mathcal{F}(\ell-s\Delta \ell) C f_2 v e^{-\mu s\Delta \ell/2}  \notag \\
& \le \frac{\gamma Cv}{6} \beta^2 (\Delta \ell)^D\sum_{s=1}^\infty s^De^{K_{\ell-s \Delta \ell} -\kappa_\beta (\ell-s\Delta \ell)} e^{-\mu s \Delta \ell/2}   \notag \\
 &\le  \frac{\gamma Cv}{6} \beta^2 (\Delta \ell)^D \mathcal{F}(\ell)  \sum_{s=1}^\infty s^D e^{-\mu s \Delta \ell/4}   \notag \\
 &\le \frac{\gamma Cv}{6} \beta^2 (\Delta \ell)^D e^{-\mu \Delta \ell /4}  \br{1+2^DD! \brr{ \max\br{1,\frac{4}{\mu \Delta \ell}}}^D} \mathcal{F}(\ell)  \notag \\
 &\le\frac{\gamma Cv \beta^2 }{6\mu^D} (\mu\Delta \ell)^D e^{-\mu \Delta \ell /4}  \br{1+2^DD!}  \mathcal{F}(\ell) \le  
  \frac{\gamma Cv\beta^2}{3}\br{\frac{8D^2}{e\mu}}^D  \mathcal{F}(\ell)  ,
\end{align}
where we use $K_{\ell'}\le K_{\ell}$ for $\ell'\le \ell$ in the second inequality,~\cite[(S.11) of Lemma~1 therein]{kuwahara2022optimal} in the third inequality, 
$4/(\mu \Delta \ell) \le 1$ in the fourth inequality, which originates from $\kappa_\beta \Delta \ell \ge 1$ and $\kappa_\beta=\min\br{ \frac{\pi \mu}{2v\beta} , \frac{\mu}{4}}$ from~\eqref{kappa_beta_Detla_ell_condition}, and $(\mu\Delta \ell)^D e^{-\mu \Delta \ell /4}\le (4D/e)^D$ from $\mu \Delta \ell\ge 4$ in the last inequality. 
In the same way, we secondly obtain using the inequality~\eqref{est_f_t_0_s_g_1case}
\begin{align}
\label{upper_bound_s^D_mathcal_F_ell-s_f_2term}
&2\gamma (\Delta \ell)^D\sum_{s=1}^\infty s^D\mathcal{F}(\ell-s\Delta \ell )f_{t_0}(s) \notag \\
&\le  
2\gamma (\Delta \ell)^Df_1 \mathcal{F}(\ell-\Delta \ell ) +
\frac{4\gamma (\Delta \ell)^D\beta}{\pi} \sum_{s=2}^\infty s^De^{K_{\ell-s\Delta \ell } -\kappa_\beta (\ell-s\Delta \ell )} e^{-2\kappa_\beta (s-2) \Delta \ell }  \notag \\
&\le  2\gamma (\Delta \ell)^D  f_1 e^{\kappa_\beta \Delta \ell }\mathcal{F}(\ell) +
\frac{4\gamma (2\Delta \ell)^D\beta}{\pi} \mathcal{F}(\ell) e^{3 \kappa_\beta\Delta \ell} \sum_{\tilde{s}=1}^\infty \tilde{s}^De^{-\kappa_\beta \tilde{s} \Delta \ell }  \notag \\
&\le  2\gamma (\Delta \ell)^D  f_1 e^{\kappa_\beta \Delta \ell }\mathcal{F}(\ell) +
\frac{4\gamma (2\Delta \ell)^D\beta}{\pi}  e^{2\kappa_\beta\Delta \ell}\br{1+2^DD! \brr{ \max\br{1,\frac{1}{\kappa_\beta\Delta \ell }}}^D} \mathcal{F}(\ell)  \notag \\
&\le 2\gamma (\Delta \ell)^D  f_1 e^{\kappa_\beta \Delta \ell }\mathcal{F}(\ell) +
\frac{4\gamma (2\Delta \ell)^D\beta}{\pi}  e^{2\kappa_\beta\Delta \ell} \br{1+2^DD! } \mathcal{F}(\ell)  \notag \\
&\le 2\gamma (\Delta \ell)^D  f_1 e^{\kappa_\beta \Delta \ell }\mathcal{F}(\ell) +  \frac{8\gamma (4D \Delta \ell)^D\beta}{\pi}  e^{2\kappa_\beta\Delta \ell}  \mathcal{F}(\ell)  ,
\end{align}
where we let $s=\tilde{s}+1$ [$(\tilde{s}+1)^D \le (2\tilde{s})^D$] and use the condition $\kappa_\beta \Delta \ell\ge 1$. 
Therefore, by combining the inequalities~\eqref{upper_bound_s^D_mathcal_F_ell-s_f_2term_0} and \eqref{upper_bound_s^D_mathcal_F_ell-s_f_2term}, we obtain
\begin{align}
\label{upper_bound_s^D_mathcal_F_ell-s_f_2term_fin}
&2\gamma (\Delta \ell)^D \sum_{s=1}^\infty s^D\mathcal{F}(\ell-s\Delta \ell ) \brr{ C f_2 v e^{-\mu s \Delta \ell /2} + f_{t_0}(s)} \notag \\
& \le 2\gamma (\Delta \ell)^D  f_1 e^{\kappa_\beta \Delta \ell }\mathcal{F}(\ell) +  \frac{8\gamma (4D \Delta \ell)^D\beta}{\pi}  e^{2\kappa_\beta\Delta \ell}  \mathcal{F}(\ell)+\frac{\gamma Cv\beta^2}{3}\br{\frac{8D^2}{e\mu}}^D  \mathcal{F}(\ell) .
 \end{align}
By applying the inequalities~\eqref{est_f_1andf_2_g_1case} and~\eqref{upper_bound_s^D_mathcal_F_ell-s_f_2term_fin} to \eqref{main_ineq:lem:Quasi-local_commutator_bound}, we obtain the first main inequality in~\eqref{main:eq:prop:Quasi-local_commutator_bound_tau} with the choice of $\mathfrak{g}_1$ as in Eq.~\eqref{main:eq:prop:Quasi-local_commutator_bound_tau_parameter_g_1and_g_2} in the case of $f(t)=g_1(t)$.

We next consider the case of $f(t)=g_2(t)$ to derive the second inequality in~\eqref{main:eq:prop:Quasi-local_commutator_bound_tau}.  
In this case, we simply choose $\Delta \ell=1$ in utilizing the inequality~\eqref{main_ineq:lem:Quasi-local_commutator_bound_Delta_eLL=1} in Lemma~\ref{lem:Quasi-local_commutator_bound}.  
Because of $\delta t < t_0=\mu s/(2v)$, we have $f_{t_0}(s)=0$ from the definition of $g_2(t)$, where $g_2(t)=0$ for $t\le \delta t$ as in Eq.~\eqref{definition_g_1t_g_2t}.
The inequality~\eqref{main_eq:connection_of_exponential_operator_trans_2} in Lemma~\ref{belief_norm_lemma} gives the parameters $f_1$ and $f_2$ as follows:
\begin{align}
\label{est_f_1andf_2_g_2case}
&f_1 \le \frac{\beta \delta t }{\pi} , \quad  f_2\le \frac{\beta \delta t^2}{2\pi} .
\end{align}
Then, we upper-bound $C f_2 v e^{-\mu s/2} + f_{t_0}(s) $ by
\begin{align} 
C f_2 v e^{-\mu s /2} + f_{t_0}(s) \le \frac{C v  \beta \delta t^2}{2\pi} e^{-\mu s/2} . 
\label{g_2_t_case_C_f_2_term}
\end{align}
By using the same calculations as~\eqref{upper_bound_s^D_mathcal_F_ell-s_f_2term_0}, we obtain
\begin{align}
\label{upper_bound_s^D_mathcal_F_ell-s_f_2term_g_2_case}
2\gamma \sum_{s=1}^\infty s^D\mathcal{F}(\ell-s) \brr{ C f_2 v e^{-\mu s/2} + f_{t_0}(s)}
 &\le \frac{\gamma C v  \beta \delta t^2}{\pi} \cdot  2^{D+1} D! (1+4/\mu)^D \mathcal{F}(\ell) \notag \\
& = \frac{\beta \delta t}{\pi} \cdot  2^{D+1} D! \gamma C v (1+4/\mu)^D  \delta t \mathcal{F}(\ell),
 \end{align}
 where we use $\max\br{1,4/\mu} \le (1+4/\mu)$. 
We therefore obtain the second inequality in~\eqref{main:eq:prop:Quasi-local_commutator_bound_tau} with the choice of $\mathfrak{g}_2$ as in Eq.~\eqref{main:eq:prop:Quasi-local_commutator_bound_tau_parameter_g_1and_g_2} by applying the inequalities~\eqref{est_f_1andf_2_g_2case} and 
\eqref{upper_bound_s^D_mathcal_F_ell-s_f_2term_g_2_case} to \eqref{main_ineq:lem:Quasi-local_commutator_bound_Delta_eLL=1}.  
This completes the proof of Proposition~\ref{prop:Quasi-local_commutator_bound_tau}. $\square$

\subsection{Proof of Proposition~\ref{prop:quasi-locality of_ad__H__tau_V}} \label{proof_prop:quasi-locality of_ad__H__tau_V}
\subsubsection{Preliminary lemmas}

We first prove several supplemental lemmas to prove the main statement. 
First of all, we prove the following lemma:
\begin{lemma} \label{lem:s_0_upper_bound}
For an arbitrary positive $\nu$, the condition 
\begin{align}
r^\nu \check{Q}(r,\tau) \le e^{-\check{\kappa}_\beta  r/2} 
\label{lem:s_0_upper_bound/cond}
\end{align}
is satisfied with 
\begin{align}
r \ge \frac{\tau+e}{2} +  \frac{32}{\check{\kappa}_\beta^2} \br{\nu^2+\check{\kappa}_1^2 \tau^2 + \frac{\check{\kappa}_\beta\check{\kappa}_0\tau  }{8}} , 
\end{align}
where $\check{Q}(r,\tau)$ was defined in Eq.~\eqref{assum_quasi_locality_of_O_first}. 
\end{lemma}

\subsubsection{Proof of Lemma~\ref{lem:s_0_upper_bound}}
We start with the expression~\eqref{assum_quasi_locality_of_O_first} of
\begin{align}
\check{Q} (\tau,r) =  e^{\check{\kappa}_0\tau +\check{\kappa}_1 \tau \log(r+\tau+e)-\check{\kappa}_\beta  r}.
\end{align}
Then, the condition~\eqref{lem:s_0_upper_bound/cond} reads 
\begin{align}
&\nu \log(r) +\check{\kappa}_0\tau  + \check{\kappa}_1 \tau \log(r+\tau+e) \le  \frac{\check{\kappa}_\beta  r}{2} .
\label{re_state_lem:s_0_upper_bound/cond}
\end{align}
For arbitrary positive $c_1$ and $c_2$, we have 
\begin{align}
\nu \log(r) +\check{\kappa}_0\tau  + \check{\kappa}_1 \tau \log(r+\tau+e) 
&=\nu \log(c_1 r) +\check{\kappa}_0\tau  + \check{\kappa}_1 \tau \log[c_2(r+\tau+e)] - \nu \log(c_1) -\check{\kappa}_1 \tau \log(c_2) \notag \\
&\le (c_1\nu + c_2 \check{\kappa}_1 \tau ) r +  \check{\kappa}_0\tau  + \check{\kappa}_1 \tau c_2(\tau+e) +\frac{\nu}{c_1} +\frac{\check{\kappa}_1 \tau}{c_2},
\end{align}
where we use $\log(x) \le x$ for $x\ge 0$. 
By choosing $c_1=\check{\kappa}_\beta /(8\nu)$ and $c_2=\check{\kappa}_\beta /(8\check{\kappa}_1 \tau)$, we reduce the above inequality to
\begin{align}
\nu \log(r) +\check{\kappa}_0\tau  + \check{\kappa}_1 \tau \log(r+\tau+e) 
&\le  \frac{\check{\kappa}_\beta  r}{4} +  \check{\kappa}_0\tau  + \frac{\check{\kappa}_\beta }{8}(\tau+e) +\frac{8\nu^2}{\check{\kappa}_\beta } +\frac{8\check{\kappa}_1^2 \tau^2}{\check{\kappa}_\beta }.
\end{align}
Therefore, the inequality~\eqref{re_state_lem:s_0_upper_bound/cond} reduces to
\begin{align}
r\ge  \frac{4}{\check{\kappa}_\beta} \brr{\check{\kappa}_0\tau  + \frac{\check{\kappa}_\beta }{8}(\tau+e) +\frac{8\nu^2}{\check{\kappa}_\beta } +\frac{8\check{\kappa}_1^2 \tau^2}{\check{\kappa}_\beta }}
=\frac{\tau+e}{2} +  \frac{32}{\check{\kappa}_\beta^2} \br{\nu^2+\check{\kappa}_1^2 \tau^2 + \frac{\check{\kappa}_\beta\check{\kappa}_0\tau  }{8}}.
\end{align}
This completes the proof. $\square$

{~}

\hrulefill{\bf [ End of Proof of Lemma~\ref{lem:s_0_upper_bound}]}

{~}

Based on Lemma~\ref{lem:s_0_upper_bound}, we further prove the following statement:
\begin{lemma} \label{tilde_Q_tau_r_lemma}
Under the definition of $\tilde{Q}(\tau,r)$ in Eq.~\eqref{def:tilde_Q_tau_r} as 
\begin{align}
\label{def:tilde_Q_tau_r_lemma}
\tilde{Q}(\tau,r):= \sum_{s< r} s^{2D-1} \bar{J}_{r-s} \check{Q}(\tau,s) + \sum_{s\ge r} s^{2D-1} \bar{J}_{0}\check{Q}(\tau,s)  ,
\end{align}
the $\tilde{Q}(\tau,r)$ is upper-bounded as 
\begin{align}
\label{upper-bound:tilde_Q_tau_r_lemma}
\tilde{Q}(\tau,r) \le 2^{2D+2} \br{\check{C}_{2,\tau}^{2D} +r^{2D}}\bar{J}_0\check{Q}(\tau,r) ,\quad \check{C}_{\nu,\tau}:= \frac{\tau+e}{2} + \frac{32}{\check{\kappa}_\beta ^2} \br{\nu^2 D^2 + \check{\kappa}_1^2 \tau^2+\frac{\check{\kappa}_\beta \check{\kappa}_0\tau}{8}},
\end{align}
where we have define $\bar{J}_x$ in Eq.~\eqref{def_short_range_long_range}, i.e., $\bar{J}_{x}:=\bar{J}_0 e^{-\mu x}$. 
\end{lemma}

{~}\\
\textit{Proof of Lemma~\ref{tilde_Q_tau_r_lemma}.}
We begin with the estimation for the summation of $  \sum_{s< r} s^{2D-1}\bar{J}_{r-s} \check{Q}(\tau,s)$, which is upper-bounded by
\begin{align}
\label{estimation_s_le_r_tilde_q_tau}
  \sum_{s< r} s^{2D-1}\bar{J}_{r-s} \check{Q}(\tau,s)   \le 
r^{2D-1}   \sum_{s< r} \bar{J}_{s} \check{Q}(\tau,r-s)  \le r^{2D-1} \check{Q} (\tau,r)   \sum_{s< r} \bar{J}_{s} e^{\check{\kappa}_\beta s} ,
\end{align} 
where we use $ \check{Q}(\tau,r-s)\le \check{Q} (\tau,r) e^{\check{\kappa}_\beta  s}$ from the form of $\check{Q}(\tau,r)$ in Eq.~\eqref{assum_quasi_locality_of_O_first}. 
Because of $\bar{J}_s=\bar{J}_0 e^{-\mu s}$, we have 
\begin{align}
 \sum_{s< r} \bar{J}_{s} e^{\check{\kappa}_\beta s} = \bar{J}_0 \sum_{s< r} e^{-(\mu-\check{\kappa}_\beta )s} \le  \frac{\bar{J}_0}{1-e^{-\mu+\check{\kappa}_\beta }}
 \le  \bar{J}_0 \br{1+\frac{1}{\mu-\check{\kappa}_\beta }} \le  \bar{J}_0 \br{1+\frac{1}{\check{\kappa}_\beta }}
 ,
\end{align} 
which reduces the inequality~\eqref{estimation_s_le_r_tilde_q_tau} to 
\begin{align}
\label{estimation_s_le_r_tilde_q_tau_final}
  \sum_{s< r} s^{2D-1}\bar{J}_{r-s} \check{Q}(\tau,s)   \le 
\bar{J}_0 \br{1+\frac{1}{\check{\kappa}_\beta }}  r^{2D-1} \check{Q} (\tau,r) ,
\end{align} 
where we use $1/(1-e^{-x})\le 1+1/x$ and $\mu-\check{\kappa}_\beta\ge 2\kappa_\beta -\check{\kappa}_\beta\ge \check{\kappa}_\beta $ from $\kappa_\beta\ge \check{\kappa}_\beta$ from the condition~\eqref{assum_quasi_locality_of_O_first}. 

We next estimate an upper bound for $ \sum_{s\ge r} s^{2D-1}\bar{J}_{0} \check{Q}(\tau,s)$. 
For this purpose, we first define $s_0$ such that for $s\ge s_0$
\begin{align}
s^{2D-1} \check{Q} (\tau,s) \le  e^{-\check{\kappa}_\beta  s /2}. 
\end{align}
Here, using Lemma~\ref{lem:s_0_upper_bound} with $\nu=2D-1$, we can ensure 
\begin{align}
\label{upper_bound_s_0_s}
s_0 \le\frac{\tau+e}{2} +  \frac{32}{\check{\kappa}_\beta ^2} \brr{(2D-1)^2 + \check{\kappa}_1^2 \tau^2+\frac{\check{\kappa}_\beta\check{\kappa}_0 \tau}{8}} \le \check{C}_{2,\tau} , 
\end{align}
where we use the definition in Eq.~\eqref{upper-bound:tilde_Q_tau_r_lemma}. 
Then, for $r\ge s_0$, we have 
\begin{align}
\label{s_ge_r_s_2D-1_upp}
 \sum_{s\ge r}  s^{2D-1}\bar{J}_0 \check{Q}(\tau,s)  
 &= \sum_{r\le s< 2r}s^{2D-1}\bar{J}_0 \check{Q}(\tau,s)+ \sum_{s\ge 2r}s^{2D-1}\bar{J}_0 \check{Q}(\tau,s) \notag \\
 & \le (2r)^{2D}\bar{J}_0 \check{Q}(\tau,r)  +  \sum_{s\ge 2r} \bar{J}_0 e^{-\check{\kappa}_\beta  s /2}  \notag \\
&\le (2r)^{2D}\bar{J}_0 \check{Q}(\tau,r)  + \frac{\bar{J}_0 e^{-\check{\kappa}_\beta  r} }{1-e^{-\check{\kappa}_\beta /2} } 
\le 2(2r)^{2D} \bar{J}_0 \check{Q}(\tau,r)   . 
\end{align}
On the other hand, for $r< s_0$, we have 
\begin{align}
 \sum_{s\ge r} s^{2D-1} \bar{J}_0\check{Q}(\tau,s)  
 &=\sum_{r\le s< s_0}s^{2D-1} \bar{J}_0\check{Q}(\tau,s)+ \sum_{s\ge s_0} s^{2D-1} \bar{J}_0 \check{Q}(\tau,s)  \notag \\
 &\le s_0^{2D}\bar{J}_0  \check{Q}(\tau,r) + 2 (2s_0)^{2D}\bar{J}_0 \check{Q}(\tau,s_0)  \le 3 (2s_0)^{2D} \bar{J}_0\check{Q}(\tau,r),
\end{align}
where we apply the upper bound in~\eqref{s_ge_r_s_2D-1_upp} with $r=s_0$ to the summation of $\sum_{s\ge s_0} s^{2D-1} \bar{J}_0 \check{Q}(\tau,s)$. 
Therefore, the following upper bound holds for general $r$:
\begin{align}
\label{estimation_s_ge_r_tilde_q_tau_final}
 \sum_{s\ge r}  s^{2D-1} \bar{J}_0\check{Q}(\tau,s)  \le  3 \cdot 2^{2D} (s_0^{2D}+r^{2D}) \bar{J}_0\check{Q}(\tau,r). 
\end{align}
Combining the inequalities~\eqref{estimation_s_le_r_tilde_q_tau_final} and~\eqref{estimation_s_ge_r_tilde_q_tau_final} and using $s_0 \le  \check{C}_{2,\tau} $ from \eqref{upper_bound_s_0_s}, we prove the main inequality~\eqref{upper-bound:tilde_Q_tau_r_lemma}.
This completes the proof. $\square$

{~}

\hrulefill{\bf [ End of Proof of Lemma~\ref{tilde_Q_tau_r_lemma}]}

{~}

\begin{lemma} \label{lem:sum_tilde_Q_tau_r}
Using Lemma~\ref{tilde_Q_tau_r_lemma}, we can obtain the following upper bound: 
\begin{align}
\label{main_ineq:sum_tilde_Q_tau_r}
\sum_{r=0}^\infty r^{D-1}  \tilde{Q}(\tau,r) \le  2^{2D+5}\check{C}_{2,\tau}^{2D} \check{C}_{3,\tau}^{3D} \bar{J}_0 ,
\end{align}
where $\check{C}_{3,\tau}$ is defined by \eqref{upper-bound:tilde_Q_tau_r_lemma}. 
\end{lemma}

{~}\\
\textit{Proof of Lemma~\ref{lem:sum_tilde_Q_tau_r}.}
We begin with the upper bound of~\eqref{upper-bound:tilde_Q_tau_r_lemma} as 
\begin{align}
r^{D-1} \tilde{Q}(\tau,r) &\le 2^{2D+2} r^{D-1}  \br{\check{C}_{2,\tau}^{2D} +r^{2D}} \bar{J}_0  \check{Q}(\tau,r)   \notag \\
&\le 2^{2D+2}\check{C}_{2,\tau}^{2D}  r^{D-1}  \br{1 +\frac{r^{2D}}{\check{C}_{2,\tau}^{2D}}} \bar{J}_0  \check{Q}(\tau,r)  \notag\\
&\le 2^{2D+2}\check{C}_{2,\tau}^{2D}  r^{D-1}  \br{ 2 r^{2D}} \bar{J}_0  \check{Q}(\tau,r) 
\le  2^{2D+3}\check{C}_{2,\tau}^{2D}  r^{3D-1} \bar{J}_0  \check{Q}(\tau,r)  .
\end{align}
Then, we introduce the parameter $r_0$ such that for $r\ge r_0$
\begin{align}
\label{cond_r_3d-1_q_1}
 r^{3D-1} Q(\tau,r) \le  e^{-\check{\kappa}_\beta  r /2} .
\end{align}
Using the notation of $r_0$, we have  
\begin{align}
\sum_{r=0}^\infty r^{D-1} \tilde{Q}(\tau,r) 
&\le \sum_{r\le r_0}2^{2D+3}\check{C}_{2,\tau}^{2D}  r^{3D-1} \bar{J}_0  \check{Q}(\tau,r)    + \sum_{r> r_0} 2^{2D+3}C_{2,\tau}^{2D}  r^{3D-1} \bar{J}_0  \check{Q}(\tau,r)  \notag \\
&\le 2^{2D+3}\check{C}_{2,\tau}^{2D} \bar{J}_0  \br{ 2r_0^{3D}  +\sum_{r> r_0}e^{-\check{\kappa}_\beta  r /2} } \notag \\
&\le 2^{2D+3}\check{C}_{2,\tau}^{2D} \bar{J}_0  \br{2 r_0^{3D} + \frac{e^{-\check{\kappa}_\beta  r_0 /2}}{1-e^{-\check{\kappa}_\beta   /2} }} 
\le 2^{2D+5}\check{C}_{2,\tau}^{2D}  r_0^{3D}  \bar{J}_0  ,
\end{align}
where we use $Q(\tau,r)\le 2$ as a trivial upper bound [see~\eqref{trivial_bound_/for_Q_tau_r}].
Finally, from Lemma~\ref{lem:s_0_upper_bound}, we can choose $r_0=\check{C}_{3,\tau}$ to satifsy the condition~\eqref{cond_r_3d-1_q_1}, which reduces the above inequality to the main inequality~\eqref{main_ineq:sum_tilde_Q_tau_r}. 
This completes the proof. $\square$

{~}

\hrulefill{\bf [ End of Proof of Lemma~\ref{lem:sum_tilde_Q_tau_r}]}

{~}

\subsubsection{Completing the proof of Proposition~\ref{prop:quasi-locality of_ad__H__tau_V}}
Throughout the proof, we omit the $\min(2,\cdots)$ notation from $\check{Q}(\tau,r)$ in Eq.~\eqref{assum_quasi_locality_of_O_first}.   
We begin with the equation of 
\begin{align}
\label{ad_H_tau/V_decomp}
\ad_{\hat{H}_\tau} (O) =  [U_\tau(  H_0 + \hat{V}_\tau )U_\tau^\dagger ,O]  
&=[ U_\tau \hat{V}_\tau U_\tau^\dagger ,O] +  [U_\tau H_0 U_\tau^\dagger ,O]  \notag \\
&=[ U_\tau \hat{V}_\tau U_\tau^\dagger ,O] +  [H_0 ,O] + [[U_\tau, H_0] U_\tau^\dagger ,O]  .
\end{align}
We then consider $\norm{[\ad_{\hat{H}_\tau} (O),u_i]}$, which is upper-bounded by the sum of 
\begin{align}
\label{0_three_term_H_tau_V_ad}
\norm{ \brr{ [ U_\tau \hat{V}_\tau U_\tau^\dagger ,O] ,u_i} },\quad \norm{ \brr{ [ H_0 ,O]  ,u_i} } , \AND \norm{ \brr{  [[U_\tau, H_0] U_\tau^\dagger ,O],u_i} } .
\end{align}

For the first term in~\eqref{0_three_term_H_tau_V_ad}, we have from Eq.~\eqref{U_tau_V_tau_mu_tau_def} 
\begin{align}
U_\tau \hat{V}_\tau U_\tau^\dagger =2 \int_0^\tau  U_\tau U_{\tau_1}^\dagger  V  U_{\tau_1}  U_\tau^\dagger d\tau_1 ,
\end{align}
which reduces $\norm{ \brr{ [ U_\tau \hat{V}_\tau U_\tau^\dagger ,O] ,u_i} }$ to 
\begin{align}
&\norm{2\int_0^\tau  \brr{ \brr{ U_\tau U_{\tau_1}^\dagger  V  U_\tau^\dagger U_{\tau_1} ,O} ,u_i}}d\tau_1 
 \le 4\int_0^\tau \br{2 \norm{V}\cdot \norm{O} \cdot \norm{ \brr{U_\tau U_{\tau_1}^\dagger,u_i}}
  +\norm{V}\cdot \norm{[O,u_i]}  +\norm{O}\cdot \norm{[V,u_i]}} d\tau_1  \notag \\
&\le 8\norm{V}\cdot \norm{O}  \int_0^\tau  \br{ \norm{[U_\tau ,u_i]} +\norm{[U_{\tau_1} ,u_i]} } d\tau_1
+  4  \norm{V} \int_0^\tau \norm{[O,u_i]} d\tau_1 +  4  \norm{O} \int_0^\tau \norm{[V,u_i]} d\tau_1  \notag \\
& \le  8\norm{V}\mathcal{N}_O \tau  \br{2 Q(\tau,r) + \frac{\check{Q}(\tau,r)+Q(0,r)}{2}}  \le 24\norm{V}\mathcal{N}_O \tau \check{Q}(\tau,r) ,
\end{align}
where we use $\norm{O}\le \mathcal{N}_O$, $Q(\tau',r)  \le Q(\tau,r)$ for $\tau'\le \tau$, and Eq.~\eqref{assumption_proof_le_0V} for $\norm{[V,u_i]}$. 
We thus obtain 
\begin{align}
\label{0_three_term_H_tau_V_ad_1st_up}
\mbox{{\bf [\textrm{1st term in \eqref{0_three_term_H_tau_V_ad}}]}} =
 \norm{ \brr{ [ U_\tau \hat{V}_\tau U_\tau^\dagger ,O] ,u_i} } \le 24\norm{V}\mathcal{N}_O \tau \check{Q}(\tau,r)  . 
\end{align}

For the second term in~\eqref{0_three_term_H_tau_V_ad}, we decompose $O$ as
\begin{align}
\label{decomp_O_mfrL_s_sum}
O= \tilde{O}_{\mfL} + \sum_{s=1}^\infty \br{\tilde{O}_{\mfL[s]} - \tilde{O}_{\mfL[s-1]} }, \quad  \tilde{O}_{\mfL[s]}:=\tilde{\tr}_{\mfL[s]^\co} (O). 
\end{align}
Note that by using the inequality~\eqref{Proof_norm_local_approx_fin} in the proof of Lemma~\ref{norm_local_approx},
Eq.~\eqref{assum_quasi_locality_of_O_first_0} gives 
\begin{align}
\label{quasi_locality_O_error_local}
\norm{\tilde{O}_{\mfL[s]} - \tilde{O}_{\mfL[s-1]} }\le \sum_{i\in \mfL[s]\setminus \mfL[s-1]} \norm{[O,u_i]} \le 
 \mathcal{N}_O |\partial (\mfL[s])| \check{Q}(\tau,s)   .
\end{align}
Then, for $i\notin \mfL[s]$ (or $\dist_{i,\mfL}=r >s$), we obtain 
\begin{align}
\brr{\brr{ H_0, \tilde{O}_{\mfL[s]} - \tilde{O}_{\mfL[s-1]} }, u_i} = \sum_{Z: Z\cap \mfL[s]\neq \emptyset,\ Z\cap \{i\} \neq \emptyset} \brr{\brr{ h_Z, \tilde{O}_{\mfL[s]} - \tilde{O}_{\mfL[s-1]} }, u_i}, 
\end{align}
and hence, by using Lemma~\ref{sum_interaction_terms}, the following inequality holds:
\begin{align}
\norm{\brr{\brr{ H_0, \tilde{O}_{\mfL[s]} - \tilde{O}_{\mfL[s-1]} }, u_i}}
&\le 4 \sum_{Z: Z\cap \mfL[s]\neq \emptyset,\ Z\cap \{i\}\neq \emptyset}  \norm{h_Z} \cdot \norm{\tilde{O}_{\mfL[s]} - \tilde{O}_{\mfL[s-1]} } \notag \\
&\le4\mathcal{N}_O |\mfL[s]|\cdot |\partial (\mfL[s])| \cdot \bar{J}_{r-s} \check{Q}(\tau,s) .
\label{0_multi:commun_H_0,V,U_x_pre1}
\end{align}
For $i\in \mfL[s]$ (or $\dist_{i,\mfL}=r \le s$), we obtain 
\begin{align}
\norm{\brr{\brr{ H_0, \tilde{O}_{\mfL[s]} - \tilde{O}_{\mfL[s-1]} }, u_i}}
&\le 4 \sum_{Z: Z\cap \mfL[s]\neq \emptyset}  \norm{h_Z} \cdot \norm{\tilde{O}_{\mfL[s]} - \tilde{O}_{\mfL[s-1]} } \le4\mathcal{N}_O |\mfL[s]|\cdot |\partial (\mfL[s])| \cdot \bar{J}_{0} \check{Q}(\tau,s) .
\label{0_multi:commun_H_0,V,U_x_pre2}
\end{align}
By using the inequality~\eqref{0_multi:commun_H_0,V,U_x_pre1} and~\eqref{0_multi:commun_H_0,V,U_x_pre2}, we take the summation as 
\begin{align}
\label{0_multi:commun_H_0,V,U_x}
\norm{\brr{ [ H_0 ,O]  ,u_i} } \le 4\mathcal{N}_O \sum_{s< r}  |\mfL[s]| \cdot |\partial (\mfL[s])| \cdot \bar{J}_{r-s} \check{Q}(\tau,s) + 4\mathcal{N}_O \sum_{s\ge  r}  |\mfL[s]| \cdot |\partial (\mfL[s])| \cdot \bar{J}_{0} \check{Q}(\tau,s) ,
\end{align}
where we let $\bar{J}_{x}=\bar{J}_0$ for $x<0$.

For the summation with respect to $s$, we have 
\begin{align}
&4\mathcal{N}_O \sum_{s< r}  |\mfL[s]| \cdot |\partial (\mfL[s])| \cdot \bar{J}_{r-s} \check{Q}(\tau,s) + 4\mathcal{N}_O \sum_{s\ge  r}  |\mfL[s]| \cdot |\partial (\mfL[s])| \cdot \bar{J}_{r-s} \check{Q}(\tau,s) \notag \\
&\le 4\mathcal{N}_O \gamma^2 |\mfL|\cdot |\partial \mfL| \br{ \sum_{s< r}  s^{2D-1}\bar{J}_{r-s}  \check{Q}(\tau,s)+ \sum_{s\ge r} s^{2D-1}\bar{J}_{0} \check{Q}(\tau,s) }\notag \\
&\le 4\mathcal{N}_O \gamma^2 |\mfL|^2 \br{ \sum_{s< r} s^{2D-1} \bar{J}_{r-s} \check{Q}(\tau,s) + \sum_{s\ge r} s^{2D-1}\bar{J}_{0} \check{Q}(\tau,s)  } ,
\label{summation_s_J_r-s_Q_0_s}
\end{align}
where we use $|\partial \mfL| \le |\mfL| $ in the last inequality.
Then, by defining 
\begin{align}
\label{def:tilde_Q_tau_r}
\tilde{Q}(\tau,r):=  \sum_{s< r} s^{2D-1} \bar{J}_{r-s} \check{Q}(\tau,s) + \sum_{s\ge r} s^{2D-1}\bar{J}_{0} \check{Q}(\tau,s)  ,
\end{align}
we obtain 
\begin{align}
\label{0_three_term_H_tau_V_ad_2nd_up}
\mbox{{\bf [\textrm{2nd term in \eqref{0_three_term_H_tau_V_ad}}]}} =
\norm{\brr{ [ H_0 ,O]  ,u_i} } \le 4\mathcal{N}_O \gamma^2 |\mfL|^2 \tilde{Q}(\tau,r) . 
\end{align}

Finally, for the third term in~\eqref{0_three_term_H_tau_V_ad}, we have
\begin{align}
\norm{ \brr{  [[U_\tau, H_0] U_\tau^\dagger ,O],u_i} } 
&\le 
2\norm{[U_\tau, H_0] } \br{\norm{O}\cdot  \norm{ [U_\tau^\dagger ,u_i]} +\norm{[O,u_i] } }
+2 \norm{O} \cdot\norm{[[U_\tau, H_0] ,u_i]}\notag \\
&\le 2\norm{[U_\tau, H_0] } \cdot\brr{ \norm{O}  Q(\tau, r) + \mathcal{N}_O   \check{Q}(\tau, r) }
+2 \norm{O} \cdot\norm{[[U_\tau, H_0] ,u_i]} \notag \\
&\le 4\mathcal{N}_O\norm{[U_\tau, H_0] } \check{Q}(\tau, r)
+2 \mathcal{N}_O \norm{[[U_\tau, H_0] ,u_i]} ,
\label{norm_brr_U_tau_H_0_U_tau_V_u_i}
\end{align}
where we use $\norm{[O,u_i] }\le  \mathcal{N}_O \check{Q}(\tau, r)$ ($\norm{O} \le \mathcal{N}_O $) and $\norm{ [U_\tau^\dagger ,u_i]}\le Q(\tau, r)\le \check{Q}(\tau, r)$ from the assumption~\eqref{assum_quasi_locality_of_O_first}.
By using a similar calculation to~\eqref{quasi_locality_O_error_local}, we obtain from the condition~\eqref{assumption_proof_le_tau} 
\begin{align}
\norm{\tilde{U}_{\tau,\mfL[s]} - \tilde{U}_{\tau,\mfL[s-1]} }\le |\partial (\mfL[s])| Q(\tau,s) \le  |\partial (\mfL[s])|  \check{Q}(\tau, s)  . 
\end{align}
Therefore, for $[[U_\tau, H_0],u_i]$ in~\eqref{norm_brr_U_tau_H_0_U_tau_V_u_i}, we can derive a similar inequality to~\eqref{0_multi:commun_H_0,V,U_x}: 
\begin{align}
\label{norm_com_U_tau_H_0_u_i}
\norm{\brr{ [U_\tau, H_0] ,u_i }}
&\le 4 \sum_{s=0}^\infty |\mfL[s]| \cdot |\partial (\mfL[s])| \cdot \bar{J}_{r-s} \check{Q}(\tau ,s) 
 \le 4\gamma^2 |\mfL|^2 \tilde{Q}(\tau,r),
\end{align}
where $\bar{J}_{r-s}= \bar{J}_0$ for $r-s\le 0$, and we use the same inequality as~\eqref{summation_s_J_r-s_Q_0_s} in the second inequality. 

We then consider the first term $4\mathcal{N}_O\norm{[U_\tau, H_0] } \check{Q}(\tau, r)$ in~\eqref{norm_brr_U_tau_H_0_U_tau_V_u_i}, which necessitates the estimation of $\norm{[U_\tau, H_0]}$.  
To estimate it, we use Lemma~\ref{lemma_commutator_generalization}.
Because of $\tr\br{ [U_\tau, H_0]}=0$, the inequality~\eqref{first_ineq:lemma_commutator_generalization} gives 
\begin{align}
\label{norm_com_U_tau_H_0_11}
\norm{[U_\tau, H_0]} 
&\le \sum _{i\in \Lambda} \sup_{u_i} \norm{[[U_\tau, H_0] , u_i] } 
= \sum_{r=0}^\infty \sum_{i: \dist_{i,\mfL}=r } \sup_{u_i} \norm{[[U_\tau, H_0] , u_i] } \notag \\
&\le 4\gamma^2 |\mfL|^2  \sum_{r=0}^\infty |\partial (\mfL[r])| \tilde{Q}(\tau,r) \le 4\gamma^3 |\mfL|^3 \sum_{r=0}^\infty r^{D-1} \tilde{Q}(\tau,r) ,
\end{align}
where we use the upper bounds of~\eqref{norm_com_U_tau_H_0_u_i} in the second inequality and $|\partial (\mfL[r])| \le  |\mfL| \gamma r^{D-1}$ in the third inequality. 
By applying the inequalities~\eqref{norm_com_U_tau_H_0_u_i} and \eqref{norm_com_U_tau_H_0_11} to~\eqref{norm_brr_U_tau_H_0_U_tau_V_u_i}, we have 
\begin{align}
\label{0_three_term_H_tau_V_ad_3rd_up}
\mbox{{\bf [\textrm{3rd term in \eqref{0_three_term_H_tau_V_ad}}]}} 
&=\norm{ \brr{  [[U_\tau, H_0] U_\tau^\dagger ,O],u_i} }  \notag \\
&\le 
 16\mathcal{N}_O \gamma^3 |\mfL|^3 \check{Q}(\tau, r)  \sum_{r=0}^\infty r^{D-1} \tilde{Q}(\tau,r) 
+8\mathcal{N}_O \gamma^2 |\mfL|^2 \tilde{Q}(\tau,r)  .
\end{align}

Finally, we combine all the upper bounds~\eqref{0_three_term_H_tau_V_ad_1st_up}, \eqref{0_three_term_H_tau_V_ad_2nd_up}, 
\eqref{0_three_term_H_tau_V_ad_3rd_up}, we have 
\begin{align}
\label{ad_H_tau_V_u_i_final_without_lemma}
&\norm{[ \ad_{\hat{H}_\tau} (O),u_i] }\le \norm{ [[ U_\tau \hat{V}_\tau U_\tau^\dagger ,O],u_i ]}+  \norm{[[H_0 ,O],u_i} +
 \norm{[[[U_\tau, H_0] U_\tau^\dagger ,O] ,u_i]}  \notag \\ 
&\le 4\mathcal{N}_O \br{6\norm{V}\tau \check{Q}(\tau,r)  + \gamma^2 |\mfL|^2 \tilde{Q}(\tau,r) + 4 \gamma^3 |\mfL|^3 \check{Q}(\tau, r)   \sum_{r=0}^\infty r^{D-1} \tilde{Q}(\tau,r) 
+2\gamma^2 |\mfL|^2 \tilde{Q}(\tau,r)}  \notag \\
&\le 4\mathcal{N}_O\brr{ \br{ 6\norm{V}\tau  + 4 \gamma^3 |\mfL|^3 \sum_{r=0}^\infty r^{D-1} \tilde{Q}(\tau,r)} \check{Q}(\tau,r)  +3  \gamma^2 |\mfL|^2 \tilde{Q}(\tau,r) } \notag \\
&\le 4\mathcal{N}_O\check{Q}(\tau,r) \br{ 6\norm{V}\tau  + 4 \gamma^3 |\mfL|^3\cdot 2^{2D+5}\check{C}_{2,\tau}^{2D} \check{C}_{3,\tau}^{3D}\bar{J}_0    +3  \gamma^2 |\mfL|^2 \cdot 2^{2D+2} \br{\check{C}_{2,\tau}^{2D} +r^{2D}}\bar{J}_0} \notag \\
&\le \mathcal{N}_O \check{Q}(\tau,r) \brr{24\norm{V}\tau  + 2^{2D+6} \gamma^2 |\mfL|^2 \bar{J}_0 \br{\check{C}_{2,\tau}^{2D} +r^{2D} +8\gamma |\mfL|\check{C}_{2,\tau}^{2D} \check{C}_{3,\tau}^{3D} }} ,
\end{align}
where we use Lemmas~\ref{tilde_Q_tau_r_lemma} and \ref{lem:sum_tilde_Q_tau_r} in the third inequality, i.e.,
\begin{align}
\label{ad_H_tau_V_u_i_final_lemma_apply}
\tilde{Q}(\tau,r) \le 2^{2D+2} \br{\check{C}_{2,\tau}^{2D} +r^{2D}}\bar{J}_0  \check{Q}(\tau,r) ,\quad 
\sum_{r=0}^\infty r^{D-1} \tilde{Q}(\tau,r) \le 2^{2D+5}\check{C}_{2,\tau}^{2D} \check{C}_{3,\tau}^{3D}\bar{J}_0 .
 \end{align}
Thus, we prove the first main inequality~\eqref{main_eq:prop:quasi-locality of_ad__H__tau_V} with the choice of $g_{\tau,r}$ as in Eq.~\eqref{def_C_V_tau_r}. 

The second main inequality~\eqref{main_eq:prop:quasi-locality of_ad__H__tau_V_2} is derived in the same way.
From Eq.~\eqref{ad_H_tau/V_decomp}, we obtain 
\begin{align}
\label{ad_H_tau/V_decomp_upp_bound}
\norm{\ad_{\hat{H}_\tau} (O)} 
&= \norm{[ U_\tau \hat{V}_\tau U_\tau^\dagger ,O] +  [H_0 ,O] + [[U_\tau, H_0] U_\tau^\dagger ,O]}   \notag \\
&\le 2 \tau  \norm{V} \cdot \norm{O} + \norm{H_0,O} + \norm{O} \cdot \norm{[U_\tau, H_0] } ,
\end{align}
where we use from Eq.~\eqref{U_tau_V_tau_mu_tau_def} 
\begin{align}
\norm{\hat{V}_\tau} \le 2\int_0^\tau  \norm{ U_{\tau_1}^\dagger  V   U_{\tau_1}}  d\tau_1\le 2 \tau\norm{V} .
\end{align}
For the estimation of $\norm{H_0,O}$, using the decomposition of Eq.~\eqref{decomp_O_mfrL_s_sum}, we have 
\begin{align}
\norm{ [ H_0 ,O]}&\le 
 2\sum_{Z: Z\cap \mfL\neq \emptyset} \norm{h_Z} \cdot \norm{\tilde{O}_{\mfL}} 
 + 2 \sum_{s=1}^\infty  \sum_{Z: Z\cap \mfL[s]\neq \emptyset}  \norm{h_Z} \cdot \norm{\tilde{O}_{\mfL[s]} - \tilde{O}_{\mfL[s-1]} } \notag \\
 &\le 2\mathcal{N}_O  \sum_{s=0}^\infty  |\mfL[s]| \cdot |\partial (\mfL[s])| \bar{J}_0 \check{Q}(\tau,s) ,
 \label{norm_H_0_O_commutator}
\end{align}
where we use $ \sum_{Z: Z\cap \mfL[s]\neq \emptyset}  \norm{h_Z} \le  \sum_{i\in \mfL[s]} \sum_{Z: Z\ni i}  \norm{h_Z} \le  \sum_{i\in \mfL[s]} \bar{J}_0= \bar{J}_0|\mfL[s]|$ from Eq.~\eqref{def:Hamiltonian}. 
The RHS of the above inequality is equal to the half of the RHS of \eqref{0_multi:commun_H_0,V,U_x} with $r=0$, and hence the inequality~\eqref{0_three_term_H_tau_V_ad_2nd_up} with $r=0$ gives
\begin{align}
\label{ad_H_tau/V_decomp_upp_bound_2}
\norm{ [ H_0 ,O]}&\le  2\mathcal{N}_O \gamma^2 |\mfL|^2 \tilde{Q}(\tau,0) \le 2^{2D+4}\mathcal{N}_O \gamma^2 |\mfL|^2 \check{C}_{2,\tau}^{2D}\bar{J}_0  , 
\end{align}
where we use $\tilde{Q}(\tau,0)  \le 2^{2D+2} \check{C}_{2,\tau}^{2D} \bar{J}_0 \check{Q}(\tau,0) \le 2^{2D+3} \check{C}_{2,\tau}^{2D} \bar{J}_0$ from Lemma~\ref{tilde_Q_tau_r_lemma} and $\check{Q}(\tau,0) \le 2$. 
Also, the norm of $\norm{[U_\tau, H_0] } $ has been upper-bounded by the inequality~\eqref{norm_com_U_tau_H_0_11}.
Then, Lemma~\ref{lem:sum_tilde_Q_tau_r} gives 
\begin{align}
\label{ad_H_tau/V_decomp_upp_bound_3}
\norm{[U_\tau, H_0] }\le  4\gamma^3 |\mfL|^3 \sum_{r=0}^\infty r^{D-1} \tilde{Q}(\tau,r)  \le 
4\gamma^3 |\mfL|^3\cdot 2^{2D+4}\check{C}_{2,\tau}^{2D} \check{C}_{3,\tau}^{3D}\bar{J}_0.
\end{align}
By applying the inequalities~\eqref{ad_H_tau/V_decomp_upp_bound_2} and \eqref{ad_H_tau/V_decomp_upp_bound_3} to \eqref{ad_H_tau/V_decomp_upp_bound}, we prove the inequality~\eqref{main_eq:prop:quasi-locality of_ad__H__tau_V_2} as follows:
\begin{align}
\label{main_eq:prop:quasi-locality of_ad__H__tau_V_2_0} 
\norm{\ad_{\hat{H}_\tau} (V)} 
&\le 2 \tau  \norm{V} \cdot \norm{O} + 2^{2D+5}\mathcal{N}_O \gamma^2 |\mfL|^2 \check{C}_{2,\tau}^{2D}\bar{J}_0 + 4 \gamma^3 |\mfL|^3\cdot 2^{2D+4}\check{C}_{2,\tau}^{2D} \check{C}_{3,\tau}^{3D}\bar{J}_0 \norm{O}\notag \\
&\le \mathcal{N}_O  \brr{2 \tau \norm{V} + 2^{2D+5}  \gamma^2 |\mfL|^2 \check{C}_{2,\tau}^{2D} \bar{J}_0 \br{ 1 +  2\gamma |\mfL|   \check{C}_{3,\tau}^{3D}  } }, 
\end{align}
where we use $\norm{O}\le \mathcal{N}_O$. 
This completes the proof of Proposition~\ref{prop:quasi-locality of_ad__H__tau_V}. $\square$

\subsection{Proof of Proposition~\ref{prop:Quasi-local_commutator_bound_tau_2}} \label{proof_prop:Quasi-local_commutator_bound_tau_2}
We begin with estimating the first term in~\eqref{main:eq:prop:Quasi-local_commutator_bound_tau__2}, i.e., 
\begin{align}
\int_{-\infty}^\infty dt  g_1(t) \int_0^t dt_1\norm{\brr{O, u_i (H_0,-t_1) } }= \int_{|t|> \delta t} dt  |g_\beta(t)|  \int_0^t dt_1\norm{\brr{O, u_i (H_0,-t_1) } }.
\end{align}
For the proof, from Corollary~\ref{corol:Quasi-local_commutator_bound}, 
we utilize Lemma~\ref{lem:Quasi-local_commutator_bound} with $f(t) =|tg_1(t)|$, where $g_1(t)$ has been defined as in Eq.~\eqref{definition_g_1t_g_2t}. 
Therefore, the parameters $f_1$ and $f_2$ in~\eqref{definition_lem:Quasi-local_commutator_bound} are upper-bounded using~\eqref{main_eq:connection_of_exponential_operator_trans_1} in Lemma~\ref{belief_norm_lemma} as follows:
\begin{align}
\label{est_f_1andf_2_g_1case_33}
&f_1=\int_{|t|>\delta t} |tg_\beta (t)| dt \le  \frac{\beta^2}{2\pi^2} \zeta(2)=  \frac{\beta^2}{12}, \notag \\
&f_2=\int_{|t|>\delta t}  |t^2 g_\beta (t)| dt \le \frac{\beta^3\zeta(3)}{2\pi^3} 
\le \frac{\beta^3}{\pi^3} ,
\end{align}
where we use $\zeta(3)\approx 1.20206$ in the second inequality. 
Also, $f_{t_0}(s)$ is given by using Eq.~\eqref{integral_t_g_beta_t_ge_delta__1}
\begin{align}
\label{integral_t_g_beta_t_ge_delta_re}
f_{t_0}(s)= \int_{|t| > t_0}  |t g_\beta (t)|  dt 
&= \frac{\beta^2}{2\pi^2} \brr{-\frac{2\pi t_0}{\beta} \log\br{1 - e^{-2\pi t_0/\beta} } + L_2\br{e^{-2\pi t_0/\beta} } } \notag \\
&\le  \frac{\beta^2}{2\pi^2} \cdot \frac{2\pi t_0}{\beta}  \log\br{e+ \frac{e\beta}{2\pi t_0}}e^{-2\pi t_0/\beta} +\frac{\beta^2}{12} e^{-2\pi t_0/\beta}\notag \\
&\le \frac{2\beta \kappa_\beta \Delta \ell}{\pi}(s-2)e^{-2\kappa_\beta (s-2)\Delta \ell  } +\frac{\beta^2}{12}  e^{-2\kappa_\beta \Delta \ell (s-2)} \for s\ge 3,
\end{align}
where we use the same analysis as in~\eqref{est_f_t_0_s_g_1case} for $-\log\br{1 - e^{-2\pi t_0/\beta} }$, $ L_2(x)\le \pi^2x/6$ for $0\le x\le 1$, 
and $2\pi t_0/\beta \ge \pi \mu (s-2) \Delta \ell / (v\beta) \ge 2\kappa_{\beta}\Delta \ell  (s-2)$.
Also, $L_s(x)$ is the polylogarithm, i.e., $L_s(x):=\sum_{m=1}^\infty x^m/m^s$.
We notice that $x \log(e+e/x) e^{-x}$ monotonically decreases with $x$ for $x\ge 1$.
For $s\le2$, we use $f_{t_0}(s)\le f_1$, where we can apply the same inequality as~\eqref{integral_t_g_beta_t_ge_delta_re} for $s=2$ since $\beta^2/12=f_1$. 

We then estimate the summation in~\eqref{main_ineq:lem:Quasi-local_commutator_bound_Delta_eLL=1}: 
\begin{align}
2\gamma (\Delta \ell)^D\sum_{s=1}^\infty s^D\mathcal{F}(\ell-s\Delta \ell) \brr{ C f_2 v e^{-\mu s\Delta \ell/2} +f_{t_0}(s)}
\end{align}
using the estimations~\eqref{est_f_1andf_2_g_1case_33} and \eqref{integral_t_g_beta_t_ge_delta_re}. 
From the calculation~\eqref{upper_bound_s^D_mathcal_F_ell-s_f_2term_0}, we obtain
\begin{align}
\label{upper_bound_s^D_mathcal_F_ell-s_f_2term_0__re}
2\gamma (\Delta \ell)^D\sum_{s=1}^\infty s^D\mathcal{F}(\ell-s\Delta \ell) C f_2 v e^{-\mu s\Delta \ell/2} 
\le  \frac{\beta^3}{\pi^3} \cdot   4\gamma Cv \br{\frac{8D^2}{e\mu}}^D  \mathcal{F}(\ell) .
\end{align}
Also, by employing the same calculation as~\eqref{upper_bound_s^D_mathcal_F_ell-s_f_2term}, we have
\begin{align}
\label{upper_bound_s^D_mathcal_F_ell-s_f_2term//re}
&2\gamma (\Delta \ell)^D\sum_{s=1}^\infty s^D\mathcal{F}(\ell-s\Delta \ell )f_{t_0}(s) \notag \\ 
&\le 2\gamma (\Delta \ell)^D \frac{\beta^2}{12} \mathcal{F}(\ell-\Delta \ell )  +  2\gamma (\Delta \ell)^D\mathcal{F}(\ell)
\sum_{s=2}^\infty s^De^{\kappa_\beta s\Delta \ell} \br{ \frac{2\beta \kappa_\beta \Delta \ell}{\pi} (s-2)  +\frac{\beta^2}{12} } e^{-2\kappa_\beta  (s-2) \Delta \ell }  \notag \\
&\le  \frac{\gamma (\Delta \ell)^D \beta^2}{6}  e^{\kappa_\beta \Delta \ell } \mathcal{F}(\ell) + 
2\gamma  (2\Delta \ell)^D\mathcal{F}(\ell) e^{3\kappa_\beta \Delta \ell }\sum_{\tilde{s}=1}^\infty  \tilde{s}^D \br{ \frac{2\beta \kappa_\beta \Delta \ell}{\pi}\tilde{s} +\frac{\beta^2}{12} } e^{-\kappa_\beta \tilde{s}\Delta \ell }
\notag \\
&\le  \frac{\gamma (\Delta \ell)^D \beta^2}{6}  e^{\kappa_\beta \Delta \ell } \mathcal{F}(\ell) + 
2\gamma  (2\Delta \ell)^D\mathcal{F}(\ell) e^{2\kappa_\beta \Delta \ell } \brr{ \frac{2\beta \kappa_\beta \Delta \ell}{\pi} \br{1+2^{D+1}(D+1)!} +
\frac{\beta^2}{12} (1+2^D D!)} \notag \\
&\le  \frac{\gamma (\Delta \ell)^D \beta^2}{6}  e^{\kappa_\beta \Delta \ell } \mathcal{F}(\ell) +  
2\gamma  (4D \Delta \ell)^D e^{2\kappa_\beta \Delta \ell }
\br {\frac{16D\beta \kappa_\beta \Delta \ell}{\pi}+\frac{\beta^2}{6} }\mathcal{F}(\ell) ,
\end{align}
where we use $\br{1+2^{D+1}(D+1)!}/(1+2^D D!)\le 2(D+1) \le 4D$ and $1+2^D D! \le 2(2D)^D$. 
Therefore, by applying the inequalities~\eqref{est_f_1andf_2_g_1case_33} and 
\eqref{upper_bound_s^D_mathcal_F_ell-s_f_2term//re} to \eqref{main_ineq:lem:Quasi-local_commutator_bound}, we prove the first main inequality~\eqref{main:eq:prop:Quasi-local_commutator_bound_tau__2} by choosing $\mathfrak{g}_3$ as in~Eq.~\eqref{definition_g_3_g_4_mathfrak}.

We next estimate the second term in Eq.~\eqref{main:eq:prop:Quasi-local_commutator_bound_tau__2}.
In this case, as in the proof of Proposition~\ref{prop:Quasi-local_commutator_bound_tau},  we simply choose $\Delta \ell=1$, where we can apply the inequality~\eqref{main_ineq:lem:Quasi-local_commutator_bound_Delta_eLL=1} in Lemma~\ref{lem:Quasi-local_commutator_bound}.  
To apply Corollary~\ref{corol:Quasi-local_commutator_bound} with $f(t) =|tg_2(t)|$ [see Eq.~\eqref{definition_g_1t_g_2t}], we estimate the parameter $f_1$ and  $f_2$ as follows:
\begin{align}
\label{est_f_1andf_2_g_2case_prop_proof}
&f_1=\int_{|t|\le \delta t} |t^2g_\beta (t)| dt \le  \frac{\beta \delta t^2}{2\pi}, \quad 
f_2=\int_{|t|\le \delta t}  |t^3 g_\beta (t)| dt \le \frac{\beta\delta t^3}{3\pi} ,
\end{align}
 where we apply the inequality \eqref{main_eq:connection_of_exponential_operator_trans_2} with $m=1$ and $m=2$, respectively.  
Then, by following the same analyses as the inequalities~\eqref{g_2_t_case_C_f_2_term} and~\eqref{upper_bound_s^D_mathcal_F_ell-s_f_2term_g_2_case}, we obtain
\begin{align} 
C f_2 v e^{-\mu s/2} + f_{t_0}(s) \le \frac{C v  \beta \delta t^3}{3\pi} e^{-\mu s/2} . 
\label{g_2_t_case_C_f_2_term___re}
\end{align}
and
\begin{align}
\label{upper_bound_s^D_mathcal_F_ell-s_f_2term_g_2_case___re}
\sum_{s=1}^\infty s^D\mathcal{F}(\ell-s) \brr{ C f_2 v e^{-\mu s/2} + f_{t_0}(s)}
 &\le \frac{2\beta \delta t^2}{3\pi} \cdot  2^{D} D! C v (1+4/\mu)^D \delta t \mathcal{F}(\ell),
 \end{align}
which yields the second inequality~\eqref{main:eq:prop:Quasi-local_commutator_bound_tau__2} from~\eqref{main_ineq:lem:Quasi-local_commutator_bound} with the parameter $\mathfrak{g}_4$ in~Eq.~\eqref{definition_g_3_g_4_mathfrak}.
This completes the proof of Proposition~\ref{prop:Quasi-local_commutator_bound_tau_2}. $\square$

\section{Refined locality estimation for entanglement Hamiltonian}

In discussing the conditional mutual information, it is not enough to ensure only the quasi-locality of the entanglement Hamiltonian as in Subtheorem~\ref{sub_thm:U_tau_u_i_commun}. 
To see the point, we consider the following entanglement Hamiltonian, 
\begin{align}
\beta \hat{H}_{\tau} = \log \br{e^{\tau (V_A + V_B)} e^{\beta H_0}e^{\tau (V_A + V_B)} }  ,
\end{align}
where $V_A$ and $V_B$ are supported on the subsets $A$ and $B$, respectively.
Here, the problem is whether or not we can ensure 
 \begin{align}
 \label{eff_AB_sum_A_B}
\beta \hat{H}_{\tau}  \overset{\text{?}}{\approx} \log \br{e^{\tau V_A} e^{\beta H_0}e^{\tau V_A} } + \log \br{e^{\tau V_B} e^{\beta H_0}e^{\tau V_B} } - \beta H_0  ,
\end{align}
if $A$ and $B$ are separated by a sufficiently long distance. 
Based on the above approximation, we can estimate the operators $H_{\tilde{\rho}_{\beta}}(A:C|B)$ in Lemma~\ref{lem:entropy_bound_1D} and $H_{\rho_\beta,\tau}(A:C|B)$ in Corollary~\ref{corol:error_est_APTP_continuity_bound_CMI}, which appear in the upper bounds for the conditional mutual information. 
 In Sec.~\ref{sec;completing_Final_proof}, we utilize this kind of approximation in completing the proof of the clustering for the conditional mutual information.

 \begin{figure}[tt]
\centering
\includegraphics[clip, scale=0.5]{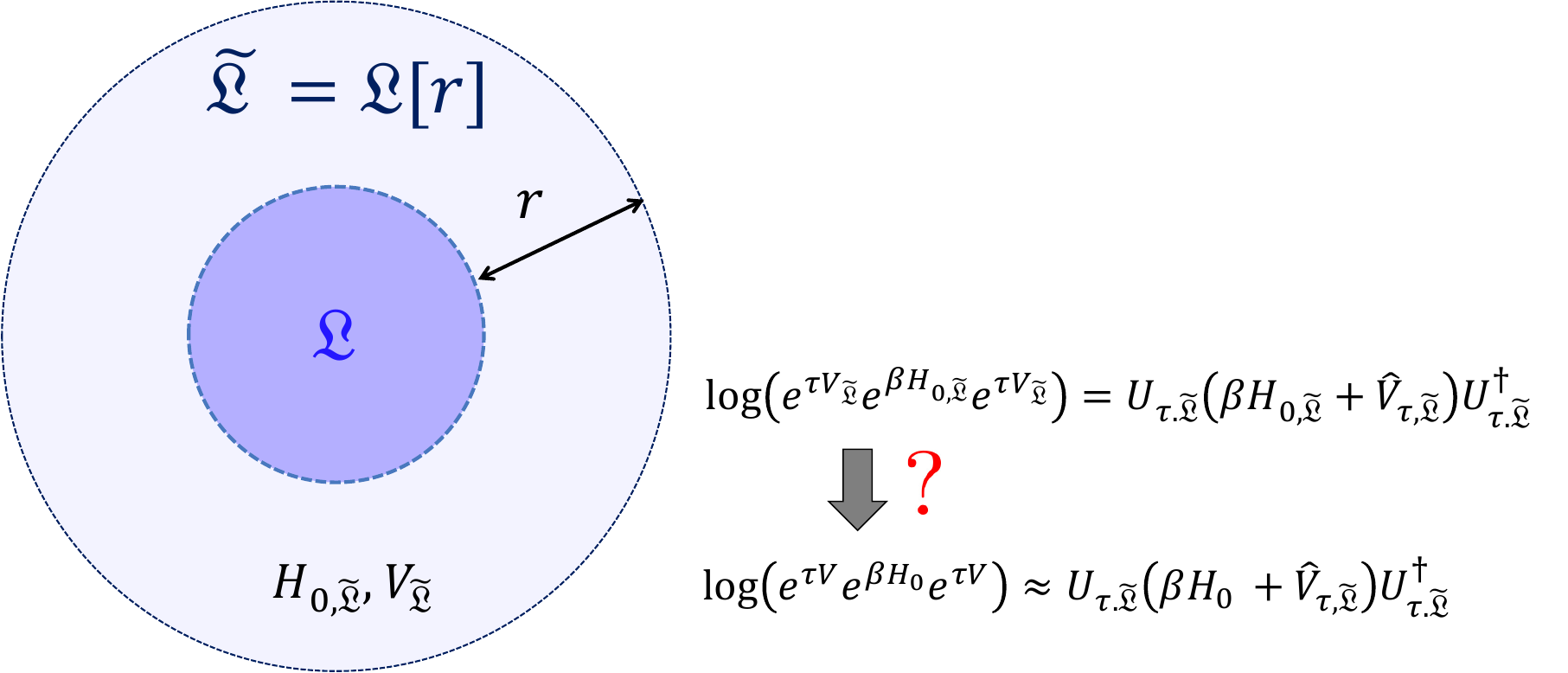}
\caption{Schematic picture to describe Theorem~\ref{thm:Refined locality estimation for effective Hamiltonian}. 
We consider the connection of the exponential operators as $e^{\tau V} e^{\beta H_{0}}e^{\tau V}=e^{\beta \hat{H}_\tau}$. 
Then, Subtheorem~\ref{sub_thm:U_tau_u_i_commun} implies that the effective interaction terms, i.e., $\hat{H}_\tau-H_0$, are localized around the region $\mfL$. However, it does not mean that the effective terms can be determined only by the subset Hamiltonian $H_{0,\mfL[r]}$. The purpose of this theorem is to ensure this point; that is, the effective terms can be approximately determined only by using the information in the region $\mfL[r]=\tilde{\mfL}$, which is given by the inequality~\eqref{error_simplest_form_thm_for_eff_ham}. 
}
\label{fig:Effective_ham_local}
\end{figure}

For the present purpose, we prove the following theorem (see Fig.~\ref{fig:Effective_ham_local}):
\begin{theorem} \label{thm:Refined locality estimation for effective Hamiltonian}
We adopt the same setup as in Subtheorem~\ref{sub_thm:U_tau_u_i_commun}. 
Let $\tilde{\mfL}$ be an extended subset from $L$ by a distance $r$, i.e., 
\begin{align}
\tilde{\mfL} := \mfL[r] .
\end{align}
Then, we construct $\hat{H}_{\tau,\tilde{\mfL}}$ using the subset Hamiltonian $H_{0,\tilde{\mfL}}$ on $\tilde{\mfL}\subset \Lambda$ as follows (see Fig.~\ref{fig:Effective_ham_local}):
\begin{align}
\hat{H}_{\tau,\tilde{\mfL}} = \log \br{e^{\tau V_{\tilde{\mfL}}} e^{\beta H_{0,\tilde{\mfL}}}e^{\tau V_{\tilde{\mfL}}}}  .
\end{align}
Using Corollary~\ref{connection_of_exponential_operator_repeat}, we can formally express $\hat{H}_{\tau,\tilde{\mfL}}$ as 
\begin{align}
&\hat{H}_{\tau,\tilde{\mfL}} = U_{\tau,\tilde{\mfL}} ( \beta H_{0,\tilde{\mfL}} + \hat{V}_{\tau,\tilde{\mfL}} ) U_{\tau,\tilde{\mfL}}^\dagger , \notag \\
&U_{\tau,\tilde{\mfL}} :=\mathcal{T} e^{- i\int_0^\tau \mC_{\tau_1,\tilde{\mfL}} d\tau_1}  ,\quad   \hat{V}_{\tau,\tilde{\mfL}} := 2\int_0^\tau U_{\tau_1,\tilde{\mfL}}^\dagger  V_{\tilde{\mfL}}U_{\tau_1,\tilde{\mfL}} d\tau_1 ,\quad 
\mC_\tau :=\frac{2}{\beta}\int_{-\infty}^\infty g_\beta(t)  V_{\tilde{\mfL}}(\hat{H}_{\tau,\tilde{\mfL}}, t)dt.
\end{align}
Then, the norm difference between  $U_\tau$ and $U_{\tau,\tilde{\mfL}}$ is upper-bounded by
\begin{align}
\norm{  U_\tau  - U_{\tau,\tilde{\mfL}} } \le  \Er (\tau,r) = \mathcal{K}_2 (1+2 \tau)  e^{\mathcal{K}_0 \tau+\mathcal{K}_1 \tau \log(r+\tau+e)-\kappa_\beta r/2} ,
\label{main_inq:thm:Refined locality estimation for effective Hamiltonian}
\end{align}
where $\mathcal{K}_0= \kappa_0$ and $\mathcal{K}_1= \kappa_1$ as in Eq.~\eqref{kappa_0_kappa_1_definition}, and $ \mathcal{K}_2$ are given by
\begin{align}
\label{def:mathcal_K_2_thm}
\mathcal{K}_2:=\Theta\br{\norm{V}^{2D+3}, |\mfL|^3, \beta^{2D^2+2D+2}, \tau^{2D+2}, r^{2D+3}}. 
\end{align}
\end{theorem}
 

{\bf Remark.} 
From the expression of $\kappa_0= \Theta(\beta^D)\norm{V} \log (\beta\norm{V}\cdot |\mfL|)$ and $\kappa_1=\Theta(\beta^D) \norm{V}$ in Eq.~\eqref{kappa_0_kappa_1_definition}, we can reduce~\eqref{main_inq:thm:Refined locality estimation for effective Hamiltonian} to the form of 
\begin{align}
\label{error_simplest_form_thm}
\norm{  U_\tau  - U_{\tau,\tilde{\mfL}} } \le   e^{ \Theta( \tau \beta^D)\norm{V} \log (\beta\norm{V}\cdot |\mfL| r \tau) -\kappa_\beta r/2} .
\end{align}
Using the above inequality, we obtain for $\log \br{e^{\tau V} e^{\beta H_0}e^{\tau V}}= U_\tau (\beta H_0 +\hat{V}_\tau) U_\tau ^\dagger=\hat{H}_\tau$
\begin{align}
\label{upper_bound_e_tau_V/tau_tilde_mfL}
&\norm{\log \br{e^{\tau V} e^{\beta H_0}e^{\tau V}} - U_{\tau,\tilde{\mfL}} (\beta H_0 +\hat{V}_{\tau,\tilde{\mfL}}) U_{\tau,\tilde{\mfL}} ^\dagger}
\le  \norm{\hat{H}_\tau - U_{\tau,\tilde{\mfL}} (\beta H_0 +\hat{V}_{\tau}) U_{\tau,\tilde{\mfL}} ^\dagger} + \norm{\hat{V}_{\tau}-\hat{V}_{\tau,\tilde{\mfL}}}     \notag \\
&= \norm{\hat{H}_\tau - U_{\tau,\tilde{\mfL}}U_\tau^\dagger \hat{H}_\tau  U_\tau U_{\tau,\tilde{\mfL}} ^\dagger}
+ 2\norm{\int_0^\tau \brr{ U_{\tau_1}^\dagger  V U_{\tau_1}- U_{\tau_1,\tilde{\mfL}}^\dagger  V_{\tilde{\mfL}}U_{\tau_1,\tilde{\mfL}} } 
 d\tau_1 } \notag \\
 &\le \norm{\brr{\hat{H}_\tau, U_\tau U_{\tau,\tilde{\mfL}}^\dagger }} + 2\tau \norm{V-V_{\tilde{\mfL}}} + 2\int_0^\tau \norm{\brr{V,  U_{\tau_1} U_{\tau_1,\tilde{\mfL}}^\dagger }}  d\tau_1  \notag \\
 &\le  \Bigl\{2 \tau  \norm{V}+ 4 \bar{J}_0 |\mfL| \brr{r^{\ast D}+ 2 \gamma |\mfL| \cdot 2^{2D+2} \br{C_{2,\tau}^{2D} +r^{\ast 2D}}}   
+  4\gamma^3 |\mfL|^3\cdot 2^{2D+5}C_{2,\tau}^{2D} C_{3,\tau}^{3D}\bar{J}_0 \Bigr\}\Er(\tau,r) \notag \\
&\quad + 2\tau\norm{V}  \Er(0,r) + 4\tau \norm{V} \Er(\tau,r) ,
\end{align}
where we use the inequality~\eqref{norm_inequ_H_0_Delta_U_tau_fibn} for $\norm{\brr{\hat{H}_\tau, U_\tau U_{\tau,\tilde{\mfL}}^\dagger }}$ ($U_\tau U_{\tau,\tilde{\mfL}}^\dagger-1=\Delta U_\tau$), the inequality~\eqref{norm_V-V_tilde_L} for $\norm{V-V_{\tilde{\mfL}}}$, and the inequality~$\norm{\brr{V,  U_{\tau_1} U_{\tau_1,\tilde{\mfL}}^\dagger }} \le 2\norm{V} \cdot \norm{U_{\tau_1} - U_{\tau_1,\tilde{\mfL}} }$. 
We note that $r^\ast=\Theta(r,\tau)$ as is defined in Eq.~\eqref{def_r^ast_in_prop:upper_bound_commutator_H_tau_dif}. 
Therefore, the same upper bound as~\eqref{error_simplest_form_thm} holds for $\norm{\log \br{e^{\tau V} e^{\beta H_0}e^{\tau V}} - U_{\tau,\tilde{\mfL}} (H_0 +\hat{V}_{\tau,\tilde{\mfL}}) U_{\tau,\tilde{\mfL}} ^\dagger}$:
\begin{align}
\label{error_simplest_form_thm_for_eff_ham}
&\norm{\log \br{e^{\tau V} e^{\beta H_0}e^{\tau V}} - U_{\tau,\tilde{\mfL}} (\beta H_0 +\hat{V}_{\tau,\tilde{\mfL}}) U_{\tau,\tilde{\mfL}} ^\dagger}
\le  e^{ \Theta( \tau \beta^D)\norm{V} \log (\beta\norm{V}\cdot |\mfL| r \tau) -\kappa_\beta r/2} .
\end{align}

Finally, when we consider the entanglement Hamiltonian $\log \br{e^{\tau V} e^{\beta H_{0,\tilde{\mfL}}}e^{\tau V}}$, we can obtain the same inequality as~\eqref{main_inq:thm:Refined locality estimation for effective Hamiltonian}, but in this case, the entanglement Hamiltonian is not strictly defined on $\tilde{\mfL}$.

\subsection{Proof of Theorem~\ref{thm:Refined locality estimation for effective Hamiltonian}}
First of all, we upper-bound $\norm{V- V_{\tilde{\mfL}}}$ using Lemma~\ref{lemma_commutator_generalization} as follows:
\begin{align}
\norm{V- V_{\tilde{\mfL}}}& \le \sum_{i\in \tilde{\mfL}^\co} \sup_{u_i}\norm{[V,u_i]}
 \le  \norm{V} \sum_{i\in \mfL[r]^\co} Q(0,\dist_{i,\mfL}) \le   \sum_{i\in \mfL[r]^\co} \norm{V} e^{-\kappa_\beta \dist_{i,\mfL}} \notag \\
 &\le  \norm{V} \cdot \gamma |\mfL| e^{\kappa_\beta/2} (2/\kappa_\beta)^D D! e^{-r\kappa_\beta/2},
\end{align}
where we use the condition~\eqref{assumption_proof_le_0V}, and in the last inequality, we calculate as  
\begin{align}
\sum_{i\in \mfL[r]^\co} Q(0,\dist_{i,\mfL}) \le \sum_{j\in \mfL} \sum_{i: \dist_{i,j} >r} e^{-\kappa_\beta \dist_{i,j}}
 \le \gamma |\mfL| e^{\kappa_\beta/2} (2/\kappa_\beta)^D D! e^{-r\kappa_\beta/2}
\end{align}
from the inequality~\eqref{ineq_sum_decay_func}. 
Therefore, choosing $\mathcal{K}_2$ such that $\gamma |\mfL| e^{\kappa_\beta/2} (2/\kappa_\beta)^D D!\le \mathcal{K}_2$, we have 
\begin{align}
\label{norm_V-V_tilde_L}
\norm{V- V_{\tilde{\mfL}}}& \le  \norm{V} \cdot \mathcal{K}_2  e^{-r\kappa_\beta/2} = \norm{V}  \Er (0,r) .  
\end{align}

We next consider the proof of the inequality~\eqref{main_inq:thm:Refined locality estimation for effective Hamiltonian}.
As in the proof of Subtheorem~\ref{sub_thm:U_tau_u_i_commun}, we rely on the inductive method. 
For $\tau=0$, the inequality trivially holds since $U_\tau=U_{\tau,\tilde{\mfL}}=\hat{1}$, and we assume the main inequality~\eqref{main_inq:thm:Refined locality estimation for effective Hamiltonian} up to a certain $\tau$ and prove the case of $\tau+d\tau$ with $d\tau\to +0$.
Then, using Eq.~\eqref{U_tau_V_tau_mu_tau_def} as 
\begin{align}
U_\tau =\mathcal{T} e^{-i\int_0^\tau \mC_{\tau_1} d\tau_1}  ,
\end{align}
we have 
\begin{align}
\norm{  U_{\tau+d\tau}  - U_{\tau+d\tau,\tilde{\mfL}}^\dagger } 
&\le \norm{e^{-i \mC_{\tau} d\tau}  U_{\tau} - e^{-i \mC_{\tau,\tilde{\mfL}} d\tau}  U_{\tau,\tilde{\mfL}} }\notag \\
&\le \norm{ \br{ e^{-i \mC_{\tau} d\tau} - e^{-i \mC_{\tau,\tilde{\mfL}} d\tau} } U_{\tau} + e^{-i \mC_{\tau,\tilde{\mfL}} d\tau} \br{ U_{\tau}- U_{\tau,\tilde{\mfL}}} }  \notag \\
&\le \norm{\mC_{\tau} -  \mC_{\tau,\tilde{\mfL}}} d\tau + \norm{ U_{\tau}- U_{\tau,\tilde{\mfL}} } .
\end{align}
Therefore, following a similar discussion to the derivation of~\eqref{aim_to_prove_statement_mC}, we aim to prove the upper bound of 
\begin{align}
\label{upp_dif_mC__mC_tau_tilde}
\norm{\mC_{\tau} -  \mC_{\tau,\tilde{\mfL}}} &= \norm{\frac{2}{\beta}\int_{-\infty}^\infty g_\beta(t) \brr{V\br{\hat{H}_\tau, t}-  V_{\tilde{\mfL}}\br{\hat{H}_{\tau,\tilde{\mfL}}, t}} dt }\notag \\
&\le  \brr{\mathcal{K}_0+ \mathcal{K}_1 \log(r+\tau+e)}\Er (\tau,r) +2 \mathcal{K}_2 e^{\mathcal{K}_0\tau+\mathcal{K}_1 \tau \log(r+\tau+e)-\kappa_\beta r/2} .
\end{align}
To achieve the above bound, it is enough to prove 
\begin{align}
\norm{\mC_{\tau} -  \mC_{\tau,\tilde{\mfL}}} \le  \brr{\mathcal{K}_0+ \mathcal{K}_1 \log(r+\tau+e)} \Er (\tau,r) +\mathcal{K}_2 \brr{Q(\tau,r/2)+ \Er (0,r)} ,
\label{goal_inequality_eff_Ham_locality}
\end{align}
which implies the inequality~\eqref{upp_dif_mC__mC_tau_tilde} because of 
\begin{align}
&e^{\mathcal{K}_0\tau+\mathcal{K}_1 \tau \log(r+\tau+e)-\kappa_\beta r/2} \ge Q(\tau,r/2)= e^{\kappa_0\tau+\kappa_1 \tau \log(r/2+\tau+e)-\kappa_\beta r/2},  \notag \\
& \mathcal{K}_2 e^{\mathcal{K}_0\tau+\mathcal{K}_1 \tau \log(r+\tau+e)-\kappa_\beta r/2} \ge  \Er (0,r)= \mathcal{K}_2 e^{-\kappa_\beta r/2} ,
\end{align}
for $\tau\ge 0$, where we use $\mathcal{K}_0= \kappa_0$ and $\mathcal{K}_1= \kappa_1$ from the definition.

For the analyses of $\norm{\mC_{\tau} -  \mC_{\tau,\tilde{\mfL}}}$, we start from the equation of
\begin{align}
&\int_{-\infty}^\infty g_\beta(t) \brr{V\br{\hat{H}_\tau, t}-  V_{\tilde{\mfL}}\br{\hat{H}_{\tau,\tilde{\mfL}}, t}} dt 
= \int_{|t|> \delta t} g_\beta(t) \brr{V\br{\hat{H}_\tau, t}dt-V_{\tilde{\mfL}}\br{\hat{H}_{\tau,\tilde{\mfL}}, t}}dt \notag \\
&\quad\quad\quad\quad\quad\quad\quad\quad\quad\quad + i\int_{|t|\le \delta t} t g_\beta(t) \int_0^1 \brrr{\brr{\ad_{\hat{H}_\tau}(V)}\br{\hat{H}_\tau,\lambda t}-\brr{\ad_{\hat{H}_{\tau,\tilde{\mfL}}}\br{V_{\tilde{\mfL}}}}\br{\hat{H}_{\tau,\tilde{\mfL}}, \lambda t}} d\lambda  dt ,
\label{int_g_beta_t_V_re_again}
\end{align}
by using Eq.~\eqref{int_g_beta_t_V}. 
We start from the equation~\eqref{start_time_O_H_tau_t}, which gives for arbitrary operators $O$ and $\tilde{O}$ 
\begin{align}
&\norm{O(\hat{H}_\tau, t) - \tilde{O}(\hat{H}_{\tau,\tilde{\mfL}}, t)} \notag \\
&=\norm{ U_\tau e^{i \br{ H_0 + \hat{V}_\tau } t }  U_\tau^\dagger  O U_\tau  e^{-i \br{ H_0 + \hat{V}_\tau } t } U_\tau^\dagger 
- U_{\tau,\tilde{\mfL}} e^{i \br{ H_{0,\tilde{\mfL}} + \hat{V}_{\tau,\tilde{\mfL}} } t }  U_{\tau,\tilde{\mfL}}^\dagger  \tilde{O} U_{\tau,\tilde{\mfL}}  e^{-i \br{ H_{0,\tilde{\mfL}} + \hat{V}_{\tau,\tilde{\mfL}} } t } U_{\tau,\tilde{\mfL}}^\dagger } \notag \\
&\le \norm{O-\tilde{O}}+ 4\norm{O} \Er(\tau,r) +\norm{  e^{i \br{ H_0 + \hat{V}_\tau } t }  U_\tau^\dagger  O U_\tau  e^{-i \br{ H_0 + \hat{V}_\tau } t } 
-  e^{i \br{ H_{0,\tilde{\mfL}} + \hat{V}_{\tau,\tilde{\mfL}} } t } U_\tau ^\dagger  O U_\tau   e^{-i \br{ H_{0,\tilde{\mfL}} + \hat{V}_{\tau,\tilde{\mfL}} } t } } ,
\label{starting point_proof_of_Theorem_main}
\end{align}
where, in the inequality, we use 
\begin{align}
&\norm{U_{\tau,\tilde{\mfL}} e^{i \br{ H_{0,\tilde{\mfL}} + \hat{V}_{\tau,\tilde{\mfL}} } t }  U_{\tau,\tilde{\mfL}}^\dagger  O U_{\tau,\tilde{\mfL}}  e^{-i \br{ H_{0,\tilde{\mfL}} + \hat{V}_{\tau,\tilde{\mfL}} } t } U_{\tau,\tilde{\mfL}}^\dagger - U_\tau e^{i \br{ H_{0,\tilde{\mfL}} + \hat{V}_{\tau,\tilde{\mfL}} } t } U_\tau ^\dagger O U_\tau   e^{-i \br{ H_{0,\tilde{\mfL}} + \hat{V}_{\tau,\tilde{\mfL}} } t } U_\tau^\dagger}   \notag \\
&\le 4\norm{O} \cdot \norm{  U_\tau -U_{\tau,\tilde{\mfL}}} \le  4\norm{O} \Er(\tau,r) .
\end{align}
In applying the inequality~\eqref{starting point_proof_of_Theorem_main} to $V(\hat{H}_\tau, t)-  V_{\tilde{\mfL}}\br{\hat{H}_{\tau,\tilde{\mfL}}, t}$ in Eq.~\eqref{int_g_beta_t_V_re_again}, the non-trivial analyses mainly arise from the estimation of the norm difference between $ e^{i \br{ H_0 + \hat{V}_\tau } t }  U_\tau^\dagger  O U_\tau  e^{-i \br{ H_0 + \hat{V}_\tau } t } $ and 
$e^{i \br{ H_{0,\tilde{\mfL}} + \hat{V}_{\tau,\tilde{\mfL}} } t } U_\tau ^\dagger  O U_\tau   e^{-i \br{ H_{0,\tilde{\mfL}} + \hat{V}_{\tau,\tilde{\mfL}} } t }$.
For this purpose, we prove the following proposition (see Sec.~\ref{sec:Proof of Proposition_prop:main_est_time_evo_comp} for the proof):

\begin{prop} \label{prop:main_est_time_evo_comp}
Let $O$ be an arbitrary quasi-local operator around $L$ such that 
\begin{align}
\norm{\brr{O, u_i}} \le \norm{O}\check{Q}(\tau,r),\quad \check{Q}(\tau,r) :=e^{\check{\kappa}_0\tau +\check{\kappa}_1 \tau \log(r+\tau+e)-\kappa_\beta r },
\label{assum_quasi_locality_of_O}
\end{align}
and 
\begin{align}
Q(\tau,r) \le \check{Q}(\tau,r),     \label{assum_quasi_locality_of_Q_tau_r}
\end{align}
where $Q(\tau,r)$ is defined by~\eqref{assumption_proof_le_tau} in Subtheorem~\ref{sub_thm:U_tau_u_i_commun}, $\dist_{i,\mfL}=r$, and $u_i$ is an arbitrary local unitary operator on the site $i$.
We then obtain for the third term in the RHS of~\eqref{starting point_proof_of_Theorem_main} 
\begin{align}
&\norm{  e^{i \br{ H_0 + \hat{V}_\tau } t }  U_\tau^\dagger  O U_\tau  e^{-i \br{ H_0 + \hat{V}_\tau } t } 
-  e^{i \br{ H_{0,\tilde{\mfL}} + \hat{V}_{\tau,\tilde{\mfL}} } t } U_\tau ^\dagger  O U_\tau   e^{-i \br{ H_{0,\tilde{\mfL}} + \hat{V}_{\tau,\tilde{\mfL}} } t } }  \notag \\
&\le 6\gamma \norm{O} \cdot |\mfL|  \br{4|t|\tau \norm{V}+1 } \Bigl \{ 2^{D+3}\br{\check{C}_{1,\tau}^{D}+r^D} \check{Q}(\tau,r/2)  
+\tilde{J}_0   |t| r^{D}\brr{ 2e^{-\mu r/8}+  \min\br{ C \gamma |\mfL| r^{D}e^{v |t|-\mu r/4},1} }   \Bigr\} \notag \\
&\quad + 4|t|\cdot \norm{O}\cdot  \norm{V}\br{\tau \Er (0,r) +2 \int_0^\tau  \Er (\tau_1,r)   d\tau_1 },
\label{main:eq:prop:main_est_time_evo_comp}
\end{align}
where the parameters $\tilde{J}_0$ and $\check{C}_{1,\tau}$ are defined in Eqs.~\eqref{sum_interaction_terms_main_eq_short_long} and \eqref{upper-bound:tilde_Q_tau_r_lemma}, respectively. 
\end{prop}

{\bf Remark.}  Using $ \Er (\tau_1,r) = \mathcal{K}_2(1+2\tau_1)e^{\mathcal{K}_0\tau+\mathcal{K}_1 \tau_1 \log(r+\tau_1+e)-\kappa_\beta r/2} $, we obtain 
\begin{align}
\int_0^\tau  \Er (\tau_1,r)   d\tau_1 \le  \mathcal{K}_2(1+2\tau) e^{\mathcal{K}_1\tau\log(r+\tau+e)-\kappa_\beta r/2} \int_0^\tau e^{\mathcal{K}_0 \tau_1}d\tau_1  
\le \frac{\Er (\tau,r) }{\mathcal{K}_0},
\end{align}
which upper-bounds the last term in the RHS of~\eqref{main:eq:prop:main_est_time_evo_comp}. 

{~} \\

We estimate the first term in the RHS of~\eqref{int_g_beta_t_V_re_again} by considering the case of $O=V$ and $\tilde{O}=V_{\tilde{\mfL}}$ in~\eqref{starting point_proof_of_Theorem_main}.
 By applying the inequality~\eqref{main:eq:prop:main_est_time_evo_comp} to \eqref{starting point_proof_of_Theorem_main}, we obtain 
\begin{align}
&\norm{V(\hat{H}_\tau, t)-V_{\tilde{\mfL}}(\hat{H}_{\tau,\tilde{\mfL}}, t)}  \notag \\
&\le  6\gamma \norm{V} \cdot |\mfL|  \br{4|t|\tau\norm{V} +1 } \Bigl \{ 2^{D+3}\br{C_{1,\tau}^{D}+r^D} Q(0,r/2)  
+\tilde{J}_0   |t| r^{D}\brr{ 2e^{-\mu r/8}+  \min\br{ C \gamma |\mfL| r^{D}e^{v |t|-\mu r/4},1} }   \Bigr\} 
\notag \\
&\quad +\norm{V-V_{\tilde{\mfL}}}+ 4\norm{V} \Er(\tau,r) + 4|t|\cdot \norm{V}^2 \br{\tau \Er(0,r) +\frac{2\Er (\tau,r) }{\mathcal{K}_0} }. 
\label{first_term_V_tau_t_estimation}
\end{align}
Using Lemma~\ref{belief_norm_lemma}, we have  
\begin{align}
&\int_{|t| > \delta t}  |g_\beta (t)|  dt \le  \frac{2\beta}{\pi}\log\br{\frac{\beta}{2\pi \delta t}} , \notag \\
&\int_{|t| > \delta t}  |tg_\beta (t)|  dt \le  \frac{\beta^2}{12} \le \frac{\beta^2}{\pi^2} ,\quad 
\int_{|t| > \delta t}  |t^2g_\beta (t)|  dt \le  \frac{\zeta(3)\beta^3}{2\pi^3} \le   \frac{\beta^3}{\pi^3}. 
\end{align}
We therefore obtain 
\begin{align}
& \int_{|t|> \delta t} |g_\beta(t)| \brr{\norm{V-V_{\tilde{\mfL}}}+ 4\norm{V} \Er(\tau,r) + 4|t| \cdot \norm{V}^2 \br{\tau \Er(0,r) +\frac{2\Er (\tau,r) }{\mathcal{K}_0}}} dt \notag \\
&\le  \frac{2\beta \norm{V}}{\pi}\log\br{\frac{\beta}{2\pi \delta t}}   \brr{\tau\Er(0,r)+ 4\Er(\tau,r)} 
+ \frac{4\beta^2 \norm{V}^2}{\pi^2}  \brr{\tau \Er(0,r) +\frac{2\Er (\tau,r) }{\mathcal{K}_0}} ,
\label{first_term_V_tau_t_estimation_1}
\end{align}
and 
\begin{align}
& \int_{|t|> \delta t} |g_\beta(t)| \cdot 6\gamma \norm{V} \cdot |\mfL|  \br{4|t| \tau \norm{V} +1 } \Bigl \{ 2^{D+3}\br{C_{1,\tau}^{D}+r^D} Q(\tau,r/2)  
+2\tilde{J}_0 |t| r^{D} e^{-\mu r/8}    \Bigr\} dt \notag \\
&\le 6\gamma \norm{V} \cdot |\mfL| \brr{ 2^{D+3}\br{C_{1,\tau}^{D}+r^D} Q(\tau,r/2) \cdot   \frac{2\beta}{\pi}\log\br{\frac{\beta}{2\pi \delta t}}
+2\tilde{J}_0  r^{D} e^{-\kappa_\beta r/2} \cdot \frac{\beta^2}{\pi^2}
} \notag \\
&\quad + 24\tau \gamma \norm{V}^2 \cdot |\mfL|   
\brr{2^{D+3}\br{C_{1,\tau}^{D}+r^D} Q(\tau,r/2)  \cdot \frac{\beta^2}{\pi^2} 
+2\tilde{J}_0r^{D} e^{-\kappa_\beta r/2} \cdot  \frac{\beta^3}{\pi^3} } \notag \\
&\le  \frac{12\cdot 2^{D+3} \gamma  \beta \norm{V} \cdot |\mfL| }{\pi} \br{\log\br{\frac{\beta}{2\pi \delta t}}+ \frac{2\tau \beta \norm{V}}{\pi}} \br{C_{1,\tau}^{D}+r^D} Q(\tau,r/2)  \notag \\
&\quad + \frac{12\gamma \beta^2 \norm{V} \cdot |\mfL| \tilde{J}_0}{\pi^2} \br{\frac{4\beta \norm{V} \tau }{\pi}+1 }  r^D e^{-\kappa_\beta r/2} \le \zeta_{\tau,r,\delta t}Q(\tau,r/2)  , 
\label{first_term_V_tau_t_estimation_2}
\end{align}
with
\begin{align}
\zeta_{\tau,r,\delta t}&:=  \frac{12\cdot 2^{D+3} \gamma  \beta \norm{V} \cdot |\mfL| }{\pi} \br{\log\br{\frac{\beta}{2\pi \delta t}}+ \frac{2\tau \beta \norm{V}}{\pi}} \br{C_{1,\tau}^{D}+r^D} + \frac{12\gamma \beta^2 \norm{V} \cdot |\mfL| \tilde{J}_0}{\pi^2} \br{\frac{4\beta \norm{V} \tau }{\pi}+1 }  r^D \notag \\
&\le \Theta\br{\norm{V}^{2D+2}, |\mfL|^2, \beta^{2D^2+2D+2}, \tau^{2D+1}, r^D, \log (\delta t^{-1})},
\end{align}
where we use the inequality~\eqref{norm_V-V_tilde_L} and $\mu /8 \ge \kappa_\beta/2$ from the definition~\eqref{definition_kappa_beta_1}, and in the last inequality, we use from Eq.~\eqref{upper-bound:tilde_Q_tau_r_lemma}
\begin{align}
C^D_{1,\tau}=\brr{ \frac{\tau+e}{2} + \frac{32}{\kappa_\beta ^2} \br{D^2 + \kappa_1^2 \tau^2+\frac{\kappa_\beta \kappa_0\tau}{8}}}^D \le \Theta\br{\norm{V}^{2D}, |\mfL|,\beta^{2D^2+2D}, \tau^{2D}}, 
\end{align}
where we use $\kappa_0= \Theta(\beta^D)\norm{V} \log (\beta\norm{V}\cdot |\mfL|)$ and $\kappa_1=\Theta(\beta^D) \norm{V}$ from Eq.~\eqref{definition_kappa_beta_1_again}.
The remaining part in Eq.~\eqref{first_term_V_tau_t_estimation} gives the integral of 
\begin{align}
\label{upper_bound_remining_v_t_tern}
& \int_{|t|> \delta t} |g_\beta(t)| \cdot 6\gamma \norm{V} \cdot |\mfL|  \br{4|t|\tau\norm{V} +1 } \tilde{J}_0   |t| r^{D}  \min\br{ C \gamma |\mfL| r^{D}e^{v |t|-\mu r/4},1}     \Bigr\} dt ,
\end{align}
which necessitates the estimation of 
\begin{align}
6\gamma \norm{V} \cdot |\mfL|  \tilde{J}_0 r^{D}  \int_{|t| > \delta t}  |t| \br{4|t|\tau\norm{V} +1 } |g_\beta (t)|  \min\br{ C \gamma |\mfL| r^{D}e^{v |t|-\mu r/4},1}   dt  . 
\label{upper_bound_remining_v_t_tern__2}
\end{align}
For this purpose, we prove the following lemma:
\begin{lemma}\label{lem:integrral_upp_eff_Ham}
For an arbitrary $m\in \mathbb{N}$ such that $m\ge 1$, we obtain the upper bound of 
\begin{align} 
\int_{|t| > \delta t}  |t^m g_\beta (t)|  \min\brr{ C \gamma |\mfL| r^{D}e^{v |t|-\mu r/4} ,1}dt  
\le  6e^{-\kappa_\beta r/2} \br{\frac{\beta}{2\pi}}^{m+1}  \br{\frac{\pi\mu r}{4v\beta}+1}^{m} \br{\frac{C \gamma |\mfL| r^{D}}{3m}+ m! }. 
\label{main_inequ_lem:integrral_upp_eff_Ham}
\end{align}
\end{lemma}

\textit{Proof of Lemma~\ref{lem:integrral_upp_eff_Ham}.}
We first let $t_0=\mu r/(8v)$ and obtain 
\begin{align}
&\int_{|t| > \delta t}  |t^m g_\beta (t)|  \min\brr{ C \gamma |\mfL| r^{D}e^{v |t|-\mu r/4} ,1}  dt  \notag \\
&\le \int_{|t| \le t_0}  |t^m g_\beta (t)|  \min\brr{C \gamma |\mfL| r^{D}e^{v |t|-\mu r/4},1}  dt + \int_{|t| > t_0}  |t^m g_\beta (t)|  \min\brr{ C \gamma |\mfL| r^{D}e^{v |t|-\mu r/4},1}  dt  \notag \\
&\le C \gamma |\mfL| r^{D}e^{v t_0-\mu r/4} \int_{|t| \le t_0}  |t^m g_\beta (t)|dt + \int_{|t| > t_0}  |t^m g_\beta (t)|   dt .
\label{main_inequ_lem:integrral_upp_eff_Ham_pr0}
\end{align}
For the first term, we have 
\begin{align} 
\label{main_inequ_lem:integrral_upp_eff_Ham_pr1}
C \gamma |\mfL| r^{D}e^{v t_0-\mu r/4} \int_{|t| \le t_0}  |t^m g_\beta (t)|dt
\le C \gamma |\mfL| r^{D}e^{-\mu r/8} \cdot \frac{2}{m} \br{\frac{\beta}{2\pi}}^{m+1}  \br{\frac{\pi\mu r}{4v\beta}}^{m},
\end{align}
where, in the first inequality, we use the inequality~\eqref{integral_t__g_beta_t_ineq} in the proof of Lemma~\ref{belief_norm_lemma}. 

For the second term, by letting $z=2\pi t/\beta$ and $z_0=2\pi t_0/\beta= \pi \mu r/(4v\beta)$, we obtain
\begin{align}
\int_{|t| > t_0}  |t^m g_\beta (t)|  dt 
&= 2\int_{z_0}^\infty  \br{\frac{\beta z}{2\pi}}^m \frac{e^{-z}}{1-e^{-z}} \frac{\beta dz}{2\pi} 
=2 \br{ \frac{\beta}{2\pi}}^{m+1} \int_{z_0}^\infty  \frac{z^m}{e^{z}-1}dz .
\end{align}
For the integral, we can derive
\begin{align} 
\int_{z_0}^\infty  \frac{z^m}{e^{z}-1}dz 
& = \sum_{s=0}^m \frac{m!}{(m-s)!} z_0^{m-s} {\rm Li}_{s+1} (e^{-z_0}) \notag\\
&\le  e^{-z_0} \brr{(z_0+1)^m + \sum_{s=1}^m \frac{m!}{(m-s)!}z_0^{m-s} \zeta(s+1)} \notag\\
&\le e^{-z_0} \brr{(z_0+1)^m +\zeta(2) \sum_{s=1}^m s! \binom{m}{s} z_0^{m-s}} \le  3 e^{-z_0} m! (z_0+1)^m ,
\end{align}
where  ${\rm Li}_s(x)$ is the polylogarithm function, i.e., ${\rm Li}_s(x):=\sum_{s=1}^\infty x^k/k^s$ [${\rm Li}_1(x)=-\log(1-x)$],  and we use the inequality of 
\begin{align} 
z_0^m{\rm Li}_{1} (e^{-z_0}) =- z_0^m  \log(1-e^{-z_0}) \le  (z_0+1)^m  e^{-z_0} 
\end{align}
and 
\begin{align} 
{\rm Li}_{s} (e^{-z_0}) = \sum_{s=1}^\infty \frac{e^{- k z_0}}{k^s} =  e^{- z_0}\sum_{s=1}^\infty \frac{e^{- (k-1) z_0}}{k^s} \le
e^{- z_0}\sum_{s=1}^\infty \frac{1}{k^s} =  e^{- z_0}\zeta(s)  \for s\ge2. 
\end{align}
We thus obtain
\begin{align}
\label{main_inequ_lem:integrral_upp_eff_Ham_pr2}
\int_{|t| > t_0}  |g_\beta (t)|  dt 
&\le  2\br{ \frac{\beta}{2\pi}}^{m+1} \cdot 3 e^{-\pi \mu r/(4v\beta)} m! \br{\frac{\pi \mu r}{4v\beta}+1}^m  . 
\end{align}
By combining the inequalities~\eqref{main_inequ_lem:integrral_upp_eff_Ham_pr1} and \eqref{main_inequ_lem:integrral_upp_eff_Ham_pr2} with~\eqref{main_inequ_lem:integrral_upp_eff_Ham_pr0}, we prove the main inequality~\eqref{main_inequ_lem:integrral_upp_eff_Ham} as follows:
\begin{align} 
&\int_{|t| > \delta t}  |t^m g_\beta (t)|  \min\brr{ C \gamma |\mfL| r^{D}e^{v |t|-\mu r/4} ,1}  dt   \notag \\
&\le  C \gamma |\mfL| r^{D}e^{-\mu r/8} \cdot\frac{2}{m} \br{\frac{\beta}{2\pi}}^{m+1}  \br{\frac{\pi\mu r}{4v\beta}}^{m} + 2\br{ \frac{\beta}{2\pi}}^{m+1} \cdot 3 e^{-\pi \mu r/(4v\beta)} m! \br{\frac{\pi \mu r}{4v\beta}+1}^m  \notag \\
&\le 6e^{-\kappa_\beta r/2} \br{\frac{\beta}{2\pi}}^{m+1}  \br{\frac{\pi\mu r}{4v\beta}+1}^{m} \br{\frac{C \gamma |\mfL| r^{D}}{3m}+ m! },
\end{align}
where we use the definition~\eqref{definition_kappa_beta_1} for $\kappa_\beta$.  
This completes the proof. $\square$

{~}

\hrulefill{\bf [ End of Proof of Lemma~\ref{lem:integrral_upp_eff_Ham}]}

{~}

Using this lemma, we obtain 
\begin{align}
&\int_{|t| > \delta t}  |t| \br{4|t|\tau\norm{V} +1 } |g_\beta (t)|  \min\br{ C \gamma |\mfL| r^{D}e^{v |t|-\mu r/4},1}   dt  \notag \\
&\le 6e^{-\kappa_\beta r/2} \brrr{ 4\tau \norm{V} \cdot \br{\frac{\beta}{2\pi}}^3   \br{\frac{\pi\mu r}{4v\beta}+1}^{2} \br{\frac{C \gamma |\mfL| r^{D}}{6}+ 2}  +
\br{\frac{\beta}{2\pi}}^2  \br{\frac{\pi\mu r}{4v\beta}+1} \br{\frac{C \gamma |\mfL| r^{D}}{3}+ 1}}\notag \\
&\le \frac{3\beta^2  }{2\pi^2} e^{-\kappa_\beta r/2}
\br{\frac{4 \tau \beta\norm{V}}{\pi} +1} \br{\frac{\pi\mu r}{4v\beta}+1}^{2}  \br{\frac{C \gamma |\mfL| r^{D}}{3}+ 2}
=\Theta(\norm{V}, |\mfL|,\beta^3, \tau,r^{D+2} )  e^{-\kappa_\beta r/2} ,
\label{first_term_V_tau_t_estimation_3}
\end{align}
which upper-bound the term~\eqref{upper_bound_remining_v_t_tern__2} as 
\begin{align}
&6\gamma \norm{V} \cdot |\mfL|  \tilde{J}_0 r^{D}  \int_{|t| > \delta t}  |t| \br{4|t|\tau\norm{V} +1 } |g_\beta (t)|  \min\br{ C \gamma |\mfL| r^{D}e^{v |t|-\mu r/4},1}   dt  \notag \\
&\le \frac{18\gamma\beta^2 \norm{V} \cdot |\mfL|  \tilde{J}_0 r^{D}  }{2\pi^2} 
\br{\frac{4 \tau \beta\norm{V}}{\pi} +1} \br{\frac{\pi\mu r}{4v\beta}+1}^{2}  \br{\frac{C \gamma |\mfL| r^{D}}{3}+ 2}
=\zeta'_{\tau,r}Q(\tau,r/2)
\label{first_term_V_tau_t_estimation_3_2}
\end{align}
with 
\begin{align}
\zeta'_{\tau,r}&:= \frac{18\gamma\beta^2 \norm{V} \cdot |\mfL|  \tilde{J}_0 r^{D}  }{2\pi^2} 
\br{\frac{4 \tau \beta\norm{V}}{\pi} +1} \br{\frac{\pi\mu r}{4v\beta}+1}^{2}  \br{\frac{C \gamma |\mfL| r^{D}}{3}+ 2} Q(\tau,r/2) \notag \\
&\le \Theta(\norm{V}^2,|\mfL|^2,\beta^3, \tau,r^{2D+2} ) .
\label{first_term_V_tau_t_estimation_3_2_bar_g}
\end{align}
By combining the three upper bounds~\eqref{first_term_V_tau_t_estimation_1}, \eqref{first_term_V_tau_t_estimation_2} and \eqref{first_term_V_tau_t_estimation_3_2} with the integral of~\eqref{first_term_V_tau_t_estimation}, we upper-bound the first term in the RHS of~\eqref{int_g_beta_t_V_re_again} as follows:
\begin{align}
&\int_{|t|> \delta t} |g_\beta(t)| \cdot \norm{V(\hat{H}_\tau, t)dt-V_{\tilde{\mfL}}(\hat{H}_{\tau,\tilde{\mfL}}, t)}dt  \notag \\
&\le  
 \frac{2\beta \norm{V}}{\pi}\log\br{\frac{\beta}{2\pi \delta t}}   \br{\tau \Er(0,r)+ 4\Er(\tau,r)}  
+ \frac{4\beta^2 \norm{V}^2}{\pi^2}  \br{\tau \Er(0,r) +\frac{2\Er (\tau,r) }{\mathcal{K}_0}} +\br{\zeta_{\tau,r,\delta t}+\zeta'_{\tau,r}}Q(\tau,r/2) \notag \\
&\zeta_{\tau,r,\delta t}+\zeta'_{\tau,r}\le 
\Theta\br{\norm{V}^{2D+2}, |\mfL|^2, \beta^{2D^2+2D+2}, \tau^{2D+1}, r^{2D+2}, \log (\delta t^{-1})}.\label{int_g_beta_t_V_re_again_final_form} 
\end{align}

We next estimate the second term in the RHS of~\eqref{int_g_beta_t_V_re_again} by considering the case of $O=\ad_{\hat{H}_\tau}(V)$ and $\tilde{O}=\ad_{\hat{H}_{\tau,\tilde{\mfL}}}(V_{\tilde{\mfL}})$ in~\eqref{starting point_proof_of_Theorem_main}.
Then, we have to estimate 
\begin{align}
\label{t_g_beta_t_integrate_ad_H_tau}
\int_{|t|\le \delta t} |t g_\beta(t)| \int_0^1 \norm{[\ad_{\hat{H}_\tau}(V)](\hat{H}_\tau,\lambda t)-[\ad_{\hat{H}_{\tau,\tilde{\mfL}}}(V_{\tilde{\mfL}})](\hat{H}_{\tau,\tilde{\mfL}}, \lambda t)} d\lambda  dt .
\end{align}
Using the inequality~\eqref{starting point_proof_of_Theorem_main} yields 
\begin{align}
\label{t_g_beta_t_integrate_ad_H_tau_error_est_1}
&\norm{[\ad_{\hat{H}_\tau}(V)](\hat{H}_\tau,\lambda t)-[\ad_{\hat{H}_{\tau,\tilde{\mfL}}}(V_{\tilde{\mfL}})](\hat{H}_{\tau,\tilde{\mfL}}, \lambda t)}   \notag \\
&\le  \norm{\ad_{\hat{H}_\tau}(V) -\ad_{\hat{H}_{\tau,\tilde{\mfL}}}(V_{\tilde{\mfL}})}+ 4\norm{\ad_{\hat{H}_\tau}(V)} \Er(\tau,r)  \notag \\
&\quad +\norm{  e^{i \br{ H_0 + \hat{V}_\tau } \lambda t }  U_\tau^\dagger \ad_{\hat{H}_\tau}(V) U_\tau  e^{-i \br{ H_0 + \hat{V}_\tau }\lambda t } 
-  e^{i \br{ H_{0,\tilde{\mfL}} + \hat{V}_{\tau,\tilde{\mfL}} }\lambda t } U_\tau ^\dagger  \ad_{\hat{H}_{\tau,\tilde{\mfL}}}(V_{\tilde{\mfL}})  U_\tau   e^{-i \br{ H_{0,\tilde{\mfL}} + \hat{V}_{\tau,\tilde{\mfL}} }\lambda t } } .
\end{align}
From the proposition~\ref{prop:quasi-locality of_ad__H__tau_V}, we obtain
\begin{align}
\label{ad_H_tau_commutator_bound}
\norm{\brr{\ad_{\hat{H}_\tau} (V) ,u_i }}\le \norm{V} g_{\tau,r}Q(\tau,r) ,
\end{align}
and 
\begin{align}
\label{ad_H_tau_commutator_norm_bound}
\norm{\ad_{\hat{H}_\tau} (V) }\le \norm{V} g'_{\tau},
\end{align}
where the parameters $g_{\tau,r}$ and $g'_{\tau}$ are estimated in Eq.~\eqref{upp_t_tau_r_g'}.
Furthermore, we need to estimate the norm difference between $\ad_{\hat{H}_\tau}(V)$ and  $\ad_{\hat{H}_{\tau,\tilde{\mfL}}}(V_{\tilde{\mfL}})$.
For this purpose, we prove the following statement (see Sec.~\ref{sec:Proof of Proposition_prop:upper_bound_commutator_H_tau_dif} for the proof):
\begin{prop} \label{prop:upper_bound_commutator_H_tau_dif}
Let us define $r^\ast$ be the minimum number such that 
\begin{align}
\label{def_r^ast_in_prop:upper_bound_commutator_H_tau_dif}
\min \brr{\Er(\tau,r), Q(\tau,r^\ast)} =  Q(\tau,r^\ast)  \quad \textrm{i.e.,} \quad  r^\ast= \Theta(\tau,r).
\end{align}
We upper-bound the norm of $\norm{ \ad_{\hat{H}_\tau}(V) - \ad_{\hat{H}_{\tau,\tilde{\mfL}}}(V_{\tilde{\mfL}})} $ by
\begin{align}
\label{main_ineq:prop:upper_bound_commutator_H_tau_dif}
\norm{ \ad_{\hat{H}_\tau}(V) - \ad_{\hat{H}_{\tau,\tilde{\mfL}}}(V_{\tilde{\mfL}})} 
\le  \norm{V} g''_{\tau,r}\Er(\tau,r) ,
\end{align}
where we define $g''_{\tau,r}$ as 
\begin{align}
g''_{\tau,r}:= 6 \brr{3\tau\norm{V}+  \tilde{J}_0  |\tilde{\mfL}|^2  +  \bar{J}_0  |\mfL| r^{\ast D} 
+ 2^{2D+3}  \bar{J}_0  \gamma |\mfL|^2   \br{C_{2,\tau}^{2D} +r^{\ast 2D} + 4\gamma^2 |\mfL|C_{2,\tau}^{2D} C_{3,\tau}^{3D} }} . 
\end{align}
\end{prop}

{\bf Remark.} Using the $\Theta$ notation, we obtain 
\begin{align}
g''_{\tau,r} = \Theta(\norm{V},|\mfL|^3,  r^{2D}, \beta^{10D}, \kappa_0^{5D}, \kappa_1^{10D},\tau^{10D}) . 
\end{align}
We again remind readers that for our purpose it is enough to ensure that $g''_{\tau,r}$ is upper-bounded by a polynomial of $\{ \norm{V},|\mfL|,  r, \beta, \kappa_0, \kappa_1,\tau$\}.

From Lemma~\ref{belief_norm_lemma}, we can derive   
\begin{align}
&\int_{|t| \le \delta t} | t^{m+1} g_\beta (t) | dt  \le  \frac{\beta}{(m+1)\pi} \delta t^{m+1} \le  \frac{\beta}{\pi} \delta t^{m+1} \quad (m\ge 0)  .
\end{align}
Applying the above inequality to~\eqref{t_g_beta_t_integrate_ad_H_tau} and \eqref{t_g_beta_t_integrate_ad_H_tau_error_est_1}, we first obtain 
\begin{align}
\label{ineq:integral_lambda_t_g_beta_1}
&\int_{|t|\le \delta t} |t g_\beta(t)| \int_0^1 \br{  \norm{\ad_{\hat{H}_\tau}(V) -\ad_{\hat{H}_{\tau,\tilde{\mfL}}}(V_{\tilde{\mfL}})}+ 4\norm{\ad_{\hat{H}_\tau}(V)} \Er(\tau,r)}   d\lambda  dt 
\le  \frac{\beta \delta t \norm{V}}{\pi} \Er(\tau,r) \br{ g''_{\tau,r} +4g'_{\tau} } .
\end{align}
Next, by using the inequality~\eqref{main:eq:prop:main_est_time_evo_comp} with the bounds of~\eqref{ad_H_tau_commutator_bound} and \eqref{ad_H_tau_commutator_norm_bound}, we upper-bound
\begin{align}
&\norm{  e^{i \br{ H_0 + \hat{V}_\tau }\lambda t }  U_\tau^\dagger  \ad_{\hat{H}_\tau}(V)  U_\tau  e^{-i \br{ H_0 + \hat{V}_\tau } \lambda t } 
-  e^{i \br{ H_{0,\tilde{\mfL}} + \hat{V}_{\tau,\tilde{\mfL}} }\lambda t } U_\tau ^\dagger  \ad_{\hat{H}_\tau}(V)  U_\tau   e^{-i \br{ H_{0,\tilde{\mfL}} + \hat{V}_{\tau,\tilde{\mfL}} }\lambda t } }  \notag \\
&\le 6\gamma \norm{V} g'_{\tau} |\mfL|  \br{4|t|\tau\norm{V} +1 } \Bigl \{ 2^{D+3}\br{C_{1,\tau}^{D}+r^D}  g_{\tau,r}Q(\tau,r/2)  
 +\tilde{J}_0   |t| r^{D}\brr{ 2e^{-\mu r/8}+  \min\br{ C \gamma |\mfL| r^{D}e^{v |t|-\mu r/4},1} }   \Bigr\} \notag \\
&\quad + 4|t| g'_{\tau} \norm{V}^2 \br{\tau \Er(0,r) + \frac{2\Er (\tau,r)}{\mathcal{K}_0} },
\end{align} 
where $0\le \lambda\le1$, and the quasi-locality of $\ad_{\hat{H}_\tau}(V)$ is given by Proposition~\ref{prop:quasi-locality of_ad__H__tau_V}. 
The integral of the above upper bounds reads 
\begin{align}
\label{ineq:integral_lambda_t_g_beta_2}
&\int_{|t|\le \delta t} |t g_\beta(t)| \int_0^1 \norm{  e^{i \br{ H_0 + \hat{V}_\tau }\lambda t }  U_\tau^\dagger  \ad_{\hat{H}_\tau}(V)  U_\tau  e^{-i \br{ H_0 + \hat{V}_\tau } \lambda t } 
-  e^{i \br{ H_{0,\tilde{\mfL}} + \hat{V}_{\tau,\tilde{\mfL}} } \lambda t } U_\tau ^\dagger  \ad_{\hat{H}_\tau}(V)  U_\tau   e^{-i \br{ H_{0,\tilde{\mfL}} + \hat{V}_{\tau,\tilde{\mfL}} }\lambda t } }d\lambda  dt  \notag \\
&\le \frac{6\beta\gamma \norm{V} g'_{\tau} |\mfL| \delta t}{\pi}  \br{4\norm{V} \tau \delta t +1 } \Bigl[ 2^{D+3}\br{C_{1,\tau}^{D}+r^D}  g_{\tau,r}Q(\tau,r/2)  
 +\tilde{J}_0  r^{D} \delta t \br{ 2e^{-\mu r/8}+  C \gamma |\mfL| r^{D}e^{v\delta t-\mu r/4}}   \Bigr] \notag \\
&\quad + \frac{4\beta \delta t^2}{\pi} g'_{\tau} \norm{V}^2\br{\tau \Er(0,r) +\frac{2\Er (\tau,r)}{\mathcal{K}_0} } \notag \\
&\le \zeta''_{\tau, r,\delta t} \delta t \norm{V} Q(\tau,r/2)  +\frac{4\beta \delta t^2}{\pi} g'_{\tau} \norm{V}^2\br{\tau \Er(0,r) +\frac{2\Er (\tau,r)}{\mathcal{K}_0}  }  ,
\end{align}
where, in the last inequality, the parameter $\zeta_{\tau, r,\delta t}$ is defined as 
\begin{align}
\zeta''_{\tau, r,\delta t}= \frac{6\beta\gamma g'_{\tau} |\mfL|}{\pi}  \br{4\norm{V} \tau \delta t +1 } \Bigl[ 2^{D+3}\br{C_{1,\tau}^{D}+r^D}  g_{\tau,r}
 +\tilde{J}_0  r^{D} \delta t \br{ 2 +  C \gamma |\mfL| r^{D}e^{v\delta t-\mu r/8}}   \Bigr] ,
\end{align}
and we use $e^{-\mu r/8} \le e^{-\kappa_\beta r/2} \le Q(\tau,r/2)$ and $e^{-\mu r/4}\le e^{-\mu r/8} e^{-\kappa_\beta r/2}\le e^{-\mu r/8} Q(\tau,r/2)$. 
Note that by using the inequality~\eqref{upp_t_tau_r_g'}, i.e., 
\begin{align}
&  g_{\tau,r}\le \Theta(\norm{V},|\mfL|^3, r^{2D}, \beta^{10D}, \kappa_0^{5D}, \kappa_1^{10D},\tau^{10D}) , \notag \\
& g'_{\tau} \le  \Theta(\norm{V},|\mfL|^3, \beta^{10D}, \kappa_0^{5D},  \kappa_1^{10D},\tau^{10D}),
\end{align}
we have 
\begin{align}
\zeta''_{\tau, r,\delta t}\le \Theta(\norm{V}^3,|\mfL|^7, r^{3D}, \beta^{22D+1}, \kappa_0^{11D}, \kappa_1^{22D},\tau^{22D+1}) .
\end{align}
Therefore, combining the inequalities~\eqref{ineq:integral_lambda_t_g_beta_1} and \eqref{ineq:integral_lambda_t_g_beta_2} with \eqref{t_g_beta_t_integrate_ad_H_tau}, we reach the desired inequality of 
\begin{align}
\label{t_g_beta_t_integrate_ad_H_tau_final}
&\int_{|t|\le \delta t} |t g_\beta(t)| \int_0^1 \norm{[\ad_{\hat{H}_\tau}(V)](\hat{H}_\tau,\lambda t)-[\ad_{\hat{H}_{\tau,\tilde{\mfL}}}(V_{\tilde{\mfL}})](\hat{H}_{\tau,\tilde{\mfL}}, \lambda t)} d\lambda  dt \notag \\
&\le \frac{\beta \delta t \norm{V}}{\pi} \Er(\tau,r) \br{ g''_{\tau,r} +3g'_{\tau} } + \zeta''_{\tau, r,\delta t} \delta t \norm{V} Q(\tau,r/2)  +\frac{4\beta \delta t^2}{\pi} g'_{\tau} \norm{V}^2\br{\tau \Er(0,r) +\frac{2\Er (\tau,r)}{\mathcal{K}_0}  } .
\end{align}
Finally, by applying the inequalities~\eqref{int_g_beta_t_V_re_again_final_form} and~\eqref{t_g_beta_t_integrate_ad_H_tau_final} to~\eqref{int_g_beta_t_V_re_again}, we obtain
\begin{align}
&\norm{\frac{2}{\beta} \int_{-\infty}^\infty g_\beta(t) \brr{V(\hat{H}_\tau, t)-  V_{\tilde{\mfL}}(\hat{H}_{\tau,\tilde{\mfL}}, t)} dt } \notag \\
&\le \frac{4 \norm{V}}{\pi}\log\br{\frac{\beta}{2\pi \delta t}}   \br{\tau \Er(0,r)+ 4\Er(\tau,r)}  
+ \frac{8\beta \norm{V}^2}{\pi^2}  \br{\tau \Er(0,r) +\frac{2\Er (\tau,r) }{\mathcal{K}_0}} +\frac{2}{\beta}\br{\zeta_{\tau,r,\delta t}+\zeta'_{\tau,r}}Q(\tau,r/2) \notag \\
&+\frac{2\delta t \norm{V}}{\pi} \Er(\tau,r) \br{ g''_{\tau,r} +3g'_{\tau} } + \frac{2}{\beta}\zeta''_{\tau, r,\delta t} \delta t \norm{V} Q(\tau,r/2)  +\frac{8 \delta t^2}{\pi} g'_{\tau} \norm{V}^2\br{\tau \Er(0,r) +\frac{2\Er (\tau,r)}{\mathcal{K}_0}   }  \notag \\
&= \frac{2\norm{V}}{\pi} \brr{8\log\br{\frac{\beta}{2\pi \delta t}}+ \frac{8\beta \norm{V}}{\pi \mathcal{K}_0}+ \delta t \br{ g''_{\tau,r} +3g'_{\tau} } +\frac{8\delta t^2 g'_{\tau} \norm{V}}{ \mathcal{K}_0} }\Er (\tau,r) \notag \\
&\quad + \frac{4\tau \norm{V}}{\pi} \brr{\log\br{\frac{\beta}{2\pi \delta t}} +\frac{2\beta \norm{V}}{\pi} + 2\delta t^2 g'_{\tau} \norm{V}} \Er (0,r)
+ \frac{2}{\beta}\br{\zeta_{\tau,r,\delta t}+\zeta'_{\tau,r}+\zeta''_{\tau, r,\delta t} \delta t \norm{V} }Q(\tau,r/2)  .
\label{upper_bound_for_inq:int_g_beta_t_V}
\end{align}
We thus obtain the upper bound of $\norm{\mC_{\tau} -  \mC_{\tau,\tilde{\mfL}}} $, which is equal to the LHS of the above inequality as in Eq.~\eqref{upp_dif_mC__mC_tau_tilde}.

\subsubsection{Completing the proof of Theorem~\ref{thm:Refined locality estimation for effective Hamiltonian}}
For the proof, we choose $\delta t$ such that 
\begin{align}
\delta t \le \frac{1}{\norm{V} }\min\br{ \frac{1}{g''_{\tau,r} +3g'_{\tau}} ,\frac{1}{\zeta''_{\tau, r,\delta t}} },
\end{align}
which reduces the inequality~\eqref{upper_bound_for_inq:int_g_beta_t_V} to 
\begin{align}
&\norm{\frac{2}{\beta} \int_{-\infty}^\infty g_\beta(t) \brr{V(\hat{H}_\tau, t)-  V_{\tilde{\mfL}}(\hat{H}_{\tau,\tilde{\mfL}}, t)} dt } \notag \\
&\le
\Theta(\norm{V}) \brr{ \log (\beta/\delta t) +  \frac{\beta \norm{V}}{\mathcal{K}_0}} \Er(\tau,r)+ \Theta(\norm{V}) \brr{ \log (\beta/\delta t) +  \beta \norm{V} } \Er(0,r)\notag \\
&+ \Theta\br{\norm{V}^{2D+2}, |\mfL|^2, \beta^{2D^2+2D+1}, \tau^{2D+1}, r^{2D+2}, \log (\delta t^{-1})}  Q(\tau,r/2)  .
\end{align}
Here, $\log(1/\delta t)$ is given by
\begin{align}
\log(\beta/\delta t) \le  \Theta(1)\log(\beta \norm{V} \cdot |\mfL|\kappa_0 \kappa_1)  +  \Theta(1) \log(r+\tau+e) \le 
 \Theta(1)\log(\beta \norm{V} \cdot |\mfL|)  +  \Theta(1) \log(r+\tau+e).
\end{align}
We note that $\kappa_0$ and $\kappa_1$ have been given by Eq.~\eqref{kappa_0_kappa_1_definition}.  
Then, in order to satisfy the inequality~\eqref{goal_inequality_eff_Ham_locality}, we have to choose 
$\mathcal{K}_0$ and $\mathcal{K}_1$ so that the following condition is satisfied:
\begin{align}
&\Theta(\norm{V}) \br{\log(\beta \norm{V}\cdot |\mfL|)  +\log(r+\tau+e)  +  \frac{\beta \norm{V}}{\mathcal{K}_0}} \Er(\tau,r) \notag \\
&+ \Theta(\norm{V}) \Bigl [\log(\beta \norm{V}\cdot |\mfL|)  +   \log(r+\tau+e)   +  \beta \norm{V} \Bigr] \Er(0,r) \notag \\
&+\Theta\br{\norm{V}^{2D+3}, |\mfL|^3, \beta^{2D^2+2D+2}, \tau^{2D+2}, r^{2D+3}}   Q(\tau,r/2)  \notag\\
&\le \brr{  \mathcal{K}_0+ \mathcal{K}_1 \log(r+\tau+e)}\Er (\tau,r) +\mathcal{K}_2 \brr{Q(\tau,r/2)+ \Er (0,r)}  .
\end{align}
Thus, the following choices satisfy the above condition. 
\begin{align}
\mathcal{K}_0 &= \max\Bigl[ \kappa_0, \Theta(\norm{V}) \log (\beta \norm{V}\cdot |\mfL|)  \Bigr]  , \quad 
\mathcal{K}_1 = \max\Bigl[ \kappa_1, \Theta(\norm{V})  \Bigr]    ,\notag \\
\mathcal{K}_2&=  \max\brr{ \Theta(\norm{V}) \br{\log(\beta \norm{V}\cdot |\mfL|)  +   \log(r+\tau+e)   +  \beta \norm{V} } ,\Theta\br{\norm{V}^{2D+3}, |\mfL|^3, \beta^{2D^2+2D+2}, \tau^{2D+2}, r^{2D+3}}  } \notag \\
&= \Theta\br{\norm{V}^{2D+3}, |\mfL|^3, \beta^{2D^2+2D+2}, \tau^{2D+2}, r^{2D+3}} ,
\label{final_condition_for_mathcal_K_012}
\end{align}
where we take $\mathcal{K}_0\ge \kappa_0$ and $\mathcal{K}_1\ge \kappa_1$  into account. 
By making the $\kappa_0$ and $\kappa_1$ sufficiently large, we can make $\mathcal{K}_0=\kappa_0$ and $\mathcal{K}_1=\kappa_1$.
We thus prove the main inequality~\eqref{main_inq:thm:Refined locality estimation for effective Hamiltonian}. 
This completes the proof of Theorem~\ref{thm:Refined locality estimation for effective Hamiltonian}. $\square$

\subsection{Proof of Proposition~\ref{prop:main_est_time_evo_comp}} \label{sec:Proof of Proposition_prop:main_est_time_evo_comp}

The purpose here is to estimate the upper bound of the norm difference of
\begin{align}
\norm{  e^{i \br{ H_0 + \hat{V}_\tau } t }  U_\tau^\dagger  O U_\tau  e^{-i \br{ H_0 + \hat{V}_\tau } t } 
-  e^{i \br{ H_{0,\tilde{\mfL}} + \hat{V}_{\tau,\tilde{\mfL}} } t } U_\tau ^\dagger  O U_\tau   e^{-i \br{ H_{0,\tilde{\mfL}} + \hat{V}_{\tau,\tilde{\mfL}} } t } } 
\end{align}
for an arbitrary operator $O$. 
To estimate it, we first prove the following lemma:
\begin{lemma} \label{second_term_estimation_lemma_unitary}
Let us define the operator function $\varepsilon_{\tau,t}(O_1)$ as follows:
\begin{align}
\varepsilon_{\tau,t}(O_1):= \norm{e^{i  H_0  t }  U_\tau^\dagger O_1 U_\tau  e^{-i  H_0  t } - 
 e^{i H_{0,\tilde{\mfL}}  t }  U_\tau^\dagger O_1 U_\tau   e^{-i H_{0,\tilde{\mfL}}  t } } ,
 \label{varepsilon_definition_tau_t}
\end{align}
where $O_1$ is an arbitrary operator.
We then obtain 
\begin{align}
\label{main:ineq:second_term_estimation_lemma_unitary}
&\norm{  e^{i \br{ H_0 + \hat{V}_\tau } t }  U_\tau^\dagger  O U_\tau  e^{-i \br{ H_0 + \hat{V}_\tau } t } 
-  e^{i \br{ H_{0,\tilde{\mfL}} + \hat{V}_{\tau,\tilde{\mfL}} } t } U_\tau ^\dagger  O U_\tau   e^{-i \br{ H_{0,\tilde{\mfL}} + \hat{V}_{\tau,\tilde{\mfL}} } t } } \notag \\
\le&\varepsilon_{\tau,t}(O)+ 4\norm{O}\br{ |t| \tau \norm{V} \Er (0,r) +2t\norm{V}  \int_0^\tau  \Er (\tau_1,r)   d\tau_1 +\int_0^\tau \int_0^{|t|}  \varepsilon_{\tau_1,x}(V) dx}  \notag \\
\le& \varepsilon_{\tau,t}(O)+ 4\norm{O} \int_0^\tau \int_0^{|t|}  \varepsilon_{\tau_1,x}(V) dx + 
 4|t|\cdot \norm{O} \cdot \norm{V}  \br{  \tau  \Er (0,r) +2 \int_0^\tau  \Er (\tau_1,r)   d\tau_1} .
\end{align}
\end{lemma}

\textit{Proof of Lemma~\ref{second_term_estimation_lemma_unitary}.}
We first note that the following equations hold 
\begin{align}
\label{interaction_picture_H_0_V_tau_tilde_L}
e^{i \br{ H_0 + \hat{V}_\tau } t }= \mathcal{T} e^{i\int_0^t  \hat{V}_\tau (H_0,x)dx} e^{i H_0  t }, \quad 
e^{i \br{ H_{0,\tilde{\mfL}} + \hat{V}_{\tau,\tilde{\mfL}} } t }= \mathcal{T} e^{i\int_0^t  \hat{V}_{\tau,\tilde{\mfL}} (H_{0,\tilde{\mfL}},x)dx} e^{i H_{0,\tilde{\mfL}}  t }.
\end{align}
Moreover, using the above equation, we consider the following decomposition of $e^{i \br{ H_0 + \hat{V}_\tau } t } $: 
\begin{align}
\label{Eq:e^H_0+V_tau_deco}
e^{i \br{ H_0 + \hat{V}_\tau } t } 
&=\br{ \mathcal{T} e^{i\int_0^t  \hat{V}_\tau (H_0,x)dx} -e^{i\int_0^t  \hat{V}_{\tau} (H_{0,\tilde{\mfL}},x)dx} +
e^{i\int_0^t  \hat{V}_{\tau} (H_{0,\tilde{\mfL}},x)dx} -   \mathcal{T} e^{i\int_0^t  \hat{V}_{\tau,\tilde{\mfL}} (H_{0,\tilde{\mfL}},x)dx}  } e^{i  H_0  t } \notag \\
&\quad + \mathcal{T} e^{i\int_0^t  \hat{V}_{\tau,\tilde{\mfL}} (H_{0,\tilde{\mfL}},x)dx}   e^{i  H_0  t } \notag \\
&=: \Delta U_{\tau} +e^{i \br{ H_{0,\tilde{\mfL}} + \hat{V}_{\tau,\tilde{\mfL}} } t }  e^{-i H_{0,\tilde{\mfL}}  t }  e^{i  H_0  t } .
\end{align}
From the above equation, we obtain 
\begin{align}
\label{upper:bound_norm_V_delta_U_tau}
&\norm{e^{i \br{ H_0 + \hat{V}_\tau } t }  U_\tau^\dagger  O U_\tau  e^{-i \br{ H_0 + \hat{V}_\tau } t } 
- e^{i \br{ H_{0,\tilde{\mfL}} + \hat{V}_{\tau,\tilde{\mfL}} } t }  e^{-i H_{0,\tilde{\mfL}}  t }  e^{i  H_0  t }  U_\tau^\dagger OU_\tau  e^{-i  H_0  t } e^{i H_{0,\tilde{\mfL}} t}
e^{-i \br{ H_{0,\tilde{\mfL}} + \hat{V}_{\tau,\tilde{\mfL}}}t}} \notag \\
&\le 2 \norm{O} \cdot \norm{\Delta U_{\tau}}  .
\end{align}

Then, using the definition of~\eqref{varepsilon_definition_tau_t}, we have 
\begin{align}
\label{upper:bound_time_H_0_tildeL_H_0diff}
&\norm{  e^{i \br{ H_{0,\tilde{\mfL}} + \hat{V}_{\tau,\tilde{\mfL}} } t }  e^{-i H_{0,\tilde{\mfL}}  t }  e^{i  H_0  t }  U_\tau^\dagger O U_\tau  e^{-i  H_0  t } e^{i H_{0,\tilde{\mfL}} t}
e^{-i \br{ H_{0,\tilde{\mfL}} + \hat{V}_{\tau,\tilde{\mfL}}}t}-
e^{i \br{ H_{0,\tilde{\mfL}} + \hat{V}_{\tau,\tilde{\mfL}} } t }  U_\tau^\dagger OU_\tau  
e^{-i \br{ H_{0,\tilde{\mfL}} + \hat{V}_{\tau,\tilde{\mfL}}}t}} \notag \\
&=\norm{  e^{i \br{ H_{0,\tilde{\mfL}} + \hat{V}_{\tau,\tilde{\mfL}} } t }  e^{-i H_{0,\tilde{\mfL}}  t }  
\br{ e^{i  H_0  t }  U_\tau^\dagger O U_\tau  e^{-i  H_0  t } - e^{i  H_{0,\tilde{\mfL}}  t }  U_\tau^\dagger OU_\tau  e^{-i  H_{0,\tilde{\mfL}} t } } e^{i H_{0,\tilde{\mfL}} t}
e^{-i \br{ H_{0,\tilde{\mfL}} + \hat{V}_{\tau,\tilde{\mfL}}}t}} \notag \\
&=\norm{e^{i  H_0  t }  U_\tau^\dagger O U_\tau  e^{-i  H_0  t } - e^{i  H_{0,\tilde{\mfL}}  t }  U_\tau^\dagger OU_\tau  e^{-i  H_{0,\tilde{\mfL}} t }} =  \varepsilon_{\tau,t}(O). 
\end{align}
By combining the two upper bounds of~\eqref{upper:bound_norm_V_delta_U_tau} and \eqref{upper:bound_time_H_0_tildeL_H_0diff}, we obtain 
\begin{align}
\label{upper:bound_norm_two_time_evo_desiredform}
\norm{e^{i \br{ H_0 + \hat{V}_\tau } t }  U_\tau^\dagger  V U_\tau  e^{-i \br{ H_0 + \hat{V}_\tau } t } 
-e^{i \br{ H_{0,\tilde{\mfL}} + \hat{V}_{\tau,\tilde{\mfL}} } t }  U_\tau^\dagger VU_\tau  
e^{-i \br{ H_{0,\tilde{\mfL}} + \hat{V}_{\tau,\tilde{\mfL}}}t}}&\le 2 \norm{O} \cdot \norm{\Delta U_{\tau}}  +\varepsilon_{\tau,t}(O) .
\end{align}

Finally, we estimate the upper bound of the norm $\norm{\Delta U_{\tau}}$. We begin with the inequality of 
\begin{align}
\label{upp_norm_Delta_U_tau}
\norm{ \Delta U_{\tau}} \le  \int_0^{|t|}  \norm{ \hat{V}_{\tau} (H_0,x) - \hat{V}_{\tau} (H_{0,\tilde{\mfL}},x)}dx+|t|\cdot \norm{ \hat{V}_\tau - \hat{V}_{\tau,\tilde{\mfL}} } ,
\end{align}
where we use the inequality~\eqref{upp_norm_Delta_U_tau_derr}.
For the estimation of the RHS of~\eqref{upp_norm_Delta_U_tau},   we first upper-bound the norm $\norm{ \hat{V}_\tau - \hat{V}_{\tau,\tilde{\mfL}} }$ as  
\begin{align}
&\norm{ \hat{V}_\tau - \hat{V}_{\tau,\tilde{\mfL}} } = 2\int_0^\tau \norm{ U_{\tau_1}^\dagger  V   U_{\tau_1}  - U_{\tau_1,\tilde{\mfL}}^\dagger  V_{\tilde{\mfL}}   U_{\tau_1,\tilde{\mfL}}} d\tau_1  \notag \\
&\le 2\tau \norm{V-V_{\tilde{\mfL}}}  + 2\int_0^\tau \br{\norm{ U_{\tau_1}^\dagger  V   U_{\tau_1} - U_{\tau_1,\tilde{\mfL}}^\dagger  V   U_{\tau_1}} +\norm{U_{\tau_1,\tilde{\mfL}}^\dagger  V   U_{\tau_1}    - U_{\tau_1,\tilde{\mfL}}^\dagger  V   U_{\tau_1,\tilde{\mfL}}}} d\tau_1 \notag \\
&\le  2\tau \norm{V-V_{\tilde{\mfL}}}  +4\norm{V}  \int_0^\tau \norm{  U_{\tau_1} -U_{\tau_1,\tilde{\mfL}}}  d\tau_1 \le 2\tau\norm{V} \Er (0,r)  +4\norm{V}  \int_0^\tau  \Er (\tau_1,r)  d\tau_1,
\end{align}
where the forms of $\hat{V}_\tau$ and $\hat{V}_{\tau,\tilde{\mfL}}$ have been defined as in Eq.~\eqref{U_tau_V_tau_mu_tau_def}.
In the same way, the error between time evolutions of $\hat{V}_{\tau} (H_0,x)$ and $\hat{V}_{\tau} (H_{0,\tilde{\mfL}},x)$ is upper-bounded as follows:
\begin{align}
& \int_0^{|t|}  \norm{ \hat{V}_{\tau} (H_0,x) - \hat{V}_{\tau} (H_{0,\tilde{\mfL}},x)}dx   \notag\\
&\le 2\int_0^\tau \int_0^{|t|}  \norm{e^{i  H_0  x } U_{\tau_1}^\dagger  V   U_{\tau_1}e^{-i  H_0  x } - e^{i H_{0,\tilde{\mfL}} x }  U_{\tau_1}^\dagger  V   U_{\tau_1}e^{-i H_{0,\tilde{\mfL}} x } }dx  =
2\int_0^\tau \int_0^{|t|}  \varepsilon_{\tau_1,x}(V)dx  ,
\end{align}
where we use the definition~\eqref{varepsilon_definition_tau_t} for $\varepsilon_{\tau_1,x}(V)$. 
By applying the above inequalities to~\eqref{upper:bound_norm_two_time_evo_desiredform}, we obtain the main inequality~\eqref{main:ineq:second_term_estimation_lemma_unitary}. $\square$

{~}

\hrulefill{\bf [ End of Proof of Lemma~\ref{second_term_estimation_lemma_unitary}]}

{~}

\begin{figure}[tt]
\centering
\includegraphics[clip, scale=0.3]{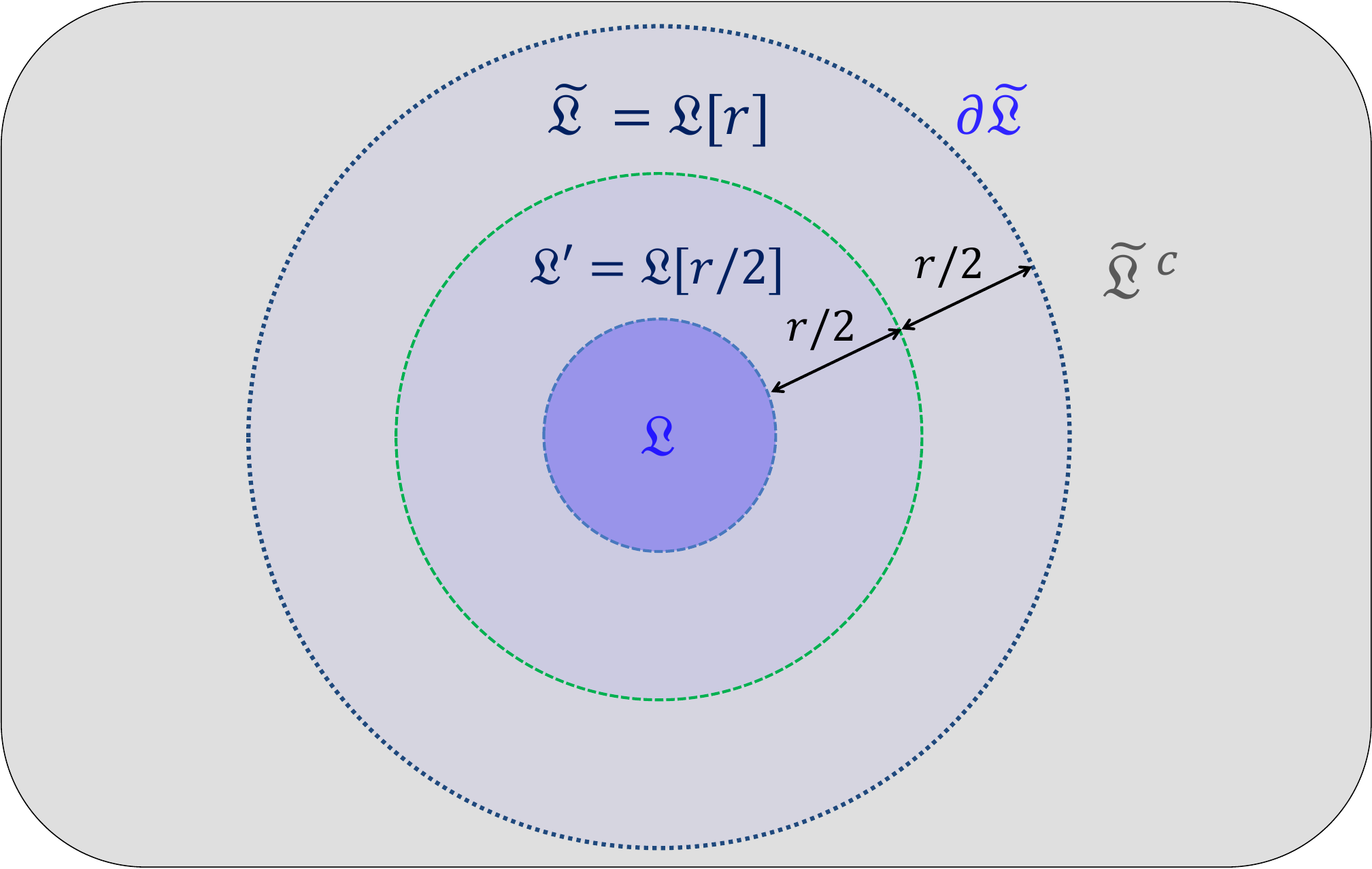}
\caption{Schematic picture of the definitions of $\mfL$, $\tilde{\mfL}$, $\mfL'$ and $\partial \tilde{\mfL}$. 
}
\label{fig_approx_eff_Hamiltonian}
\end{figure}

Then, we have to estimate the function $\varepsilon_{\tau,t}(O)$ in Eq.~\eqref{varepsilon_definition_tau_t}. 
For this purpose, we prove the following statement:
\begin{lemma} \label{lemm:varepsilon:upper bound}
The parameter $\varepsilon_{\tau,t}(O)$ is upper-bounded by 
\begin{align}
\varepsilon_{\tau,t}(O)\le 8\norm{O} \cdot \norm{U_\tau-\tilde{\tr}_{\tilde{\mfL}^{'\co}}(U_\tau)}  + 4\norm{O -\tilde{\tr}_{\tilde{\mfL}^{'\co}}(O)} +
6\norm{O} \cdot \norm{U_{\partial \tilde{\mfL},t}- \tilde{\tr}_{\tilde{\mfL}'}(U_{\partial \tilde{\mfL},t})}  
\label{main_eq:lemm:varepsilon:upper bound}
\end{align}
with
\begin{align}
U_{\partial \tilde{\mfL},t}:= e^{-iH_0t} e^{i (H_{0,\tilde{\mfL}}+H_{0,\tilde{\mfL}^\co})t}=
e^{-iH_0t} e^{i H_{0,\tilde{\mfL}}t}  e^{i H_{0,\tilde{\mfL}^\co}t} , \label{def:U_partial_tilde_L_,t}
\end{align}
where we define the subset $\tilde{\mfL}'$ as (see also Fig.~\ref{fig_approx_eff_Hamiltonian})
\begin{align}
\label{tilde_L'_definition}
\tilde{\mfL}':= \mfL[r/2] . 
\end{align}
\end{lemma}

\textit{Proof of Lemma~\ref{lemm:varepsilon:upper bound}.}
We first consider an arbitrary operator $O'$ and decompose the time evolved operator $O(H_0,t)$ using the definition~\eqref{def:U_partial_tilde_L_,t} as  
\begin{align}
O' (H_0,t) 
 &= e^{iH_{0,\tilde{\mfL}}t} e^{iH_{0,\tilde{\mfL}^\co}t}  U_{\partial \tilde{\mfL},t}^\dagger O' U_{\partial \tilde{\mfL},t} e^{-iH_{0,\tilde{\mfL}^\co}t}  e^{-iH_{0,\tilde{\mfL}}t},
 \label{def:U_partial_tilde_L_,t__re}
\end{align}
which yields
\begin{align}
&\norm{ O' (H_0,t) -  O' (H_{0,\tilde{\mfL}},t) }  \notag \\
& = \norm{e^{iH_{0,\tilde{\mfL}}t} e^{iH_{0,\tilde{\mfL}^\co}t} U_{\partial \tilde{\mfL},t}^\dagger \brr{ O', U_{\partial \tilde{\mfL},t}} e^{-iH_{0,\tilde{\mfL}^\co}t}  e^{-iH_{0,\tilde{\mfL}}t} 
 + e^{iH_{0,\tilde{\mfL}}t} e^{iH_{0,\tilde{\mfL}^\co}t}  O' e^{-iH_{0,\tilde{\mfL}^\co}t}  e^{-iH_{0,\tilde{\mfL}}t} - e^{iH_{0,\tilde{\mfL}}t}O' e^{-iH_{0,\tilde{\mfL}}t}} \notag \\
&\le  \norm{\brr{ O', U_{\partial \tilde{\mfL},t}}}   
 + \norm{ e^{iH_{0,\tilde{\mfL}}t} e^{iH_{0,\tilde{\mfL}^\co}t}\brr{O', e^{-iH_{0,\tilde{\mfL}^\co}t}}  e^{-iH_{0,\tilde{\mfL}}t}  } 
 \le  \norm{\brr{ O', U_{\partial \tilde{\mfL},t}}} +  \norm{\brr{O',e^{-iH_{0,\tilde{\mfL}^\co}t}}} .
 \end{align}
By applying the above inequality to $O'=U_\tau^\dagger OU_\tau $, we upper-bound 
$\varepsilon_{\tau,t}(O)$ as 
\begin{align}
\varepsilon_{\tau,t}(O)&= \norm{\brr{ U_\tau^\dagger O U_\tau}(H_0,t)  - 
\brr{ U_\tau^\dagger O U_\tau}(H_{0,\tilde{\mfL}},t)}  \notag \\
&\le \norm{\brr{  U_\tau^\dagger OU_\tau, U_{\partial \tilde{\mfL},t}}} +  \norm{\brr{ U_\tau^\dagger OU_\tau,e^{-iH_{0,\tilde{\mfL}^\co}t}}} \notag \\
&\le 2\norm{O} \br{\norm{\brr{ U_\tau, U_{\partial \tilde{\mfL},t}}} +  \norm{\brr{U_\tau,e^{-iH_{0,\tilde{\mfL}^\co}t}}} } 
+ 
\norm{\brr{O, U_{\partial \tilde{\mfL},t}}} +  \norm{\brr{O,e^{-iH_{0,\tilde{\mfL}^{\co}}t}}} \notag \\
&\le 2\norm{O} \cdot \norm{\brr{ U_\tau, U_{\partial \tilde{\mfL},t}}}  +4\norm{O} \cdot \norm{U_\tau- \tilde{\tr}_{\tilde{\mfL}^{\co}}(U_\tau)} 
+ 
\norm{\brr{O, U_{\partial \tilde{\mfL},t}}} + 2\norm{O- \tilde{\tr}_{\tilde{\mfL}^{\co}}(O)}  ,
\label{upper_bound//varepsilon}
\end{align}
where we use $\brr{\tilde{\tr}_{\tilde{\mfL}^{\co}}(U_\tau),e^{-iH_{0,\tilde{\mfL}^\co}t}}=\brr{\tilde{\tr}_{\tilde{\mfL}^{\co}}(O),e^{-iH_{0,\tilde{\mfL}^\co}t}}=0$
since $\tilde{\tr}_{\tilde{\mfL}^{\co}}(U_\tau)$ and $\tilde{\tr}_{\tilde{\mfL}^{\co}}(O)$ are supported on the sebset $\tilde{\mfL}$ ($\tilde{\mfL} \cap \tilde{\mfL}^\co=\emptyset$ as in Fig.~\ref{fig_approx_eff_Hamiltonian}).
To estimate the norm of $\norm{\brr{ U_\tau, U_{\partial \tilde{\mfL},t}}}$, we approximate $U_\tau$ and $U_{\partial \tilde{\mfL},t}$ 
onto the subsets $\tilde{\mfL}'$ and $\tilde{\mfL}^{'\co}$, respectively.
We then obtain 
\begin{align}
\norm{\brr{ U_\tau, U_{\partial \tilde{\mfL},t}}} 
&\le 
\norm{\brr{ U_\tau-\tilde{\tr}_{\tilde{\mfL}^{'\co}}(U_\tau), U_{\partial \tilde{\mfL},t}}} 
+\norm{\brr{ \tilde{\tr}_{\tilde{\mfL}^{'\co}}(U_\tau), U_{\partial \tilde{\mfL},t}}} \notag \\
&\le 2 \norm{U_\tau-\tilde{\tr}_{\tilde{\mfL}^{'\co}}(U_\tau)} 
+\norm{\brr{ \tilde{\tr}_{\tilde{\mfL}^{'\co}}(U_\tau), U_{\partial \tilde{\mfL},t}}} \notag \\
&\le 2 \norm{U_\tau-\tilde{\tr}_{\tilde{\mfL}^{'\co}}(U_\tau)} 
+\norm{\brr{ \tilde{\tr}_{\tilde{\mfL}^{'\co}}(U_\tau), U_{\partial \tilde{\mfL},t}- \tilde{\tr}_{\tilde{\mfL}'}(U_{\partial \tilde{\mfL},t})}} \notag \\
&\le 2 \norm{U_\tau-\tilde{\tr}_{\tilde{\mfL}^{'\co}}(U_\tau)} 
+2 \norm{U_{\partial \tilde{\mfL},t}- \tilde{\tr}_{\tilde{\mfL}'}(U_{\partial \tilde{\mfL},t})} ,
\label{upper_bound//varepsilon_0}
\end{align}
where in the last inequality we use $\tilde{\tr}_{\tilde{\mfL}^{'\co}}(O_1) \le \norm{O_1}$ for an arbitrary $O_1$.
We can derive a similar inequality for $\norm{\brr{O, U_{\partial \tilde{\mfL},t}}} $.
By applying the upper bound~\eqref{upper_bound//varepsilon_0} to the inequality~\eqref{upper_bound//varepsilon}, we prove the main inequality~\eqref{main_eq:lemm:varepsilon:upper bound} as follows: 
\begin{align}
\varepsilon_{\tau,t}
&\le  4\norm{O} \cdot \norm{U_\tau-\tilde{\tr}_{\tilde{\mfL}^{'\co}}(U_\tau)} +
4\norm{O} \cdot \norm{U_{\partial \tilde{\mfL},t}- \tilde{\tr}_{\tilde{\mfL}'}(U_{\partial \tilde{\mfL},t})}   +4\norm{O} \cdot \norm{U_\tau- \tilde{\tr}_{\tilde{\mfL}^{\co}}(U_\tau)}  \notag \\
&\quad 
+ 2\norm{O- \tilde{\tr}_{\tilde{\mfL}^{'\co}}(O)}
+2\norm{O} \cdot \norm{U_{\partial \tilde{\mfL},t}- \tilde{\tr}_{\tilde{\mfL}'}(U_{\partial \tilde{\mfL},t})} 
+ 2 \norm{O -\tilde{\tr}_{\tilde{\mfL}^{\co}}(O)} 
  \notag \\
&\le 8\norm{O} \cdot \norm{U_\tau-\tilde{\tr}_{\tilde{\mfL}^{'\co}}(U_\tau)}  + 4\norm{O -\tilde{\tr}_{\tilde{\mfL}^{'\co}}(O)} +
6\norm{O} \cdot \norm{U_{\partial \tilde{\mfL},t}- \tilde{\tr}_{\tilde{\mfL}'}(U_{\partial \tilde{\mfL},t})} .
\end{align}
This completes the proof. $\square$

{~}

\hrulefill{\bf [ End of Proof of Lemma~\ref{lemm:varepsilon:upper bound}]}

{~}

From Lemma~\ref{lemm:varepsilon:upper bound}, we aim to estimate the terms in the RHS of \eqref{main_eq:lemm:varepsilon:upper bound} to derive the upper bound of $\varepsilon_{\tau,t}(O)$.
We first estimate the norm of $\norm{U_\tau-\tilde{\tr}_{\tilde{\mfL}^{'\co}}(U_\tau)} $ and $\norm{O-\tilde{\tr}_{\tilde{\mfL}^{'\co}}(O)}$. 
Using the inequality~\eqref{basic_ansatz_m_general_X_start}, we have 
\begin{align}
\label{ineq_U_tau_tilde_L_U_tau'}
\norm{U_\tau-\tilde{\tr}_{\tilde{\mfL}^{'\co}}(U_\tau)} \le 
  \sum _{i\in \tilde{\mfL}^{'\co}} \sup_{u_i} \norm{ [U_\tau, u_i ]}\le 
  \sum _{i\in \tilde{\mfL}^{'\co}} Q(\dist_{i,\mfL},\tau)
  &\le \sum_{s=r/2+1}^\infty \sum _{i\in \partial \mfL[s]} Q(s,\tau) \notag \\
  &\le  \gamma |\mfL| \sum_{s=r/2+1}^\infty s^{D-1} Q(s,\tau) ,
\end{align}
where we use the fact that $\tilde{\mfL}^{'\co}= \mfL[r/2]^\co$ from $\tilde{\mfL}'=\mfL[r/2]$ as in Eq.~\eqref{tilde_L'_definition}. 
Note that $|\partial \mfL[s]|\le s^{D-1} |\partial \mfL| \le s^{D-1} |\mfL| $ using the $\gamma$ in the inequality~\eqref{def:parameter_gamma}. 
By employing the same analyses as the derivation of~\eqref{estimation_s_ge_r_tilde_q_tau_final}, we obtain  
\begin{align}
\label{estimation_s_ge_r_tilde_q_tau_final_again_re}
 \sum_{s> r/2}  s^{D-1} Q(\tau,s)  \le  3 \cdot 2^D \brr{C_{1,\tau}^{D}+(r/2)^D} Q(\tau,r/2)\le  2^{D+2} \br{C_{1,\tau}^{D}+r^D} Q(\tau,r/2). 
\end{align}
By combining the inequalities~\eqref{ineq_U_tau_tilde_L_U_tau'} and \eqref{estimation_s_ge_r_tilde_q_tau_final_again_re}, the following inequality holds:
\begin{align}
\label{U_tau_err_r_estimation}
\norm{U_\tau-\tilde{\tr}_{\tilde{\mfL}^{'\co}}(U_\tau)} \le 2^{D+2} \gamma  |\mfL| \br{C_{1,\tau}^{D}+r^D} Q(\tau,r/2).
\end{align}
Following the above analyses, we can derive the same inequality for $\norm{O -\tilde{\tr}_{\tilde{\mfL}^{'\co}}(O)}$ using the assumption of~\eqref{assum_quasi_locality_of_O}:
 \begin{align}
\label{U_tau_err_r_estimation2}
\norm{O -\tilde{\tr}_{\tilde{\mfL}^{'\co}}(O)} \le 2^{D+2} \gamma \norm{O}\cdot |\mfL|  \br{\check{C}_{1,\tau}^{D}+r^D} \check{Q}(\tau,r/2), 
\end{align}
where we adopt a similar definition to $\check{C}_{1,\tau}$ in Eq.~\eqref{upper-bound:tilde_Q_tau_r_lemma} for $\check{C}_{\nu,\tau}$, i.e., $\check{C}_{\nu,\tau}:= \frac{\tau+e}{2} + 32\br{\nu^2 D^2 + \check{\kappa}_1^2 \tau^2+\check{\kappa}_\beta \check{\kappa}_0\tau /8}/\kappa_\beta^2$.

We second estimate the norm of $\norm{U_{\partial \tilde{\mfL},t}- \tilde{\tr}_{\tilde{\mfL}'}(U_{\partial \tilde{\mfL},t})}$.
From Eq.~\eqref{def:U_partial_tilde_L_,t}, we have 
\begin{align}
&U_{\partial \tilde{\mfL},t}:= e^{-iH_0t} e^{i (H_{0,\tilde{\mfL}}+H_{0,\tilde{\mfL}^\co})t}=\mathcal{T} e^{-i  \int_0^t \partial h_{\tilde{\mfL}}(H_0,-x) dx},
 \label{def:U_partial_tilde_L_,t_2} \\
&\partial h_{\tilde{\mfL}}= \sum_{Z:Z \cap \tilde{\mfL}\neq \emptyset,\ Z \cap \tilde{\mfL}^\co \neq \emptyset} h_Z,
\end{align}
where the second notation comes from the definition of Eq.~\eqref{def:Ham_surface}. 
From the inequality~\eqref{local_approxiamtion_O_X_X_brr_r}, we can derive 
\begin{align}
\norm{U_{\partial \tilde{\mfL},t}- \tilde{\tr}_{\tilde{\mfL}'}(U_{\partial \tilde{\mfL},t})} 
\le  \sup_{U_{\tilde{\mfL}'}} \norm{ \brr{U_{\partial \tilde{\mfL},t}, U_{\tilde{\mfL}'}} }
\le  \int_0^{|t|} \sup_{U_{\tilde{\mfL}'}} \norm{ \brr{ \partial h_{\tilde{\mfL}}(H_0,-x) , U_{\tilde{\mfL}'}} } dx. 
 \label{def:U_partial_tilde_L_,t_2_3} 
\end{align}

We then decompose $\partial h_{\tilde{\mfL}}$ as $\partial h_{\tilde{\mfL}}^{(1)} + \partial h_{\tilde{\mfL}}^{(2)}$ with 
\begin{align}
\label{def_partial_h_tilde_mfL_1_2}
\partial h_{\tilde{\mfL}}^{(1)} := \sum_{Z:Z \cap \tilde{\mfL}[-r/4]\neq \emptyset,\ Z \cap \tilde{\mfL}^\co \neq \emptyset} h_Z, \quad  \partial h_{\tilde{\mfL}}^{(2)} :=\partial h_{\tilde{\mfL}}-\partial h_{\tilde{\mfL}}^{(1)} ,
\end{align}
where $\partial h_{\tilde{\mfL}}^{(2)}$ is supported on $\tilde{\mfL}[-r/4]^\co=\tilde{\mfL}'[r/4]^\co$ from the definition. 
By applying Lemma~\ref{sum_interaction_terms} to the norms of $\partial h_{\tilde{\mfL}}^{(1)}$ and $\partial h_{\tilde{\mfL}}^{(2)}$, we obtain
\begin{align}
\label{def_partial_h_tilde_mfL_1_2/uuper_boud}
\norm{\partial h_{\tilde{\mfL}}^{(1)}} \le  |\partial (\tilde{\mfL}[-r/4])| \mathcal{J}_{r/4}\le \tilde{J}_0  |\partial \tilde{\mfL}| e^{-\mu r/8} ,
\quad \norm{\partial h_{\tilde{\mfL}}^{(2)}} \le  |\partial \tilde{\mfL}| \mathcal{J}_{0}=\tilde{J}_0  |\partial \tilde{\mfL}|, 
\end{align}
which reduces the inequality~\eqref{def:U_partial_tilde_L_,t_2_3} to 
\begin{align}
\norm{U_{\partial \tilde{\mfL},t}- \tilde{\tr}_{\tilde{\mfL}'}(U_{\partial \tilde{\mfL},t})} 
&\le 2|t| \tilde{J}_0   |\partial \tilde{\mfL}| e^{-\mu r/8}+ \int_0^{|t|} \sup_{U_{\tilde{\mfL}'}} \norm{ \brr{ h^{(2)}_{\partial \tilde{\mfL}}(H_0,-x) , U_{\tilde{\mfL}'}} } dx \notag \\
&\le 2|t| \tilde{J}_0   |\partial \tilde{\mfL}| e^{-\mu r/8}+ 
\int_0^{|t|} \tilde{J}_0  |\partial \tilde{\mfL}| \cdot \min\brr{ C |\partial \tilde{\mfL}'|\br{e^{v |t|}-1} e^{-\mu r/4},1} dx \notag \\
&\le \gamma |\mfL| r^{D} \tilde{J}_0 |t| \brr{ 2e^{-\mu r/8}+  \min\br{ C \gamma |\mfL| r^D e^{v |t|-\mu r/4},1} },
 \label{def:U_partial_tilde_L_,t_2_4} 
\end{align}
where we use $|\partial \tilde{\mfL}'|= |\partial (\mfL[r/2])| \le \gamma |\mfL| r^D$ apply the Lieb--Robinson bound~\eqref{Lieb--Robinson_main_short} in the second inequality.

By combining the inequalities~\eqref{U_tau_err_r_estimation}, \eqref{U_tau_err_r_estimation2} and \eqref{def:U_partial_tilde_L_,t_2_4} with~\eqref{main_eq:lemm:varepsilon:upper bound}
in Lemma~\ref{lemm:varepsilon:upper bound}, we obtain 
\begin{align}
 \varepsilon_{\tau,t}(O) &\le \gamma \norm{O} \cdot |\mfL|  \Bigl \{ 2^{D+5}\br{C_{1,\tau}^{D}+r^D} Q(\tau,r/2) + 2^{D+4}\br{\check{C}_{1,\tau}^{D}+r^D} \check{Q}(\tau,r/2)  
 \notag\\
&\quad +6r^{D}\tilde{J}_0 |t| \brr{ 2e^{-\mu r/8}+  \min\br{ C \gamma |\mfL| r^{D}e^{v |t|-\mu r/4},1} } \Bigr\} \notag \\
&\le 6\gamma \norm{O} \cdot |\mfL|  \Bigl \{ 2^{D+3}\br{\check{C}_{1,\tau}^{D}+r^D} \check{Q}(\tau,r/2)  
+\tilde{J}_0   |t| r^{D}\brr{ 2e^{-\mu r/8}+  \min\br{ C \gamma |\mfL| r^{D}e^{v |t|-\mu r/4},1} }   \Bigr\} ,
\label{main:ineq:second_term_estimation_lemma_unitary_before}
\end{align}
where we use $Q(\tau,r/2)\le \check{Q}(\tau,r/2)$ from the assumption. 

By applying the inequality~\eqref{main:ineq:second_term_estimation_lemma_unitary_before}, 
we upper-bound the terms in~\eqref{main:ineq:second_term_estimation_lemma_unitary} that include $\epsilon_{\tau,t}$ as 
\begin{align}
&\varepsilon_{\tau,t}(O)+ 4\norm{O} \int_0^\tau \int_0^{|t|}  \varepsilon_{\tau_1,x}(V)dx  \notag  \\
\le& 6\gamma \norm{O} \cdot |\mfL|  \br{4|t| \tau \norm{V} +1 } \Bigl \{ 2^{D+3}\br{\check{C}_{1,\tau}^{D}+r^D} \check{Q}(\tau,r/2)  
+\tilde{J}_0   |t| r^{D}\brr{ 2e^{-\mu r/8}+  \min\br{ C \gamma |\mfL| r^{D}e^{v |t|-\mu r/4},1} }   \Bigr\} , \notag 
\end{align}
where, in considering the integrals of $\int_0^\tau \int_0^{|t|}  \varepsilon_{\tau_1,x}(V)dx$,  we use the property that the RHS of~\eqref{main:ineq:second_term_estimation_lemma_unitary_before} monotonically increases with $\tau$ and $|t|$.
By combining the above inequality with~\eqref{main:ineq:second_term_estimation_lemma_unitary} in Lemma~\ref{second_term_estimation_lemma_unitary}, 
we prove the main inequality~\eqref{main:eq:prop:main_est_time_evo_comp}. 
This completes the proof of Proposition~\ref{prop:main_est_time_evo_comp}. $\square$

\subsection{Proof of Proposition~\ref{prop:upper_bound_commutator_H_tau_dif}} \label{sec:Proof of Proposition_prop:upper_bound_commutator_H_tau_dif}

The basic proof relies on the proof of Proposition~\ref{prop:quasi-locality of_ad__H__tau_V}. 
We start from the simple upper bound of
\begin{align}
\label{starting@point_Proof of proposition_com__0}
\norm{ \ad_{\hat{H}_\tau}(V) - \ad_{\hat{H}_{\tau,\tilde{\mfL}}}(V_{\tilde{\mfL}})} 
&\le \norm{\brr{ \hat{H}_\tau,V - V_{\tilde{\mfL}}}}+\norm{\brr{\hat{H}_\tau-\hat{H}_{\tau,\tilde{\mfL}}, V_{\tilde{\mfL}}}}.
\end{align}

First, from the expressions of $\hat{H}_\tau=U_\tau (  H_0 + \hat{V}_{\tau} ) U_\tau^\dagger$ and $\hat{H}_{\tau,\tilde{\mfL}}=U_{\tau,\tilde{\mfL}} (  H_{0,\tilde{\mfL}} + \hat{V}_{\tau,\tilde{\mfL}} ) U_{\tau,\tilde{\mfL}}^\dagger$, we have 
\begin{align}
\label{bar_H_tau_hat_H_tau_tilde_mfl_up_comm}
&\norm{ \brr{ \hat{H}_\tau - \hat{H}_{\tau,\tilde{\mfL}}, V_{\tilde{\mfL}}}}
= \norm{ \brr{ \hat{H}_\tau   -U_{\tau,\tilde{\mfL}} U_\tau^\dagger U_\tau (  H_{0,\tilde{\mfL}} + \hat{V}_{\tau,\tilde{\mfL}} ) U_\tau^\dagger  U_\tau U_{\tau,\tilde{\mfL}}^\dagger   , V_{\tilde{\mfL}}}} 
 \notag \\
&\le 2\norm{V} \cdot \norm{\hat{H}_\tau   -U_{\tau,\tilde{\mfL}} U_\tau^\dagger \hat{H}_\tau  U_\tau U_{\tau,\tilde{\mfL}}^\dagger   } 
+2\norm{V}\cdot  \norm{\hat{V}_{\tau}- \hat{V}_{\tau,\tilde{\mfL}} } 
+ \norm{ \brr{U_{\tau,\tilde{\mfL}} \br{H_0-H_{0,\tilde{\mfL}}}  U_{\tau,\tilde{\mfL}}^\dagger ,V_{\tilde{\mfL}} } }
\notag \\
&\le 2\norm{V} \cdot  \norm{ \brr{ U_\tau   U_{\tau,\tilde{\mfL}}^\dagger ,  \hat{H}_\tau }} 
+12\tau \norm{V}^2 \Er(\tau,r) +\norm{V} \cdot  \norm{\brr{U_{\tau,\tilde{\mfL}}, \widehat{H_{0,\tilde{\mfL}^\co}}}}
 +\norm{\brr{\widehat{H_{0,\tilde{\mfL}^\co}},V_{\tilde{\mfL}}}},
\end{align}
where we use the definition~\eqref{def:Hamiltonian_subset_L} for $\widehat{H_{0,\tilde{\mfL}^\co}}$ and  the similar inequality to~\eqref{upper_bound_e_tau_V/tau_tilde_mfL} in deriving 
\begin{align}
 \norm{\hat{V}_{\tau}-\hat{V}_{\tau,\tilde{\mfL}}}  
 \le 2\tau \norm{V-V_{\tilde{\mfL}}} + 2\int_0^\tau \norm{\brr{V,  U_{\tau_1} U_{\tau_1,\tilde{\mfL}}^\dagger }}  d\tau_1 \le  2\tau\norm{V}
 \brr{ \Er(0,r) + 2\Er(\tau,r)}
 \le 6\tau \norm{V} \Er(\tau,r) .
\end{align}
Here, we use the definition of $\widehat{H_{0,\tilde{\mfL}^\co}}= H_{0,\tilde{\mfL}^\co} + \partial h_{\tilde{\mfL}}$ and define the similar decomposition to Eq.~\eqref{def_partial_h_tilde_mfL_1_2}, i.e., 
\begin{align}
\partial h_{\tilde{\mfL}}= \partial h_{\tilde{\mfL}}^{(1)}+\partial h_{\tilde{\mfL}}^{(2)},
\quad \partial h_{\tilde{\mfL}}^{(1)}:= \sum_{Z:Z \cap \mfL[r/2]\neq \emptyset,\ Z \cap \tilde{\mfL}^\co \neq \emptyset} h_Z, \quad  \partial h_{\tilde{\mfL}}^{(2)} :=\partial h_{\tilde{\mfL}}-\partial h_{\tilde{\mfL}}^{(1)},
\end{align}
where $h_{\tilde{\mfL}}^{(2)}$ is supported on the subset $\mfL[r/2]^\co$. 
Using the inequality~\eqref{sec_ineq:lemma_commutator_generalization} in Lemma~\ref{lemma_commutator_generalization},  we have 
\begin{align}
\label{bar_H_tau_hat_H_tau_tilde_mfl_up_comm_2}
&\norm{V} \cdot  \norm{\brr{U_{\tau,\tilde{\mfL}}, \widehat{H_{0,\tilde{\mfL}^\co}}}}
 +\norm{\brr{\widehat{H_{0,\tilde{\mfL}^\co}},V_{\tilde{\mfL}}}}
 = \norm{V} \cdot  \norm{\brr{U_{\tau,\tilde{\mfL}},\partial h_{\tilde{\mfL}} }}
 +\norm{\brr{\partial h_{\tilde{\mfL}} ,V_{\tilde{\mfL}}}}  \notag\\
&\le \norm{V} \cdot  \norm{\brr{U_{\tau,\tilde{\mfL}},\partial h_{\tilde{\mfL}}^{(1)} }}
 +\norm{\brr{\partial h_{\tilde{\mfL}}^{(1)} ,V_{\tilde{\mfL}}}}  
 +\norm{V} \cdot  \norm{\brr{U_{\tau,\tilde{\mfL}},\partial h_{\tilde{\mfL}}^{(2)} }}
 +\norm{\brr{\partial h_{\tilde{\mfL}}^{(2)} ,V_{\tilde{\mfL}}}}   \notag \\
 &\le 4 \norm{V} \cdot   \norm{\partial h_{\tilde{\mfL}}^{(1)}} + \norm{\partial h_{\tilde{\mfL}}^{(2)}}\sum_{i\in \mfL[r/2]^\co \setminus \tilde{\mfL}}  \sup_{u_i}\br{\norm{V} \cdot \norm{\brr{U_{\tau,\tilde{\mfL}}, u_i}}+\norm{\brr{V_{\tilde{\mfL}},u_i}}}  \notag \\
 &\le 4 \norm{V} \tilde{J}_0  |\partial \tilde{\mfL}| \brr{ e^{-\mu r/4} +  |\tilde{\mfL}|   \frac{Q(\tau,r/2)+ Q(0,r/2)}{4}}
 \le  6\norm{V}  \tilde{J}_0  |\tilde{\mfL}|^2   Q(\tau,r/2)  , 
\end{align}
where we use the similar inequality to~\eqref{def_partial_h_tilde_mfL_1_2/uuper_boud} and the upper bounds of 
$\norm{\brr{U_{\tau,\tilde{\mfL}}, u_i}}\le Q(\tau,\dist_{i,\mfL})$ and $\norm{\brr{V_{\tilde{\mfL}},u_i}} \le \norm{V} Q(0,\dist_{i,\mfL})$.

By applying the inequalities~\eqref{bar_H_tau_hat_H_tau_tilde_mfl_up_comm} and \eqref{bar_H_tau_hat_H_tau_tilde_mfl_up_comm_2}  to~\eqref{starting@point_Proof of proposition_com__0}, 
We obtain the upper bound of
\begin{align}
\label{starting@point_Proof of proposition_com}
\norm{ \ad_{\hat{H}_\tau}(V) - \ad_{\hat{H}_{\tau,\tilde{\mfL}}}(V_{\tilde{\mfL}})} 
&\le \norm{\brr{ \hat{H}_\tau,V - V_{\tilde{\mfL}}}}+2\norm{V} \cdot  \norm{ \brr{ U_\tau   U_{\tau,\tilde{\mfL}}^\dagger ,  \hat{H}_\tau }} 
+6 \norm{V} \brr{ \tau\Er(\tau,r)+  \tilde{J}_0  |\tilde{\mfL}|^2   Q(\tau,r/2) }
\notag \\
&\le \norm{ \brr{ \Delta V  ,\hat{H}_\tau}} + 2\norm{V}\cdot \norm{ \brr{   \Delta U_\tau,  \hat{H}_\tau }} 
+ 6 \norm{V} \br{ 2\tau\norm{V} +  \tilde{J}_0  |\tilde{\mfL}|^2  } \Er(\tau,r) ,
\end{align}
where we define $\Delta V:= V - V_{\tilde{\mfL}}$ and $ \Delta U_\tau :=U_\tau U_{\tau,\tilde{\mfL}}^\dagger  -1$ with $\norm{\Delta U_\tau }\le \norm{ U_\tau - U_{\tau,\tilde{\mfL}}}$. 

For the norm of $\Delta V$ and $\Delta U_\tau$, we obtain 
\begin{align}
\norm{ \brr{ V - V_{\tilde{\mfL}} ,u_i}} 
\le 2 \min \br{\norm{V - V_{\tilde{\mfL}}}, \norm{V} Q(0,\ell)} \le 2\norm{V} \min \brr{\Er(\tau,r),  Q(\tau,\ell)}   ,
\end{align}
and
\begin{align}
\norm{ \brr{ U_\tau - U_{\tau,\tilde{\mfL}},u_i}} 
\le 2\min \br{\norm{U_\tau - U_{\tau,\tilde{\mfL}}}, Q(\tau,\ell)} \le 2\min \brr{\Er(\tau,r),  Q(\tau,\ell)}  ,
\end{align}
where $\dist_{i,\mfL}=\ell$ and we use the ansatz~\eqref{main_inq:thm:Refined locality estimation for effective Hamiltonian}, $Q(0,\ell)\le Q(\tau,\ell)$ for the first upper bound and Subtheorem~\ref{sub_thm:U_tau_u_i_commun} for $U_\tau$ and $U_{\tau,\tilde{\mfL}}$ in the second upper bound.
We can ensure that the commutators of $[U_\tau,u_i]$ and $[U_{\tau,\tilde{\mfL}},u_i]$ satisfy the same upper bound from Subtheorem~\ref{sub_thm:U_tau_u_i_commun}.

We then estimate the norms of $\norm{ \brr{ \Delta V  ,\hat{H}_\tau}} $ and $\norm{ \brr{   \Delta U_\tau,  \hat{H}_\tau }}$. 
Using the inequality~\eqref{ad_H_tau/V_decomp_upp_bound}, we have 
\begin{align}
\label{ad_H_tau/V_decomp_upp_bound_re_00}
\norm{\ad_{\hat{H}_\tau} (\Delta V)} 
&\le 2 \tau  \norm{V} \cdot \norm{\Delta V} + \norm{H_0,\Delta V} + \norm{\Delta V} \cdot \norm{[U_\tau, H_0] } .
\end{align}
For the third term in the above inequality, we have already obtained the upper bound in~\eqref{ad_H_tau/V_decomp_upp_bound_3} as 
\begin{align}
\label{ad_H_tau/V_decomp_upp_bound_3_re_}
\norm{[U_\tau, H_0] }\le  4\gamma^3 |\mfL|^3 \sum_{r=0}^\infty r^{D-1} \tilde{Q}(\tau,r)  \le 
4\gamma^3 |\mfL|^3\cdot 2^{2D+4}C_{2,\tau}^{2D} C_{3,\tau}^{3D}\bar{J}_0,
\end{align}
where we can let $\check{Q}(\tau,r)$ be $Q(\tau,r)$ in~\eqref{ad_H_tau/V_decomp_upp_bound_3} because the estimation of $\norm{[U_\tau, H_0] }$ does not depend on the choice of $O$ in Proposition~\ref{prop:quasi-locality of_ad__H__tau_V}.
By using $\norm{\Delta V}\le \norm{V} \Er(0,r) \le \norm{V} \Er(\tau,r)$, we reduce the inequality~\eqref{ad_H_tau/V_decomp_upp_bound_re_00} to 
\begin{align}
\label{ad_H_tau/V_decomp_upp_bound_re_}
\norm{\ad_{\hat{H}_\tau} (\Delta V)} 
&\le2\norm{V}  \Er(\tau,r) \br{\tau  \norm{V} + 2^{2D+5} \gamma^3 |\mfL|^3C_{2,\tau}^{2D} C_{3,\tau}^{3D}\bar{J}_0 }
+  \norm{H_0,\Delta V}  .
\end{align}
Thus, the remaining task is refining the estimation of $\norm{H_0,\Delta V}$. 

For this purpose, we adopt the decomposition of $\Delta V$ as 
\begin{align}
\Delta V= \Delta \tilde{V}_{\mfL[r^\ast]} + \sum_{s=1}^\infty \br{ \Delta \tilde{V}_{\mfL[r^\ast+s]} -  \Delta \tilde{V}_{\mfL[r^\ast+s-1]} }, \quad 
\Delta \tilde{V}_{\mfL[r^\ast+s]}:=\tilde{\tr}_{\mfL[r^\ast+s]^\co} (\Delta V), 
\end{align}
where $r^\ast$ is chosen appropriately afterward.
We then obtain a similar inequality to~\eqref{quasi_locality_O_error_local} as 
\begin{align}
\label{quasi_locality_O_Delta_V_error_local}
\norm{\Delta \tilde{V}_{\mfL[r^\ast+s]} -  \Delta \tilde{V}_{\mfL[r^\ast+s-1]} }
&\le \sum_{i\in \mfL[r^\ast+s] \setminus \mfL[r^\ast+s-1]} \norm{[\Delta V,u_i]}  \notag \\
&\le  2 |\partial (\mfL[r^\ast+s])|  \cdot \norm{V} \min \brr{\Er(\tau,r),  Q(\tau,r^\ast+s)}  .
\end{align}
From the above inequality, we have a similar inequality to~\eqref{norm_H_0_O_commutator} as follows:
\begin{align}
\norm{ [ H_0 ,\Delta V]}&\le 
 2\sum_{Z: Z\cap \mfL[r^\ast]\neq \emptyset} \norm{h_Z} \cdot \norm{\Delta \tilde{V}_{\mfL[r^\ast]}} 
 + 2 \sum_{s=1}^\infty  \sum_{Z: Z\cap \mfL[r^\ast+s]\neq \emptyset}  \norm{h_Z} \cdot \norm{\Delta \tilde{V}_{\mfL[r^\ast+s]} - \Delta \tilde{V}_{\mfL[r^\ast+s-1]} } \notag \\
 &\le 2 \bar{J}_0|\mfL[r^\ast]| \cdot \norm{\Delta V} + 
 4\norm{V}  \sum_{s=0}^\infty  |\mfL[r^\ast+s]| \cdot |\partial (\mfL[r^\ast+s])| \bar{J}_0 \min \brr{ \Er(\tau,r) , Q(\tau,r^\ast+s)} \notag \\
 &\le 2 \bar{J}_0 \norm{V} r^{\ast D} |\mfL|  \Er(\tau,r) +
 4\bar{J}_0 \norm{V} \gamma |\mfL|^2 \sum_{s=1}^\infty  (r^\ast+s)^{2D-1}\min \brr{\Er(\tau,r), Q(\tau,r^\ast+s)} .
 \label{norm_inequ_H_0_Delta_V}
\end{align}
We can prove a similar lemma to Lemma~\ref{tilde_Q_tau_r_lemma}:
\begin{lemma} \label{tilde_Q_tau_r_lemma_again}
Let $r^\ast$ be the minimum number such that 
\begin{align}
\label{def_r^ast_1}
\min \brr{\Er(\tau,r), Q(\tau,r^\ast)} =  Q(\tau,r^\ast)  \longrightarrow  r^\ast= \Theta(\tau,r),
\end{align}
where we use the $\Theta$ notation in Eq.~\eqref{Theta_notation_def}.
Then, we obtain 
\begin{align}
\label{upper-bound:tilde_Q_tau_r_lemma_again}
 \sum_{s=1}^\infty  (r^\ast+s)^{2D-1}\min \brr{ \norm{V - V_{\tilde{\mfL}}}, Q(\tau,r^\ast+s)} 
 &\le  2^{2D+2} \br{C_{2,\tau}^{2D} +r^{\ast 2D}}Q(\tau ,r^\ast) \notag \\
 &\le  2^{2D+2} \br{C_{2,\tau}^{2D} +r^{\ast 2D}}\Er(\tau,r) , 
\end{align}
where we use $Q(\tau ,r^\ast)\le \Er(\tau,r)$ from Eq.~\eqref{def_r^ast_1} and adopt the definition of $C_{\nu,\tau}$ in Eq.~\eqref{def_C_V_tau_r_without_check}. 
\end{lemma}

By applying Lemma~\ref{tilde_Q_tau_r_lemma_again} to the inequality~\eqref{norm_inequ_H_0_Delta_V}, we obtain 
\begin{align}
\norm{ [ H_0 ,\Delta V]}&\le 
 2 \bar{J}_0 \norm{V} \cdot  |\mfL| \Er(\tau,r) \brr{r^{\ast D}+ 2 \gamma |\mfL| \cdot 2^{2D+2} \br{C_{2,\tau}^{2D} +r^{\ast 2D}}}  ,
 \label{norm_inequ_H_0_Delta_V_fibn}
\end{align}
which reduces the inequality~\eqref{ad_H_tau/V_decomp_upp_bound_re_} to 
\begin{align}
&\norm{\ad_{\hat{H}_\tau} (\Delta V)}  \notag \\
&\le 2\norm{V}  \Er(\tau,r) \brr{\tau  \norm{V} + 2^{2D+5} \gamma^3 |\mfL|^3C_{2,\tau}^{2D} C_{3,\tau}^{3D}\bar{J}_0 
+  \bar{J}_0  |\mfL| r^{\ast D}+ 2^{2D+3}  \bar{J}_0  \gamma |\mfL|^2   \br{C_{2,\tau}^{2D} +r^{\ast 2D}}} .
\end{align}
The same analyses yield 
\begin{align}
&\norm{\ad_{\hat{H}_\tau} (\Delta U_\tau)}  \notag \\
&\le  2\Er(\tau,r) \brr{\tau  \norm{V} + 2^{2D+5} \gamma^3 |\mfL|^3C_{2,\tau}^{2D} C_{3,\tau}^{3D}\bar{J}_0 
+  \bar{J}_0  |\mfL| r^{\ast D}+ 2^{2D+3}  \bar{J}_0  \gamma |\mfL|^2   \br{C_{2,\tau}^{2D} +r^{\ast 2D}}} .
 \label{norm_inequ_H_0_Delta_U_tau_fibn}
\end{align}
By applying the above two inequalities~\eqref{norm_inequ_H_0_Delta_V_fibn} and~\eqref{norm_inequ_H_0_Delta_U_tau_fibn} to 
\eqref{starting@point_Proof of proposition_com}, we can derive the main inequality as 
\begin{align}
&\norm{ \ad_{\hat{H}_\tau}(V) - \ad_{\hat{H}_{\tau,\tilde{\mfL}}}(V_{\tilde{\mfL}})} 
\le \norm{ \brr{ \Delta V  ,\hat{H}_\tau}} + 2\norm{V}\cdot \norm{ \brr{   \Delta U_\tau,  \hat{H}_\tau }} + 6 \norm{V} \br{ 2\tau \norm{V}+  \tilde{J}_0  |\tilde{\mfL}|^2  } \Er(\tau,r) \notag \\
&\le 6\norm{V}  \Er(\tau,r)  \brr{3\tau\norm{V}+  \tilde{J}_0  |\tilde{\mfL}|^2  +  \bar{J}_0  |\mfL| r^{\ast D} 
+ 2^{2D+3}  \bar{J}_0  \gamma |\mfL|^2   \br{C_{2,\tau}^{2D} +r^{\ast 2D} + 4\gamma^2 |\mfL|C_{2,\tau}^{2D} C_{3,\tau}^{3D} }}  . 
\end{align}
This completes the proof of Proposition~\ref{prop:upper_bound_commutator_H_tau_dif}. $\square$

\subsubsection{Proof of Lemma~\ref{tilde_Q_tau_r_lemma_again}}
From the definition of $r^\ast$ in Eq.~\eqref{def_r^ast_1}, we obtain 
\begin{align}
 \sum_{s=1}^\infty  (r^\ast+s)^{2D-1}\min \brr{ \norm{V - V_{\tilde{\mfL}}}, Q(\tau,r^\ast+s)} 
 =  \sum_{s=1}^\infty  (r^\ast+s)^{2D-1} Q(\tau,r^\ast+s) = \sum_{s>r^\ast}  s^{2D-1} Q(\tau,s) .
\end{align}
We next estimate an upper bound for $ \sum_{s> r^\ast} s^{2D-1} Q(\tau,s)$.
The upper bound can be derived by the same calculations for the derivation of~\eqref{estimation_s_ge_r_tilde_q_tau_final}, which gives 
\begin{align}
\label{estimation_s_ge_r_tilde_q_tau_final_again}
 \sum_{s> r^\ast}  s^{2D-1}  Q(\tau,s)  \le  3 \cdot 2^{2D} (s_0^{\ast 2D}+r^{\ast2D}) Q(\tau ,r^\ast), 
\end{align}
where $s_0^\ast$ is defined by 
 \begin{align}
s^{2D-1} Q(\tau,s) \le  e^{-\kappa_\beta  s /2}. 
\end{align}
for $s\ge s_0^\ast$. Note that  Lemma~\ref{lem:s_0_upper_bound} with $\nu=2D-1$ gives 
\begin{align}
\label{upper_bound_s_0_s_ast}
s_0^\ast \le \frac{\tau+e}{2}+\frac{32}{\kappa_\beta^2} \brr{(2D-1)^2 + \kappa_1^2 \tau^2+\frac{\kappa_\beta\kappa_0\tau}{8}} \le C_{2,\tau} . 
\end{align}
By applying~\eqref{upper_bound_s_0_s_ast} to the inequality~\eqref{estimation_s_ge_r_tilde_q_tau_final_again}, we prove the main inequality~\eqref{upper-bound:tilde_Q_tau_r_lemma_again}. This completes the proof. $\square$

{~}

\hrulefill{\bf [ End of Proof of Lemma~\ref{tilde_Q_tau_r_lemma_again}]}

{~}

\section{Improved bound for 1D cases} \label{sec:Improved bound for 1D cases}

In Theorem~\ref{thm:Refined locality estimation for effective Hamiltonian}, the norm error between the unitary operator $U_\tau$ and $U_{\tau,\tilde{\mfL}}$ is given by~\eqref{error_simplest_form_thm}:
\begin{align}
\label{U_tay_U_tauilde_mafl}
\norm{  U_\tau  - U_{\tau,\tilde{\mfL}} } \le  \exp \brr{ \Theta(\tau \beta^D) \norm{V} \log (\beta\norm{V}\cdot |\mfL| r \tau) -\kappa_\beta r/2},
\end{align}
which yields the behaviour of the RHS as $e^{\tilde{\mathcal{O}}(\tau \beta^D \norm{V}) - \Theta(r/\beta)}$ from $\kappa_{\beta}\propto 1/\beta$.
Therefore, when we apply the above inequality to the BP formalism in one-dimensional systems as in~\eqref{BP_formalism_exponential_connection_eq_log}, i.e., $$ \log\br{\tilde{\Phi}_{XY} e^{\beta \br{\tilde{H}^\ast_{X}+H_{L[\ell]^\co}}} \tilde{\Phi}_{XY}^\dagger}
 =\log\brr{ 
\mathcal{T} e^{\int_0^1 \tilde{\phi}_{XY,\tau}d\tau} e^{\beta \br{\tilde{H}^\ast_{X}+H_{L[\ell]^\co}}} \br{\mathcal{T} e^{\int_0^1 \tilde{\phi}_{XY,\tau}d\tau} }^\dagger} ,$$ we have $\norm{V}=\orderof{\beta}$ from $\norm{\tilde{\phi}_{XY,\tau}}\le \beta\norm{\partial h_{L[\ell]}}=\orderof{\beta}$ [see also Eq.~\eqref{X:L_brr_r_Y:L_brr_r_^co_belief_prop}] and $\tau=1$ in~\eqref{U_tay_U_tauilde_mafl}, and hence, the RHS reduces to $e^{\tilde{\mathcal{O}}(\beta^2) - \orderof{r/\beta}}$. 
This leads to the correlation length of the CMI in the form of $\orderof{\beta^{3}}$. 
Note that the operator $V$ is now quasi-local around the surface region of $X$ (or $Y$), and hence we can take $|\mfL|=|\partial X|=1$.  
 
In the present section, we aim to improve the dependence of $\beta$ by adopting a different ansatz in Subtheorem~\ref{sub_thm:U_tau_u_i_commun}.
In detail, we prove the RHS as in $e^{\tilde{\mathcal{O}}(\beta) - \orderof{r/\beta}}$, which is expected to be optimal since it gives qualitatively the same behavior as for the quasi-locality of the belief propagation operator [see the inequality~\eqref{main_ineq:corol:high_dimensional_applicaton_bp}].
We prove the following theorem:

\begin{theorem} \label{sub_thm:U_tau_u_i_commun_1D}
Let us adopt the same setup as in Theorem~\ref{thm:Refined locality estimation for effective Hamiltonian}. 
Then, we prove 
\begin{align}
\norm{  U_\tau  - U_{\tau,\tilde{\mfL}} } \le  \Er^{(1)} (\tau,r) =\brr{1+2 \tau \mathcal{K}_2} e^{\kappa_0^{(1)} \tau+\kappa_1^{(1)} \tau \log(r+\tau+e)-\kappa_\beta r/2} ,
\label{main_inq:thm:Refined locality estimation for effective Hamiltonian1D}
\end{align}
where $\mathcal{K}_2$ has been defined in Eq.~\eqref{def:mathcal_K_2_thm} and $\kappa_0^{(1)}$ and $\kappa_1^{(1)}$ are defined in similar ways to Eq.~\eqref{kappa_0_kappa_1_definition}: 
 \begin{align}
&\kappa_0^{(1)} = \Theta(1)\norm{V} \log (\beta\norm{V}\cdot |\mfL|)  , \quad \kappa_1^{(1)} =\Theta(1) \norm{V}  . 
\label{definition_kappa_beta_1_1D}
\end{align}
\end{theorem}

{\bf Remark.} Under the choice as in~\eqref{BP_formalism_exponential_connection_eq_log} for the BP formalism,
we have $V\to \tilde{\phi}_{XY,\tau}$ ($\norm{V}=\orderof{\beta}$), $\tau\to1$ and $|\mfL|=|\partial X|=\orderof{1}$.
We thus reduce the upper bound~\eqref{main_inq:thm:Refined locality estimation for effective Hamiltonian1D} to 
\begin{align}
\norm{  U_{\tau=1}  - U_{\tau=1,\tilde{\mfL}} } \le   e^{\Theta(\beta) \log(\beta r)-\kappa_\beta r/2} ,
\end{align}
where we let $\tilde{\mfL}=\partial{X}[r]$. 
We, therefore, obtain the desired upper bound. 

Using the same analyses for the derivation of~\eqref{error_simplest_form_thm_for_eff_ham}, we can obtain 
\begin{align}
\label{error_simplest_form_thm_for_eff_ham_1D}
&\norm{\log \br{\tilde{\Phi}_{XY} e^{\beta \br{\tilde{H}^\ast_{X}+H_{L[\ell]^\co}}} \tilde{\Phi}_{XY}^\dagger} - U_{\tau=1,\tilde{\mfL}} \brr{\beta \br{\tilde{H}^\ast_{X}+H_{L[\ell]^\co}}+\hat{\tilde{\Phi}}_{\tau=1,\tilde{\mfL}} }U_{\tau=1,\tilde{\mfL}} ^\dagger}
\le e^{\Theta(\beta) \log(\beta r)-\kappa_\beta r/2} ,
\end{align}
where we use Eq.~\eqref{connection_of_exponential_operator_repeat_def_tau_dependent}, which define $\hat{\tilde{\Phi}}_{\tau=1,\tilde{\mfL}}$ as 
\begin{align}
\label{hat_tilde_Phi_1_tilde_mfL_def}
\hat{\tilde{\Phi}}_{\tau=1,\tilde{\mfL}}  := 2\int_0^1 U_{\tau_1,\tilde{\mfL}}^\dagger  \tilde{\phi}_{XY,\tau_1}   U_{\tau_1,\tilde{\mfL}} d\tau_1 .
\end{align}

%

\subsection{Proof of Theorem~\ref{sub_thm:U_tau_u_i_commun_1D}} 
We follow the same proof techniques for Subtheorem~\ref{sub_thm:U_tau_u_i_commun} and Theorem~\ref{thm:Refined locality estimation for effective Hamiltonian}.
We start with the improvement of Subtheorem~\ref{sub_thm:U_tau_u_i_commun}.
In the improvement, we adopt a different ansatz to characterize the quasi-locality of the unitary operator $U_\tau$ so that we can utilize Lemma~\ref{lem:Quasi-local_commutator_bound2}.  
For this purpose, let $u_{X}$ be an arbitrary unitary operator defined on the subset $X$ separated from the subset $\mfL$ by distance $r$.  
We aim to prove an upper bound of the commutator norm in the form of 
\begin{align}
\label{upper_bound_U_tau_u_i__re}
\norm{[U_\tau,u_X]}  \le Q^{(1)} (\tau,r) = e^{\kappa_0^{(1)} \tau+\kappa_1^{(1)} \tau \log(r+\tau+e)-\kappa_\beta r}  \quad (\dist_{X,\mfL}=r). 
\end{align}

Almost all the proof techniques are applied to the ansatz~\eqref{upper_bound_U_tau_u_i__re}.
The differences arise from the estimations of the parameters $\mathfrak{g}_1$ and $\mathfrak{g}_3$ in Eqs.~\eqref{main:eq:prop:Quasi-local_commutator_bound_tau_parameter_g_1and_g_2} and \eqref{main:eq:prop:Quasi-local_commutator_bound_tau__2}, respectively.
For the modifications of Propositions~\ref{prop:Quasi-local_commutator_bound_tau} and \ref{prop:Quasi-local_commutator_bound_tau_2}, we utilize Lemma~\ref{lem:Quasi-local_commutator_bound2} instead of Lemma~\ref{lem:Quasi-local_commutator_bound}. 
We then prove the following proposition: 

\begin{prop}\label{prop:Quasi-local_commutator_bound_tau_improve_1D}
Let us adopt the same setup as in Propositions~\ref{prop:Quasi-local_commutator_bound_tau} and~\ref{prop:Quasi-local_commutator_bound_tau_2}. 
We also choose $\mathcal{F}(\ell)$ as 
\begin{align}
\label{choice_of_nathcal_F_ell_1D}
\mathcal{F}(\ell)= \exp\br{ K_\ell- \kappa_\beta \ell} , 
\end{align}
where $K_\ell$ monotonically increases with $\ell$. 
We then obtain
\begin{align}
\label{main:eq:prop:Quasi-local_commutator_bound_tau_1DD}
\int_{-\infty}^\infty g_1(t) \norm{[O,u_X(H_0,t)]}dt \le  \mathfrak{g}_1^{(1)}  \mathcal{F}(\ell)  , 
\AND 
&\int_{-\infty}^\infty g_1(t) \int_0^t \norm{[O,u_X(H_0,t_1)]} dt_1 dt\le  \mathfrak{g}_3^{(1)}   \mathcal{F}(\ell)   , 
\end{align}
with $g_1(t)$ defined in Eq.~\eqref{definition_g_1t_g_2t}, i.e.,
\begin{align}
g_1(t) =  \begin{cases}  
| g_\beta(t)| &\for |t| \ge \delta t,  \\
0 & \for |t| < \delta t, 
\end{cases} 
\end{align}
where $\mathfrak{g}_1^{(1)}$ and $\mathfrak{g}_3^{(1)}$ are defined as  
\begin{align}
\label{main:eq:prop:modified_ver_g_1and_g_3}
&\mathfrak{g}_1^{(1)} :=\beta \brr{ \frac{41 C  |\partial \mfL| }{\pi v} 
+\frac{4\br{1+4 e^{2\kappa_\beta \Delta \ell}}}{\pi} \log\br{\frac{\beta}{2\pi \delta t}}  }  , \notag \\
&\mathfrak{g}_3^{(1)} := \beta^2  \brr{ \frac{1}{6} + \frac{703 C  |\partial \mfL|}{\pi v^2 \beta}  
+ 2e^{\kappa_\beta \Delta \ell} \br{  \frac{3\mu \Delta \ell}{\pi v \beta}
 + \frac{1}{6} + \frac{1}{6} e^{\kappa_\beta \Delta \ell} }} .
\end{align}
Here, the parameter $\Delta \ell$ can be arbitrarily chosen such that $\kappa_\beta \Delta \ell \ge 1$ (or $1/(\kappa_\beta \Delta \ell)\le 1$).
\end{prop}

On the estimations of $g_{\tau,r}$, $g'_{\tau}$, $\mathfrak{g}_2$, and $\mathfrak{g}_4$ in Propositions~\ref{prop:Quasi-local_commutator_bound_tau}, \ref{prop:quasi-locality of_ad__H__tau_V}, and \ref{prop:Quasi-local_commutator_bound_tau_2}, we use the previous loose estimate based on the ansatz for $\norm{[U_\tau,u_i]}$. 
The upper bound for $\norm{[U_\tau,u_X]}$ is trivially applied to the commutator with a local unitary operator $u_i$, i.e., $\norm{[U_\tau,u_i]}  \le Q^{(1)} (\tau,r)$ for $\dist_{i,\mfL}=r$.
On the other hand, from the local commutator bound $\norm{[U_\tau,u_i]}$, one can upper-bound $\norm{[U_\tau,u_X]}$ using the following general lemma:
\begin{lemma} \label{lema_from_u_i_to_u_X}
We adopt a similar setup to Proposition~\ref{prop:Quasi-local_commutator_bound_tau}. 
Let $O$ be an arbitrary operator with $\norm{[O,u_i]}  \le \mathcal{F}(\ell) = e^{K_\ell -\kappa_\beta \ell} $ for $\forall i \in \Lambda$ with $\ell=\dist_{i,\mfL}$, we obtain the upper bound of
\begin{align}
\label{main_inequality:lema_from_u_i_to_u_X}
\norm{[U_\tau,u_X]}  \le 2^{D+2} \gamma   |\mfL|    \br{\ell^{D} +{s^\ast}^{D}  }\mathcal{F}(\ell) ,
\end{align}
with $s^\ast$ the solution of the following equation
\begin{align}
\label{s_ast_lema_from_u_i_to_u_X}
s^\ast = \frac{2}{\kappa_\beta} \brr{ K_{s^\ast} + (D-1) \log (s^\ast)},
\end{align}
where $\mathcal{F}(\ell) $ monotonically decreases with $\ell$.
\end{lemma}

{\bf Remark.} If $K_\ell \le  \Theta[\log(\ell)] + \poly( \norm{V},\beta, |\mfL|,\tau)$, the solution of Eq.~\eqref{s_ast_lema_from_u_i_to_u_X} gives 
\begin{align}
s^\ast =\poly( \norm{V},\beta, |\mfL|,\tau) ,
\end{align}
and hence, we can obtain 
\begin{align}
\label{main_inequality:lema_from_u_i_to_u_X__02}
\norm{[U_\tau,u_X]}\le 2^{D+2} \gamma   |\mfL|    \br{\ell^{D} +{s^\ast}^{D}  }\mathcal{F}(\ell)  \le \poly( \norm{V},\beta, |\mfL|,\tau) \mathcal{F}(\ell) .
\end{align}
The norm increases by $\poly( \norm{V},\beta, |\mfL|,\tau)$ can be cancelled by

\textit{Proof of Lemma~\ref{lema_from_u_i_to_u_X}.}
First, using the inequality~\eqref{sec_ineq:lemma_commutator_generalization} in Lemma~\ref{lemma_commutator_generalization}, we can derive 
\begin{align}
\label{prooof_1:lema_from_u_i_to_u_X}
\norm{[O,u_X]}  \le \sum_{i\in X}\sup_{u_i} \norm{[O,u_i]} \le 
\sum_{i\in X}  \mathcal{F}(\dist_{i,\mfL}) \le  \sum_{i\in \mfL[\ell]^\co}  \mathcal{F}(\dist_{i,\mfL}) .
\end{align}
We then obtain 
\begin{align}
\label{prooof_2:lema_from_u_i_to_u_X}
\sum_{i\in \mfL[\ell]^\co} \mathcal{F}(\dist_{i,\mfL}) \le \sum_{j\in \mfL} \sum_{i: \dist_{i,j}>\ell} \mathcal{F}(\dist_{i,j})
\le \sum_{j\in \mfL} \sum_{q=1}^\infty \partial (j[\ell+q])  \mathcal{F}(\ell+q) 
\le |\mfL| \gamma   \sum_{s>\ell} s^{D-1} \mathcal{F}(s) .
\end{align}
We now rely on similar analyses to the proof of  Lemma~\ref{tilde_Q_tau_r_lemma}.
We define $s^\ast$ as the constant such that 
\begin{align}
(s^\ast)^{D-1} \mathcal{F}(s^\ast) =  (s^\ast)^{D-1} e^{K_{s^\ast} -\kappa_\beta s^\ast}   = e^{-\kappa_\beta s^\ast/2} ,
\end{align}
which reduces to the equation~\eqref{s_ast_lema_from_u_i_to_u_X}.
In the case where $\ell \ge s^\ast$, we have 
\begin{align}
\sum_{s>\ell} s^{D-1} \mathcal{F}(s) 
&\le \sum_{\ell < s\le 2\ell} s^{D-1} \mathcal{F}(s) + \sum_{s> 2\ell} s^{D-1} \mathcal{F}(s) \notag \\
&\le (2\ell)^{D} \mathcal{F}(\ell) + \sum_{s> 2\ell} e^{-\kappa_\beta \ell/2} 
=  (2\ell)^{D} \mathcal{F}(\ell) + \frac{e^{-K_\ell} }{1-e^{-\kappa_\beta }  }\mathcal{F}(\ell) .
\end{align}
In the case where $\ell < s^\ast$, we have 
\begin{align}
\sum_{s>\ell} s^{D-1} \mathcal{F}(s) 
&\le \sum_{\ell < s\le s^\ast} s^{D-1} \mathcal{F}(s) + \sum_{s> s^\ast} s^{D-1} \mathcal{F}(s) \notag \\
&\le (s^\ast)^{D}  \mathcal{F}(\ell) + \sum_{s> s^\ast} e^{-\kappa_\beta \ell/2}
 \le (s^\ast)^{D}  \mathcal{F}(\ell) +(2s^\ast)^{D} \mathcal{F}(s^\ast) + \frac{e^{-K_{s^\ast}} }{1-e^{-\kappa_\beta }  }\mathcal{F}(s^\ast)   \notag \\
&\le \br{ 2 (2s^\ast)^{D} +\frac{e^{-K_{s^\ast}} }{1-e^{-\kappa_\beta }}}  \mathcal{F}(\ell) 
\end{align}
By combining the above two upper bounds, the following inequality always holds:
\begin{align}
\label{prooof_3:lema_from_u_i_to_u_X}
\sum_{s>\ell} s^{D-1} \mathcal{F}(s) 
&\le 2^{D+2}  \br{\ell^{D} +  {s^\ast}^{D}  }\mathcal{F}(\ell) .
\end{align}
By applying the inequalities~\eqref{prooof_2:lema_from_u_i_to_u_X} and~\eqref{prooof_3:lema_from_u_i_to_u_X} to 
\eqref{prooof_1:lema_from_u_i_to_u_X}, we prove the main inequality~\eqref{main_inequality:lema_from_u_i_to_u_X}.
This completes the proof. $\square$

{~}

\hrulefill{\bf [ End of Proof of Lemma~\ref{lema_from_u_i_to_u_X}]}

{~}

 By using Lemma~\ref{lema_from_u_i_to_u_X}, we have 
\begin{align}
\int_{-\infty}^\infty g_2(t) \norm{[O,u_X(H_0,\lambda t)]}dt 
&\le  \sum_{i\in X}\sup_{u_i}\int_{|t|\le \delta t} |tg_\beta(t)| \cdot \norm{[O,u_i(H_0,\lambda t)]}dt  \notag \\
&\le  \mathfrak{g}_2   \delta t \sum_{i\in X} \mathcal{F}(\dist_{i,\mfL}) 
\le  \mathfrak{g}_2 \delta \tilde{t}\mathcal{F}(\ell) ,
\end{align}
where we let $\delta \tilde{t}:= 2^{D+2} \gamma   |\mfL|    \br{\ell^{D} +{s^\ast}^{D}  } \delta t\le \poly( \norm{V},\beta, |\mfL|,\tau)\delta t$.
Therefore, by replacing the $\delta t$ by $\delta \tilde{t}$, we can obtain the same upper bound for the summation including $ |tg_\beta(t)|$, i.e., 
in the inequalities~\eqref{norm_V_tau_t_u_i_com_2_fin} and \eqref{norm_V_tau_t_u_i_com_3_up_fin}.
Here, we can obtain the same estimations for $\mathfrak{g}_2$ and $\mathfrak{g}_4$ in Propositions~\ref{prop:Quasi-local_commutator_bound_tau} and \ref{prop:Quasi-local_commutator_bound_tau_2} as $\mathfrak{g}_2,\mathfrak{g}_4\le \Theta(\beta)$.

The remaining part of the proof is the same, and we can replace the inequality~\eqref{condition_kappa_1and_kpappa_2} with
\begin{align}
\label{condition_kappa_1and_kpappa_2_1D}
 \frac{2\norm{V}}{\beta}\brr{ 4\mathfrak{g}_1 ^{(1)}  + \mathfrak{g}_2 +  \frac{6}{\kappa_0^{(1)}}    \br{\mathfrak{g}_3^{(1)}\norm{V} +  \frac{\mathfrak{g}_4}{9g'_{\tau}}  }} \le \kappa_0^{(1)} + \kappa_1^{(1)}   \log(r+\tau+e) ,
\end{align}
where we adopt the same choice for $\delta \tilde{t}$ as in Eq.~\eqref{choice_fo_delta_t}, i.e., $
\delta \tilde{t}^{-1} =\poly(\norm{V},|\mfL|, \beta, \kappa_0^{(1)}, \kappa_1^{(1)},\tau)$, which also implies $\delta t=\poly(\norm{V},|\mfL|, \beta, \kappa_0^{(1)}, \kappa_1^{(1)},\tau)$. 
Moreover, using Eq.~\eqref{main:eq:prop:modified_ver_g_1and_g_3} in Proposition~\ref{prop:Quasi-local_commutator_bound_tau_improve_1D}, 
we can replace the inequality~\eqref{ineq:mathfrak_g_1_2_3_4_g} by 
\begin{align}
&\mathfrak{g}_{1}^{(1)} \le \Theta(\beta)\log (\beta/\delta t)  , 
\quad\mathfrak{g}_4\le \mathfrak{g}_{2}\le  \Theta(\beta) , \quad \mathfrak{g}_3^{(1)}  \le  \Theta(\beta^{2}).
\end{align}
Therefore, the choices of~\eqref{definition_kappa_beta_1_1D} satisfy the inequality~\eqref{condition_kappa_1and_kpappa_2_1D}. 
We thus achieve the improvement of Subtheorem~\ref{sub_thm:U_tau_u_i_commun} to the form of~\eqref{upper_bound_U_tau_u_i__re} in one-dimensional systems.

Finally, we will prove the main statement~\eqref{main_inq:thm:Refined locality estimation for effective Hamiltonian1D}, which improves Theorem~\ref{thm:Refined locality estimation for effective Hamiltonian}. 
On this point, we do not need any modification. 
To see it, we start with the final conditions for $\{\mathcal{K}_0,\mathcal{K}_1,\mathcal{K}_2\}$ in the form of \eqref{final_condition_for_mathcal_K_012}. 
Now, we denote the parameters by $\{\mathcal{K}_0^{(1)},\mathcal{K}_1^{(1)},\mathcal{K}_2\}$ in one-dimensional systems, which should satisfy
\begin{align}
\mathcal{K}_0^{(1)} &= \max\Bigl[ \kappa_0^{(1)}, \Theta(\norm{V}) \log (\beta \norm{V}\cdot|\mfL|)  \Bigr]  , \quad 
\mathcal{K}_1^{(1)} = \max\Bigl[ \kappa_1^{(1)}, \Theta(\norm{V})  \Bigr]   ,\notag \\
\mathcal{K}_2&=  \max\brr{ \Theta(\norm{V}) \br{\log(\beta \norm{V}\cdot |\mfL|)  +   \log(r+\tau+e)   +  \beta \norm{V} } ,\Theta\br{\norm{V}^{2D+3}, |\mfL|^3, \beta^{2D^2+2D+2}, \tau^{2D+2}, r^{2D+3}}  } \notag \\
&= \Theta\br{\norm{V}^{2D+3}, |\mfL|^3, \beta^{2D^2+2D+2}, \tau^{2D+2}, r^{2D+3}}  .
\end{align}
By applying the forms of $\kappa_0^{(1)} $ and $\kappa_1^{(1)}$, i.e., $\kappa_0^{(1)} = \Theta(1)\norm{V} \log (\beta\norm{V}\cdot |\mfL|)$ and $\kappa_1^{(1)} =\Theta(1) \norm{V}$, we ensure that the choices $\mathcal{K}_0^{(1)}=\kappa_0^{(1)}$ and $\mathcal{K}_1^{(1)}=\kappa_1^{(1)}$ satisfy the above conditions. 
Therefore, we prove the main inequality~\eqref{main_inq:thm:Refined locality estimation for effective Hamiltonian1D}
This completes the proof of Theorem~\ref{sub_thm:U_tau_u_i_commun_1D}. $\square$

\subsection{Proof of Proposition~\ref{prop:Quasi-local_commutator_bound_tau_improve_1D}}
The proof strategy is the same as those for Propositions~\ref{prop:Quasi-local_commutator_bound_tau} and \ref{prop:Quasi-local_commutator_bound_tau_2}. 
First of all, because of the condition $\kappa_\beta \Delta \ell \ge 1$ (or $1/(\kappa_\beta \Delta \ell)\le 1$), we have 
\begin{align}
\label{ell_condl_inequality_g13}
\frac{\pi \mu}{2v\beta} \Delta \ell \ge 1 \AND  \frac{\mu \Delta \ell}{4} \ge 1 . 
\end{align}
We recall that $\kappa_\beta$ has been defined as $\kappa_\beta := \min\br{ \frac{\pi \mu}{2v\beta} , \frac{\mu}{4}}$ in Eq.~\eqref{definition_kappa_beta_1}.

We start from the former part in~\eqref{main:eq:prop:modified_ver_g_1and_g_3}. 
Using the inequality~\eqref{main_eq:connection_of_exponential_operator_trans_1} in Lemma~\ref{belief_norm_lemma}, the parameters $f_1$ and $f_2$ in~\eqref{definition_lem:Quasi-local_commutator_bound} immediately given by
\begin{align}
\label{est_f_1andf_2_g_1case_re_ag}
&f_1= \int_{-\infty}^\infty g_1(t) dt \le  \frac{2\beta}{\pi}\log\br{\frac{\beta}{2\pi \delta t}}, \notag \\
&f_2= \int_{|t| \le t_0}  |t| g_1(t) dt \le  \frac{\beta}{2\pi} t_0^{2} \le \frac{\beta}{2\pi} \br{\frac{\mu s \Delta \ell}{2v}}^{2} ,
\end{align}
where, in the second inequality, we use the inequality~\eqref{main_eq:connection_of_exponential_operator_trans_2} by replacing $\delta t$ with $t_0$.
Note that $t_0$ has been defined as $t_0=[\mu/(2v)]\Delta \ell(s-2)$, where we can arbitrarily control the value of $\Delta \ell$.  
For $f_{t_0}(s)$, we can utilize the same inequality as~\eqref{est_f_t_0_s_g_1case}: 
\begin{align}
\label{est_f_t_0_s_g_1case_re_ag}
f_{t_0}(s)
\le  \frac{\beta}{\pi}  \log\br{e+ \frac{e\beta}{2\pi t_0}}e^{-2\pi t_0/\beta} 
\le  \frac{\beta}{\pi}  \log\br{e+ \frac{e}{2}  }e^{-2\kappa_\beta (s-2)\Delta \ell} \le 
\frac{2\beta}{\pi} e^{-2\kappa_\beta (s-2)\Delta \ell} 
\end{align}
for $s\ge 3$, 
where we use $2\pi t_0/\beta= [\pi \mu/(v\beta)] (s-2)\Delta \ell \ge  2\kappa_\beta (s-2)\Delta \ell \ge 2(s-2)$. 
For $s\le 2$, we can simply obtain $f_{t_0}(s)=f_1$ since $t_0\le0$.  

Then, by applying the upper bounds for $f_2$ and $f_{t_0}(s)$ to~\eqref{main_ineq:lem:Quasi-local_commutator_bound2}, we first obtain 
 \begin{align}
 \label{upper_bound_s^D_mathcal_F_ell-s_f_2term_0_re_ag}
&2\sum_{s=1}^\infty \mathcal{F}(\ell-s\Delta \ell) C|\partial (\mfL[\ell])| f_2 v e^{-\mu s \Delta \ell/2} 
\le  4C v |\partial \mfL|  \sum_{s=1}^\infty e^{K_{\ell-s\Delta \ell} -\kappa_\beta (\ell-s\Delta \ell)} \cdot \frac{\beta}{2\pi} \br{\frac{\mu \Delta \ell}{2v}}^{2}  s^2 e^{-\mu s \Delta \ell/2}  \notag \\
&\le 4C v |\partial \mfL| \frac{\beta}{2\pi} \br{\frac{\mu \Delta \ell}{2v}}^{2}  \mathcal{F}(\ell)    \sum_{s=1}^\infty  s^2 e^{\kappa_\beta s \Delta \ell-\mu s \Delta \ell/2}  
\le 4C v |\partial \mfL|   \frac{\beta}{2\pi} \br{\frac{\mu \Delta \ell}{2v}}^{2}   \mathcal{F}(\ell) \sum_{s=1}^\infty  s^2 e^{-\mu s \Delta \ell/4}   \notag \\
&\le 4C  v |\partial \mfL|  \frac{\beta}{2\pi} \br{\frac{\mu \Delta \ell}{2v}}^{2}   \mathcal{F}(\ell)  e^{-\mu \Delta \ell/4}  \br{1+2^2 \cdot 2! \brr{ \max\br{1,\frac{4}{\mu \Delta \ell}}}^3}  \notag \\
&\le   \frac{2C  v |\partial \mfL|  \beta}{\pi}  \mathcal{F}(\ell) e^{-\mu \Delta \ell/4}  \br{\frac{\mu \Delta \ell}{2v}}^{2}  \br{1+2^3}= 
\frac{41 C  |\partial \mfL|  \beta}{\pi v}  \mathcal{F}(\ell) ,
\end{align}
where we use $|\partial (\mfL[\ell])| \le 2 |\partial \mfL| $ in one-dimensional systems in the first inequality,~\cite[(S.11) of Lemma~1 therein]{kuwahara2022optimal} in the fourth inequality, and the following inequality for the last inequality:
 \begin{align}
e^{-\mu \Delta \ell/4}  (\mu \Delta \ell)^{2}\le 8.66146 \cdots  \le 9 ,
\end{align}
which is derived under the condition of $\mu \Delta \ell \ge 4$ in \eqref{ell_condl_inequality_g13}. 
In the same way, we secondly obtain 
 \begin{align}
 \label{upper_bound_s^D_mathcal_F_ell-s_f_2term_re_ag}
2\sum_{s=1}^\infty \mathcal{F}(\ell-s\Delta \ell)  f_{t_0}(s) 
&\le 
2 \mathcal{F}(\ell)\brr{ f_1e^{\kappa_\beta \Delta \ell}  \br{ 1+e^{\kappa_\beta \Delta \ell}} 
+ \sum_{s=3}^\infty e^{\kappa_\beta s \Delta \ell}\cdot  \frac{2\beta}{\pi}  e^{-2\kappa_\beta (s-2)\Delta \ell} } \notag \\
&\le2 \mathcal{F}(\ell)\brr{ 2 e^{2\kappa_\beta \Delta \ell} \cdot \frac{2\beta}{\pi}\log\br{\frac{\beta}{2\pi \delta t}}
+ \frac{2\beta}{\pi}  \frac{e^{\kappa_\beta \Delta \ell}}{1-e^{-\kappa_\beta \Delta \ell}} }  \notag \\
&\le \frac{16 \beta}{\pi} \log\br{\frac{\beta}{2\pi \delta t}} e^{2\kappa_\beta \Delta \ell}   \mathcal{F}(\ell) ,
\end{align}
where we use $\br{1-e^{-\kappa_\beta \Delta \ell}}^{-1} \le 2$ from $\kappa_\beta \Delta \ell\ge 1$. 
Therefore, by combining the inequalities~\eqref{est_f_1andf_2_g_1case_re_ag}, \eqref{upper_bound_s^D_mathcal_F_ell-s_f_2term_0_re_ag}
 and \eqref{upper_bound_s^D_mathcal_F_ell-s_f_2term_re_ag}, we obtain
\begin{align}
\label{upper_bound_s^D_mathcal_F_ell-s_f_2term_fin_re_ag}
&2f_1 \mathcal{F}(\ell) + 2\sum_{s=1}^\infty \mathcal{F}(\ell-s\Delta \ell) \brr{ C|\partial (\mfL[\ell])| f_2 v e^{-\mu s \Delta \ell /2} + f_{t_0}(s)} \notag \\
& \le 
\beta  \mathcal{F}(\ell) \brr{ \frac{41 C  |\partial \mfL| }{\pi v} 
+\frac{4\br{1+4 e^{2\kappa_\beta \Delta \ell}}}{\pi} \log\br{\frac{\beta}{2\pi \delta t}}   }.
 \end{align}
We thus prove the first main inequality~\eqref{main:eq:prop:Quasi-local_commutator_bound_tau_1DD} with $\mathfrak{g}_1^{(1)}$ given by Eq.~\eqref{main:eq:prop:modified_ver_g_1and_g_3} from~\eqref{main_ineq:lem:Quasi-local_commutator_bound2} in Lemma~\ref{lem:Quasi-local_commutator_bound2}.

We next proceed to the latter part. The proof itself is quite similar to the proof for the former part. 
For the proof, from Corollary~\ref{corol:Quasi-local_commutator_bound}, 
we use Lemma~\ref{lem:Quasi-local_commutator_bound2} by replacing $f(t) =|tg_1(t)|$, where $g_1(t)$ has been defined in the same way as in Eq.~\eqref{definition_g_1t_g_2t}. 
Therefore, the parameters $f_1$ and $f_2$ in~\eqref{definition_lem:Quasi-local_commutator_bound} are upper-bounded using~\eqref{main_eq:connection_of_exponential_operator_trans_1} in Lemma~\ref{belief_norm_lemma} as follows:
\begin{align}
\label{est_f_1andf_2_g_1case_33_re_ag}
&f_1=\int_{|t|>\delta t} |tg_\beta (t)| dt \le  \frac{\beta^2}{2\pi^2} \zeta(2)=  \frac{\beta^2}{12}, \notag \\
&f_2=\int_{|t|\le t_0}  |t^2 g_\beta (t)| dt \le \frac{\beta t_0^3}{3\pi} 
=\frac{\beta}{3\pi} \br{\frac{\mu s \Delta \ell}{2v}}^3 ,
\end{align}
where we use the inequality~\eqref{main_eq:connection_of_exponential_operator_trans_2} in the second inequality.
Then, by following the same analyses as~\eqref{integral_t_g_beta_t_ge_delta_re} and \eqref{est_f_t_0_s_g_1case_re_ag}, we can obtain
\begin{align}
\label{integral_t_g_beta_t_ge_delta_re__agg}
f_{t_0}(s):= \int_{|t| > t_0}  |t g_\beta (t)|  dt 
&\le  \frac{\beta t_0}{\pi}\log\br{e+ \frac{e\beta}{2\pi t_0}}e^{-2\pi t_0/\beta} +\frac{\beta^2}{12} e^{-2\pi t_0/\beta}\notag \\
&\le  \brr{ \frac{\mu \beta }{\pi v} (s-2) \Delta \ell   +\frac{\beta^2}{12}} e^{-2\kappa_\beta (s-2)\Delta \ell} 
\end{align}
for $s\ge 3$, and we have $f_{t_0}(s)\le f_1$ for $s=1,2$.

Then, from the above $f_2$ and $f_{t_0}(s)$, we upper-bound the second term in the RHS of~\eqref{main_ineq:lem:Quasi-local_commutator_bound2}. 
Following a similar analysis to ~\eqref{upper_bound_s^D_mathcal_F_ell-s_f_2term_0_re_ag}, we obtain 
 \begin{align}
 \label{mathcal_F_ell-s_f_2term_0_re_ag_g_3}
&2\sum_{s=1}^\infty \mathcal{F}(\ell-s\Delta \ell) C|\partial (\mfL[\ell])| f_2 v e^{-\mu s \Delta \ell/2} 
\le  4C v |\partial \mfL| \frac{\beta}{3\pi} \br{\frac{\mu \Delta \ell}{2v}}^{3}  \mathcal{F}(\ell)    \sum_{s=1}^\infty  s^3 e^{\kappa_\beta s \Delta \ell-\mu s \Delta \ell/2}   \notag \\
&\le 4C v |\partial \mfL| \frac{\beta}{3\pi} \br{\frac{\mu \Delta \ell}{2v}}^{3}  \mathcal{F}(\ell)   e^{-\mu \Delta \ell/4}  \br{1+2^3 \cdot 3! \brr{ \max\br{1,\frac{4}{\mu \Delta \ell}}}^4}  \notag \\
&\le   4C v |\partial \mfL| \frac{\beta}{3\pi} \br{\frac{\mu \Delta \ell}{2v}}^{3}  \mathcal{F}(\ell)   e^{-\mu \Delta \ell/4} \br{1+2^3 \cdot 3! }  \le
\frac{703 C  |\partial \mfL|  \beta}{\pi v^2}  \mathcal{F}(\ell)  ,
\end{align}
where, in the last inequality, we use 
 \begin{align}
e^{-\mu \Delta \ell/4} (\mu \Delta \ell)^3  \le86.0321\cdots 
\end{align}
from $\mu \Delta \ell\ge 4$ as in~\eqref{ell_condl_inequality_g13}. 
Also, by replacing the function $f_{t_0}(s)$ in~\eqref{upper_bound_s^D_mathcal_F_ell-s_f_2term_re_ag} by the form of~\eqref{integral_t_g_beta_t_ge_delta_re__agg}, we have the following inequality: 
 \begin{align}
 \label{s^D_mathcal_F_ell-s_f_2term_re_ag_g_3}
2\sum_{s=1}^\infty \mathcal{F}(\ell-s\Delta \ell)  f_{t_0}(s) 
&\le 
2\mathcal{F}(\ell)\brrr{ f_1e^{\kappa_\beta \Delta \ell}  \br{ 1+e^{\kappa_\beta \Delta \ell}} 
+ \sum_{s=3}^\infty e^{\kappa_\beta s \Delta \ell}\cdot \brr{ \frac{\mu \beta }{\pi v} (s-2) \Delta \ell   +\frac{\beta^2}{12}} e^{-2\kappa_\beta (s-2)\Delta \ell} }\notag \\
&\le2 \beta^2 \mathcal{F}(\ell) e^{\kappa_\beta \Delta \ell} \br{  \frac{3\mu \Delta \ell}{\pi v \beta}
 + \frac{1}{6} + \frac{1}{6} e^{\kappa_\beta \Delta \ell} },
 \end{align}
where we use the following inequality: 
 \begin{align} 
&\sum_{s=3}^\infty e^{\kappa_\beta s \Delta \ell}\cdot \brr{ \frac{\mu \beta }{\pi v} (s-2) \Delta \ell   +\frac{\beta^2}{12}} e^{-2\kappa_\beta (s-2)\Delta \ell}  \notag \\
&\le \frac{\mu \beta \Delta \ell}{\pi v} e^{2\kappa_\beta \Delta \ell} 
\sum_{s=1}^\infty s e^{- \kappa_\beta s \Delta \ell} + \frac{\beta^2}{12}e^{4\kappa_\beta \Delta \ell}   \sum_{s=3}^\infty  e^{- \kappa_\beta s \Delta \ell} \notag \\
&\le \frac{\mu \beta \Delta \ell}{\pi v} e^{\kappa_\beta \Delta \ell} 
 \br{1+2 \brr{ \max\br{1,\frac{1}{\kappa_\beta \Delta \ell}}}^2}  
 + \frac{\beta^2}{12} \frac{e^{\kappa_\beta \Delta \ell}}{1-e^{- \kappa_\beta \Delta \ell} }   
 \le \br{  \frac{3\mu \beta \Delta \ell}{\pi v}
 + \frac{\beta^2}{6} }    e^{\kappa_\beta \Delta \ell} .
 \end{align}
Note that in the above inequality, we use $1/(\kappa_\beta \Delta \ell) \le 1$ ($\kappa_\beta \Delta \ell\ge 1$). 
 By using the above inequalities~\eqref{mathcal_F_ell-s_f_2term_0_re_ag_g_3} and~\eqref{s^D_mathcal_F_ell-s_f_2term_re_ag_g_3} with $f_1\le \beta^2/12$, we obtain 
\begin{align}
\label{upper_bound_s^D_mathcal_F_ell-s_f_2term_fin_re_ag___g3}
&2f_1 \mathcal{F}(\ell) + 2\sum_{s=1}^\infty \mathcal{F}(\ell-s\Delta \ell) \brr{ C|\partial (\mfL[\ell])| f_2 v e^{-\mu s \Delta \ell /2} + f_{t_0}(s)} \notag \\
& \le 
\beta^2  \mathcal{F}(\ell) \brr{ \frac{1}{6} + \frac{703 C  |\partial \mfL|}{\pi v^2 \beta}  
+ 2e^{\kappa_\beta \Delta \ell} \br{  \frac{3\mu \Delta \ell}{\pi v \beta}
 + \frac{1}{6} + \frac{1}{6} e^{\kappa_\beta \Delta \ell} }}.
 \end{align}
Therefore, by combining the inequalities~\eqref{est_f_1andf_2_g_1case_re_ag}, \eqref{upper_bound_s^D_mathcal_F_ell-s_f_2term_0_re_ag}
 and \eqref{upper_bound_s^D_mathcal_F_ell-s_f_2term_re_ag}, we obtain the second inequality in \eqref{main:eq:prop:Quasi-local_commutator_bound_tau_1DD} with $\mathfrak{g}_3^{(1)}$ provided by Eq.~\eqref{main:eq:prop:modified_ver_g_1and_g_3}.
This completes the proof of Proposition~\ref{prop:Quasi-local_commutator_bound_tau_improve_1D}. $\square$

\section{Clustering theorem for conditional mutual information at arbitrary temperatures}\label{sec;completing_Final_proof}

\subsection{High-dimensional systems}

We now have all the ingredients to prove the clustering theorem for conditional mutual information.
We begin with the general dimensional cases, which can be treated by the PTP formalism in Sec.~\ref{sec:PTP_formalism}.
We prove the following theorem on the decay of the conditional mutual information:
\begin{theorem} \label{CMI_decay_PTP_general_D}
Using $\Theta$ notation in Eq.~\eqref{Theta_notation_def}, we obtain 
 \begin{align}
\mI_{\rho_\beta}(A:C|B) \le 
   \mathcal{D}_{AC}   e^{-R/\Theta\brr{\beta^{D+1}\log \br{R}} + \Theta(1)\log(R)}    
 \label{main:ineq:CMI_decay_PTP_general_D}
\end{align}
for an arbitrary tripartition of the total system $\Lambda=A\sqcup B\sqcup C$,
where $R=\dist_{A,C}$, and $\mathcal{D}_{AC}$ has been defined as the Hilbert space dimension on the subset $AC$.
\end{theorem}

{\bf Remark.} From the upper bound, the conditional mutual information decays beyond the distance $\tilde{\mathcal{O}}\brr{\beta^{D+1} \log(\mathcal{D}_{AC})}$, and the asymptotic decay is given by $e^{-R/\log(R)}$. 
Therefore, when $|A|$ and $|C|$ is order of $1$, we have 
 \begin{align}
\mI_{\rho_\beta}(A:C|B) \le e^{-R/\Theta\brr{\beta^{D+1}\log \br{R}}} .
\end{align}
This gives the pairwise Markov structure in general quantum Gibbs states, while we cannot ensure the local and the global Markov structures, which require $\max(|A|,|C|)=\orderof{|\Lambda|}$.  

The RHS of the inequality~\eqref{main:ineq:CMI_decay_PTP_general_D} has the dependence of $e^{\Theta\br{|AC| }}$, which originates from the coefficient $\mathcal{D}_{AC}=e^{\Theta(|AC|)}$.
This condition is necessary so that the approximate PTP operator $P_\tau$ has a sufficiently small error to ensure the exact partial trace. 
In the proof, it appears in the inequality~\eqref{ineq:corol:error_est_APTP_continuity_bound_CMI_again}. 
As will be proven in the subsequent section~\ref{sec:One-dimensional case: improved bound}, we can resolve this point in one-dimensional cases.  

%

\subsubsection{Proof of Theorem~\ref{CMI_decay_PTP_general_D}}

\begin{figure}[tt]
\centering
\includegraphics[clip, scale=0.4]{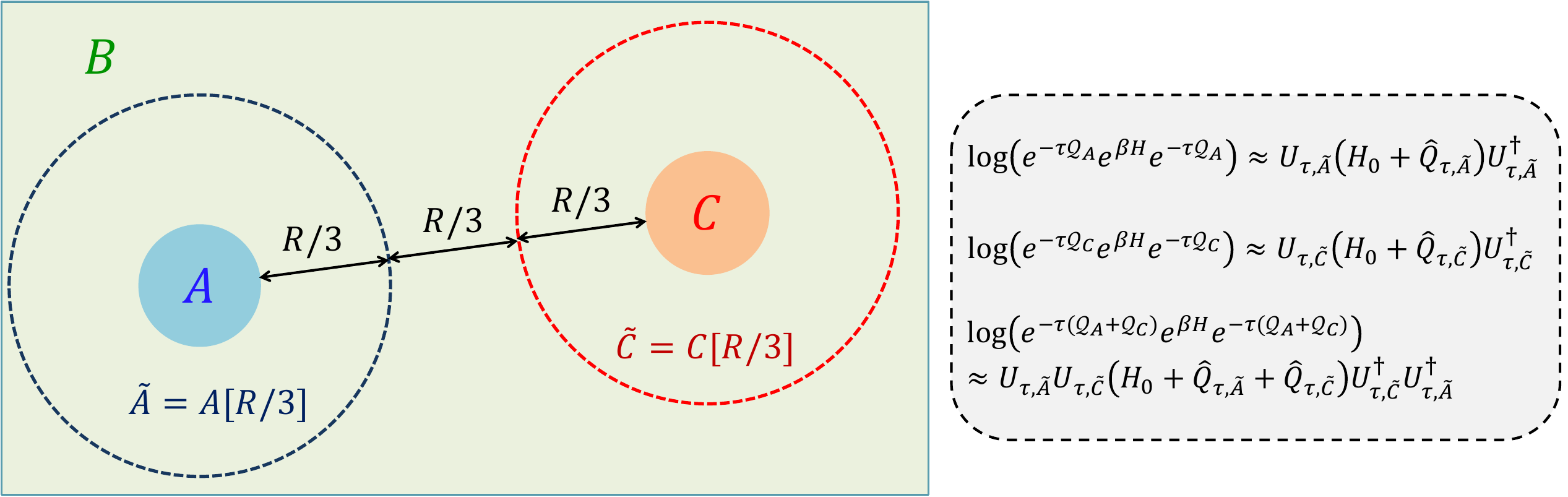}
\caption{Clustering theorem using the PTP formalism. We consider the approximate PTP operator $e^{\mQ_A}$ and $e^{\mQ_C}$ for the subsets $A$ and $C$, respectively. 
Using Theorem~\ref{thm:Refined locality estimation for effective Hamiltonian}, we can ensure that the effects of the PTP operators of $e^{\mQ_A}$ and $e^{\mQ_C}$ are approximately localized around $A$ and $C$.
We thus consider the approximations onto $\tilde{A}=A[R/3]$ and $\tilde{C}=C[R/3]$ and utilize  Lemma~\ref{lem:error_est_APTP_continuity_bound} to derive the upper bound for the conditional mutual information.
}
\label{fig:CMI_clustering_high}
\end{figure}

By applying Lemma~\ref{lem:error_est_APTP_continuity_bound} and Corollary~\ref{corol:error_est_APTP_continuity_bound_CMI} to the quantum Gibbs state $\rho_\beta$, we upper-bound the conditional mutual information (see also Fig.~\ref{fig:CMI_clustering_high}). 
For the convenience of readers, we show the statement again:

{~}\\
{\bf Lemma~\ref{lem:error_est_APTP_continuity_bound} and Corollary~\ref{corol:error_est_APTP_continuity_bound_CMI} (restatement).}
\textit{Let $\rho_{\beta,L^\co,\tau}$ be defined as 
 \begin{align}
 \label{definition_rho_L^co_tau_PTP_again}
\rho_{\beta,L^\co,\tau}:=  \frac{\mP_{L,\tau} \rho_{\beta} \mP_{L,\tau}}{\tr\br{\mP_{L,\tau} \rho_{\beta} \mP_{L,\tau}}} 
=\frac{e^{-\tau \mQ_L}\rho_{\beta} e^{-\tau \mQ_L}}{\tr\br{e^{-\tau \mQ_L} \rho_{\beta} e^{-\tau \mQ_L} }}  .
\end{align}
Then, for an arbitrary tripartition of $\Lambda=A\sqcup B \sqcup C$, the quantum conditional mutual information $\mI_{\rho_\beta}(A:C|B)$ is upper-bounded as follows:
\begin{align}
\label{ineq:corol:error_est_APTP_continuity_bound_CMI_again}
\mI_{\rho_\beta}(A:C|B) \le  8  e^{-\tau } \mathcal{D}_{AC} \br{\mathcal{\chi}_{\tau,\rho_\beta,AB} +\mathcal{\chi}_{\tau,\rho_\beta,BC} + \mathcal{\chi}_{\tau,\rho_\beta,B}} 
 + \norm{H_{\rho_\beta, \tau}(A:C|B)} ,
\end{align}
where we define 
 \begin{align}
\mathcal{\chi}_{\tau,\rho_\beta,L^\co}:=  \sup_{u_L} \norm{\brr{\log\br{\rho_{\beta,L^\co,\tau}}, u_L}} + \int_{\tau}^\infty   e^{\tau-\tau_1}\sup_{u_L} \norm{\brr{\log\br{\rho_{\beta,L^\co,\tau_1}}, u_L}} d\tau_1 .
\end{align}
and 
 \begin{align}
 \label{def_H_rho_beta_tau_A:C|B}
H_{\rho_\beta,\tau}(A:C|B):=- \log(\rho_{\beta,AB,\tau}) - \log(\rho_{\beta,BC,\tau}) +\log(\rho_{\beta,ABC,\tau})+\log(\rho_{\beta,B,\tau}) .
\end{align}
}

To estimate the RHS of the inequality~\eqref{ineq:corol:error_est_APTP_continuity_bound_CMI_again}, we begin with upper-bounding $\mathcal{\chi}_{\tau,\rho_\beta,L^\co}$. 
We prove the following lemma:
\begin{lemma}  \label{lemma:upp_bound_chi/tau/rho_beta_Lco}
Using the inequality~\eqref{main_eq:prop:quasi-locality of_ad__H__tau_V_2} with $O\to u_L$ and $V\to \mQ_L$ ($\norm{\mQ_L}=1$) in Proposition~\ref{prop:quasi-locality of_ad__H__tau_V}, we can upper-bound the parameter $\mathcal{\chi}_{\tau,\rho_\beta,L^\co}$ as follows:
 \begin{align}
\mathcal{\chi}_{\tau,\rho_\beta,L^\co}\le  \Theta (\beta^{10D}, \tau^{10D+1},|L|^3)  ,
\end{align}
where we use the $\Theta$ notation in Eq.~\eqref{Theta_notation_def}. 
\end{lemma}

\textit{Proof of Lemma~\ref{lemma:upp_bound_chi/tau/rho_beta_Lco}.}
We first note that 
\begin{align}
\log\br{\rho_{\beta,L^\co,\tau}}=  \log \br{\frac{e^{-\tau \mQ_L}\rho_{\beta} e^{-\tau \mQ_L}}{\tr\br{e^{-\tau \mQ_L} \rho_{\beta} e^{-\tau \mQ_L} }} } = \log \br{e^{-\tau \mQ_L}e^{\beta H}e^{-\tau \mQ_L}} - \log \brr{Z_\beta \tr\br{e^{-\tau \mQ_L} \rho_{\beta} e^{-\tau \mQ_L} }} .
\end{align}
The first term characterizes the effective Hamiltonian $\hat{H}_\tau$ in Subtheorem~\ref{sub_thm:U_tau_u_i_commun}.
Because the operator $u_L$ is strictly localized on the subset $L$, we can apply Proposition~\ref{prop:quasi-locality of_ad__H__tau_V} to 
$\ad_{\hat{H}_\tau}(u_L)$.
We then obtain the upper bound as in~\eqref{main_eq:prop:quasi-locality of_ad__H__tau_V_2}: 
 \begin{align}
\sup_{u_L} \norm{\brr{\log\br{\rho_{\beta,L^\co,\tau_1}}, u_L}} \le g'_{\tau_1},
\end{align}
where $g'_{\tau_1}=2  \tau_1 + 2^{2D+5} \gamma^2 |L|^2C_{2,\tau_1}^{2D} \bar{J}_0 \br{ 1 +  8\gamma |L| C_{3,\tau_1}^{3D}  }$, and we adopt the definition~\eqref{def_C_V_tau_r_without_check} for $C_{\nu,\tau}$, i.e., $C_{\nu,\tau_1}:=\frac{\tau_1+e}{2}+ \frac{32}{\kappa_\beta^2}  \br{\nu^2 D^2 + \kappa_1^2 \tau_1^2+\frac{\kappa_\beta\kappa_0\tau_1}{8}}$. 
Using the upper bound in Ref.~\cite{pinelis2020exact} for the incomplete Gamma function, we obtain 
 \begin{align}
 \label{incomplete_gamma_upp}
 \int_{\tau}^\infty   e^{\tau-\tau_1} \tau_1^{m-1} d\tau_1 \le \frac{\br{\tau+\Gamma(m + 1)^{1/(m - 1)}}^m-\tau^m}{m \Gamma(m + 1)^{1/(m - 1)}}
 \le \frac{\br{\tau+m+1}^m}{2m}
\end{align}
for $m\ge 2$, where $\Gamma(m)$ is the Gamma function.  For $m=1$, we can trivially obtain $ \int_{\tau}^\infty   e^{\tau-\tau_1} d\tau_1=1$. 
Here, $g'_{\tau_1}$ is described by a $(10D)$th order polynomial with respect to $\tau$. 
We therefore obtain 
 \begin{align}
 \int_{\tau}^\infty   e^{\tau-\tau_1} g'_{\tau_1} d\tau_1 = \Theta (\beta^{10D}, \tau^{10D+1},|L|^3) ,
\end{align}
where we use~\eqref{incomplete_gamma_upp} with $m\le 10D+1=\orderof{1}$.
This completes the proof. $\square$ 

{~}

\hrulefill{\bf [ End of Proof of Lemma~\ref{lemma:upp_bound_chi/tau/rho_beta_Lco}]}

{~}

By applying Lemma~\ref{lemma:upp_bound_chi/tau/rho_beta_Lco} to~\eqref{ineq:corol:error_est_APTP_continuity_bound_CMI_again}, we have 
\begin{align}
\label{8_e^-tau_mathcal_D_AC_upp}
8  e^{-\tau } \mathcal{D}_{AC} \br{\mathcal{\chi}_{\tau,\rho_\beta,AB} +\mathcal{\chi}_{\tau,\rho_\beta,BC} + \mathcal{\chi}_{\tau,\rho_\beta,B}}
&\le  e^{-\tau } \mathcal{D}_{AC} \Theta (\beta^{10D}, \tau^{10D+1},|AC|^3)  \notag \\
&\le   \mathcal{D}_{AC}  e^{-\tau/2 + \Theta(1)\log(\beta|AC|)}   ,
\end{align} 
where we use $\Theta (\tau^{10D+1}) e^{-\tau} \le \Theta(1) e^{-\tau/2}$. 
Then, the remaining task is to estimate the norm of  $H_{\rho_\beta,\tau}(A:C|B)$ in Eq.~\eqref{def_H_rho_beta_tau_A:C|B}.
We now prove the following lemma, which is derived from Theorem~\ref{thm:Refined locality estimation for effective Hamiltonian}
\begin{lemma}  \label{lemma:upp_bound_norm_CMI_Hamiltonian}
The norm of the operator $H_{\rho_\beta,\tau}(A:C|B)$ in Eq.~\eqref{def_H_rho_beta_tau_A:C|B} is upper-bounded by
\begin{align}
\norm{H_{\rho_\beta,\tau}(A:C|B) }
 &\le 6e^{-\tau}+e^{ \Theta( \tau \beta^D) \log \br{\beta |AC| R \tau} -\kappa_\beta R/6}+\Theta(|AC|,R^{2D},\beta) e^{-4\kappa_\beta R/3} \notag \\
&\le 6e^{-\tau}+ e^{ \Theta( \tau \beta^D) \log \br{\beta |AC| R \tau} -\kappa_\beta R/6} . 
  \label{main_ineq:lemma:upp_bound_norm_CMI_Hamiltonian}
\end{align} 
\end{lemma}

\textit{Proof of Lemma~\ref{lemma:upp_bound_norm_CMI_Hamiltonian}.}
We first define $H'_{\rho_\beta,\tau}(A:C|B)$ by slightly modifying the definition~\eqref{def_H_rho_beta_tau_A:C|B} as 
 \begin{align}
 \label{def_H_rho_beta_tau_A:C|B'ver}
&H'_{\rho_\beta,\tau}(A:C|B)
:=
- \log(e^{-\tau \mQ_C}\rho_{\beta} e^{-\tau \mQ_C}) 
- \log(e^{-\tau \mQ_A}\rho_{\beta} e^{-\tau \mQ_A}) 
+ \log(\rho_{\beta})
+\log(e^{-\tau (\mQ_A+\mQ_C)}\rho_{\beta}  e^{-\tau (\mQ_A+\mQ_C)}) \notag \\
&=- \log(e^{-\tau \mQ_C}e^{\beta H}  e^{-\tau \mQ_C}) 
- \log(e^{-\tau \mQ_A}e^{\beta H}  e^{-\tau \mQ_A}) 
+\beta H 
+\log(e^{-\tau (\mQ_A+\mQ_C)}e^{\beta H}  e^{-\tau (\mQ_A+\mQ_C)}),
\end{align}
where the definition removes the normalization factors in Eq.~\eqref{definition_rho_L^co_tau_PTP_again}. 
Note that $\mQ_{(ABC)^\co}=\mQ_{\emptyset} = \hat{0}$. 
Using the inequality~\eqref{main:eq:lem:error_est_APTP} in Lemma~\ref{lem:error_est_APTP}, we obtain 
\begin{align}
&\norm{e^{-\tau \mQ_L }\rho_{\beta} e^{-\tau \mQ_L } - \mP_{L}\rho_{\beta} \mP_{L} }_1 \le 2 e^{-\tau} 
\longrightarrow 
\tr \br{e^{-\tau \mQ_L }\rho_{\beta}  e^{-\tau \mQ_L }} \le  1+ 2 e^{-\tau} 
\end{align}
for an arbitrary $L \subseteq \Lambda$.
Therefore, from the inequality of $\log(1+x)\le x$ for $x\ge0$, we obtain 
 \begin{align}
 \label{def_H_rho_beta_tau_A:C|B'ver_dif}
&\norm{ H'_{\rho_\beta,\tau}(A:C|B) -H_{\rho_\beta,\tau}(A:C|B) }
\le 6e^{-\tau}.
\end{align}
In the following, we aim to estimate the upper bound for $\norm{H'_{\rho_\beta,\tau}(A:C|B)}$ instead of $\norm{H_{\rho_\beta,\tau}(A:C|B) }$ relying on Theorem~\ref{thm:Refined locality estimation for effective Hamiltonian}.

For the convenience of readers, we first restate Theorem~\ref{thm:Refined locality estimation for effective Hamiltonian} with the use of the inequalities~\eqref{error_simplest_form_thm} and \eqref{error_simplest_form_thm_for_eff_ham}:

{~}\\
{\bf Theorem~\ref{thm:Refined locality estimation for effective Hamiltonian} (restatement).}
\textit{Let $\tilde{\mfL}$ be an extended subset from $\mfL$ by a distance $r$, i.e., 
\begin{align}
\tilde{\mfL} := \mfL[r] .
\end{align}
Then, we construct $\hat{H}_{\tau,\tilde{\mfL}}$ using the subset Hamiltonian $H_{0,\tilde{\mfL}}$ on $\tilde{\mfL}\subset \Lambda$ as follows:
\begin{align}
\label{def_hat_H_tau_tilde_mfL}
\hat{H}_{\tau,\tilde{\mfL}} = \log \br{e^{\tau V_{\tilde{\mfL}}} e^{\beta H_{0,\tilde{\mfL}}}e^{\tau V_{\tilde{\mfL}}}}  .
\end{align}
\begin{align}
\norm{  U_\tau  - U_{\tau,\tilde{\mfL}} } \le   e^{ \Theta( \tau \beta^D)\norm{V} \log (\beta\norm{V}\cdot |\mfL| r \tau) -\kappa_\beta r/2}  ,
\label{main_inq:thm:Refined locality estimation for effective Hamiltonian_re}
\end{align}
and 
\begin{align}
&\norm{\log \br{e^{\tau V} e^{\beta H_0}e^{\tau V}} - U_{\tau,\tilde{L}} (\beta H_0 +\hat{V}_{\tau,\tilde{\mfL}}) U_{\tau,\tilde{\mfL}} ^\dagger}
\le  e^{ \Theta( \tau \beta^D)\norm{V} \log (\beta\norm{V}\cdot |\mfL| r \tau) -\kappa_\beta r/2} .
\label{error_simplest_form_thm_for_eff_ham_rre}
\end{align}
}

To utilize the theorem with $H_0=H$ and $V=-\mQ_A$ ($\mfL=A$), $V=-\mQ_C$ ($\mfL=C$), $V=-\mQ_A-\mQ_C$ ($\mfL=A\sqcup C$), we first define $U_{\tau,\tilde{A}}$ and $U_{\tau,\tilde{C}}$ by the following effective Hamiltonian 
\begin{align}
&\hat{H}_{\tau,\tilde{A}} = \log \br{e^{-\tau \mQ_A} e^{\beta H_{\tilde{A}}}e^{-\tau \mQ_A} }
= U_{\tau,\tilde{A}}\brr{\beta H_{\tilde{A}} + \tau \mQ_{\tau,\tilde{A}}} U_{\tau,\tilde{A}}^\dagger , \notag \\
&\mQ_{\tau,\tilde{A}}= -2 \int_0^\tau U_{\tau_1,\tilde{A}}^\dagger \mQ_A U_{\tau_1,\tilde{A}}
\end{align}
and 
\begin{align}
&\hat{H}_{\tau,\tilde{C}} = \log \br{e^{-\tau \mQ_C}  e^{\beta H_{\tilde{C}}}e^{-\tau \mQ_C}}  
= U_{\tau,\tilde{C}}\brr{\beta H_{\tilde{C}} + \tau \mQ_{\tau,\tilde{C}}} U_{\tau,\tilde{C}}^\dagger , \notag \\
&\mQ_{\tau,\tilde{C}}= -2 \int_0^\tau U_{\tau_1,\tilde{C}}^\dagger \mQ_C U_{\tau_1,\tilde{C}} ,
\end{align}
where we define $\tilde{A}=A[R/3]$ and $\tilde{C}=C[R/3]$ and use the expression in Corollary~\ref{connection_of_exponential_operator_repeat}. 
Then, from the inequality~\eqref{error_simplest_form_thm_for_eff_ham_rre} in Theorem~\ref{thm:Refined locality estimation for effective Hamiltonian}, we have 
\begin{align}
&\norm{\log\br{e^{-\tau \mQ_A}e^{\beta H} e^{-\tau \mQ_A}} - U_{\tau,\tilde{A}}\brr{\beta H + \tau \mQ_{\tau,\tilde{A}}} U_{\tau,\tilde{A}}^\dagger} 
\le e^{ \Theta( \tau \beta^D) \log (\beta \cdot |A| R \tau) -\kappa_\beta R/6}
, \notag \\
&\norm{\log\br{e^{-\tau \mQ_C}e^{\beta H} e^{-\tau \mQ_C}}
-  U_{\tau,\tilde{C}}\brr{\beta H + \tau \mQ_{\tau,\tilde{C}}} U_{\tau,\tilde{C}}^\dagger}
\le e^{ \Theta( \tau \beta^D) \log (\beta \cdot |C| R \tau) -\kappa_\beta R/6}, \notag \\
&\norm{\log\br{e^{-\tau(\mQ_A+\mQ_C)}e^{\beta H} e^{-\tau(\mQ_A+\mQ_C)}} 
-  U_{\tau,\tilde{A}} U_{\tau,\tilde{C}}\brr{\beta H +\tau \mQ_{\tau,\tilde{A}}+ \tau \mQ_{\tau,\tilde{C}}} U_{\tau,\tilde{C}}^\dagger U_{\tau,\tilde{A}}^\dagger}  \notag \\
&\le e^{ \Theta( \tau \beta^D) \log \br{\beta |AC| R \tau} -\kappa_\beta R/6},
\end{align}
where, in the third inequality, we use $e^{\beta H_{0,\tilde{A} \sqcup \tilde{C}}}= e^{\beta H_{0,\tilde{A}}} \otimes e^{\beta H_{0,\tilde{C}}} $, which reduces Eq.~\eqref{def_hat_H_tau_tilde_mfL} to
\begin{align}
\log \br{e^{-\tau(\mQ_A+\mQ_C)} e^{\beta H_{0,\tilde{A} \sqcup \tilde{C}}} e^{-\tau(\mQ_A+\mQ_C)}} 
= \log \br{e^{-\tau\mQ_A} e^{\beta H_{0,\tilde{A}}} e^{-\tau \mQ_A}} +  \log \br{e^{-\tau\mQ_C} e^{\beta H_{0,\tilde{C}}} e^{-\tau \mQ_C}}   .
\end{align}
Therefore, by defining $H''_{\rho_\beta,\tau}(A:C|B)$
 \begin{align}
 \label{def_H_rho_beta_tau_A:C|B''ver}
H''_{\rho_\beta,\tau}(A:C|B)
:=&
- U_{\tau,\tilde{A}}\brr{\beta H + \tau \mQ_{\tau,\tilde{A}}} U_{\tau,\tilde{A}}^\dagger
- U_{\tau,\tilde{C}}\brr{\beta H + \tau \mQ_{\tau,\tilde{C}}} U_{\tau,\tilde{C}}^\dagger \notag \\
&+\beta H 
+U_{\tau,\tilde{A}} U_{\tau,\tilde{C}}\brr{\beta H +\tau \mQ_{\tau,\tilde{A}}+ \tau \mQ_{\tau,\tilde{C}}} U_{\tau,\tilde{C}}^\dagger U_{\tau,\tilde{A}}^\dagger ,
\end{align}
we have 
 \begin{align}
 \label{def_H_rho_beta_tau_A:C|B''ver_ver'_dif}
\norm{H'_{\rho_\beta,\tau}(A:C|B)-H''_{\rho_\beta,\tau}(A:C|B)}
\le  e^{ \Theta( \tau \beta^D) \log \br{\beta |AC| R \tau} -\kappa_\beta R/6}.
\end{align}

We finally estimate the norm of $\norm{H''_{\rho_\beta,\tau}(A:C|B)}$. If there are no interactions between the regions $\tilde{A}$ and $\tilde{C}$, i.e., $H=H_{\tilde{A},\tilde{B}}+H_{\tilde{B},\tilde{C}}$ [$\tilde{B}=\Lambda\setminus (\tilde{A}\sqcap \tilde{C})$], we can immediately obtain $\norm{H''_{\rho_\beta,\tau}(A:C|B)}=0$. Therefore, by letting $h_{\tilde{A},\tilde{C}}$ be the interaction term as $\sum_{Z: Z\cap \tilde{A}=\emptyset ,Z\cap \tilde{C}=\emptyset} h_Z$, we have 
 \begin{align}
  \label{def_H_rho_A:C|B''ver_norm}
\norm{H''_{\rho_\beta,\tau}(A:C|B)}\le 4\beta \norm{h_{\tilde{A},\tilde{C}}} \le 4\beta \bar{J}_0 |\tilde{A}| \cdot |\tilde{C}| e^{-\mu R/3} 
\le \Theta(|AC|,R^{2D},\beta) e^{-4\kappa_\beta R/3}  ,
\end{align}
where we use the inequality~\eqref{sum_interaction_terms_main_eq} in Lemma~\ref{sum_interaction_terms} and $\mu\ge 4\kappa_\beta$ from Eq.~\eqref{definition_kappa_beta_1}.

By combining the inequalities~\eqref{def_H_rho_beta_tau_A:C|B'ver_dif}, \eqref{def_H_rho_beta_tau_A:C|B''ver_ver'_dif} and \eqref{def_H_rho_A:C|B''ver_norm}, 
 \begin{align}
& \norm{H_{\rho_\beta,\tau}(A:C|B) } \notag \\
&\le \norm{ H'_{\rho_\beta,\tau}(A:C|B) -H_{\rho_\beta,\tau}(A:C|B) } + \norm{H'_{\rho_\beta,\tau}(A:C|B)-H''_{\rho_\beta,\tau}(A:C|B)} + \norm{H''_{\rho_\beta,\tau}(A:C|B)} \notag \\
&\le 6e^{-\tau}+e^{ \Theta( \tau \beta^D) \log \br{\beta |AC| R \tau} -\kappa_\beta R/6}+\Theta(|AC|,R^{2D},\beta) e^{-4\kappa_\beta R/3} .
\end{align}
We thus prove the main inequality. This completes the proof. $\square$

{~}

\hrulefill{\bf [ End of Proof of Lemma~\ref{lemma:upp_bound_norm_CMI_Hamiltonian}]}

{~}

By applying the inequalities~\eqref{8_e^-tau_mathcal_D_AC_upp} and~\eqref{main_ineq:lemma:upp_bound_norm_CMI_Hamiltonian} to~\eqref{ineq:corol:error_est_APTP_continuity_bound_CMI_again}, we finally obtain 
\begin{align}
\label{ineq:bound_CMI_fin}
\mI_{\rho_\beta}(A:C|B) 
&\le \mathcal{D}_{AC}  e^{-\tau/2 + \Theta(1)\log(\beta |AC|)} + e^{ \Theta( \tau \beta^D) \log \br{\beta |AC| R \tau} -\kappa_\beta R/6}  \notag\\
 &\le \mathcal{D}_{AC}  e^{-\tau/2 + \Theta(1)\log(\tau R)} + e^{ \Theta( \tau \beta^D) \log \br{\tau R} -\kappa_\beta R/6} ,\end{align}
which holds for an arbitrary positive $\tau>0$, where we use the fact that the inequality is meaningful only for $\tau \ge  \Theta(|AC|)$ and $R\ge \Theta(\beta)$.
Therefore, by choosing $\tau$ such that 
\begin{align}
&\Theta( \tau \beta^D) \log \br{\tau R} -\frac{\kappa_\beta R}{6} = - \frac{\kappa_\beta R}{12} \notag \\
&\longrightarrow  \tau=\frac{R}{\Theta\br{\beta^{D+1}} \log \br{R} } \ge \frac{R}{\Theta\br{\beta^{D+1}} \log \br{R} }  ,
\end{align}
which reduces the inequality~\eqref{ineq:bound_CMI_fin} to 
\begin{align}
\label{Final_ineq_CMI_decayhigh_dim}
\mI_{\rho_\beta}(A:C|B) \le 
  \mathcal{D}_{AC}   e^{-R/\Theta\brr{\beta^{D+1}\log \br{R}} + \Theta(1)\log(R)}  + e^{-\kappa_\beta R/12}  .  
\end{align}
Because the first term is more dominant than the second term, we prove the main inequality~\eqref{main:ineq:CMI_decay_PTP_general_D}.
This completes the proof of Theorem~\ref{CMI_decay_PTP_general_D}. $\square$

\subsection{One-dimensional case: improved bound} \label{sec:One-dimensional case: improved bound}

We now consider an improved bound for the one-dimensional cases. 
The result in the previous section can be applied to one-dimensional systems, but it is meaningless when $A$ or $C$ is macroscopically large, i.e., $|A|=|C|=\orderof{|\Lambda|}$. 
In one-dimensional systems, we can resolve this drawback by using the BP formalism as in Sec.~\ref{sec:BP_formalism}.  
We now aim to prove the following theorem:

\begin{theorem} \label{CMI_decay_PTP_general_1D_improve}
Let us define $A\sqcup B \sqcup C=\Lambda$ and let $A_0$ and $C_0$ be subsystems such that $A_0 \subseteq A$ and $C_0\subseteq C$.
Then, for an arbitrary quantum Gibbs state on one dimension, the conditional mutual information $\mI_{\rho_\beta}(A_0:C_0|B) $ is upper-bounded by
 \begin{align}
\mI_{\rho_\beta}(A_0:C_0|B) \le  e^{\Theta(\beta) \log(R)-\kappa_\beta R/10}   ,
 \label{main:ineq:CMI_decay_PTP_1D/}
\end{align}
where $R=\dist_{A,C}$. 
Note that $\kappa_\beta$ has been defined as in Eq.~\eqref{definition_kappa_beta_1}. 
\end{theorem}

{\bf Remark.} The inequality gives the exponential decay of the conditional mutual information as $e^{\Theta(\beta) \log(R)- R/\Theta(\beta)}$ because of $\kappa_\beta =\orderof{1/\beta}$. Hence, the conditional mutual information has a correlation length of $\tilde{\mathcal{O}}(\beta^2)$. 
As has been mentioned, this scaling is the same as the length scale of the quantum belief propagation as was given in Corollary~\ref{corol:high_dimensional_applicaton_bp}.

Also, in 1D cases, the clustering of the CMI can be derived for arbitrary contiguous subsystems, i.e., $A_0\sqcup B \sqcup C_0\subset \Lambda$.
However, when the conditional region $B$ is disconnected (non-contiguous), there exists a counterexample where the decay of the CMI no longer holds.
For example, let us consider the 1D cluster Hamiltonian as~\cite{PhysRevA.80.022316,PhysRevLett.103.020506} 
 \begin{align}
H=- \sum_{i\in \Lambda} \sigma_{i-1}^z \otimes  \sigma_i^x  \otimes \sigma_{i+1}^z 
\end{align}
with $\{\sigma^x,\sigma^y,\sigma^z\}$ the Pauli matrices, where the periodic boundary condition is adopted. 
Here the quantum Gibbs state $e^{\beta H}$ is close to the cluster state up to an error of $1/\poly(|\Lambda|)$ for $\beta =\Omega [ \log (|\Lambda|)]$. 
On the other hand, if we decompose the total system into $\Lambda_{\rm odd}=\{1,3,5,\ldots\}$ and $\Lambda_{\rm even}=\{2,4,6,\ldots\}$, the CMI $\mI(A_0:C_0|B_0)$ ($A_0\sqcup B_0\sqcup C_0=\Lambda_{\rm odd}$) has an infinite correlation length for the CMI.
Thus, the subsystem states $\rho_{\beta,\Lambda_{\rm odd}}$ have a long-range CMI.

In comparison with the previous bound in Ref.~\cite{kato2016quantum}, the present upper bound is better in the following sense. 
First, the decay is given by the exponential function instead of the subexponential function $e^{-\Theta(\sqrt{R})}$.
Second, the correlation length is exponentially improved from $e^{\orderof{\beta}}$ to $\tilde{\mathcal{O}}(\beta^2)$.
The primary reason for this improvement is that we do not rely on the clustering theorem for the bipartite correlation function, which should have a correlation length of $e^{\orderof{\beta}}$.   
Third, from our derivation, we can ensure the quasi-locality of the entanglement Hamiltonian for an approximate quantum Gibbs state [see Eq.~\eqref{def_tilde_rho_beta_1D} below] after the partial trace operation.
However, to derive the quasi-locality for the exact quantum Gibbs state, 
we need a significant leap from the current analyses (see Sec.~\ref{section:Quasi-locality of the true effective Hamiltonian}).

Finally, we also mention the recent works~\cite{10.1063/5.0085358,svetlichnyy2022matrix} that prove the exponential decay of the conditional mutual information for the matrix product states (or the matrix product density matrices) under the two assumptions: i) the translation invariance, ii) existence of a constant gap in the transfer matrix\footnote{In addition, there is a coefficient that is proportional to the minimum eigenvalue of the target density matrix $\rho$; specifically, the coefficient $Q$ in Ref.~\cite[Theorem~1]{svetlichnyy2022matrix} is proportional to $\lambda_{\min}^{-3/2}$, where $\lambda_{\min}$ is the minimum eigenvalue of $\rho$. Hence, it is only applicable to quantum states with low purity.}.  
While proving these conditions is still challenging, it is of further interest to investigate the decay of conditional mutual information within the class of tensor network states, as opposed to quantum Gibbs states.
In general, the gap of the transfer matrix is related to the decay rate of the bipartite correlation function~\cite{Hern_ndez_Santana_2015}, and hence the gap is expected to be as small as $e^{-\Omega(\beta)}$. From this aspect, the CMI decay in the Gibbs state may have a qualitatively different property from the matrix product states. 

\subsubsection{Proof of Theorem~\ref{CMI_decay_PTP_general_1D_improve}}

Using the monotonicity of the conditional mutual information ~(see Refs.~\cite[Proposition 3 therein]{doi:10.1063/1.1643788} and \cite[Ineq. (1.1)]{berta2015renyi}), we prove 
 \begin{align}
 \label{monotonicity_CMI_winter}
\mI_{\rho_\beta}(A_0:C_0|B) \le  \mI_{\rho_\beta}(A:C|B) \for \forall A_0 \subseteq A,\quad C_0 \subseteq C. 
\end{align}
Hence, we only have to estimate the upper bound of $\mI_{\rho_\beta}(A:C|B) $.  

\begin{figure}[tt]
\centering
\includegraphics[clip, scale=0.41]{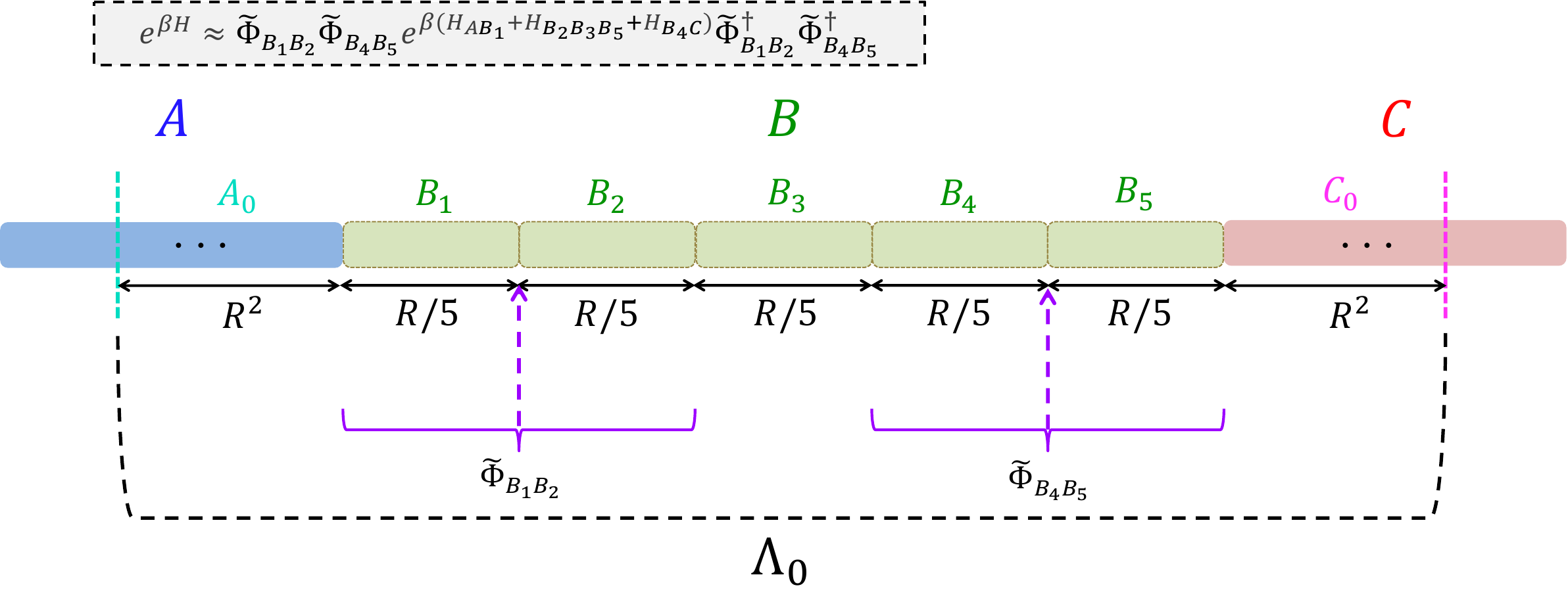}
\caption{Proof of the clustering theorem for the conditional mutual information in one dimension.
To apply the BP formalism (see Sec.~\ref{sec:BP_formalism}), we decompose the subsystem $B$ into 5 pieces as $\{B_1,B_2,B_3,B_4,B_5\}$ and approximate the effects of the partial traces for $A$ and $C$ onto $B_1B_2$ and $B_3B_4$, respectively [see also~\eqref{log_AB_U_B1B2_U'_B3B4_dif_up} and \eqref{log_AB_U_B1B2_U'_B3B4_dif_up_other_ones}].
Also, to upper-bound the conditional mutual information, we rely on Proposition~\ref{prop:bp_error_est_effective}.
As a drawback of the proposition, it includes the dependence on $\log(\mathcal{D}_{AC})=\orderof{|A|+|C|}$. 
To obtain meaningful bound in the limit of $|A|,|C|\to \infty$, we utilize the decomposition of \eqref{key_decomposition_for_norm_eff_Ham_dif_apply_1D}, which allows us to use Proposition~\ref{prop:bp_error_est_effective} for a smaller system $\Lambda_0=A_0\sqcup B \sqcup C_0$, where $A_0$ and $C_0$ are subsets of $A$ and $C$, respectively, and have the length of $R^2$. 
By combining all the techniques together, we prove Theorem~\ref{CMI_decay_PTP_general_1D_improve}. 
}
\label{fig:CMI_clustering_1D}
\end{figure}

For this purpose, we consider the following partition of the total system (see Fig.~\ref{fig:CMI_clustering_1D}):
\begin{align}
\Lambda = A \sqcup B \sqcup C=  A \sqcup B_1 \sqcup B_2 \sqcup B_3 \sqcup B_4\sqcup B_5 \sqcup C ,
\end{align}
where $|B_1|=|B_2|=|B_3|=|B_4|=|B_5|=R/5$. 
In Eqs~\eqref{X:L_brr_r_Y:L_brr_r_^co} and~\eqref{X:L_brr_r_Y:L_brr_r_^co_belief_prop}, we choose $L=AC$, $X=B_1B_4$, and $Y=B_2B_3$, 
and the following decomposition holds from the belief propagation:
\begin{align}
\label{BP_decompo_E_beta_H}
e^{\beta H} = \Phi_{\partial h_{B_5C}} \Phi_{\partial h_{AB_1}}e^{\beta \br{H_{AB_1}+ H_{B_2B_3B_4}+ H_{B_5C}}}\Phi_{\partial h_{AB_1}}^\dagger  \Phi_{\partial h_{B_5C}}^\dagger  ,
\end{align}
where the belief propagation operators $\Phi_{\partial h_{AB_1}}$ and $ \Phi_{\partial h_{B_5C}}$ are defined as follows:
\begin{align}
&e^{\beta \br{H_{AB_1B_2B_3B_4}+ H_{B_5C}}}= \Phi_{\partial h_{AB_1}}e^{\beta \br{H_{AB_1}+ H_{B_2B_3B_4} + H_{B_5C}}}\Phi_{\partial h_{AB_1}}^\dagger  ,\notag \\
&e^{\beta H} = \Phi_{\partial h_{B_5C}} e^{\beta \br{H_{AB_1B_2B_3B_4}+ H_{B_5C}}}  \Phi_{\partial h_{B_5C}}^\dagger  .
\label{BP_A_B1_B2_B3_B4_B5_C//}
\end{align}
Using Lemma~\ref{lem:BP_error_approx_local} and Corollary~\ref{corol:high_dimensional_applicaton_bp}, we approximate $\Phi_{\partial h_{AB_1}}$ and $\Phi_{\partial h_{B_5C}}$ by 
$\tilde{\Phi}_{B_1 B_2}$ and $\tilde{\Phi}_{B_4B_5}$, respectively, where they are given by
\begin{align}
\tilde{\Phi}_{B_1 B_2}= \mathcal{T} \exp \br{\int_0^1 \tilde{\phi}_{B_1B_2,\tau} d\tau}, \quad 
\tilde{\Phi}_{B_4B_5}= \mathcal{T} \exp \br{\int_0^1 \tilde{\phi}_{B_4B_5,\tau} d\tau},
\label{approx_tilde_Phi_B1_B4}
\end{align}
such that
\begin{align}
\label{main_ineq::lem:BP_error_approx_local}
\norm{\phi_{\partial h_{AB_1},\tau}-\tilde{\phi}_{B_1B_2,\tau}  }\le \bar{\phi}_{\beta, 1} e^{-\kappa_\beta R/5} 
,\quad 
\norm{\phi_{\partial h_{B_5C},\tau}-\tilde{\phi}_{B_4B_5,\tau}  }\le \bar{\phi}_{\beta, 1} e^{-\kappa_\beta R/5} ,
\end{align}
where we use $|\partial(AB_1)| = |\partial(B_5C)|=1$ in the definition~\eqref{def:bar_phi_beta_partial_L} for $\bar{\phi}_{\beta, |\partial L|}$. 
Using the truncated belief propagation operators in~\eqref{approx_tilde_Phi_B1_B4}, we approximate $e^{\beta H}$ by $e^{\beta \tilde{H}_\Lambda}$ as in
\begin{align}
e^{\beta \tilde{H}_\Lambda} := \tilde{\Phi}_{B_4B_5}\tilde{\Phi}_{B_1 B_2} e^{\beta \br{H_{AB_1}+ H_{B_2B_3B_4}+ H_{B_5C}}}\tilde{\Phi}_{B_1 B_2}^\dagger  \tilde{\Phi}_{B_4B_5}^\dagger  ,\quad 
\tilde{\rho}_\beta =\frac{e^{\beta \tilde{H}_\Lambda}}{\tilde{Z}_\beta} , 
\label{def_tilde_rho_beta_1D}
\end{align}
where we define $\tilde{Z}_\beta:=\tr\br{e^{\beta \tilde{H}_\Lambda} }$.
Note that the operator $\tilde{\phi}_{B_1B_2,\tau}$ (resp. $\tilde{\phi}_{B_4B_5,\tau}$) is quasi-local around the joint between $B_1$ and $B_2$ (resp. $B_3$ and $B_4$).

We now utilize Lemma~\ref{lem:entropy_bound_1D} for the approximate Gibbs state $\tilde{\rho}_\beta$, where the original definition for $\tilde{\rho}_\beta$ in Eq.~\eqref{def_tilde_rho_beta_delta} is now replaced by the one in Eq.~\eqref{def_tilde_rho_beta_1D}.
 For the convenience of readers, we show it again so that the statement meets our current purpose:

{~}\\
{\bf Lemma~\ref{lem:entropy_bound_1D} (restatement).}
\textit{Let us define $H_{\rho}(A:C|B)$ as 
\begin{align}
H_{\rho}(A:C|B) =-\log(\rho_{AB}) -\log(\rho_{BC})  + \log(\rho_{ABC}) + \log(\rho_{B})
\end{align}
for an arbitrary quantum state $\rho$. 
Then, we obtain 
\begin{align}
\label{main_ineq;lem:entropy_bound_1D_rrrreee}
\mI_{\rho_\beta}(A:C|B) 
\le &\norm{ H_{\tilde{\rho}_{\beta}}(A:C|B) }+4\beta\norm{H - \tilde{H}_\Lambda} +4\delta_{\beta,R/5,AB_1} +4\delta_{\beta,R/5,B_5C} ,
\end{align}
where we use the form of $\tilde{H}_\lambda$ in Eq.~\eqref{def_tilde_rho_beta_1D} and define $\delta_{\beta,\ell,L}$ by~\eqref{norm_ineq:tr_e_beta_H/00} as follows:
\begin{align}
\label{def_delta_beta_ell__rreee}
\delta_{\beta,\ell,L}:=13  \bar{\phi}_{\beta, |\partial L[\ell]|}  \norm{\partial h_{L[\ell]}}e^{2\beta \norm{\partial h_{L[\ell]}}-\kappa_\beta (\ell-1)} .
\end{align}}

First of all, we can obtain for an arbitrary concatenated region $L \subset \Lambda$
\begin{align}
\delta_{\beta,R/5,L} \le 
  \Theta(\beta^2 )e^{\Theta(\beta) -\kappa_\beta R/5}
\end{align}
for an arbitrary connected region $L \subset \Lambda$,   
where we use for $\bar{\phi}_{\beta,2}$ from Eq.~\eqref{def:bar_phi_beta_partial_L} 
\begin{align}
\label{orderof_phi_Beta_1_beta^2}
\bar{\phi}_{\beta,2} := 8\beta \gamma\tilde{J}_0   e^{\mu/2} 
 \brr{1+\frac{4\beta \gamma Cv }{7} \br{\frac{4D}{e\mu}}^{D} +\frac{8}{\pi^2}  \log\br{e+\frac{e}{\kappa_\beta}} \frac{e^{\kappa_\beta }}{e^{\kappa_\beta }-1}} \le \Theta\br{\beta^2}.
\end{align}
Therefore, we obtain the upper bound for $4\delta_{\beta,R/5,AB_1} +4\delta_{\beta,R/5,B_5C}$ in the inequality~\eqref{main_ineq;lem:entropy_bound_1D_rrrreee}.

In the following, we aim to estimate the norms of $\norm{ H_{\tilde{\rho}_{\beta}}(A:C|B)}$ and the norm
\begin{align}
\label{defini_tilde_Delta_R}
\varepsilon(R):=\beta \norm{H - \tilde{H}_\Lambda}=  \norm{\beta H- \log \br{\tilde{\Phi}_{B_4B_5}\tilde{\Phi}_{B_1 B_2} e^{\beta \br{H_{AB_1}+ H_{B_2B_3B_4}+ H_{B_5C}}}\tilde{\Phi}_{B_1 B_2}^\dagger \tilde{\Phi}_{B_4B_5}^\dagger }}, 
\end{align}
 separately. 
We first prove the following lemma for $\norm{ H_{\tilde{\rho}_{\beta}}(A:C|B)}$:
\begin{lemma} \label{Lemm:norm_H_tile_rho_beta}
Under the definition of $\tilde{\rho}_{\beta}$ as in Eq.~\eqref{def_tilde_rho_beta_1D}, we upper-bound 
\begin{align}
\norm{ H_{\tilde{\rho}_{\beta}}(A:C|B) } \le  e^{\Theta(\beta) \log(\beta R)-\kappa_\beta R/10} + 4\beta \bar{J}_0 |B_2| \cdot |B_4| e^{-\mu R/5} .
 \label{main:inq_Lemm:norm_H_tile_rho_beta}
\end{align}
Note that the second term in the RHS of the above inequality can be absorbed to the first term since $\mu\ge 4\kappa_\beta$ from Eq.~\eqref{definition_kappa_beta_1} and $4\beta \bar{J}_0 |B_2| \cdot |B_4| \le e^{\Theta(1) \log(\beta R)}$. 
\end{lemma}

{~}

\textit{Proof of Lemma~\ref{Lemm:norm_H_tile_rho_beta}.}
Throughout the proof, we let $\tilde{Z}_\beta=1$ in Eq.~\eqref{def_tilde_rho_beta_1D} for simplicity. 
For the proof, we need to consider the entanglement Hamiltonian $\log(\tilde{\rho}_{\beta,AB})$,  $\log(\tilde{\rho}_{\beta,BC})$,  $\log(\tilde{\rho}_{\beta,ABC})$ and $\log(\tilde{\rho}_{\beta,B})$ which is given by using Eq.~\eqref{partial_trace_tr_L_H_LX}:
\begin{align}
&\log(\tilde{\rho}_{\beta,AB}) 
= \log\br{ \tilde{\Phi}_{B_4B_5}\tilde{\Phi}_{B_1 B_2} e^{\beta \br{H_{AB_1}+ H_{B_2B_3B_4}+ \tilde{H}^\ast_{B_5}}}\tilde{\Phi}_{B_1 B_2}^\dagger \tilde{\Phi}_{B_4B_5}^\dagger} , \notag \\
&\log(\tilde{\rho}_{\beta,BC})
= \log\br{ \tilde{\Phi}_{B_4B_5}\tilde{\Phi}_{B_1 B_2} e^{\beta \br{\tilde{H}^\ast_{B_1}+ H_{B_2B_3B_4}+ H_{B_5C}}}\tilde{\Phi}_{B_1 B_2}^\dagger \tilde{\Phi}_{B_4B_5}^\dagger}  , \notag \\
&\log(\tilde{\rho}_{\beta,ABC}) = \log\br{ \tilde{\Phi}_{B_4B_5}\tilde{\Phi}_{B_1 B_2} e^{\beta \br{H_{AB_1}+ H_{B_2B_3B_4}+ H_{B_5C}}}\tilde{\Phi}_{B_1 B_2}^\dagger \tilde{\Phi}_{B_4B_5}^\dagger} , \notag \\
&\log(\tilde{\rho}_{\beta,B}) = \log\br{ \tilde{\Phi}_{B_4B_5}\tilde{\Phi}_{B_1 B_2} e^{\beta \br{\tilde{H}^\ast_{B_1}+ H_{B_2B_3B_4}+ \tilde{H}^\ast_{B_5}}}\tilde{\Phi}_{B_1 B_2}^\dagger \tilde{\Phi}_{B_4B_5}^\dagger}.
\label{log_rho_AB_BC_ABC_B}
\end{align}

By adopting the expression of Eq.~\eqref{hat_tilde_Phi_1_tilde_mfL_def}, 
we here define the unitary operators $U_{\tau, B_1B_2}$, $U_{\tau, B_4B_5}$, $U'_{\tau, B_1B_2}$, $U'_{\tau, B_4B_5}$ and $\hat{\tilde{\Phi}}_{\tau,B_1 B_2}$, $\hat{\tilde{\Phi}}_{\tau,B_1 B_2}'$, $\hat{\tilde{\Phi}}_{\tau,B_4B_5}$, $\hat{\tilde{\Phi}}_{\tau,B_4B_5}'$ as follows:
\begin{align}
& \log\br{\tilde{\Phi}_{B_1 B_2} e^{\beta \br{H_{B_1}+ H_{B_2}}}\tilde{\Phi}_{B_1 B_2}^\dagger }
 =U_{B_1B_2}  \brr{\beta \br{H_{B_1}+ H_{B_2}}+ \hat{\tilde{\Phi}}_{B_1 B_2} }U_{B_1B_2}^\dagger , \notag \\
 & \log\br{\tilde{\Phi}_{B_4B_5} e^{\beta \br{H_{B_4}+ H_{B_5}}}\tilde{\Phi}_{B_4B_5}^\dagger }
 =U_{B_4B_5}  \brr{\beta \br{H_{B_4}+ H_{B_5}}+ \hat{\tilde{\Phi}}_{B_4B_5} }U_{B_4B_5}^\dagger , \notag \\
& \log\br{\tilde{\Phi}_{B_1 B_2} e^{\beta \br{H^\ast_{B_1}+ H_{B_2}}}\tilde{\Phi}_{B_1 B_2}^\dagger }
 =U'_{B_1B_2}  \brr{\beta \br{H^\ast_{B_1}+ H_{B_2}}+ \hat{\tilde{\Phi}}'_{B_1 B_2} }U_{B_1B_2}^{'\dagger} , \notag \\
 & \log\br{\tilde{\Phi}_{B_4B_5} e^{\beta \br{H_{B_4}+ H^\ast_{B_5}}}\tilde{\Phi}_{B_4B_5}^\dagger}
 =U'_{B_4B_5}  \brr{\beta \br{H_{B_4}+ H^\ast_{B_5}}+ \hat{\tilde{\Phi}}'_{B_4B_5} }U_{B_4B_5}^{'\dagger}.
 \label{def_U_U'_B1B2_B4B5}
\end{align}
where we omit the index $\tau=1$ and simply denote $U_{\tau=1, B_1B_2}$ by $U_{B_1B_2}$ for example.
Then, from the inequality~\eqref{error_simplest_form_thm_for_eff_ham_1D} in Theorem~\ref{sub_thm:U_tau_u_i_commun_1D}, we approximate $\log(\rho_{AB})$ in Eq.~\eqref{log_rho_AB_BC_ABC_B} by 
\begin{align}
&\norm{\log(\tilde{\rho}_{\beta,AB}) - U_{B_1B_2} U'_{B_4B_5} \brr{\beta \br{H_{AB_1}+ H_{B_2B_3B_4}+ \tilde{H}^\ast_{B_5}} 
+ \hat{\tilde{\Phi}}_{B_1 B_2}+\hat{\tilde{\Phi}}'_{B_4B_5} } U_{B_4B_5}^{'\dagger}  U_{B_1B_2}^\dagger} \notag \\
&\le  e^{\Theta(\beta) \log(\beta R)-\kappa_\beta R/10} ,
\label{log_AB_U_B1B2_U'_B3B4_dif_up}
\end{align}
where we choose $\mfL=(\partial B_1) \cup (\partial B_4)$ and $\tilde{\mfL}=B_1B_2 B_4B_5$, and the length $r$ in~\eqref{error_simplest_form_thm_for_eff_ham_1D} is chosen as $R/5$. 
The same upper bounds as~\eqref{log_AB_U_B1B2_U'_B3B4_dif_up} hold for the norm differences of
\begin{align}
&\norm{\log(\tilde{\rho}_{\beta,BC}) - U'_{B_1B_2} U_{B_4B_5} \brr{\beta \br{\tilde{H}_{B_1}^\ast+ H_{B_2B_3B_4}+ \tilde{H}_{B_5C}} 
+ \hat{\tilde{\Phi}}'_{B_1 B_2}+\hat{\tilde{\Phi}}_{B_4B_5} } U_{B_4B_5}^\dagger  U_{B_1B_2}^{'\dagger}} \notag \\
&\norm{\log(\tilde{\rho}_{\beta,ABC}) - U_{B_1B_2} U_{B_4B_5} \brr{\beta \br{H_{AB_1}+ H_{B_2B_3B_4}+ H_{B_5C}} 
+ \hat{\tilde{\Phi}}_{B_1 B_2}+\hat{\tilde{\Phi}}_{B_4B_5} } U_{B_4B_5}^{\dagger}  U_{B_1B_2}^{\dagger}} \notag \\
&\norm{\log(\tilde{\rho}_{\beta,B}) - U'_{B_1B_2} U'_{B_4B_5} \brr{\beta \br{\tilde{H}_{B_1}^\ast+ H_{B_2B_3B_4}+ \tilde{H}^\ast_{B_5}} 
+ \hat{\tilde{\Phi}}'_{B_1 B_2}+\hat{\tilde{\Phi}}'_{B_4B_5} } U_{B_4B_5}^{'\dagger}  U_{B_1B_2}^{'\dagger}} . 
\label{log_AB_U_B1B2_U'_B3B4_dif_up_other_ones}
\end{align}
Relying on similar analyses to~\eqref{def_H_rho_beta_tau_A:C|B''ver}, \eqref{def_H_rho_beta_tau_A:C|B''ver_ver'_dif} and \eqref{def_H_rho_A:C|B''ver_norm}, we can obtain
\begin{align}
\norm{ H_{\tilde{\rho}_{\beta}}(A:C|B) } \le  e^{\Theta(\beta) \log(\beta R)-\kappa_\beta R/10} + 4\beta \bar{J}_0 |B_2| \cdot |B_4| e^{-\mu R/5} ,
\end{align}
which proves the main inequality~\eqref{main:inq_Lemm:norm_H_tile_rho_beta}.
This completes the proof. $\square$

{~}

\hrulefill{\bf [ End of Proof of Lemma~\ref{Lemm:norm_H_tile_rho_beta}]}

{~}

We next estimate the norm~\eqref{defini_tilde_Delta_R} in $\varepsilon(R)$ by utilizing Proposition~\ref{prop:bp_error_est_effective}. 
We prove the following statement: 
\begin{lemma}\label{lem:beta_H_log_tilde..._up}
Under the definition of Eq.~\eqref{defini_tilde_Delta_R} for $\varepsilon(R)$, we obtain the upper bound of 
\begin{align}
\label{main_ineq_lem:beta_H_log_tilde..._up}
\varepsilon(R)=\beta \norm{H - \tilde{H}_\Lambda}\le e^{\Theta(\beta) \log(\beta R)-\kappa_\beta R/10}  . 
\end{align}
\end{lemma}

\textit{Proof of Lemma~\ref{lem:beta_H_log_tilde..._up}.}
Here, we adopt the notation of $H_0$ as 
\begin{align}
H_0=H_{AB_1}+ H_{B_2B_3B_4}+ H_{B_5C}.
\end{align}
Using the notation, we obtain from Eq.~\eqref{BP_decompo_E_beta_H}  
\begin{align}
e^{\beta H}=\Phi_{\partial h_{B_5C}} \Phi_{\partial h_{AB_1}}e^{\beta H_0}\Phi_{\partial h_{AB_1}}  \Phi_{\partial h_{B_5C}} 
\end{align}
Then, we start with applying the decomposition of~\eqref{key_decomposition_for_norm_eff_Ham_dif} to Eq.~\eqref{defini_tilde_Delta_R}:
\begin{align}
&\norm{\beta H- \log \br{\tilde{\Phi}_{B_4B_5}\tilde{\Phi}_{B_1 B_2} e^{\beta \br{H_{AB_1}+ H_{B_2B_3B_4}+ H_{B_5C}}}\tilde{\Phi}_{B_1 B_2}^\dagger \tilde{\Phi}_{B_4B_5}^\dagger }} \notag \\
&=\norm{\log \br{\Phi_{\partial h_{B_5C}} \Phi_{\partial h_{AB_1}}e^{\beta H_0}\Phi_{\partial h_{AB_1}}^\dagger  \Phi_{\partial h_{B_5C}}^\dagger }
 - \log \br{\tilde{\Phi}_{B_4B_5}\tilde{\Phi}_{B_1 B_2} e^{\beta H_0}\tilde{\Phi}_{B_1 B_2}^\dagger \tilde{\Phi}_{B_4B_5}^\dagger } }  \notag \\
& \le \norm{\log \br{\Phi_{\partial h_{B_5C}} \Phi_{\partial h_{AB_1}}e^{\beta H_{0,\Lambda_0} }\Phi_{\partial h_{AB_1}}^\dagger  \Phi_{\partial h_{B_5C}}^\dagger }
 - \log \br{\tilde{\Phi}_{B_4B_5}\tilde{\Phi}_{B_1 B_2} e^{\beta H_{0,\Lambda_0} }\tilde{\Phi}_{B_1 B_2}^\dagger \tilde{\Phi}_{B_4B_5}^\dagger } }   \notag \\
 &\quad+ \biggl \| \log \br{\Phi_{\partial h_{B_5C}} \Phi_{\partial h_{AB_1}}e^{\beta H_0}\Phi_{\partial h_{AB_1}}^\dagger  \Phi_{\partial h_{B_5C}}^\dagger }
 - \log \br{\tilde{\Phi}_{B_4B_5}\tilde{\Phi}_{B_1 B_2} e^{\beta H_0}\tilde{\Phi}_{B_1 B_2}^\dagger \tilde{\Phi}_{B_4B_5}^\dagger }   \notag \\
 &\quad\quad - \brr{\log \br{\Phi_{\partial h_{B_5C}} \Phi_{\partial h_{AB_1}}e^{\beta H_{0,\Lambda_0} }\Phi_{\partial h_{AB_1}}^\dagger  \Phi_{\partial h_{B_5C}}^\dagger }
 - \log \br{\tilde{\Phi}_{B_4B_5}\tilde{\Phi}_{B_1 B_2} e^{\beta H_{0,\Lambda_0} }\tilde{\Phi}_{B_1 B_2}^\dagger \tilde{\Phi}_{B_4B_5}^\dagger } }   \biggr \| \notag \\
&\le
\norm{\log \br{\Phi_{\partial h_{B_5C}} \Phi_{\partial h_{AB_1}}e^{\beta H_{0,\Lambda_0} }\Phi_{\partial h_{AB_1}}^\dagger  \Phi_{\partial h_{B_5C}}^\dagger }
 - \log \br{\tilde{\Phi}_{B_4B_5}\tilde{\Phi}_{B_1 B_2} e^{\beta H_{0,\Lambda_0} }\tilde{\Phi}_{B_1 B_2}^\dagger \tilde{\Phi}_{B_4B_5}^\dagger } }   \notag \\
 &\quad+ \norm{\beta \widehat{H_{0,\Lambda_0^\co}}+\log \br{\tilde{\Phi}_{B_4B_5}\tilde{\Phi}_{B_1 B_2} e^{\beta H_{0,\Lambda_0} }\tilde{\Phi}_{B_1 B_2}^\dagger \tilde{\Phi}_{B_4B_5}^\dagger }- 
  \log \br{\tilde{\Phi}_{B_4B_5}\tilde{\Phi}_{B_1 B_2} e^{\beta H_0}\tilde{\Phi}_{B_1 B_2}^\dagger \tilde{\Phi}_{B_4B_5}^\dagger }} \notag \\
 &\quad + \norm{ 
\beta \widehat{H_{0,\Lambda_0^\co}}+ \log \br{\Phi_{\partial h_{B_5C}} \Phi_{\partial h_{AB_1}}e^{\beta H_{0,\Lambda_0} }\Phi_{\partial h_{AB_1}}^\dagger  \Phi_{\partial h_{B_5C}}^\dagger }
- \log \br{\Phi_{\partial h_{B_5C}} \Phi_{\partial h_{AB_1}}e^{\beta H_0}\Phi_{\partial h_{AB_1}}^\dagger  \Phi_{\partial h_{B_5C}}^\dagger }}
 ,
 \label{key_decomposition_for_norm_eff_Ham_dif_apply_1D}
 \end{align}
 where we define the subsets $A_0$, $C_0$ (see also Fig.~\ref{fig:CMI_clustering_1D}) as follows:
\begin{align}
\label{def_subsets_A_0_C_0}
&A_0  := \{i \in A| \dist_{i,B} \le R^2\} ,\quad C_0  := \{i \in C| \dist_{i,B} \le R^2\} , \quad  \Lambda_0:=  A_0 \sqcup B \sqcup C_0 \subset \Lambda. 
\end{align}
Also, we use the notation~\eqref{def:Ham_surface} for $\widehat{H_{0,\Lambda_0^\co}}$, which gives 
$H_{0,\Lambda_0}+\widehat{H_{0,\Lambda_0^\co}}=H_0$. 
 
For the first term in the RHS of~\eqref{key_decomposition_for_norm_eff_Ham_dif_apply_1D},  we apply Proposition~\ref{prop:bp_error_est_effective} or the inequality~\eqref{main_ineq:lema:bp_error_est_rougth}. 
For this purpose, we define $\Phi_{\partial h_{A_0 B_1}}$ and $\Phi_{\partial h_{B_5C_0}}$ in similar ways to Eq.~\eqref{BP_A_B1_B2_B3_B4_B5_C//}:
\begin{align}
&e^{\beta \br{H_{A_0B_1B_2B_3B_4}+ H_{B_5C_0}}}= \Phi_{\partial h_{A_0B_1}}e^{\beta \br{H_{A_0B_1}+ H_{B_2B_3B_4} + H_{B_5C}}}\Phi_{\partial h_{A_0B_1}}^\dagger  ,\notag \\
&e^{\beta H} = \Phi_{\partial h_{B_5C_0}} e^{\beta \br{H_{A_0B_1B_2B_3B_4}+ H_{B_5C_0}}}  \Phi_{\partial h_{B_5C_0}}^\dagger  ,
\end{align}
and
\begin{align}
\Phi_{\partial h_{A_0B_1}}= \mathcal{T} \exp \br{\int_0^1 \phi_{\partial h_{A_0B_1},\tau} d\tau}, \quad 
\Phi_{\partial h_{B_5C_0}}= \mathcal{T} \exp \br{\int_0^1 \phi_{\partial h_{B_5C_0},\tau} d\tau},
\end{align}
which also yields
\begin{align}
e^{\beta H_{\Lambda_0}}= \Phi_{\partial h_{B_5C_0}} \Phi_{\partial h_{A_0B_1}}e^{\beta H_{0,\Lambda_0}}\Phi_{\partial h_{A_0B_1}}^\dagger   \Phi_{\partial h_{B_5C_0}}^\dagger   .
\end{align}
We then obtain a similar inequality to~\eqref{key_decomposition_for_norm_eff_Ham_dif__alt} as 
\begin{align}
&\norm{\log \br{\Phi_{\partial h_{B_5C}} \Phi_{\partial h_{AB_1}}e^{\beta H_{0,\Lambda_0} }\Phi_{\partial h_{AB_1}}^\dagger  \Phi_{\partial h_{B_5C}}^\dagger }
 - \log \br{\tilde{\Phi}_{B_4B_5}\tilde{\Phi}_{B_1 B_2} e^{\beta H_{0,\Lambda_0} }\tilde{\Phi}_{B_1 B_2}^\dagger \tilde{\Phi}_{B_4B_5}^\dagger } }  \notag \\
\le& \norm{ \log \br{\Phi_{\partial h_{B_5C_0}} \Phi_{\partial h_{A_0B_1}}e^{\beta H_{0,\Lambda_0} }\Phi_{\partial h_{A_0B_1}}^\dagger   \Phi_{\partial h_{B_5C_0}}^\dagger  }-\log \br{\Phi_{\partial h_{B_5C}} \Phi_{\partial h_{AB_1}}e^{\beta H_{0,\Lambda_0} }\Phi_{\partial h_{AB_1}}^\dagger  \Phi_{\partial h_{B_5C}}^\dagger }}
 \notag \\
 &+ \norm{\log \br{\Phi_{\partial h_{B_5C_0}} \Phi_{\partial h_{A_0B_1}}e^{\beta H_{0,\Lambda_0} }\Phi_{\partial h_{A_0B_1}}^\dagger   \Phi_{\partial h_{B_5C_0}}^\dagger  }- \log \br{\tilde{\Phi}_{B_4B_5}\tilde{\Phi}_{B_1 B_2} e^{\beta H_{0,\Lambda_0} }\tilde{\Phi}_{B_1 B_2}^\dagger \tilde{\Phi}_{B_4B_5}^\dagger } }
\end{align}
We note that because of $|A_0|=|C_0|=R^2$, we have 
\begin{align}
\label{main_ineq::lem:BP_error_approx_local_A_0_C_0}
\norm{\phi_{\partial h_{AB_1},\tau}-\phi_{\partial h_{A_0B_1},\tau}}\le e^{-\Theta(R^2/\beta)} 
,\quad 
\norm{\phi_{\partial h_{B_5C},\tau}-\partial h_{B_5C_0,\tau}}\le e^{-\Theta(R^2/\beta)} ,
\end{align}
by employing similar calculations to Lemma~\ref{lem:BP_error_approx_local}. 

We now choose as 
\begin{align}
&\mA \to H_{0,\Lambda_0}, \quad  \Phi_{\mB} \to \Phi_{\partial h_{B_5C_0}} \Phi_{\partial h_{A_0B_1}}  ,\quad 
\tilde{\Phi}_{\mB}\to \Phi_{\partial h_{B_5C}} \Phi_{\partial h_{AB_1}}\Or  \tilde{\Phi}_{B_4B_5}\tilde{\Phi}_{B_1 B_2},  \notag \\
&\mB\to \partial h_{A_0B_1}+ \partial h_{B_5C_0},\quad   \delta \to  2 \bar{\phi}_{\beta, 1} e^{-\kappa_\beta R/5} + e^{-\Theta(R^2/\beta)} ,
\end{align}
where we use the upper bounds~\eqref{main_ineq::lem:BP_error_approx_local} and \eqref{main_ineq::lem:BP_error_approx_local_A_0_C_0} for
the parameter $\delta$. 
Using $\norm{\partial h_{A_0B_1}}+\norm{ \partial h_{B_5C_0}}=\orderof{1}$, the inequality~\eqref{main_ineq:lema:bp_error_est_rougth} now reduces to
 \begin{align}
\label{main_ineq:lema:bp_error_est_rougth_1D_CMI}
&\norm{\log \br{\Phi_{\partial h_{B_5C}} \Phi_{\partial h_{AB_1}}e^{\beta H_{0,\Lambda_0} }\Phi_{\partial h_{AB_1}}^\dagger  \Phi_{\partial h_{B_5C}}^\dagger }
 - \log \br{\tilde{\Phi}_{B_4B_5}\tilde{\Phi}_{B_1 B_2} e^{\beta H_{0,\Lambda_0} }\tilde{\Phi}_{B_1 B_2}^\dagger \tilde{\Phi}_{B_4B_5}^\dagger } } \notag\\
 &\le e^{\Theta(\beta) \log\br{\beta \norm{H_{0,\Lambda_0}}}  } \bar{\phi}_{\beta, 1} e^{-\kappa_\beta R/5}  
 = e^{\Theta(\beta) \log\br{\beta R}  } e^{-\kappa_\beta R/5}  ,
\end{align}
where we use $\bar{\phi}_{\beta,1}\le \Theta(\beta^2)$ in Eq.~\eqref{orderof_phi_Beta_1_beta^2} and 
\begin{align}
\norm{H_{0,\Lambda_0}}\le \sum_{i\in \Lambda_0} \sum_{Z:Z\ni i} \norm{h_Z} \le \bar{J}_0 |\Lambda_0| = \Theta(R^2) .
\end{align}

For the second and the third terms in the RHS of~\eqref{key_decomposition_for_norm_eff_Ham_dif_apply_1D},  we apply 
Theorem~\ref{sub_thm:U_tau_u_i_commun_1D} or the inequality~\eqref{error_simplest_form_thm_for_eff_ham_1D}. 
We consider the second term, and the third term can be treated in the same way. 
Using the definition of $U_{B_1,B_2}$ and $U_{B_4,B_5}$ as in Eq.~\eqref{def_U_U'_B1B2_B4B5}, 
we can derive the second inequality in \eqref{log_AB_U_B1B2_U'_B3B4_dif_up_other_ones}:
\begin{align} 
&\norm{ \log \br{\tilde{\Phi}_{B_4B_5}\tilde{\Phi}_{B_1 B_2} e^{\beta H_0}\tilde{\Phi}_{B_1 B_2}^\dagger \tilde{\Phi}_{B_4B_5}^\dagger} - U_{B_1B_2} U_{B_4B_5} \brr{\beta H_0 + \hat{\tilde{\Phi}}_{B_1 B_2}+\hat{\tilde{\Phi}}_{B_4B_5} } U_{B_4B_5}^{\dagger} U_{B_1B_2}^{\dagger}} \notag \\
&\le e^{\Theta(\beta) \log(\beta R)-\kappa_\beta R/10},  \notag \\
&\norm{ \log \br{\tilde{\Phi}_{B_4B_5}\tilde{\Phi}_{B_1 B_2} e^{\beta H_{0,\Lambda_0} }\tilde{\Phi}_{B_1 B_2}^\dagger \tilde{\Phi}_{B_4B_5}^\dagger} - U_{B_1B_2} U_{B_4B_5} \brr{\beta  H_{0,\Lambda_0} + \hat{\tilde{\Phi}}_{B_1 B_2}+\hat{\tilde{\Phi}}_{B_4B_5} } U_{B_4B_5}^{\dagger} U_{B_1B_2}^{\dagger}}\notag \\
&\le e^{\Theta(\beta) \log(\beta R)-\kappa_\beta R/10}.
\end{align}
By using the above inequality, we obtain 
\begin{align}
&\norm{\beta \widehat{H_{0,\Lambda_0^\co}}+\log \br{\tilde{\Phi}_{B_4B_5}\tilde{\Phi}_{B_1 B_2} e^{\beta H_{0,\Lambda_0} }\tilde{\Phi}_{B_1 B_2}^\dagger \tilde{\Phi}_{B_4B_5}^\dagger }- 
\log \br{\tilde{\Phi}_{B_4B_5}\tilde{\Phi}_{B_1 B_2} e^{\beta H_0}\tilde{\Phi}_{B_1 B_2}^\dagger \tilde{\Phi}_{B_4B_5}^\dagger }}  \notag \\
&\le   2 e^{\Theta(\beta) \log(\beta R)-\kappa_\beta R/10}  
+ \norm{\beta \widehat{H_{0,\Lambda_0^\co}} +\beta  U_{B_1B_2} U_{B_4B_5}  \br{ H_0 -H_{0,\Lambda_0}}   U_{B_4B_5}^{\dagger} U_{B_1B_2}^{\dagger}} \notag \\
&\le  2 e^{\Theta(\beta) \log(\beta R)-\kappa_\beta R/10}  
+\beta  \norm{\brr{ \widehat{H_{0,\Lambda_0^\co}} , U_{B_1B_2} U_{B_4B_5}}}  \notag \\
&\le  2 e^{\Theta(\beta) \log(\beta R)-\kappa_\beta R/10} +2 \beta \tilde{J}_0 e^{-\mu R^2/2} ,
\label{second_term_upp_ineq_decomp_beta_H_0_Lambda_0}
\end{align}
where we use $H_0 -H_{0,\Lambda_0}=\widehat{H_{0,\Lambda_0^\co}}$ from Eq.~\eqref{def:Hamiltonian_subset_L}, and the norm of the commutator 
$\norm{\brr{ \widehat{H_{0,\Lambda_0^\co}} , U_{B_1B_2} U_{B_4B_5}}}$ is upper-bounded by using the inequality~\eqref{sum_interaction_terms_main_eq_short_long} in Lemma~\ref{sum_interaction_terms}:
\begin{align}
\norm{\brr{ \widehat{H_{0,\Lambda_0^\co}} , U_{B_1B_2} U_{B_4B_5}}} 
&\le 2 \sum_{Z: Z\cap \Lambda_0^\co\neq \emptyset, \ Z\cap (B_1B_2B_4B_5) \neq \emptyset} \norm{h_Z} \notag \\
&\le 2 \sum_{Z: Z\cap B[R^2]\neq \emptyset, \ Z\cap B \neq \emptyset} \norm{h_Z} \le  |\partial B| \tilde{J}_0 e^{-\mu R^2/2} = 2\tilde{J}_0 e^{-\mu R^2/2}  .
\end{align}
Note that $\dist_{\Lambda_0^\co,B} =R^2$ from the definition~\eqref{def_subsets_A_0_C_0}. 
We can obtain the same inequality for the third term in the RHS of~\eqref{key_decomposition_for_norm_eff_Ham_dif_apply_1D}. 

By applying the inequalities~\eqref{main_ineq:lema:bp_error_est_rougth_1D_CMI} and \eqref{second_term_upp_ineq_decomp_beta_H_0_Lambda_0} to \eqref{key_decomposition_for_norm_eff_Ham_dif_apply_1D}, we obtain 
\begin{align}
&\norm{\log \br{\Phi_{\partial h_{B_5C}} \Phi_{\partial h_{AB_1}}e^{\beta H_0}\Phi_{\partial h_{AB_1}}^\dagger  \Phi_{\partial h_{B_5C}}^\dagger }
 - \log \br{\tilde{\Phi}_{B_4B_5}\tilde{\Phi}_{B_1 B_2} e^{\beta H_0}\tilde{\Phi}_{B_1 B_2}^\dagger \tilde{\Phi}_{B_4B_5}^\dagger } }  \notag \\
& \le e^{\Theta(\beta) \log\br{\beta R}  } e^{-\kappa_\beta R/5} + 4 e^{\Theta(\beta) \log(\beta R)-\kappa_\beta R/10} +4 \beta \tilde{J}_0 e^{-\mu R^2/2}
= e^{\Theta(\beta) \log(\beta R)-\kappa_\beta R/10}  ,
 \end{align}
which yields the main inequality~\eqref{main_ineq_lem:beta_H_log_tilde..._up}. 
This completes the proof. $\square$

{~}

\hrulefill{\bf [ End of Proof of Lemma~\ref{lem:beta_H_log_tilde..._up}]}

{~}

Finally by applying the inequalities in Lemmas~\ref{Lemm:norm_H_tile_rho_beta} and~\ref{lem:beta_H_log_tilde..._up} to~\eqref{main_ineq;lem:entropy_bound_1D_rrrreee}, we finally upper-bound the conditional mutual information as follows:
\begin{align}
\mI_{\rho_\beta}(A:C|B) 
\le e^{\Theta(\beta) \log(\beta R)-\kappa_\beta R/10} + 4\beta \bar{J}_0 |B_2| \cdot |B_4| e^{-\mu R/5} 
+ e^{\Theta(\beta) \log(\beta R)-\kappa_\beta R/10}   + \Theta(\beta^2 )e^{\Theta(\beta) -\kappa_\beta R/5} , \notag 
\end{align}
which reduces to the desired inequality~\eqref{main:ineq:CMI_decay_PTP_1D/}, where we use $R\ge \Theta(\beta)$ [or $\log(\beta R) \le \Theta(1) \log(R)$] since the inequality is meaningless for $R\le \Theta(\beta)$. 
This completes the proof of Theorem~\ref{CMI_decay_PTP_general_1D_improve}. $\square$

\section{Relation to other information measures} \label{sec:Relation to other information measures}

\subsection{Clustering of Entanglement of Formation (EoF) at arbitrary temperatures}

As one application of the clustering theorem for the CMI, we show the clustering of quantum entanglement. 
In the previous results~\cite{PhysRevX.12.021022}, the exponential clustering of the positive-partial-transpose (PPT) relative entanglement~\cite{PhysRevLett.87.217902,PhysRevA.66.032310,PhysRevA.78.032310,Girard_2014} has been proved.
However, the PPT class cannot capture the existence of the bound entanglement~\cite{PhysRevLett.80.5239,PhysRevLett.82.1056}.   
To analyze the genuine bipartite entanglement, we consider the Entanglement of Formation (EoF) since it gives the upper bound for other entanglement measures~\cite{PhysRevA.99.042304} such as the relative entanglement~\cite{PhysRevLett.78.2275}, the entanglement cost~\cite{PhysRevA.54.3824}, the squashed entanglement~\cite{doi:10.1063/1.1643788}, etc.
The EoF for an arbitrary bipartite quantum state $\rho_{AB}$ is defined as follows:
\begin{align}
\label{def_Entanglement_form_corr}
E_F(\rho_{AB}) &:= \inf_{\{p_s, \ket{\psi_{s,AB}}\}} 
\sum_{s}p_s  S_{\ket{\psi_{s,AB}}} (A:B) ,
\end{align}
where $S_{\ket{\psi_{s,AB}}}(A)$ is the von Neumann entropy for the reduced density matrix on the subset $A$, respectively. 
The convex roof $\inf_{\{p_s, \ket{\psi_{s,AB}}\}}$ is taken for an arbitrary decomposition $\rho=\sum_s p_s \ket{\psi_{s,AB}}\bra{\psi_{s,AB}}$ with $p_s>0$.

In high-dimensional cases, we can prove the following corollary as the direct consequence of Theorem~\ref{CMI_decay_PTP_general_D}:
\begin{corol}[High dimensional cases] \label{corol:ent_F_high}
Let $A$ and $B$ be subsystems that are separated by a distance $R$.
Then, for the reduced density matrix $\rho_{\beta,AB}$, the EoF obeys the following clustering theorem:
\begin{align}
\label{corol:ent_F_high_ineq_main}
E_F(\rho_{\beta,AB}) \le \mathcal{D}_{AB} e^{-R/\Theta\brr{\beta^{D+1}\log \br{R}} + \Theta(1)\log(R)}  . 
\end{align}
\end{corol}

{\bf Remark.}
The upper bound is meaningful only when the subsystem sizes of $A$ and $B$ do not depend on the system size $|\Lambda|$.
The same constraints were also imposed in the previous result~\cite{PhysRevX.12.021022}. 
On the other hand, the temperature dependence of the entanglement length becomes worse in comparison with the previous one, 
where correlation length for the PPT relative entanglement was given by $\orderof{\beta}$.
From this point, we still have room to further improve the inequality~\eqref{corol:ent_F_high_ineq_main}. 

{~}

\textit{Proof of Corollary~\ref{corol:ent_F_high}.}
We first introduce the continuity inequality for the EoF~\cite[Corollary~4 therein]{Winter2016}.
Given two bipartite states $\rho_{AB}$ and $\sigma_{AB}$ such that $\delta=\norm{\rho_{AB}-\sigma_{AB}}_1/2$,  the EoF for $\rho_{AB}$ is upper-bounded by
 \begin{align}
 \label{minimum_distance_SEP_continuity}
E_F(\rho_{AB}) \le E_F(\sigma_{AB}) + \bar{\delta} \log\brr{\max(\mathcal{D}_A,\mathcal{D}_B)} + (1+\bar{\delta}) \br{\frac{\bar{\delta}}{1+\bar{\delta}}} ,
\end{align} 
where $ \bar{\delta}:=\sqrt{(\delta/2)(2-\delta/2)}\le \delta^{1/2}$ and $h(x)=-x \log(x) - (1-x)\log(1-x)$ ($0<x<1$). 
By using $\delta \le 1$ and $h(x)\le x \log(3/x)$, we simplify the inequality~\eqref{minimum_distance_SEP_continuity} to 
 \begin{align}
 \label{minimum_distance_SEP_continuity2}
E_F(\rho_{AB}) \le E_F(\sigma_{AB}) + \delta^{1/2} \brrr{ \log\brr{\max(\mathcal{D}_A,\mathcal{D}_B)} + 2 \log\br{3 \delta^{-1/2}}} .
\end{align}
 
Using the minimum distance $\delta_{\rho_{\beta,AB}}$ in Eq.~\eqref{minimum_distance_SEP} between the reduced density matrix $\rho_{\beta,AB}$ and separable (i.e., non-entangled) states, 
we prove the upper bound for the EoF $E_F(\rho_{\beta,AB})$ as follows:
 \begin{align}
 \label{minimum_distance_SEP_E_F}
E_F(\rho_{\beta,AB}) \le \delta_{\rho_{\beta,AB}}^{1/2} \brrr{ \log\brr{\max(\mathcal{D}_A,\mathcal{D}_B)} + 2 \log\br{3 \delta_{\rho_{\beta,AB}}^{-1/2}} }.
\end{align}
Note that the separable state $\sigma_{AB}$ satisfy $E_F(\sigma_{AB})=0$.
For the quantity $\delta_{\rho_{\beta,AB}}$, by applying Theorem~\ref{CMI_decay_PTP_general_D} to the inequality~\eqref{minimum_distance_SEP_squash_relation}, we obtain 
\begin{align}
\label{minimum_distance_SEP_squash_relation_CMI_beta}
\delta_{\rho_{\beta,AB}}& \le 
2 \min(\mathcal{D}_A,\mathcal{D}_B) \sqrt{\mathcal{D}_{AB}}   e^{-R/\Theta\brr{\beta^{D+1}\log \br{R}} + \Theta(1)\log(R)}  ,
\end{align}
which reduces the inequality~\eqref{minimum_distance_SEP_E_F} to the main inequality~\eqref{corol:ent_F_high_ineq_main}. 
This completes the proof of Corollary~\ref{corol:ent_F_high}. $\square$

{~}

In the one-dimensional case, as in the previous work~\cite{PhysRevX.12.021022}, we can remove the dependence on the Hilbert space dimension $\mathcal{D}_{AB}$. 
We prove the following proposition:
 \begin{prop}[One dimensional cases] \label{corol:ent_F_one}
Let us adopt the same setup as in Corollary~\ref{corol:ent_F_high} for one-dimensional systems.
We then obtain 
\begin{align}
\label{corol:ent_F_one_ineq_main}
E_F(\rho_{\beta,AB}) \le  e^{\Theta(\beta \log(\beta)) - \kappa_\beta^2 R/[81\log(d_0)]}  ,
\end{align}
where $d_0$ is the dimension of the local Hilbert space.  
\end{prop}

{\bf Remark.}
In the previous bound~\cite[Theorem~12]{PhysRevX.12.021022} for the PPT relative entanglement, there exists an additional coefficient of $\orderof{|\Lambda|}$, which spoils the bound in the thermodynamic limit. 
A prominent advantage of the current clustering theorem~\eqref{corol:ent_F_one_ineq_main} is that we can remove the $|\Lambda|$ dependence. 
To achieve this, we cannot rely solely on the continuity inequality~\eqref{minimum_distance_SEP_continuity}, which yields the additional coefficient of $ \log\brr{\max(\mathcal{D}_A,\mathcal{D}_B)} = \orderof{|\Lambda|}$.  
An essential technique here is the iterative use of the approximate recovery map to reconstruct the total quantum Gibbs state (see Lemma~\ref{lemm:recovery_map_error_M} below).  
The approximation error is controlled by the CMI decay in Theorem~\ref{CMI_decay_PTP_general_1D_improve} with the combination of the Fawzi-Renner theorem~\cite{Fawzi2015}.

\subsubsection{Proof of Proposition~\ref{corol:ent_F_one}}

\begin{figure}[tt]
\centering
\includegraphics[clip, scale=0.41]{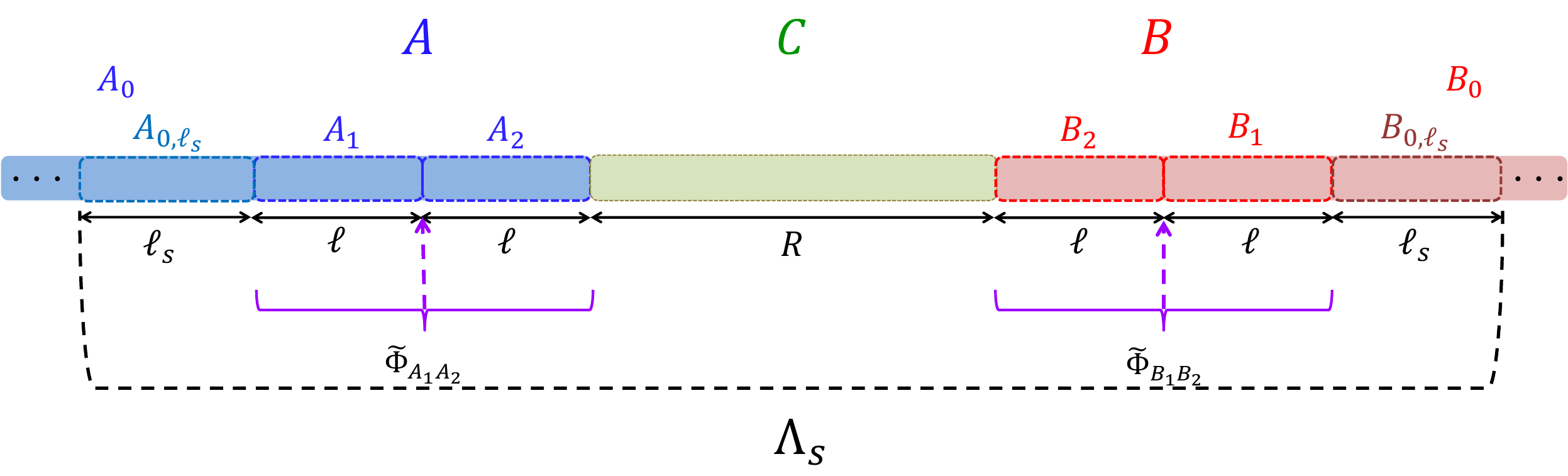}
\caption{Setup of the proof of Proposition~\ref{prop:O_X_imaginary_time}.
We decompose the subsets $A$ and $B$ into three pieces as in Eq.~\eqref{A_decomp_A_0to2}, i.e., $A=A_0\sqcup A_1\sqcup A_2 $ and $B=B_0\sqcup B_1\sqcup B_2 $ respectively. Here, the sizes of the subsets $A_1$ and $A_2$ (or $B_1$ and $B_2$) are set to be equal to $\ell$, which will be chosen afterward such that it minimizes the RHS of~\eqref{norm_Delta_rho_beta_AB}. 
Then, the closeness $\delta_{\rho_{\beta,AB}}$ between the separable state and $\rho_{AB}$ is connected to that of $e^{\beta H_{A_2 C B_2}}$ as in Eq.~\eqref{connect_rho'beta_sigma_beta}. 
In the inequality~\eqref{norm_Delta_rho_beta_AB}, we can prove $\delta_{\rho_{\beta,AB}}\lesssim e^{-R/\Theta(\beta)}$. 
Unfortunately, only the bound on $\delta_{\rho_{\beta,AB}}$ is not enough to derive the main inequality because of the $\log\brr{\max(\mathcal{D}_A,\mathcal{D}_B)} $ dependence in the continuity inequality~\eqref{minimum_distance_SEP_continuity2}. 
This dependence cannot be ignored in the thermodynamic limit of $|\Lambda|\to \infty$. 
To address this point, we extend the left and the right end regions little by little as in Eq.~\eqref{mathfrak_A_B_define}, where the subsets $\{\Lambda_s\}_s$ extend with $s$ so that $\Lambda_s \to \Lambda$ in the limit of $s\to\infty$. 
Using the recovery map (see Lemma~\ref{lemm:recovery_map_error_M}), we take into account the influences coming from the edges, which finally leads to the inequality~\eqref{mathfrak_A_B_s_upper_bound__infty} without $|\Lambda|$ dependence. 
}
\label{fig:Entanglement_clustering}
\end{figure}

For the proof, we rely on a similar proof technique to~\cite[Proof of Theorem~12]{PhysRevX.12.021022}.
We first decompose the subsystems $A$ and $B$ into three pieces, respectively (see Fig.~\ref{fig:Entanglement_clustering}).
We let the decomposition of $A$ be
\begin{align}
\label{A_decomp_A_0to2}
A=A_0\sqcup A_1\sqcup A_2 ,\quad |A_1|=|A_2|=\ell ,
\end{align}
where we adopt the same decomposition for $B$.

We then use the belief propagation operator to obtain a similar decomposition to Eq.~\eqref{BP_decompo_E_beta_H} as 
\begin{align}
e^{\beta H}= \Phi_{\partial h_{A_0A_1}}  \Phi_{\partial h_{B_1B_0}} e^{\beta (H_{A_0A_1} + H_{A_2 C B_2} +H_{B_1B_0})} \Phi_{\partial h_{B_1B_0}}^\dagger \Phi_{\partial h_{A_0A_1}}^\dagger .
\end{align}
By approximating the belief propagation operators $\Phi_{\partial h_{A_0A_1}}$ and $ \Phi_{\partial h_{B_1B_0}}$ onto $A_1A_2$ and $B_1B_2$, respectively, we have a similar approximation to Eq.~\eqref{def_tilde_rho_beta_1D} as follows:
\begin{align}
&e^{\beta \tilde{H}'}:= \tilde{\Phi}_{A_1A_2}   \tilde{\Phi}_{B_1B_2} e^{\beta (H_{A_0A_1} + H_{A_2 C B_2} +H_{B_1B_0})}  \tilde{\Phi}_{B_1B_2} \tilde{\Phi}_{A_1A_2}^\dagger, \notag \\
&\tilde{\rho}'_\beta:= \frac{e^{\beta \tilde{H}'}}{\tilde{Z}'_\beta} ,\quad \tilde{Z}'_\beta:=\tr\br{e^{\beta \tilde{H}'}}.
\label{tilde_rho'_beta_def}
\end{align}
Using Corollary~\ref{corol:high_dimensional_applicaton_bp}, the approximation error between $e^{\beta H}$ and $e^{\beta \tilde{H}'}$ is upper-bounded by
\begin{align}
\frac{1}{Z_\beta}\norm{ e^{\beta H} - e^{\beta \tilde{H}'}}_1 \le e^{\Theta(\beta) - \kappa_\beta \ell} 
\longrightarrow \norm{\rho_{\beta,AB}- \tilde{\rho}'_{\beta,AB} }_1\le e^{\Theta(\beta) - \kappa_\beta \ell} 
,
\label{e-beta_H-e_beta_tilde_H'}
\end{align}
where we use $\norm{\rho_{\beta,AB}- \tilde{\rho}'_{\beta,AB} }_1 \le \frac{1}{Z_\beta}\norm{ e^{\beta H} - e^{\beta \tilde{H}'}}_1 
+ \abs{\frac{\tilde{Z}'_\beta}{Z_\beta} - 1}$ and 
\begin{align}
\tilde{Z}'_\beta=\tr\br{e^{\beta \tilde{H}'}} = Z_\beta + \tr\br{e^{\beta \tilde{H}'}- e^{\beta H} } \le   
Z_\beta \br{1+ \frac{1}{Z_\beta} \norm{e^{\beta \tilde{H}'}- e^{\beta H} } } \le 
Z_\beta \br{1+ e^{\Theta(\beta) - \kappa_\beta \ell}} .
\end{align}

%

We then consider the quantum state $\sigma_\beta$ defined by
\begin{align}
 \sigma_\beta = \frac{e^{\beta (H_{A_0A_1} + H_{A_2 C B_2} +H_{B_1B_0})}}{\tilde{Z}''_{\beta}} , \quad \tilde{Z}''_\beta=\tr\br{e^{\beta (H_{A_0A_1} + H_{A_2 C B_2} +H_{B_1B_0})}}.
 \end{align}
Note that from Eq.~\eqref{tilde_rho'_beta_def} we have 
\begin{align}
\label{connect_rho'beta_sigma_beta}
\tilde{\rho}'_\beta = \frac{\tilde{Z}''_\beta}{\tilde{Z}'_\beta}\tilde{\Phi}_{A_1A_2}   \tilde{\Phi}_{B_1B_2}   \sigma_\beta\tilde{\Phi}_{B_1B_2} \tilde{\Phi}_{A_1A_2}^\dagger .
\end{align} 
Then, by writing the reduced density matrix of $ \sigma_{\beta,AB}= \sigma_{\beta,A_0A_1} \otimes  \sigma_{\beta,A_2B_2} \otimes \sigma_{\beta,B_1B_0}$, we obtain 
\begin{align}
\label{sigma_beta_AB_decomp}
&\sigma_{\beta,A_2B_2} = \sigma^{({\rm SEP})}_{\beta,A_2B_2} + \Delta \sigma_{\beta,A_2B_2}  , \quad
\sigma_{\beta,AB} = \sigma^{({\rm SEP})}_{\beta,AB} +  \Delta \sigma_{\beta,AB} , \notag \\
&{\rm with} \quad 
\sigma^{({\rm SEP})}_{\beta,AB} = \sigma_{\beta,A_0A_1} \otimes\sigma^{({\rm SEP})}_{\beta,A_2B_2} \otimes \sigma_{\beta,B_1B_0} \AND 
  \Delta \sigma_{\beta,AB} =\sigma_{\beta,A_0A_1} \otimes\Delta \sigma_{\beta,A_2B_2}\otimes \sigma_{\beta,B_1B_0}. 
 \end{align}
Note that $ \sigma_{\beta,A_0A_1} \otimes \sigma^{({\rm SEP})}_{\beta,A_2B_2} \otimes \sigma_{\beta,B_1B_0}$ is also the separable state. For the norm of $\Delta \sigma_{\beta,AB} $, we can utilize the inequality~\eqref{minimum_distance_SEP_squash_relation} and Theorem~\ref{CMI_decay_PTP_general_1D_improve} to derive 
\begin{align}
\label{minimum_distance_SEP_squash_relation_1D_gibbs}
\delta_{\sigma_{\beta,AB}}=\norm{\Delta \sigma_{\beta,AB} }_1 = \norm{\Delta \sigma_{\beta,A_2B_2} }_1 
&\le 4 \min(\mathcal{D}_{A_2},\mathcal{D}_{B_2}) e^{\Theta(\beta) \log(R)-\kappa_\beta R/20}   \notag \\
&\le 4 e^{\Theta(\beta) \log(R)-\kappa_\beta R/20 +\ell \log(d_0)} ,
\end{align}
where we use $\dist_{A_2,B_2}=R$ and $\mathcal{D}_{A_2}=\mathcal{D}_{B_2} \le d_0^\ell$. 

In a similar way to Eq.~\eqref{sigma_beta_AB_decomp}, we decompose $\tilde{\rho}'_{\beta,AB}$ as 
\begin{align}
\tilde{\rho}'_{\beta,AB}= \tilde{\rho}^{'({\rm SEP})}_{\beta,AB}+  \Delta \tilde{\rho}'_{\beta,AB} . 
\end{align}
By applying the inequality~\eqref{minimum_distance_SEP_squash_relation_1D_gibbs} to Eq.~\eqref{connect_rho'beta_sigma_beta}, we obtain 
\begin{align}
\label{eq_Delta_tilde_rho'_beta_AB}
\delta_{\rho'_{\beta,AB}}= \norm{\Delta \tilde{\rho}'_{\beta,AB}}_1
&\le  \frac{\tilde{Z}''_\beta}{\tilde{Z}'_\beta} \norm{\tilde{\Phi}_{A_1A_2}   \tilde{\Phi}_{B_1B_2} \Delta \sigma_{\beta,AB}  \tilde{\Phi}_{B_1B_2} \tilde{\Phi}_{A_1A_2}^\dagger }_1 \notag\\
&\le \norm{\tilde{\Phi}_{A_1A_2}^{-1}}^2\norm{\tilde{\Phi}_{B_1B_2}^{-1}}^2 \norm{\tilde{\Phi}_{A_1A_2}}^2\norm{\tilde{\Phi}_{B_1B_2}}^2  \cdot 4 e^{\Theta(\beta) \log(R)-\kappa_\beta R/20 +\ell \log(d_0)}  \notag \\
&\le e^{\Theta(\beta) \log(R)-\kappa_\beta R/20 +\ell \log(d_0)}  
\end{align}
where we use the inequality~\eqref{sup_Def:Phi_0_phi_norm_Phi_norm}, which yields the inequality as $ \norm{\tilde{\Phi}_{A_1A_2}}\le e^{\Theta(\beta)}$, and 
\begin{align}
 \tilde{Z}''_\beta=\tr\br{e^{\beta (H_{A_0A_1} + H_{A_2 C B_2} +H_{B_1B_0})} }
 &\le \norm{\tilde{\Phi}_{A_1A_2}^{-1}}^2\norm{\tilde{\Phi}_{B_1B_2}^{-1}}^2  \tilde{Z}'_\beta.
\end{align}
By applying the bound~\eqref{eq_Delta_tilde_rho'_beta_AB} to~\eqref{e-beta_H-e_beta_tilde_H'}, we reach the upper bound of 
\begin{align}
\label{norm_Delta_rho_beta_AB}
\delta_{\rho_{\beta,AB}}=\norm{\Delta \rho_{\beta,AB}}_1 \le e^{\Theta(\beta) \log(R)-\kappa_\beta R/20 +\ell \log(d_0)}  +e^{\Theta(\beta) - \kappa_\beta \ell} 
\end{align}
for the decomposition of 
$
\rho_{\beta,AB}= \rho^{({\rm SEP})}_{\beta,AB}+  \Delta \rho_{\beta,AB} . 
$
By choosing $\ell$ as $\ell=\kappa_\beta R/[40\log(d_0)]$, we have 
\begin{align}
\label{norm_Delta_rho_beta_AB___02}
\delta_{\rho_{\beta,AB}} \le e^{\Theta(\beta \log(\beta))- \kappa_\beta^2 R/[40\log(d_0)]} .
\end{align}

From the upper bound~\eqref{minimum_distance_SEP_E_F}, we can immediately derive an upper bound as 
\begin{align}
E_F(\rho_{\beta,AB}) \lesssim  \log\brr{\max(\mathcal{D}_A,\mathcal{D}_B)} \delta_{\rho_{\beta,AB}}^{1/2} ,
\end{align}
but it is infinitely large in the limit of $|A|,|B| \to \infty$. 
To obtain a better bound, we define the contiguous subsystems in $A_0$ and $B_0$ as $\{A_{0,\ell_s}\}_s$ and $\{B_{0,\ell_s}\}_s$, respectively, where  
$|A_{0,\ell_s}|=|B_{0,\ell_s}|=\ell_s$.
We also denote 
\begin{align}
&\mathfrak{A}_s=A_{0,\ell_s} \sqcup A_1\sqcup A_2 ,\quad \mathfrak{B}_s=B_2\sqcup B_1 \sqcup B_{0,\ell_s} , \notag \\
&\Delta \mathfrak{A}_s = \mathfrak{A}_s \setminus \mathfrak{A}_{s-1}, \quad  \Delta \mathfrak{B}_s = \mathfrak{B}_s \setminus \mathfrak{B}_{s-1} ,\notag \\
&\Lambda'_s = \mathfrak{A}_s \sqcup C \sqcup  \mathfrak{B}_{s-1}, \quad \Lambda_s = \mathfrak{A}_s \sqcup C \sqcup  \mathfrak{B}_s . 
\label{mathfrak_A_B_define}
\end{align}
We choose the length $\ell_s$ as 
\begin{align}
\label{ell_s_choice_1}
\ell_s =R^{s+1} \longrightarrow  \max\brr{ \log(\mathcal{D}_{\mathfrak{A}_s}) , \log(\mathcal{D}_{\mathfrak{B}_s}) } = (R^{s+1} +2\ell )\log(d_0)  .
\end{align}
To derive the upper bound for $E_F(\rho_{\beta,AB})$, we need to estimate $E_F(\rho_{\beta,\mathfrak{A}_\infty\mathfrak{B}_\infty})$.

We start from considering the EoF of $\rho_{\beta,\mathfrak{A}_1\mathfrak{B}_1}$, which also satisfies 
\begin{align}
\label{norm_Delta_rho_beta_mathfrakA_1B_1}
\delta_{\rho_{\beta,\mathfrak{A}_1\mathfrak{B}_1}} = \norm{\Delta \rho_{\beta,\mathfrak{A}_1\mathfrak{B}_1} }\le \norm{\Delta \rho_{\beta,AB}}_1 \le e^{\Theta(\beta \log(\beta))- \kappa_\beta^2 R/[40\log(d_0)]}  
\end{align}
for the decomposition of 
\begin{align}
\rho_{\beta,\mathfrak{A}_1\mathfrak{B}_1}= \tr_{\mathfrak{A}_1^\co} \tr_{\mathfrak{B}_1^\co} \br{\rho_{\beta,AB}}=
\rho^{({\rm SEP})}_{\beta,\mathfrak{A}_1\mathfrak{B}_1}+  \Delta \rho_{\beta,\mathfrak{A}_1\mathfrak{B}_1}  .
\end{align}
Note that the partial trace does not increase the norm of operators. 
By applying the continuity inequality~\eqref{minimum_distance_SEP_E_F} with $\delta_{\rho_{\beta,AB}} \to \norm{\Delta \rho_{\beta,\mathfrak{A}_1\mathfrak{B}_1} }$ and~\eqref{ell_s_choice_1} to~\eqref{norm_Delta_rho_beta_mathfrakA_1B_1}, we have 
\begin{align}
E_F(\rho_{\beta,\mathfrak{A}_1\mathfrak{B}_1}) 
&\le e^{\Theta(\beta \log(\beta))+ \Theta(1)\log(R\log(d_0)) - \kappa_\beta^2 R/[80\log(d_0)]} \notag \\
&\le e^{\Theta(\beta \log(\beta)) - \kappa_\beta^2 R/[81\log(d_0)]} ,
\label{E_F_rho_beta_mathcal_A_1_B_1}
\end{align}
where we use $\Theta(1)\log[R\log(d_0)] - c  \kappa_\beta^2 R \le \Theta(\log(\beta))$ for an arbitrary $c>0$.

In the next step, we consider $E_F(\rho_{\beta,\mathfrak{A}_2\mathfrak{B}_2})$. 
For this purpose, we prove the following lemma:
\begin{lemma} \label{lemm:recovery_map_error_M}
Let us consider a recovery map that transforms $\rho_{\beta,\Lambda_s}$ to $\rho_{\beta,\Lambda_{s+1}}$. 
Then, there exists local completely-positive-trace-preserving (CPTP) maps $\mathcal{M}_{\mathfrak{A}_s\to \mathfrak{A}_{s+1}}$ and $\mathcal{M}_{\mathfrak{B}_s\to \mathfrak{B}_{s+1}}$ such that 
\begin{align}
\norm{\rho_{\beta,\Lambda_{s+1}}-\mathcal{M}_{\mathfrak{B}_s\to \mathfrak{B}_{s+1}} \mathcal{M}_{\mathfrak{A}_s\to \mathfrak{A}_{s+1}}(\rho_{\beta,\Lambda_s})}_1 \le e^{\Theta(\beta) \log(\ell_s)-\kappa_\beta \ell_s/10} ,
\label{lemm:recovery_map_error_M/main_ineq}
\end{align}
where $\ell_s$ is defined in Eqs.~\eqref{mathfrak_A_B_define} and~\eqref{ell_s_choice_1}, which gives $|\mathfrak{A}_s| \ge \ell_s$ and $|\mathfrak{B}_s| \ge \ell_s$.
\end{lemma}

 \begin{figure}[tt]
\centering
\includegraphics[clip, scale=0.33]{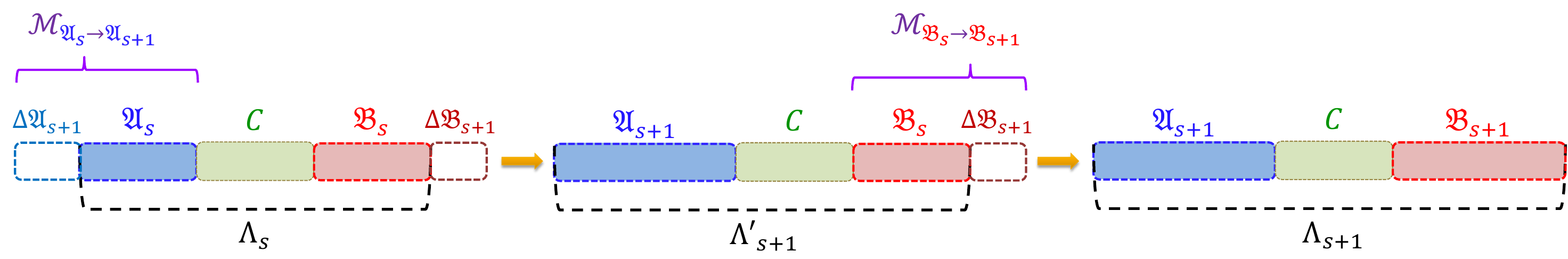}
\caption{Recovery from $\rho_{\beta,\Lambda_s}$ to $\rho_{\beta,\Lambda_{s+1}}$ in Lemma~\ref{lemm:recovery_map_error_M}. 
We take the two steps: i) recovery from $\rho_{\beta,\Lambda_s}$ to $\rho_{\beta,\Lambda'_{s+1}}$, ii) recovery from $\rho_{\beta,\Lambda'_{s+1}}$ to $\rho_{\beta,\Lambda_{s+1}}$, where $\Lambda'_{s+1}=\mathfrak{A}_{s+1} \sqcup \Lambda_s$. 
We approximate the recovery maps so that they only act on $ \mathfrak{A}_s$ and $\mathfrak{B}_s$, respectively. 
Under these approximated local maps, i.e., $\mathcal{M}_{\mathfrak{A}_s\to \mathfrak{A}_{s+1}}$ and $\mathcal{M}_{\mathfrak{B}_s\to \mathfrak{B}_{s+1}}$, the entanglement cannot increase due to the monotonicity. 
}
\label{fig:Recovery_Lambda_s.pdf}
\end{figure}

\textit{Proof of Lemma~\ref{lemm:recovery_map_error_M}.}
We construct the recovery map in two steps: i) $\rho_{\beta,\Lambda_s}$ to $\rho_{\beta,\Lambda'_{s+1}}$ and 
(ii) $\rho_{\beta,\Lambda'_{s+1}}$ to $\rho_{\beta,\Lambda_{s+1}}$ (see Fig.~\ref{fig:Recovery_Lambda_s.pdf}).  
We first consider the recovery map (i). 
From Theorem~\ref{CMI_decay_PTP_general_1D_improve}, we can ensure
\begin{align}
\mI_{\rho_{\beta}}(\Delta \mathfrak{A}_{s+1},C\mathfrak{B}_s|\mathfrak{A}_s) \le e^{\Theta(\beta) \log(|\mathfrak{A}_s|)-\kappa_\beta |\mathfrak{A}_s|/10}.
\end{align}
By using the Fawzi-Renner theorem~\cite[Ineq. (6) therein]{Fawzi2015} (see also~\cite{PhysRevLett.115.050501,doi101098rspa20150623,Junge2018}), 
we can construct a recovery map $\mathcal{M}_{\mathfrak{A}_s\to \mathfrak{A}_{s+1}}$ such that   
\begin{align}
\label{eq:M_A_map_1}
\norm{\mathcal{M}_{\mathfrak{A}_s\to \mathfrak{A}_{s+1}}(\rho_{\beta,\Lambda_s}) - \rho_{\beta,\Lambda'_{s+1}}}^2_1 \le  \log(2) \mI_{\rho_{\beta}}(\Delta \mathfrak{A}_{s+1},C\mathfrak{B}_s|\mathfrak{A}_s)
\le e^{\Theta(\beta) \log(\ell_s)-\kappa_\beta \ell_s/10},
\end{align}
where we use $\ell_s\le |\mathfrak{A}_s| \le \Theta(1) \ell_s$ from $\mathfrak{A}_s \supset A_{0,\ell_s} $ as in Eq.~\eqref{mathfrak_A_B_define}. 
In the same way, we consider the recovery map (ii). Because of  
\begin{align}
\mI_{\rho_{\beta}}(\Delta \mathfrak{B}_{s+1},\mathfrak{A}_{s+1} C|\mathfrak{B}_s) \le e^{\Theta(\beta) \log(\ell_s)-\kappa_\beta\ell_s/10},
\end{align}
we have 
\begin{align}
\label{eq:M_B_map_2}
\norm{\mathcal{M}_{\mathfrak{B}_s\to \mathfrak{B}_{s+1}}(\rho_{\beta,\Lambda'_{s+1}}) - \rho_{\beta,\Lambda_{s+1}}}^2_1 \le e^{\Theta(\beta) \log(\ell_s)-\kappa_\beta \ell_s/10} .
\end{align}
By combining the inequality~\eqref{eq:M_A_map_1} and \eqref{eq:M_B_map_2}, we prove the main inequality~\eqref{lemm:recovery_map_error_M/main_ineq}, where we use the fact that the CPTP map $\mathcal{M}_{\mathfrak{A}_s\to \mathfrak{A}_{s+1}}$ and $\mathcal{M}_{\mathfrak{B}_s\to \mathfrak{B}_{s+1}}$ does not increase the norm.  This completes the proof. $\square$

{~}

\hrulefill{\bf [ End of Proof of Lemma~\ref{lemm:recovery_map_error_M}]}

{~}

Using the Lemma~\ref{lemm:recovery_map_error_M} with $s=1$, we can derive 
\begin{align}
\norm{\rho_{\beta,\Lambda_{2}}-\mathcal{M}_{\mathfrak{B}_1\to \mathfrak{B}_{2}} \mathcal{M}_{\mathfrak{A}_1\to \mathfrak{A}_{2}}(\rho_{\beta,\Lambda_1})}_1 \le e^{\Theta(\beta) \log(\ell_1)-\kappa_\beta \ell_1/10} ,
\end{align}
which yields by taking the partial trace with respect to the subset $C$ 
\begin{align}
\label{mathfrak_A_B_2_upper_bound}
\norm{\rho_{\beta,\mathfrak{A}_2\mathfrak{B}_2}-\mathcal{M}_{\mathfrak{B}_1\to \mathfrak{B}_{2}} \mathcal{M}_{\mathfrak{A}_1\to \mathfrak{A}_{2}}(\rho_{\beta,\mathfrak{A}_1\mathfrak{B}_1})}_1 \le e^{\Theta(\beta) \log(\ell_1)-\kappa_\beta \ell_1/10} .
\end{align}
By using the continuity inequality~\eqref{minimum_distance_SEP_continuity2} and the monotonicity of the entanglement, i.e.,
\begin{align}
E_F\br{\mathcal{M}_{\mathfrak{B}_1\to \mathfrak{B}_{2}} \mathcal{M}_{\mathfrak{A}_1\to \mathfrak{A}_{2}}(\rho_{\beta,\mathfrak{A}_1\mathfrak{B}_1})} 
\le  E_F(\rho_{\beta,\mathfrak{A}_1\mathfrak{B}_1}), 
\end{align}
we obtain 
\begin{align}
E_F(\rho_{\beta,\mathfrak{A}_2\mathfrak{B}_2})
&\le E_F(\rho_{\beta,\mathfrak{A}_1\mathfrak{B}_1})+ e^{\Theta(\beta) \log(\ell_1)-\kappa_\beta \ell_1/20} 
\log\br{\mathcal{D}_{\Lambda_2}} \notag \\
&\le E_F(\rho_{\beta,\mathfrak{A}_1\mathfrak{B}_1})+ \Theta(\ell_2) e^{\Theta(\beta) \log(\ell_1)-\kappa_\beta \ell_1/20} .
\end{align}

In the same way, we can derive the same inequality as~\eqref{mathfrak_A_B_2_upper_bound} for $\rho_{\beta,\mathfrak{A}_3\mathfrak{B}_3}$: 
\begin{align}
\label{mathfrak_A_B_2_upper_bound__3}
\norm{\rho_{\beta,\mathfrak{A}_3\mathfrak{B}_3}-\mathcal{M}_{\mathfrak{B}_2\to \mathfrak{B}_{3}} \mathcal{M}_{\mathfrak{A}_2\to \mathfrak{A}_{3}}(\rho_{\beta,\mathfrak{A}_2\mathfrak{B}_2})}_1 \le e^{\Theta(\beta) \log(\ell_2)-\kappa_\beta \ell_2/20} ,
\end{align}
and hence 
\begin{align}
E_F(\rho_{\beta,\mathfrak{A}_3\mathfrak{B}_3})
&\le E_F(\rho_{\beta,\mathfrak{A}_2\mathfrak{B}_2})+ \Theta(\ell_3) e^{\Theta(\beta) \log(\ell_2)-\kappa_\beta \ell_2/20} .
\end{align}
By repeating the same process, we finally obtain 
\begin{align}
\label{mathfrak_A_B_s_upper_bound__infty}
E_F(\rho_{\beta,\mathfrak{A}_{\infty}\mathfrak{B}_{\infty}})
=E_F(\rho_{\beta,AB})
&\le E_F(\rho_{\beta,\mathfrak{A}_1\mathfrak{B}_1})+ \sum_{j=1}^\infty \Theta(\ell_{j+1}) e^{\Theta(\beta) \log(\ell_j)-\kappa_\beta \ell_j/20} \notag\\
&\le e^{\Theta(\beta \log(\beta)) - \kappa_\beta^2 R/[81\log(d_0)]} + e^{\Theta(\beta) \log(R) - \kappa_\beta \Theta(R^2)},
\end{align}
where we use the inequality~\eqref{E_F_rho_beta_mathcal_A_1_B_1} for $E_F(\rho_{\beta,\mathfrak{A}_1\mathfrak{B}_1}) $
 and the definition of $\ell_s$ in Eq.~\eqref{ell_s_choice_1}, i.e., $\ell_s =R^{s+1}$.
 We thus prove the main inequality~\eqref{corol:ent_F_one_ineq_main}.
 This completes the proof of Proposition~\ref{corol:ent_F_one}. $\square$

\subsection{Clustering of information distribution: strong 1D thermal area law}

\begin{figure}[tt]
\centering
\includegraphics[clip, scale=0.33]{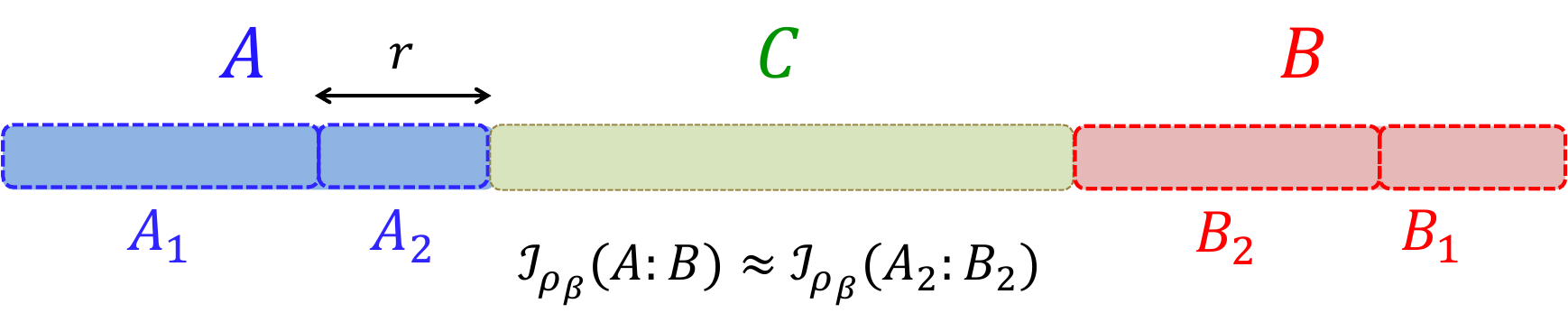}
\caption{Setup of Corollary~\ref{corol:mutual_info_decomp}.
When we consider the mutual information between the subsets $A$ and $B$, the shared information between $A$ and $B$ is approximately localized around the boundary regions $A_2$ and $B_2$ with an exponential tails as $e^{-r/\Theta(\beta)}$, where $r=\min(|A_2|,|B_2|)$. 
This implies a strong version of the area law; that is, the total amount of information is not only proportional to the surface region of $A$ (or $B$) but also exponentially localized around the boundary. 
}
\label{fig:Clustering_mi.pdf}
\end{figure}

As a relevant information measure, we consider the mutual information as 
\begin{align}
\label{mutual_info_defnition1}
\mI_{\rho_\beta}(A:B):=S(\rho_{\beta,A}) + S(\rho_{\beta,B}) -S(\rho_{\beta,AB}) = S(\rho_{\beta,AB}|\rho_{\beta,A} \otimes \rho_{\beta,B}).
\end{align}
When $A$ and $B$ are connected as $A\sqcup B=\Lambda$, the mutual information is known to obey the area law as follows~\cite{PhysRevLett.100.070502,PhysRevX.11.011047}: 
\begin{align}
\mI_{\rho_\beta}(A:B) = \tO\br{\beta^{2/3}} ,
\end{align}
where the exponent $2/3$ may be further improved~\cite{PhysRevX.11.011047}. 
The area law implies that the total amount of the shared information between $A$ and $B$ is proportional to the surface region $\partial A$. 
Here, we aim to prove that almost all the information localizes around the surface region between $A$ and $B$.
In detail, we decompose $A=A_1 \sqcup A_2$ and $B = B_2 \sqcup B_1$ with $A_2$ and $B_2$ neighboring to each other and prove 
\begin{align}
\mI_{\rho_\beta}(A:B) \approx \mI_{\rho_\beta}(A_2:B_2),
\end{align}
where the approximation error depends on the size of $A_2$ and $B_2$ and approaches to zero for $|A_2|\to \infty$ and $|B_2|\to \infty$ (i.e., $A_1=B_1=\emptyset$).
We, in general, prove the following lemma:
\begin{corol} \label{corol:mutual_info_decomp}
Let us consider arbitrary subsets $A$ and $B$ and decompose $A=A_1 \sqcup A_2$ and $B = B_2 \sqcup B_1$ such that $\dist_{A,B}=\dist_{A_2,B_2}$ (see Fig.~\ref{fig:Clustering_mi.pdf}). Also, we define $C$ be the intermediate region between $A$ and $B$, i.e., $|C|=\dist_{A,B}$. Then, we obtain 
 \begin{align}
\label{lem:mutual_info_decomp_ineq/main}
\mI_{\rho_\beta}(A:B)
&\le e^{\Theta(\beta) \log(r)-\kappa_\beta r/10}  + \mI_{\rho_\beta}(A_2:B_2) ,
\end{align}
where we set $\min(|A_2|,|B_2|)=r$. Note that the above bound does not depend on the size of $|C|$. 
\end{corol}

\textit{Proof of Corollary~\ref{corol:mutual_info_decomp}.}
From the definition of the CMI as $\mI_{\rho_\beta}(A:B|C)=\mI_{\rho_\beta}(A:BC) - \mI_{\rho_\beta}(A:C)$,
we have 
 \begin{align}
\mI_{\rho_\beta}(A:B)
&=\mI_{\rho_\beta}(A_1A_2:B_2 B_1) \notag \\
&= \mI_{\rho_\beta}(A_1A_2:B_1|B_2)+  \mI_{\rho_\beta}(A_1:B_2|A_2) + \mI_{\rho_\beta}(A_2:B_2) \notag \\
&\le \mI_{\rho_\beta}(AC:B_1|B_2)+  \mI_{\rho_\beta}(A_1:BC|A_2) + \mI_{\rho_\beta}(A_2:B_2) ,
\end{align}
where we use the inequality~\eqref{monotonicity_CMI_winter}. 
By applying Theorem~\ref{CMI_decay_PTP_general_1D_improve} to the above inequality, we prove the main inequality~\eqref{lem:mutual_info_decomp_ineq/main}.
This completes the proof. $\square$

{~}

\hrulefill{\bf [ End of Proof of Corollary~\ref{corol:mutual_info_decomp}]}

{~}

Furthermore, Corollary~\ref{corol:mutual_info_decomp} immediately leads to the exponential clustering of the mutual information.
\begin{corol} \label{mutual_info_corol}
Let us assume that the correlation function is given by
\begin{align}
\label{def_corr_decay_assymp}
{\rm Cor}_{\rho_\beta} (O_A,O_B) \le \norm{O_A} \cdot  \norm{O_B}  {\rm Cor}(R) ,\quad \dist_{A,B}=R, 
\end{align}
for arbitrary operator $O_A,O_B$, 
where ${\rm Cor}_{\rho_\beta} (O_A,O_B)=\tr (\rho_\beta O_A O_B) - \tr (\rho_\beta O_A) \tr (\rho_\beta O_B)$.
Then, we obtain 
\begin{align}
\label{mutual_info_corol_main_ineq}
\mI_{\rho_\beta} (A:B) &\le \br{\frac{\log[1/{\rm Cor}(R)]}{\log(d_0)}}^{\Theta(\beta)} [{\rm Cor}(R)]^{\frac{\kappa_\beta}{100\log(d_0)}}  \notag \\
&\le [{\rm Cor}(R)]^{1/\Theta(\beta)} ,
\end{align}
where we assume $d_0=\orderof{1}$. 
\end{corol}

{\bf Remark.} Using Araki's result~\cite{Araki1969} or its generalization~\cite{Perez-Garcia2023}, we have ${\rm Cor}(R) =e^{-\Omega(R)}$ for translation invariant infinite systems, where the temperature dependence is ignored. 
 In a more general setup, a recent result~\cite{kimura2024clustering} has proven ${\rm Cor}(R) =e^{-e^{-\Theta(\beta)}\Omega(\sqrt{R})}$ including the $\beta$ dependence. 
By using it, we prove in general 
\begin{align}
\mI_{\rho_\beta} (A:B) &\le \exp\brr{-e^{-\Theta(\beta)}\Omega\br{\sqrt{R}}}
\end{align}
for arbitrary short-range interacting systems.
A similar clustering theorem for the mutual information has also been derived in Ref.~\cite{bluhm2021exponential}.
An advantage of our approach is its ability to extend to more general classes of interactions, e.g., long-range interactions. 

The result implies the closeness between the reduced density matrix $\rho_{\beta,AB}$ and a product state, i.e., 
\begin{align}
\norm{\rho_{\beta,AB}- \rho_{\beta,A} \otimes \rho_{\beta,B}}_1 \le \sqrt{2\mI_{\rho_\beta}(A:B)},
\end{align}
where we use the Pinsker's inequality for Eq.~\eqref{mutual_info_defnition1} as $\norm{\rho- \sigma}_1\le \sqrt{2S(\rho|\sigma)}$.
Hence, the clustering of the mutual information imposes a stronger constraint than the entanglement clustering.
On the other hand, we emphasize that the correlation length of the mutual information is as large as $e^{\Theta(\beta)}$, while the correlation length of the entanglement is at most $\orderof{\beta^2}$ as in Proposition~\ref{corol:ent_F_one}.

{~}

\textit{Proof of Corollary~\ref{mutual_info_corol}.}
We first use Corollary~\ref{corol:mutual_info_decomp} to obtain 
 \begin{align}
 \label{mutual_info_CMI_upper}
\mI_{\rho_\beta}(A:B)
&\le e^{\Theta(\beta) \log(r)-\kappa_\beta r/10}  + \mI_{\rho_\beta}(A_2:B_2) ,
\end{align}
where we let $|A_2|=|B_2|=r$.
We then upper-bound the mutual information $\mI_{\rho_\beta}(A_2:B_2)$ using the correlation function~\eqref{def_corr_decay_assymp}. 
By expanding the reduced density matrix $\rho_{\beta ,A_2B_2}$ using the operator bases $\{P_{A_2,s}\}_{s=1}^{\mathcal{D}_{A_2}^2}$ and $\{P_{B_2,s}\}_{s=1}^{\mathcal{D}_{B_2}^2}$, we decompose
 \begin{align}
\rho_{\beta ,A_2B_2}= \sum_{s,s'} \lambda_{s,s'} P_{A_2,s} \otimes P_{B_2,s'} ,
\end{align}
where we choose $P_{A_2,s}$ and $P_{B_2,s'}$ such that $\norm{P_{A_2,s}}_1\le 2$ and $\norm{P_{B_2,s'}}_1\le 2$\footnote{For example, it is possible by choosing $P_{A_2,s} = \ket{x} \bra{x'} + {\rm h.c.}$ with $\{\ket{x}\}$ orthonormal bases on the Hilbert space for $A_2$.}. 
By using the assumption~\eqref{def_corr_decay_assymp}, we have 
 \begin{align}
\abs{\lambda_{s,s'} -  \bar{\lambda}_s  \bar{\lambda}'_{s'} } \le   {\rm Cor}(R),
\end{align}
where $\bar{\lambda}_s:=\tr\br{P_{A_2,s} \rho_{\beta ,A_2B_2}}$ and $\bar{\lambda}'_{s'}:=\tr\br{P_{B_2,s'} \rho_{\beta ,A_2B_2}}$.
We therefore obtain 
 \begin{align}
\norm{\rho_{\beta ,A_2B_2} - \sum_{s}\bar{\lambda}_s  P_{A_2,s} \otimes  \sum_{s'}\bar{\lambda}'_{s'}   P_{B_2,s'}}_1
& \le 
 \sum_{s,s'} \abs{\lambda_{s,s'}- \bar{\lambda}_s \bar{\lambda}'_{s'} } \cdot \norm{P_{A_2,s} \otimes P_{B_2,s'}}_1 \notag \\
&\le 4  \mathcal{D}_{A_2}^2 \mathcal{D}_{B_2}^2 {\rm Cor}(R) \le 4 d_0^{4r}  {\rm Cor}(R) ,
 \end{align}
where we use the local Hilbert dimension $d_0$ and the inequality $\norm{P_{A_2,s} \otimes P_{B_2,s'}}_1\le \norm{P_{A_2,s}}_1 \norm{P_{B_2,s'}}_1\le 4$. 
By using the continuity inequality for the mutual information as in Ref.~\cite[Remark 1 therein]{10.1063/1.4987135}, we prove 
 \begin{align}
\mI_{\rho_\beta}(A_2:B_2) \le \Theta(1) d_0^{5r}  {\rm Cor}(R).
\end{align}
Here, the parameter $r$ can be adjusted to achieve any desired power of ${\rm Cor}(R)$ in the right-hand side of the above inequality; for example, $[{\rm Cor}(R)]^\eta$ with $0 < \eta < 1$. For simplicity, we choose $r = \log[1/{\rm Cor}(R)]/[10 \log(d_0)]$ so that the right-hand side is proportional to $\sqrt{{\rm Cor}(R)}$:
 \begin{align}
 \label{mutual_upp_A_2_B_2}
\mI_{\rho_\beta}(A_2:B_2) \le \Theta(1) \sqrt{{\rm Cor}(R)}.
\end{align}
By combining the inequality~\eqref{mutual_info_CMI_upper} and \eqref{mutual_upp_A_2_B_2}, we arrive at the inequality of 
 \begin{align}
 \label{mutual_info_CMI_upper_22}
\mI_{\rho_\beta}(A:B)
&\le \br{\frac{\log[1/{\rm Cor}(R)]}{10\log(d_0)}}^{\Theta(\beta)} [{\rm Cor}(R)]^{\frac{\kappa_\beta}{100\log(d_0)}}  + \Theta(1) \sqrt{{\rm Cor}(R)} ,
\end{align}
which reduces to the main inequality~\eqref{mutual_info_corol_main_ineq}.
This completes the proof of Corollary~\ref{mutual_info_corol}. $\square$

\section{Quasi-locality of the true entanglement Hamiltonian} \label{section:Quasi-locality of the true effective Hamiltonian}

In Sec.~\ref{sec:One-dimensional case: improved bound} for the 1D CMI, we have shown that the approximate quantum Gibbs state $\tilde{\rho}_\beta$ 
[see Eq.~\eqref{def_tilde_rho_beta_1D}]
satisfies the quasi-locality of entanglement Hamiltonians on the subsets $AB$, $BC$ and $B$ as in Eq.~\eqref{log_rho_AB_BC_ABC_B}. 
The inequality implies that the effective interactions of $\log(\tilde{\rho}_{\beta,L})$ for $\forall L\subset \Lambda$ is localized around the boundary $\partial L$ with an exponential tail of $e^{-\Theta(R/\beta)}$.

Our fundamental question is whether the quasi-locality of the effective interactions is proved for the true entanglement Hamiltonian of $\rho_{\beta,B}$ instead of $\tilde{\rho}_{\beta,B}$, or $\tr_{AC}\br{e^{\beta \tilde{H}_\Lambda}}$, that is,
\begin{align}
\log \br{ \rho_{\beta,B} }  \overset{?}{\approx } \log \br{\tilde{\rho}_{\beta,B}}.
\end{align}
We note that in general for given $\rho$ and $\tilde{\rho}$, we cannot ensure $\log(\rho) \approx \log(\tilde{\rho})$ only from the norm error of $\norm{\rho-\tilde{\rho}}_1$ even in the classical cases [see the inequality~\eqref{classical_log_sigma_rho} below].
The quasi-locality of the true entanglement Hamiltonian is a stronger concept than the decay of the CMI and has several applications, e.g., in Hamiltonian learning of 1D quantum systems.  

Using the inequality~\eqref{main_ineq_lem:beta_H_log_tilde..._up} in Lemma~\ref{lem:beta_H_log_tilde..._up}, we can ensure 
\begin{align}
\beta \norm{H   -  \tilde{H}_\Lambda}  \lesssim e^{-\Theta(R/\beta)}.
\end{align}
That is, the approximate density matrix $\tilde{\rho}_{\beta}$ has a similar global Hamiltonian.
We then have the following fundamental question:
\begin{center}
\textit {``Can we prove closeness of the entanglement Hamiltonians between $e^{\beta H}$ and $\tilde{\rho}_{\beta}$\text{?}''}
\end{center}
We address the problem in the 1D cases where the interactions are of a finite range. 
Precisely speaking, we can prove the following theorem:
\begin{theorem} \label{thm:1D_effective_Ham}
Let us assume that the one-dimensional Hamiltonian only has finite-range interactions as follows:
\begin{align}
\label{k-local_asummp_finite/range}
H=\sum_{Z: \diam(Z)\le k} h_Z ,\quad  \sum_{Z:Z\ni i } \norm{h_Z} \le J_0,
\end{align}
where $\diam(Z)$ is defined as $\diam(Z)=\max_{i,i'\in Z} (\dist_{i,i'})$. 
Then, for an arbitrary concatenate subset $B$ ($\subset \Lambda$) such that $|B|=R$, we can prove 
\begin{align}
\label{main_ineq/thm:1D_effective_Ham}
\norm{\log(\rho_{\beta,B})- \log(\tilde{\rho}_{\beta,B})}\le e^{\xi_\beta -R/\Theta(\beta)},
\end{align}
where we define $\xi_\beta$ as a doubly exponential function of $\beta$, i.e., $\xi_\beta=e^{e^{\Theta(\beta)}}$. 
\end{theorem}

The proof is deferred to Subsections~\ref{sec:Continuity inequality for logarithmic operators}, \ref{Proof of Theorem_thm_1D_effective_Ham} and~\ref{Proof of prop:O_X_imaginary_time}. 
The proof relies on the following two steps:
\begin{enumerate}
\item{} We first prove the closeness of two density matrices $\rho$ and $\sigma$ under the condition of the small relative error (see Definition~\ref{def:relative error}). The statement is summarized in Theorem~\ref{thm:refined continuity} in Sec.~\ref{sec:Continuity inequality for logarithmic operators}.

\item{} We then prove an upper bound for the relative error between $\rho_{\beta,B}$ and  $\tilde{\rho}_{\beta,B}$, which will be given in Subtheorem~\ref{subthm:est_relative_error}, which will be proven in Sec.~\ref{Proof of prop:O_X_imaginary_time}.
Theorem~\ref{thm:1D_effective_Ham} is straightforwardly derived by a simple combination of Theorem~\ref{thm:refined continuity} and Subtheorem~\ref{subthm:est_relative_error} as in Sec.~\ref{Proof of Theorem_thm_1D_effective_Ham}. 
 Here, the quasi-locality of the imaginary-time evolution plays a crucial role (see Lemma~\ref{lem:error_estimation_BP_imag} and Proposition~\ref{prop:O_X_imaginary_time}).
To ensure the quasi-locality by the imaginary time evolution (Lemma~\ref{lem:imaginary_time_LR}), we need the condition of the finite interaction length in the Hamiltonian\footnote{In general, the quasi-locality of the imaginary time evolution breaks down~\cite{10.1063/1.4936209,Perez-Garcia2023} even when the interaction decay is exponential.}. 
\end{enumerate}

From the inequality~\eqref{log_AB_U_B1B2_U'_B3B4_dif_up} we have 
\begin{align}
\label{main_ineq/thm:1D_effective_Ham_approx}
&\norm{\log(\tilde{\rho}_{\beta,B}) - \beta (\tilde{H}_{B_1B_2} +H_{B_3} + \tilde{H}_{B_4B_5}) }  \le e^{\Theta(\beta) \log(\beta R)-\kappa_\beta R/10}  ,
\end{align}
where we denote $\tilde{H}_{B_1B_2} +H_{B_3} + \tilde{H}_{B_4B_5}$ as follows [see also~\eqref{log_AB_U_B1B2_U'_B3B4_dif_up_other_ones}]: 
\begin{align}
\label{main_ineq/thm:1D_effective_Ham_notation_til}
\beta (\tilde{H}_{B_1B_2} +H_{B_3} + \tilde{H}_{B_4B_5})
= U'_{B_1B_2} U'_{B_4B_5} \brr{\beta \br{\tilde{H}_{B_1}^\ast+ H_{B_2B_3B_4}+ \tilde{H}^\ast_{B_5}} 
+ \hat{\tilde{\Phi}}'_{B_1 B_2}+\hat{\tilde{\Phi}}'_{B_4B_5} } U_{B_4B_5}^{'\dagger}  U_{B_1B_2}^{'\dagger} .
\end{align}
By combining the inequalities~\eqref{main_ineq/thm:1D_effective_Ham} and \eqref{main_ineq/thm:1D_effective_Ham_approx}, we obtain 
\begin{align}
\label{main_ineq/thm:1D_effective_Ham____2}
\norm{\log(\rho_{\beta,B})- \beta (\tilde{H}_{B_1B_2} +H_{B_3} + \tilde{H}_{B_4B_5}) }\le e^{\xi_\beta -R/\Theta(\beta)} + e^{\Theta(\beta) \log(\beta R)-\kappa_\beta R/10}. 
\end{align}
From the above bound, we can ensure that the true effective interaction terms are localized around the boundary up to the distance of $\xi_\beta$. 
Although mathematical tools to access the true entanglement Hamiltonian are scarce so far, we believe that a refined analytical technique may improve the temperature dependence to a similar form to the CMI decay, i.e., $e^{\Theta(\beta) \log(\beta R)-\kappa_\beta R/10}$.

{~}

From Theorem~\ref{thm:1D_effective_Ham}, the approximation of the effective interaction terms up to a distance $R/5$ from the boundary $\partial B$ is considered. As a more convenient statement, we prove the following corollary, which argues the quasi-locality of the effective interaction around the boundary of $B$

 \begin{figure}[tt]
\centering
\includegraphics[clip, scale=0.4]{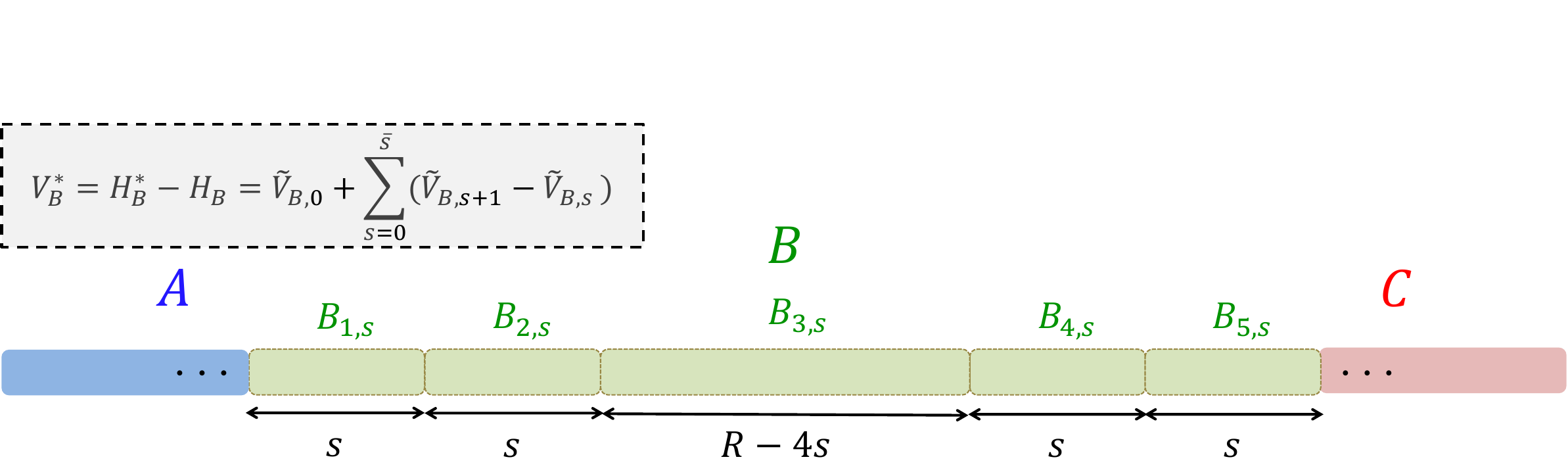}
\caption{Setup of Corollary~\ref{corol:1D_Ham_eff_short}.
We construct effective interactions $\{\tilde{V}_{B,s} \}_{s=0}^{\bar{s}+1}$, each which has an interaction length at most $2s + k$ as in Eq.~\eqref{def:tilde_V_B_ells}. By expanding $V_B^\ast$ as in Eq.~\eqref{V_B_ast_decompsition}, we can prove the exponential decay of the effective interaction $V_B^\ast$. 
}
\label{fig:Short_range_eff_int}
\end{figure}

{~}

\begin{corol} \label{corol:1D_Ham_eff_short}
Let $V_B^\ast$ be the effective interaction defined as the difference between the entanglement Hamiltonian $H_B^\ast$ and the subset Hamiltonian $H_B$:
\begin{align}
 \label{def:1D_Ham_eff_short}
V_B^\ast := H_B^\ast - H_B  =\frac{1}{\beta} \log\brr{ \tr_{AC} \br{e^{\beta H}}} - H_B ,
\end{align}
where $B^\co=AC$. 
Then, the effective interaction $V_B^\ast$ is quasi-local around the boundary in the sense that
\begin{align}
 \label{main:1D_Ham_eff_short}
\norm{\brr{ V_B^\ast , u_i}} \le \Theta(\beta) e^{\xi_\beta -d_{i,AC}/\Theta(\beta)}.
\end{align}
The inequality means that the effective interactions decay exponentially with the distance from the boundary. 
\end{corol}

{~}

\textit{Proof of Corollary~\ref{corol:1D_Ham_eff_short}.} 
For the proof, we adopt a similar decomposition to Fig.~\ref{fig:CMI_clustering_1D} and 
decompose the subsystem $B$ as $B_{1,s}$, $B_{2,s}$, $B_{3,s}$, $B_{4,s}$ and $B_{5,s}$ such that (see Fig.~\ref{fig:Short_range_eff_int})
 \begin{align}
B_{1,s}=B_{2,s}=B_{4,s}=B_{5,s}= s ,\quad B_{3,s}= |B| - 4s = R-4s ,\quad  s \le \bar{s}:=R/4.
\end{align}
Then, using the notation of Eq.~\eqref{main_ineq/thm:1D_effective_Ham_notation_til},]
we define the approximate effective interaction $\tilde{V}_{B,s}$ as 
\begin{align}
\label{def:tilde_V_B_ells}
\tilde{V}_{B,s} &:= \br{\tilde{H}_{B_{1,s}B_{2,s}} +H_{B_{3,s}} + \tilde{H}_{B_{4,s}B_{5,s}}} -H_B \notag \\
&=\br{\tilde{H}_{B_{1,s}B_{2,s}}  -H_{B_{1,s}B_{2,s}} }+  \br{\tilde{H}_{B_{4,s}B_{5,s}}  -H_{B_{4,s}B_{5,s}} }  + h_{\d B_{3,s}} ,
\end{align}
where we use the notation of Eq.~\eqref{def:Ham_surface}, and $h_{\d B_{3,s}}$ is localized around the boundary of $B_{3,s}$ up to the distance $k$. 
We also let $\tilde{V}_{B,\bar{s}+1}=V_B^\ast $.
Note that the operator $\tilde{V}_{B,s}$ is supported on $(B_{1,s}B_{2,s})[k] \cup (B_{4,s}B_{5,s})[k]$, where the boundary interaction between $B_{3,s}$ and $B_{1,s}B_{2,s}B_{4,s}B_{5,s}$ is taken into account. 

Using the notation of $\tilde{V}_{B,s}$, we rewrite the effective interaction $V_B^\ast $ as 
\begin{align}
\label{V_B_ast_decompsition}
V_B^\ast  =\tilde{V}_{B,0}+ \sum_{s=0}^{\bar{s}} \br{ \tilde{V}_{B,s+1}- \tilde{V}_{B,s} } .
\end{align}
Note that $\br{ \tilde{V}_{B,s+1}- \tilde{V}_{B,s} }$ is supported to the boundary region of $B$ up to an distance $2(s+1)+k$ (see Fig.~\ref{fig:Short_range_eff_int}). 
This means 
\begin{align}
\brr{\tilde{V}_{B,s+1}- \tilde{V}_{B,s} ,u_i} = 0 \for 2(s+1) < \dist_{i,AC}-k ,
\end{align}
where the condition  is satisfied for $s < \dist_{i,AC}/2 - 1 -k/2$. 
We therefore prove 
\begin{align}
\label{commutator_u/iu_V/B}
\brr{ V_B^\ast ,u_i}  = \sum_{s\ge \dist_{i,AC}/2 - 1 -k/2}^{\bar{s}} \brr{ \tilde{V}_{B,s+1}- \tilde{V}_{B,s},u_i } .
\end{align}

From the inequality~\eqref{main_ineq/thm:1D_effective_Ham____2}, we can derive 
\begin{align}
\label{main_ineq/thm:1D_effective_Ham2__re}
\norm{V_B^\ast -\tilde{V}_{B,s} }\le e^{\xi_\beta -s/\Theta(\beta)}  ,
\end{align}
which also yields the upper bound of 
\begin{align}
\label{commutator_u/iu_V/B_diff}
\norm{ \brr{ \tilde{V}_{B,s+1}- \tilde{V}_{B,s},u_i}} \le 4 e^{\xi_\beta -s/\Theta(\beta)}  .
\end{align}
By applying the inequality~\eqref{commutator_u/iu_V/B_diff} to~\eqref{commutator_u/iu_V/B}, we obtain the main inequality as follows:
\begin{align}
\label{commutator_u/iu_V/B_upp_1}
\norm{\brr{ V_B^\ast ,u_i} } \le \sum_{s\ge\dist_{i,AC}/2 - 1 -k/2}^{\bar{s}} 4 e^{\xi_\beta -s/\Theta(\beta)}  
\le \Theta(\beta) e^{\xi_\beta -d_{i,AC}/\Theta(\beta)}  . 
\end{align}
This completes the proof of Corollary~\ref{corol:1D_Ham_eff_short}. $\square$

\subsection{1D Hamiltonian learning with log sample complexity} \label{sec:1D Hamiltonian learning with poly-log sample complexity}

Here, we show an efficient Hamiltonian learning in one-dimensional systems by applying the quasi-locality of the true effective interactions.
By combining Theorem~\ref{thm:1D_effective_Ham} and Proposition~\ref{prop:Norm of the effective Hamiltonian} with the method in Ref.~\cite{anshu2021efficient}, we prove the following corollary:
\begin{corol} \label{corol:1D_Ham_learning}
Let us adopt the same setup as in Theorem~\ref{thm:1D_effective_Ham}.
Then, the sufficient number of copies of the quantum Gibbs state to learn the Hamiltonian $H$ up to an error $\epsilon$ is given by 
\begin{align}
\label{corol:1D_Ham_learning/eq1}
N=e^{\xi_\beta}   (1/\epsilon)^{\Theta(\beta^2)} \log(|\Lambda|) 
\end{align}
with $\xi_\beta=e^{e^{\Theta(\beta)}}$, where the success probability is larger than $0.99$ as in Corollary~\ref{corol:effective_Ham_Learning}, and the error is measured by 
  \begin{align}
\epsilon= \max_{Z\subset \Lambda} \norm{h_Z- \tilde{h}_Z} ,
\end{align}
with $\{\tilde{h}_Z\}_{Z\subset \Lambda}$ the reconstructed interactions. 
The time complexity for the learning is $|\Lambda| e^{\beta \xi_\beta}   (1/\epsilon)^{\Theta(\beta)}$. 
\end{corol}

{\bf Remark.} 
The corollary provides not only sample-efficient but also time-efficient Hamiltonian learning. 
On the time efficiency, the recent result achieved the polynomial time complexity in arbitrary dimensions, including infinite-dimensional graphs~\cite{bakshi2023learning,narayanan2024}. 
Although achieving the logarithmic sample complexity is still an active open research area, we believe that the logarithmic sample complexity should be derived without relying on the effective Hamiltonian theory\footnote{Private communication with Quynh The Nguyen.}.

{~}

\textit{Proof of Corollary~\ref{corol:1D_Ham_learning}.}
We adopt the same setup as in Ref.~\cite{Anshu_2021}, and the proof wholly relies on Ref.~\cite{anshu2021efficient}, which treats the learning of commuting Hamiltonians.
First, we consider a set of concatenated regions $\{L_s\}_{s=1}^{n}$ such that $|L_s|=R$ and $n =\Theta( |\Lambda|)$.
For each of $\{L_s\}_{s=1}^{n}$, we define the center region with the length $R/5$ as $\tilde{L}_s$. 

Let us define the operator bases on $L_s$ as $\{\hat{E}_{L_s,j}\}_{j=1}^{\mathcal{D}_{L_s}^2}$ for $s\in [n]$.
Then, the set of $\{e_{L_s,j}\}_{j=1}^{\mathcal{D}_{L_s}^2}$, defined by
\begin{align}
e_{L_s,j} = \tr \br{\hat{E}_{L_s,j} \rho_{\beta, L_s}} , 
\end{align}
characterizes the reduced density matrices $\{\rho_{\beta, L_s}\}_{s=1}^n$.
That is, the total number of parameters is given by $n\mathcal{D}_{L_s}^2$.
We now denote $e'_{L_s,j}$ by the measurement average of $\hat{E}_{L_s,j}$ for the copies of the quantum Gibbs state. 
Using Refs.~\cite{PhysRevX.10.031064,PhysRevLett.124.100401,Huang2020}, the sufficient number of samples $N$ to ensure
\begin{align}
\label{error_est_measure}
\norm{e_{L_s,j} - e'_{L_s,j} } \le \epsilon_0 \for \forall s,j
\end{align}
is given by
\begin{align}
\label{sample_compcl_N}
N = \frac{\Theta(1)}{\epsilon_0^2} \log\br{\frac{n \mathcal{D}_{L_s}^2}{\delta}} ,
\end{align}
where the success probability is at least $1-\delta$.

Under the condition~\eqref{error_est_measure}, the reconstructed quantum states $\{\rho'_{\beta, L_s}\}_{s=1}^n$ satisfies 
\begin{align}
\label{error_est_measure_2}
\norm{\rho_{\beta, L_s} - \rho'_{\beta, L_s}}_1 \le \mathcal{D}_{L_s}^2 \epsilon_0 \le e^{\Theta(R)} \epsilon_0.
\end{align}
Hence, using the inequality~\eqref{main_effective_Ham_Learning_2} with $\lambda_{\min}=e^{-\Theta(\beta) R}$ from Proposition~\ref{prop:Norm of the effective Hamiltonian}, we have 
\begin{align}
\label{error_est_measure_3}
\norm{\log\br{\rho_{\beta, L_s}} - \log\br{\rho'_{\beta, L_s}}} \le e^{\Theta(\beta) R} \epsilon_0.
\end{align}
Therefore, from Theorem~\ref{thm:1D_effective_Ham}, the subset Hamiltonian on the center region $\tilde{L}_s$ can be estimated up to the error of
\begin{align}
e^{\Theta(\beta) R} \epsilon_0 + e^{\xi_\beta -R/\Theta(\beta)} ,
\end{align}
where the second term comes from the effective interaction terms due to the partial trace.  
To make it smaller than the desired error $\epsilon$, we have to let
\begin{align}
R= \xi_\beta + \Theta(\beta) \log(1/\epsilon),  \quad 
\epsilon_0=e^{-\Theta(\beta) R}\epsilon = e^{-\Theta(\beta) \xi_\beta}  (1/\epsilon)^{-\Theta(\beta^2)}\epsilon =
e^{-\xi_\beta}   \epsilon^{\Theta(\beta^2)} ,
\end{align}
where we let $\beta \xi_\beta =\beta e^{e^{\Theta(\beta)}} = e^{e^{\Theta(\beta)+\log\log(\beta)}} =e^{e^{\Theta(\beta)}} $. 
By applying the above choice and $\delta=0.01$ to Eq.~\eqref{sample_compcl_N}, we prove the statement~\eqref{corol:1D_Ham_learning/eq1}. 
Also, the necessary time to construct the Hamiltonian is at most $n\cdot \poly( \mathcal{D}_{L_s})= ne^{\Theta(R)} \le |\Lambda| e^{ \xi_\beta} (1/\epsilon)^{\Theta(\beta)}$. 
This completes the proof of Corollary~\ref{corol:1D_Ham_learning}. $\square$

\subsection{Continuity inequality for logarithmic operators} \label{sec:Continuity inequality for logarithmic operators}

In this section, we aim to prove the following statements to relate the relative error to the continuity inequality of the operator logarithm.
We first define the relative error as follows:
\begin{definition}[Relative error] \label{def:relative error}
We define the relative error $\delta_{\rm R}(\rho,\sigma)$ between the density matrices $\rho$ and $\sigma$ as 
\begin{align}
\delta_{\rm R}(\rho,\sigma):= \sup_{\ket{\psi}} \frac{|\bra{\psi} \rho-\sigma \ket{\psi}|}{\bra{\psi} \rho \ket{\psi}} ,
\label{eq:def:relative error}
\end{align}
where the $\sup_{\ket{\psi}}$ is taken for all the set of quantum states $\ket{\psi}$. 
\end{definition}

{\bf Remark.} When we simply consider $\sup_{\ket{\psi}}|\bra{\psi} \rho-\sigma \ket{\psi}|$, it reduces to the standard operator norm as $\norm{\rho-\sigma}$. 
By using the minimum eigenvalue of $\rho$ as $\lambda_{\min}(\rho)$, we can upper bound 
\begin{align}
\label{ineq:delta_rho_sigma_norm}
\delta_{\rm R}(\rho,\sigma) \le \frac{1}{\lambda_{\min}(\rho)} \sup_{\ket{\psi}}|\bra{\psi} \rho-\sigma \ket{\psi}|=  \frac{\norm{\rho-\sigma}}{\lambda_{\min}(\rho)} .
\end{align}

We note that we usually obtain $\delta_{\rm R}(\rho,\sigma)\neq\delta_{\rm R}(\sigma,\rho)$. 
However, the two quantities are related to each other.
To see it, we use the definition~\eqref{eq:def:relative error} to provide
\begin{align}
\brr{1-\delta_{\rm R}(\rho,\sigma)} \bra{\psi} \rho \ket{\psi} \le     \bra{\psi} \sigma \ket{\psi} \le \brr{1+\delta_{\rm R}(\rho,\sigma)} \bra{\psi} \rho \ket{\psi}  ,
\end{align}
and hence 
\begin{align}
\frac{\bra{\psi} \sigma \ket{\psi} }{1+\delta_{\rm R}(\rho,\sigma)} \le     \bra{\psi} \rho \ket{\psi} \le \frac{\bra{\psi} \sigma \ket{\psi} }{1-\delta_{\rm R}(\rho,\sigma)}   .
\end{align}
Then, for $\delta_{\rm R}(\rho,\sigma)\le 1/2$, we have 
\begin{align}
\bra{\psi} \sigma \ket{\psi}  [1-\delta_{\rm R}(\rho,\sigma)] \le     \bra{\psi} \rho \ket{\psi} \le \bra{\psi} \sigma \ket{\psi} [1+2 \delta_{\rm R}(\rho,\sigma)]   ,
\end{align}
which yields for $\delta_{\rm R}(\sigma,\rho)$ 
\begin{align}
\label{delta_R_sigma_rho_up_rho_sigma}
\delta_{\rm R}(\sigma,\rho) \le 2 \delta_{\rm R}(\rho,\sigma) \for  \delta_{\rm R}(\rho,\sigma) \le \frac{1}{2}  .
\end{align}

On the relative error, one can immediately prove the following lemma:
\begin{lemma} \label{relative_error_local}
A small relative error between $\rho$ and $\sigma$ also implies a small relative error between their reduced density matrices:
\begin{align}
\label{relative_error_local/mainieq}
\delta_{\rm R}(\rho_L,\sigma_L) \le \delta_{\rm R}(\rho,\sigma) ,
\end{align}
where $\rho_L$ and $\sigma_L$ are reduced density matrices on a subset $L \subset \Lambda$. 
\end{lemma}

\textit{Proof of Lemma~\ref{relative_error_local}.}
For the proof, we let $\ket{\psi_L}$ be an arbitrary quantum state on the subset $L$ and $\{\ket{x_{L^\co}}\}_x$ be the orthonormal bases on the subset $L^\co$. We then obtain 
\begin{align}
\label{relative_error_loca_eq1}
\abs{ \sum_x \bra{\psi_L,x_{L^\co}} \rho-\sigma \ket{\psi_L,x_{L^\co}}} = 
\abs{\bra{\psi_L} \rho_L -\sigma_L \ket{\psi_L}} , 
\end{align}
where we use $\sum_x \bra{x_{L^\co}} \rho-\sigma \ket{x_{L^\co}}= \tr_{L^\co} (\rho-\sigma)$. 
Also, by using the global relative error $\delta_{\rm R}(\rho,\sigma)$, we have 
\begin{align}
\label{relative_error_loca_eq2}
\abs{ \sum_x \bra{\psi_L,x_{L^\co}} \rho-\sigma \ket{\psi_L,x_{L^\co}}} \le 
\sum_x\abs{\bra{\psi_L,x_{L^\co}} \rho-\sigma \ket{\psi_L,x_{L^\co}}} 
&\le \delta_{\rm R}(\rho,\sigma)\sum_x \bra{\psi_L,x_{L^\co}} \rho \ket{\psi_L,x_{L^\co}} \notag \\
&= \delta_{\rm R}(\rho,\sigma) \bra{\psi_L} \rho \ket{\psi_L} .  
\end{align}
By applying the inequality~\eqref{relative_error_loca_eq2} to Eq.~\eqref{relative_error_loca_eq1}, we arrive at the inequality of 
\begin{align}
\label{relative_error_loca_eq3}
\frac{\abs{\bra{\psi_L} \rho_L -\sigma_L \ket{\psi_L}}}{\bra{\psi_L} \rho \ket{\psi_L} } \le  \delta_{\rm R}(\rho,\sigma) 
\end{align}
for $\forall \psi_L$, which yields the desired inequality~\eqref{relative_error_local/mainieq}.
This completes the proof. $\square$ 

{~}

\hrulefill{\bf [ End of Proof of Lemma~\ref{relative_error_local}]}

{~}

Under the assumption of the relative error, one can prove the following continuity inequality for logarithmic operators (see Appendix~\ref{Theorem_proof:thm:refined continuity} for the proof):
\begin{theorem} \label{thm:refined continuity}
Under the condition that 
\begin{align}
 \varepsilon=\max\brr{ \delta_{\rm R}(\rho,\sigma), \delta_{\rm R}(\sigma,\rho) }  \le \frac{1}{2},
\label{cond:thm:refined continuity_0}
\end{align}
we obtain the upper bound as 
 \begin{align}
\norm{\log(\sigma)- \log(\rho)}\le  \varepsilon \brr{ \frac{4\log [2\lambda^{-1}_{\min}(\rho)] }{\pi} \log\br{\frac{e\log [2\lambda^{-1}_{\min}(\rho)]}{2\pi }} +23 } , 
\label{main_ineq:thm:refined continuity_0}
\end{align}
where $\lambda^{-1}_{\min}(\rho)$ is the minimum eigenvalue of $\rho$. 
\end{theorem}

{\bf Remark.}
To satisfy the condition~\eqref{cond:thm:refined continuity_0}, it is enough to ensure 
$$\min \br{ \delta_{\rm R}(\rho,\sigma), \delta_{\rm R}(\sigma,\rho) } \le \frac{1}{4}$$
from the inequality~\eqref{delta_R_sigma_rho_up_rho_sigma}. 
In practically important cases, the inverse minimum eigenvalue $\lambda^{-1}_{\min}(\rho)$ is at most of ${\rm Poly}(\mathcal{D}_\Lambda)$, where $\mathcal{D}_\Lambda$ is the Hilbert space dimension. Then, the upper bound is roughly given by $\norm{\log(\sigma)- \log(\rho)}\lesssim  \varepsilon \log(\mathcal{D}_\Lambda)\log\log(\mathcal{D}_\Lambda)$.  
 
In the classical case or the commuting case of $[\rho,\sigma]=0$, the inequality is rather trivial because $\sigma$ and $\rho$ have simultaneous eigenstates. By letting the eigenvalues of $\rho$ and $\sigma$ be $\{\rho_m\}_m$ and $\{\sigma_m\}_m$, respectively, we have 
 \begin{align}
 \label{classical_log_sigma_rho}
\norm{\log(\sigma)- \log(\rho)}= \max_m \abs{\log (\rho_m) - \log (\sigma_m) } =  \max_m \abs{\log \br{\frac{\sigma_m}{\rho_m}} }\le \log(1+\varepsilon) \le \varepsilon , 
\end{align}
where we use the condition~\eqref{cond:thm:refined continuity_0} in the second inequality. 
Therefore, the dependence on the minimum eigenvalue $\lambda_{\min}(\rho)$ only appears in the quantum setups. 
 
If we use only the norm distance $\norm{\rho-\sigma}$, we obtain from~\eqref{ineq:delta_rho_sigma_norm}:
\begin{align}
\norm{\log(\sigma)- \log(\rho)}\lesssim  \frac{{\rm  Polylog} [\lambda^{-1}_{\min}(\rho)] }{\lambda_{\min}(\rho)} \norm{\rho-\sigma} ,
\end{align}
which requires $\norm{\rho-\sigma} \ll \lambda_{\min}(\rho)$ for a good approximation. 

Finally, we emphasize that the converse of Theorem~\ref{thm:refined continuity} is not true\footnote{In the commuting cases, from the relations in~\eqref{classical_log_sigma_rho}, the small relative error and the closeness of the operator logarithms are equivalent.}. 
That is, the closeness of the logarithmic operators does not imply a small relative error, i.e., 
\begin{align}
\norm{\log(\sigma)- \log(\rho)} \ll 1   \xrightarrow{\textrm{Not imply}}   \delta_{\rm R}(\rho,\sigma) \ll 1 .
\end{align}
The counterexample is given by the following case in a $1$ qubit system:
\begin{align}
\log(\sigma)= J\sigma_z+ \epsilon \sigma_x ,\quad \log(\rho)= J\sigma_z  \quad (J>0), 
\end{align}
where $\sigma_x$ and $\sigma_z$ are the Pauli matrix with $\sigma_z \ket{0}=\ket{0}$ and $\sigma_z \ket{1}=-\ket{1}$.
Then, we have $\norm{\log(\sigma)- \log(\rho)}=\epsilon$, whereas we have 
\begin{align}
\delta_{\rm R}(\rho,\sigma)  \ge \frac{\bra{1} \sigma- \rho \ket{1}} {\bra{1}\rho \ket{1}}
&= e^J \abs{\cosh(J) - \cosh\br{\sqrt{J^2+\epsilon^2}} - \sinh(J) + \frac{J\sinh\br{\sqrt{J^2+\epsilon^2}}}{\sqrt{J^2+\epsilon^2}}} \notag \\
&=\frac{\abs{e^J \sinh(J)-J }}{2J^2} \epsilon^2  + \orderof{\epsilon^4} , 
\end{align}
which can be arbitrarily large in the limit of $J\to \infty$.
From this point, it is an important mathematical open problem to identify the necessary and efficient condition to get the small logarithmic error $\norm{\log(\sigma)- \log(\rho)} \ll 1 $. 

In applying Theorem~\ref{thm:refined continuity} to quantum Gibbs states, a technical challenge is to estimate the relative error between $e^{\beta H}$ and $e^{\beta \tilde{H}_\Lambda}$.
We prove the following subtheorem: 
\begin{subtheorem} \label{subthm:est_relative_error}
Let us adopt the same setup as in Theorem~\ref{thm:1D_effective_Ham}. 
Then, we obtain the upper bound for $\delta_{\rm R} (e^{\beta H}, e^{\beta \tilde{H}_\Lambda})$ as follows:
\begin{align}
\label{relative_error_finite_/range_fin}
\delta_{\rm R} \br{e^{\beta H}, e^{\beta \tilde{H}_\Lambda}} \le \exp\brr{\xi_\beta -R/\Theta(\beta)} ,
\end{align} 
where $\xi_\beta=e^{e^{\Theta(\beta)}}$, and $R$ is defined as the size of $B$ [see the definition~\eqref{def_tilde_rho_beta_1D__re} below].
\end{subtheorem}

\subsection{Proof of Theorem~\ref{thm:1D_effective_Ham}} \label{Proof of Theorem_thm_1D_effective_Ham}
 
Here, we prove Theorem~\ref{thm:1D_effective_Ham} based on Theorem~\ref{thm:refined continuity} and Subtheorem~\ref{subthm:est_relative_error}. 
We first apply Theorem~\ref{thm:refined continuity} to the operators $\rho_{\beta,B}$ and $\tilde{\rho}_{\beta,B}$.
To utilize the inequality~\eqref{main_ineq:thm:refined continuity_0}, we need to estimate the minimum eigenvalue of $\rho_{\beta,B}$, which we denote by $\lambda_{\min}$.
By using the inequality~\eqref{main_ineq_prop:Norm of the effective Hamiltonian} in Proposition~\ref{prop:Norm of the effective Hamiltonian}, we can immediately obtain
\begin{align}
\log(1/\lambda_{\min}) = \orderof{\beta |B|}  ,
\end{align} 
where we use the fact that the Hamiltonian is assumed to be $k$-local as in~\eqref{k-local_asummp_finite/range}.  
Also, from Lemma~\ref{relative_error_local}, we can ensure that the relative error is upper-bounded as 
\begin{align}
\delta_{\rm R} \brr{\tr_{AC} \br{e^{\beta H}}, \tr_{AC} \br{e^{\beta \tilde{H}_\Lambda}}}\le \delta_{\rm R} \br{e^{\beta H}, e^{\beta \tilde{H}_\Lambda}} .
\end{align} 
Then, by using Theorem~\ref{thm:refined continuity}, we prove 
 \begin{align}
\norm{\log(\rho_{\beta,B})- \log(\tilde{\rho}_{\beta,B})}
&\le  \Theta(1 )  \delta_{\rm R} \br{e^{\beta H}, e^{\beta \tilde{H}_\Lambda}} (\beta |B|)^2 ,
\end{align}
where we use $\beta |B| \log(\beta |B|) \le (\beta |B|)^2$ from $x\log(x) \le x^2$ for $x>0$.
By applying the upper bound~\eqref{relative_error_finite_/range_fin} for $\delta_{\rm R} \br{e^{\beta H}, e^{\beta \tilde{H}_\Lambda}}$ to the above inequality, 
we prove 
 \begin{align}
\norm{\log(\rho_{\beta,B})- \log(\tilde{\rho}_{\beta,B})}
&\le  e^{\xi_\beta -R/\Theta(\beta) +\Theta(1) \log(\beta R)} ,
\end{align}
where we use $|B|=R$. 
For $R\ge \xi_\beta$, we have $-R/\Theta(\beta) +\Theta(1) \log(\beta R) \le -R/\Theta(\beta)$, which gives the main inequality~\eqref{main_ineq/thm:1D_effective_Ham}. 
This completes the proof of Theorem~\ref{thm:1D_effective_Ham}. $\square$

\subsection{Proof of Subtheorem~\ref{subthm:est_relative_error}: estimation of the relative error}

For simplicity of the notations, we write
\begin{align}
\label{notation_v_1_v_2}
&H_0=H_{AB_1}+ H_{B_2B_3B_4}+ H_{B_5C}, \notag \\
&v_1= \partial h_{AB_1} ,\quad v_2=\partial h_{B_5C} ,
\end{align}
where $H=H_0+v_1+v_2$.
For the convenience of readers, we show the definition~\eqref{def_tilde_rho_beta_1D} of $e^{\beta \tilde{H}_\Lambda}$ again with the decomposition of $e^{\beta H}$ in Eq.~\eqref{BP_decompo_E_beta_H}: 
\begin{align}
&e^{\beta H} = \Phi_{\partial h_{B_5C}} \Phi_{\partial h_{AB_1}}e^{\beta H_0}\Phi_{\partial h_{AB_1}}^\dagger  \Phi_{\partial h_{B_5C}}^\dagger  , \notag \\
&e^{\beta \tilde{H}_\Lambda} =\tilde{\Phi}_{B_4B_5}\tilde{\Phi}_{B_1 B_2} e^{\beta H_0}\tilde{\Phi}_{B_1 B_2}^\dagger  \tilde{\Phi}_{B_4B_5}^\dagger  , 
\label{def_tilde_rho_beta_1D__re}
\end{align}
with
\begin{align}
&e^{\beta \br{H_0+ v_1}}= \Phi_{\partial h_{AB_1}}e^{\beta H_0}\Phi_{\partial h_{AB_1}}^\dagger  ,\notag \\
&e^{\beta \br{H_0+ v_1+v_2}} = \Phi_{\partial h_{B_5C}} e^{\beta \br{H_0+ v_1}}  \Phi_{\partial h_{B_5C}}^\dagger  .
\label{BP_A_B1_B2_B3_B4_B5_C//__re}
\end{align}

Here, the approximation  $\tilde{\Phi}_{B_1 B_2}$ and $\tilde{\Phi}_{B_4B_5}$ are defined by the local approximation of the belief propagation operators $\Phi_{\partial h_{AB_1}}$ and $\Phi_{\partial h_{B_5C}}$, respectively: 
\begin{align}
\tilde{\Phi}_{B_1B_2}= \mathcal{T} \exp \br{\int_0^1 \tilde{\phi}_{B_1B_2,\tau} d\tau}, \quad 
\tilde{\Phi}_{B_4B_5}= \mathcal{T} \exp \br{\int_0^1 \tilde{\phi}_{B_4B_5,\tau} d\tau},
\label{approx_tilde_Phi_B1_B4__re}
\end{align}
where $\tilde{\phi}_{B_1B_2,\tau}$ and $\tilde{\phi}_{B_4B_5,\tau}$ are explicitly given by
\begin{align}
\label{sup_Def:Phi_0_phie_re}
&\tilde{\phi}_{B_1B_2,\tau}= \frac{\beta}{2}  \int_{-\infty}^\infty f_\beta(t) \tilde{\tr}_{(B_1B_2)^\co }\brr{v_1(H_0+ \tau v_1,t) }dt,  \notag \\
&\tilde{\phi}_{B_4B_5,\tau}= \frac{\beta}{2}  \int_{-\infty}^\infty f_\beta(t)\tilde{\tr}_{(B_4B_5)^\co }\brr{v_2(H_0+v_1+ \tau v_2,t)} dt.
\end{align}
Note that after the partial traces $ \tilde{\tr}_{(B_1B_2)^\co}$ and $ \tilde{\tr}_{(B_4B_5)^\co }$, the operators are supported on $B_1B_2$ and $B_4B_5$, respectively.

To derive the relative error between $e^{\beta H}$ and $e^{\beta \tilde{H}_\Lambda}$ in Eq.~\eqref{def_tilde_rho_beta_1D__re}, we use the following inequality which holds for an arbitrary quantum state $\ket{\psi}$:
\begin{align}
&| \bra{\psi} e^{\beta H}  -e^{\beta \tilde{H}_\Lambda} \ket{\psi} | \notag \\
&=
| \bra{\psi} \Phi_{\partial h_{B_5C}} \Phi_{\partial h_{AB_1}}  e^{\beta H_0/2} \br{ 1  - e^{-\beta H_0/2} \Phi_{\partial h_{AB_1}}^{-1} \Phi_{\partial h_{B_5C}}^{-1}e^{\beta \tilde{H}_\Lambda}  \Phi_{\partial h_{B_5C}}^{-1} \Phi_{\partial h_{AB_1}}^{-1} e^{-\beta H_0/2} } e^{\beta H_0/2}\Phi_{\partial h_{AB_1}}^\dagger  \Phi_{\partial h_{B_5C}}^\dagger  \ket{\psi} | \notag \\
&\le  \bra{\psi} e^{\beta H}  \ket{\psi} \norm{ 1  - \mathcal{W}\mathcal{W}^\dagger} ,
\end{align}
where we define $\mathcal{W}$ as follows:
\begin{align}
\mathcal{W}:= e^{-\beta H_0/2} \Phi_{\partial h_{AB_1}}^{-1} \Phi_{\partial h_{B_5C}}^{-1} \tilde{\Phi}_{B_4B_5}\tilde{\Phi}_{B_1 B_2} e^{\beta H_0/2}.
\end{align}
By decomposing $1  - \mathcal{W}\mathcal{W}^\dagger = 1  - \mathcal{W} +\mathcal{W}(1- \mathcal{W}^\dagger)$, we obtain 
\begin{align}
\label{relative_error_finite_/range}
\delta_{\rm R} \br{e^{\beta H} , e^{\beta \tilde{H}_\Lambda}} \le  \norm{ 1  - \mathcal{W}\mathcal{W}^\dagger} \le 
\norm{1  - \mathcal{W}} \br{1+\norm{\mathcal{W}} }.
\end{align} 
To estimate the norm of $\norm{1  - \mathcal{W}}$, we use the following lemma:
\begin{lemma} \label{lem:error_estimation_BP_imag}
Let $A_x$ and $B_x$ be arbitrary operators that depend on $x$ with $0\le x\le x_0$.
We then define an arbitrary parameterization of the Hamiltonian as $H_x$ ($0\le x\le x_0$) such that $H_{x=0}=H_0$ and $\norm{dH_x/dx} < \infty$. 
We then obtain 
\begin{align}
&\norm{ e^{-\beta H_0/2}  \br{\mathcal{T} e^{\int_0^{x_0} B_x dx} }^{-1} \mathcal{T} e^{\int_0^{x_0} A_x dx} e^{\beta H_0/2} -1} \notag \\
&\le e^{\int_0^{x_0} \norm{e^{-\beta H_x/2} A_xe^{\beta H_x/2}} +  \norm{e^{-\beta H_x/2} B_x e^{\beta H_x/2}} + \norm{\Delta H_x}  dx} \int_0^{x_0} \norm{e^{-\beta H_x/2}(A_x - B_x ) e^{\beta H_x/2}}dx ,
\label{lem:error_estimation_BP_imag/main}
\end{align}
where we define $\Delta H_x$ as 
\begin{align}
\label{Def_H_X_effctive_Ham}
\Delta H_x :=  \beta \int_0^{\beta} e^{-z H_x/2} \frac{dH_x}{dx} e^{z H_x/2} dz  .
\end{align} 
\end{lemma}

\textit{Proof of Lemma~\ref{lem:error_estimation_BP_imag}.}
We first write
\begin{align}
\mathcal{T} e^{\int_0^{x_0} A_x dx} e^{\beta H_0/2} 
= e^{A_{x_M}/N}   \cdots e^{A_{x_2}/N}e^{A_{x_1}/N}  e^{\beta H_0/2}  + \orderof{N^{-2}}   ,
\end{align} 
with $x_m=m/N$ for $1\le m \le Nx_0=: M$. We then discretize $H_x$ by $\{H_{x_1},H_{x_2} ,\ldots,H_{x_M}\}$ and obtain 
\begin{align}
&e^{A_{x_M}/N} \cdots e^{A_{x_2}/N}e^{A_{x_1}/N}  e^{\beta H_0/2}  \notag \\
&=  e^{\beta H_{x_M}/2}  \br{e^{-\beta H_{x_M}/2} e^{A_{x_M}/N} e^{\beta H_{x_M}/2}}  \br{e^{-\beta H_{x_M}/2}  e^{\beta H_{x_{M-1}}/2}} \notag \\
&\cdots  \br{e^{-\beta H_{x_2}/2} e^{A_{x_2}/N}  e^{\beta H_{x_2}/2}}   \br{e^{-\beta H_{x_2}/2}  e^{\beta H_{x_1}/2}} \br{e^{-\beta H_{x_1}/2} e^{A_{x_1}/N}  e^{\beta H_{x_1}/2}}\br{e^{-\beta H_{x_1}/2}  e^{\beta H_0/2}} .
\end{align} 
For general $m$, we let $H_{x_m} -H_{x_{m-1}} =(1/N) (dH_x/dx)_{x=x_{m-1}}+\orderof{N^{-2}}$ and reduce 
\begin{align}
& \br{e^{-\beta H_{x_m}/2} e^{A_{x_m}/N}  e^{\beta H_{x_m}/2}}   \br{e^{-\beta H_{x_m}/2}  e^{\beta H_{x_{m-1}}/2}} \notag \\
&=\br{ 1 + \frac{1}{N}e^{-\beta H_{x_m}/2}  A_{x_m}e^{\beta H_{x_m}/2}}   \br{1- \frac{\beta}{2N} \int_0^{\beta} e^{-z H_{x_{m-1}}/2}  \br{\frac{dH_x}{dx}}_{x=x_{m-1}} e^{z H_{x_{m-1}}/2} dz } + \orderof{N^{-2}} \notag \\
&= \exp\brr{ \frac{1}{N} \br{ e^{-\beta H_{x_m}/2}  A_{x_m}e^{\beta H_{x_m}/2}-  \frac{1}{2}\Delta H_{x_{m-1}}}}+ \orderof{N^{-2}} ,
\end{align} 
where $\Delta H_x$ has been defined in Eq.~\eqref{Def_H_X_effctive_Ham}.

By combining the above calculations, we prove 
\begin{align}
\label{mathcal_T_A_x/exp}
\mathcal{T} e^{\int_0^{x_0} A_x dx} e^{\beta H_0/2} 
= e^{\beta H_{x_0}/2} \mathcal{T}  e^{\int_0^{x_0} \br{e^{-\beta H_x/2}  A_x e^{\beta H_x/2}- \Delta H_x/2} }dx
\end{align} 
In the same way, we obtain 
\begin{align}
\label{mathcal_T_B_x/exp}
e^{-\beta H_0/2}  \br{ \mathcal{T} e^{\int_0^{x_0} B_x dx}}^{-1}
=  \tilde{\mathcal{T}} e^{-\int_0^{x_0} \br{e^{-\beta H_x/2}  B_x e^{\beta H_x/2}+ \Delta H_x/2 }dx} e^{-\beta H_{x_0}/2} ,
\end{align} 
where we let the time-ordering operator $\tilde{\mathcal{T}}$ be
\begin{align}
\br{ \mathcal{T} e^{\int_0^{x_0} B_x dx}}^{-1}
=\tilde{\mathcal{T}} e^{-\int_0^{x_0} B_x dx} 
= e^{-B_{x_1}/N}e^{-B_{x_2}/N}   \cdots e^{-B_{x_M}/N}  + \orderof{N^{-2}}   
\end{align} 
with $N\to \infty$. 

From Eqs.~\eqref{mathcal_T_A_x/exp} and \eqref{mathcal_T_B_x/exp}, we can derive 
\begin{align}
&e^{-\beta H_0/2}  \br{ \mathcal{T} e^{\int_0^{x_0} B_x dx}}^{-1}  \mathcal{T} e^{\int_0^{x_0} A_x dx} e^{\beta H_0/2}  \notag \\
&=  \tilde{\mathcal{T}} e^{-\int_0^{x_0} \br{e^{-\beta H_x/2}  B_x e^{\beta H_x/2}+ \Delta H_x/2} dx}  \mathcal{T} e^{\int_0^{x_0} \br{e^{-\beta H_x/2}  A_x e^{\beta H_x/2}- \Delta H_x/2} dx} .
\end{align}
To estimate the norm of the above operator, we define the operator $G_z$ as  
\begin{align}
G_z := \tilde{\mathcal{T}} e^{-\int_0^{z} \br{ e^{-\beta H_x/2}  B_x e^{\beta H_x/2}+ \Delta H_x /2} dx}  \mathcal{T} e^{\int_0^{z} \br{e^{-\beta H_x/2}  A_x e^{\beta H_x/2}- \Delta H_x /2}dx}-1 ,
\end{align}
which depends on the parameter $z$. 
Here, we aim to derive the upper bound for $\norm{G_{x_0}}$. 
We then calculate the derivative of 
\begin{align}
\frac{d}{dz} G_z = \tilde{\mathcal{T}} e^{-\int_0^{z} \br{e^{-\beta H_x/2}  B_x e^{\beta H_x/2}+ \Delta H_x/2} dx} 
\brr{e^{-\beta H_z/2} \br{A_z- B_z} e^{\beta H_z/2}}   \mathcal{T} e^{\int_0^{z} \br{e^{-\beta H_x/2}  A_x e^{\beta H_x/2}- \Delta H_x/2} dx}, 
\end{align}
and obtain 
\begin{align}
\norm{\frac{d}{dz} G_z} \le e^{\int_0^{z} \norm{e^{-\beta H_x/2} A_xe^{\beta H_x/2}} +  \norm{e^{-\beta H_x/2} B_x e^{\beta H_x/2}} + \norm{\Delta H_x}  dx} \cdot \norm{e^{-\beta H_z/2}(A_z - B_z ) e^{\beta H_z/2}} .
\end{align}
By applying the above inequality to 
\begin{align}
\norm{G_{x_0}}  \le  \int_0^{x_0} \norm{\frac{d}{dx} G_x} dx,
\end{align}
we prove the main inequality~\eqref{lem:error_estimation_BP_imag/main}.
This completes the proof. $\square$

{~}

\hrulefill{\bf [ End of Proof of Lemma~\ref{lem:error_estimation_BP_imag}]}

{~}

\begin{figure}[tt]
\centering
\includegraphics[clip, scale=0.41]{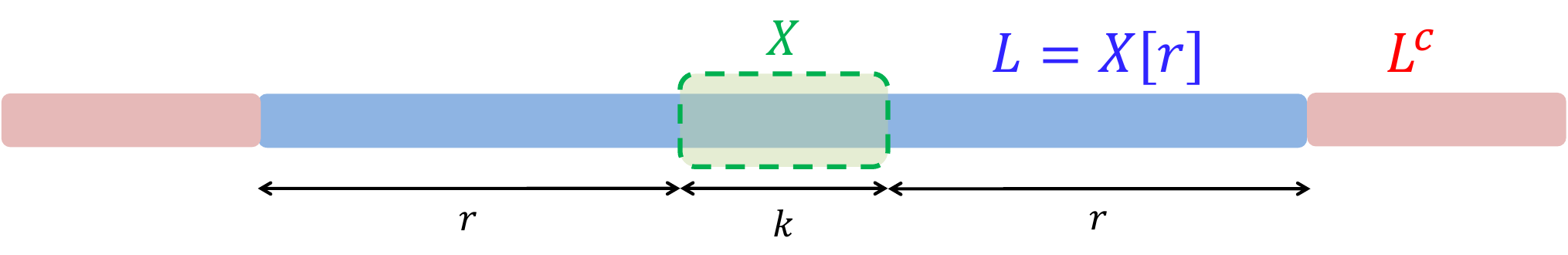}
\caption{Setup of Proposition~\ref{prop:O_X_imaginary_time}.
We consider a concatenated subset $X$ such that $|X|\le k$. We then consider the time evolution of $O_X(H,t)$ and its local approximation onto the region $X[r]$, which is also denoted by $L$. The approximated operator is defined by $\tilde{\tr}_{L^\co} [O_X(H,t)]$, whose approximation error has been estimated by Lemma~\ref{norm_local_approx} with the use of the Lieb--Robinson bound. 
In Proposition~\ref{prop:O_X_imaginary_time}, we aim to estimate the amplification of the error by the imaginary time evolution~\eqref{norm_e_-beta_H_O_X_e^beta_H_main}. 
The statement plays a key role in proving Subtheorem~\ref{subthm:est_relative_error} through the estimation of $\norm{1  - \mathcal{W}}$ in~\eqref{norm_1-mathcak_W}.
}
\label{fig:BP_imaginary}
\end{figure}

Here, we adopt the Hamiltonian $H_{\tau_1,\tau_2}$ as 
\begin{align}
H_{0,\tau}: =  H_0+ \tau v_1,\quad H_{1,\tau}:=H_0+v_1+ \tau v_2 , 
\end{align}
which reduce Eq.~\eqref{sup_Def:Phi_0_phie_re} to
\begin{align}
\label{sup_Def:Phi_0_phie_re2}
&\tilde{\phi}_{B_1B_2,\tau}= \frac{\beta}{2}  \int_{-\infty}^\infty f_\beta(t) \tilde{\tr}_{(B_1B_2)^\co }\brr{v_1(H_{0,\tau},t) }dt,  \quad \tilde{\phi}_{B_4B_5,\tau}= \frac{\beta}{2}  \int_{-\infty}^\infty f_\beta(t)\tilde{\tr}_{(B_4B_5)^\co }\brr{v_2(H_{1,\tau},t)} dt.
\end{align}
Then, by applying Lemma~\ref{lem:error_estimation_BP_imag} to 
\begin{align}
\mathcal{W}-1=e^{-\beta H_0/2} \Phi_{\partial h_{AB_1}}^{-1} \Phi_{\partial h_{B_5C}}^{-1} \tilde{\Phi}_{B_4B_5}\tilde{\Phi}_{B_1 B_2} e^{\beta H_0/2}-1,
\end{align}
we obtain the upper bound of
\begin{align}
\label{norm_1-mathcak_W}
\norm{1  - \mathcal{W}} 
\le &e^{\bar{w}_{B_1B_2} + \bar{w}_{B_4B_5}+  \bar{w}_0} 
\int_0^1  \norm{ e^{-\beta H_{0,\tau}/2} \br{\phi_{\partial h_{AB_1},\tau}-\tilde{\phi}_{B_1B_2,\tau} }e^{\beta H_{0,\tau}/2} }  d\tau \notag \\
&+e^{\bar{w}_{B_1B_2} + \bar{w}_{B_4B_5}+\bar{w}_0}  \int_0^1 \norm{ e^{-\beta H_{1,\tau}/2} \br{\phi_{\partial h_{B_5C},\tau}-\tilde{\phi}_{B_4B_5,\tau} }e^{\beta H_{1,\tau}/2} }    d\tau  ,
\end{align}
where $\bar{w}_{B_1B_2}$, $\bar{w}_{B_4B_5}$ and $\bar{w}_0$, are defined as follows:
\begin{align}
\label{bar_w_B1B2_B4_B5}
& \bar{w}_{B_1B_2} := \int_{0}^1 \br{ \norm{ e^{-\beta H_{0,\tau}/2} \phi_{\partial h_{AB_1},\tau} e^{\beta H_{0,\tau}/2} }  
 + \norm{ e^{-\beta H_{0,\tau}/2} \tilde{\phi}_{B_1B_2,\tau} e^{\beta H_{0,\tau}/2} }  }  d\tau, \notag \\
&\bar{w}_{B_4B_5} := \int_{0}^1  \br{\norm{ e^{-\beta H_{1,\tau}/2} \phi_{\partial h_{B_5C},\tau} e^{\beta H_{1,\tau}/2} }  
 + \norm{ e^{-\beta H_{1,\tau}/2} \tilde{\phi}_{B_4B_5,\tau} e^{\beta H_{1,\tau}/2} } } d\tau, \notag \\
 & \bar{w}_0:=  \beta \int_{0}^1  \int_0^{\beta} \br{\norm{e^{-z H_{0,\tau}/2} v_1 e^{z H_{0,\tau}/2} }+ \norm{e^{-z H_{1,\tau}/2} v_2 e^{z H_{1,\tau}/2} } } dz   d\tau.
\end{align}
Using Eq.~\eqref{sup_Def:Phi_0_phie_re2}, we further obtain an upper bound of 
\begin{align}
& \norm{ e^{-\beta H_{0,\tau}/2} \br{\phi_{\partial h_{AB_1},\tau}-\tilde{\phi}_{B_1B_2,\tau} }e^{\beta H_{0,\tau}/2} }   \notag \\
 &\le  \frac{\beta}{2}  \int_{-\infty}^\infty f_\beta(t) \norm{ e^{-\beta H_{0,\tau}/2} \brrr{v_1(H_{0,\tau},t)  -\tilde{\tr}_{(B_1B_2)^\co }\brr{v_1(H_{0,\tau},t) }  }e^{\beta H_{0,\tau}/2} }  dt,
 \label{imaginary_Bp_error}
 \end{align}
where we can derive a similar bound for $\norm{ e^{-\beta H_0/2} \br{\phi_{\partial h_{B_5C},\tau}-\tilde{\phi}_{B_4B_5,\tau} }e^{\beta H_0/2} }$.  
We can estimate the integral in Eq.~\eqref{imaginary_Bp_error} using the following general proposition:
\begin{prop}\label{prop:O_X_imaginary_time}
Let $O_X$ be an arbitrary operator on a subset $X$ with $|X|\le k$, where we assume $\norm{O_X}=1$. 
Then, for an arbitrary Hamiltonian with finite-range interactions as in Eq.~\eqref{k-local_asummp_finite/range},  we have 
\begin{align}
&\norm{e^{-\beta H/2} \brrr{O_X(H,t) - \tilde{\tr}_{L^\co} \br{O_X(H,t)}} e^{\beta H/2} } \le \exp\br{e^{e^{\Theta(\beta)}}} \min\br{1, e^{-\mu r/3+vt}},
\label{norm_e_-beta_H_O_X_e^beta_H_main}
\end{align}
where $L=X[r]$ (see Fig.~\ref{fig:BP_imaginary} for the setup). 
Moreover, as long as the RHS of~\eqref{norm_e_-beta_H_O_X_e^beta_H_main} is smaller than $1$, i.e., $r=e^{e^{\Omega(\beta)}}$\footnote{The condition is utilized in the inequality~\eqref{lastt_ineq_Prop50} for the proof below.}, 
we also prove the upper bound of 
\begin{align}
&\norm{e^{-\beta H/2} O_X(H,t) e^{\beta H/2} } \le e^{e^{\Theta(\beta)}}, \quad 
\norm{e^{-\beta H/2}\tilde{\tr}_{L^\co} \br{O_X(H,t)} e^{\beta H/2} } \le e^{e^{\Theta(\beta)}}. 
\label{norm_e_-beta_H_O_X_e^beta_H_main__23}
\end{align}
We defer the proof to the subsequent subsection (Sec.~\ref{Proof of prop:O_X_imaginary_time}).
\end{prop}

{~}

By applying the above proposition to~\eqref{imaginary_Bp_error} with
\begin{align}
&O_X \to v_1 = \sum_{\substack{Z: \diam(Z)\le k  \\   Z\cap B_1 \neq \emptyset,\ Z\cap B_2 \neq \emptyset}} h_Z, \notag \\
&H \to H_{0,\tau}  , \quad L\to B_1B_2,\quad r\to R/5 -k,
\end{align}
we reduce the inequality~\eqref{imaginary_Bp_error} to
\begin{align}
& \norm{ e^{-\beta H_0/2} \br{\phi_{\partial h_{AB_1},\tau}-\tilde{\phi}_{B_1B_2,\tau} }e^{\beta H_0/2} }  \le  \frac{\beta}{2} \exp\br{e^{e^{\Theta(\beta)}}}  \int_{-\infty}^\infty f_\beta(t) \min\br{1, e^{-\mu R/15+\mu k/3+ vt}} dt  ,
 \label{imaginary_Bp_error__2}
 \end{align}
Note that the notations $v_1$ and $v_2$ were given in Eq.~\eqref{notation_v_1_v_2}.
Using a similar decomposition to~\eqref{integral_ft_ge_delta_t_1D}, we obtain the following inequality by choosing $t_0$ such that $ e^{-\mu R/15+vt} = e^{-\mu R/30}$, or $t_0=\mu R/(30v)$: 
\begin{align}
 \int_{-\infty}^\infty f_\beta(t) \min\br{1, e^{-\mu R/15+\mu k/3+vt}} dt  
\le  e^{-\mu R/30+\mu k/3}  \int_{|t|\le t_0}  f_\beta(t)  dt + \int_{|t| \ge t_0}  f_\beta(t)  dt  
\le e^{-R/\Theta(\beta)},
\end{align}
which reduces the inequality~\eqref{imaginary_Bp_error__2} to 
\begin{align}
 \label{imaginary_Bp_error__1_fin}
& \norm{ e^{-\beta H_0/2} \br{\phi_{\partial h_{AB_1},\tau}-\tilde{\phi}_{B_1B_2,\tau} }e^{\beta H_0/2} } 
  \le  \exp\br{e^{e^{\Theta(\beta)}}} e^{-R/\Theta(\beta)} ,
\end{align}
where we absorb the coefficient $(\beta/2)$ into $\exp\br{e^{e^{\Theta(\beta)}}}$. 
We obtain the same inequality for the norm of $\norm{ e^{-\beta H_0/2} \br{\phi_{\partial h_{B_5C},\tau}-\tilde{\phi}_{B_4B_5,\tau} }e^{\beta H_0/2} }$.

Also, by using the inequality~\eqref{norm_e_-beta_H_O_X_e^beta_H_main__23}, we can upper-bound $ \bar{w}_{B_1B_2}$, $\bar{w}_{B_4B_5}$ and $\bar{w}_0$ in Eq.~\eqref{bar_w_B1B2_B4_B5} by $e^{e^{\Theta(\beta)}}$, which yields 
\begin{align}
\label{bar_w_B1B2_B4_B5_ffin}
e^{ \bar{w}_{B_1B_2} + \bar{w}_{B_4B_5}+\bar{w}_0} \le \exp\br{e^{e^{\Theta(\beta)}}} . 
\end{align}
By applying the inequalities~\eqref{imaginary_Bp_error__1_fin} and \eqref{bar_w_B1B2_B4_B5_ffin} to \eqref{norm_1-mathcak_W}, we prove 
\begin{align}
\norm{1  - \mathcal{W}} 
\le \exp\br{e^{e^{\Theta(\beta)}}} e^{-R/\Theta(\beta)} ,
\end{align}
which reduces the inequality~\eqref{relative_error_finite_/range} to the desired one in~\eqref{relative_error_finite_/range_fin}.
This completes the proof of Subtheorem~\ref{subthm:est_relative_error}. $\square$

\subsection{Proof of Proposition~\ref{prop:O_X_imaginary_time}} \label{Proof of prop:O_X_imaginary_time}

\subsubsection{Preliminaries}
Here, we consider the one-dimensional Hamiltonian with finite-range interactions as follows:
\begin{align}
H=\sum_{Z: \diam(Z)\le k} h_Z ,\quad  \sum_{Z:Z\ni i } \norm{h_Z} \le J_0,
\end{align}
where $\diam(Z)$ is defined as $\diam(Z)=\max_{i,j\in Z} (\dist_{i,j})$. 
We consider the imaginary-time evolution for an arbitrary $O_X$ with $\diam(X)\le k$ and estimate 
 \begin{align}
e^{-\tau H} O_X e^{\tau H} . 
\end{align}
We first obtain the following lemma, which is a direct consequence of Ref.~\cite[Corollary~1 therein]{kuwahara2018polynomialtime}.
\begin{lemma} \label{lem:imaginary_time_norm}
For  the norm of imaginary-time evolved operator $\norm{ e^{-\tau H} O_X e^{\tau H} } $, we can prove the following upper bound:
  \begin{align}
\norm{ e^{-\tau H} O_X e^{\tau H} } 
&\le \sum_{m=0}^\infty \frac{\tau^m}{m!} \norm{\ad^m_{H}(O_X)}
 \le 
2 m_0 e^{m_0} \norm{O_X} \le e^{e^{\Theta(\tau)}}   \norm{O_X}   ,
\label{main_ineq_lem:imaginary_time_norm}
\end{align}
where $m_0=e^{ 4e^2 k J_0 \tau} $. 
\end{lemma}

\textit{Proof of Lemma~\ref{lem:imaginary_time_norm}.}
We start from the statement in Ref.~\cite[Corollary~1 therein]{kuwahara2018polynomialtime} as follows:
 \begin{align}
\norm{\ad_H^m (O_X) } 
& \le \norm{O_X} \br{6kJ_0}^m \br{\frac{m+\tilde{l}_0}{W[e(m+\tilde{l}_0)]}}^{m+\tilde{l}_0-\frac{m+\tilde{l}_0}{W[e(m+\tilde{l}_0)]}} \notag \\
&\le \norm{O_X} \br{\frac{6kJ_0m}{1.602\log(m)}}^m \for \forall X\subset \Lambda \quad (|X| \le k) ,
\end{align}
where $\tilde{l}_0=|X|/k=1$ and $m\ge 2$ is assumed. 
By using the above inequality, we have 
 \begin{align}
 \frac{\tau^m}{m!} \norm{\ad^m_{H}(O_X)} \le  \br{\frac{e\tau}{m}}^m  \norm{\ad^m_{H}(O_X)}\le
\begin{cases} 
 e^{m_0}  &\for m< m_0=e^{ 4e^2 k J_0 \tau} , \\
e^{-m} & \for m\ge m_0 ,
\end{cases}
 \end{align}
which yields 
 \begin{align}
\norm{ e^{-\tau H} O_X e^{\tau H} } 
&\le \sum_{m=0}^\infty \frac{\tau^m}{m!} \norm{\ad^m_{H}(O_X)}
 \le 
\norm{O_X}  \br{m_0 e^{m_0} + \frac{e^{-m_0}}{1-e^{-1}}} .
\end{align}
We thus prove the main inequality~\eqref{main_ineq_lem:imaginary_time_norm}. 
This completes the proof. $\square$

{~}

\hrulefill{\bf [ End of Proof of Lemma~\ref{lem:imaginary_time_norm}]}

{~}

Also, we can prove the imaginary Lieb--Robinson bound as follows (see Ref.~\cite[Lemma~3 therein]{kuwahara2018polynomialtime}):
\begin{lemma} \label{lem:imaginary_time_LR}
We can approximate $e^{-\tau H} O_X e^{\tau H}$ by $\tilde{O}^{(\tau)}_{X[r]}$ with the error of 
  \begin{align}
\norm { e^{-\tau H} O_X e^{\tau H} -\tilde{O}^{(\tau)}_{X[r]}} 
 \le  \frac{\zeta_r^{\lceil r/k \rceil}}{1-\zeta_r} \le e^{-\mu r + e^{\Theta(\tau)}} ,
\end{align}
where $\zeta_r=6ek\tau/\log(\lceil r/k \rceil)$, and $\tilde{O}^{(\tau)}_{X[r]}$ is constructed from the truncation of the expansion of 
$e^{-\tau H} O_X e^{\tau H} \le \sum_{m=0}^\infty \frac{(-\tau)^m}{m!} \ad^m_{H}(O_X)$, and $\mu$ can be arbitrarily chosen. 
\end{lemma} 
In the proof below, we fully utilize the above two lemmas.

\subsubsection{Proof of Proposition~\ref{prop:O_X_imaginary_time}}

%

We begin with defining $\tilde{O}$ as 
\begin{align}
\label{def_tilde_O_x}
\tilde{O}:= e^{-\beta H/2} O_X e^{\beta H/2} ,
\end{align}
which gives $e^{-\beta H/2} O_X(H,t) e^{\beta H/2} = \tilde{O}(H,t)$.
We aim to upper-bound the norm of  
\begin{align}
&\norm{e^{-\beta H/2} \brrr{O_X(H,t) - \tilde{\tr}_{L^\co} \br{O_X(H,t)}} e^{\beta H/2} }  \notag \\
&= \exp\br{e^{e^{\Theta(\beta)}}} \norm{\tilde{O}(H,t) - e^{-\beta H/2} \tilde{\tr}_{L^\co} \brr{O_X(H,t)} e^{\beta H/2} } .
\label{norm_e_-beta_H_O_X_e^beta_H}
\end{align}
A difficulty stems from the fact that $e^{-\beta H/2} \tilde{\tr}_{L^\co} \brr{O_X(H,t)} e^{\beta H/2} \neq \tilde{\tr}_{L^\co} \brr{\tilde{O}(H,t)}$ because $e^{-\beta H/2}$ cannot be inserted in $\tilde{\tr}_{L^\co} \br{\cdots}$. 

In the following, we aim to upper-bound
\begin{align}
\label{norm_e-beta_H/2_tr_L^co_target}
 \norm{e^{-\beta H/2} \tilde{\tr}_{L^\co} \brr{O_X(H,t)} e^{\beta H/2} -   \tilde{\tr}_{L^\co} \brr{\tilde{O}(H,t)} } .
\end{align}
We introduce the operator $W_\beta$ as
\begin{align}
W_\beta=  e^{-\beta H/2}  e^{\beta (H_L+ H_{L^\co})/2} ,
\end{align}
where $H_L$ and $H_{L^\co}$ are subset Hamiltonian of $H$ on $L$ and $L^\co$, respectively. 
Using the above notation, we have 
\begin{align}
\label{tilde_L_tr_e^-beta_H}
e^{-\beta H/2} \tilde{\tr}_{L^\co} \brr{O_X(H,t)} e^{\beta H/2} 
& =W_\beta \tilde{\tr}_{L^\co} \br{e^{-\beta (H_L+ H_{L^\co})/2} O_X(H,t) e^{\beta (H_L+ H_{L^\co})/2}  }W_\beta^{-1}  . 
\end{align}
Furthermore, we calculate as 
\begin{align}
\label{tilde_L_tr_e^-beta_H_without}
\tilde{\tr}_{L^\co} 
\brr{e^{-\beta (H_L+ H_{L^\co})/2} O_X(H,t) e^{\beta (H_L+ H_{L^\co})/2}  }&=\tilde{\tr}_{L^\co} \brr{W_\beta^{-1} e^{-\beta H/2} O_X(H,t)  e^{\beta H/2}W_\beta} \notag \\
&=\tilde{\tr}_{L^\co} \brr{W_\beta^{-1} \tilde{O}(H,t) W_\beta} \notag \\
&=\tilde{\tr}_{L^\co} \br{W_\beta^{-1} \brr{ \tilde{O}(H,t), W_\beta}} + 
\tilde{\tr}_{L^\co} \brr{\tilde{O}(H,t) } .
\end{align}
By combining Eqs.~\eqref{tilde_L_tr_e^-beta_H} and \eqref{tilde_L_tr_e^-beta_H_without}, we have 
\begin{align}
e^{-\beta H/2} \tilde{\tr}_{L^\co} \brr{O_X(H,t)} e^{\beta H/2} 
& =W_\beta \brrr{\tilde{\tr}_{L^\co} \br{W_\beta^{-1} \brr{ \tilde{O}(H,t), W_\beta}} + 
\tilde{\tr}_{L^\co} \brr{\tilde{O}(H,t) }}W_\beta^{-1}  ,
\end{align}
which yields 
\begin{align}
\label{tilde_L_tr_e^-beta_H_iupper}
&\norm{e^{-\beta H/2} \tilde{\tr}_{L^\co} \brr{O_X(H,t)} e^{\beta H/2} - \tilde{\tr}_{L^\co} \brr{\tilde{O}(H,t) }} \notag \\
&\le \norm{W_\beta} \cdot \norm{W_\beta^{-1}}^2\cdot \norm{\brr{ \tilde{O}(H,t), W_\beta}}
+\norm{W_\beta^{-1}}\cdot \norm{\brr{W_\beta, \tilde{\tr}_{L^\co} \br{\tilde{O}(H,t) } }} \notag \\
&\le \exp\br{e^{e^{\Theta(\beta)}}} \br{\norm{\brr{ \tilde{O}(H,t), W_\beta}}+ \norm{\brr{W_\beta, \tilde{\tr}_{L^\co} \br{\tilde{O}(H,t) } } }} ,
\end{align}
where we use the following inequality to derive $\norm{W_\beta}= \norm{ e^{-\beta H/2}  e^{\beta (H_L+ H_{L^\co})/2}}\le \exp\br{e^{e^{\Theta(\beta)}}}$:
\begin{align}
\label{imaginary_decomp_e_beta_H_H}
 &\norm{e^{-\beta H/2}e^{\beta (H_L+ H_{L^\co})/2}} 
 =  \norm{e^{-\beta H/2} e^{\beta H/2} e^{-(\beta/2) \int_0^1 e^{-\beta xH/2} \partial h_L e^{\beta xH/2}} dx} \notag \\
 &\le \exp \br{\frac{\beta}{2} \int_0^1 \sum_{Z:Z\cap L \neq \emptyset, \ Z\cap L^\co \neq \emptyset} \norm{e^{-\beta xH/2} h_Z e^{\beta xH/2}} dx}   \le \exp\br{e^{e^{\Theta(\beta)}}} 
\end{align}
with $\partial h_L=H- (H_L+ H_{L^\co})= \sum_{Z:Z\cap L \neq \emptyset, \ Z\cap L^\co \neq \emptyset} h_Z$.

In the following, by using the imaginary Lieb--Robinson bound in Lemma~\ref{lem:imaginary_time_LR}, we approximate $\tilde{O}$ in Eq.~\eqref{def_tilde_O_x} onto the subsets $X[r/3]$:
\begin{align}
\label{local_approx_tilde_O}
\norm{\tilde{O} - \tilde{O}_{X[r/3]}} \le e^{-\mu r/3  + e^{\Theta(\beta)}} .
\end{align}
In the same way, we approximate $W_\beta$ onto the subset $L^\co[r/3]=(X[2r/3])^\co$, where the approximation error is given by
\begin{align}
\norm{ W_\beta - W_{\beta, L^\co[r/3]} } \le\exp\br{e^{e^{\Theta(\beta)}}}  e^{-\mu r/3  + e^{\Theta(\beta)}}  \le \exp\br{e^{e^{\Theta(\beta)}}} e^{-\mu r/3} , 
\end{align}
where we apply Lemma~\ref{lem:imaginary_time_LR} to $e^{-\beta x H/2} \partial h_Le^{\beta x H/2}$ and use the inequality~\eqref{upp_norm_Delta_U_tau_derr} for 
\begin{align}
&W_{\beta}=  e^{-\beta H/2}  e^{\beta (H_L+ H_{L^\co})/2} = \mathcal{T} e^{-(\beta/2) \int_0^1 e^{-\beta x H/2} \partial h_Le^{\beta x H/2} dx} , \notag \\
&W_{\beta, L^\co[r/3]}=  \mathcal{T} e^{-(\beta/2) \int_0^1 \br{\partial h_L}^{(\beta x/2)}_{L^\co[r/3]} dx}
\end{align}
to derive 
\begin{align}
\norm{ W_\beta - W_{\beta, L^\co[r/3]} } \le e^{(\beta/2)\int_0^1 \norm{e^{-\beta x H/2} \partial h_Le^{\beta x H/2}}dx} \int_0^1 \norm{ e^{-\beta x H/2} \partial h_Le^{\beta x H/2}- \br{\partial h_L}^{(\beta x/2)}_{L^\co[r/3]}} dx .
\end{align}
Here, the operator $\br{\partial h_L}^{(\beta x/2)}_{L^\co[r/3]}$ is an approximation of $e^{-\beta x H/2} \partial h_Le^{\beta x H/2}$ onto $L^\co[r/3]$ as has been defined in Lemma~\ref{lem:imaginary_time_LR}. 
Note that $\norm{e^{-\beta x H/2} \partial h_Le^{\beta x H/2}}\le e^{e^{\Theta(\beta)}} $ can be derived from Lemma~\ref{lem:imaginary_time_norm}.

By using the above upper bounds to the RHS of~\eqref{tilde_L_tr_e^-beta_H_iupper}, we have 
\begin{align}
\label{tilde_L_tr_e^-beta_H_iupper__1}
&\norm{e^{-\beta H/2} \tilde{\tr}_{L^\co} \br{O_X(H,t)} e^{\beta H/2} - \tilde{\tr}_{L^\co} \br{\tilde{O}(H,t) }} \notag \\
&\le \exp\br{e^{e^{\Theta(\beta)}}}   \Bigl( \norm{\brr{  \tilde{O}_{X[r/3]}(H,t), W_{\beta, L^\co[r/3]}}} + \norm{\brr{W_{\beta, L^\co[r/3]}, \tilde{\tr}_{L^\co} \br{ \tilde{O}_{X[r/3]}(H,t) } } } +e^{-\mu r/3 }\Bigr).
\end{align}
Because of $X[2r/3] \cap L^\co[r/3]= \emptyset $ from $X[r]=:L$, we obtain 
 \begin{align}
 \label{tilde_L_tr_e^-beta_H_iupper/2}
& \norm{\brr{  \tilde{O}_{X[r/3]}(H,t), W_{\beta, L^\co[r/3]}}} + \norm{\brr{W_{\beta, L^\co[r/3]}, \tilde{\tr}_{L^\co} \br{ \tilde{O}_{X[r/3]}(H,t) } } }  \notag \\
&\le 4 \norm{W_{\beta, L^\co[r/3]}} \cdot \norm{\tilde{O}_{X[r/3]}(H,t) - \tilde{\tr}_{X[2r/3]^\co}\br{\tilde{O}_{X[r/3]}(H,t)}},
\end{align} 
where we use that $ \tilde{\tr}_{X[2r/3]^\co}\brr{\tilde{O}_{X[r/3]}(H,t)}$ is supported on the subset $X[2r/3]$. 
Finally,  by using the Lieb--Robinson bound, we have 
 \begin{align}
  \label{tilde_L_tr_e^-beta_H_iupper/23}
\norm{\tilde{O}_{X[r/3]}(H,t) - \tilde{\tr}_{X[2r/3]^\co}\br{\tilde{O}_{X[r/3]}(H,t)}} 
&\le \norm{\tilde{O}_{X[r/3]}} \min\br{2, e^{-\mu r/3+vt+\Theta(1)} } \notag \\
&\le e^{e^{\Theta(\beta)}}   \min\br{1, e^{-\mu r/3+vt}}.
\end{align} 
By combining the inequalities~\eqref{tilde_L_tr_e^-beta_H_iupper__1}, \eqref{tilde_L_tr_e^-beta_H_iupper/2} and~\eqref{tilde_L_tr_e^-beta_H_iupper/23}, we arrive at the inequality of 
\begin{align}
\label{tilde_L_tr_e^-beta_H_iupper_fin}
&\norm{e^{-\beta H/2} \tilde{\tr}_{L^\co} \brr{O_X(H,t)} e^{\beta H/2} - \tilde{\tr}_{L^\co} \brr{\tilde{O}(H,t) }} \le
\exp\br{e^{e^{\Theta(\beta)}}}   \min\br{1, e^{-\mu r/3+vt}} , 
\end{align}
which gives the upper bound for the target quantity~\eqref{norm_e-beta_H/2_tr_L^co_target}. 
It reduces the inequality~\eqref{norm_e_-beta_H_O_X_e^beta_H} to 
\begin{align}
\label{norm_e-beta_H/2_tr_L^co}
&\norm{e^{-\beta H/2} \brrr{O_X(H,t) - \tilde{\tr}_{L^\co} \brr{O_X(H,t)}} e^{\beta H/2} }  \notag \\
&\le \exp\br{e^{e^{\Theta(\beta)}}}\brr{ \norm{\tilde{O}(H,t) - \tilde{\tr}_{L^\co} \br{\tilde{O}(H,t) }} + \min\br{1, e^{-\mu r/3+vt}}} .
\end{align}

We finally estimate the norm of $\norm{\tilde{O}(H,t) - \tilde{\tr}_{L^\co} \brr{\tilde{O}(H,t) }}$. 
From the inequalities~\eqref{local_approx_tilde_O} and~\eqref{tilde_L_tr_e^-beta_H_iupper/23}, we can also derive 
\begin{align}
\label{norm_e-beta_H/2_tr_L^co_Otil_H_t}
&\norm{\tilde{O}(H,t) - \tilde{\tr}_{L^\co} \brr{\tilde{O}(H,t) }  } \le 2\norm{\tilde{O} - \tilde{O}_{X[r/3]}} + \norm{\tilde{O}_{X[r/3]}(H,t) - \tilde{\tr}_{L^\co} \brr{\tilde{O}_{X[r/3]}(H,t) }  } \notag \\
&\le  e^{e^{\Theta(\beta)} -\mu r/3} +    \norm{\tilde{O}_{X[r/3]}(H,t) - \tilde{\tr}_{X[2r/3]^\co} \brr{\tilde{O}_{X[r/3]}(H,t) }  } \notag \\
&\le e^{e^{\Theta(\beta)}} \min\br{1, e^{-\mu r/3+vt}}  .
\end{align}
By combining the two inequalities~\eqref{norm_e-beta_H/2_tr_L^co} and \eqref{norm_e-beta_H/2_tr_L^co_Otil_H_t}, 
we prove the first main inequality~\eqref{norm_e_-beta_H_O_X_e^beta_H_main}.

For the proof of the other main inequalities in~\eqref{norm_e_-beta_H_O_X_e^beta_H_main__23}, we immediately prove 
\begin{align}
\norm{e^{-\beta H/2} O_X(H,t) e^{\beta H/2} } 
&\le 
 \norm{e^{-\beta H/2} O_X e^{\beta H/2}}  \le e^{e^{\Theta(\beta)}}  .
\end{align}
Moreover, using the inequality~\eqref{norm_e_-beta_H_O_X_e^beta_H_main}, we have  
\begin{align}
\label{lastt_ineq_Prop50}
&\norm{e^{-\beta H/2}  \tilde{\tr}_{L^\co} \brr{O_X(H,t)} e^{\beta H/2} } \notag \\
&\le \norm{e^{-\beta H/2} O_X(H,t) e^{\beta H/2} } + \norm{e^{-\beta H/2} O_X(H,t) e^{\beta H/2}- e^{-\beta H/2}  \tilde{\tr}_{L^\co} \brr{O_X(H,t)} e^{\beta H/2} }  \notag \\
&\le e^{e^{\Theta(\beta)}} + 1 \le  e^{e^{\Theta(\beta)}}  ,
\end{align}
where we use the assumption that the second term in the second line is smaller than $1$. 
We thus prove the second main inequalities in~\eqref{norm_e_-beta_H_O_X_e^beta_H_main__23}.
This completes the proof of Proposition~\ref{prop:O_X_imaginary_time}. $\square$

%

\appendix

\widetext{

\section{Another proof for Lemma~\ref{connection_of_exponential_operator}} 
\label{Another proof for Lemma_connection_of_exponential_operator} 

We show another proof for Lemma~\ref{connection_of_exponential_operator}, which plays a key role in our analyses and has been proved based on the quantum belief propagation technique in Sec.~\ref{Sec:Quantum Belief propagation}. 
Here, we utilize the derivation of the operator logarithm in Ref.~\cite{adlertaylor,haber2018notes}, which gives 
 \begin{align}
 \label{dx_d_log_sigma_x_delta_rho}
\frac{d}{dx}\log \br{\rho+ x \delta \rho} = \int_0^\infty \frac{1}{\rho_x + z \hat{1}} \delta \rho \frac{1}{\rho_x + z \hat{1}} dz , 
\end{align}
where we let $\rho_x := \rho+ x \delta \rho$. 
The purpose here is that we cannot refine the effective Hamiltonian theory based on the above perturbation method. 
For the convenience of readers, we show the statement of Lemma~\ref{connection_of_exponential_operator} again:
\\{~}\\
{\bf Restatement of Lemma~\ref{connection_of_exponential_operator}.}
\textit{ 
For arbitrary operators in the form of 
\begin{align}
 e^{\epsilon  \mB} e^{\beta \mA} e^{\epsilon  \mB} ,
\end{align}
we obtain the logarithm as 
\begin{align}
\label{main_eq:connection_of_exponential_operator_rre}
&\log( e^{\epsilon \mB} e^{\beta \mA} e^{\epsilon \mB} ) =\beta  e^{-2i \epsilon \mathcal{C}} \mA e^{2i \epsilon \mathcal{C}} + 2\epsilon  \mB + \orderof{\epsilon^2}, \\
&\mathcal{C} := \frac{1}{\beta}\int_{-\infty}^\infty g_\beta(t) e^{i\mA t} \mB e^{-i\mA t}dt,
\end{align}
where $g_\beta(t)$ has been defined in the context of the belief propagation~\eqref{def_g_t_exp_decay}.
}
{~}\\

For the proof, we let
\begin{align}
\rho= e^{\beta \mA} ,\quad \delta \rho= \brrr{ e^{\beta \mA} , \mB} ,\quad  x=\epsilon, 
\end{align}
which reduces Eq.~\eqref{dx_d_log_sigma_x_delta_rho} to
 \begin{align}
\log \br{e^{\beta \mA} + \epsilon \brrr{ e^{\beta \mA} , \mB}} =\log \br{e^{\beta \mA}}+ \epsilon\int_0^\infty \frac{1}{e^{\beta \mA} + z \hat{1}} \brrr{ e^{\beta \mA} , \mB} \frac{1}{e^{\beta \mA} + z \hat{1}} dz
+\orderof{\epsilon^2} ,  \label{proof_other1_connection_of_exponential_operator}
\end{align}
where $\brrr{ e^{\beta \mA} , \mB}=e^{\beta \mA} \mB +  \mB e^{\beta \mA}$. 
Using the spectral decomposition of $\mB$ by the eigenspace of $\mA$, we obtain 
\begin{align}
\mB = \sum_{l,m} \bra{m} \mB\ket{l} \ket{l} \bra{m},
\end{align}
where $\mA\ket{m}=a_m \ket{m}$. Then, we have 
 \begin{align}
\int_0^\infty \frac{1}{e^{\beta \mA} + z \hat{1}} \brrr{ e^{\beta \mA} , \mB} \frac{1}{e^{\beta \mA} + z \hat{1}} dz
&= \sum_{l,m} \int_0^\infty \frac{1}{e^{\beta a_l} + z } \br{e^{\beta a_l} + e^{\beta a_{m}}}\bra{l} \mB\ket{m} \ket{l} \bra{m} \frac{1}{e^{\beta a_m} + z}  dz  \notag \\
&=  \sum_{l,m} \frac{\beta (a_l-a_m)}{e^{\beta a_l} -e^{\beta a_m} } \br{e^{\beta a_l} + e^{\beta a_{m}}}\bra{l} \mB\ket{m} \ket{l} \bra{m}  \notag \\
&=   \sum_{l,m} \frac{\beta (a_l-a_m)}{\tanh\brr{\beta (a_l-a_m)/2}} \bra{l} \mB\ket{m} \ket{l} \bra{m}   \notag \\
&=\frac{\beta \ad_{\mA}}{\tanh\br{\beta \ad_{\mA}/2}}  \sum_{l,m} \bra{l} \mB\ket{m} \ket{l} \bra{m} = \frac{\beta \ad_{\mA}}{\tanh\br{\beta \ad_{\mA}/2}}  \mB,
\label{proof_other2_connection_of_exponential_operator}
\end{align}
where we use $\ad_{\mA}(\ket{l} \bra{m})= a_l-a_m$. 
Using $\mB_\omega = \sum_{i,j} \bra{a_i} \mB \ket{a_j} \delta(a_i-a_j-\omega) \ket{a_i}\bra{a_j}$ and Eq.~\eqref{B_omega_B_eq_} as 
\begin{align}
\mB_\omega=\frac{1}{2\pi} \int_{-\infty}^\infty \mB(\mA,t) e^{-i\omega t} dt,\quad 
\mB= \int_{-\infty}^\infty  \mB_\omega d\omega, 
\end{align}
we obtain 
\begin{align}
\frac{\beta \ad_{\mA}}{\tanh\brr{\beta \ad_{\mA}/2}}  \mB=  2\int_{-\infty}^\infty \frac{\beta \omega/2}{\tanh\br{\beta \omega /2}}   \mB_\omega d\omega 
= 2\int_{-\infty}^\infty  dt \mB(\mA,t) \frac{1}{2\pi} \int_{-\infty}^\infty \frac{\beta \omega/2}{\tanh\br{\beta \omega /2}}e^{-i\omega t}   d\omega  .
\label{proof_other3_connection_of_exponential_operator}
\end{align}
To perform the Fourier transform, we use the decomposition of 
\begin{align}
\frac{\beta \omega/2}{\tanh\br{\beta \omega /2}}
= 1+ \br{\frac{\beta \omega/2}{\tanh\br{\beta \omega /2}}-1}
= 1+ \frac{1}{\omega}\br{\frac{\beta \omega/2}{\tanh\br{\beta \omega /2}}-1} \omega .
\end{align}
Therefore, from $ \omega \mB_\omega = \ad_{\mA} (\mB_\omega)$, we obtain 
\begin{align}
\int_{-\infty}^\infty \frac{\beta \omega/2}{\tanh\br{\beta \omega /2}}   \mB_\omega d\omega 
&= \mB+ \ad_{\mA}  \int_{-\infty}^\infty  \frac{1}{\omega}\br{\frac{\beta \omega/2}{\tanh\br{\beta \omega /2}}-1}\mB_\omega d\omega   \notag \\
&=  \mB+ \ad_{\mA}  \int_{-\infty}^\infty dt \mB(\mA,t)  \frac{1}{2\pi} \int_{-\infty}^\infty  \frac{1}{\omega}\br{\frac{\beta \omega/2}{\tanh\br{\beta \omega /2}}-1}e^{-i\omega t}   d\omega \notag \\
&=      \mB+ i \ad_{\mA}  \int_{-\infty}^\infty g_\beta(t)  \mB(\mA,t) dt ,
=   \mB+ i\beta  \ad_{\mA} (\mathcal{C}),
\label{proof_other4_connection_of_exponential_operator}
\end{align}
where we use Eq.~\eqref{definition_of_g_beta_t} for $ g_\beta(t)$ and the definition of $\mathcal{C}$ in Eq.~\eqref{main_eq:connection_of_exponential_operator_rre}.

By combining Eqs~\eqref{proof_other1_connection_of_exponential_operator}, \eqref{proof_other2_connection_of_exponential_operator}, \eqref{proof_other3_connection_of_exponential_operator}, and \eqref{proof_other4_connection_of_exponential_operator}, we obtain the main inequality~\eqref{main_eq:connection_of_exponential_operator_rre} as follows:
 \begin{align}
\log \br{e^{\beta \mA} + \epsilon \brrr{ e^{\beta \mA} , \mB}} =\beta \mA+ 2\epsilon \mB + 2 i\beta \epsilon \ad_{\mA} (\mathcal{C}) + \orderof{\epsilon^2}
= \beta e^{-2 i\epsilon \mathcal{C}} \mA e^{2 i\epsilon \mathcal{C}} +2\epsilon \mB  + \orderof{\epsilon^2}. 
\end{align}
This completes the proof of Lemma~\ref{connection_of_exponential_operator}. $\square$

\section{Exponential $\tau\norm{V}$ dependence}

From our analyses, we observe that the non-locality grows with $e^{\tau \norm{V}}$ as has been shown in Subtheorem~\ref{sub_thm:U_tau_u_i_commun}.
This point is a critical obstacle to improving the current theorem from the pairwise Markov property to the global Markov property, where we choose the size of the subset as large as the system size. 
In this case, the dependence $e^{\tau \norm{V}}$ yields the exponential factor as $e^{\orderof{|AC|}}$ in Theorem~\ref{CMI_decay_PTP_general_D}.

For this purpose, one might consider that proof techniques in Subtheorem~\ref{sub_thm:U_tau_u_i_commun} might be further refined.
However, we can provide an explicit example of an exponentially enhanced effective interaction term due to a connection of quasi-local exponential operators.
In detail, one can prove the following statement: 
\begin{claim}
Let $H$ be a one-dimensional Hamiltonian that has short-range interactions as in Eq.~\eqref{def_short_range_long_range}. 
Then, under an appropriate choice of the local operator $V$, the effective interaction for 
$
\log\br{e^{V} e^{H} e^V}
$
can be exponentially amplified as $e^{\orderof{\norm{V}}}$. Here, we can include the $\tau$ by simply replacing $V \to  \tau  V'$.
\end{claim}
To show an example of the statement, let $H_0$ be a Hamiltonian that consists of only one interaction term as   
\begin{align}
\label{H_0_counter_exp_0}
H_0= - 2v_0 \sigma_{1,z} +4 v_0  \varepsilon\sigma_{1,x} \otimes \sigma_{\ell,x}  , \quad   \varepsilon= e^{-\orderof{\ell}} ,
\end{align}
where $\{\sigma_{i,x},\sigma_{i,y},\sigma_{i,z}\}_{i\in \Lambda}$ is the Pauli matrix on the site $i$.
We note that the Hamiltonian $H_0$ satisfies the condition of short-range interaction~\eqref{def_short_range_long_range} as long as $v_0 = \Poly(\ell)$. 
We then choose $V$ as 
\begin{align}
V=\norm{V} \sigma_{1,z} ,
\end{align}
and define $H_{\rm eff}$ as 
\begin{align}
H_{\rm eff}=  \log (e^{V} e^{H_0} e^{V})  .
\end{align}
Then, the Hamiltonian $H$ is essentially a two-qubit Hamiltonian; hence, we can analytically calculate its form.
The analytic form is quite complicated, but we can approximate it as 
\begin{align}
H_{\rm eff} \approx  \br{-2v_0+ 2\norm{V} }\sigma_{1,z}  + \varepsilon e^{2\norm{V}}\sigma_{1,x} \otimes \sigma_{\ell,x} \for \norm{V} \lesssim v_0 .
\end{align}
We thus obtain the exponential enhancement of the long-range interaction between the sites $1$ and $\ell$.

The mechanism behind the above analysis is as follows: To make the point more transparent, let us choose $\norm{V}=v_0$, where $v_0$ appears in $H_0$.
First of all, we find that the Hamiltonian $H_0$ is approximately given by
\begin{align}
H_0 \approx \log\br{e^{ \varepsilon\sigma_{1,x} \otimes \sigma_{\ell,x} } e^{-2v_0 \sigma_{1,z} } e^{ \varepsilon\sigma_{1,x} \otimes \sigma_{\ell,x} } },
\end{align}
and hence the Hamiltonian $H_{\rm eff}$ is given by
\begin{align}
H_{\rm eff}&\approx 
\log\br{e^{v_0\sigma_{1,z}}  e^{ \varepsilon\sigma_{1,x} \otimes \sigma_{\ell,x} } e^{-2v_0 \sigma_{1,z} } e^{ \varepsilon\sigma_{1,x} \otimes \sigma_{\ell,x} } e^{v_0\sigma_{1,z}}  } \notag \\
&=\log\brr{ \exp\br{ \varepsilon  e^{v_0\sigma_{1,z}} \sigma_{1,x} e^{-v_0\sigma_{1,z}} \otimes \sigma_{\ell,x} }
 \exp\br{ \varepsilon e^{-v_0 \sigma_{1,z} } \sigma_{1,x}  e^{v_0 \sigma_{1,z} } \otimes \sigma_{\ell,x} }  } ,
 \label{imaginary_tm_evo_counter_ex}
\end{align}
where we use $e^{v_0\sigma_{1,z}}  e^{ \varepsilon\sigma_{1,x} \otimes \sigma_{\ell,x} } e^{-v_0 \sigma_{1,z} } =
\exp\brr{\varepsilon e^{v_0\sigma_{1,z}} \sigma_{1,x}e^{-v_0 \sigma_{1,z} } \otimes \sigma_{\ell,x} }$.
Now, we have 
\begin{align}
&\norm{e^{\pm v_0\sigma_{1,z}} \sigma_{1,x}e^{\mp v_0 \sigma_{1,z} } } =\norm{\cosh(2v_0) \sigma_{1,x}  \pm i \sinh(2v_0) \sigma_{1,y}} ,
\end{align}
which grows exponentially with $v_0$. 
In this way, if $e^{V} e^{H_0} e^{V}$ includes imaginary time evolutions for $V$, it induces the exponential growth of the amplitude of the effective interactions.  

The model~\eqref{H_0_counter_exp_0} itself is far from the practical many-body systems.
However, this mechanism may be feasible in more natural quantum many-body systems. 
Therefore, to avoid exponential growth, we have to impose additional constraints on the form of the operator $V$\footnote{Indeed, if we choose $V$ as the PTP operator, the imaginary time evolution as in Eq.~\eqref{imaginary_tm_evo_counter_ex} may not appear since the original Hamiltonian $H_0$ does not act on the ancilla subspace and cannot be expressed in a form including $V$. }.

\section{Continuity bound on the logarithm of operators based on relative error}

\subsection{Setup and Main result}
For the readers' convenience, we show the setup again so that the section can be closed independently.

Without loss of generality, we can let two operators $\rho$ and $\sigma$ be density matrices with $\tr(\rho)=1$ and $\tr(\sigma)=1$.
In general, we get the difference between the logarithms of $c_1 \rho$ and $c_2 \sigma$ as 
\begin{align}
\log(c_1 \rho) - \log(c_2 \sigma) = \log(c_1) - \log(c_2) +  \log(\rho) - \log(\sigma) .
\end{align}
We restate the definition of the relative error as follows:

{~}

{\bf Definition~\ref{def:relative error}} [Relative error]. \textit{
We define the relative error $\delta_{\rm R}(\rho,\sigma)$ between $\rho$ and $\sigma$ as 
\begin{align}
\delta_{\rm R}(\rho,\sigma):= \sup_{\ket{\psi}} \frac{|\bra{\psi} \rho-\sigma \ket{\psi}|}{\bra{\psi} \rho \ket{\psi}} ,
\end{align}
where the $\sup_{\ket{\psi}}$ is taken for all the set of quantum states $\ket{\psi}$.}

Our purpose now is to derive a continuity inequality for $\norm{\log(\rho) - \log(\sigma)}$ based on the relative error of $\delta_{\rm R}(\rho,\sigma)$, which has been given as follows:

{~}

{\bf Theorem~\ref{thm:refined continuity}.} \textit{
Under the condition that 
\begin{align}
 \varepsilon=\max\brr{ \delta_{\rm R}(\rho,\sigma), \delta_{\rm R}(\sigma,\rho) }  \le \frac{1}{2},
\label{cond:thm:refined continuity}
\end{align}
we obtain the upper bound as 
 \begin{align}
\norm{\log(\sigma)- \log(\rho)}\le  \varepsilon \brr{ \frac{4\log [2\lambda^{-1}_{\min}(\rho)] }{\pi} \log\br{\frac{e\log [2\lambda^{-1}_{\min}(\rho)]}{2\pi }} +23 } , 
\label{main_ineq:thm:refined continuity}
\end{align}
where $\lambda^{-1}_{\min}(\rho)$ is the minimum eigenvalue of $\rho$.}


\subsection{Proof of Theorem~\ref{thm:refined continuity}}\label{Theorem_proof:thm:refined continuity}

Throughout the proof, we denote $\sigma-\rho$ by $\delta \rho$, i.e., $\delta \rho:=\sigma-\rho$.
We first note that the condition~\eqref{cond:thm:refined continuity} immediately gives
 \begin{align}
 \label{relative_error_cond}
|\bra{\psi} \delta \rho \ket{\psi}| \le \varepsilon \bra{\psi} \rho \ket{\psi},   \quad |\bra{\psi} \delta \rho \ket{\psi}| \le \varepsilon \bra{\psi} \sigma \ket{\psi}
\end{align}
from Definition~\ref{def:relative error}. 
For the proof, we start with the equation of  
 \begin{align}
 \label{log_dif_ryo:sigma}
\log(\sigma)- \log(\rho)= \int_0^1  \frac{d}{dx}\log \br{\rho+ x \delta \rho}   dx .
\end{align}
Using Ref.~\cite{adlertaylor}\footnote{In the reference, there is a typo in the first equation (1) therein. 
The second term in the RHS of the equation should be $t\int_0^\infty \frac{1}{A+zI} B \frac{1}{A+zI} dz $ 
instead of $t\int_0^\infty \frac{1}{B+zI} A \frac{1}{B+zI} dz $.}, we have 
 \begin{align}
 \label{dx_d_log_sigma_x_delta_rho_2}
\frac{d}{dx}\log \br{\rho+ x \delta \rho} = \int_0^\infty \frac{1}{\rho_x + z \hat{1}} \delta \rho \frac{1}{\rho_x + z \hat{1}} dz , 
\quad \rho_x := \rho+ x \delta \rho. 
\end{align}
We note that the same inequality has been used in Eq.~\eqref{dx_d_log_sigma_x_delta_rho}.

We now define 
 \begin{align}
\rho_x = e^{H_x}  = \sum_{m} e^{E_m} \ket{E_m} \bra{E_m},
\end{align}
and obtain the following bound:
\begin{align}
\label{delta_rho_relative_error_rho_x}
\frac{|\bra{\psi} \delta \rho \ket{\psi}|}{\bra{\psi}\rho_x \ket{\psi}}\le \varepsilon  \longrightarrow
|\bra{\psi} \delta \rho \ket{\psi}| \le \varepsilon  \bra{\psi}\rho_x \ket{\psi}
\end{align}
for an arbitrary quantum state $\ket{\psi}$. 
It is immediately obtained from the condition~\eqref{relative_error_cond} by using 
 \begin{align}
|\bra{\psi} \delta \rho \ket{\psi}| \le x \varepsilon \bra{\psi} \rho \ket{\psi}+(1-x) \varepsilon  \bra{\psi} \sigma \ket{\psi}  = \varepsilon  \bra{\psi} \rho_x \ket{\psi} . 
\end{align}

We then calculate
 \begin{align}
\int_0^\infty \frac{1}{\rho_x + z \hat{1}} \delta \rho \frac{1}{\rho_x + z \hat{1}} dz 
&=\sum_{m,n}  \int_0^\infty \frac{1}{e^{E_m}  + z } \bra{E_m} \delta \rho \ket{E_n} \frac{1}{e^{E_n} + z } \ket{E_m}\bra{E_n} dz \notag \\
&= \sum_{m,n}  \frac{E_m-E_n}{e^{E_m}-e^{E_n}}  \bra{E_m} \delta \rho \ket{E_n}  \ket{E_m}\bra{E_n}.
\end{align}
Note that for arbitrary quantum state $\ket{\psi}=\sum_m a_m \ket{E_m}$, we have from the inequality~\eqref{delta_rho_relative_error_rho_x}
 \begin{align}
\abs{\Braket{\psi | \int_0^\infty \frac{1}{\rho_x + z \hat{1}} \delta \rho \frac{1}{\rho_x + z \hat{1}} dz | \psi}}
&\le \sum_{m,n}  \frac{E_m-E_n}{e^{E_m}-e^{E_n}}  \abs{ \bra{E_m} \delta \rho \ket{E_n}} \cdot \abs{a_m} \cdot \abs{a_n} \notag \\
&\le  \sum_{m,n}  \frac{E_m-E_n}{e^{E_m}-e^{E_n}} \frac{\bra{E_m} \delta \rho \ket{E_m} + \bra{E_n} \delta \rho \ket{E_n} }{2} \cdot \abs{a_m} \cdot \abs{a_n} \notag \\
&\le  \varepsilon  \sum_{m,n}\frac{E_m-E_n}{e^{E_m}-e^{E_n}} \frac{\bra{E_m} \rho_x  \ket{E_m} + \bra{E_n} \rho_x \ket{E_n} }{2} \cdot \abs{a_m} \cdot \abs{a_n} \notag \\
&= \varepsilon  \sum_{m,n}  \frac{E_m-E_n}{e^{E_m}-e^{E_n}} \frac{e^{E_m}+ e^{E_n} }{2} \cdot \abs{a_m} \cdot \abs{a_n},
\end{align}
where we use $\frac{E_m-E_n}{e^{E_m}-e^{E_n}}\ge 0$ for arbitrary $E_m,E_n\in \mathbb{Z}$ and the Cauthy--Schwarz inequality as 
  \begin{align}
 \abs{ \bra{E_m} \delta \rho \ket{E_n}}  \le \sqrt{\bra{E_m} \delta \rho \ket{E_m} \bra{E_n} \delta \rho \ket{E_n}} \le \frac{\bra{E_m} \delta \rho \ket{E_m} +\bra{E_n} \delta \rho \ket{E_n}}{2}.
\end{align}
Therefore, by defining the operator $\bar{\mathcal{R}}$ as 
 \begin{align}
 \label{bar_mathcal_R_deff}
\bar{\mathcal{R}}:=  \sum_{m,n} \frac{E_m-E_n}{e^{E_m}-e^{E_n}} \frac{e^{E_m}+ e^{E_n} }{2}\ket{E_m}\bra{E_n},
\end{align}
we have 
 \begin{align}
 \label{Main_ineq_var_epsilon_rho}
\norm{\int_0^\infty \frac{1}{\rho_x + z \hat{1}} \delta \rho \frac{1}{\rho_x + z \hat{1}} dz }\le   \varepsilon  \norm{\bar{\mathcal{R}}} = \varepsilon 
 \sup_{P} \abs{ \tr \br{ P \bar{\mathcal{R}}} } , 
\end{align}
where $P$ is an arbitrary projector onto a quantum state. 
For the trace of $\tr \br{ P \bar{\mathcal{R}}}$, we aim to prove the following lemma:
\begin{lemma}\label{tr_bar_mathcal_R_deff_P}
Under the definition of Eq.~\eqref{bar_mathcal_R_deff} for $\bar{\mathcal{R}}$, we obtain the upper bound of 
 \begin{align}
\abs{ \tr\br{ \bar{\mathcal{R}}P  } }
\le \frac{4\norm{H_x}}{\pi} \log\br{\frac{e\norm{H_x}}{2\pi }} +23,
\label{tr_bar_mathcal_R_deff_P/main_ineq}
\end{align}
for an arbitrary projection $P$. 
\end{lemma}

By applying Lemma~\ref{tr_bar_mathcal_R_deff_P} to the inequality~\eqref{Main_ineq_var_epsilon_rho}, we have 
 \begin{align}
 \label{Main_ineq_var_epsilon_rho_fin}
\norm{\int_0^\infty \frac{1}{\rho_x + z \hat{1}} \delta \rho \frac{1}{\rho_x + z \hat{1}} dz }\le   \varepsilon  \norm{\bar{\mathcal{R}}} = \varepsilon \brr{ \frac{4\norm{H_x}}{\pi} \log\br{\frac{e\norm{H_x}}{2\pi }} +23 } .
\end{align}
Here we have $\norm{H_x}=\log [\lambda^{-1}_{\min}(\rho_x)]$ since all the eigenvalues of $\rho_x$ are smaller than or equal to $1$.
We can also obtain the lower bound of 
 \begin{align}
\inf_{\ket{\psi}} \bra{\psi} \rho_x \ket{\psi} 
&= \inf_{\ket{\psi}}\br{ \bra{\psi} \rho \ket{\psi} + x \bra{\psi} \sigma- \rho \ket{\psi}} \notag \\
&\ge  \inf_{\ket{\psi}}\br{ \bra{\psi} \rho \ket{\psi} - x |\bra{\psi} \sigma- \rho \ket{\psi}|} \notag \\
&\ge \inf_{\ket{\psi}}\br{ \bra{\psi} \rho \ket{\psi} - x \varepsilon \bra{\psi} \rho \ket{\psi}}  \ge  \frac{1}{2} \lambda_{\min}(\rho),
\end{align}
where we use $1-x\varepsilon \ge 1-\varepsilon \ge 1/2$ from $\varepsilon \le 1/2$. 
We thus obtain $\norm{H_x}\le \log [2\lambda^{-1}_{\min}(\rho)]$ for $0\le x\le 1$.

By combining Eqs.~\eqref{log_dif_ryo:sigma} and \eqref{dx_d_log_sigma_x_delta_rho_2} with the inequality~\eqref{Main_ineq_var_epsilon_rho_fin} and $\norm{H_x}\le \log [2\lambda^{-1}_{\min}(\rho)]$, we prove the main inequality~\eqref{main_ineq:thm:refined continuity}. 
This completes the proof of Theorem~\ref{thm:refined continuity}. $\square$

\subsubsection{Proof of Lemma~\ref{tr_bar_mathcal_R_deff_P}}
We first rewrite $\bar{\mathcal{R}}$ as 
 \begin{align}
 \label{bar_mathcal_R_P_spectral}
 \bar{\mathcal{R}} 
 &= \sum_{m,n}\frac{E_m-E_n}{e^{E_m}-e^{E_n}}  \frac{e^{E_m}+ e^{E_n} }{2}  \ket{E_m}\bra{E_n}
= \frac{\ad_{H_x}\br{e^{\ad_{H_x}}+ 1}}{2\br{e^{\ad_{H_x}}-1}}, 
\end{align}
where we use $ f(E_m-E_n) \ket{E_m}\bra{E_n}= f(\ad_{H_x}) \ket{E_m}\bra{E_n}$ because of $\ad_{H_x} \ket{E_m}\bra{E_n}=(E_m-E_n) \ket{E_m}\bra{E_n}$.
We thus obtain 
 \begin{align}
\bar{\mathcal{R}}  P =\frac{\ad_{H_x}\br{e^{\ad_{H_x}}+ 1}}{2\br{e^{\ad_{H_x}}-1}} P = 
\frac{\ad_{H_x}/2}{\tanh\br{\ad_{H_x}/2}} P.
\end{align}
We second adopt the spectral decomposition as in Ref.~\cite[Appendix~A therein]{PhysRevX.12.021022}
\begin{align}
\label{P_omega_B_eq_}
&P_\omega = \sum_{m,n} \bra{E_m} P \ket{E_n} \delta(E_m-E_n-\omega) \ket{E_m}\bra{E_n}  \notag \\
&\longrightarrow
P_\omega=\frac{1}{2\pi} \int_{-\infty}^\infty P(H_x,t) e^{-i\omega t} dt,\quad 
P= \int_{-\infty}^\infty  P_\omega d\omega.
\end{align}
By applying Eq.~\eqref{P_omega_B_eq_} to Eq.~\eqref{bar_mathcal_R_P_spectral} with $\ad_{H_x} P_\omega= \omega P_\omega$, we obtain 
 \begin{align}
 \label{bar_mathcal_R_P_spectral_integral}
 \bar{\mathcal{R}}P  
&=  \int_{-\infty}^\infty   \frac{\omega/2}{\tanh(\omega/2)}  P_\omega d\omega =   \int_{-\infty}^\infty \brr{\frac{1}{\omega}\br{  \frac{\omega/2}{\tanh(\omega/2)}-1} \omega P_\omega + P_\omega} d\omega \notag \\
&=P+ \frac{1}{2\pi} \int_{-\infty}^\infty \ad_{H_x}[ P(H_x,t)]  \int_{-\infty}^\infty  \frac{1}{\omega}\br{  \frac{\omega/2}{\tanh(\omega/2)}-1} e^{-i\omega t} d\omega dt          \notag \\
&= P+ i \int_{-\infty}^\infty g(t) \ad_{H_x}[ P(H_x,t)] dt  ,
\end{align}
where $g(t)$ is defined by the Fourier transform of
 \begin{align}
 \label{g_t_definition/}
g(t) =  \frac{1}{2\pi i} \int_{-\infty}^\infty  \frac{1}{\omega}\br{  \frac{\omega/2}{\tanh(\omega/2)}-1} e^{-i\omega t} d\omega
={\rm sign}(t) \frac{e^{-2\pi|t|}}{1-e^{-2\pi|t|}}. 
\end{align}

The remaining task is to estimate the upper bound of 
 \begin{align}
 \label{integrals_upp_bound_0}
\abs{ \tr\br{ \bar{\mathcal{R}}P  } }
\le  1+  \norm{ \int_{-\infty}^\infty g(t) \ad_{H_x}[ P(H_x,t)] dt }_1 ,
\end{align}
where we use $\tr(P)=1$ from the condition.  We first note that the following simple upper bound as 
 \begin{align}
 \norm{ \int_{-\infty}^\infty g(t) \ad_{H_x}[ P(H_x,t)] dt }_1 \le 
\int_{-\infty}^\infty |g(t)| \cdot \norm{\ad_{H_x}(P)}_1dt  \le 
2 \norm{H_x} \int_{-\infty}^\infty |g(t)| dt 
\end{align}
can NOT be used since the integral of $|g(t)|$ is divergent because of $|g(t)| \propto 1/|t| $ for $|t|\ll1$. 
Note that the H\"older inequality gives $\norm{\ad_{H_x}(P)}_1\le 2\norm{H_x} \cdot \norm{P}_1= 2\norm{H_x} $.

To avoid the divergence, we adopt the prescription in Ref.~\cite[(D. 83) and (D.84)]{PhysRevX.12.021022}.
We first note that the following decomposition holds for an arbitrary operator $O$ in general: 
\begin{align}
O(H_x,t) = O + \int_0^1 \frac{d}{d\lambda}O(H_x,\lambda t) d\lambda = 
 O +it \int_0^1\ad_{H_x} [O(H_x,\lambda t)] d\lambda  .
\end{align}
By applying it to $\ad_{H_x}[ P(H_x,t)] = [\ad_{H_x}(P)](H_x,t)$, we have 
\begin{align}
\ad_{H_x}[ P(H_x,t)]=\ad_{H_x}(P) +it \int_0^1\ad^2_{H_x} [P(H_x,\lambda t)] d\lambda  .
\end{align}
Here, we utilize the above decomposition for the integrals~\eqref{integrals_upp_bound_0} in $|t|\le \delta t$, which yields 
 \begin{align}
\int_{-\infty}^\infty g(t) \ad_{H_x}[ P(H_x,t)] dt 
=  \int_{|t| > \delta t} g(t) \ad_{H_x}[ P(H_x,t)] dt + \int_{|t| \le \delta t} it  g(t)\int_0^1\ad^2_{H_x} [P(H_x,\lambda t)] d\lambda dt ,
\end{align}
where we use $\int_{|t| \le \delta t}g(t)=0$ since $g(t)$ in Eq.~\eqref{g_t_definition/} is an odd function. 
We thus prove 
 \begin{align}
 \label{norm_g_t_int_H_x_P_x}
 \norm{ \int_{-\infty}^\infty g(t) \ad_{H_x}[ P(H_x,t)] dt }_1
\le 2 \norm{H_x} \int_{|t| > \delta t} | g(t)| dt +4 \norm{H_x}^2 \int_{|t| \le \delta t} |t  g(t)| dt ,
\end{align}
where we use $\norm{\ad_{H_x}(P)}_1\le  2\norm{H_x}$ and $\norm{\ad^2_{H_x}(P)}_1\le 4\norm{H_x}$ from $\norm{P}_1=\tr(P)=1$. 

Using Lemma~\ref{belief_norm_lemma} with $\beta=1$, we have 
 \begin{align}
 \int_{|t| > \delta t} | g(t)| dt \le \frac{2}{\pi} \log\br{\frac{1}{2\pi \delta t}} ,\quad 
  \int_{|t| \le \delta t} |t g(t)| dt \le \frac{\delta t}{\pi},
\end{align}
where we assume $\delta t \le 1/(4\pi)$. 
By applying the above bound to the inequality~\eqref{norm_g_t_int_H_x_P_x}, we derive 
 \begin{align}
 \label{norm_g_t_int_H_x_P_x_fin}
 \norm{ \int_{-\infty}^\infty g(t) \ad_{H_x}[ P(H_x,t)] dt }_1
\le \frac{4\norm{H_x}}{\pi} \log\br{\frac{1}{2\pi \delta t}} +\frac{4\delta t}{\pi} \norm{H_x}^2  .
\end{align}
Finally, by choosing $\delta t$ as $\min\brr{1/(4\pi), 1/\norm{H_x}}$, we obtain 
 \begin{align}
 \norm{ \int_{-\infty}^\infty g(t) \ad_{H_x}[ P(H_x,t)] dt }_1
&\le \begin{cases}
8\log(2)+16 &\for \norm{H_x} \le 4\pi \\
 \frac{4\norm{H_x}}{\pi} \log\br{\frac{e\norm{H_x}}{2\pi }}   &\for \norm{H_x}> 4\pi 
\end{cases} \notag \\
&\le \max\brr{ 22 ,   \frac{4\norm{H_x}}{\pi} \log\br{\frac{e\norm{H_x}}{2\pi }} }.
\end{align}
Therefore, from the inequality~\eqref{integrals_upp_bound_0}, we prove the inequality of
 \begin{align}
\abs{ \tr\br{ \bar{\mathcal{R}}P  } }
\le 1+ \max\brr{ 22 , \frac{4\norm{H_x}}{\pi} \log\br{\frac{e\norm{H_x}}{2\pi }} },
\end{align}
which reduces to the main inequality~\eqref{tr_bar_mathcal_R_deff_P/main_ineq}.
This completes the proof of Lemma~\ref{tr_bar_mathcal_R_deff_P}. $\square$

\bibliography{Quantum_Markov}

%% file: CMI_decay_main_arxiv.bbl
\providecommand{\noopsort}[1]{}\providecommand{\singleletter}[1]{#1}%
\begin{thebibliography}{183}%
\makeatletter
\providecommand \@ifxundefined [1]{%
 \@ifx{#1\undefined}
}%
\providecommand \@ifnum [1]{%
 \ifnum #1\expandafter \@firstoftwo
 \else \expandafter \@secondoftwo
 \fi
}%
\providecommand \@ifx [1]{%
 \ifx #1\expandafter \@firstoftwo
 \else \expandafter \@secondoftwo
 \fi
}%
\providecommand \natexlab [1]{#1}%
\providecommand \enquote  [1]{``#1''}%
\providecommand \bibnamefont  [1]{#1}%
\providecommand \bibfnamefont [1]{#1}%
\providecommand \citenamefont [1]{#1}%
\providecommand \href@noop [0]{\@secondoftwo}%
\providecommand \href [0]{\begingroup \@sanitize@url \@href}%
\providecommand \@href[1]{\@@startlink{#1}\@@href}%
\providecommand \@@href[1]{\endgroup#1\@@endlink}%
\providecommand \@sanitize@url [0]{\catcode `\\12\catcode `\$12\catcode
  `\&12\catcode `\#12\catcode `\^12\catcode `\_12\catcode `\%12\relax}%
\providecommand \@@startlink[1]{}%
\providecommand \@@endlink[0]{}%
\providecommand \url  [0]{\begingroup\@sanitize@url \@url }%
\providecommand \@url [1]{\endgroup\@href {#1}{\urlprefix }}%
\providecommand \urlprefix  [0]{URL }%
\providecommand \Eprint [0]{\href }%
\providecommand \doibase [0]{https://doi.org/}%
\providecommand \selectlanguage [0]{\@gobble}%
\providecommand \bibinfo  [0]{\@secondoftwo}%
\providecommand \bibfield  [0]{\@secondoftwo}%
\providecommand \translation [1]{[#1]}%
\providecommand \BibitemOpen [0]{}%
\providecommand \bibitemStop [0]{}%
\providecommand \bibitemNoStop [0]{.\EOS\space}%
\providecommand \EOS [0]{\spacefactor3000\relax}%
\providecommand \BibitemShut  [1]{\csname bibitem#1\endcsname}%
\let\auto@bib@innerbib\@empty
\bibitem [{\citenamefont {Araki}(1969)}]{Araki1969}%
  \BibitemOpen
  \bibfield  {author} {\bibinfo {author} {\bibfnamefont {H.}~\bibnamefont
  {Araki}},\ }\bibfield  {title} {\bibinfo {title} {{\it Gibbs states of a one
  dimensional quantum lattice}},\ }\href {https://doi.org/10.1007/BF01645134}
  {\bibfield  {journal} {\bibinfo  {journal} {Communications in Mathematical
  Physics}\ }\textbf {\bibinfo {volume} {14}},\ \bibinfo {pages} {120}
  (\bibinfo {year} {1969})}\BibitemShut {NoStop}%
\bibitem [{\citenamefont {Park}\ and\ \citenamefont {Yoo}(1995)}]{Park1995}%
  \BibitemOpen
  \bibfield  {author} {\bibinfo {author} {\bibfnamefont {Y.~M.}\ \bibnamefont
  {Park}}\ and\ \bibinfo {author} {\bibfnamefont {H.~J.}\ \bibnamefont {Yoo}},\
  }\bibfield  {title} {\bibinfo {title} {{\it Uniqueness and clustering
  properties of Gibbs states for classical and quantum unbounded spin
  systems}},\ }\href {https://doi.org/10.1007/BF02178359} {\bibfield  {journal}
  {\bibinfo  {journal} {Journal of Statistical Physics}\ }\textbf {\bibinfo
  {volume} {80}},\ \bibinfo {pages} {223} (\bibinfo {year} {1995})}\BibitemShut
  {NoStop}%
\bibitem [{\citenamefont {Ueltschi}(2004)}]{ueltschi2004cluster}%
  \BibitemOpen
  \bibfield  {author} {\bibinfo {author} {\bibfnamefont {D.}~\bibnamefont
  {Ueltschi}},\ }\bibfield  {title} {\bibinfo {title} {{\it Cluster expansions
  and correlation functions}},\ }\href@noop {} {\bibfield  {journal} {\bibinfo
  {journal} {Moscow Mathematical Journal}\ }\textbf {\bibinfo {volume} {4}},\
  \bibinfo {pages} {511} (\bibinfo {year} {2004})}\BibitemShut {NoStop}%
\bibitem [{\citenamefont {Kliesch}\ \emph {et~al.}(2014)\citenamefont
  {Kliesch}, \citenamefont {Gogolin}, \citenamefont {Kastoryano}, \citenamefont
  {Riera},\ and\ \citenamefont {Eisert}}]{PhysRevX.4.031019}%
  \BibitemOpen
  \bibfield  {author} {\bibinfo {author} {\bibfnamefont {M.}~\bibnamefont
  {Kliesch}}, \bibinfo {author} {\bibfnamefont {C.}~\bibnamefont {Gogolin}},
  \bibinfo {author} {\bibfnamefont {M.~J.}\ \bibnamefont {Kastoryano}},
  \bibinfo {author} {\bibfnamefont {A.}~\bibnamefont {Riera}},\ and\ \bibinfo
  {author} {\bibfnamefont {J.}~\bibnamefont {Eisert}},\ }\bibfield  {title}
  {\bibinfo {title} {{\it Locality of Temperature}},\ }\href
  {https://doi.org/10.1103/PhysRevX.4.031019} {\bibfield  {journal} {\bibinfo
  {journal} {Phys. Rev. X}\ }\textbf {\bibinfo {volume} {4}},\ \bibinfo {pages}
  {031019} (\bibinfo {year} {2014})}\BibitemShut {NoStop}%
\bibitem [{\citenamefont {Fr{\"o}hlich}\ and\ \citenamefont
  {Ueltschi}(2015)}]{frohlich2015some}%
  \BibitemOpen
  \bibfield  {author} {\bibinfo {author} {\bibfnamefont {J.}~\bibnamefont
  {Fr{\"o}hlich}}\ and\ \bibinfo {author} {\bibfnamefont {D.}~\bibnamefont
  {Ueltschi}},\ }\bibfield  {title} {\bibinfo {title} {{\it Some properties of
  correlations of quantum lattice systems in thermal equilibrium}},\ }\href
  {https://doi.org/10.1063/1.4921305} {\bibfield  {journal} {\bibinfo
  {journal} {Journal of Mathematical Physics}\ }\textbf {\bibinfo {volume}
  {56}},\ \bibinfo {pages} {053302} (\bibinfo {year} {2015})}\BibitemShut
  {NoStop}%
\bibitem [{\citenamefont {Hastings}(2004)}]{PhysRevB.69.104431}%
  \BibitemOpen
  \bibfield  {author} {\bibinfo {author} {\bibfnamefont {M.~B.}\ \bibnamefont
  {Hastings}},\ }\bibfield  {title} {\bibinfo {title} {{\it Lieb-Schultz-Mattis
  in higher dimensions}},\ }\href {https://doi.org/10.1103/PhysRevB.69.104431}
  {\bibfield  {journal} {\bibinfo  {journal} {Phys. Rev. B}\ }\textbf {\bibinfo
  {volume} {69}},\ \bibinfo {pages} {104431} (\bibinfo {year}
  {2004})}\BibitemShut {NoStop}%
\bibitem [{ref()}]{ref:Hastings2004-Markov}%
  \BibitemOpen
  \bibfield  {title} {\bibinfo {title} {{\it Locality in Quantum and Markov
  Dynamics on Lattices and Networks}},\ }\href@noop {} {\ }\BibitemShut
  {NoStop}%
\bibitem [{\citenamefont {Hastings}\ and\ \citenamefont
  {Koma}(2006)}]{ref:Hastings2006-ExpDec}%
  \BibitemOpen
  \bibfield  {author} {\bibinfo {author} {\bibfnamefont {M.}~\bibnamefont
  {Hastings}}\ and\ \bibinfo {author} {\bibfnamefont {T.}~\bibnamefont
  {Koma}},\ }\bibfield  {title} {\bibinfo {title} {{\it Spectral Gap and
  Exponential Decay of Correlations}},\ }\href
  {https://doi.org/10.1007/s00220-006-0030-4} {\bibfield  {journal} {\bibinfo
  {journal} {Communications in Mathematical Physics}\ }\textbf {\bibinfo
  {volume} {265}},\ \bibinfo {pages} {781} (\bibinfo {year}
  {2006})}\BibitemShut {NoStop}%
\bibitem [{\citenamefont {Nachtergaele}\ and\ \citenamefont
  {Sims}(2006)}]{ref:Nachtergaele2006-LR}%
  \BibitemOpen
  \bibfield  {author} {\bibinfo {author} {\bibfnamefont {B.}~\bibnamefont
  {Nachtergaele}}\ and\ \bibinfo {author} {\bibfnamefont {R.}~\bibnamefont
  {Sims}},\ }\bibfield  {title} {\bibinfo {title} {{\it Lieb-Robinson Bounds
  and the Exponential Clustering Theorem}},\ }\href
  {https://doi.org/10.1007/s00220-006-1556-1} {\bibfield  {journal} {\bibinfo
  {journal} {Communications in Mathematical Physics}\ }\textbf {\bibinfo
  {volume} {265}},\ \bibinfo {pages} {119} (\bibinfo {year}
  {2006})}\BibitemShut {NoStop}%
\bibitem [{\citenamefont {Amico}\ \emph {et~al.}(2008)\citenamefont {Amico},
  \citenamefont {Fazio}, \citenamefont {Osterloh},\ and\ \citenamefont
  {Vedral}}]{RevModPhys.80.517}%
  \BibitemOpen
  \bibfield  {author} {\bibinfo {author} {\bibfnamefont {L.}~\bibnamefont
  {Amico}}, \bibinfo {author} {\bibfnamefont {R.}~\bibnamefont {Fazio}},
  \bibinfo {author} {\bibfnamefont {A.}~\bibnamefont {Osterloh}},\ and\
  \bibinfo {author} {\bibfnamefont {V.}~\bibnamefont {Vedral}},\ }\bibfield
  {title} {\bibinfo {title} {{\it Entanglement in many-body systems}},\ }\href
  {https://doi.org/10.1103/RevModPhys.80.517} {\bibfield  {journal} {\bibinfo
  {journal} {Rev. Mod. Phys.}\ }\textbf {\bibinfo {volume} {80}},\ \bibinfo
  {pages} {517} (\bibinfo {year} {2008})}\BibitemShut {NoStop}%
\bibitem [{\citenamefont {Latorre}\ and\ \citenamefont
  {Riera}(2009)}]{Latorre_2009}%
  \BibitemOpen
  \bibfield  {author} {\bibinfo {author} {\bibfnamefont {J.~I.}\ \bibnamefont
  {Latorre}}\ and\ \bibinfo {author} {\bibfnamefont {A.}~\bibnamefont
  {Riera}},\ }\bibfield  {title} {\bibinfo {title} {{\it A short review on
  entanglement in quantum spin systems}},\ }\href
  {https://doi.org/10.1088/1751-8113/42/50/504002} {\bibfield  {journal}
  {\bibinfo  {journal} {Journal of Physics A: Mathematical and Theoretical}\
  }\textbf {\bibinfo {volume} {42}},\ \bibinfo {pages} {504002} (\bibinfo
  {year} {2009})}\BibitemShut {NoStop}%
\bibitem [{\citenamefont {Laflorencie}(2016)}]{LAFLORENCIE20161}%
  \BibitemOpen
  \bibfield  {author} {\bibinfo {author} {\bibfnamefont {N.}~\bibnamefont
  {Laflorencie}},\ }\bibfield  {title} {\bibinfo {title} {{\it Quantum
  entanglement in condensed matter systems}},\ }\href
  {https://doi.org/https://doi.org/10.1016/j.physrep.2016.06.008} {\bibfield
  {journal} {\bibinfo  {journal} {Physics Reports}\ }\textbf {\bibinfo {volume}
  {646}},\ \bibinfo {pages} {1} (\bibinfo {year} {2016})},\ \bibinfo {note}
  {quantum entanglement in condensed matter systems}\BibitemShut {NoStop}%
\bibitem [{\citenamefont {Cirac}\ \emph {et~al.}(2021)\citenamefont {Cirac},
  \citenamefont {P\'erez-Garc\'{\i}a}, \citenamefont {Schuch},\ and\
  \citenamefont {Verstraete}}]{RevModPhys.93.045003}%
  \BibitemOpen
  \bibfield  {author} {\bibinfo {author} {\bibfnamefont {J.~I.}\ \bibnamefont
  {Cirac}}, \bibinfo {author} {\bibfnamefont {D.}~\bibnamefont
  {P\'erez-Garc\'{\i}a}}, \bibinfo {author} {\bibfnamefont {N.}~\bibnamefont
  {Schuch}},\ and\ \bibinfo {author} {\bibfnamefont {F.}~\bibnamefont
  {Verstraete}},\ }\bibfield  {title} {\bibinfo {title} {Matrix product states
  and projected entangled pair states: Concepts, symmetries, theorems},\ }\href
  {https://doi.org/10.1103/RevModPhys.93.045003} {\bibfield  {journal}
  {\bibinfo  {journal} {Rev. Mod. Phys.}\ }\textbf {\bibinfo {volume} {93}},\
  \bibinfo {pages} {045003} (\bibinfo {year} {2021})}\BibitemShut {NoStop}%
\bibitem [{\citenamefont {Eisert}\ \emph {et~al.}(2010)\citenamefont {Eisert},
  \citenamefont {Cramer},\ and\ \citenamefont {Plenio}}]{RevModPhys.82.277}%
  \BibitemOpen
  \bibfield  {author} {\bibinfo {author} {\bibfnamefont {J.}~\bibnamefont
  {Eisert}}, \bibinfo {author} {\bibfnamefont {M.}~\bibnamefont {Cramer}},\
  and\ \bibinfo {author} {\bibfnamefont {M.~B.}\ \bibnamefont {Plenio}},\
  }\bibfield  {title} {\bibinfo {title} {{\it Colloquium: Area laws for the
  entanglement entropy}},\ }\href {https://doi.org/10.1103/RevModPhys.82.277}
  {\bibfield  {journal} {\bibinfo  {journal} {Rev. Mod. Phys.}\ }\textbf
  {\bibinfo {volume} {82}},\ \bibinfo {pages} {277} (\bibinfo {year}
  {2010})}\BibitemShut {NoStop}%
\bibitem [{\citenamefont {Hastings}(2007{\natexlab{a}})}]{Hastings_2007}%
  \BibitemOpen
  \bibfield  {author} {\bibinfo {author} {\bibfnamefont {M.~B.}\ \bibnamefont
  {Hastings}},\ }\bibfield  {title} {\bibinfo {title} {{\it An area law for
  one-dimensional quantum systems}},\ }\href
  {https://doi.org/10.1088/1742-5468/2007/08/p08024} {\bibfield  {journal}
  {\bibinfo  {journal} {Journal of Statistical Mechanics: Theory and
  Experiment}\ }\textbf {\bibinfo {volume} {2007}},\ \bibinfo {pages} {P08024}
  (\bibinfo {year} {2007}{\natexlab{a}})}\BibitemShut {NoStop}%
\bibitem [{\citenamefont {Arad}\ \emph {et~al.}(2013)\citenamefont {Arad},
  \citenamefont {Kitaev}, \citenamefont {Landau},\ and\ \citenamefont
  {Vazirani}}]{arad2013area}%
  \BibitemOpen
  \bibfield  {author} {\bibinfo {author} {\bibfnamefont {I.}~\bibnamefont
  {Arad}}, \bibinfo {author} {\bibfnamefont {A.}~\bibnamefont {Kitaev}},
  \bibinfo {author} {\bibfnamefont {Z.}~\bibnamefont {Landau}},\ and\ \bibinfo
  {author} {\bibfnamefont {U.}~\bibnamefont {Vazirani}},\ }\bibfield  {title}
  {\bibinfo {title} {{\it An area law and sub-exponential algorithm for 1D
  systems}},\ }\href@noop {} {\bibfield  {journal} {\bibinfo  {journal} {arXiv
  preprint arXiv:1301.1162}\ } (\bibinfo {year} {2013})},\ \Eprint
  {https://arxiv.org/abs/arXiv:1301.1162} {arXiv:1301.1162} \BibitemShut
  {NoStop}%
\bibitem [{\citenamefont {Anshu}\ \emph {et~al.}(2022)\citenamefont {Anshu},
  \citenamefont {Arad},\ and\ \citenamefont
  {Gosset}}]{10.1145/3519935.3519962}%
  \BibitemOpen
  \bibfield  {author} {\bibinfo {author} {\bibfnamefont {A.}~\bibnamefont
  {Anshu}}, \bibinfo {author} {\bibfnamefont {I.}~\bibnamefont {Arad}},\ and\
  \bibinfo {author} {\bibfnamefont {D.}~\bibnamefont {Gosset}},\ }\bibfield
  {title} {\bibinfo {title} {An area law for 2d frustration-free spin
  systems},\ }in\ \href {https://doi.org/10.1145/3519935.3519962} {\emph
  {\bibinfo {booktitle} {Proceedings of the 54th Annual ACM SIGACT Symposium on
  Theory of Computing}}},\ \bibinfo {series and number} {STOC 2022}\ (\bibinfo
  {publisher} {Association for Computing Machinery},\ \bibinfo {address} {New
  York, NY, USA},\ \bibinfo {year} {2022})\ p.\ \bibinfo {pages}
  {12–18}\BibitemShut {NoStop}%
\bibitem [{\citenamefont {Wolf}\ \emph {et~al.}(2008)\citenamefont {Wolf},
  \citenamefont {Verstraete}, \citenamefont {Hastings},\ and\ \citenamefont
  {Cirac}}]{PhysRevLett.100.070502}%
  \BibitemOpen
  \bibfield  {author} {\bibinfo {author} {\bibfnamefont {M.~M.}\ \bibnamefont
  {Wolf}}, \bibinfo {author} {\bibfnamefont {F.}~\bibnamefont {Verstraete}},
  \bibinfo {author} {\bibfnamefont {M.~B.}\ \bibnamefont {Hastings}},\ and\
  \bibinfo {author} {\bibfnamefont {J.~I.}\ \bibnamefont {Cirac}},\ }\bibfield
  {title} {\bibinfo {title} {{\it Area Laws in Quantum Systems: Mutual
  Information and Correlations}},\ }\href
  {https://doi.org/10.1103/PhysRevLett.100.070502} {\bibfield  {journal}
  {\bibinfo  {journal} {Phys. Rev. Lett.}\ }\textbf {\bibinfo {volume} {100}},\
  \bibinfo {pages} {070502} (\bibinfo {year} {2008})}\BibitemShut {NoStop}%
\bibitem [{\citenamefont {Kuwahara}\ \emph {et~al.}(2021)\citenamefont
  {Kuwahara}, \citenamefont {Alhambra},\ and\ \citenamefont
  {Anshu}}]{PhysRevX.11.011047}%
  \BibitemOpen
  \bibfield  {author} {\bibinfo {author} {\bibfnamefont {T.}~\bibnamefont
  {Kuwahara}}, \bibinfo {author} {\bibfnamefont {A.~M.}\ \bibnamefont
  {Alhambra}},\ and\ \bibinfo {author} {\bibfnamefont {A.}~\bibnamefont
  {Anshu}},\ }\bibfield  {title} {\bibinfo {title} {{\it Improved Thermal Area
  Law and Quasilinear Time Algorithm for Quantum Gibbs States}},\ }\href
  {https://doi.org/10.1103/PhysRevX.11.011047} {\bibfield  {journal} {\bibinfo
  {journal} {Phys. Rev. X}\ }\textbf {\bibinfo {volume} {11}},\ \bibinfo
  {pages} {011047} (\bibinfo {year} {2021})}\BibitemShut {NoStop}%
\bibitem [{\citenamefont {Bernigau}\ \emph {et~al.}(2015)\citenamefont
  {Bernigau}, \citenamefont {Kastoryano},\ and\ \citenamefont
  {Eisert}}]{Bernigau_2015}%
  \BibitemOpen
  \bibfield  {author} {\bibinfo {author} {\bibfnamefont {H.}~\bibnamefont
  {Bernigau}}, \bibinfo {author} {\bibfnamefont {M.~J.}\ \bibnamefont
  {Kastoryano}},\ and\ \bibinfo {author} {\bibfnamefont {J.}~\bibnamefont
  {Eisert}},\ }\bibfield  {title} {\bibinfo {title} {{\it Mutual information
  area laws for thermal free fermions}},\ }\href
  {https://doi.org/10.1088/1742-5468/2015/02/p02008} {\bibfield  {journal}
  {\bibinfo  {journal} {Journal of Statistical Mechanics: Theory and
  Experiment}\ }\textbf {\bibinfo {volume} {2015}},\ \bibinfo {pages} {P02008}
  (\bibinfo {year} {2015})}\BibitemShut {NoStop}%
\bibitem [{\citenamefont {Lemm}\ and\ \citenamefont
  {Siebert}(2023)}]{Lemm2023thermalarealaw}%
  \BibitemOpen
  \bibfield  {author} {\bibinfo {author} {\bibfnamefont {M.}~\bibnamefont
  {Lemm}}\ and\ \bibinfo {author} {\bibfnamefont {O.}~\bibnamefont {Siebert}},\
  }\bibfield  {title} {\bibinfo {title} {{\it Thermal {A}rea {L}aw for
  {L}attice {B}osons}},\ }\href {https://doi.org/10.22331/q-2023-08-16-1083}
  {\bibfield  {journal} {\bibinfo  {journal} {{Quantum}}\ }\textbf {\bibinfo
  {volume} {7}},\ \bibinfo {pages} {1083} (\bibinfo {year} {2023})}\BibitemShut
  {NoStop}%
\bibitem [{\citenamefont {Verstraete}\ \emph {et~al.}(2004)\citenamefont
  {Verstraete}, \citenamefont {Garc\'{\i}a-Ripoll},\ and\ \citenamefont
  {Cirac}}]{PhysRevLett.93.207204}%
  \BibitemOpen
  \bibfield  {author} {\bibinfo {author} {\bibfnamefont {F.}~\bibnamefont
  {Verstraete}}, \bibinfo {author} {\bibfnamefont {J.~J.}\ \bibnamefont
  {Garc\'{\i}a-Ripoll}},\ and\ \bibinfo {author} {\bibfnamefont {J.~I.}\
  \bibnamefont {Cirac}},\ }\bibfield  {title} {\bibinfo {title} {{\it Matrix
  Product Density Operators: Simulation of Finite-Temperature and Dissipative
  Systems}},\ }\href {https://doi.org/10.1103/PhysRevLett.93.207204} {\bibfield
   {journal} {\bibinfo  {journal} {Phys. Rev. Lett.}\ }\textbf {\bibinfo
  {volume} {93}},\ \bibinfo {pages} {207204} (\bibinfo {year}
  {2004})}\BibitemShut {NoStop}%
\bibitem [{\citenamefont {Schollw\"ock}(2005)}]{RevModPhys.77.259}%
  \BibitemOpen
  \bibfield  {author} {\bibinfo {author} {\bibfnamefont {U.}~\bibnamefont
  {Schollw\"ock}},\ }\bibfield  {title} {\bibinfo {title} {{\it The
  density-matrix renormalization group}},\ }\href
  {https://doi.org/10.1103/RevModPhys.77.259} {\bibfield  {journal} {\bibinfo
  {journal} {Rev. Mod. Phys.}\ }\textbf {\bibinfo {volume} {77}},\ \bibinfo
  {pages} {259} (\bibinfo {year} {2005})}\BibitemShut {NoStop}%
\bibitem [{\citenamefont {Landau}\ \emph {et~al.}(2015)\citenamefont {Landau},
  \citenamefont {Vazirani},\ and\ \citenamefont
  {Vidick}}]{landau2015polynomial}%
  \BibitemOpen
  \bibfield  {author} {\bibinfo {author} {\bibfnamefont {Z.}~\bibnamefont
  {Landau}}, \bibinfo {author} {\bibfnamefont {U.}~\bibnamefont {Vazirani}},\
  and\ \bibinfo {author} {\bibfnamefont {T.}~\bibnamefont {Vidick}},\
  }\bibfield  {title} {\bibinfo {title} {{\it A polynomial time algorithm for
  the ground state of one-dimensional gapped local Hamiltonians}},\ }\href
  {https://doi.org/10.1038/nphys3345} {\bibfield  {journal} {\bibinfo
  {journal} {Nature Physics}\ }\textbf {\bibinfo {volume} {11}},\ \bibinfo
  {pages} {566} (\bibinfo {year} {2015})}\BibitemShut {NoStop}%
\bibitem [{\citenamefont {Arad}\ \emph {et~al.}(2017)\citenamefont {Arad},
  \citenamefont {Landau}, \citenamefont {Vazirani},\ and\ \citenamefont
  {Vidick}}]{Arad2017}%
  \BibitemOpen
  \bibfield  {author} {\bibinfo {author} {\bibfnamefont {I.}~\bibnamefont
  {Arad}}, \bibinfo {author} {\bibfnamefont {Z.}~\bibnamefont {Landau}},
  \bibinfo {author} {\bibfnamefont {U.}~\bibnamefont {Vazirani}},\ and\
  \bibinfo {author} {\bibfnamefont {T.}~\bibnamefont {Vidick}},\ }\bibfield
  {title} {\bibinfo {title} {{\it Rigorous RG Algorithms and Area Laws for Low
  Energy Eigenstates in 1D}},\ }\href
  {https://doi.org/10.1007/s00220-017-2973-z} {\bibfield  {journal} {\bibinfo
  {journal} {Communications in Mathematical Physics}\ }\textbf {\bibinfo
  {volume} {356}},\ \bibinfo {pages} {65} (\bibinfo {year} {2017})}\BibitemShut
  {NoStop}%
\bibitem [{\citenamefont {Kuwahara}\ and\ \citenamefont
  {Saito}(2022)}]{PhysRevX.12.021022}%
  \BibitemOpen
  \bibfield  {author} {\bibinfo {author} {\bibfnamefont {T.}~\bibnamefont
  {Kuwahara}}\ and\ \bibinfo {author} {\bibfnamefont {K.}~\bibnamefont
  {Saito}},\ }\bibfield  {title} {\bibinfo {title} {{\it Exponential Clustering
  of Bipartite Quantum Entanglement at Arbitrary Temperatures}},\ }\href
  {https://doi.org/10.1103/PhysRevX.12.021022} {\bibfield  {journal} {\bibinfo
  {journal} {Phys. Rev. X}\ }\textbf {\bibinfo {volume} {12}},\ \bibinfo
  {pages} {021022} (\bibinfo {year} {2022})}\BibitemShut {NoStop}%
\bibitem [{\citenamefont {Chen}\ \emph {et~al.}(2010)\citenamefont {Chen},
  \citenamefont {Gu},\ and\ \citenamefont {Wen}}]{PhysRevB.82.155138}%
  \BibitemOpen
  \bibfield  {author} {\bibinfo {author} {\bibfnamefont {X.}~\bibnamefont
  {Chen}}, \bibinfo {author} {\bibfnamefont {Z.-C.}\ \bibnamefont {Gu}},\ and\
  \bibinfo {author} {\bibfnamefont {X.-G.}\ \bibnamefont {Wen}},\ }\bibfield
  {title} {\bibinfo {title} {{\it Local unitary transformation, long-range
  quantum entanglement, wave function renormalization, and topological
  order}},\ }\href {https://doi.org/10.1103/PhysRevB.82.155138} {\bibfield
  {journal} {\bibinfo  {journal} {Phys. Rev. B}\ }\textbf {\bibinfo {volume}
  {82}},\ \bibinfo {pages} {155138} (\bibinfo {year} {2010})}\BibitemShut
  {NoStop}%
\bibitem [{\citenamefont {Wen}(2017)}]{RevModPhys.89.041004}%
  \BibitemOpen
  \bibfield  {author} {\bibinfo {author} {\bibfnamefont {X.-G.}\ \bibnamefont
  {Wen}},\ }\bibfield  {title} {\bibinfo {title} {{\it Colloquium: Zoo of
  quantum-topological phases of matter}},\ }\href
  {https://doi.org/10.1103/RevModPhys.89.041004} {\bibfield  {journal}
  {\bibinfo  {journal} {Rev. Mod. Phys.}\ }\textbf {\bibinfo {volume} {89}},\
  \bibinfo {pages} {041004} (\bibinfo {year} {2017})}\BibitemShut {NoStop}%
\bibitem [{\citenamefont {Christandl}\ and\ \citenamefont
  {Winter}(2004)}]{doi:10.1063/1.1643788}%
  \BibitemOpen
  \bibfield  {author} {\bibinfo {author} {\bibfnamefont {M.}~\bibnamefont
  {Christandl}}\ and\ \bibinfo {author} {\bibfnamefont {A.}~\bibnamefont
  {Winter}},\ }\bibfield  {title} {\bibinfo {title} {{\it “Squashed
  entanglement”: An additive entanglement measure}},\ }\href
  {https://doi.org/10.1063/1.1643788} {\bibfield  {journal} {\bibinfo
  {journal} {Journal of Mathematical Physics}\ }\textbf {\bibinfo {volume}
  {45}},\ \bibinfo {pages} {829} (\bibinfo {year} {2004})}\BibitemShut
  {NoStop}%
\bibitem [{\citenamefont {Berta}\ \emph
  {et~al.}(2015{\natexlab{a}})\citenamefont {Berta}, \citenamefont
  {Seshadreesan},\ and\ \citenamefont {Wilde}}]{berta2015renyi}%
  \BibitemOpen
  \bibfield  {author} {\bibinfo {author} {\bibfnamefont {M.}~\bibnamefont
  {Berta}}, \bibinfo {author} {\bibfnamefont {K.~P.}\ \bibnamefont
  {Seshadreesan}},\ and\ \bibinfo {author} {\bibfnamefont {M.~M.}\ \bibnamefont
  {Wilde}},\ }\bibfield  {title} {\bibinfo {title} {{\it R{\'e}nyi
  generalizations of the conditional quantum mutual information}},\ }\href
  {https://doi.org/10.1063/1.4908102} {\bibfield  {journal} {\bibinfo
  {journal} {Journal of Mathematical Physics}\ }\textbf {\bibinfo {volume}
  {56}},\ \bibinfo {pages} {022205} (\bibinfo {year}
  {2015}{\natexlab{a}})}\BibitemShut {NoStop}%
\bibitem [{\citenamefont {Sutter}(2018)}]{sutter2018approximate}%
  \BibitemOpen
  \bibfield  {author} {\bibinfo {author} {\bibfnamefont {D.}~\bibnamefont
  {Sutter}},\ }\bibfield  {title} {\bibinfo {title} {{\it Approximate quantum
  Markov chains}},\ }\href@noop {} {\bibfield  {journal} {\bibinfo  {journal}
  {arXiv preprint arXiv:1802.05477}\ } (\bibinfo {year} {2018})},\ \Eprint
  {https://arxiv.org/abs/arXiv:1802.05477} {arXiv:1802.05477} \BibitemShut
  {NoStop}%
\bibitem [{\citenamefont {Kitaev}\ and\ \citenamefont
  {Preskill}(2006)}]{PhysRevLett.96.110404}%
  \BibitemOpen
  \bibfield  {author} {\bibinfo {author} {\bibfnamefont {A.}~\bibnamefont
  {Kitaev}}\ and\ \bibinfo {author} {\bibfnamefont {J.}~\bibnamefont
  {Preskill}},\ }\bibfield  {title} {\bibinfo {title} {{\it Topological
  Entanglement Entropy}},\ }\href
  {https://doi.org/10.1103/PhysRevLett.96.110404} {\bibfield  {journal}
  {\bibinfo  {journal} {Phys. Rev. Lett.}\ }\textbf {\bibinfo {volume} {96}},\
  \bibinfo {pages} {110404} (\bibinfo {year} {2006})}\BibitemShut {NoStop}%
\bibitem [{\citenamefont {Levin}\ and\ \citenamefont
  {Wen}(2006)}]{PhysRevLett.96.110405}%
  \BibitemOpen
  \bibfield  {author} {\bibinfo {author} {\bibfnamefont {M.}~\bibnamefont
  {Levin}}\ and\ \bibinfo {author} {\bibfnamefont {X.-G.}\ \bibnamefont
  {Wen}},\ }\bibfield  {title} {\bibinfo {title} {{\it Detecting Topological
  Order in a Ground State Wave Function}},\ }\href
  {https://doi.org/10.1103/PhysRevLett.96.110405} {\bibfield  {journal}
  {\bibinfo  {journal} {Phys. Rev. Lett.}\ }\textbf {\bibinfo {volume} {96}},\
  \bibinfo {pages} {110405} (\bibinfo {year} {2006})}\BibitemShut {NoStop}%
\bibitem [{\citenamefont {Kato}\ \emph {et~al.}(2016)\citenamefont {Kato},
  \citenamefont {Furrer},\ and\ \citenamefont {Murao}}]{PhysRevA.93.022317}%
  \BibitemOpen
  \bibfield  {author} {\bibinfo {author} {\bibfnamefont {K.}~\bibnamefont
  {Kato}}, \bibinfo {author} {\bibfnamefont {F.}~\bibnamefont {Furrer}},\ and\
  \bibinfo {author} {\bibfnamefont {M.}~\bibnamefont {Murao}},\ }\bibfield
  {title} {\bibinfo {title} {{\it Information-theoretical analysis of
  topological entanglement entropy and multipartite correlations}},\ }\href
  {https://doi.org/10.1103/PhysRevA.93.022317} {\bibfield  {journal} {\bibinfo
  {journal} {Phys. Rev. A}\ }\textbf {\bibinfo {volume} {93}},\ \bibinfo
  {pages} {022317} (\bibinfo {year} {2016})}\BibitemShut {NoStop}%
\bibitem [{\citenamefont {Ding}\ \emph {et~al.}(2016)\citenamefont {Ding},
  \citenamefont {Hayden},\ and\ \citenamefont {Walter}}]{Ding2016}%
  \BibitemOpen
  \bibfield  {author} {\bibinfo {author} {\bibfnamefont {D.}~\bibnamefont
  {Ding}}, \bibinfo {author} {\bibfnamefont {P.}~\bibnamefont {Hayden}},\ and\
  \bibinfo {author} {\bibfnamefont {M.}~\bibnamefont {Walter}},\ }\bibfield
  {title} {\bibinfo {title} {{\it Conditional mutual information of bipartite
  unitaries and scrambling}},\ }\href {https://doi.org/10.1007/JHEP12(2016)145}
  {\bibfield  {journal} {\bibinfo  {journal} {Journal of High Energy Physics}\
  }\textbf {\bibinfo {volume} {2016}},\ \bibinfo {pages} {145} (\bibinfo {year}
  {2016})}\BibitemShut {NoStop}%
\bibitem [{\citenamefont {Iyoda}\ and\ \citenamefont
  {Sagawa}(2018)}]{PhysRevA.97.042330}%
  \BibitemOpen
  \bibfield  {author} {\bibinfo {author} {\bibfnamefont {E.}~\bibnamefont
  {Iyoda}}\ and\ \bibinfo {author} {\bibfnamefont {T.}~\bibnamefont {Sagawa}},\
  }\bibfield  {title} {\bibinfo {title} {{\it Scrambling of quantum information
  in quantum many-body systems}},\ }\href
  {https://doi.org/10.1103/PhysRevA.97.042330} {\bibfield  {journal} {\bibinfo
  {journal} {Phys. Rev. A}\ }\textbf {\bibinfo {volume} {97}},\ \bibinfo
  {pages} {042330} (\bibinfo {year} {2018})}\BibitemShut {NoStop}%
\bibitem [{\citenamefont {Bertini}\ and\ \citenamefont
  {Piroli}(2020)}]{PhysRevB.102.064305}%
  \BibitemOpen
  \bibfield  {author} {\bibinfo {author} {\bibfnamefont {B.}~\bibnamefont
  {Bertini}}\ and\ \bibinfo {author} {\bibfnamefont {L.}~\bibnamefont
  {Piroli}},\ }\bibfield  {title} {\bibinfo {title} {{\it Scrambling in random
  unitary circuits: Exact results}},\ }\href
  {https://doi.org/10.1103/PhysRevB.102.064305} {\bibfield  {journal} {\bibinfo
   {journal} {Phys. Rev. B}\ }\textbf {\bibinfo {volume} {102}},\ \bibinfo
  {pages} {064305} (\bibinfo {year} {2020})}\BibitemShut {NoStop}%
\bibitem [{\citenamefont {Monaco}\ \emph {et~al.}(2023)\citenamefont {Monaco},
  \citenamefont {Innocenti}, \citenamefont {Cilluffo}, \citenamefont
  {Chisholm}, \citenamefont {Lorenzo},\ and\ \citenamefont
  {Palma}}]{LoMonaco_2023}%
  \BibitemOpen
  \bibfield  {author} {\bibinfo {author} {\bibfnamefont {G.~L.}\ \bibnamefont
  {Monaco}}, \bibinfo {author} {\bibfnamefont {L.}~\bibnamefont {Innocenti}},
  \bibinfo {author} {\bibfnamefont {D.}~\bibnamefont {Cilluffo}}, \bibinfo
  {author} {\bibfnamefont {D.~A.}\ \bibnamefont {Chisholm}}, \bibinfo {author}
  {\bibfnamefont {S.}~\bibnamefont {Lorenzo}},\ and\ \bibinfo {author}
  {\bibfnamefont {G.~M.}\ \bibnamefont {Palma}},\ }\bibfield  {title} {\bibinfo
  {title} {{\it Quantum scrambling via accessible tripartite information}},\
  }\href {https://doi.org/10.1088/2058-9565/accd92} {\bibfield  {journal}
  {\bibinfo  {journal} {Quantum Science and Technology}\ }\textbf {\bibinfo
  {volume} {8}},\ \bibinfo {pages} {035006} (\bibinfo {year}
  {2023})}\BibitemShut {NoStop}%
\bibitem [{\citenamefont {Gullans}\ and\ \citenamefont
  {Huse}(2020)}]{PhysRevX.10.041020}%
  \BibitemOpen
  \bibfield  {author} {\bibinfo {author} {\bibfnamefont {M.~J.}\ \bibnamefont
  {Gullans}}\ and\ \bibinfo {author} {\bibfnamefont {D.~A.}\ \bibnamefont
  {Huse}},\ }\bibfield  {title} {\bibinfo {title} {{\it Dynamical Purification
  Phase Transition Induced by Quantum Measurements}},\ }\href
  {https://doi.org/10.1103/PhysRevX.10.041020} {\bibfield  {journal} {\bibinfo
  {journal} {Phys. Rev. X}\ }\textbf {\bibinfo {volume} {10}},\ \bibinfo
  {pages} {041020} (\bibinfo {year} {2020})}\BibitemShut {NoStop}%
\bibitem [{\citenamefont {Ippoliti}\ \emph {et~al.}(2021)\citenamefont
  {Ippoliti}, \citenamefont {Gullans}, \citenamefont {Gopalakrishnan},
  \citenamefont {Huse},\ and\ \citenamefont {Khemani}}]{PhysRevX.11.011030}%
  \BibitemOpen
  \bibfield  {author} {\bibinfo {author} {\bibfnamefont {M.}~\bibnamefont
  {Ippoliti}}, \bibinfo {author} {\bibfnamefont {M.~J.}\ \bibnamefont
  {Gullans}}, \bibinfo {author} {\bibfnamefont {S.}~\bibnamefont
  {Gopalakrishnan}}, \bibinfo {author} {\bibfnamefont {D.~A.}\ \bibnamefont
  {Huse}},\ and\ \bibinfo {author} {\bibfnamefont {V.}~\bibnamefont
  {Khemani}},\ }\bibfield  {title} {\bibinfo {title} {{\it Entanglement Phase
  Transitions in Measurement-Only Dynamics}},\ }\href
  {https://doi.org/10.1103/PhysRevX.11.011030} {\bibfield  {journal} {\bibinfo
  {journal} {Phys. Rev. X}\ }\textbf {\bibinfo {volume} {11}},\ \bibinfo
  {pages} {011030} (\bibinfo {year} {2021})}\BibitemShut {NoStop}%
\bibitem [{\citenamefont {Koh}\ \emph {et~al.}(2023)\citenamefont {Koh},
  \citenamefont {Sun}, \citenamefont {Motta},\ and\ \citenamefont
  {Minnich}}]{Koh2023}%
  \BibitemOpen
  \bibfield  {author} {\bibinfo {author} {\bibfnamefont {J.~M.}\ \bibnamefont
  {Koh}}, \bibinfo {author} {\bibfnamefont {S.-N.}\ \bibnamefont {Sun}},
  \bibinfo {author} {\bibfnamefont {M.}~\bibnamefont {Motta}},\ and\ \bibinfo
  {author} {\bibfnamefont {A.~J.}\ \bibnamefont {Minnich}},\ }\bibfield
  {title} {\bibinfo {title} {{\it Measurement-induced entanglement phase
  transition on a superconducting quantum processor with mid-circuit
  readout}},\ }\href {https://doi.org/10.1038/s41567-023-02076-6} {\bibfield
  {journal} {\bibinfo  {journal} {Nature Physics}\ }\textbf {\bibinfo {volume}
  {19}},\ \bibinfo {pages} {1314} (\bibinfo {year} {2023})}\BibitemShut
  {NoStop}%
\bibitem [{\citenamefont {Hayden}\ \emph {et~al.}(2013)\citenamefont {Hayden},
  \citenamefont {Headrick},\ and\ \citenamefont
  {Maloney}}]{PhysRevD.87.046003}%
  \BibitemOpen
  \bibfield  {author} {\bibinfo {author} {\bibfnamefont {P.}~\bibnamefont
  {Hayden}}, \bibinfo {author} {\bibfnamefont {M.}~\bibnamefont {Headrick}},\
  and\ \bibinfo {author} {\bibfnamefont {A.}~\bibnamefont {Maloney}},\
  }\bibfield  {title} {\bibinfo {title} {{\it Holographic mutual information is
  monogamous}},\ }\href {https://doi.org/10.1103/PhysRevD.87.046003} {\bibfield
   {journal} {\bibinfo  {journal} {Phys. Rev. D}\ }\textbf {\bibinfo {volume}
  {87}},\ \bibinfo {pages} {046003} (\bibinfo {year} {2013})}\BibitemShut
  {NoStop}%
\bibitem [{\citenamefont {Rota}(2016)}]{Rota2016}%
  \BibitemOpen
  \bibfield  {author} {\bibinfo {author} {\bibfnamefont {M.}~\bibnamefont
  {Rota}},\ }\bibfield  {title} {\bibinfo {title} {{\it Tripartite information
  of highly entangled states}},\ }\href
  {https://doi.org/10.1007/JHEP04(2016)075} {\bibfield  {journal} {\bibinfo
  {journal} {Journal of High Energy Physics}\ }\textbf {\bibinfo {volume}
  {2016}},\ \bibinfo {pages} {75} (\bibinfo {year} {2016})}\BibitemShut
  {NoStop}%
\bibitem [{\citenamefont {Mari\ifmmode~\acute{c}\else \'{c}\fi{}}\ and\
  \citenamefont {Fagotti}(2023)}]{PhysRevB.108.L161116}%
  \BibitemOpen
  \bibfield  {author} {\bibinfo {author} {\bibfnamefont {V.}~\bibnamefont
  {Mari\ifmmode~\acute{c}\else \'{c}\fi{}}}\ and\ \bibinfo {author}
  {\bibfnamefont {M.}~\bibnamefont {Fagotti}},\ }\bibfield  {title} {\bibinfo
  {title} {{\it Universality in the tripartite information after global
  quenches}},\ }\href {https://doi.org/10.1103/PhysRevB.108.L161116} {\bibfield
   {journal} {\bibinfo  {journal} {Phys. Rev. B}\ }\textbf {\bibinfo {volume}
  {108}},\ \bibinfo {pages} {L161116} (\bibinfo {year} {2023})}\BibitemShut
  {NoStop}%
\bibitem [{\citenamefont {Mari{\'{c}}}\ and\ \citenamefont
  {Fagotti}(2023)}]{Maric2023}%
  \BibitemOpen
  \bibfield  {author} {\bibinfo {author} {\bibfnamefont {V.}~\bibnamefont
  {Mari{\'{c}}}}\ and\ \bibinfo {author} {\bibfnamefont {M.}~\bibnamefont
  {Fagotti}},\ }\bibfield  {title} {\bibinfo {title} {{\it Universality in the
  tripartite information after global quenches: (generalised) quantum XY
  models}},\ }\href {https://doi.org/10.1007/JHEP06(2023)140} {\bibfield
  {journal} {\bibinfo  {journal} {Journal of High Energy Physics}\ }\textbf
  {\bibinfo {volume} {2023}},\ \bibinfo {pages} {140} (\bibinfo {year}
  {2023})}\BibitemShut {NoStop}%
\bibitem [{\citenamefont {Petz}(1986)}]{Petz1986}%
  \BibitemOpen
  \bibfield  {author} {\bibinfo {author} {\bibfnamefont {D.}~\bibnamefont
  {Petz}},\ }\bibfield  {title} {\bibinfo {title} {{\it Sufficient subalgebras
  and the relative entropy of states of a von Neumann algebra}},\ }\href
  {https://doi.org/10.1007/BF01212345} {\bibfield  {journal} {\bibinfo
  {journal} {Communications in Mathematical Physics}\ }\textbf {\bibinfo
  {volume} {105}},\ \bibinfo {pages} {123} (\bibinfo {year}
  {1986})}\BibitemShut {NoStop}%
\bibitem [{\citenamefont {Fawzi}\ and\ \citenamefont
  {Renner}(2015)}]{Fawzi2015}%
  \BibitemOpen
  \bibfield  {author} {\bibinfo {author} {\bibfnamefont {O.}~\bibnamefont
  {Fawzi}}\ and\ \bibinfo {author} {\bibfnamefont {R.}~\bibnamefont {Renner}},\
  }\bibfield  {title} {\bibinfo {title} {{\it Quantum Conditional Mutual
  Information and Approximate Markov Chains}},\ }\href
  {https://doi.org/10.1007/s00220-015-2466-x} {\bibfield  {journal} {\bibinfo
  {journal} {Communications in Mathematical Physics}\ }\textbf {\bibinfo
  {volume} {340}},\ \bibinfo {pages} {575} (\bibinfo {year}
  {2015})}\BibitemShut {NoStop}%
\bibitem [{\citenamefont {Brand\~ao}\ \emph {et~al.}(2015)\citenamefont
  {Brand\~ao}, \citenamefont {Harrow}, \citenamefont {Oppenheim},\ and\
  \citenamefont {Strelchuk}}]{PhysRevLett.115.050501}%
  \BibitemOpen
  \bibfield  {author} {\bibinfo {author} {\bibfnamefont {F.~G. S.~L.}\
  \bibnamefont {Brand\~ao}}, \bibinfo {author} {\bibfnamefont {A.~W.}\
  \bibnamefont {Harrow}}, \bibinfo {author} {\bibfnamefont {J.}~\bibnamefont
  {Oppenheim}},\ and\ \bibinfo {author} {\bibfnamefont {S.}~\bibnamefont
  {Strelchuk}},\ }\bibfield  {title} {\bibinfo {title} {{\it Quantum
  Conditional Mutual Information, Reconstructed States, and State
  Redistribution}},\ }\href {https://doi.org/10.1103/PhysRevLett.115.050501}
  {\bibfield  {journal} {\bibinfo  {journal} {Phys. Rev. Lett.}\ }\textbf
  {\bibinfo {volume} {115}},\ \bibinfo {pages} {050501} (\bibinfo {year}
  {2015})}\BibitemShut {NoStop}%
\bibitem [{\citenamefont {Wilde}(2015)}]{doi101098rspa20150338}%
  \BibitemOpen
  \bibfield  {author} {\bibinfo {author} {\bibfnamefont {M.~M.}\ \bibnamefont
  {Wilde}},\ }\bibfield  {title} {\bibinfo {title} {{\it Recoverability in
  quantum information theory}},\ }\href
  {https://doi.org/10.1098/rspa.2015.0338} {\bibfield  {journal} {\bibinfo
  {journal} {Proceedings of the Royal Society A: Mathematical, Physical and
  Engineering Sciences}\ }\textbf {\bibinfo {volume} {471}},\ \bibinfo {pages}
  {20150338} (\bibinfo {year} {2015})}\BibitemShut {NoStop}%
\bibitem [{\citenamefont {Junge}\ \emph {et~al.}(2018)\citenamefont {Junge},
  \citenamefont {Renner}, \citenamefont {Sutter}, \citenamefont {Wilde},\ and\
  \citenamefont {Winter}}]{Junge2018}%
  \BibitemOpen
  \bibfield  {author} {\bibinfo {author} {\bibfnamefont {M.}~\bibnamefont
  {Junge}}, \bibinfo {author} {\bibfnamefont {R.}~\bibnamefont {Renner}},
  \bibinfo {author} {\bibfnamefont {D.}~\bibnamefont {Sutter}}, \bibinfo
  {author} {\bibfnamefont {M.~M.}\ \bibnamefont {Wilde}},\ and\ \bibinfo
  {author} {\bibfnamefont {A.}~\bibnamefont {Winter}},\ }\bibfield  {title}
  {\bibinfo {title} {{\it Universal Recovery Maps and Approximate Sufficiency
  of Quantum Relative Entropy}},\ }\href
  {https://doi.org/10.1007/s00023-018-0716-0} {\bibfield  {journal} {\bibinfo
  {journal} {Annales Henri Poincar{\'e}}\ }\textbf {\bibinfo {volume} {19}},\
  \bibinfo {pages} {2955} (\bibinfo {year} {2018})}\BibitemShut {NoStop}%
\bibitem [{\citenamefont {Sutter}\ and\ \citenamefont
  {Renner}(2018)}]{Sutter2018}%
  \BibitemOpen
  \bibfield  {author} {\bibinfo {author} {\bibfnamefont {D.}~\bibnamefont
  {Sutter}}\ and\ \bibinfo {author} {\bibfnamefont {R.}~\bibnamefont
  {Renner}},\ }\bibfield  {title} {\bibinfo {title} {{\it Necessary Criterion
  for Approximate Recoverability}},\ }\href
  {https://doi.org/10.1007/s00023-018-0715-1} {\bibfield  {journal} {\bibinfo
  {journal} {Annales Henri Poincar{\'e}}\ }\textbf {\bibinfo {volume} {19}},\
  \bibinfo {pages} {3007} (\bibinfo {year} {2018})}\BibitemShut {NoStop}%
\bibitem [{\citenamefont {Hammersley}\ and\ \citenamefont
  {Clifford}(1971)}]{HammersleyClifford1971}%
  \BibitemOpen
  \bibfield  {author} {\bibinfo {author} {\bibfnamefont {J.~M.}\ \bibnamefont
  {Hammersley}}\ and\ \bibinfo {author} {\bibfnamefont {P.}~\bibnamefont
  {Clifford}},\ }\bibfield  {title} {\bibinfo {title} {{\it Markov fields on
  finite graphs and lattices}},\ }\href@noop {} {\bibfield  {journal} {\bibinfo
   {journal} {Unpublished manuscript}\ }\textbf {\bibinfo {volume} {46}}
  (\bibinfo {year} {1971})}\BibitemShut {NoStop}%
\bibitem [{\citenamefont {Leifer}\ and\ \citenamefont
  {Poulin}(2008)}]{Leifer2008}%
  \BibitemOpen
  \bibfield  {author} {\bibinfo {author} {\bibfnamefont {M.}~\bibnamefont
  {Leifer}}\ and\ \bibinfo {author} {\bibfnamefont {D.}~\bibnamefont
  {Poulin}},\ }\bibfield  {title} {\bibinfo {title} {{\it Quantum Graphical
  Models and Belief Propagation}},\ }\href
  {https://doi.org/10.1016/j.aop.2007.10.001} {\bibfield  {journal} {\bibinfo
  {journal} {Annals of Physics}\ }\textbf {\bibinfo {volume} {323}},\ \bibinfo
  {pages} {1899 } (\bibinfo {year} {2008})}\BibitemShut {NoStop}%
\bibitem [{\citenamefont {Brown}\ and\ \citenamefont
  {Poulin}(2012)}]{brown2012quantum}%
  \BibitemOpen
  \bibfield  {author} {\bibinfo {author} {\bibfnamefont {W.}~\bibnamefont
  {Brown}}\ and\ \bibinfo {author} {\bibfnamefont {D.}~\bibnamefont {Poulin}},\
  }\bibfield  {title} {\bibinfo {title} {{\it Quantum Markov networks and
  commuting Hamiltonians}},\ }\href@noop {} {\bibfield  {journal} {\bibinfo
  {journal} {arXiv preprint arXiv:1206.0755}\ } (\bibinfo {year} {2012})},\
  \Eprint {https://arxiv.org/abs/arXiv:1206.0755} {arXiv:1206.0755}
  \BibitemShut {NoStop}%
\bibitem [{\citenamefont {Jouneghani}\ \emph {et~al.}(2014)\citenamefont
  {Jouneghani}, \citenamefont {Babazadeh}, \citenamefont {Bayramzadeh},\ and\
  \citenamefont {Movla}}]{Jouneghani2014}%
  \BibitemOpen
  \bibfield  {author} {\bibinfo {author} {\bibfnamefont {F.~G.}\ \bibnamefont
  {Jouneghani}}, \bibinfo {author} {\bibfnamefont {M.}~\bibnamefont
  {Babazadeh}}, \bibinfo {author} {\bibfnamefont {R.}~\bibnamefont
  {Bayramzadeh}},\ and\ \bibinfo {author} {\bibfnamefont {H.}~\bibnamefont
  {Movla}},\ }\bibfield  {title} {\bibinfo {title} {{\it Investigation of
  Commuting Hamiltonian in Quantum Markov Network}},\ }\href
  {https://doi.org/10.1007/s10773-014-2042-8} {\bibfield  {journal} {\bibinfo
  {journal} {International Journal of Theoretical Physics}\ }\textbf {\bibinfo
  {volume} {53}},\ \bibinfo {pages} {2521} (\bibinfo {year}
  {2014})}\BibitemShut {NoStop}%
\bibitem [{\citenamefont {Kato}\ and\ \citenamefont
  {Brand{\~a}o}(2019)}]{kato2016quantum}%
  \BibitemOpen
  \bibfield  {author} {\bibinfo {author} {\bibfnamefont {K.}~\bibnamefont
  {Kato}}\ and\ \bibinfo {author} {\bibfnamefont {F.~G. S.~L.}\ \bibnamefont
  {Brand{\~a}o}},\ }\bibfield  {title} {\bibinfo {title} {{\it Quantum
  Approximate Markov Chains are Thermal}},\ }\bibfield  {journal} {\bibinfo
  {journal} {Communications in Mathematical Physics}\ }\href
  {https://doi.org/10.1007/s00220-019-03485-6} {10.1007/s00220-019-03485-6}
  (\bibinfo {year} {2019})\BibitemShut {NoStop}%
\bibitem [{\citenamefont {Kuwahara}\ \emph {et~al.}(2020)\citenamefont
  {Kuwahara}, \citenamefont {Kato},\ and\ \citenamefont
  {Brand\~ao}}]{PhysRevLett.124.220601}%
  \BibitemOpen
  \bibfield  {author} {\bibinfo {author} {\bibfnamefont {T.}~\bibnamefont
  {Kuwahara}}, \bibinfo {author} {\bibfnamefont {K.}~\bibnamefont {Kato}},\
  and\ \bibinfo {author} {\bibfnamefont {F.~G. S.~L.}\ \bibnamefont
  {Brand\~ao}},\ }\bibfield  {title} {\bibinfo {title} {{\it Clustering of
  Conditional Mutual Information for Quantum Gibbs States above a Threshold
  Temperature}},\ }\href {https://doi.org/10.1103/PhysRevLett.124.220601}
  {\bibfield  {journal} {\bibinfo  {journal} {Phys. Rev. Lett.}\ }\textbf
  {\bibinfo {volume} {124}},\ \bibinfo {pages} {220601} (\bibinfo {year}
  {2020})}\BibitemShut {NoStop}%
\bibitem [{\citenamefont {Kuwahara}\ \emph {et~al.}(2025)\citenamefont
  {Kuwahara}, \citenamefont {Kato},\ and\ \citenamefont
  {Brand\~ao}}]{Erratum2025}%
  \BibitemOpen
  \bibfield  {author} {\bibinfo {author} {\bibfnamefont {T.}~\bibnamefont
  {Kuwahara}}, \bibinfo {author} {\bibfnamefont {K.}~\bibnamefont {Kato}},\
  and\ \bibinfo {author} {\bibfnamefont {F.~G. S.~L.}\ \bibnamefont
  {Brand\~ao}},\ }\bibfield  {title} {\bibinfo {title} {Erratum: Clustering of
  conditional mutual information for quantum gibbs states above a threshold
  temperature [phys. rev. lett. 124, 220601 (2020)]},\ }\href
  {https://doi.org/10.1103/PhysRevLett.134.199901} {\bibfield  {journal}
  {\bibinfo  {journal} {Phys. Rev. Lett.}\ }\textbf {\bibinfo {volume} {134}},\
  \bibinfo {pages} {199901} (\bibinfo {year} {2025})}\BibitemShut {NoStop}%
\bibitem [{\citenamefont {Castelnovo}\ and\ \citenamefont
  {Chamon}(2008)}]{PhysRevB.78.155120}%
  \BibitemOpen
  \bibfield  {author} {\bibinfo {author} {\bibfnamefont {C.}~\bibnamefont
  {Castelnovo}}\ and\ \bibinfo {author} {\bibfnamefont {C.}~\bibnamefont
  {Chamon}},\ }\bibfield  {title} {\bibinfo {title} {{\it Topological order in
  a three-dimensional toric code at finite temperature}},\ }\href
  {https://doi.org/10.1103/PhysRevB.78.155120} {\bibfield  {journal} {\bibinfo
  {journal} {Phys. Rev. B}\ }\textbf {\bibinfo {volume} {78}},\ \bibinfo
  {pages} {155120} (\bibinfo {year} {2008})}\BibitemShut {NoStop}%
\bibitem [{\citenamefont {Hastings}(2011)}]{PhysRevLett.107.210501}%
  \BibitemOpen
  \bibfield  {author} {\bibinfo {author} {\bibfnamefont {M.~B.}\ \bibnamefont
  {Hastings}},\ }\bibfield  {title} {\bibinfo {title} {{\it Topological Order
  at Nonzero Temperature}},\ }\href
  {https://doi.org/10.1103/PhysRevLett.107.210501} {\bibfield  {journal}
  {\bibinfo  {journal} {Phys. Rev. Lett.}\ }\textbf {\bibinfo {volume} {107}},\
  \bibinfo {pages} {210501} (\bibinfo {year} {2011})}\BibitemShut {NoStop}%
\bibitem [{\citenamefont {Eldar}\ and\ \citenamefont {Harrow}(2017)}]{8104078}%
  \BibitemOpen
  \bibfield  {author} {\bibinfo {author} {\bibfnamefont {L.}~\bibnamefont
  {Eldar}}\ and\ \bibinfo {author} {\bibfnamefont {A.~W.}\ \bibnamefont
  {Harrow}},\ }\bibfield  {title} {\bibinfo {title} {{\it Local Hamiltonians
  Whose Ground States Are Hard to Approximate}},\ }in\ \href
  {https://doi.org/10.1109/FOCS.2017.46} {\emph {\bibinfo {booktitle} {2017
  IEEE 58th Annual Symposium on Foundations of Computer Science (FOCS)}}}\
  (\bibinfo {year} {2017})\ pp.\ \bibinfo {pages} {427--438}\BibitemShut
  {NoStop}%
\bibitem [{\citenamefont {Anshu}\ \emph {et~al.}(2023)\citenamefont {Anshu},
  \citenamefont {Breuckmann},\ and\ \citenamefont
  {Nirkhe}}]{10.1145/3564246.3585114}%
  \BibitemOpen
  \bibfield  {author} {\bibinfo {author} {\bibfnamefont {A.}~\bibnamefont
  {Anshu}}, \bibinfo {author} {\bibfnamefont {N.~P.}\ \bibnamefont
  {Breuckmann}},\ and\ \bibinfo {author} {\bibfnamefont {C.}~\bibnamefont
  {Nirkhe}},\ }\bibfield  {title} {\bibinfo {title} {{\it NLTS Hamiltonians
  from Good Quantum Codes}},\ }in\ \href
  {https://doi.org/10.1145/3564246.3585114} {\emph {\bibinfo {booktitle}
  {Proceedings of the 55th Annual ACM Symposium on Theory of Computing}}},\
  \bibinfo {series and number} {STOC 2023}\ (\bibinfo  {publisher} {Association
  for Computing Machinery},\ \bibinfo {address} {New York, NY, USA},\ \bibinfo
  {year} {2023})\ p.\ \bibinfo {pages} {1090–1096}\BibitemShut {NoStop}%
\bibitem [{\citenamefont {Ackley}\ \emph {et~al.}(1985)\citenamefont {Ackley},
  \citenamefont {Hinton},\ and\ \citenamefont {Sejnowski}}]{ACKLEY1985147}%
  \BibitemOpen
  \bibfield  {author} {\bibinfo {author} {\bibfnamefont {D.~H.}\ \bibnamefont
  {Ackley}}, \bibinfo {author} {\bibfnamefont {G.~E.}\ \bibnamefont {Hinton}},\
  and\ \bibinfo {author} {\bibfnamefont {T.~J.}\ \bibnamefont {Sejnowski}},\
  }\bibfield  {title} {\bibinfo {title} {{\it A learning algorithm for
  boltzmann machines}},\ }\href
  {https://doi.org/https://doi.org/10.1016/S0364-0213(85)80012-4} {\bibfield
  {journal} {\bibinfo  {journal} {Cognitive Science}\ }\textbf {\bibinfo
  {volume} {9}},\ \bibinfo {pages} {147} (\bibinfo {year} {1985})}\BibitemShut
  {NoStop}%
\bibitem [{\citenamefont {Salakhutdinov}\ and\ \citenamefont
  {Hinton}(2009)}]{pmlr-v5-salakhutdinov09a}%
  \BibitemOpen
  \bibfield  {author} {\bibinfo {author} {\bibfnamefont {R.}~\bibnamefont
  {Salakhutdinov}}\ and\ \bibinfo {author} {\bibfnamefont {G.}~\bibnamefont
  {Hinton}},\ }\bibfield  {title} {\bibinfo {title} {{\it Deep Boltzmann
  Machines}},\ }in\ \href
  {https://proceedings.mlr.press/v5/salakhutdinov09a.html} {\emph {\bibinfo
  {booktitle} {Proceedings of the Twelth International Conference on Artificial
  Intelligence and Statistics}}},\ \bibinfo {series} {Proceedings of Machine
  Learning Research}, Vol.~\bibinfo {volume} {5},\ \bibinfo {editor} {edited
  by\ \bibinfo {editor} {\bibfnamefont {D.}~\bibnamefont {van Dyk}}\ and\
  \bibinfo {editor} {\bibfnamefont {M.}~\bibnamefont {Welling}}}\ (\bibinfo
  {publisher} {PMLR},\ \bibinfo {address} {Hilton Clearwater Beach Resort,
  Clearwater Beach, Florida USA},\ \bibinfo {year} {2009})\ pp.\ \bibinfo
  {pages} {448--455}\BibitemShut {NoStop}%
\bibitem [{\citenamefont {Anshu}\ \emph
  {et~al.}(2021{\natexlab{a}})\citenamefont {Anshu}, \citenamefont
  {Arunachalam}, \citenamefont {Kuwahara},\ and\ \citenamefont
  {Soleimanifar}}]{Anshu_2021}%
  \BibitemOpen
  \bibfield  {author} {\bibinfo {author} {\bibfnamefont {A.}~\bibnamefont
  {Anshu}}, \bibinfo {author} {\bibfnamefont {S.}~\bibnamefont {Arunachalam}},
  \bibinfo {author} {\bibfnamefont {T.}~\bibnamefont {Kuwahara}},\ and\
  \bibinfo {author} {\bibfnamefont {M.}~\bibnamefont {Soleimanifar}},\
  }\bibfield  {title} {\bibinfo {title} {{\it Sample-efficient learning of
  interacting quantum systems}},\ }\href
  {https://doi.org/10.1038/s41567-021-01232-0} {\bibfield  {journal} {\bibinfo
  {journal} {Nature Physics}\ }\textbf {\bibinfo {volume} {17}},\ \bibinfo
  {pages} {931} (\bibinfo {year} {2021}{\natexlab{a}})}\BibitemShut {NoStop}%
\bibitem [{\citenamefont {Anshu}\ \emph
  {et~al.}(2021{\natexlab{b}})\citenamefont {Anshu}, \citenamefont
  {Arunachalam}, \citenamefont {Kuwahara},\ and\ \citenamefont
  {Soleimanifar}}]{anshu2021efficient}%
  \BibitemOpen
  \bibfield  {author} {\bibinfo {author} {\bibfnamefont {A.}~\bibnamefont
  {Anshu}}, \bibinfo {author} {\bibfnamefont {S.}~\bibnamefont {Arunachalam}},
  \bibinfo {author} {\bibfnamefont {T.}~\bibnamefont {Kuwahara}},\ and\
  \bibinfo {author} {\bibfnamefont {M.}~\bibnamefont {Soleimanifar}},\
  }\bibfield  {title} {\bibinfo {title} {{\it Efficient learning of commuting
  Hamiltonians on lattices}},\ }\href
  {https://anuraganshu.seas.harvard.edu/sites/g/files/omnuum6156/files/anshu/files/learning_commuting_hamiltonian.pdf}
  {\bibfield  {journal} {\bibinfo  {journal} {Electronic notes}\ }\textbf
  {\bibinfo {volume} {25}} (\bibinfo {year} {2021}{\natexlab{b}})}\BibitemShut
  {NoStop}%
\bibitem [{\citenamefont {Poulin}\ and\ \citenamefont
  {Wocjan}(2009)}]{PhysRevLett.103.220502}%
  \BibitemOpen
  \bibfield  {author} {\bibinfo {author} {\bibfnamefont {D.}~\bibnamefont
  {Poulin}}\ and\ \bibinfo {author} {\bibfnamefont {P.}~\bibnamefont
  {Wocjan}},\ }\bibfield  {title} {\bibinfo {title} {{\it Sampling from the
  Thermal Quantum Gibbs State and Evaluating Partition Functions with a Quantum
  Computer}},\ }\href {https://doi.org/10.1103/PhysRevLett.103.220502}
  {\bibfield  {journal} {\bibinfo  {journal} {Phys. Rev. Lett.}\ }\textbf
  {\bibinfo {volume} {103}},\ \bibinfo {pages} {220502} (\bibinfo {year}
  {2009})}\BibitemShut {NoStop}%
\bibitem [{\citenamefont {Brand{\~a}o}\ and\ \citenamefont
  {Kastoryano}(2019)}]{Brandao2019}%
  \BibitemOpen
  \bibfield  {author} {\bibinfo {author} {\bibfnamefont {F.~G. S.~L.}\
  \bibnamefont {Brand{\~a}o}}\ and\ \bibinfo {author} {\bibfnamefont {M.~J.}\
  \bibnamefont {Kastoryano}},\ }\bibfield  {title} {\bibinfo {title} {{\it
  Finite Correlation Length Implies Efficient Preparation of Quantum Thermal
  States}},\ }\href {https://doi.org/10.1007/s00220-018-3150-8} {\bibfield
  {journal} {\bibinfo  {journal} {Communications in Mathematical Physics}\
  }\textbf {\bibinfo {volume} {365}},\ \bibinfo {pages} {1} (\bibinfo {year}
  {2019})}\BibitemShut {NoStop}%
\bibitem [{\citenamefont {Kim}(2016)}]{kim2016markovian}%
  \BibitemOpen
  \bibfield  {author} {\bibinfo {author} {\bibfnamefont {I.~H.}\ \bibnamefont
  {Kim}},\ }\href@noop {} {\bibinfo {title} {{\it Markovian Marignals}}}
  (\bibinfo {year} {2016}),\ \Eprint {https://arxiv.org/abs/1609.08579}
  {arXiv:1609.08579 [quant-ph]} \BibitemShut {NoStop}%
\bibitem [{\citenamefont {Kim}(2021)}]{PhysRevX.11.021039}%
  \BibitemOpen
  \bibfield  {author} {\bibinfo {author} {\bibfnamefont {I.~H.}\ \bibnamefont
  {Kim}},\ }\bibfield  {title} {\bibinfo {title} {{\it Entropy Scaling Law and
  the Quantum Marginal Problem}},\ }\href
  {https://doi.org/10.1103/PhysRevX.11.021039} {\bibfield  {journal} {\bibinfo
  {journal} {Phys. Rev. X}\ }\textbf {\bibinfo {volume} {11}},\ \bibinfo
  {pages} {021039} (\bibinfo {year} {2021})}\BibitemShut {NoStop}%
\bibitem [{\citenamefont {Lauritzen}(1996)}]{lauritzen1996graphical}%
  \BibitemOpen
  \bibfield  {author} {\bibinfo {author} {\bibfnamefont {S.~L.}\ \bibnamefont
  {Lauritzen}},\ }\href@noop {} {\emph {\bibinfo {title} {{\it Graphical
  models}}}},\ Vol.~\bibinfo {volume} {17}\ (\bibinfo  {publisher} {Clarendon
  Press},\ \bibinfo {year} {1996})\BibitemShut {NoStop}%
\bibitem [{\citenamefont {Koller}\ and\ \citenamefont
  {Friedman}(2009)}]{koller2009probabilistic}%
  \BibitemOpen
  \bibfield  {author} {\bibinfo {author} {\bibfnamefont {D.}~\bibnamefont
  {Koller}}\ and\ \bibinfo {author} {\bibfnamefont {N.}~\bibnamefont
  {Friedman}},\ }\href@noop {} {\emph {\bibinfo {title} {{\it Probabilistic
  graphical models: principles and techniques}}}}\ (\bibinfo  {publisher} {MIT
  press},\ \bibinfo {year} {2009})\BibitemShut {NoStop}%
\bibitem [{Sup()}]{Supplement_CMI}%
  \BibitemOpen
  \href@noop {} {}See Supplemental Material [url] for the technical details,
  which includes
  Refs.~\cite{Kastoryano2016,kim2017markovian,lami2018nonclassical,berta2023entanglement,Nachtergaele2006,PhysRevLett.114.157201,chen2019finite,PhysRevX.10.031010,PhysRevLett.127.160401,Scharf_1988,10.1145/2897518.2897585,Phd_kuwahara,10129917,Lindblad1974,Uhlmann1977,10.5555/2871378.2871383,doi:10.1063/1.2044667,schur1923uber,Horn1954,10.5555/2011326.2011331,kuwahara2022optimal,pinelis2020exact,PhysRevA.80.022316,PhysRevLett.103.020506,Hern_ndez_Santana_2015,Winter2016,doi101098rspa20150623,bluhm2021exponential,10.1063/1.4987135,10.1063/1.4936209,PhysRevX.10.031064,PhysRevLett.124.100401,Huang2020,kuwahara2018polynomialtime,adlertaylor,haber2018notes}.\BibitemShut
  {Stop}%
\bibitem [{\citenamefont {Svetlichnyy}\ and\ \citenamefont
  {Kennedy}(2022)}]{10.1063/5.0085358}%
  \BibitemOpen
  \bibfield  {author} {\bibinfo {author} {\bibfnamefont {P.}~\bibnamefont
  {Svetlichnyy}}\ and\ \bibinfo {author} {\bibfnamefont {T.~A.~B.}\
  \bibnamefont {Kennedy}},\ }\bibfield  {title} {\bibinfo {title} {{\it Decay
  of quantum conditional mutual information for purely generated finitely
  correlated states}},\ }\href {https://doi.org/10.1063/5.0085358} {\bibfield
  {journal} {\bibinfo  {journal} {Journal of Mathematical Physics}\ }\textbf
  {\bibinfo {volume} {63}},\ \bibinfo {pages} {072201} (\bibinfo {year}
  {2022})}\BibitemShut {NoStop}%
\bibitem [{\citenamefont {Svetlichnyy}\ \emph {et~al.}(2024)\citenamefont
  {Svetlichnyy}, \citenamefont {Mittal},\ and\ \citenamefont
  {Kennedy}}]{svetlichnyy2022matrix}%
  \BibitemOpen
  \bibfield  {author} {\bibinfo {author} {\bibfnamefont {P.}~\bibnamefont
  {Svetlichnyy}}, \bibinfo {author} {\bibfnamefont {S.}~\bibnamefont
  {Mittal}},\ and\ \bibinfo {author} {\bibfnamefont {T.~A.~B.}\ \bibnamefont
  {Kennedy}},\ }\bibfield  {title} {\bibinfo {title} {{\it Matrix product
  states and the decay of quantum conditional mutual information}},\ }\href
  {https://doi.org/10.1063/5.0152063} {\bibfield  {journal} {\bibinfo
  {journal} {Journal of Mathematical Physics}\ }\textbf {\bibinfo {volume}
  {65}},\ \bibinfo {pages} {022201} (\bibinfo {year} {2024})}\BibitemShut
  {NoStop}%
\bibitem [{\citenamefont {Chen}\ \emph {et~al.}(2023)\citenamefont {Chen},
  \citenamefont {Kato},\ and\ \citenamefont {Brandão}}]{chen2023matrix}%
  \BibitemOpen
  \bibfield  {author} {\bibinfo {author} {\bibfnamefont {C.-F.}\ \bibnamefont
  {Chen}}, \bibinfo {author} {\bibfnamefont {K.}~\bibnamefont {Kato}},\ and\
  \bibinfo {author} {\bibfnamefont {F.~G. S.~L.}\ \bibnamefont {Brandão}},\
  }\href@noop {} {\bibinfo {title} {{\it Matrix Product Density Operators: when
  do they have a local parent Hamiltonian?}}} (\bibinfo {year} {2023}),\
  \Eprint {https://arxiv.org/abs/2010.14682} {arXiv:2010.14682} \BibitemShut
  {NoStop}%
\bibitem [{\citenamefont {Kimura}\ and\ \citenamefont
  {Kuwahara}(2024)}]{kimura2024clustering}%
  \BibitemOpen
  \bibfield  {author} {\bibinfo {author} {\bibfnamefont {Y.}~\bibnamefont
  {Kimura}}\ and\ \bibinfo {author} {\bibfnamefont {T.}~\bibnamefont
  {Kuwahara}},\ }\href@noop {} {\bibinfo {title} {{\it Clustering theorem in 1D
  long-range interacting systems at arbitrary temperatures}}} (\bibinfo {year}
  {2024}),\ \Eprint {https://arxiv.org/abs/2403.11431} {arXiv:2403.11431
  [quant-ph]} \BibitemShut {NoStop}%
\bibitem [{DK_()}]{DK_com}%
  \BibitemOpen
  \href@noop {} {\bibinfo  {journal} {Private communication with Donghoon Kim}\
  }\BibitemShut {NoStop}%
\bibitem [{\citenamefont {Gondolf}\ \emph {et~al.}(2024)\citenamefont
  {Gondolf}, \citenamefont {Scalet}, \citenamefont {de~Alarcon}, \citenamefont
  {Alhambra},\ and\ \citenamefont {Capel}}]{gondolf2024conditional}%
  \BibitemOpen
\bibfield  {journal} {  }\bibfield  {author} {\bibinfo {author} {\bibfnamefont
  {P.}~\bibnamefont {Gondolf}}, \bibinfo {author} {\bibfnamefont {S.~O.}\
  \bibnamefont {Scalet}}, \bibinfo {author} {\bibfnamefont {A.~R.}\
  \bibnamefont {de~Alarcon}}, \bibinfo {author} {\bibfnamefont {A.~M.}\
  \bibnamefont {Alhambra}},\ and\ \bibinfo {author} {\bibfnamefont
  {A.}~\bibnamefont {Capel}},\ }\href@noop {} {\bibinfo {title} {{\it
  Conditional Independence of 1D Gibbs States with Applications to Efficient
  Learning}}} (\bibinfo {year} {2024}),\ \Eprint
  {https://arxiv.org/abs/2402.18500} {arXiv:2402.18500} \BibitemShut {NoStop}%
\bibitem [{\citenamefont {Belavkin}(1982)}]{Belavkin1982}%
  \BibitemOpen
  \bibfield  {author} {\bibinfo {author} {\bibfnamefont {S.~P.}\ \bibnamefont
  {Belavkin}, \bibfnamefont {V.~P.}},\ }\bibfield  {title} {\bibinfo {title}
  {{\it $C^\ast$-algebraic generalization of relative entropy and entropy}},\
  }\href {http://eudml.org/doc/76163} {\bibfield  {journal} {\bibinfo
  {journal} {Annales de l'I.H.P. Physique théorique}\ }\textbf {\bibinfo
  {volume} {37}},\ \bibinfo {pages} {51} (\bibinfo {year} {1982})}\BibitemShut
  {NoStop}%
\bibitem [{\citenamefont {Horodecki}\ \emph {et~al.}(1998)\citenamefont
  {Horodecki}, \citenamefont {Horodecki},\ and\ \citenamefont
  {Horodecki}}]{PhysRevLett.80.5239}%
  \BibitemOpen
  \bibfield  {author} {\bibinfo {author} {\bibfnamefont {M.}~\bibnamefont
  {Horodecki}}, \bibinfo {author} {\bibfnamefont {P.}~\bibnamefont
  {Horodecki}},\ and\ \bibinfo {author} {\bibfnamefont {R.}~\bibnamefont
  {Horodecki}},\ }\bibfield  {title} {\bibinfo {title} {{\it Mixed-State
  Entanglement and Distillation: Is there a ``Bound'' Entanglement in
  Nature?}},\ }\href {https://doi.org/10.1103/PhysRevLett.80.5239} {\bibfield
  {journal} {\bibinfo  {journal} {Phys. Rev. Lett.}\ }\textbf {\bibinfo
  {volume} {80}},\ \bibinfo {pages} {5239} (\bibinfo {year}
  {1998})}\BibitemShut {NoStop}%
\bibitem [{\citenamefont {Horodecki}\ \emph {et~al.}(1999)\citenamefont
  {Horodecki}, \citenamefont {Horodecki},\ and\ \citenamefont
  {Horodecki}}]{PhysRevLett.82.1056}%
  \BibitemOpen
  \bibfield  {author} {\bibinfo {author} {\bibfnamefont {P.}~\bibnamefont
  {Horodecki}}, \bibinfo {author} {\bibfnamefont {M.}~\bibnamefont
  {Horodecki}},\ and\ \bibinfo {author} {\bibfnamefont {R.}~\bibnamefont
  {Horodecki}},\ }\bibfield  {title} {\bibinfo {title} {{\it Bound Entanglement
  Can Be Activated}},\ }\href {https://doi.org/10.1103/PhysRevLett.82.1056}
  {\bibfield  {journal} {\bibinfo  {journal} {Phys. Rev. Lett.}\ }\textbf
  {\bibinfo {volume} {82}},\ \bibinfo {pages} {1056} (\bibinfo {year}
  {1999})}\BibitemShut {NoStop}%
\bibitem [{\citenamefont {Audenaert}\ \emph {et~al.}(2001)\citenamefont
  {Audenaert}, \citenamefont {Eisert}, \citenamefont {Jan\'e}, \citenamefont
  {Plenio}, \citenamefont {Virmani},\ and\ \citenamefont
  {De~Moor}}]{PhysRevLett.87.217902}%
  \BibitemOpen
  \bibfield  {author} {\bibinfo {author} {\bibfnamefont {K.}~\bibnamefont
  {Audenaert}}, \bibinfo {author} {\bibfnamefont {J.}~\bibnamefont {Eisert}},
  \bibinfo {author} {\bibfnamefont {E.}~\bibnamefont {Jan\'e}}, \bibinfo
  {author} {\bibfnamefont {M.~B.}\ \bibnamefont {Plenio}}, \bibinfo {author}
  {\bibfnamefont {S.}~\bibnamefont {Virmani}},\ and\ \bibinfo {author}
  {\bibfnamefont {B.}~\bibnamefont {De~Moor}},\ }\bibfield  {title} {\bibinfo
  {title} {{\it Asymptotic Relative Entropy of Entanglement}},\ }\href
  {https://doi.org/10.1103/PhysRevLett.87.217902} {\bibfield  {journal}
  {\bibinfo  {journal} {Phys. Rev. Lett.}\ }\textbf {\bibinfo {volume} {87}},\
  \bibinfo {pages} {217902} (\bibinfo {year} {2001})}\BibitemShut {NoStop}%
\bibitem [{\citenamefont {Audenaert}\ \emph {et~al.}(2002)\citenamefont
  {Audenaert}, \citenamefont {De~Moor}, \citenamefont {Vollbrecht},\ and\
  \citenamefont {Werner}}]{PhysRevA.66.032310}%
  \BibitemOpen
  \bibfield  {author} {\bibinfo {author} {\bibfnamefont {K.}~\bibnamefont
  {Audenaert}}, \bibinfo {author} {\bibfnamefont {B.}~\bibnamefont {De~Moor}},
  \bibinfo {author} {\bibfnamefont {K.~G.~H.}\ \bibnamefont {Vollbrecht}},\
  and\ \bibinfo {author} {\bibfnamefont {R.~F.}\ \bibnamefont {Werner}},\
  }\bibfield  {title} {\bibinfo {title} {{\it Asymptotic relative entropy of
  entanglement for orthogonally invariant states}},\ }\href
  {https://doi.org/10.1103/PhysRevA.66.032310} {\bibfield  {journal} {\bibinfo
  {journal} {Phys. Rev. A}\ }\textbf {\bibinfo {volume} {66}},\ \bibinfo
  {pages} {032310} (\bibinfo {year} {2002})}\BibitemShut {NoStop}%
\bibitem [{\citenamefont {Miranowicz}\ and\ \citenamefont
  {Ishizaka}(2008)}]{PhysRevA.78.032310}%
  \BibitemOpen
  \bibfield  {author} {\bibinfo {author} {\bibfnamefont {A.}~\bibnamefont
  {Miranowicz}}\ and\ \bibinfo {author} {\bibfnamefont {S.}~\bibnamefont
  {Ishizaka}},\ }\bibfield  {title} {\bibinfo {title} {{\it Closed formula for
  the relative entropy of entanglement}},\ }\href
  {https://doi.org/10.1103/PhysRevA.78.032310} {\bibfield  {journal} {\bibinfo
  {journal} {Phys. Rev. A}\ }\textbf {\bibinfo {volume} {78}},\ \bibinfo
  {pages} {032310} (\bibinfo {year} {2008})}\BibitemShut {NoStop}%
\bibitem [{\citenamefont {Girard}\ \emph {et~al.}(2014)\citenamefont {Girard},
  \citenamefont {Gour},\ and\ \citenamefont {Friedland}}]{Girard_2014}%
  \BibitemOpen
  \bibfield  {author} {\bibinfo {author} {\bibfnamefont {M.~W.}\ \bibnamefont
  {Girard}}, \bibinfo {author} {\bibfnamefont {G.}~\bibnamefont {Gour}},\ and\
  \bibinfo {author} {\bibfnamefont {S.}~\bibnamefont {Friedland}},\ }\bibfield
  {title} {\bibinfo {title} {{\it On convex optimization problems in quantum
  information theory}},\ }\href
  {https://doi.org/10.1088/1751-8113/47/50/505302} {\bibfield  {journal}
  {\bibinfo  {journal} {Journal of Physics A: Mathematical and Theoretical}\
  }\textbf {\bibinfo {volume} {47}},\ \bibinfo {pages} {505302} (\bibinfo
  {year} {2014})}\BibitemShut {NoStop}%
\bibitem [{\citenamefont {Vidal}\ and\ \citenamefont
  {Werner}(2002)}]{PhysRevA.65.032314}%
  \BibitemOpen
  \bibfield  {author} {\bibinfo {author} {\bibfnamefont {G.}~\bibnamefont
  {Vidal}}\ and\ \bibinfo {author} {\bibfnamefont {R.~F.}\ \bibnamefont
  {Werner}},\ }\bibfield  {title} {\bibinfo {title} {{\it Computable measure of
  entanglement}},\ }\href {https://doi.org/10.1103/PhysRevA.65.032314}
  {\bibfield  {journal} {\bibinfo  {journal} {Phys. Rev. A}\ }\textbf {\bibinfo
  {volume} {65}},\ \bibinfo {pages} {032314} (\bibinfo {year}
  {2002})}\BibitemShut {NoStop}%
\bibitem [{\citenamefont {Brand{\~a}o}\ \emph {et~al.}(2011)\citenamefont
  {Brand{\~a}o}, \citenamefont {Christandl},\ and\ \citenamefont
  {Yard}}]{Brandao2011}%
  \BibitemOpen
  \bibfield  {author} {\bibinfo {author} {\bibfnamefont {F.~G. S.~L.}\
  \bibnamefont {Brand{\~a}o}}, \bibinfo {author} {\bibfnamefont
  {M.}~\bibnamefont {Christandl}},\ and\ \bibinfo {author} {\bibfnamefont
  {J.}~\bibnamefont {Yard}},\ }\bibfield  {title} {\bibinfo {title} {{\it
  Faithful Squashed Entanglement}},\ }\href
  {https://doi.org/10.1007/s00220-011-1302-1} {\bibfield  {journal} {\bibinfo
  {journal} {Communications in Mathematical Physics}\ }\textbf {\bibinfo
  {volume} {306}},\ \bibinfo {pages} {805} (\bibinfo {year}
  {2011})}\BibitemShut {NoStop}%
\bibitem [{\citenamefont {Li}\ and\ \citenamefont {Winter}(2014)}]{Li2014}%
  \BibitemOpen
  \bibfield  {author} {\bibinfo {author} {\bibfnamefont {K.}~\bibnamefont
  {Li}}\ and\ \bibinfo {author} {\bibfnamefont {A.}~\bibnamefont {Winter}},\
  }\bibfield  {title} {\bibinfo {title} {{\it Relative Entropy and Squashed
  Entanglement}},\ }\href {https://doi.org/10.1007/s00220-013-1871-2}
  {\bibfield  {journal} {\bibinfo  {journal} {Communications in Mathematical
  Physics}\ }\textbf {\bibinfo {volume} {326}},\ \bibinfo {pages} {63}
  (\bibinfo {year} {2014})}\BibitemShut {NoStop}%
\bibitem [{\citenamefont {Jeng}\ \emph {et~al.}(2019)\citenamefont {Jeng},
  \citenamefont {Tserkis}, \citenamefont {Haw}, \citenamefont {Chrzanowski},
  \citenamefont {Janousek}, \citenamefont {Ralph}, \citenamefont {Lam},\ and\
  \citenamefont {Assad}}]{PhysRevA.99.042304}%
  \BibitemOpen
  \bibfield  {author} {\bibinfo {author} {\bibfnamefont {H.}~\bibnamefont
  {Jeng}}, \bibinfo {author} {\bibfnamefont {S.}~\bibnamefont {Tserkis}},
  \bibinfo {author} {\bibfnamefont {J.~Y.}\ \bibnamefont {Haw}}, \bibinfo
  {author} {\bibfnamefont {H.~M.}\ \bibnamefont {Chrzanowski}}, \bibinfo
  {author} {\bibfnamefont {J.}~\bibnamefont {Janousek}}, \bibinfo {author}
  {\bibfnamefont {T.~C.}\ \bibnamefont {Ralph}}, \bibinfo {author}
  {\bibfnamefont {P.~K.}\ \bibnamefont {Lam}},\ and\ \bibinfo {author}
  {\bibfnamefont {S.~M.}\ \bibnamefont {Assad}},\ }\bibfield  {title} {\bibinfo
  {title} {{\it Entanglement properties of a measurement-based entanglement
  distillation experiment}},\ }\href
  {https://doi.org/10.1103/PhysRevA.99.042304} {\bibfield  {journal} {\bibinfo
  {journal} {Phys. Rev. A}\ }\textbf {\bibinfo {volume} {99}},\ \bibinfo
  {pages} {042304} (\bibinfo {year} {2019})}\BibitemShut {NoStop}%
\bibitem [{\citenamefont {Vedral}\ \emph {et~al.}(1997)\citenamefont {Vedral},
  \citenamefont {Plenio}, \citenamefont {Rippin},\ and\ \citenamefont
  {Knight}}]{PhysRevLett.78.2275}%
  \BibitemOpen
  \bibfield  {author} {\bibinfo {author} {\bibfnamefont {V.}~\bibnamefont
  {Vedral}}, \bibinfo {author} {\bibfnamefont {M.~B.}\ \bibnamefont {Plenio}},
  \bibinfo {author} {\bibfnamefont {M.~A.}\ \bibnamefont {Rippin}},\ and\
  \bibinfo {author} {\bibfnamefont {P.~L.}\ \bibnamefont {Knight}},\ }\bibfield
   {title} {\bibinfo {title} {{\it Quantifying Entanglement}},\ }\href
  {https://doi.org/10.1103/PhysRevLett.78.2275} {\bibfield  {journal} {\bibinfo
   {journal} {Phys. Rev. Lett.}\ }\textbf {\bibinfo {volume} {78}},\ \bibinfo
  {pages} {2275} (\bibinfo {year} {1997})}\BibitemShut {NoStop}%
\bibitem [{\citenamefont {Bennett}\ \emph {et~al.}(1996)\citenamefont
  {Bennett}, \citenamefont {DiVincenzo}, \citenamefont {Smolin},\ and\
  \citenamefont {Wootters}}]{PhysRevA.54.3824}%
  \BibitemOpen
  \bibfield  {author} {\bibinfo {author} {\bibfnamefont {C.~H.}\ \bibnamefont
  {Bennett}}, \bibinfo {author} {\bibfnamefont {D.~P.}\ \bibnamefont
  {DiVincenzo}}, \bibinfo {author} {\bibfnamefont {J.~A.}\ \bibnamefont
  {Smolin}},\ and\ \bibinfo {author} {\bibfnamefont {W.~K.}\ \bibnamefont
  {Wootters}},\ }\bibfield  {title} {\bibinfo {title} {{\it Mixed-state
  entanglement and quantum error correction}},\ }\href
  {https://doi.org/10.1103/PhysRevA.54.3824} {\bibfield  {journal} {\bibinfo
  {journal} {Phys. Rev. A}\ }\textbf {\bibinfo {volume} {54}},\ \bibinfo
  {pages} {3824} (\bibinfo {year} {1996})}\BibitemShut {NoStop}%
\bibitem [{\citenamefont {Li}\ and\ \citenamefont
  {Haldane}(2008)}]{PhysRevLett.101.010504}%
  \BibitemOpen
  \bibfield  {author} {\bibinfo {author} {\bibfnamefont {H.}~\bibnamefont
  {Li}}\ and\ \bibinfo {author} {\bibfnamefont {F.~D.~M.}\ \bibnamefont
  {Haldane}},\ }\bibfield  {title} {\bibinfo {title} {{\it Entanglement
  Spectrum as a Generalization of Entanglement Entropy: Identification of
  Topological Order in Non-Abelian Fractional Quantum Hall Effect States}},\
  }\href {https://doi.org/10.1103/PhysRevLett.101.010504} {\bibfield  {journal}
  {\bibinfo  {journal} {Phys. Rev. Lett.}\ }\textbf {\bibinfo {volume} {101}},\
  \bibinfo {pages} {010504} (\bibinfo {year} {2008})}\BibitemShut {NoStop}%
\bibitem [{\citenamefont {Peschel}\ and\ \citenamefont
  {Eisler}(2009)}]{Peschel_2009}%
  \BibitemOpen
  \bibfield  {author} {\bibinfo {author} {\bibfnamefont {I.}~\bibnamefont
  {Peschel}}\ and\ \bibinfo {author} {\bibfnamefont {V.}~\bibnamefont
  {Eisler}},\ }\bibfield  {title} {\bibinfo {title} {{\it Reduced density
  matrices and entanglement entropy in free lattice models}},\ }\href
  {https://doi.org/10.1088/1751-8113/42/50/504003} {\bibfield  {journal}
  {\bibinfo  {journal} {Journal of Physics A: Mathematical and Theoretical}\
  }\textbf {\bibinfo {volume} {42}},\ \bibinfo {pages} {504003} (\bibinfo
  {year} {2009})}\BibitemShut {NoStop}%
\bibitem [{\citenamefont {Bilgin}\ and\ \citenamefont
  {Poulin}(2010)}]{PhysRevB.81.054106}%
  \BibitemOpen
  \bibfield  {author} {\bibinfo {author} {\bibfnamefont {E.}~\bibnamefont
  {Bilgin}}\ and\ \bibinfo {author} {\bibfnamefont {D.}~\bibnamefont
  {Poulin}},\ }\bibfield  {title} {\bibinfo {title} {{\it Coarse-grained belief
  propagation for simulation of interacting quantum systems at all
  temperatures}},\ }\href {https://doi.org/10.1103/PhysRevB.81.054106}
  {\bibfield  {journal} {\bibinfo  {journal} {Phys. Rev. B}\ }\textbf {\bibinfo
  {volume} {81}},\ \bibinfo {pages} {054106} (\bibinfo {year}
  {2010})}\BibitemShut {NoStop}%
\bibitem [{\citenamefont {Parisen~Toldin}\ and\ \citenamefont
  {Assaad}(2018)}]{PhysRevLett.121.200602}%
  \BibitemOpen
  \bibfield  {author} {\bibinfo {author} {\bibfnamefont {F.}~\bibnamefont
  {Parisen~Toldin}}\ and\ \bibinfo {author} {\bibfnamefont {F.~F.}\
  \bibnamefont {Assaad}},\ }\bibfield  {title} {\bibinfo {title} {{\it
  Entanglement Hamiltonian of Interacting Fermionic Models}},\ }\href
  {https://doi.org/10.1103/PhysRevLett.121.200602} {\bibfield  {journal}
  {\bibinfo  {journal} {Phys. Rev. Lett.}\ }\textbf {\bibinfo {volume} {121}},\
  \bibinfo {pages} {200602} (\bibinfo {year} {2018})}\BibitemShut {NoStop}%
\bibitem [{\citenamefont {Eisler}\ \emph {et~al.}(2020)\citenamefont {Eisler},
  \citenamefont {Giulio}, \citenamefont {Tonni},\ and\ \citenamefont
  {Peschel}}]{Eisler_2020}%
  \BibitemOpen
  \bibfield  {author} {\bibinfo {author} {\bibfnamefont {V.}~\bibnamefont
  {Eisler}}, \bibinfo {author} {\bibfnamefont {G.~D.}\ \bibnamefont {Giulio}},
  \bibinfo {author} {\bibfnamefont {E.}~\bibnamefont {Tonni}},\ and\ \bibinfo
  {author} {\bibfnamefont {I.}~\bibnamefont {Peschel}},\ }\bibfield  {title}
  {\bibinfo {title} {{\it Entanglement Hamiltonians for non-critical quantum
  chains}},\ }\href {https://doi.org/10.1088/1742-5468/abb4da} {\bibfield
  {journal} {\bibinfo  {journal} {Journal of Statistical Mechanics: Theory and
  Experiment}\ }\textbf {\bibinfo {volume} {2020}},\ \bibinfo {pages} {103102}
  (\bibinfo {year} {2020})}\BibitemShut {NoStop}%
\bibitem [{\citenamefont {Kokail}\ \emph {et~al.}(2021)\citenamefont {Kokail},
  \citenamefont {van Bijnen}, \citenamefont {Elben}, \citenamefont
  {Vermersch},\ and\ \citenamefont {Zoller}}]{Kokail2021}%
  \BibitemOpen
  \bibfield  {author} {\bibinfo {author} {\bibfnamefont {C.}~\bibnamefont
  {Kokail}}, \bibinfo {author} {\bibfnamefont {R.}~\bibnamefont {van Bijnen}},
  \bibinfo {author} {\bibfnamefont {A.}~\bibnamefont {Elben}}, \bibinfo
  {author} {\bibfnamefont {B.}~\bibnamefont {Vermersch}},\ and\ \bibinfo
  {author} {\bibfnamefont {P.}~\bibnamefont {Zoller}},\ }\bibfield  {title}
  {\bibinfo {title} {{\it Entanglement Hamiltonian tomography in quantum
  simulation}},\ }\href {https://doi.org/10.1038/s41567-021-01260-w} {\bibfield
   {journal} {\bibinfo  {journal} {Nature Physics}\ }\textbf {\bibinfo {volume}
  {17}},\ \bibinfo {pages} {936} (\bibinfo {year} {2021})}\BibitemShut
  {NoStop}%
\bibitem [{\citenamefont {Zache}\ \emph {et~al.}(2022)\citenamefont {Zache},
  \citenamefont {Kokail}, \citenamefont {Sundar},\ and\ \citenamefont
  {Zoller}}]{Zache2022entanglement}%
  \BibitemOpen
  \bibfield  {author} {\bibinfo {author} {\bibfnamefont {T.~V.}\ \bibnamefont
  {Zache}}, \bibinfo {author} {\bibfnamefont {C.}~\bibnamefont {Kokail}},
  \bibinfo {author} {\bibfnamefont {B.}~\bibnamefont {Sundar}},\ and\ \bibinfo
  {author} {\bibfnamefont {P.}~\bibnamefont {Zoller}},\ }\bibfield  {title}
  {\bibinfo {title} {{\it Entanglement {S}pectroscopy and probing the
  {L}i-{H}aldane {C}onjecturein {T}opological {Q}uantum {M}atter}},\ }\href
  {https://doi.org/10.22331/q-2022-04-27-702} {\bibfield  {journal} {\bibinfo
  {journal} {{Quantum}}\ }\textbf {\bibinfo {volume} {6}},\ \bibinfo {pages}
  {702} (\bibinfo {year} {2022})}\BibitemShut {NoStop}%
\bibitem [{\citenamefont {Talkner}\ and\ \citenamefont
  {H\"anggi}(2020)}]{RevModPhys.92.041002}%
  \BibitemOpen
  \bibfield  {author} {\bibinfo {author} {\bibfnamefont {P.}~\bibnamefont
  {Talkner}}\ and\ \bibinfo {author} {\bibfnamefont {P.}~\bibnamefont
  {H\"anggi}},\ }\bibfield  {title} {\bibinfo {title} {{\it Colloquium:
  Statistical mechanics and thermodynamics at strong coupling: Quantum and
  classical}},\ }\href {https://doi.org/10.1103/RevModPhys.92.041002}
  {\bibfield  {journal} {\bibinfo  {journal} {Rev. Mod. Phys.}\ }\textbf
  {\bibinfo {volume} {92}},\ \bibinfo {pages} {041002} (\bibinfo {year}
  {2020})}\BibitemShut {NoStop}%
\bibitem [{\citenamefont {Cresser}\ and\ \citenamefont
  {Anders}(2021)}]{PhysRevLett.127.250601}%
  \BibitemOpen
  \bibfield  {author} {\bibinfo {author} {\bibfnamefont {J.~D.}\ \bibnamefont
  {Cresser}}\ and\ \bibinfo {author} {\bibfnamefont {J.}~\bibnamefont
  {Anders}},\ }\bibfield  {title} {\bibinfo {title} {{\it Weak and Ultrastrong
  Coupling Limits of the Quantum Mean Force Gibbs State}},\ }\href
  {https://doi.org/10.1103/PhysRevLett.127.250601} {\bibfield  {journal}
  {\bibinfo  {journal} {Phys. Rev. Lett.}\ }\textbf {\bibinfo {volume} {127}},\
  \bibinfo {pages} {250601} (\bibinfo {year} {2021})}\BibitemShut {NoStop}%
\bibitem [{\citenamefont {Trushechkin}\ \emph {et~al.}(2022)\citenamefont
  {Trushechkin}, \citenamefont {Merkli}, \citenamefont {Cresser},\ and\
  \citenamefont {Anders}}]{10.1116/5.0073853}%
  \BibitemOpen
  \bibfield  {author} {\bibinfo {author} {\bibfnamefont {A.~S.}\ \bibnamefont
  {Trushechkin}}, \bibinfo {author} {\bibfnamefont {M.}~\bibnamefont {Merkli}},
  \bibinfo {author} {\bibfnamefont {J.~D.}\ \bibnamefont {Cresser}},\ and\
  \bibinfo {author} {\bibfnamefont {J.}~\bibnamefont {Anders}},\ }\bibfield
  {title} {\bibinfo {title} {{\it Open quantum system dynamics and the mean
  force Gibbs state}},\ }\href {https://doi.org/10.1116/5.0073853} {\bibfield
  {journal} {\bibinfo  {journal} {AVS Quantum Science}\ }\textbf {\bibinfo
  {volume} {4}},\ \bibinfo {pages} {012301} (\bibinfo {year}
  {2022})}\BibitemShut {NoStop}%
\bibitem [{\citenamefont {Burke}\ \emph {et~al.}(2023)\citenamefont {Burke},
  \citenamefont {Nakerst},\ and\ \citenamefont {Haque}}]{burke2023structure}%
  \BibitemOpen
  \bibfield  {author} {\bibinfo {author} {\bibfnamefont {P.~C.}\ \bibnamefont
  {Burke}}, \bibinfo {author} {\bibfnamefont {G.}~\bibnamefont {Nakerst}},\
  and\ \bibinfo {author} {\bibfnamefont {M.}~\bibnamefont {Haque}},\
  }\href@noop {} {\bibinfo {title} {{\it Structure of the Hamiltonian of mean
  force}}} (\bibinfo {year} {2023}),\ \Eprint
  {https://arxiv.org/abs/2311.10427} {arXiv:2311.10427 [quant-ph]} \BibitemShut
  {NoStop}%
\bibitem [{Foo()}]{Foot2}%
  \BibitemOpen
  \href@noop {} {}This can happen when the temperature is sufficiently low. For
  example, consider a Hamiltonian whose ground state is a product state (e.g.,
  the XX model with magnetic fields). At low temperatures, numerical
  simulations indicate that non-local interactions can be induced by tracing
  out the spin at the right end (or the left end). In this example, the CMI is
  small at low temperatures because the quantum Gibbs state is close to a
  product state.\BibitemShut {Stop}%
\bibitem [{\citenamefont {Hakoshima}\ \emph {et~al.}(2024)\citenamefont
  {Hakoshima}, \citenamefont {Endo}, \citenamefont {Yamamoto}, \citenamefont
  {Matsuzaki},\ and\ \citenamefont {Yoshioka}}]{hakoshima2024localized}%
  \BibitemOpen
  \bibfield  {author} {\bibinfo {author} {\bibfnamefont {H.}~\bibnamefont
  {Hakoshima}}, \bibinfo {author} {\bibfnamefont {S.}~\bibnamefont {Endo}},
  \bibinfo {author} {\bibfnamefont {K.}~\bibnamefont {Yamamoto}}, \bibinfo
  {author} {\bibfnamefont {Y.}~\bibnamefont {Matsuzaki}},\ and\ \bibinfo
  {author} {\bibfnamefont {N.}~\bibnamefont {Yoshioka}},\ }\href@noop {}
  {\bibinfo {title} {{\it Localized Virtual Purification}}} (\bibinfo {year}
  {2024}),\ \Eprint {https://arxiv.org/abs/2308.13500} {arXiv:2308.13500}
  \BibitemShut {NoStop}%
\bibitem [{\citenamefont {Hastings}(2007{\natexlab{b}})}]{PhysRevB.76.201102}%
  \BibitemOpen
  \bibfield  {author} {\bibinfo {author} {\bibfnamefont {M.~B.}\ \bibnamefont
  {Hastings}},\ }\bibfield  {title} {\bibinfo {title} {{\it Quantum belief
  propagation: An algorithm for thermal quantum systems}},\ }\href
  {https://doi.org/10.1103/PhysRevB.76.201102} {\bibfield  {journal} {\bibinfo
  {journal} {Phys. Rev. B}\ }\textbf {\bibinfo {volume} {76}},\ \bibinfo
  {pages} {201102} (\bibinfo {year} {2007}{\natexlab{b}})}\BibitemShut
  {NoStop}%
\bibitem [{\citenamefont {Kim}(2012)}]{PhysRevB.86.245116}%
  \BibitemOpen
  \bibfield  {author} {\bibinfo {author} {\bibfnamefont {I.~H.}\ \bibnamefont
  {Kim}},\ }\bibfield  {title} {\bibinfo {title} {{\it Perturbative analysis of
  topological entanglement entropy from conditional independence}},\ }\href
  {https://doi.org/10.1103/PhysRevB.86.245116} {\bibfield  {journal} {\bibinfo
  {journal} {Phys. Rev. B}\ }\textbf {\bibinfo {volume} {86}},\ \bibinfo
  {pages} {245116} (\bibinfo {year} {2012})}\BibitemShut {NoStop}%
\bibitem [{\citenamefont {P{\'e}rez-Garc{\'i}a}\ and\ \citenamefont
  {P{\'e}rez-Hern{\'a}ndez}(2023)}]{Perez-Garcia2023}%
  \BibitemOpen
  \bibfield  {author} {\bibinfo {author} {\bibfnamefont {D.}~\bibnamefont
  {P{\'e}rez-Garc{\'i}a}}\ and\ \bibinfo {author} {\bibfnamefont
  {A.}~\bibnamefont {P{\'e}rez-Hern{\'a}ndez}},\ }\bibfield  {title} {\bibinfo
  {title} {{\it Locality Estimates for Complex Time Evolution in 1D}},\ }\href
  {https://doi.org/10.1007/s00220-022-04573-w} {\bibfield  {journal} {\bibinfo
  {journal} {Communications in Mathematical Physics}\ }\textbf {\bibinfo
  {volume} {399}},\ \bibinfo {pages} {929} (\bibinfo {year}
  {2023})}\BibitemShut {NoStop}%
\bibitem [{\citenamefont {Bakshi}\ \emph
  {et~al.}(2024{\natexlab{a}})\citenamefont {Bakshi}, \citenamefont {Liu},
  \citenamefont {Moitra},\ and\ \citenamefont {Tang}}]{bak2023}%
  \BibitemOpen
  \bibfield  {author} {\bibinfo {author} {\bibfnamefont {A.}~\bibnamefont
  {Bakshi}}, \bibinfo {author} {\bibfnamefont {A.}~\bibnamefont {Liu}},
  \bibinfo {author} {\bibfnamefont {A.}~\bibnamefont {Moitra}},\ and\ \bibinfo
  {author} {\bibfnamefont {E.}~\bibnamefont {Tang}},\ }\bibfield  {title}
  {\bibinfo {title} {{\it Learning Quantum Hamiltonians at Any Temperature in
  Polynomial Time}},\ }in\ \href {https://doi.org/10.1145/3618260.3649619}
  {\emph {\bibinfo {booktitle} {Proceedings of the 56th Annual ACM Symposium on
  Theory of Computing}}},\ \bibinfo {series and number} {STOC 2024}\ (\bibinfo
  {publisher} {Association for Computing Machinery},\ \bibinfo {address} {New
  York, NY, USA},\ \bibinfo {year} {2024})\ p.\ \bibinfo {pages}
  {1470–1477}\BibitemShut {NoStop}%
\bibitem [{\citenamefont {Haah}\ \emph {et~al.}(2024)\citenamefont {Haah},
  \citenamefont {Kothari},\ and\ \citenamefont {Tang}}]{Haah2024}%
  \BibitemOpen
  \bibfield  {author} {\bibinfo {author} {\bibfnamefont {J.}~\bibnamefont
  {Haah}}, \bibinfo {author} {\bibfnamefont {R.}~\bibnamefont {Kothari}},\ and\
  \bibinfo {author} {\bibfnamefont {E.}~\bibnamefont {Tang}},\ }\bibfield
  {title} {\bibinfo {title} {{\it Learning quantum Hamiltonians from
  high-temperature Gibbs states and real-time evolutions}},\ }\href
  {https://doi.org/10.1038/s41567-023-02376-x} {\bibfield  {journal} {\bibinfo
  {journal} {Nature Physics}\ }\textbf {\bibinfo {volume} {20}},\ \bibinfo
  {pages} {1027} (\bibinfo {year} {2024})}\BibitemShut {NoStop}%
\bibitem [{\citenamefont {Narayanan}(2024)}]{narayanan2024}%
  \BibitemOpen
  \bibfield  {author} {\bibinfo {author} {\bibfnamefont {S.}~\bibnamefont
  {Narayanan}},\ }\href {https://arxiv.org/abs/2407.04540} {\bibinfo {title}
  {{\it Improved algorithms for learning quantum Hamiltonians, via flat
  polynomials}}} (\bibinfo {year} {2024}),\ \Eprint
  {https://arxiv.org/abs/2407.04540} {arXiv:2407.04540 [quant-ph]} \BibitemShut
  {NoStop}%
\bibitem [{\citenamefont {Lieb}\ and\ \citenamefont
  {Robinson}(1972)}]{ref:LR-bound72}%
  \BibitemOpen
  \bibfield  {author} {\bibinfo {author} {\bibfnamefont {E.}~\bibnamefont
  {Lieb}}\ and\ \bibinfo {author} {\bibfnamefont {D.}~\bibnamefont
  {Robinson}},\ }\bibfield  {title} {\bibinfo {title} {{\it The finite group
  velocity of quantum spin systems}},\ }\href
  {https://doi.org/10.1007/BF01645779} {\bibfield  {journal} {\bibinfo
  {journal} {Communications in Mathematical Physics}\ }\textbf {\bibinfo
  {volume} {28}},\ \bibinfo {pages} {251} (\bibinfo {year} {1972})}\BibitemShut
  {NoStop}%
\bibitem [{\citenamefont {Bravyi}\ \emph {et~al.}(2006)\citenamefont {Bravyi},
  \citenamefont {Hastings},\ and\ \citenamefont
  {Verstraete}}]{PhysRevLett.97.050401}%
  \BibitemOpen
  \bibfield  {author} {\bibinfo {author} {\bibfnamefont {S.}~\bibnamefont
  {Bravyi}}, \bibinfo {author} {\bibfnamefont {M.~B.}\ \bibnamefont
  {Hastings}},\ and\ \bibinfo {author} {\bibfnamefont {F.}~\bibnamefont
  {Verstraete}},\ }\bibfield  {title} {\bibinfo {title} {{\it Lieb-Robinson
  Bounds and the Generation of Correlations and Topological Quantum Order}},\
  }\href {https://doi.org/10.1103/PhysRevLett.97.050401} {\bibfield  {journal}
  {\bibinfo  {journal} {Phys. Rev. Lett.}\ }\textbf {\bibinfo {volume} {97}},\
  \bibinfo {pages} {050401} (\bibinfo {year} {2006})}\BibitemShut {NoStop}%
\bibitem [{\citenamefont {Blanes}\ \emph {et~al.}(2009)\citenamefont {Blanes},
  \citenamefont {Casas}, \citenamefont {Oteo},\ and\ \citenamefont
  {Ros}}]{BLANES2009151}%
  \BibitemOpen
  \bibfield  {author} {\bibinfo {author} {\bibfnamefont {S.}~\bibnamefont
  {Blanes}}, \bibinfo {author} {\bibfnamefont {F.}~\bibnamefont {Casas}},
  \bibinfo {author} {\bibfnamefont {J.}~\bibnamefont {Oteo}},\ and\ \bibinfo
  {author} {\bibfnamefont {J.}~\bibnamefont {Ros}},\ }\bibfield  {title}
  {\bibinfo {title} {{\it The Magnus expansion and some of its applications}},\
  }\href {https://doi.org/https://doi.org/10.1016/j.physrep.2008.11.001}
  {\bibfield  {journal} {\bibinfo  {journal} {Physics Reports}\ }\textbf
  {\bibinfo {volume} {470}},\ \bibinfo {pages} {151} (\bibinfo {year}
  {2009})}\BibitemShut {NoStop}%
\bibitem [{\citenamefont {Bukov}\ \emph {et~al.}(2015)\citenamefont {Bukov},
  \citenamefont {D'Alessio},\ and\ \citenamefont
  {Polkovnikov}}]{doi:10.1080/00018732.2015.1055918}%
  \BibitemOpen
  \bibfield  {author} {\bibinfo {author} {\bibfnamefont {M.}~\bibnamefont
  {Bukov}}, \bibinfo {author} {\bibfnamefont {L.}~\bibnamefont {D'Alessio}},\
  and\ \bibinfo {author} {\bibfnamefont {A.}~\bibnamefont {Polkovnikov}},\
  }\bibfield  {title} {\bibinfo {title} {{\it Universal high-frequency behavior
  of periodically driven systems: from dynamical stabilization to Floquet
  engineering}},\ }\href {https://doi.org/10.1080/00018732.2015.1055918}
  {\bibfield  {journal} {\bibinfo  {journal} {Advances in Physics}\ }\textbf
  {\bibinfo {volume} {64}},\ \bibinfo {pages} {139} (\bibinfo {year}
  {2015})}\BibitemShut {NoStop}%
\bibitem [{\citenamefont {Kuwahara}\ \emph {et~al.}(2016)\citenamefont
  {Kuwahara}, \citenamefont {Mori},\ and\ \citenamefont
  {Saito}}]{KUWAHARA201696}%
  \BibitemOpen
  \bibfield  {author} {\bibinfo {author} {\bibfnamefont {T.}~\bibnamefont
  {Kuwahara}}, \bibinfo {author} {\bibfnamefont {T.}~\bibnamefont {Mori}},\
  and\ \bibinfo {author} {\bibfnamefont {K.}~\bibnamefont {Saito}},\ }\bibfield
   {title} {\bibinfo {title} {{\it Floquet-Magnus theory and generic transient
  dynamics in periodically driven many-body quantum systems}},\ }\href
  {https://doi.org/10.1016/j.aop.2016.01.012} {\bibfield  {journal} {\bibinfo
  {journal} {Annals of Physics}\ }\textbf {\bibinfo {volume} {367}},\ \bibinfo
  {pages} {96 } (\bibinfo {year} {2016})}\BibitemShut {NoStop}%
\bibitem [{\citenamefont {Mori}\ \emph {et~al.}(2016)\citenamefont {Mori},
  \citenamefont {Kuwahara},\ and\ \citenamefont
  {Saito}}]{PhysRevLett.116.120401}%
  \BibitemOpen
  \bibfield  {author} {\bibinfo {author} {\bibfnamefont {T.}~\bibnamefont
  {Mori}}, \bibinfo {author} {\bibfnamefont {T.}~\bibnamefont {Kuwahara}},\
  and\ \bibinfo {author} {\bibfnamefont {K.}~\bibnamefont {Saito}},\ }\bibfield
   {title} {\bibinfo {title} {{\it Rigorous Bound on Energy Absorption and
  Generic Relaxation in Periodically Driven Quantum Systems}},\ }\href
  {https://doi.org/10.1103/PhysRevLett.116.120401} {\bibfield  {journal}
  {\bibinfo  {journal} {Phys. Rev. Lett.}\ }\textbf {\bibinfo {volume} {116}},\
  \bibinfo {pages} {120401} (\bibinfo {year} {2016})}\BibitemShut {NoStop}%
\bibitem [{\citenamefont {Hastings}(2022)}]{hastings2022doublebracket}%
  \BibitemOpen
  \bibfield  {author} {\bibinfo {author} {\bibfnamefont {M.~B.}\ \bibnamefont
  {Hastings}},\ }\href {https://arxiv.org/abs/2201.07141} {\bibinfo {title}
  {{\it On Lieb-Robinson Bounds for the Double Bracket Flow}}} (\bibinfo {year}
  {2022}),\ \Eprint {https://arxiv.org/abs/2201.07141} {arXiv:2201.07141
  [quant-ph]} \BibitemShut {NoStop}%
\bibitem [{\citenamefont {Alicki}\ and\ \citenamefont
  {Fannes}(2004)}]{Alicki_2004}%
  \BibitemOpen
  \bibfield  {author} {\bibinfo {author} {\bibfnamefont {R.}~\bibnamefont
  {Alicki}}\ and\ \bibinfo {author} {\bibfnamefont {M.}~\bibnamefont
  {Fannes}},\ }\bibfield  {title} {\bibinfo {title} {{\it Continuity of quantum
  conditional information}},\ }\href
  {https://doi.org/10.1088/0305-4470/37/5/l01} {\bibfield  {journal} {\bibinfo
  {journal} {Journal of Physics A: Mathematical and General}\ }\textbf
  {\bibinfo {volume} {37}},\ \bibinfo {pages} {L55} (\bibinfo {year}
  {2004})}\BibitemShut {NoStop}%
\bibitem [{\citenamefont {Arad}\ \emph {et~al.}(2016)\citenamefont {Arad},
  \citenamefont {Kuwahara},\ and\ \citenamefont {Landau}}]{Arad_2016}%
  \BibitemOpen
  \bibfield  {author} {\bibinfo {author} {\bibfnamefont {I.}~\bibnamefont
  {Arad}}, \bibinfo {author} {\bibfnamefont {T.}~\bibnamefont {Kuwahara}},\
  and\ \bibinfo {author} {\bibfnamefont {Z.}~\bibnamefont {Landau}},\
  }\bibfield  {title} {\bibinfo {title} {{\it Connecting global and local
  energy distributions in quantum spin models on a lattice}},\ }\href
  {https://doi.org/10.1088/1742-5468/2016/03/033301} {\bibfield  {journal}
  {\bibinfo  {journal} {Journal of Statistical Mechanics: Theory and
  Experiment}\ }\textbf {\bibinfo {volume} {2016}},\ \bibinfo {pages} {033301}
  (\bibinfo {year} {2016})}\BibitemShut {NoStop}%
\bibitem [{\citenamefont {Castelnovo}\ and\ \citenamefont
  {Chamon}(2007)}]{PhysRevB.76.184442}%
  \BibitemOpen
  \bibfield  {author} {\bibinfo {author} {\bibfnamefont {C.}~\bibnamefont
  {Castelnovo}}\ and\ \bibinfo {author} {\bibfnamefont {C.}~\bibnamefont
  {Chamon}},\ }\bibfield  {title} {\bibinfo {title} {{\it Entanglement and
  topological entropy of the toric code at finite temperature}},\ }\href
  {https://doi.org/10.1103/PhysRevB.76.184442} {\bibfield  {journal} {\bibinfo
  {journal} {Phys. Rev. B}\ }\textbf {\bibinfo {volume} {76}},\ \bibinfo
  {pages} {184442} (\bibinfo {year} {2007})}\BibitemShut {NoStop}%
\bibitem [{\citenamefont {Mazáč}\ and\ \citenamefont
  {Hamma}(2012)}]{MAZAC20122096}%
  \BibitemOpen
  \bibfield  {author} {\bibinfo {author} {\bibfnamefont {D.}~\bibnamefont
  {Mazáč}}\ and\ \bibinfo {author} {\bibfnamefont {A.}~\bibnamefont
  {Hamma}},\ }\bibfield  {title} {\bibinfo {title} {{\it Topological order,
  entanglement, and quantum memory at finite temperature}},\ }\href
  {https://doi.org/https://doi.org/10.1016/j.aop.2012.05.004} {\bibfield
  {journal} {\bibinfo  {journal} {Annals of Physics}\ }\textbf {\bibinfo
  {volume} {327}},\ \bibinfo {pages} {2096} (\bibinfo {year}
  {2012})}\BibitemShut {NoStop}%
\bibitem [{\citenamefont {Lee}\ and\ \citenamefont
  {Vidal}(2013)}]{PhysRevA.88.042318}%
  \BibitemOpen
  \bibfield  {author} {\bibinfo {author} {\bibfnamefont {Y.~A.}\ \bibnamefont
  {Lee}}\ and\ \bibinfo {author} {\bibfnamefont {G.}~\bibnamefont {Vidal}},\
  }\bibfield  {title} {\bibinfo {title} {{\it Entanglement negativity and
  topological order}},\ }\href {https://doi.org/10.1103/PhysRevA.88.042318}
  {\bibfield  {journal} {\bibinfo  {journal} {Phys. Rev. A}\ }\textbf {\bibinfo
  {volume} {88}},\ \bibinfo {pages} {042318} (\bibinfo {year}
  {2013})}\BibitemShut {NoStop}%
\bibitem [{\citenamefont {Gabbrielli}\ \emph {et~al.}(2018)\citenamefont
  {Gabbrielli}, \citenamefont {Smerzi},\ and\ \citenamefont
  {Pezz{\`e}}}]{Gabbrielli2018}%
  \BibitemOpen
  \bibfield  {author} {\bibinfo {author} {\bibfnamefont {M.}~\bibnamefont
  {Gabbrielli}}, \bibinfo {author} {\bibfnamefont {A.}~\bibnamefont {Smerzi}},\
  and\ \bibinfo {author} {\bibfnamefont {L.}~\bibnamefont {Pezz{\`e}}},\
  }\bibfield  {title} {\bibinfo {title} {{\it Multipartite Entanglement at
  Finite Temperature}},\ }\href {https://doi.org/10.1038/s41598-018-31761-3}
  {\bibfield  {journal} {\bibinfo  {journal} {Scientific Reports}\ }\textbf
  {\bibinfo {volume} {8}},\ \bibinfo {pages} {15663} (\bibinfo {year}
  {2018})}\BibitemShut {NoStop}%
\bibitem [{\citenamefont {Shapourian}\ and\ \citenamefont
  {Ryu}(2019)}]{Shapourian_2019}%
  \BibitemOpen
  \bibfield  {author} {\bibinfo {author} {\bibfnamefont {H.}~\bibnamefont
  {Shapourian}}\ and\ \bibinfo {author} {\bibfnamefont {S.}~\bibnamefont
  {Ryu}},\ }\bibfield  {title} {\bibinfo {title} {{\it Finite-temperature
  entanglement negativity of free fermions}},\ }\href
  {https://doi.org/10.1088/1742-5468/ab11e0} {\bibfield  {journal} {\bibinfo
  {journal} {Journal of Statistical Mechanics: Theory and Experiment}\ }\textbf
  {\bibinfo {volume} {2019}},\ \bibinfo {pages} {043106} (\bibinfo {year}
  {2019})}\BibitemShut {NoStop}%
\bibitem [{\citenamefont {Lu}\ \emph {et~al.}(2020)\citenamefont {Lu},
  \citenamefont {Hsieh},\ and\ \citenamefont
  {Grover}}]{PhysRevLett.125.116801}%
  \BibitemOpen
  \bibfield  {author} {\bibinfo {author} {\bibfnamefont {T.-C.}\ \bibnamefont
  {Lu}}, \bibinfo {author} {\bibfnamefont {T.~H.}\ \bibnamefont {Hsieh}},\ and\
  \bibinfo {author} {\bibfnamefont {T.}~\bibnamefont {Grover}},\ }\bibfield
  {title} {\bibinfo {title} {{\it Detecting Topological Order at Finite
  Temperature Using Entanglement Negativity}},\ }\href
  {https://doi.org/10.1103/PhysRevLett.125.116801} {\bibfield  {journal}
  {\bibinfo  {journal} {Phys. Rev. Lett.}\ }\textbf {\bibinfo {volume} {125}},\
  \bibinfo {pages} {116801} (\bibinfo {year} {2020})}\BibitemShut {NoStop}%
\bibitem [{\citenamefont {Wu}\ \emph {et~al.}(2020)\citenamefont {Wu},
  \citenamefont {Lu}, \citenamefont {Chung}, \citenamefont {Kao},\ and\
  \citenamefont {Grover}}]{PhysRevLett.125.140603}%
  \BibitemOpen
  \bibfield  {author} {\bibinfo {author} {\bibfnamefont {K.-H.}\ \bibnamefont
  {Wu}}, \bibinfo {author} {\bibfnamefont {T.-C.}\ \bibnamefont {Lu}}, \bibinfo
  {author} {\bibfnamefont {C.-M.}\ \bibnamefont {Chung}}, \bibinfo {author}
  {\bibfnamefont {Y.-J.}\ \bibnamefont {Kao}},\ and\ \bibinfo {author}
  {\bibfnamefont {T.}~\bibnamefont {Grover}},\ }\bibfield  {title} {\bibinfo
  {title} {{\it Entanglement Renyi Negativity across a Finite Temperature
  Transition: A Monte Carlo Study}},\ }\href
  {https://doi.org/10.1103/PhysRevLett.125.140603} {\bibfield  {journal}
  {\bibinfo  {journal} {Phys. Rev. Lett.}\ }\textbf {\bibinfo {volume} {125}},\
  \bibinfo {pages} {140603} (\bibinfo {year} {2020})}\BibitemShut {NoStop}%
\bibitem [{\citenamefont {Lu}\ and\ \citenamefont
  {Grover}(2020)}]{PhysRevResearch.2.043345}%
  \BibitemOpen
  \bibfield  {author} {\bibinfo {author} {\bibfnamefont {T.-C.}\ \bibnamefont
  {Lu}}\ and\ \bibinfo {author} {\bibfnamefont {T.}~\bibnamefont {Grover}},\
  }\bibfield  {title} {\bibinfo {title} {{\it Structure of quantum entanglement
  at a finite temperature critical point}},\ }\href
  {https://doi.org/10.1103/PhysRevResearch.2.043345} {\bibfield  {journal}
  {\bibinfo  {journal} {Phys. Rev. Research}\ }\textbf {\bibinfo {volume}
  {2}},\ \bibinfo {pages} {043345} (\bibinfo {year} {2020})}\BibitemShut
  {NoStop}%
\bibitem [{\citenamefont {Kim}\ \emph {et~al.}(2024)\citenamefont {Kim},
  \citenamefont {Kuwahara},\ and\ \citenamefont {Saito}}]{kim2024thermal}%
  \BibitemOpen
  \bibfield  {author} {\bibinfo {author} {\bibfnamefont {D.}~\bibnamefont
  {Kim}}, \bibinfo {author} {\bibfnamefont {T.}~\bibnamefont {Kuwahara}},\ and\
  \bibinfo {author} {\bibfnamefont {K.}~\bibnamefont {Saito}},\ }\href@noop {}
  {\bibinfo {title} {{\it Thermal Area Law in Long-Range Interacting Systems}}}
  (\bibinfo {year} {2024}),\ \Eprint {https://arxiv.org/abs/2404.04172}
  {arXiv:2404.04172} \BibitemShut {NoStop}%
\bibitem [{\citenamefont {Achutha}\ \emph {et~al.}(2025)\citenamefont
  {Achutha}, \citenamefont {Kim}, \citenamefont {Kimura},\ and\ \citenamefont
  {Kuwahara}}]{PhysRevLett.134.190404}%
  \BibitemOpen
  \bibfield  {author} {\bibinfo {author} {\bibfnamefont {R.}~\bibnamefont
  {Achutha}}, \bibinfo {author} {\bibfnamefont {D.}~\bibnamefont {Kim}},
  \bibinfo {author} {\bibfnamefont {Y.}~\bibnamefont {Kimura}},\ and\ \bibinfo
  {author} {\bibfnamefont {T.}~\bibnamefont {Kuwahara}},\ }\bibfield  {title}
  {\bibinfo {title} {{\it Provably Efficient Simulation of 1D Long-Range
  Interacting Systems at Any Temperature}},\ }\href
  {https://doi.org/10.1103/PhysRevLett.134.190404} {\bibfield  {journal}
  {\bibinfo  {journal} {Phys. Rev. Lett.}\ }\textbf {\bibinfo {volume} {134}},\
  \bibinfo {pages} {190404} (\bibinfo {year} {2025})}\BibitemShut {NoStop}%
\bibitem [{\citenamefont {Miyake}(2003)}]{PhysRevA.67.012108}%
  \BibitemOpen
  \bibfield  {author} {\bibinfo {author} {\bibfnamefont {A.}~\bibnamefont
  {Miyake}},\ }\bibfield  {title} {\bibinfo {title} {{\it Classification of
  multipartite entangled states by multidimensional determinants}},\ }\href
  {https://doi.org/10.1103/PhysRevA.67.012108} {\bibfield  {journal} {\bibinfo
  {journal} {Phys. Rev. A}\ }\textbf {\bibinfo {volume} {67}},\ \bibinfo
  {pages} {012108} (\bibinfo {year} {2003})}\BibitemShut {NoStop}%
\bibitem [{\citenamefont {Horodecki}\ \emph {et~al.}(2009)\citenamefont
  {Horodecki}, \citenamefont {Horodecki}, \citenamefont {Horodecki},\ and\
  \citenamefont {Horodecki}}]{RevModPhys.81.865}%
  \BibitemOpen
  \bibfield  {author} {\bibinfo {author} {\bibfnamefont {R.}~\bibnamefont
  {Horodecki}}, \bibinfo {author} {\bibfnamefont {P.}~\bibnamefont
  {Horodecki}}, \bibinfo {author} {\bibfnamefont {M.}~\bibnamefont
  {Horodecki}},\ and\ \bibinfo {author} {\bibfnamefont {K.}~\bibnamefont
  {Horodecki}},\ }\bibfield  {title} {\bibinfo {title} {{\it Quantum
  entanglement}},\ }\href {https://doi.org/10.1103/RevModPhys.81.865}
  {\bibfield  {journal} {\bibinfo  {journal} {Rev. Mod. Phys.}\ }\textbf
  {\bibinfo {volume} {81}},\ \bibinfo {pages} {865} (\bibinfo {year}
  {2009})}\BibitemShut {NoStop}%
\bibitem [{\citenamefont {Szalay}(2015)}]{PhysRevA.92.042329}%
  \BibitemOpen
  \bibfield  {author} {\bibinfo {author} {\bibfnamefont {S.}~\bibnamefont
  {Szalay}},\ }\bibfield  {title} {\bibinfo {title} {{\it Multipartite
  entanglement measures}},\ }\href {https://doi.org/10.1103/PhysRevA.92.042329}
  {\bibfield  {journal} {\bibinfo  {journal} {Phys. Rev. A}\ }\textbf {\bibinfo
  {volume} {92}},\ \bibinfo {pages} {042329} (\bibinfo {year}
  {2015})}\BibitemShut {NoStop}%
\bibitem [{\citenamefont {Xie}\ and\ \citenamefont
  {Eberly}(2021)}]{PhysRevLett.127.040403}%
  \BibitemOpen
  \bibfield  {author} {\bibinfo {author} {\bibfnamefont {S.}~\bibnamefont
  {Xie}}\ and\ \bibinfo {author} {\bibfnamefont {J.~H.}\ \bibnamefont
  {Eberly}},\ }\bibfield  {title} {\bibinfo {title} {{\it Triangle Measure of
  Tripartite Entanglement}},\ }\href
  {https://doi.org/10.1103/PhysRevLett.127.040403} {\bibfield  {journal}
  {\bibinfo  {journal} {Phys. Rev. Lett.}\ }\textbf {\bibinfo {volume} {127}},\
  \bibinfo {pages} {040403} (\bibinfo {year} {2021})}\BibitemShut {NoStop}%
\bibitem [{\citenamefont {Beckey}\ \emph {et~al.}(2021)\citenamefont {Beckey},
  \citenamefont {Gigena}, \citenamefont {Coles},\ and\ \citenamefont
  {Cerezo}}]{PhysRevLett.127.140501}%
  \BibitemOpen
  \bibfield  {author} {\bibinfo {author} {\bibfnamefont {J.~L.}\ \bibnamefont
  {Beckey}}, \bibinfo {author} {\bibfnamefont {N.}~\bibnamefont {Gigena}},
  \bibinfo {author} {\bibfnamefont {P.~J.}\ \bibnamefont {Coles}},\ and\
  \bibinfo {author} {\bibfnamefont {M.}~\bibnamefont {Cerezo}},\ }\bibfield
  {title} {\bibinfo {title} {{\it Computable and Operationally Meaningful
  Multipartite Entanglement Measures}},\ }\href
  {https://doi.org/10.1103/PhysRevLett.127.140501} {\bibfield  {journal}
  {\bibinfo  {journal} {Phys. Rev. Lett.}\ }\textbf {\bibinfo {volume} {127}},\
  \bibinfo {pages} {140501} (\bibinfo {year} {2021})}\BibitemShut {NoStop}%
\bibitem [{\citenamefont {Ma}\ \emph {et~al.}(2024)\citenamefont {Ma},
  \citenamefont {Li},\ and\ \citenamefont {Shang}}]{MA2024}%
  \BibitemOpen
  \bibfield  {author} {\bibinfo {author} {\bibfnamefont {M.}~\bibnamefont
  {Ma}}, \bibinfo {author} {\bibfnamefont {Y.}~\bibnamefont {Li}},\ and\
  \bibinfo {author} {\bibfnamefont {J.}~\bibnamefont {Shang}},\ }\bibfield
  {title} {\bibinfo {title} {{\it Multipartite entanglement measures: a
  review}},\ }\bibfield  {journal} {\bibinfo  {journal} {Fundamental Research}\
  }\href {https://doi.org/https://doi.org/10.1016/j.fmre.2024.03.031}
  {https://doi.org/10.1016/j.fmre.2024.03.031} (\bibinfo {year}
  {2024})\BibitemShut {NoStop}%
\bibitem [{\citenamefont {Poilblanc}(2010)}]{PhysRevLett.105.077202}%
  \BibitemOpen
  \bibfield  {author} {\bibinfo {author} {\bibfnamefont {D.}~\bibnamefont
  {Poilblanc}},\ }\bibfield  {title} {\bibinfo {title} {{\it Entanglement
  Spectra of Quantum Heisenberg Ladders}},\ }\href
  {https://doi.org/10.1103/PhysRevLett.105.077202} {\bibfield  {journal}
  {\bibinfo  {journal} {Phys. Rev. Lett.}\ }\textbf {\bibinfo {volume} {105}},\
  \bibinfo {pages} {077202} (\bibinfo {year} {2010})}\BibitemShut {NoStop}%
\bibitem [{\citenamefont {Li}\ \emph {et~al.}(2024)\citenamefont {Li},
  \citenamefont {Huang}, \citenamefont {Ding}, \citenamefont {Meng},
  \citenamefont {Wang},\ and\ \citenamefont {Yan}}]{PhysRevB.109.195169}%
  \BibitemOpen
  \bibfield  {author} {\bibinfo {author} {\bibfnamefont {C.}~\bibnamefont
  {Li}}, \bibinfo {author} {\bibfnamefont {R.-Z.}\ \bibnamefont {Huang}},
  \bibinfo {author} {\bibfnamefont {Y.-M.}\ \bibnamefont {Ding}}, \bibinfo
  {author} {\bibfnamefont {Z.~Y.}\ \bibnamefont {Meng}}, \bibinfo {author}
  {\bibfnamefont {Y.-C.}\ \bibnamefont {Wang}},\ and\ \bibinfo {author}
  {\bibfnamefont {Z.}~\bibnamefont {Yan}},\ }\bibfield  {title} {\bibinfo
  {title} {{\it Relevant long-range interaction of the entanglement Hamiltonian
  emerges from a short-range gapped system}},\ }\href
  {https://doi.org/10.1103/PhysRevB.109.195169} {\bibfield  {journal} {\bibinfo
   {journal} {Phys. Rev. B}\ }\textbf {\bibinfo {volume} {109}},\ \bibinfo
  {pages} {195169} (\bibinfo {year} {2024})}\BibitemShut {NoStop}%
\bibitem [{\citenamefont {Giudici}\ \emph {et~al.}(2018)\citenamefont
  {Giudici}, \citenamefont {Mendes-Santos}, \citenamefont {Calabrese},\ and\
  \citenamefont {Dalmonte}}]{PhysRevB.98.134403}%
  \BibitemOpen
  \bibfield  {author} {\bibinfo {author} {\bibfnamefont {G.}~\bibnamefont
  {Giudici}}, \bibinfo {author} {\bibfnamefont {T.}~\bibnamefont
  {Mendes-Santos}}, \bibinfo {author} {\bibfnamefont {P.}~\bibnamefont
  {Calabrese}},\ and\ \bibinfo {author} {\bibfnamefont {M.}~\bibnamefont
  {Dalmonte}},\ }\bibfield  {title} {\bibinfo {title} {{\it Entanglement
  Hamiltonians of lattice models via the Bisognano-Wichmann theorem}},\ }\href
  {https://doi.org/10.1103/PhysRevB.98.134403} {\bibfield  {journal} {\bibinfo
  {journal} {Phys. Rev. B}\ }\textbf {\bibinfo {volume} {98}},\ \bibinfo
  {pages} {134403} (\bibinfo {year} {2018})}\BibitemShut {NoStop}%
\bibitem [{\citenamefont {Chen}\ \emph {et~al.}(2025)\citenamefont {Chen},
  \citenamefont {Anshu},\ and\ \citenamefont {Nguyen}}]{chen2025_learn}%
  \BibitemOpen
  \bibfield  {author} {\bibinfo {author} {\bibfnamefont {C.-F.}\ \bibnamefont
  {Chen}}, \bibinfo {author} {\bibfnamefont {A.}~\bibnamefont {Anshu}},\ and\
  \bibinfo {author} {\bibfnamefont {Q.~T.}\ \bibnamefont {Nguyen}},\ }\href
  {https://arxiv.org/abs/2504.02706} {\bibinfo {title} {{\it Learning quantum
  Gibbs states locally and efficiently}}} (\bibinfo {year} {2025}),\ \Eprint
  {https://arxiv.org/abs/2504.02706} {arXiv:2504.02706 [quant-ph]} \BibitemShut
  {NoStop}%
\bibitem [{\citenamefont {Lee}\ \emph {et~al.}(2024)\citenamefont {Lee},
  \citenamefont {Oh}, \citenamefont {Wong}, \citenamefont {Chen},\ and\
  \citenamefont {Jiang}}]{lee2024universalspreadingconditionalmutual}%
  \BibitemOpen
  \bibfield  {author} {\bibinfo {author} {\bibfnamefont {S.-u.}\ \bibnamefont
  {Lee}}, \bibinfo {author} {\bibfnamefont {C.}~\bibnamefont {Oh}}, \bibinfo
  {author} {\bibfnamefont {Y.}~\bibnamefont {Wong}}, \bibinfo {author}
  {\bibfnamefont {S.}~\bibnamefont {Chen}},\ and\ \bibinfo {author}
  {\bibfnamefont {L.}~\bibnamefont {Jiang}},\ }\bibfield  {title} {\bibinfo
  {title} {{\it Universal Spreading of Conditional Mutual Information in Noisy
  Random Circuits}},\ }\href {https://doi.org/10.1103/PhysRevLett.133.200402}
  {\bibfield  {journal} {\bibinfo  {journal} {Phys. Rev. Lett.}\ }\textbf
  {\bibinfo {volume} {133}},\ \bibinfo {pages} {200402} (\bibinfo {year}
  {2024})}\BibitemShut {NoStop}%
\bibitem [{\citenamefont {Zhang}\ and\ \citenamefont
  {Gopalakrishnan}(2024)}]{zhang2024nonlocalgrowthquantumconditional}%
  \BibitemOpen
  \bibfield  {author} {\bibinfo {author} {\bibfnamefont {Y.}~\bibnamefont
  {Zhang}}\ and\ \bibinfo {author} {\bibfnamefont {S.}~\bibnamefont
  {Gopalakrishnan}},\ }\bibfield  {title} {\bibinfo {title} {{\it Nonlocal
  growth of quantum conditional mutual information under decoherence}},\ }\href
  {https://doi.org/10.1103/PhysRevA.110.032426} {\bibfield  {journal} {\bibinfo
   {journal} {Phys. Rev. A}\ }\textbf {\bibinfo {volume} {110}},\ \bibinfo
  {pages} {032426} (\bibinfo {year} {2024})}\BibitemShut {NoStop}%
\bibitem [{\citenamefont {Sang}\ and\ \citenamefont
  {Hsieh}(2025)}]{sang2024stabilitymixedstatequantumphases}%
  \BibitemOpen
  \bibfield  {author} {\bibinfo {author} {\bibfnamefont {S.}~\bibnamefont
  {Sang}}\ and\ \bibinfo {author} {\bibfnamefont {T.~H.}\ \bibnamefont
  {Hsieh}},\ }\bibfield  {title} {\bibinfo {title} {{\it Stability of
  Mixed-State Quantum Phases via Finite Markov Length}},\ }\href
  {https://doi.org/10.1103/PhysRevLett.134.070403} {\bibfield  {journal}
  {\bibinfo  {journal} {Phys. Rev. Lett.}\ }\textbf {\bibinfo {volume} {134}},\
  \bibinfo {pages} {070403} (\bibinfo {year} {2025})}\BibitemShut {NoStop}%
\bibitem [{\citenamefont {Kato}\ and\ \citenamefont
  {Kuwahara}(2025)}]{kato2025}%
  \BibitemOpen
  \bibfield  {author} {\bibinfo {author} {\bibfnamefont {K.}~\bibnamefont
  {Kato}}\ and\ \bibinfo {author} {\bibfnamefont {T.}~\bibnamefont
  {Kuwahara}},\ }\href {https://arxiv.org/abs/2504.02235} {\bibinfo {title}
  {{\it Clustering of Conditional Mutual Information via Quantum
  Belief-Propagation Channels}}} (\bibinfo {year} {2025}),\ \Eprint
  {https://arxiv.org/abs/2504.02235} {arXiv:2504.02235 [quant-ph]} \BibitemShut
  {NoStop}%
\bibitem [{\citenamefont {Chen}\ and\ \citenamefont {Rouzé}(2025)}]{chen2025}%
  \BibitemOpen
  \bibfield  {author} {\bibinfo {author} {\bibfnamefont {C.-F.}\ \bibnamefont
  {Chen}}\ and\ \bibinfo {author} {\bibfnamefont {C.}~\bibnamefont {Rouzé}},\
  }\href {https://arxiv.org/abs/2504.02208} {\bibinfo {title} {{\it Quantum
  Gibbs states are locally Markovian}}} (\bibinfo {year} {2025}),\ \Eprint
  {https://arxiv.org/abs/2504.02208} {arXiv:2504.02208 [quant-ph]} \BibitemShut
  {NoStop}%
\bibitem [{\citenamefont {Kastoryano}\ and\ \citenamefont
  {Brand{\~a}o}(2016)}]{Kastoryano2016}%
  \BibitemOpen
  \bibfield  {author} {\bibinfo {author} {\bibfnamefont {M.~J.}\ \bibnamefont
  {Kastoryano}}\ and\ \bibinfo {author} {\bibfnamefont {F.~G. S.~L.}\
  \bibnamefont {Brand{\~a}o}},\ }\bibfield  {title} {\bibinfo {title} {{\it
  Quantum Gibbs Samplers: The Commuting Case}},\ }\href
  {https://doi.org/10.1007/s00220-016-2641-8} {\bibfield  {journal} {\bibinfo
  {journal} {Communications in Mathematical Physics}\ }\textbf {\bibinfo
  {volume} {344}},\ \bibinfo {pages} {915} (\bibinfo {year}
  {2016})}\BibitemShut {NoStop}%
\bibitem [{\citenamefont {Kim}(2017)}]{kim2017markovian}%
  \BibitemOpen
  \bibfield  {author} {\bibinfo {author} {\bibfnamefont {I.~H.}\ \bibnamefont
  {Kim}},\ }\bibfield  {title} {\bibinfo {title} {{\it Markovian Matrix Product
  Density Operators: Efficient computation of global entropy}},\ }\href@noop {}
  {\bibfield  {journal} {\bibinfo  {journal} {arXiv preprint arXiv:1709.07828}\
  } (\bibinfo {year} {2017})},\ \Eprint
  {https://arxiv.org/abs/arXiv:1709.07828} {arXiv:1709.07828} \BibitemShut
  {NoStop}%
\bibitem [{\citenamefont {Lami}(2018)}]{lami2018nonclassical}%
  \BibitemOpen
  \bibfield  {author} {\bibinfo {author} {\bibfnamefont {L.}~\bibnamefont
  {Lami}},\ }\href@noop {} {\bibinfo {title} {{\it Non-classical correlations
  in quantum mechanics and beyond}}} (\bibinfo {year} {2018}),\ \Eprint
  {https://arxiv.org/abs/1803.02902} {arXiv:1803.02902 [quant-ph]} \BibitemShut
  {NoStop}%
\bibitem [{\citenamefont {Berta}\ and\ \citenamefont
  {Tomamichel}(2024)}]{berta2023entanglement}%
  \BibitemOpen
  \bibfield  {author} {\bibinfo {author} {\bibfnamefont {M.}~\bibnamefont
  {Berta}}\ and\ \bibinfo {author} {\bibfnamefont {M.}~\bibnamefont
  {Tomamichel}},\ }\bibfield  {title} {\bibinfo {title} {{\it Entanglement
  Monogamy via Multivariate Trace Inequalities}},\ }\href
  {https://doi.org/10.1007/s00220-023-04920-5} {\bibfield  {journal} {\bibinfo
  {journal} {Communications in Mathematical Physics}\ }\textbf {\bibinfo
  {volume} {405}},\ \bibinfo {pages} {29} (\bibinfo {year} {2024})}\BibitemShut
  {NoStop}%
\bibitem [{\citenamefont {Nachtergaele}\ \emph {et~al.}(2006)\citenamefont
  {Nachtergaele}, \citenamefont {Ogata},\ and\ \citenamefont
  {Sims}}]{Nachtergaele2006}%
  \BibitemOpen
  \bibfield  {author} {\bibinfo {author} {\bibfnamefont {B.}~\bibnamefont
  {Nachtergaele}}, \bibinfo {author} {\bibfnamefont {Y.}~\bibnamefont
  {Ogata}},\ and\ \bibinfo {author} {\bibfnamefont {R.}~\bibnamefont {Sims}},\
  }\bibfield  {title} {\bibinfo {title} {{\it Propagation of Correlations in
  Quantum Lattice Systems}},\ }\href
  {https://doi.org/10.1007/s10955-006-9143-6} {\bibfield  {journal} {\bibinfo
  {journal} {Journal of Statistical Physics}\ }\textbf {\bibinfo {volume}
  {124}},\ \bibinfo {pages} {1} (\bibinfo {year} {2006})}\BibitemShut {NoStop}%
\bibitem [{\citenamefont {Foss-Feig}\ \emph {et~al.}(2015)\citenamefont
  {Foss-Feig}, \citenamefont {Gong}, \citenamefont {Clark},\ and\ \citenamefont
  {Gorshkov}}]{PhysRevLett.114.157201}%
  \BibitemOpen
  \bibfield  {author} {\bibinfo {author} {\bibfnamefont {M.}~\bibnamefont
  {Foss-Feig}}, \bibinfo {author} {\bibfnamefont {Z.-X.}\ \bibnamefont {Gong}},
  \bibinfo {author} {\bibfnamefont {C.~W.}\ \bibnamefont {Clark}},\ and\
  \bibinfo {author} {\bibfnamefont {A.~V.}\ \bibnamefont {Gorshkov}},\
  }\bibfield  {title} {\bibinfo {title} {{\it Nearly Linear Light Cones in
  Long-Range Interacting Quantum Systems}},\ }\href
  {https://doi.org/10.1103/PhysRevLett.114.157201} {\bibfield  {journal}
  {\bibinfo  {journal} {Phys. Rev. Lett.}\ }\textbf {\bibinfo {volume} {114}},\
  \bibinfo {pages} {157201} (\bibinfo {year} {2015})}\BibitemShut {NoStop}%
\bibitem [{\citenamefont {Chen}\ and\ \citenamefont
  {Lucas}(2019)}]{chen2019finite}%
  \BibitemOpen
  \bibfield  {author} {\bibinfo {author} {\bibfnamefont {C.-F.}\ \bibnamefont
  {Chen}}\ and\ \bibinfo {author} {\bibfnamefont {A.}~\bibnamefont {Lucas}},\
  }\bibfield  {title} {\bibinfo {title} {{\it Finite Speed of Quantum
  Scrambling with Long Range Interactions}},\ }\href
  {https://doi.org/10.1103/PhysRevLett.123.250605} {\bibfield  {journal}
  {\bibinfo  {journal} {Phys. Rev. Lett.}\ }\textbf {\bibinfo {volume} {123}},\
  \bibinfo {pages} {250605} (\bibinfo {year} {2019})}\BibitemShut {NoStop}%
\bibitem [{\citenamefont {Kuwahara}\ and\ \citenamefont
  {Saito}(2020)}]{PhysRevX.10.031010}%
  \BibitemOpen
  \bibfield  {author} {\bibinfo {author} {\bibfnamefont {T.}~\bibnamefont
  {Kuwahara}}\ and\ \bibinfo {author} {\bibfnamefont {K.}~\bibnamefont
  {Saito}},\ }\bibfield  {title} {\bibinfo {title} {{\it Strictly Linear Light
  Cones in Long-Range Interacting Systems of Arbitrary Dimensions}},\ }\href
  {https://doi.org/10.1103/PhysRevX.10.031010} {\bibfield  {journal} {\bibinfo
  {journal} {Phys. Rev. X}\ }\textbf {\bibinfo {volume} {10}},\ \bibinfo
  {pages} {031010} (\bibinfo {year} {2020})}\BibitemShut {NoStop}%
\bibitem [{\citenamefont {Tran}\ \emph {et~al.}(2021)\citenamefont {Tran},
  \citenamefont {Guo}, \citenamefont {Baldwin}, \citenamefont {Ehrenberg},
  \citenamefont {Gorshkov},\ and\ \citenamefont
  {Lucas}}]{PhysRevLett.127.160401}%
  \BibitemOpen
  \bibfield  {author} {\bibinfo {author} {\bibfnamefont {M.~C.}\ \bibnamefont
  {Tran}}, \bibinfo {author} {\bibfnamefont {A.~Y.}\ \bibnamefont {Guo}},
  \bibinfo {author} {\bibfnamefont {C.~L.}\ \bibnamefont {Baldwin}}, \bibinfo
  {author} {\bibfnamefont {A.}~\bibnamefont {Ehrenberg}}, \bibinfo {author}
  {\bibfnamefont {A.~V.}\ \bibnamefont {Gorshkov}},\ and\ \bibinfo {author}
  {\bibfnamefont {A.}~\bibnamefont {Lucas}},\ }\bibfield  {title} {\bibinfo
  {title} {{\it Lieb-Robinson Light Cone for Power-Law Interactions}},\ }\href
  {https://doi.org/10.1103/PhysRevLett.127.160401} {\bibfield  {journal}
  {\bibinfo  {journal} {Phys. Rev. Lett.}\ }\textbf {\bibinfo {volume} {127}},\
  \bibinfo {pages} {160401} (\bibinfo {year} {2021})}\BibitemShut {NoStop}%
\bibitem [{\citenamefont {Scharf}(1988)}]{Scharf_1988}%
  \BibitemOpen
  \bibfield  {author} {\bibinfo {author} {\bibfnamefont {R.}~\bibnamefont
  {Scharf}},\ }\bibfield  {title} {\bibinfo {title} {{\it The
  Campbell-Baker-Hausdorff expansion for classical and quantum kicked
  dynamics}},\ }\href {https://doi.org/10.1088/0305-4470/21/9/017} {\bibfield
  {journal} {\bibinfo  {journal} {Journal of Physics A: Mathematical and
  General}\ }\textbf {\bibinfo {volume} {21}},\ \bibinfo {pages} {2007}
  (\bibinfo {year} {1988})}\BibitemShut {NoStop}%
\bibitem [{\citenamefont {Haah}\ \emph {et~al.}(2016)\citenamefont {Haah},
  \citenamefont {Harrow}, \citenamefont {Ji}, \citenamefont {Wu},\ and\
  \citenamefont {Yu}}]{10.1145/2897518.2897585}%
  \BibitemOpen
  \bibfield  {author} {\bibinfo {author} {\bibfnamefont {J.}~\bibnamefont
  {Haah}}, \bibinfo {author} {\bibfnamefont {A.~W.}\ \bibnamefont {Harrow}},
  \bibinfo {author} {\bibfnamefont {Z.}~\bibnamefont {Ji}}, \bibinfo {author}
  {\bibfnamefont {X.}~\bibnamefont {Wu}},\ and\ \bibinfo {author}
  {\bibfnamefont {N.}~\bibnamefont {Yu}},\ }\bibfield  {title} {\bibinfo
  {title} {{\it Sample-optimal tomography of quantum states}}\ }(\bibinfo
  {publisher} {Association for Computing Machinery},\ \bibinfo {address} {New
  York, NY, USA},\ \bibinfo {year} {2016})\ p.\ \bibinfo {pages}
  {913–925}\BibitemShut {NoStop}%
\bibitem [{\citenamefont {Kuwahara}(2015)}]{Phd_kuwahara}%
  \BibitemOpen
  \bibfield  {author} {\bibinfo {author} {\bibfnamefont {T.}~\bibnamefont
  {Kuwahara}},\ }\bibfield  {title} {\bibinfo {title} {{\it Fundamental
  inequalities in quantum many-body systems}},\ }\href
  {http://hatano-lab.iis.u-tokyo.ac.jp/thesis/dron2014/thesis_kuwahara.pdf}
  {\bibfield  {journal} {\bibinfo  {journal} {PhD thesis, University of Tokyo}\
  } (\bibinfo {year} {2015})}\BibitemShut {NoStop}%
\bibitem [{\citenamefont {Bluhm}\ \emph {et~al.}(2023)\citenamefont {Bluhm},
  \citenamefont {Capel}, \citenamefont {Gondolf},\ and\ \citenamefont
  {P\'erez-Hern\'andez}}]{10129917}%
  \BibitemOpen
  \bibfield  {author} {\bibinfo {author} {\bibfnamefont {A.}~\bibnamefont
  {Bluhm}}, \bibinfo {author} {\bibfnamefont {A.}~\bibnamefont {Capel}},
  \bibinfo {author} {\bibfnamefont {P.}~\bibnamefont {Gondolf}},\ and\ \bibinfo
  {author} {\bibfnamefont {A.}~\bibnamefont {P\'erez-Hern\'andez}},\ }\bibfield
   {title} {\bibinfo {title} {{\it Continuity of Quantum Entropic Quantities
  via Almost Convexity}},\ }\href {https://doi.org/10.1109/tit.2023.3277892}
  {\bibfield  {journal} {\bibinfo  {journal} {IEEE Transactions on Information
  Theory}\ }\textbf {\bibinfo {volume} {69}},\ \bibinfo {pages} {5869–5901}
  (\bibinfo {year} {2023})}\BibitemShut {NoStop}%
\bibitem [{\citenamefont {Lindblad}(1974)}]{Lindblad1974}%
  \BibitemOpen
  \bibfield  {author} {\bibinfo {author} {\bibfnamefont {G.}~\bibnamefont
  {Lindblad}},\ }\bibfield  {title} {\bibinfo {title} {{\it Expectations and
  entropy inequalities for finite quantum systems}},\ }\href
  {https://doi.org/10.1007/BF01608390} {\bibfield  {journal} {\bibinfo
  {journal} {Communications in Mathematical Physics}\ }\textbf {\bibinfo
  {volume} {39}},\ \bibinfo {pages} {111} (\bibinfo {year} {1974})}\BibitemShut
  {NoStop}%
\bibitem [{\citenamefont {Uhlmann}(1977)}]{Uhlmann1977}%
  \BibitemOpen
  \bibfield  {author} {\bibinfo {author} {\bibfnamefont {A.}~\bibnamefont
  {Uhlmann}},\ }\bibfield  {title} {\bibinfo {title} {{\it Relative entropy and
  the Wigner-Yanase-Dyson-Lieb concavity in an interpolation theory}},\ }\href
  {https://doi.org/10.1007/BF01609834} {\bibfield  {journal} {\bibinfo
  {journal} {Communications in Mathematical Physics}\ }\textbf {\bibinfo
  {volume} {54}},\ \bibinfo {pages} {21} (\bibinfo {year} {1977})}\BibitemShut
  {NoStop}%
\bibitem [{\citenamefont {Berta}\ \emph
  {et~al.}(2015{\natexlab{b}})\citenamefont {Berta}, \citenamefont {Lemm},\
  and\ \citenamefont {Wilde}}]{10.5555/2871378.2871383}%
  \BibitemOpen
  \bibfield  {author} {\bibinfo {author} {\bibfnamefont {M.}~\bibnamefont
  {Berta}}, \bibinfo {author} {\bibfnamefont {M.}~\bibnamefont {Lemm}},\ and\
  \bibinfo {author} {\bibfnamefont {M.~M.}\ \bibnamefont {Wilde}},\ }\bibfield
  {title} {\bibinfo {title} {{\it Monotonicity of Quantum Relative Entropy and
  Recoverability}},\ }\href@noop {} {\bibfield  {journal} {\bibinfo  {journal}
  {Quantum Info. Comput.}\ }\textbf {\bibinfo {volume} {15}},\ \bibinfo {pages}
  {1333–1354} (\bibinfo {year} {2015}{\natexlab{b}})}\BibitemShut {NoStop}%
\bibitem [{\citenamefont {Audenaert}\ and\ \citenamefont
  {Eisert}(2005)}]{doi:10.1063/1.2044667}%
  \BibitemOpen
  \bibfield  {author} {\bibinfo {author} {\bibfnamefont {K.~M.~R.}\
  \bibnamefont {Audenaert}}\ and\ \bibinfo {author} {\bibfnamefont
  {J.}~\bibnamefont {Eisert}},\ }\bibfield  {title} {\bibinfo {title} {{\it
  Continuity bounds on the quantum relative entropy}},\ }\href
  {https://doi.org/10.1063/1.2044667} {\bibfield  {journal} {\bibinfo
  {journal} {Journal of Mathematical Physics}\ }\textbf {\bibinfo {volume}
  {46}},\ \bibinfo {pages} {102104} (\bibinfo {year} {2005})},\ \Eprint
  {https://arxiv.org/abs/https://doi.org/10.1063/1.2044667}
  {https://doi.org/10.1063/1.2044667} \BibitemShut {NoStop}%
\bibitem [{\citenamefont {Schur}(1923)}]{schur1923uber}%
  \BibitemOpen
  \bibfield  {author} {\bibinfo {author} {\bibfnamefont {I.}~\bibnamefont
  {Schur}},\ }\bibfield  {title} {\bibinfo {title} {{\it Uber eine Klasse von
  Mittelbildungen mit Anwendungen auf die Determinantentheorie}},\ }\href@noop
  {} {\bibfield  {journal} {\bibinfo  {journal} {Sitzungsberichte der Berliner
  Mathematischen Gesellschaft}\ }\textbf {\bibinfo {volume} {22}},\ \bibinfo
  {pages} {51} (\bibinfo {year} {1923})}\BibitemShut {NoStop}%
\bibitem [{\citenamefont {Horn}(1954)}]{Horn1954}%
  \BibitemOpen
  \bibfield  {author} {\bibinfo {author} {\bibfnamefont {A.}~\bibnamefont
  {Horn}},\ }\bibfield  {title} {\bibinfo {title} {{\it Doubly Stochastic
  Matrices and the Diagonal of a Rotation Matrix}},\ }\href
  {https://doi.org/10.2307/2372705} {\bibfield  {journal} {\bibinfo  {journal}
  {American Journal of Mathematics}\ }\textbf {\bibinfo {volume} {76}},\
  \bibinfo {pages} {620} (\bibinfo {year} {1954})},\ \bibinfo {note} {full
  publication date: Jul., 1954}\BibitemShut {NoStop}%
\bibitem [{\citenamefont {Nielsen}\ and\ \citenamefont
  {Vidal}(2001)}]{10.5555/2011326.2011331}%
  \BibitemOpen
  \bibfield  {author} {\bibinfo {author} {\bibfnamefont {M.~A.}\ \bibnamefont
  {Nielsen}}\ and\ \bibinfo {author} {\bibfnamefont {G.}~\bibnamefont
  {Vidal}},\ }\bibfield  {title} {\bibinfo {title} {{\it Majorization and the
  Interconversion of Bipartite States}},\ }\href@noop {} {\bibfield  {journal}
  {\bibinfo  {journal} {Quantum Info. Comput.}\ }\textbf {\bibinfo {volume}
  {1}},\ \bibinfo {pages} {76–93} (\bibinfo {year} {2001})}\BibitemShut
  {NoStop}%
\bibitem [{\citenamefont {Kuwahara}\ \emph {et~al.}(2024)\citenamefont
  {Kuwahara}, \citenamefont {Vu},\ and\ \citenamefont
  {Saito}}]{kuwahara2022optimal}%
  \BibitemOpen
  \bibfield  {author} {\bibinfo {author} {\bibfnamefont {T.}~\bibnamefont
  {Kuwahara}}, \bibinfo {author} {\bibfnamefont {T.~V.}\ \bibnamefont {Vu}},\
  and\ \bibinfo {author} {\bibfnamefont {K.}~\bibnamefont {Saito}},\ }\bibfield
   {title} {\bibinfo {title} {{\it Effective light cone and digital quantum
  simulation of interacting bosons}},\ }\href
  {https://doi.org/10.1038/s41467-024-46501-7} {\bibfield  {journal} {\bibinfo
  {journal} {Nature Communications}\ }\textbf {\bibinfo {volume} {15}},\
  \bibinfo {pages} {2520} (\bibinfo {year} {2024})}\BibitemShut {NoStop}%
\bibitem [{\citenamefont {Pinelis}(2020)}]{pinelis2020exact}%
  \BibitemOpen
  \bibfield  {author} {\bibinfo {author} {\bibfnamefont {I.}~\bibnamefont
  {Pinelis}},\ }\bibfield  {title} {\bibinfo {title} {{\it Exact lower and
  upper bounds on the incomplete gamma function}},\ }\href
  {https://doi.org/dx.doi.org/10.7153/mia-2020-23-95} {\bibfield  {journal}
  {\bibinfo  {journal} {Mathematical Inequalities and Applications}\ }\textbf
  {\bibinfo {volume} {23}},\ \bibinfo {pages} {1261} (\bibinfo {year}
  {2020})}\BibitemShut {NoStop}%
\bibitem [{\citenamefont {Skr\o{}vseth}\ and\ \citenamefont
  {Bartlett}(2009)}]{PhysRevA.80.022316}%
  \BibitemOpen
  \bibfield  {author} {\bibinfo {author} {\bibfnamefont {S.~O.}\ \bibnamefont
  {Skr\o{}vseth}}\ and\ \bibinfo {author} {\bibfnamefont {S.~D.}\ \bibnamefont
  {Bartlett}},\ }\bibfield  {title} {\bibinfo {title} {{\it Phase transitions
  and localizable entanglement in cluster-state spin chains with Ising
  couplings and local fields}},\ }\href
  {https://doi.org/10.1103/PhysRevA.80.022316} {\bibfield  {journal} {\bibinfo
  {journal} {Phys. Rev. A}\ }\textbf {\bibinfo {volume} {80}},\ \bibinfo
  {pages} {022316} (\bibinfo {year} {2009})}\BibitemShut {NoStop}%
\bibitem [{\citenamefont {Doherty}\ and\ \citenamefont
  {Bartlett}(2009)}]{PhysRevLett.103.020506}%
  \BibitemOpen
  \bibfield  {author} {\bibinfo {author} {\bibfnamefont {A.~C.}\ \bibnamefont
  {Doherty}}\ and\ \bibinfo {author} {\bibfnamefont {S.~D.}\ \bibnamefont
  {Bartlett}},\ }\bibfield  {title} {\bibinfo {title} {{\it Identifying Phases
  of Quantum Many-Body Systems That Are Universal for Quantum Computation}},\
  }\href {https://doi.org/10.1103/PhysRevLett.103.020506} {\bibfield  {journal}
  {\bibinfo  {journal} {Phys. Rev. Lett.}\ }\textbf {\bibinfo {volume} {103}},\
  \bibinfo {pages} {020506} (\bibinfo {year} {2009})}\BibitemShut {NoStop}%
\bibitem [{\citenamefont {Hern{\'{a}}ndez-Santana}\ \emph
  {et~al.}(2015)\citenamefont {Hern{\'{a}}ndez-Santana}, \citenamefont {Riera},
  \citenamefont {Hovhannisyan}, \citenamefont {Perarnau-Llobet}, \citenamefont
  {Tagliacozzo},\ and\ \citenamefont {Ac{\'{\i}}n}}]{Hern_ndez_Santana_2015}%
  \BibitemOpen
  \bibfield  {author} {\bibinfo {author} {\bibfnamefont {S.}~\bibnamefont
  {Hern{\'{a}}ndez-Santana}}, \bibinfo {author} {\bibfnamefont
  {A.}~\bibnamefont {Riera}}, \bibinfo {author} {\bibfnamefont {K.~V.}\
  \bibnamefont {Hovhannisyan}}, \bibinfo {author} {\bibfnamefont
  {M.}~\bibnamefont {Perarnau-Llobet}}, \bibinfo {author} {\bibfnamefont
  {L.}~\bibnamefont {Tagliacozzo}},\ and\ \bibinfo {author} {\bibfnamefont
  {A.}~\bibnamefont {Ac{\'{\i}}n}},\ }\bibfield  {title} {\bibinfo {title}
  {{\it Locality of temperature in spin chains}},\ }\href
  {https://doi.org/10.1088/1367-2630/17/8/085007} {\bibfield  {journal}
  {\bibinfo  {journal} {New Journal of Physics}\ }\textbf {\bibinfo {volume}
  {17}},\ \bibinfo {pages} {085007} (\bibinfo {year} {2015})}\BibitemShut
  {NoStop}%
\bibitem [{\citenamefont {Winter}(2016)}]{Winter2016}%
  \BibitemOpen
  \bibfield  {author} {\bibinfo {author} {\bibfnamefont {A.}~\bibnamefont
  {Winter}},\ }\bibfield  {title} {\bibinfo {title} {{\it Tight Uniform
  Continuity Bounds for Quantum Entropies: Conditional Entropy, Relative
  Entropy Distance and Energy Constraints}},\ }\href
  {https://doi.org/10.1007/s00220-016-2609-8} {\bibfield  {journal} {\bibinfo
  {journal} {Communications in Mathematical Physics}\ }\textbf {\bibinfo
  {volume} {347}},\ \bibinfo {pages} {291} (\bibinfo {year}
  {2016})}\BibitemShut {NoStop}%
\bibitem [{\citenamefont {Sutter}\ \emph {et~al.}(2016)\citenamefont {Sutter},
  \citenamefont {Fawzi},\ and\ \citenamefont {Renner}}]{doi101098rspa20150623}%
  \BibitemOpen
  \bibfield  {author} {\bibinfo {author} {\bibfnamefont {D.}~\bibnamefont
  {Sutter}}, \bibinfo {author} {\bibfnamefont {O.}~\bibnamefont {Fawzi}},\ and\
  \bibinfo {author} {\bibfnamefont {R.}~\bibnamefont {Renner}},\ }\bibfield
  {title} {\bibinfo {title} {{\it Universal recovery map for approximate Markov
  chains}},\ }\href {https://doi.org/10.1098/rspa.2015.0623} {\bibfield
  {journal} {\bibinfo  {journal} {Proceedings of the Royal Society A:
  Mathematical, Physical and Engineering Sciences}\ }\textbf {\bibinfo {volume}
  {472}},\ \bibinfo {pages} {20150623} (\bibinfo {year} {2016})}\BibitemShut
  {NoStop}%
\bibitem [{\citenamefont {Bluhm}\ \emph {et~al.}(2022)\citenamefont {Bluhm},
  \citenamefont {Capel},\ and\ \citenamefont
  {P{\'{e}}rez-Hern{\'{a}}ndez}}]{bluhm2021exponential}%
  \BibitemOpen
  \bibfield  {author} {\bibinfo {author} {\bibfnamefont {A.}~\bibnamefont
  {Bluhm}}, \bibinfo {author} {\bibfnamefont {{\'{A}}.}~\bibnamefont {Capel}},\
  and\ \bibinfo {author} {\bibfnamefont {A.}~\bibnamefont
  {P{\'{e}}rez-Hern{\'{a}}ndez}},\ }\bibfield  {title} {\bibinfo {title} {{\it
  Exponential decay of mutual information for {G}ibbs states of local
  {H}amiltonians}},\ }\href {https://doi.org/10.22331/q-2022-02-10-650}
  {\bibfield  {journal} {\bibinfo  {journal} {{Quantum}}\ }\textbf {\bibinfo
  {volume} {6}},\ \bibinfo {pages} {650} (\bibinfo {year} {2022})}\BibitemShut
  {NoStop}%
\bibitem [{\citenamefont {Shirokov}(2017)}]{10.1063/1.4987135}%
  \BibitemOpen
  \bibfield  {author} {\bibinfo {author} {\bibfnamefont {M.~E.}\ \bibnamefont
  {Shirokov}},\ }\bibfield  {title} {\bibinfo {title} {{{\it Tight uniform
  continuity bounds for the quantum conditional mutual information, for the
  Holevo quantity, and for capacities of quantum channels}}},\ }\href
  {https://doi.org/10.1063/1.4987135} {\bibfield  {journal} {\bibinfo
  {journal} {Journal of Mathematical Physics}\ }\textbf {\bibinfo {volume}
  {58}},\ \bibinfo {pages} {102202} (\bibinfo {year} {2017})}\BibitemShut
  {NoStop}%
\bibitem [{\citenamefont {Bouch}(2015)}]{10.1063/1.4936209}%
  \BibitemOpen
  \bibfield  {author} {\bibinfo {author} {\bibfnamefont {G.}~\bibnamefont
  {Bouch}},\ }\bibfield  {title} {\bibinfo {title} {{\it Complex-time
  singularity and locality estimates for quantum lattice systems}},\ }\href
  {https://doi.org/10.1063/1.4936209} {\bibfield  {journal} {\bibinfo
  {journal} {Journal of Mathematical Physics}\ }\textbf {\bibinfo {volume}
  {56}},\ \bibinfo {pages} {123303} (\bibinfo {year} {2015})}\BibitemShut
  {NoStop}%
\bibitem [{\citenamefont {Bonet-Monroig}\ \emph {et~al.}(2020)\citenamefont
  {Bonet-Monroig}, \citenamefont {Babbush},\ and\ \citenamefont
  {O'Brien}}]{PhysRevX.10.031064}%
  \BibitemOpen
  \bibfield  {author} {\bibinfo {author} {\bibfnamefont {X.}~\bibnamefont
  {Bonet-Monroig}}, \bibinfo {author} {\bibfnamefont {R.}~\bibnamefont
  {Babbush}},\ and\ \bibinfo {author} {\bibfnamefont {T.~E.}\ \bibnamefont
  {O'Brien}},\ }\bibfield  {title} {\bibinfo {title} {{\it Nearly Optimal
  Measurement Scheduling for Partial Tomography of Quantum States}},\ }\href
  {https://doi.org/10.1103/PhysRevX.10.031064} {\bibfield  {journal} {\bibinfo
  {journal} {Phys. Rev. X}\ }\textbf {\bibinfo {volume} {10}},\ \bibinfo
  {pages} {031064} (\bibinfo {year} {2020})}\BibitemShut {NoStop}%
\bibitem [{\citenamefont {Cotler}\ and\ \citenamefont
  {Wilczek}(2020)}]{PhysRevLett.124.100401}%
  \BibitemOpen
  \bibfield  {author} {\bibinfo {author} {\bibfnamefont {J.}~\bibnamefont
  {Cotler}}\ and\ \bibinfo {author} {\bibfnamefont {F.}~\bibnamefont
  {Wilczek}},\ }\bibfield  {title} {\bibinfo {title} {{\it Quantum Overlapping
  Tomography}},\ }\href {https://doi.org/10.1103/PhysRevLett.124.100401}
  {\bibfield  {journal} {\bibinfo  {journal} {Phys. Rev. Lett.}\ }\textbf
  {\bibinfo {volume} {124}},\ \bibinfo {pages} {100401} (\bibinfo {year}
  {2020})}\BibitemShut {NoStop}%
\bibitem [{\citenamefont {Huang}\ \emph {et~al.}(2020)\citenamefont {Huang},
  \citenamefont {Kueng},\ and\ \citenamefont {Preskill}}]{Huang2020}%
  \BibitemOpen
  \bibfield  {author} {\bibinfo {author} {\bibfnamefont {H.-Y.}\ \bibnamefont
  {Huang}}, \bibinfo {author} {\bibfnamefont {R.}~\bibnamefont {Kueng}},\ and\
  \bibinfo {author} {\bibfnamefont {J.}~\bibnamefont {Preskill}},\ }\bibfield
  {title} {\bibinfo {title} {{\it Predicting many properties of a quantum
  system from very few measurements}},\ }\href
  {https://doi.org/10.1038/s41567-020-0932-7} {\bibfield  {journal} {\bibinfo
  {journal} {Nature Physics}\ }\textbf {\bibinfo {volume} {16}},\ \bibinfo
  {pages} {1050} (\bibinfo {year} {2020})}\BibitemShut {NoStop}%
\bibitem [{\citenamefont {Kuwahara}\ and\ \citenamefont
  {Saito}(2018)}]{kuwahara2018polynomialtime}%
  \BibitemOpen
  \bibfield  {author} {\bibinfo {author} {\bibfnamefont {T.}~\bibnamefont
  {Kuwahara}}\ and\ \bibinfo {author} {\bibfnamefont {K.}~\bibnamefont
  {Saito}},\ }\href@noop {} {\bibinfo {title} {{\it Polynomial-time Classical
  Simulation for One-dimensional Quantum Gibbs States}}} (\bibinfo {year}
  {2018}),\ \Eprint {https://arxiv.org/abs/1807.08424} {arXiv:1807.08424
  [quant-ph]} \BibitemShut {NoStop}%
\bibitem [{\citenamefont {Adler}()}]{adlertaylor}%
  \BibitemOpen
  \bibfield  {author} {\bibinfo {author} {\bibfnamefont {S.~L.}\ \bibnamefont
  {Adler}},\ }\href
  {https://www.ias.edu/sites/default/files/sns/files/1-matrixlog_tex(1).pdf}
  {\bibinfo {title} {{\it Taylor expansion and derivative formulas for matrix
  logarithms}}}\BibitemShut {NoStop}%
\bibitem [{\citenamefont {Haber}(2018)}]{haber2018notes}%
  \BibitemOpen
  \bibfield  {author} {\bibinfo {author} {\bibfnamefont {H.~E.}\ \bibnamefont
  {Haber}},\ }\bibfield  {title} {\bibinfo {title} {{\it Notes on the matrix
  exponential and logarithm}},\ }\href
  {http://scipp.ucsc.edu/~haber/webpage/MatrixExpLog.pdf} {\bibfield  {journal}
  {\bibinfo  {journal} {Santa Cruz Institute for Particle Physics, University
  of California: Santa Cruz, CA, USA}\ } (\bibinfo {year} {2018})}\BibitemShut
  {NoStop}%
\bibitem [{\citenamefont {Kuwahara}\ and\ \citenamefont
  {Saito}(2021)}]{kuwahara2020absence}%
  \BibitemOpen
  \bibfield  {author} {\bibinfo {author} {\bibfnamefont {T.}~\bibnamefont
  {Kuwahara}}\ and\ \bibinfo {author} {\bibfnamefont {K.}~\bibnamefont
  {Saito}},\ }\bibfield  {title} {\bibinfo {title} {{\it Absence of Fast
  Scrambling in Thermodynamically Stable Long-Range Interacting Systems}},\
  }\href {https://doi.org/10.1103/PhysRevLett.126.030604} {\bibfield  {journal}
  {\bibinfo  {journal} {Phys. Rev. Lett.}\ }\textbf {\bibinfo {volume} {126}},\
  \bibinfo {pages} {030604} (\bibinfo {year} {2021})}\BibitemShut {NoStop}%
\bibitem [{\citenamefont {Bakshi}\ \emph
  {et~al.}(2024{\natexlab{b}})\citenamefont {Bakshi}, \citenamefont {Liu},
  \citenamefont {Moitra},\ and\ \citenamefont {Tang}}]{bakshi2023learning}%
  \BibitemOpen
  \bibfield  {author} {\bibinfo {author} {\bibfnamefont {A.}~\bibnamefont
  {Bakshi}}, \bibinfo {author} {\bibfnamefont {A.}~\bibnamefont {Liu}},
  \bibinfo {author} {\bibfnamefont {A.}~\bibnamefont {Moitra}},\ and\ \bibinfo
  {author} {\bibfnamefont {E.}~\bibnamefont {Tang}},\ }\bibfield  {title}
  {\bibinfo {title} {{\it Learning Quantum Hamiltonians at Any Temperature in
  Polynomial Time}},\ }in\ \href {https://doi.org/10.1145/3618260.3649619}
  {\emph {\bibinfo {booktitle} {Proceedings of the 56th Annual ACM Symposium on
  Theory of Computing}}},\ \bibinfo {series and number} {STOC 2024}\ (\bibinfo
  {publisher} {Association for Computing Machinery},\ \bibinfo {address} {New
  York, NY, USA},\ \bibinfo {year} {2024})\ p.\ \bibinfo {pages}
  {1470–1477}\BibitemShut {NoStop}%
\end{thebibliography}%
